\documentclass[12pt,a4paper,oneside]{memoir}
\usepackage[left=2.8cm,right=2.8cm,top=2.8cm,bottom=2.8cm]{geometry}
\usepackage[OT2,T1]{fontenc}
\usepackage{amsmath}
\usepackage{amsthm}
\usepackage{stmaryrd}

\usepackage{amssymb}
\usepackage{mathrsfs}
\usepackage{mathtools}
\usepackage[usenames,dvipsnames]{color}
\usepackage{xcolor}
\usepackage{appendix}
\usepackage{bm}
\usepackage{graphicx}
\usepackage{here}
\usepackage{balance}
\DeclareMathAlphabet{\mathscrbf}{OMS}{mdugm}{b}{n}

\setsecnumdepth{subsection}
\chapterstyle{ger}

\makepagestyle{Contents}
\makeevenhead{Contents}{\small{\it{Contents}}}{}{\thepage}
\makeoddhead{Contents}{\small{\it{Contents}}}{}{\thepage}

\makepagestyle{Foreword}
\makeevenhead{Foreword}{\small{\it{Foreword}}}{}{\thepage}
\makeoddhead{Foreword}{\small{\it{Foreword}}}{}{\thepage}

\makepagestyle{Regular}
\makeevenhead{Regular}{{\it{\leftmark}}}{}{\thepage}
\makeoddhead{Regular}{{\it{\rightmark}}}{}{\thepage}

\makepagestyle{Bibliography}
\makeevenhead{Bibliography}{\small{\it{Bibliography}}}{}{\thepage}
\makeoddhead{Bibliography}{\small{\it{Bibliography}}}{}{\thepage}

\makepagestyle{Index}
\makeevenhead{Index}{\small{\it{Index}}}{}{\thepage}
\makeoddhead{Index}{\small{\it{Index}}}{}{\thepage}

\nouppercaseheads

\usepackage[pagestyles,raggedright]{titlesec}
\usepackage{titletoc}

\newcommand{\secmark}{}
\newcommand{\marktotoc}[1]{\renewcommand{\secmark}{#1}}
\newenvironment{advanced}{
  \renewcommand{\secmark}{*}%
  \addtocontents{toc}{\protect\marktotoc{*}}
}
{\addtocontents{toc}{\protect\marktotoc{}}}

\titleformat{\section}{\normalfont\Large\bfseries}
{\makebox[1.5em][l]{\llap{\secmark}\thesection}}{0.4em}{}
\titlecontents{section}[3.7em]{}
{\contentslabel[\llap{\secmark}\thecontentslabel]{2.3em}}
{\hspace*{-2.3em}}{\titlerule*{.}\contentspage}

\titleformat{\subsection}{\normalfont\large\bfseries}
{\makebox[2.3em][l]{\llap{\secmark}\thesubsection}}{0.4em}{}
\titlecontents{subsection}[6.7em]{}
{\small\contentslabel[\llap{\secmark}\thecontentslabel]{3.1em}}
{\hspace*{-2.3em } }{\titlerule*{.}\contentspage}

\usepackage{makeidx}
\makeindex

\usepackage{cite}

\usepackage[backref=page,colorlinks,allcolors=blue]{hyperref}
\renewcommand*{\backref}[1]{}
\renewcommand*{\backrefalt}[4]{
    \ifcase #1 (Not cited in the text.)
    \or        (Cited on page~#2.)
    \else      (Cited on pages~#2.)
    \fi}

\let\OLDthebibliography\thebibliography
\renewcommand\thebibliography[1]{
  \OLDthebibliography{#1}
  \setlength{\parskip}{0pt}
  \setlength{\itemsep}{3.6pt plus 0.3ex}
}

\makeatletter
\renewcommand\part{%
  \if@openright
    \cleardoublepage
  \else
    \clearpage
  \fi
  \thispagestyle{empty}%
  \if@twocolumn
    \onecolumn
    \@tempswatrue
  \else
    \@tempswafalse
  \fi
  \null\vfil
  \secdef\@part\@spart}
\makeatother

\newcommand\cyr{%
  \renewcommand\rmdefault{wncyr}%
  \renewcommand\sfdefault{wncyss}%
  \renewcommand\encodingdefault{OT2}%
  \normalfont
  \selectfont}
\DeclareTextFontCommand{\textcyr}{\cyr}

\def\i{\index}

\renewcommand\it[1]{\textit{#1}}
\renewcommand\refeq[1]{\stackrel{\text{(\ref{#1})}}{=}}

\newcommand{\be}{\begin{equation}}
\newcommand{\ee}{\end{equation}}
\newcommand{\bea}{\begin{align}}
\newcommand{\eea}{\end{align}}
\newcommand{\nn}{\nonumber}
\newcommand{\bmm}{\begin{pmatrix}}
\newcommand{\emm}{\end{pmatrix}}
\newcommand{\bra}{\left<}
\newcommand{\ket}{\right>}
\newcommand{\Ad}{\text{Ad}}
\newcommand{\ad}{\text{ad}}
\newcommand{\BMS}{\text{BMS}_3}
\newcommand{\bms}{\mathfrak{bms}_3}
\newcommand{\CM}{C^{\infty}({\mathcal M})}
\newcommand{\demi}{\frac{1}{2}}
\newcommand{\der}{\partial}
\newcommand{\Diff}{\text{Diff}(S^1)}
\newcommand{\Diffc}{\widetilde{\text{Diff}}{}^+(S^1)}
\newcommand{\Diffp}{\text{Diff}^+(S^1)}

\newcommand{\gca}{\mathfrak{gca}_2}
\newcommand{\hAd}{\widehat{\text{Ad}}}
\newcommand{\had}{\widehat{\text{ad}}}
\newcommand{\hBMS}{\widehat{\text{BMS}}{}_3}
\newcommand{\hbms}{\widehat{\mathfrak{bms}}{}_3}
\newcommand{\hDiff}{\widehat{\text{Diff}}(S^1)}
\newcommand{\hVect}{\widehat{\text{Vect}}(S^1)}
\newcommand{\phii}{\varphi}
\newcommand{\PSL}{\text{PSL}(2,\mathbb{R})}
\newcommand{\SL}{\text{SL}(2,\mathbb{R})}
\renewcommand{\sl}{\mathfrak{sl}(2,\mathbb{R})}
\newcommand{\SO}{\text{SO}}

\newcommand{\sooh}{\mathfrak{so}(2,2)}
\newcommand{\SU}{\text{SU}}
\newcommand{\su}{\mathfrak{su}}
\newcommand{\Tr}{\text{Tr}}
\newcommand{\un}{\text{U}(1)}
\newcommand{\Vect}{\text{Vect}(S^1)}

\newcommand{\vir}{\mathfrak{vir}}

\newcommand{\bbk}{\textbf{k}}

\newcommand{\bbp}{\textbf{p}}
\newcommand{\bbq}{\textbf{q}}

\newcommand{\bbv}{\textbf{v}}
\newcommand{\bbw}{\textbf{w}}
\newcommand{\bbx}{\textbf{x}}

\newcommand{\cA}{{\mathcal A}}
\newcommand{\cB}{{\mathcal B}}
\newcommand{\cC}{{\mathcal C}}
\newcommand{\cD}{{\mathcal D}}
\newcommand{\cE}{{\mathcal E}}
\newcommand{\cF}{{\mathcal F}}
\newcommand{\cG}{{\mathcal G}}
\newcommand{\cH}{{\mathcal H}}
\newcommand{\cI}{{\mathcal I}}
\newcommand{\cJ}{{\mathcal J}}
\newcommand{\cK}{{\mathcal K}}
\newcommand{\cL}{{\mathcal L}}
\newcommand{\cM}{{\mathcal M}}
\newcommand{\cN}{{\mathcal N}}
\newcommand{\cO}{{\mathcal O}}
\newcommand{\cP}{{\mathcal P}}
\newcommand{\cQ}{{\mathcal Q}}
\newcommand{\cR}{{\mathcal R}}
\newcommand{\cS}{{\mathcal S}}
\newcommand{\cT}{{\mathcal T}}
\newcommand{\cU}{{\mathcal U}}
\newcommand{\cV}{{\mathcal V}}
\newcommand{\cW}{{\mathcal W}}
\newcommand{\cX}{{\mathcal X}}

\newcommand{\cZ}{{\mathcal Z}}

\newcommand{\hp}{{\hat p}}
\newcommand{\hq}{{\hat q}}

\newcommand{\hA}{{\hat A}}

\newcommand{\hG}{{\widehat{G}}{}}

\newcommand{\hU}{{\hat U}}

\newcommand{\mg}{\mathfrak{g}}
\newcommand{\mh}{\mathfrak{h}}

\newcommand{\hmg}{{\widehat{\mathfrak{g}}}{}}

\newcommand{\sC}{\mathscr{C}}

\newcommand{\sH}{\mathscr{H}}
\newcommand{\sI}{\mathscr{I}}

\newcommand{\sR}{\mathscr{R}}
\newcommand{\sS}{\mathscr{S}}
\newcommand{\sT}{\mathscr{T}}

\newcommand{\CC}{\mathbb{C}}

\newcommand{\HH}{\mathbb{H}}
\newcommand{\II}{\mathbb{I}}

\newcommand{\NN}{\mathbb{N}}

\newcommand{\PP}{\mathbb{P}}

\newcommand{\RR}{\mathbb{R}}

\newcommand{\VV}{\mathbb{V}}
\newcommand{\WW}{\mathbb{W}}

\newcommand{\ZZ}{\mathbb{Z}}

\newcommand{\sfB}{\mathsf{B}}
\newcommand{\sfC}{\mathsf{C}}

\newcommand{\sfK}{\mathsf{K}}

\newcommand{\sfM}{\mathsf{M}}

\newcommand{\sfS}{\mathsf{S}}
\newcommand{\sfT}{\mathsf{T}}

\newcommand{\sfW}{\mathsf{W}}

\newcommand{\sfb}{\mathsf{b}}
\newcommand{\sfc}{\mathsf{c}}
\newcommand{\sfd}{\mathsf{d}}

\newcommand{\sfff}{\mathsf{f}}
\newcommand{\sfg}{\mathsf{g}}
\newcommand{\sfh}{\mathsf{h}}
\newcommand{\sfi}{\mathsf{i}}

\newcommand{\sfk}{\mathsf{k}}

\newcommand{\sfp}{\mathsf{p}}

\newcommand{\sfs}{\mathsf{s}}
\newcommand{\sft}{\mathsf{t}}

\newcommand{\sfw}{\mathsf{w}}

\newcommand{\bbalpha}{{\boldsymbol\alpha}}
\newcommand{\bbbeta}{{\boldsymbol\beta}}

\newcommand{\bbomega}{{\boldsymbol\omega}}

\newcommand{\bbOmega}{{\boldsymbol\Omega}}

\begin{document}

\frontmatter

~\\
~\\
\hrule
\begin{center}
{\fontsize{50}{60}\selectfont{\bfseries{\scshape{BMS Particles}}}}\\[.3cm]
{\fontsize{25}{30}\selectfont{\bfseries{\scshape{in Three Dimensions}}}}
\end{center}
\hrule
~\\

\begin{center}
\large{\bfseries{\scshape{Blagoje Oblak$^*$}}}
\end{center}
~\\
~\\

\begin{figure}[H]
\begin{center}
\includegraphics[width=0.50\textwidth]{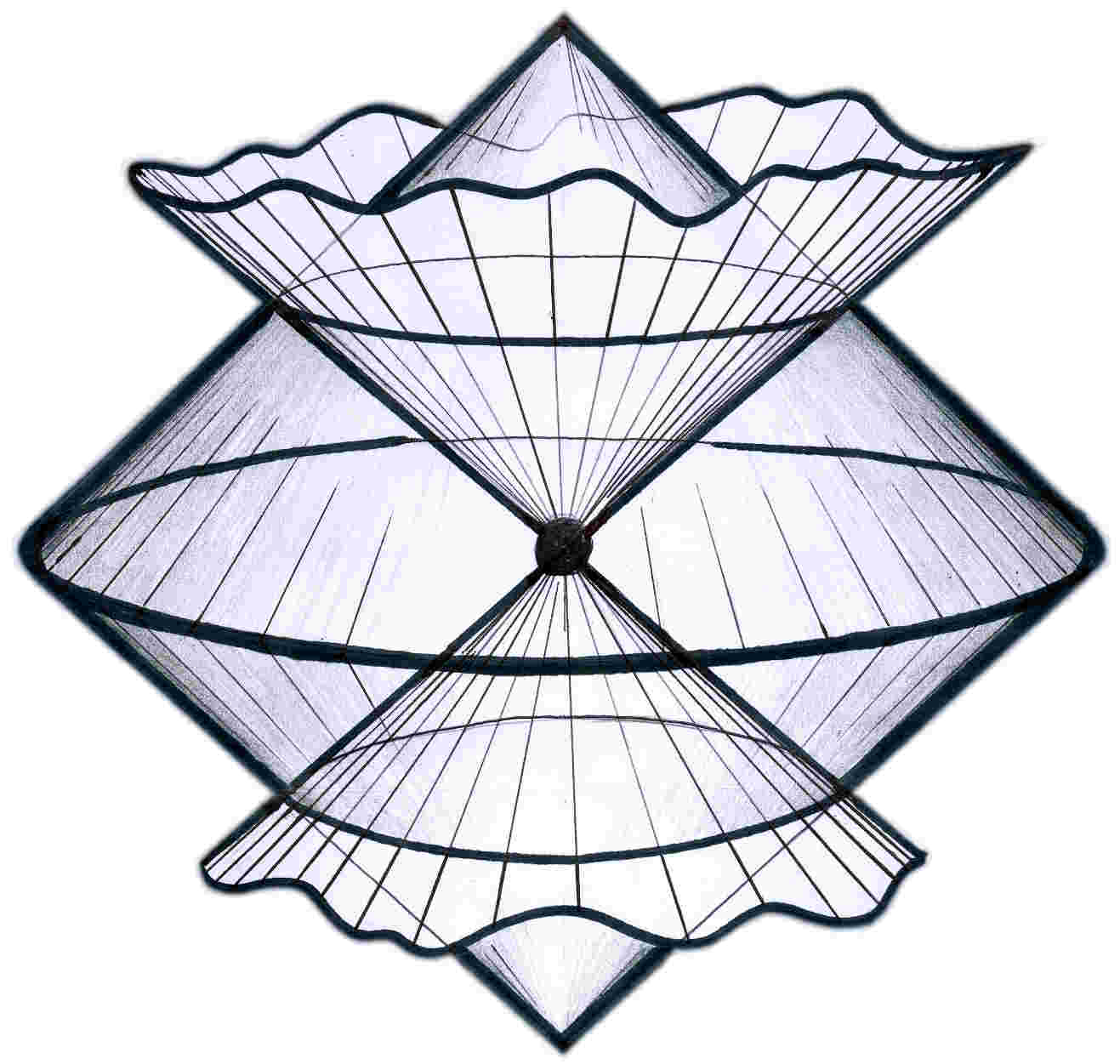}
\end{center}
\end{figure}

\begin{center}
\begin{minipage}{.9\textwidth}
\begin{center}{\bfseries{Abstract}}\end{center}
This thesis is devoted to the group-theoretic aspects of three-dimensional quantum gravity on 
Anti-de Sitter and Minkowskian backgrounds. In particular we describe the relation between unitary 
representations of asymptotic symmetry groups and gravitational perturbations around a space-time metric. In 
the asymptotically flat case this leads to \it{BMS particles}, representing standard relativistic
particles dressed with gravitational degrees of freedom accounted for by coadjoint orbits of the 
Virasoro group. Their thermodynamics are described by BMS characters, which coincide with gravitational 
one-loop partition functions. We also extend these considerations to higher-spin theories and supergravity.
\end{minipage}
\end{center}

\vfill
\noindent
\mbox{}
\raisebox{-3\baselineskip}{%
  \parbox{\textwidth}{\mbox{}\hrulefill\\[-4pt]}}
{\scriptsize$^*$ Current e-mail: boblak@phys.ethz.ch}

\thispagestyle{empty}


\newpage
~
\thispagestyle{empty}


\newpage

\begin{center}
~\\
~\\
~\\
{\scshape{Universit\'e Libre de Bruxelles}}\\
~\\
{\scshape{facult\'e des sciences}}\\
~\\
{\scshape{service de physique th\'eorique et math\'ematique}}\\
~\\
~\\
~\\
~\\
~\\
~\\
~\\
{\Huge{\bfseries{\scshape{BMS Particles}}}}\\[.3cm]
{\Huge{\bfseries{\scshape{in Three Dimensions}}}}
~\\
~\\
~\\
{\large{\bfseries{\scshape{Blagoje Oblak}}}}\\
~\\
~\\
~\\
~\\
~\\
~\\
~\\
~\\
\it{Th\`ese pr\'esent\'ee en vue de l'obtention du titre de Docteur en Sciences}\\
~\\
\it{R\'ealis\'ee sous la direction du} Prof.\ Glenn {\scshape{Barnich}}\\
~\\
\it{Ann\'ee acad\'emique 2015-2016}\\
~\\
~\\
~\\
\scshape{--- electronic version ---}
\end{center}

\thispagestyle{empty}


\newpage
~\\
~\\
~\\
\noindent This thesis was defended behind closed doors on June 22nd, and publicly on June 24th, in front of 
the following jury at \it{Universit\'e Libre de Bruxelles}:
\begin{center}
\begin{tabular}{rl}
President: & $\!\!\!$Prof.~Marc \textsc{Henneaux}\\[.2cm]
Secretary: & $\!\!\!$Prof.~Petr \textsc{Tinyakov}\\[.2cm]
Advisor: & $\!\!\!$Prof.~Glenn \textsc{Barnich}
\end{tabular}
\end{center}
~\\
Prof.~Guillaume \textsc{Bossard},\\
\indent\indent\indent\indent\indent\it{Ecole Polytechnique, Paris} (France)\\[.2cm]
Prof.~St\'ephane \textsc{Detournay},\\
\indent\indent\indent\indent\indent\it{Universit\'e Libre de Bruxelles} (Belgium)\\[.2cm]
Prof.~Axel \textsc{Kleinschmidt},\\
\indent\indent\indent\indent\indent\it{Albert Einstein Institut, Potsdam} (Germany)

\vfill

{\small{

\noindent The work presented in this thesis was carried out in the group of Mathematical Physics of 
Fundamental Interactions at \it{Universit\'e 
Libre de Bruxelles} (Belgium), and at the Department of Applied Mathematics and Theoretical Physics of 
the University of Cambridge (United Kingdom).
It was supported by the \it{Fonds de la Recherche Scientifique -- F.N.R.S.}, of which the 
author is a 
Research Fellow (\it{aspirant}), by the Wiener-Anspach Foundation, and by the International 
Solvay Institutes.\\

\noindent Printed in June 2016 by \it{Presses Universitaires de Bruxelles ASBL} --- Avenue Franklin
Roosevelt 50, C.P.\ 149, B-1050, Belgium.\\

\noindent Typeset in \LaTeX~with the {\textsf{memoir}} class.\\

~\\
~\\
~\\
~

}}

\thispagestyle{empty}


\newpage
\begin{flushright}
~\\
~\\
~\\
~\\
~\\
~\\
\large{{\cyr Mojoj najmilijoj porodici,\\
s mnogo ljubavi.}}
\end{flushright}
\thispagestyle{empty}


\newpage
~
\thispagestyle{empty}


\pagestyle{Contents}
\renewcommand\cftsubsectionfont{\small}
\newpage

\addtocontents{toc}{\unexpanded{\unexpanded{{%
  \noindent\itshape
  Note: sections marked with an asterisk may be skipped on a first reading.
  \par\bigskip
}}}}

\tableofcontents*

\chapter*{Foreword}
\addcontentsline{toc}{chapter}{Foreword}
\markboth{}{\small{\chaptername~\thechapter. Foreword}}
\pagestyle{Foreword}

This thesis collects thoughts and results that originate from a four-year long research 
project in theoretical physics. The main topic is representation theory and its application to quantum 
gravity, in particular in the context of BMS symmetry.

\section*{How to read this thesis}
\addcontentsline{toc}{section}{How to read this thesis}

The text consists of three parts:
\begin{center}
\begin{tabular}{ll}
Part I: & Group theory;\\
Part II: & Virasoro symmetry and AdS$_3$/CFT$_2$;\\
Part III: & BMS symmetry in three dimensions.
\end{tabular}
\end{center}
It is written in such a way that each part can be read more or less independently of the others, although
the later parts do depend on background material presented in the earlier ones; the logical flow of chapters 
is explained in section \ref{i5} below. A few 
sections are marked with an asterisk; they contain somewhat more advanced material that may be skipped 
without affecting the reading of the main track.\\

The original contributions of this thesis are based on the following 
publications:
\begin{itemize}
\item
G.~Barnich and B.~Oblak, ``{Holographic positive energy theorems in
  three\--di\-men\-sio\-nal gravity},'' {\em Class. Quant. Grav.} \textbf{31} (2014)
  152001,
\href{http://www.arXiv.org/abs/1403.3835}{\texttt{1403.3835}}.
\item
G.~Barnich and B.~Oblak, ``{Notes on the BMS group in three dimensions: I.
  Induced representations},'' {\em JHEP} \textbf{06} (2014) 129,
\href{http://www.arXiv.org/abs/1403.5803}{\texttt{1403.5803}}.
\item
G.~Barnich and B.~Oblak, ``{Notes on the BMS group in three dimensions: II.
  Coadjoint representation},'' {\em JHEP} \textbf{03} (2015) 033,
\href{http://www.arXiv.org/abs/1502.00010}{\texttt{1502.00010}}.
\item
B.~Oblak, ``{Characters of the BMS Group in Three Dimensions},'' {\em Commun.
  Math. Phys.} \textbf{340} (2015), no.~1, 413--432,
\href{http://www.arXiv.org/abs/1502.03108}{\texttt{1502.03108}}.
\item
G.~Barnich, H.~A. Gonz\'alez, A.~Maloney, and B.~Oblak, ``{One-loop partition
  function of three-dimensional flat gravity},'' {\em JHEP} \textbf{04} (2015)
  178,
\href{http://www.arXiv.org/abs/1502.06185}{\texttt{1502.06185}}.
\item
B.~Oblak, ``{From the Lorentz Group to the Celestial Sphere},''
\newblock \it{Notes de la Septi\`eme BSSM}, U.L.B.~(2015).
\newblock
\href{http://www.arXiv.org/abs/1508.00920}{\texttt{1508.00920}}.
\item
A.~Campoleoni, H.~A. Gonz\'alez, B.~Oblak, and M.~Riegler, ``{Rotating Higher
  Spin Partition Functions and Extended BMS Symmetries},'' {\em JHEP} \textbf{04}
  (2016) 034,
\href{http://www.arXiv.org/abs/1512.03353}{\texttt{1512.03353}}.
\item
H.~Afshar, S.~Detournay, D.~Grumiller, and B.~Oblak, ``{Near-Horizon Geometry
  and Warped Conformal Symmetry},'' {\em JHEP} \textbf{03} (2016) 187,
\href{http://www.arXiv.org/abs/1512.08233}{\texttt{1512.08233}}.
\item
A.~Campoleoni, H.~A. Gonz\'alez, B.~Oblak, and M.~Riegler, ``{BMS Modules in
  Three Dimensions},'' {\em Int. J. Mod. Phys.} \textbf{A31} (2016), no.~12,
  1650068,
\href{http://www.arXiv.org/abs/1603.03812}{\texttt{1603.03812}}.
\end{itemize}

\section*{Acknowledgements}
\addcontentsline{toc}{section}{Acknowledgements}

\subsection*{Professional community}

This thesis could not have been completed without the help and support of a number of people. First and 
foremost I am indebted to my supervisor, Prof.\ Glenn Barnich, for suggesting the topic in the first 
place and guiding me through its completion. Working with Glenn has been both a delight and a challenge; the 
former for his contagious passion for physics, and the latter for his sharp critical mind and 
unwavering skepticism, leading to numerous lively discussions about the nature of our work and of 
science altogether. Our complementary approaches have made our collaboration all the more 
fruitful; I am grateful to him for preserving my complete freedom while guiding me along the thesis. He has 
been a teacher, a friend, and a model of 
independence and scientific integrity that I hope to emulate myself.\\

I also wish to thank my other collaborators. The earliest ones in the history of my Ph.D.\ 
are Sophie De Buyl, St\'ephane Detournay and Antonin Rovai, with whom I spent a few weeks in Harvard in the 
Spring of 2013. I am grateful to them for their friendship and our many relaxed discussions about physics 
and other matters, including world economy, Belgian movies and American food. 
Later on my path crossed that of Hern\'an Gonz\'alez, starting with several ultra 
local journal clubs on quantum gravity, and ending with common projects. I am 
grateful to him for sharing his enthusiasm and almost convincing me that three-dimensional gravity 
is, in fact, gravity. My gratitude also goes to Hamid Afshar, Daniel Grumiller, Alexander 
Maloney, Max Riegler, and especially Andrea Campoleoni, for fruitful and enjoyable scientific 
collaborations; I hope there will be many more in the future.\\

The friendly environment at the physics department of the \it{Universit\'e Libre de Bruxelles} has 
also 
been a great help in completing this thesis. In particular I am grateful to my office neighbour, Laura 
Donnay, as well as Pujian Mao and Marco Fazzi, for enjoyable discussions during coffee breaks or lunches. I 
also thank the other 
students, postdocs and professors in the service of mathematical physics for making my days as a Ph.D.\ 
student in Brussels as pleasant as can be; this includes Riccardo Argurio, Andr\'es 
Collinucci, Geoffrey Comp\`ere, Laure-Anne Douxchamps, Simone Giacomelli, Gaston Giribet, Paolo Gregori, Marc 
Henneaux, Victor 
Lekeu, Arnaud Lepage-Jutier, Andrea 
Marzolla, Roberto Oliveri, Arash Ranjbar, Waldemar Schulgin, Shyam Sunder Gopalakrishnan, and C\'eline 
Zwikel. 
In addition I wish to thank 
my colleagues outside of Brussels --- Harold Erbin, Lucien Heurtier, Jules Lamers, Ruben Monten, Ali Seraj,
and Ellen Van der Woerd --- for our numerous delightful interactions. Finally, the 
logistics and organization of pretty much anything at the service of mathematical physics would be 
impossible without the precious help and efficiency of the administrative staff of the service and 
the Solvay Institutes --- Dominique Bogaerts, Fabienne De Neyn, Marie-France Rogge, Isabelle Van Geet, 
and Chantal Verrier. To them, thank you for your unshakeable goodwill in the face of the 
fiercest of administrative 
challenges.\\

Since October 
2015 I have had the chance to meet a number of new colleagues at the Deparment of Applied Mathematics 
and Theoretical Physics of the University of Cambridge. This would not have been possible without the support 
of Harvey Reall, whom I wish to thank warmly for this incredible opportunity. I am 
also grateful to Siavash Golkar, Shahar Hadar, Sasha Hajnal-Corob, Kai Roehrig, Arnab Rudra, Joshua Schiffrin 
and Piotr Tourkine for making my time there a daily enjoyment.\\

Still on the international side, I would like to thank Matthias Gaberdiel and Bianca Dittrich for giving me 
the opportunity to give seminars at their respective institutions. Both of these visits were a delight, and I 
had great pleasure in sharing some thoughts about my research with them and their colleagues.\\

Finally, I am grateful to Guillaume Bossard, Axel Kleinschmidt and Petr Tinyakov (in addition to Glenn 
Barnich, St\'ephane Detournay and Marc Henneaux who have already been cited) for accepting to be part of my 
thesis jury, and for their many questions and suggestions at the private and public defences.

\subsection*{Family and friends}

On a more private side I must mention the help and unconditional support provided by my family, 
and in particular my parents, Tijana and Du\v{s}an Oblak. Their contribution to this work is invisible to the 
naked eye, but it is actually so all-pervasive that it is hard to tell what would have 
become of me if I hadn't had such an amazing team behind my back. I am grateful to them for all the love they 
have given me and in particular for their support (both moral and practical) during the last year. I also 
wish 
to thank my cousins Sofija and Milena Stevanovi\'c, my aunt 
Ksenija Stevanovi\'c and my grandmother Vera Vujadinovi\'c for always being there for me when it counts and 
for the many beautiful moments we have had the chance to share.\\

My friends have also contributed greatly to this thesis, often unknowingly. Above all I wish to thank David 
Alaluf, Mitia Duerinckx, Jihane Elyahyioui, Geoffrey Mullier, Olmo Nieto-Silleras, and Roxane Verdikt for our 
many 
adventures together, including holidays and unforgettable parties (in the Balkans and elsewhere). As a 
matter of fact, Mitia Duerinckx has been a great help in understanding some of the mathematics used in this 
thesis, 
always patiently replying to my many e-mails with questions about Hilbert spaces, measures and the like. More 
generally, I am grateful to my friends from the mathematics department of ULB for their friendship and 
sometimes professional interactions, including Charel Antony, C\'edric De Groote, Julien Meyer and Patrick 
Weber. Finally, I am indebted to Thierry Maerschalk for helping me 
out long ago with some cute \LaTeX ~tricks, many of which were used in this thesis.\\

On the other side of the Channel I wish to thank the flurry of people whom I have had the chance to meet 
during the last year of my Ph.D.\, and who have allowed me to take helpful breaks away from physics. This 
includes Abhimanyu Chandra, Robert Cochrane, Marius Leonhardt, Du\v{s}an Perovi\'{c}, Frank Schindler, Bianca 
Schor, and Aaron Wienkers. I hope to meet them again in the future, despite our living in different corners 
of the world.\\

Finally, I wish to thank Vanessa Drianne, who has been in an entangled state between Brussels and Cambridge 
for most of the past year and whose support has been critical for my well-being in the last 
stages of writing the thesis.

\subsection*{Physics and art}

At this point I would like to thank two professors who have been most important for my development, both as a 
scientist and as a person. The first is my high school physics teacher, Emmanuel Thiran, who first managed to 
show me a glimpse of the beauty of Nature and the thrill of lifting its veil. While indirect, his influence 
pervades the entirety of my approach to physics and guides me to this very day. The second is Michel 
Laurent, my piano teacher. Since more than a decade he has 
been showing me the subtleties of music, but in truth his teaching extends far beyond that. While difficult 
to express in words, the conceptions of art and beauty that he has conveyed to me have had a 
great influence on me, and hence on this thesis.

\subsection*{Financial and logistic support}

To conclude I wish to thank the institutions who have provided me with financial and logistic 
support throughout my Ph.D. Above all I am grateful to the \it{Fonds de la Recherche Scientifique} 
-- F.N.R.S.\ for the grant (number FC-95570) that has allowed me to make a living out of the most delightful 
of occupations. On the other hand the financial 
support allowing me to work at the university of Cambridge during the last academic year has been granted to 
me by the \it{Fondation Wiener-Anspach}. I gratefully acknowledge their support in this marvelous experience, 
and wish to thank particularly Nicole Bosmans for her help with logistics.

\mainmatter

\chapter{Introduction}
\markboth{}{\small{\chaptername~\thechapter. Introduction}}
\pagestyle{Regular}

The quantization of gravity is one of the long-standing puzzles of theoretical physics. The purpose of this 
thesis is to study certain aspects of the problem that can be studied on the sole basis of symmetries, 
without any assumptions on the underlying microscopic theory. In this 
introduction we 
describe this strategy in some more detail, starting in section \ref{i1} with a broad overview of 
asymptotic 
symmetries in general and Bondi-Metzner-Sachs (BMS) symmetry in particular. We then introduce the 
distinction between global and extended BMS groups in section \ref{i2}. Section \ref{i3} is 
devoted to a lightning review of AdS/CFT and its putative Minkowskian counterpart. Finally, in section 
\ref{i4} we describe the relation between BMS symmetry and soft graviton degrees of freedom. Section \ref{i5} 
contains a general presentation of the upcoming chapters and describes their logical flow.

\section{Asymptotic BMS symmetry}
\label{i1}

The notion of symmetry is a cornerstone of physics and mathematics.\i{symmetry} A system is 
\it{symmetric} if there exists a set of transformations that leave it invariant, i.e.\ that preserve its 
structure. In physical terms, saying that a system has symmetries is really saying 
that there exist certain transformations that can be performed without affecting the outcome of 
experiments. For instance, translational symmetry\i{translation} is the statement 
that the 
result of an experiment does not depend on where one carries it out. By construction, the set of symmetry 
transformations of a system forms a \it{group}, so the mathematical tool used in the study of symmetries is 
group theory.\\

In this thesis we shall be concerned with symmetries of gravitational systems, that is, changes of 
coordinates that can be applied to space-time and that leave invariant the 
large-distance behaviour of the gravitational field. They are known as 
\it{asymptotic symmetries}\i{asymptotic symmetry} and can be thought of as a generalization of Poincar\'e 
symmetry for systems endowed with a weak gravitational 
field. In other words, these 
symmetries are those 
one would observe by looking at a gravitational system ``from far away''. In that context, the 
general type of question that we will ask is the following: given the asymptotic symmetries of a 
gravitational system, what are their physical implications? In particular, how do these symmetries 
affect one's intuition about particle physics?\\

Asymptotic symmetries of gravitational systems have been studied for about 
fifty years by now. Their first appearance in the literature is also the one 
that motivates the present work. Indeed, it was observed in the sixties by Bondi, van der 
Burg, Metzner \cite{Bondi:1960jsa,Bondi:1962px} and Sachs \cite{Sachs:1962zza,Sachs:1962wk} that the 
presence of gravitation in an 
asymptotically flat space-time leads to a symmetry group that is much, 
\it{much} larger than standard Poincar\'e. The group that they found turned out to be an 
infinite-dimensional extension of the Poincar\'e group, and is known today as the 
\it{Bondi-Metzner-Sachs group}, or \it{BMS 
group} for short.\i{BMS group}\\

The BMS group\i{BMS group} considered by the authors of 
\cite{Bondi:1962px,Sachs:1962zza,Sachs:1962wk} consists of two pieces: the first is the standard 
Lorentz group\i{Lorentz group} of special relativity, and the second is an infinite-dimensional 
Abelian group of so-called \it{supertranslations}.\i{supertranslation}\footnote{The terminology of 
``super-things'' here has nothing to do with supersymmetry: ``super-object'' simply means that a 
certain object, which one is familiar with in the finite-dimensional context of special relativity, gets 
extended in an infinite-dimensional way in the BMS group.}
In abstract mathematical notation, its structure\i{BMS group} can be written 
symbolically as
\be
\text{BMS}=\text{Lorentz}\ltimes\text{Supertranslations}.
\label{BMS}
\ee
The notation $\ltimes$ used here means that elements of the BMS group are pairs consisting of a 
Lorentz transformation and a supertranslation, and that Lorentz transformations act non-trivially 
on supertranslations. In the same way, the Poincar\'e group\i{Poincar\'e group} is
\be
\text{Poincar\'e}=\text{Lorentz}\ltimes\text{Translations}.
\label{poin}
\ee
The latter is a subgroup of BMS: the group of space-time translations is contained 
in the infinite-dimensional group of supertranslations.\\

Groups of the form (\ref{BMS}) or (\ref{poin}) are known as \it{semi-direct 
products}.\i{semi-direct product} They are ubiquitous in physics, and many of the 
conclusions of this thesis rely on this structure.

\section{Global BMS and extended BMS}
\label{i2}

In this section we introduce Bondi coordinates to explain briefly how BMS symmetry emerges from an asymptotic 
analysis. We then describe the distinction between ``global'' and ``extended'' BMS transformations.

\subsection*{Bondi coordinates}

Consider 
Minkowski space-time\i{Minkowski space}, endowed with inertial coordinates $x^{\mu}$ 
in 
terms of which the metric reads
\be
ds^2=\eta_{\mu\nu}dx^{\mu}dx^{\nu},\qquad
\text{with }\;(\eta_{\mu\nu})=\text{diag}(-1,+1,+1,+1).
\label{mink}
\ee
Now suppose we wish to study, say, outgoing massless particles sent by an observer located at the 
spatial origin. For this purpose we 
introduce \it{retarded Bondi coordinates}\i{Bondi coordinates}\i{coordinates!Bondi}
\be
r\equiv\left[x^ix^i\right]^{1/2},
\qquad
z\equiv\frac{x^1+ix^2}{r+x^3},
\qquad
u\equiv x^0-r.
\label{bondi}
\ee
Here $r$ is a space-like radial coordinate, $z$ is a stereographic coordinate\i{stereographic 
coordinate} on the sphere of radius $r$ (such that the north and south poles respectively correspond to $z=0$ 
and 
$z=\infty$), and $u$ is known as \it{retarded time}.\i{retarded time} In these 
coordinates the Minkowski metric (\ref{mink}) reads
\be
ds^2=
-du^2-2\,dudr+r^2\frac{4dzd\bar z}{(1+z\bar z)^2}
\label{minkB}
\ee
and the world line of an outgoing massless particle (moving away from the origin) is of the form 
$u=\text{const.}$, $z=\text{const.}$:

\begin{figure}[H]
\centering
\includegraphics[width=0.40\textwidth]{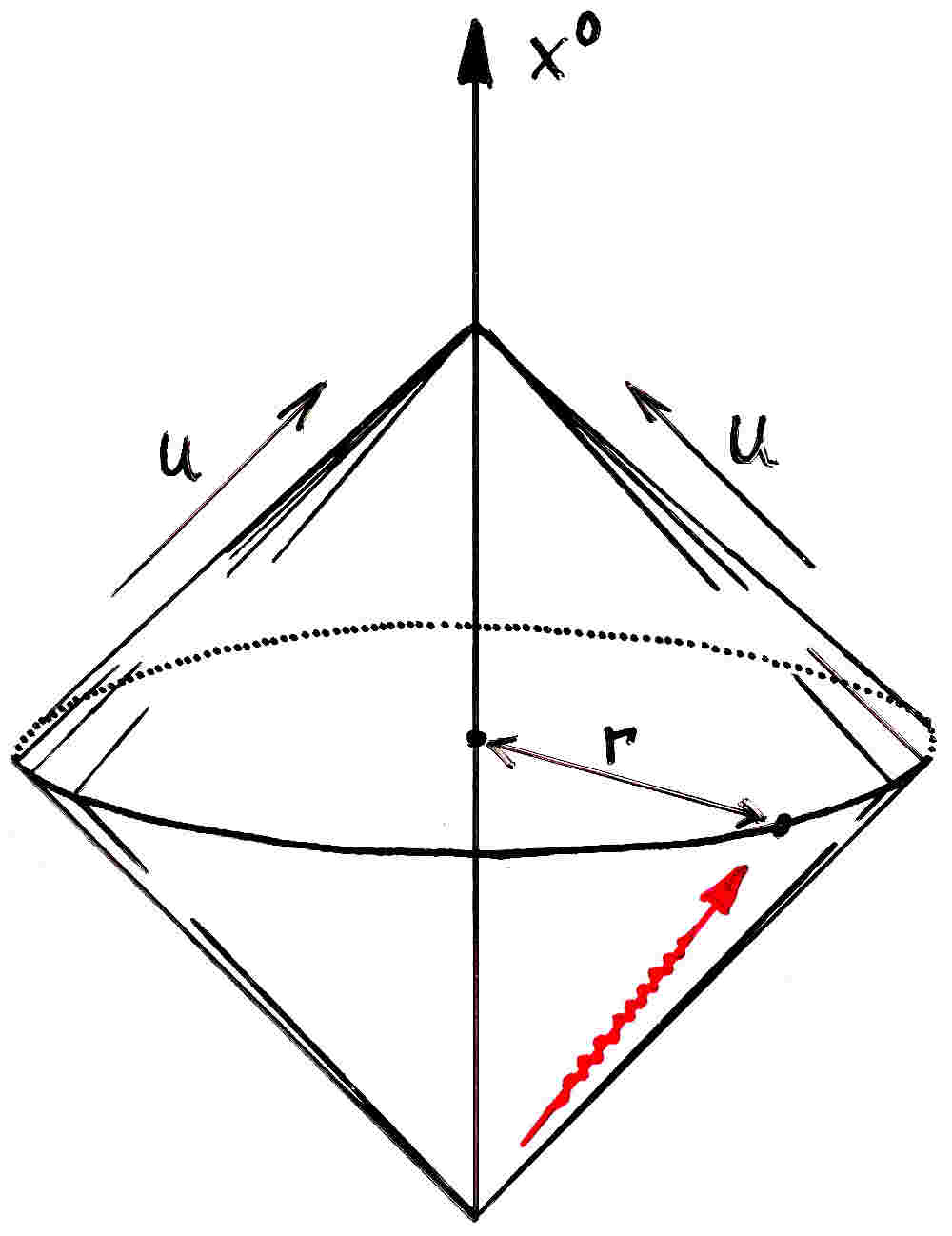}
\caption{The coordinates $u$ and $r$ in space-time. 
The time coordinate $x^0$ points 
upwards. The wavy red line represents an outgoing radial massless particle emitted at $r=0$ and moving to 
some non-zero distance $r$ away from the observer at $r=0$; the particle moves along one of the 
generators of the light cone 
given by $u=\text{const}$. The drawing is three-dimensional, so the circle of radius $r$ in 
this picture would actually be a sphere (spanned by the coordinate $z$) in a four-dimensional 
space-time.\label{BondiC}}
\end{figure}

In terms of Bondi coordinates, the region reached by massless particles emitted at some moment from the 
origin $r=0$ is a sphere at null infinity ($r\rightarrow+\infty$) spanned by the complex coordinate $z$, 
called a 
(future) \it{celestial sphere}.\i{celestial sphere} There is one such sphere for each value of 
retarded time $u$; 
the succession of all possible celestial spheres is a manifold $\RR\times S^2$ located at 
$r\rightarrow+\infty$ and known as \it{future null infinity}.\i{null infinity} It is the region where all 
outgoing massless radiation ``escapes'' out of space-time; it is the upper null cone of the Penrose 
diagram\i{Penrose diagram} of Minkowski space-time.

\begin{figure}[H]
\centering
\includegraphics[width=0.40\textwidth]{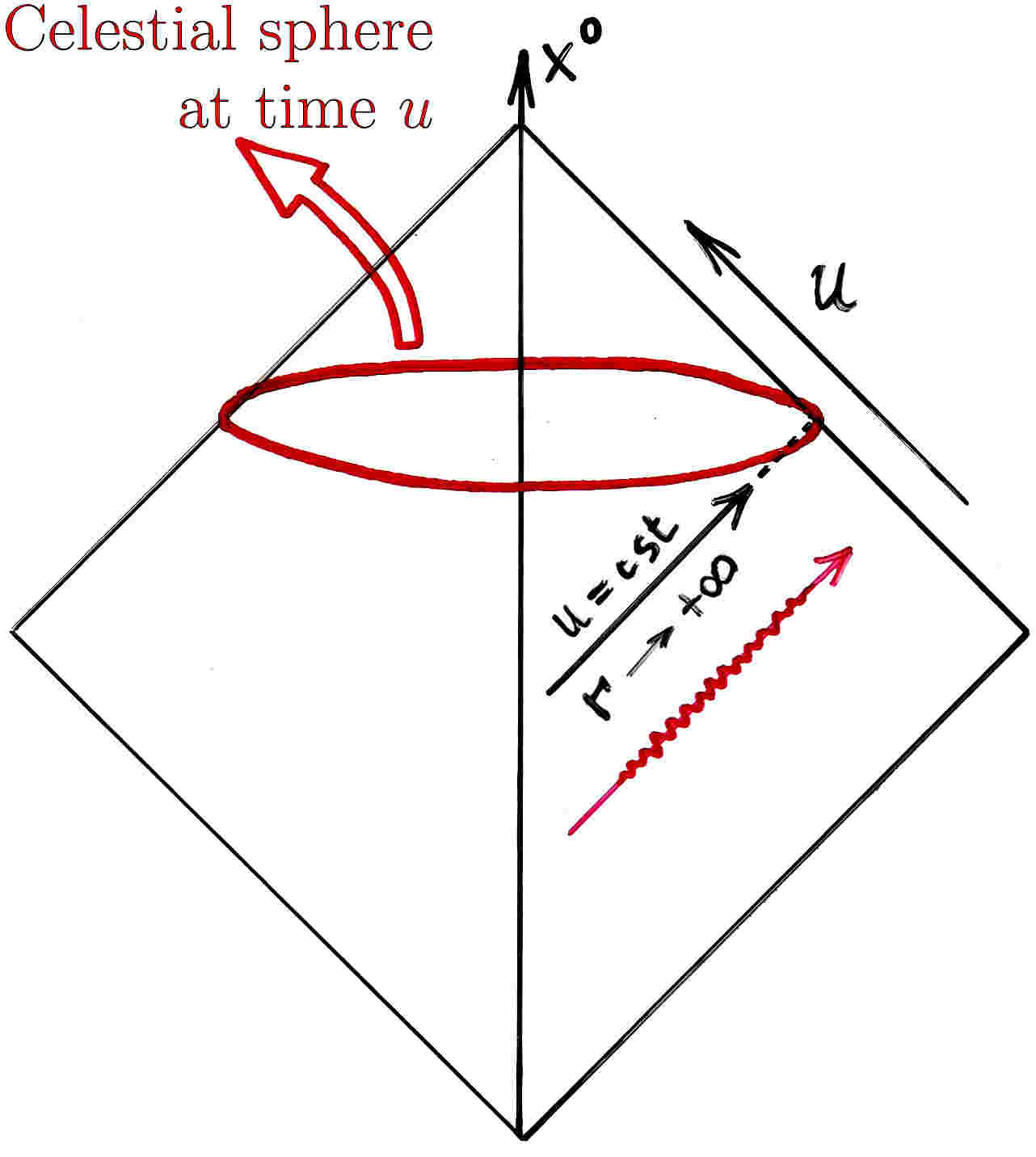}
\caption{A representation of celestial spheres on the Penrose diagram of Minkowski space. As in 
fig.~\ref{BondiC}, the wavy red line represents an outgoing radial light ray. The drawing is 
three-dimensional, so the red circle at the top of the picture would really be a sphere --- a celestial 
sphere --- in a four-dimensional space-time. Future null infinity is the cone $\RR\times S^2$ on the upper 
half of the image, spanned by $u$ and $z$.\label{penrose}}
\end{figure}

There exists a parallel construction of Bondi coordinates which is convenient for the study of 
\it{past} null infinity. These are \it{advanced Bondi coordinates} $(r,z,v)$,\i{Bondi 
coordinates}\i{coordinates!Bondi}\i{advanced Bondi coordinates} defined 
in terms of inertial coordinates $x^{\mu}$ exactly as in (\ref{bondi}) up to a sign difference in retarded 
time: $v=x^0+r$. The spheres at $r\rightarrow+\infty$ then are past celestial spheres 
and they foliate past null infinity (spanned by $v$ and $z$) in slices of constant time. 
In particular, all incoming massless particles originate from past null infinity.

\subsection*{Asymptotic flatness and the BMS group}

We now have the tools needed to introduce BMS symmetry. First, one declares that a space-time 
manifold is \it{asymptotically flat}\i{asymptotic flatness} at, say, future null 
infinity, if it admits local 
coordinates $(r,z,u)$ such that, as $r$ goes to infinity with $u$ finite, the metric takes the form 
(\ref{minkB}) up to subleading corrections. These coordinates need not be defined globally --- all that is 
needed is that they span a neighbourhood of future null infinity. Also, there is a precise definition of what 
is 
meant by ``subleading corrections''; these are alterations of the Minkowski metric (\ref{minkB}) that 
typically decay as inverse powers of $r$ at infinity, but the allowed powers themselves are constrained in a 
specific way. These constraints are motivated by physical considerations and they are part of the definition 
of ``asymptotic flatness''. (We will not deal with 
these subtleties for now, but we shall display them in section \ref{sebmsboco} in the 
three-dimensional case.)\\

The notion of asymptotic flatness allows one to define the associated asymptotic symmetry group. The latter 
consists, roughly speaking, of diffeomorphisms of space-time that preserve 
the asymptotic behaviour of the metric.\footnote{More precisely, the asymptotic symmetry group is the 
quotient of the group of diffeomorphisms that preserve the asymptotic behaviour of the metric by its normal 
subgroup consisting of so-called trivial diffeomorphisms. We will return to this in section 
\ref{seGeTri}.} 
Bondi \it{et al}.~\cite{Bondi:1962px,Sachs:1962zza,Sachs:1962wk} found that there are two families of such 
diffeomorphisms:
\begin{itemize}
\item The first family consists of Lorentz transformations. Their effect at null infinity is that of 
conformal 
transformations\i{Mobius transformation@M\"obius transformation} of celestial spheres (i.e.\ M\"obius 
transformations) given in 
terms of the stereographic coordinate $z$ of (\ref{bondi}) by
\be
z\mapsto\frac{az+b}{cz+d}+\cO(1/r),
\qquad
\begin{pmatrix}
a & b \\ c & d
\end{pmatrix}
\in\text{SL}(2,\CC).
\label{lor}
\ee
This property relies on the isomorphism\footnote{$\text{O}(3,1)$ is the Lorentz group in four 
dimensions 
and $\text{SO}(3,1)^{\uparrow}$ is its largest connected subgroup.} 
$\text{SO}(3,1)^{\uparrow}\cong\text{SL}(2,\CC)/\ZZ_2$, which expresses Lorentz transformations in terms of 
$\text{SL}(2,\CC)$ matrices.\i{SL$(2,\CC)$}\i{Lorentz group} Lorentz 
transformations 
also act on the coordinates $r$ and $u$ at infinity by angle-dependent rescalings, but this subtlety is 
unimportant at this stage.
\item The second family consists of angle-dependent translations of retarded time,\i{supertranslation}
\be
u\mapsto u+\alpha(z,\bar z),
\label{supert}
\ee
where $\alpha(z,\bar z)$ is any (smooth) real function on the sphere. In this language, Poincar\'e 
space-time 
translations are reproduced by functions $\alpha$ which are linear combinations of the functions
\be
\nn
1\,,\qquad\frac{1-z\bar z}{1+z\bar z}\,,
\qquad\frac{z+\bar z}{1+z\bar z}\,,\qquad\frac{i(z-\bar z)}{1+z\bar z}\,.
\ee
(In terms of polar coordinates $\theta$ and $\phii$, this corresponds to the 
spherical harmonics\i{spherical harmonic} $Y_{00}(\theta,\phii)$ and $Y_{1,m}(\theta,\phii)$ with 
$m=-1,0,1$.) This is the main surprise discovered by Bondi \it{et al}. It states that asymptotic 
symmetries (as opposed to isometries) enhance the Poincar\'e group to an 
infinite-dimensional group with an infinite-dimensional Abelian normal subgroup consisting of 
transformations (\ref{supert}). These transformations are the 
\it{supertranslations}\i{supertranslation} alluded to in eq.\ (\ref{BMS}).
\end{itemize}

\subsection*{Extended BMS}

The group of asymptotic symmetry transformations (\ref{lor}) and (\ref{supert}) is the original BMS 
group discovered in \cite{Bondi:1962px,Sachs:1962zza,Sachs:1962wk}.\i{BMS group} It consists of globally 
well-defined, invertible transformations of null infinity, so from now on we call it the 
\it{global} BMS group.\i{global BMS group} This slight terminological alteration is rooted in one of the most 
intriguing aspects 
of BMS symmetry. Indeed, in their work, Bondi \it{et al}.~observed that asymptotic 
symmetries include conformal transformations (\ref{lor}), but in principle one may even include 
transformations generated by arbitrary 
(generally singular) conformal Killing vector fields\i{conformal Killing vector} on the celestial 
spheres. Only six of those vector fields generate 
the invertible M\"obius transformations 
(\ref{lor}); the remaining ones are singular. Upon 
including these extra generators, the global conformal transformations (\ref{lor}) are enhanced to 
arbitrary local conformal transformations
\be
z\mapsto f(z)+\cO(1/r),
\label{locc}
\ee
where $f(z)$ is any meromorphic function. Despite their singularities, these transformations do preserve the 
asymptotic behaviour of the metric and may therefore qualify as asymptotic symmetries, at least 
infinitesimally.\\

The extension of the BMS group obtained by replacing Lorentz transformations by local 
conformal transformations (\ref{locc}) is called the \it{extended} BMS group.\footnote{Strictly speaking, 
local conformal transformations do not span a group but a semi-group, and the same applies to 
extended BMS. This abuse of terminology is pretty common, and it will be inconsequential for the 
discussion of this introduction.}\i{BMS group!extended}\i{extended BMS group} In that context, local 
conformal transformations of 
celestial spheres are known as \it{superrotations}\i{superrotation} and should be thought of as an 
infinite-dimensional extension of Lorentz transformations, in the same way that supertranslations extend 
space-time translations. In the notation of (\ref{BMS}) and (\ref{poin}), the extended BMS group looks like
\be
\label{ebms}
\boxed{
\Big.
\text{Extended BMS}
=
\text{Superrotations}\ltimes\text{Supertranslations}
}
\ee
where now both factors of the semi-direct product are infinite-dimensional.\\

It was recently suggested by Barnich and Troessaert \cite{Barnich:2009se,Barnich:2010eb} that 
extended (as opposed to global) BMS symmetry is the true, physically relevant symmetry of 
asymptotically 
flat gravitational systems in four dimensions (see also footnote 17 
of \cite{Banks:2003vp}).\label{btABBA} This proposal is motivated by a similar symmetry 
enhancement\i{symmetry enhancement} occurring 
in 
two-dimensional conformally-invariant systems: while their global symmetry algebra is finite-dimensional, 
they 
turn out to enjoy a much richer infinite-dimensional symmetry. This observation first appeared in 
a seminal paper by Belavin, Polyakov and Zamolodchikov \cite{Belavin:1984vu} and triggered the development of 
two-dimensional conformal field theory (CFT).\i{conformal field theory}\\

Thus, the truly thrilling aspect of extended BMS symmetry is the prospect of applying conformal 
field-theoretic techniques to gravitational phenomena in four dimensions. \label{bt} This reduction from four 
to two 
dimensions is reminiscent of holograms, and indeed the notion of ``holography'' in quantum gravity is one of 
the main motivations that led to these considerations.

\section{Holography}
\label{i3}

The elementary concept of holography in quantum gravity is simple: it is the statement that gravitational 
phenomena 
occurring in a certain space-time manifold can be described equivalently in terms of some lower-dimensional, 
``dual'' theory. This idea is originally due to 't Hooft \cite{'tHooft:1993gx} and Susskind 
\cite{Susskind:1994vu}, who were led to it by model-independent considerations.\i{holography} In particular, 
holography is 
compatible with the Bekenstein-Hawking entropy formula \cite{Bekenstein:1973ur,Hawking:1974sw},\i{black 
hole} according to which the entropy of a black hole is 
proportional to the area of its horizon. (The keyword here is ``area'', as opposed to the ``volume'' expected 
on the basis of standard thermodynamics.)

\subsection*{AdS/CFT}

In practice, the first genuine illustration of holography in a concrete model of quantum gravity --- namely 
string theory --- was exhibited by Maldacena \cite{Maldacena:1997re}, initiating what has come to be known as 
the Anti-de 
Sitter/Conformal Field Theory (AdS/CFT) correspondence.\i{AdS/CFT} The latter states, in a 
nutshell, that (quantum) gravity on a $D$-dimensional asymptotically Anti-de Sitter space-time is dual to a 
$(D-1)$-dimensional conformal field theory. The CFT may be seen as living on 
the boundary of AdS, that is, at spatial infinity, and is supposed to capture all the information on 
gravitational observables. This is a statement of \it{duality}, where two completely different 
theories contain the same physical information. While there is (as yet) no proof of the full equivalence, a 
substantial amount of checks have been carried out to confirm that gravity on AdS and a suitable CFT on its 
boundary do indeed produce the same physical predictions.\\

The case of a three-dimensional bulk space-time ($D=3$) is especially important for our 
purposes. In that context the first hint of a holographic duality actually dates back to the eighties, 
when Brown 
and Henneaux \cite{Brown:1986nw} noticed that the asymptotic symmetries of AdS$_3$ gravity are 
infinite-dimensional. Analogously to the Minkowskian setting studied two decades earlier by Bondi \it{et 
al}., Brown and Henneaux found that asymptotic symmetries enhance the usual AdS$_3$ isometry 
algebra $\mathfrak{so}(2,2)$ in an infinite-dimensional way and span the algebra of local conformal 
transformations (\ref{locc}) in two dimensions. In addition, the conserved charges generating 
these symmetries turn out to satisfy a centrally extended algebra, with a central charge proportional to 
the AdS radius measured in Planck units\i{Brown-Henneaux central charge}, now known as the \it{Brown-Henneaux 
central charge}. The latter is the one key parameter specifying the 
putative two-dimensional CFT dual to gravity on AdS$_3$. For instance, it was used in 
\cite{Strominger:1997eq} to show that the entropy of
black holes 
in three dimensions \cite{Banados:1992wn,Banados:1992gq} can be reproduced by a purely 
conformal field-theoretic computation.\\

The proposal that BMS symmetry might account for gravitational 
physics in asymptotically flat space-times is similar in spirit to AdS/CFT. The problem is that 
most known holographic constructions rely on the key assumption that the bulk space-time is endowed with 
a 
negative cosmological constant, 
i.e.~that it is of the AdS type. This leads to a natural question: how should one deal with holography in 
flat space?

\subsection*{Flat space holography}

In the context of AdS/CFT, the dual theory of gravity is a CFT; in particular, even without full knowledge of 
the dual theory, one can at least hope to make sense of it by relying on the well understood 
consequences of conformal invariance. By contrast, in asymptotically flat space-times, the concept of a 
``dual theory'' is unclear, partly due to poorly understood symmetries and partly because (in dimension four 
or higher) gravitational waves cross the null boundary of space-time; in fact, one may ask whether flat space 
holography makes any sense to begin with. In view of this pessimistic omen, a safe approach to the problem is 
to 
avoid 
unnecessary assumptions and rely solely on the one known property of asymptotically flat gravity, namely BMS 
symmetry.\i{flat space holography}\i{holography!in flat space} Indeed, whatever flat space holography means, 
if a dual theory exists, then it must be invariant 
under a certain version of BMS.\\

The most interesting 
incarnation of BMS symmetry is the four-dimensional one, since it is relevant to macroscopic gravitational 
waves. 
Unfortunately, the structure of the 
extended BMS (semi-)group in four dimensions is very poorly understood (despite recent progress 
\cite{Barnich:2011mi,Strominger:2013jfa}). In short, this structure appears to be such that standard group 
theory fails to apply. One is thus 
led to study toy models that capture the key features of BMS symmetry without the complications of a 
four-dimensional world.\\

In this thesis we argue that the \it{BMS group in three dimensions} \cite{Ashtekar:1996cd}, or 
BMS$_3$,\i{BMS$_3$ 
group} provides such a toy model. We shall see that it displays the extended structure (\ref{ebms}) in a 
simplified 
and controlled setting, and successfully accounts for many aspects of three-dimensional asymptotically flat 
gravity, both classically and quantum-mechanically. The BMS$_3$ group is the main actor of this work and we 
will use it to develop our intuition on flat space holography in general, including the four-dimensional case.

\subsection*{Holography as an Erlangen programme}
\label{suserlangen}

Aside from the study of quantum gravity in asymptotically flat space-times, this thesis puts a strong accent 
on the relation between group theory and physics. Most, if not all, of the topics that we will encounter in 
both AdS$_3$/CFT$_2$ and flat space holography follow from the properties 
of suitable groups --- Virasoro and BMS$_3$, respectively. For 
instance, the phase space of gravity will turn out to coincide with the space of the coadjoint representation 
of its asymptotic symmetry group, and its quantization will produce families of unitary 
representations of that group. In this sense, 
three-dimensional gravity and its ``holographic'' properties can be reformulated as statements in group 
theory.\\

In hindsight this observation is not too surprising. Indeed, Klein's Erlangen programme \cite{Klein} 
posits that geometric statements can be recast in the language of group theory.\i{Erlangen programme} 
This point of view has led to numerous developments in mathematics 
throughout the twentieth century (including e.g.\ the work of Poincar\'e on special relativity). 
Since 
general relativity is essentially the dynamics of pseudo-Riemannian geometry, it is natural that the 
programme should apply to it as well provided one identifies the correct symmetry group. In particular, 
holography may sometimes be seen as an Erlangen programme in disguise.\footnote{See e.g.\ the Wikipedia 
page \url{https://en.wikipedia.org/wiki/Erlangen_program}.}

\section{BMS particles and soft gravitons}
\label{i4}

The symmetry representation theorem of Wigner \cite{wigner1931} states that the symmetry group of any 
quantum-mechanical system\i{symmetry representation thm} 
acts on the corresponding Hilbert space by unitary transformations.\i{symmetry!and unitarity} Accordingly, a 
natural first step in the study of BMS symmetry is the 
construction and classification of its irreducible, unitary representations.\\

Since the Poincar\'e group is a subgroup of BMS, it provides a first rough picture of what one
should expect from BMS representations. Indeed, irreducible unitary representations of 
Poincar\'e 
are, by definition, \it{particles}:\i{particle} they are classified by their mass\i{mass} and spin\i{spin} 
\cite{Wigner:1939cj} and their Hilbert space accounts for the available one-particle 
states. The BMS group is expected to generalize this notion in a way that incorporates certain 
gravitational effects. It should describe the quantum states of a particle, plus some extra degrees of 
freedom 
accounting for the fact that BMS is only an \it{asymptotic}, rather than an exact, symmetry group. Guided by 
this picture, we introduce the following terminology:\i{BMS particle}
\be
\text{\it{A BMS particle is an irreducible, unitary representation of the BMS group.}}
\nn
\ee
This thesis is devoted to the description and classification of such particles in three space-time 
dimensions.\\

Recent developments in the study of BMS symmetry provide a simple interpretation for BMS 
particles. Indeed, it was observed in \cite{Strominger:2013jfa,He:2014laa} that, in four dimensions, the 
statement of 
supertranslation-invariance of the gravitational $S$-matrix is equivalent to Weinberg's soft graviton theorem 
\cite{Weinberg:1965nx}.\i{soft graviton} This result was subsequently generalized to include superrotations 
\cite{Cachazo:2014fwa,Kapec:2014opa}, producing a subleading term in the soft graviton expansion of the 
$S$-matrix. The bottom line of these considerations is that BMS symmetry describes the soft sector of 
gravity, that is, 
the one consisting of infinite-wavelength gravitational degrees of freedom. The
interpretation of BMS particles follows: they are particles (in the standard sense) dressed with soft 
gravitons. Dressed particles are indeed ubiquitous in the quantization of gauge theories 
\cite{Bloch:1937pw,Dirac:1955uv,Chung:1965zza,Kibble:1968zza,Kibble:1969ip,Kibble:1969ep,Kibble:1969kd,
Kulish:1970ut,Zwanziger:1973if}, and 
this in itself is not a new result. What \it{is} new, however, is the fact that this dressing is accounted 
for by a symmetry principle that generalizes Poincar\'e; this is the key content of the relation between 
BMS symmetry and soft theorems.\\

Accordingly, the classification of BMS particles that we expose in this thesis may be thought of as a 
classification of all possible ways to dress a Poincar\'e particle with soft gravitons. A word of caution is 
in order: since we will be working in three space-time dimensions, the gravitational field will have no local 
degrees of freedom so there will be no genuine gravitons. \label{pagedisc} In particular, the name ``soft 
graviton'' is ambiguous, as there is no actual graviton whose zero-energy limit would be a soft particle. 
\i{gravity in 3D} However, asymptotic symmetries precisely account for soft 
graviton degrees of freedom, so we shall adopt the viewpoint that any system with non-trivial asymptotic 
symmetries does indeed have non-trivial soft degrees of freedom. This amounts to using the words ``soft 
graviton'' as a synonym for the more standard ``topological'' or ``boundary degree of freedom''.
In particular, three-dimensional 
gravitational systems generally do have highly non-trivial asymptotic symmetries 
\cite{Brown:1986nw,Barnich:2006av} and 
therefore 
possess soft degrees of freedom in this sense. In this language, the statement that three-dimensional gravity 
has no bulk degrees of freedom turns into the fact that the \it{only} non-trivial degrees of freedom of 
three-dimensional gravity are soft.

\paragraph{Remark.} Unitary representations of the globally well-defined BMS group (\ref{BMS}) have already 
been classified by McCarthy and others in \cite{McCarthy01,McCarthy00,McCarthy317}, and it was indeed 
suggested in \cite{Mccarthy:1972ry} that BMS symmetry is relevant to particle physics in that it provides 
a better definition of the notion of ``particle''. However, these representations 
appear to miss the fact that supertranslations create soft gravitons when acting on the 
vacuum, which is crucial for the application of BMS symmetry to soft theorems. In 
this sense the understanding of BMS particles in four dimensions is still an open problem; it suggests that 
some extension of (\ref{BMS}) is necessary if representations of BMS are to reflect reality. 
We shall comment further on this issue in section \ref{sebmspar}.

\section{Plan of the thesis}
\label{i5}

We now describe the topics studied in this thesis. The latter is divided in three parts, devoted respectively 
to group theory in quantum mechanics, to the Virasoro group, and to the BMS$_3$ group.

\subsection*{Quantum symmetries}

The first part of the thesis deals with the implementation of symmetries in quantum mechanics through 
projective unitary representations, which are worked out in detail for the Poincar\'e groups and the Bargmann 
groups. It consists of four chapters.\\

Quantum symmetries generally act in a projective way, which is to say that the group operation of the 
underlying symmetry group is represented up to certain constant phases. The presence of such phases is 
captured by \it{central extensions} of the symmetry group. Accordingly, chapter \ref{c1} is devoted to 
central extensions and to the more general notion of group and Lie algebra cohomology. Chapter \ref{c1b} then 
explains how one can build Hilbert spaces of wavefunctions on a homogeneous space endowed with a unitary 
action of 
a symmetry group. This involves the important notion of \it{induced representations}, which we discuss in 
detail.\\

As an application, in chapter \ref{c2bis} we describe the irreducible unitary representations of semi-direct 
products of the general form (\ref{BMS}) or (\ref{poin}). As it turns out, \it{all} these representations are 
induced 
representations and consist of wavefunctions on a \it{momentum orbit}. This
provides an exhaustive classification of unitary representations for such groups. We illustrate these 
considerations with the Poincar\'e group (in any space-time dimension) and with its non-relativistic 
counterpart, the Bargmann group, corresponding respectively to relativistic and Galilean particles.\\

Finally, chapter \ref{c3} describes the relation between classical and quantum symmetries through 
geometric quantization. In a nutshell this relation is obtained by defining a space of wavefunctions on what 
is known as a \it{coadjoint orbit} of a symmetry group. For semi-direct products this 
approach reproduces the classification of representations by momentum orbits and leads to a group-theoretic 
version of the world line formalism.

\paragraph{Remark.} The tools used in chapters \ref{c1} to \ref{c2bis} rely on elementary group theory; we 
refer 
for instance to \cite{rotman1999} for an introduction. The language of chapter \ref{c3}, on the 
other hand, relies more heavily on differential and symplectic geometry; see 
e.g.\ \cite{abraham1978foundations,lee2003} 
for some background material.

\subsection*{Virasoro symmetry and AdS$_3$ gravity}

The second part of the thesis deals with the Virasoro group and its application to three-dimensional gravity 
on Anti-de Sitter backgrounds. It consists of three chapters. The material exposed in part II relies in a 
crucial way on 
chapter \ref{c1} and to a lesser extent on chapter \ref{c3}, but is independent of the considerations of 
chapters \ref{c1b} and \ref{c2bis}.\\

Chapter \ref{c4} is devoted to the construction of the Virasoro group as a central extension of the group of 
diffeomorphisms of the circle and introduces its coadjoint representation. The latter coincides with the 
transformation law of stress tensors in two-dimensional conformal field theory. In chapter \ref{c4bis} we 
classify the orbits of this action, i.e.\ the coadjoint orbits of the Virasoro group, and observe that they 
look roughly like infinite-dimensional cousins of Poincar\'e momentum orbits.\\

In chapter \ref{AdS3} we show how Virasoro symmetry emerges in AdS$_3$ gravity with Brown-Henneaux 
boundary conditions, after explaining some basic notions on asymptotic symmetries in general. We also show 
that the phase space of AdS$_3$ gravity is embedded as a hyperplane at constant central charge in the space 
of the coadjoint representation of two copies of the Virasoro group. As an application we relate 
highest-weight representations of the Virasoro algebra to the quantization of gravitational boundary degrees 
of freedom.

\subsection*{BMS$_3$ symmetry and gravity in flat space}

The third and last part of the thesis is devoted to three-dimensional BMS symmetry and contains most of the 
original contributions of this work. It consists of three chapters, plus a conclusion. The material 
presented in part III relies crucially on the content of parts I and II.\\

In chapter \ref{c6} we introduce BMS$_3$ symmetry by way of an asymptotic analysis of Brown-Henneaux 
type applied to Minkowskian backgrounds, and show that the resulting algebra of surface charges has a 
classical central extension. We then put 
this observation on firm mathematical ground by defining rigorously the BMS$_3$ group and its central 
extension. We also show that the phase space of asymptotically flat gravity is a hyperplane at fixed central 
charges embedded in the space of the coadjoint representation of BMS$_3$ 
\cite{Barnich:2014zoa,Barnich:2015uva}.\\

Chapter \ref{c7} is devoted to the quantization of BMS$_3$ symmetry, i.e.\ to its irreducible unitary 
representations \cite{Barnich:2014kra}. In the language 
introduced above, each representation is a \it{BMS$_3$ particle}. We show that the supermomentum orbits that 
classify these particles coincide with coadjoint orbits of the Virasoro group and describe the resulting 
Hilbert spaces of one-particle states. This leads in particular to the interpretation of BMS$_3$ particles as 
particles dressed with gravitational degrees of freedom.\\

Finally, chapter \ref{c8} deals with rotating one-loop partition functions of quantum fields in flat space at 
finite temperature. Each partition function takes the form of an exponential of Poincar\'e characters. In 
three space-time dimensions and for a massless field with spin two, the combination of characters is 
precisely such that the whole partition function coincides with the character of a unitary representation of 
the BMS$_3$ group \cite{Oblak:2015sea,Barnich:2015mui}. For higher spins in three dimensions we similarly 
obtain characters of flat non-linear $\cW_N$ algebras \cite{Campoleoni:2015qrh}. Along the way we describe 
unitary representations of these algebras \cite{Campoleoni:2016vsh} and show that they differ qualitatively 
from earlier proposals in the literature. We end by describing certain supersymmetric extensions of the 
BMS$_3$ group, their representations, and their characters.

\paragraph{Remark.} The group-theoretic methods developed in this thesis apply to essentially any 
symmetry group involving the Virasoro group. In particular one can use this approach to derive the 
transformation laws of the stress tensor of a warped conformal field theory for all values of its three 
central charges. Since these 
considerations are somewhat out of our main line of thought we will not review them in this thesis and refer 
instead to \cite{Afshar:2015wjm}, where they were used to derive a Cardy-like formula 
for the entropy of Rindler backgrounds.

\renewcommand{\afterpartskip}{} 
\part*{Part I\\[.3cm]
Quantum symmetries}
\addcontentsline{toc}{part}{I~\ Quantum symmetries}
\begin{center}
\begin{minipage}{.9\textwidth}
~\\
~\\
~\\
~\\
~\\
In this part we describe symmetry groups in quantum mechanics along three related lines of thought. First we 
argue that the action of symmetry transformations in quantum mechanics is unitary up to phases, which leads 
to central extensions. Then we show how to build concrete unitary representations using the method of induced 
representations, which we apply to the description of relativistic particles. Finally we describe the general 
relation between unitary representations and homogeneous spaces through geometric quantization.
\end{minipage}
\end{center}
\newpage
~
\thispagestyle{empty}

\chapter{Quantum mechanics and central extensions}
\label{c1}
\markboth{}{\small{\chaptername~\thechapter. Central extensions}}

In this short chapter we discuss the implementation of symmetries in a quantum-mechanical 
context. For definiteness and simplicity we assume throughout that these symmetries span a Lie group.
We start in 
section \ref{s1.1} with a brief review of the symmetry representation theorem of Wigner and show how quantum 
mechanics gives rise to projective unitary representations. The problem of classifying such 
representations then leads to sections 
\ref{s1.2} and \ref{gcoho}, respectively devoted to Lie algebra cohomology and group cohomology. The 
presentation is inspired by \cite{weinberg1995,ovsienko2004projective,khesin2008geometry,guieu2007algebre}; 
see also \cite{Tuynman}.

\section{Symmetries and projective representations}
\label{s1.1}

In this section we review the interplay between quantum mechanics and symmetries. After 
a brief general reminder on the formalism of quantum theory, we state the symmetry representation theorem 
which justifies the study of unitary representations of groups and Lie algebras. We also show how the fact 
that quantum states are rays in a Hilbert space (rather than individual vectors) leads to 
projective representations, hence to central extensions. We end with a discussion of topological 
central extensions, while algebraic central extensions are postponed to 
section \ref{s1.2}.

\subsection{Quantum mechanics}

\paragraph{Definition.} A (complex) \it{Hilbert space} $\sH$\i{Hilbert space} is a vector 
space over $\CC$ endowed with a Hermitian form
\be
\langle\cdot|\cdot\rangle:
\sH\times\sH\rightarrow\CC:
(\Phi,\Psi)\mapsto\langle\Phi|\Psi\rangle,
\label{scal}
\ee
such that the norm of a vector $\Psi$ be $\sqrt{\langle\Psi|\Psi\rangle}$, and such that the resulting 
normed vector space be complete.\footnote{Recall that a metric space is \it{complete} if 
any Cauchy sequence converges.\i{complete metric space}} We take the scalar product (\ref{scal}) to be linear 
in 
its second argument and antilinear in the first one.\\

Note that our 
notation is \it{not} the standard Dirac notation of bras and kets:\i{bra}\i{ket} a vector in $\sH$ is 
denoted as $\Psi$ (not $|\Psi\rangle$), and its dual is the linear form $\langle\Psi|\cdot\rangle$ on $\sH$. 
Accordingly, the 
Hermitian conjugate 
$\hA^{\dagger}$ of a linear operator $\hA$ is defined by
\be
\langle\Phi|\hA^{\dagger}\Psi\rangle\equiv\langle\hA\Phi|\Psi\rangle
\quad\text{for all }\Phi,\Psi\in\sH.
\label{hermAKU}
\ee
An 
operator $\hA$ is Hermitian (or self-adjoint)\footnote{We will not take into 
account issues related to the domains of operators.} if $\hA^{\dagger}=\hA$.\i{Hermitian 
operator}\i{self-adjoint operator}\\

Now consider a quantum system whose space of states is a Hilbert space 
$\sH$. A pure \it{quantum state} of the system is a ray\i{quantum state}\i{ray}\i{Hilbert space!ray}
in $\sH$, that is, a one-dimensional subspace
\be
[\Psi]
=
\left\{z\Psi|z\in\CC\right\}
\label{ray}
\ee
where $\Psi$ is some non-zero state vector. The vanishing vector does not represent a quantum 
state, so the set of mutually inequivalent pure states is the projective space\i{projective 
space} $\PP\sH=(\sH\backslash0)/\CC$.
It is the set of one-dimensional subspaces of $\sH$. 
Stated differently, the set of distinct states in $\sH$ is the 
quotient of the unit sphere in $\sH$ by the equivalence relation
\be
\label{ia}
\Psi\sim e^{i\theta}\Psi
\quad\text{for all }\theta\in\RR.
\ee
We shall denote by $[\Psi]$ the resulting equivalence class of $\Psi$. For 
example, in a two-level system where $\sH=\CC^2$, the set of inequivalent states is $\CC P^1\cong S^2$.\\

Now let the system be in a state $[\Psi]$. If $\hA$ is an observable and if $\lambda$ 
is one of its eigenvalues with eigenvector $\Phi$ say, the probability of finding the value $\lambda$ is
\be
\label{prob}
\text{Prob}(\lambda,\hA,[\Psi])
=
\frac{\left|\langle\Phi|\Psi\rangle\right|^2}{\langle\Phi|\Phi\rangle\langle\Psi|\Psi\rangle}.
\ee
(We are assuming for simplicity that the eigenvalue $\lambda$ is not degenerate.) Note that this expression 
is independent, as it should, of the choice of both the representative $\Psi$ of the state 
$[\Psi]$, and the eigenvector $\Phi$.

\paragraph{Remark.} In quantum mechanics, one generally assumes that the Hilbert space is \it{separable}, 
i.e.\ that it admits a countable basis.\i{separable Hilbert space} Any such
space is isometric to the space $\ell^2(\NN)$ of 
square-integrable sequences of complex numbers --- so there really exists only \it{one} infinite-dimensional 
separable Hilbert space. This is not to say that all separable Hilbert spaces describe 
the same quantum system, because the definition of a system also involves the set of observables that act on 
it --- and identical Hilbert spaces may well come with very different operator algebras.

\subsection{Symmetry representation theorem}
\label{symrep}

\subsubsection*{Symmetry groups}

A \it{symmetry}\i{symmetry} is a transformation of a system that leaves it invariant. In particular, the set 
of symmetries of a system always contains the identity transformation, and any symmetry transformation is 
invertible. In addition the composition of any two symmetry transformations is itself a symmetry, and 
composition is associative. Put together, these properties imply that\i{group}
\be
\text{\it{the set of symmetries of any system forms a group.}}
\nn
\ee
Accordingly, the framework suited for the study of symmetries is \it{group theory}.\\

In this thesis we will be concerned with \it{Lie groups}, consisting of symmetry transformations that depend 
smoothly on a certain number of real parameters. This number is the \it{dimension} of the group. In part I of 
the thesis, all Lie groups are finite-dimensional.

\paragraph{Remark.} The notion of symmetry can be relaxed in such a way that not all pairs of symmetry 
transformations are allowed to be composed together. The resulting set of symmetry transformations then spans 
a \it{groupoid} rather than a group (see e.g.\ \cite{Weinstein96,crainic2011lectures}). This 
relaxed notion of 
symmetry is relevant to gauge theories \cite{Barnich:2010xq}, and in particular to BMS symmetry in four 
dimensions \cite{Barnich:2011mi}. However, standard group theory suffices for all symmetry considerations in 
three-dimensional gravity (and in particular for BMS$_3$), so we will not deal with groupoids in this thesis.

\subsubsection*{Symmetries in quantum mechanics}

Consider a quantum Hilbert space of states $\sH$. In these terms a symmetry is a bijection 
$\PP\sH\rightarrow\PP\sH:[\Psi]\mapsto\cS([\Psi])$ that preserves the 
probabilities (\ref{prob}). Equivalently, if we represent rays in $\sH$ by normalized 
vectors subject to the identification (\ref{ia}), a symmetry transformation $\cS$ must be such that 
\be
\left|\langle\Phi|\Psi\rangle\right|
=
\left|\langle\Phi'|\Psi'\rangle\right|
\label{unit}
\ee
for all normalized vectors $\Phi$, $\Psi$, $\Phi'$, $\Psi'$ such that $\Phi'\in\cS([\Phi])$ and 
$\Psi'\in\cS([\Psi])$. The key result on symmetries in quantum mechanics is the following \cite{wigner1931}:

\paragraph{Symmetry representation theorem.} Let $\cS:\PP\sH\rightarrow\PP\sH$ be an invertible 
transformation satisfying property (\ref{unit}).\i{symmetry representation thm} Then it takes the form
$\cS([\Psi])=[\hU\cdot\Psi]$, where $\hU$ is either a linear, unitary operator so that
\be
\hU\cdot\left(\lambda\Phi+\mu\Psi\right)
=
\lambda\,\hU\cdot\Phi+\mu\,\hU\cdot\Psi
\qquad\text{and}\qquad
\langle\hU\cdot\Phi|\hU\cdot\Psi\rangle
=
\langle\Phi|\Psi\rangle\,,
\nn
\ee
or an antilinear, antiunitary 
operator\i{antilinear}\i{antiunitary} so that
\be
\hU\cdot\left(\lambda\Phi+\mu\Psi\right)
=
\bar\lambda\,\hU\cdot\Phi+\bar\mu\,\hU\cdot\Psi
\qquad\text{and}\qquad
\langle\hU\cdot\Phi|\hU\cdot\Psi\rangle
=
\langle\Psi|\Phi\rangle
\nn
\ee
for all $\lambda,\mu\in\CC$ and all $\Phi,\Psi\in\sH$.
A proof of this theorem can be found in chapter 2 (appendix A) of \cite{weinberg1995}.\\

Note that symmetries represented by antiunitary\i{antiunitary} operators only arise when the symmetry group 
is 
disconnected. For 
example, in Lorentz-invariant theories, time-reversal is always represented in an antiunitary way (see 
e.g.~\cite{peskin1995introduction}). In this work we will restrict attention to connected symmetry 
groups, 
in which case all symmetry operators are linear and unitary. In particular they satisfy 
$\hU^{\dagger}=\hU^{-1}$, where Hermitian conjugation is defined by (\ref{hermAKU}).

\subsection{Projective representations}

The symmetry representation theorem implies that all (connected) symmetry groups are represented 
unitarily in a quantum-mechanical system, and thus motivates the study of unitary representations in 
general. Let us first recall the basics:

\paragraph{Definition.} A \it{representation}\i{representation} of a group $G$ in a vector space $\sH$ is a 
homomorphism\footnote{Throughout this thesis representations of groups are denoted by the letters $\cR$, 
$\cS$, $\cT$, etc. The letter $G$ denotes a group whose elements are written $f$, $g$, $h$, etc. The identity 
in $G$ is denoted $e$.}
\be
\cT:G\rightarrow\text{GL}(\sH):
g\mapsto\cT[g]
\nn
\ee
where $\text{GL}(\sH)$ is the group of invertible linear transformations of $\sH$. When $\sH$ is a Hilbert 
space, the 
representation is \it{unitary} if $\cT[g]$ is a unitary operator for each $g\in G$.\i{unitary 
representation}\i{representation!unitary}\\

In quantum mechanics the notion of symmetry as a transformation that satisfies (\ref{unit}) leads to a key 
subtlety. Let us call $\cT[f]$ the unitary operator that represents a symmetry transformation $f$ belonging 
to some group $G$. Then, because a quantum state is really an 
equivalence class (\ref{ray}) of vectors 
in $\sH$, there is no need to require $\cT$ to be a homomorphism; rather, all we need is that the ray 
of $\cT[f]\cdot\cT[g]\cdot\Phi$ coincides with that of $\cT[f\cdot g]\cdot\Phi$ 
(for all $f,g\in G$ and any $\Phi\in\sH$). Accordingly, $\cT$ must really be a unitary representation \it{up 
to a phase},
\be
\cT[f]\cdot\cT[g]
=
e^{i\sfC(f,g)}\,\cT[f\cdot g]
\qquad\text{for }f,g\in G,
\label{cent}
\ee
where $\sfC$ is some real function on $G\times G$. In more abstract terms, $\cT$ must define a group action 
on 
the projective space $\PP\sH$, which is to say that the map
\be
[\cT]:G\rightarrow\text{GL}(\sH)/\CC^*:f\mapsto\big[\cT[f]\big]
\label{ppro}
\ee
is a homomorphism. Here $\text{GL}(\sH)/\CC^*$ is the projective group\i{projective group} of $\sH$, i.e. the 
quotient of 
the linear group of $\sH$ by its normal subgroup consisting of multiples of the identity. For any operator 
$\cO$ in $\text{GL}(\sH)$, the symbol $[\cO]$ denotes its class in the projective 
group. Throughout this thesis, any map $\cT$ satisfying this property will be called a \it{projective 
representation}.\i{projective representation}\i{representation!projective} In quantum mechanics, symmetries 
are represented 
by \it{unitary} projective representations, i.e.~projective representations whose operators are unitary.\\

From now 
on, if we wish to stress 
that a 
representation is \it{not} projective, we will call it \it{exact}.\i{exact 
representation} Quantum mechanics tells us 
that exact representations are overrated: the truly important ones are generally projective. This seemingly 
anecdotal observation is at the core of the richest aspects of the representation theory of the Virasoro 
algebra, and it will also play a key role for BMS$_3$ particles. For 
instance, all interesting two-dimensional conformal field theories are such that the conformal group is 
represented projectively in their Hilbert space, and how exactly this phenomenon takes place is measured 
by the central charge. For this reason, this whole chapter is devoted to
the various ways in which projective effects occur; they are accounted for by group and Lie 
algebra cohomology.

\paragraph{Remark.} Since we are focussing on Lie groups, the representations of interest are 
\it{continuous}\i{continuous representation}\i{representation!continuous} 
in the sense that the map $G\times\sH\rightarrow\sH:(f,\Psi)\mapsto\cT[f]\cdot\Psi$ is continuous. From now 
on it is understood that all representations are continuous.

\subsection{Central extensions}
\label{centext}

The function $\sfC$ appearing in (\ref{cent}) is not completely arbitrary. Indeed, the product (\ref{cent}) 
must be associative in the sense that 
$\cT[f]\cdot\left(\cT[g]\cdot\cT[h]\right)
=
\left(\cT[f]\cdot\cT[g]\right)\cdot\cT[h]$ for all group elements $f,g,h$, so that\i{cocycle condition}
\be
\sfC(f,gh)+\sfC(g,h)
=
\sfC(fg,h)+\sfC(f,g)
\quad\text{for all }f,g,h\in G.
\label{cocy}
\ee
Any function $\sfC:G\times G\rightarrow\RR$ satisfying this requirement is known as a (real) 
\it{two-cocycle}, 
and the 
condition itself is known as the \it{cocycle condition}. Given any 
such function one can define a new group
\be
\hG\equiv G\times\RR
\label{gcen}
\ee
whose elements are pairs $(f,\lambda)$, endowed with a group operation\i{central extension}\i{group!central 
extension}
\be
\boxed{\Big.
(f,\lambda)\cdot(g,\mu)
=
\big(f\cdot g,\lambda+\mu+\sfC(f,g)\big).
}
\label{groupop}
\ee
The group (\ref{gcen}) is called a \it{central extension} of the group $G$.\i{central extension!of group} 
We will study this notion in much greater detail in section \ref{gcoho}. For now let us only work out 
the basic consequences of this structure and its relation to representation theory.

\subsubsection*{Projective versus exact representations}

Property (\ref{cent}) says that $\cT$ is an exact unitary representation of the centrally extended 
group (\ref{gcen}), provided one represents the pair $(f,\lambda)$ by $e^{i\lambda}\cT[f]$. In other words, 
exact representations 
are not overrated after all: we may view any projective representation of $G$ as an exact
(i.e.~non-projective) representation of a central extension $\hG$ of $G$, and the problem of classifying 
projective unitary representations of $G$ boils down to that of classifying \it{exact} unitary 
representations of its central extensions.\\

The question then is whether $G$ 
admits central extensions to begin with. For any group, an obvious type of central 
extension always exists. Namely, suppose $\sfK$ is a real function on $G$ and define $\sfC:G\times 
G\rightarrow\RR$ by
\be
\sfC(f,g)\equiv \sfK(fg)-\sfK(f)-\sfK(g).
\label{triv}
\ee
This automatically satisfies condition (\ref{cocy}). A two-cocycle of that form is said to 
be \it{trivial}.\i{central extension!trivial}\i{trivial central extension} In 
particular, if the cocycle in (\ref{cent}) is trivial, it can be 
absorbed by defining $\tilde\cT[f]\equiv e^{i\sfK(f)}\cT[f]$, which is an exact representation of $G$. Thus, 
what we wish to know is not quite whether $G$ admits 
two-cocycles at all (since trivial ones are always available), but rather whether it admits \it{non-trivial} 
two-cocycles. If yes, it admits genuine projective representations, whose phases cannot be absorbed by a 
mere redefinition.\\

This question leads to group (and Lie algebra) cohomology, studied in detail in sections \ref{s1.2} and 
\ref{gcoho}. 
For now we simply 
point out that central extensions may arise via two distinct mechanisms. The first is \it{algebraic} in that 
it follows from the local group structure of $G$, or equivalently from the commutation relations of its Lie 
algebra. In short, in some cases, the Lie algebra $\mg$ of $G$ can be enlarged into a bigger algebra 
$\hmg$ which contains extra generators commuting with those of $\mg$ (see eq.\ (\ref{tatb}) 
below). The group corresponding to this enlarged algebra then is a central extension of $G$. The second 
mechanism is \it{topological} in the sense 
that it is due to the global structure of $G$. We now describe this topological mechanism in some more detail.

\subsection{Topological central extensions}
\label{susefamitop}

If 
the group $G$ is not simply connected\i{central extension!topological} (i.e.\ its fundamental group is 
non-trivial), there exist closed 
paths 
in $G$ that cannot be continuously deformed into a point.\i{simply connected}\i{group!simply connected} Let 
$\gamma:[0,1]\rightarrow G$ be such a path, starting and ending at some group element $f$ so that 
$\gamma(0)=\gamma(1)=f$. Suppose we are given a (continuous) projective unitary representation $\cT$ of $G$, 
and consider the path
\be
\cT\circ\gamma:[0,1]\rightarrow\text{GL}(\sH):t\mapsto\cT[\gamma(t)]
\nn
\ee
in the space of unitary operators on $\sH$. Since $\cT$ is projective, the fact that $\gamma$ is a closed 
path does \it{not} imply that $\cT\circ\gamma$ is closed: in general
$\cT[\gamma(0)]$ and $\cT[\gamma(1)]$ differ by a $\gamma$-dependent 
phase, $\cT[\gamma(1)]=e^{i\phi(\gamma)}\cT[\gamma(0)]$.\\

Owing to the fact that the map $\cT$ is continuous, the phase $\phi(\gamma)$ only depends
on the homotopy class of $\gamma$. In 
addition, if $\gamma_1$ and $\gamma_2$ are two closed paths starting at $f$, we can concatenate them into a 
single path $\gamma_1\cdot\gamma_2$ (which is $\gamma_1$ at double speed followed by $\gamma_2$ at 
double speed); the phase $\phi$ must be compatible with this operation in the sense that 
$e^{i\phi(\gamma_1)}\cdot e^{i\phi(\gamma_2)}=e^{i\phi(\gamma_1\cdot\gamma_2)}$.\i{topological central 
extension}\i{central extension!topological} Thus, 
any one-dimensional unitary representation of the fundamental group of $G$, multiplying an exact 
unitary representation of $G$, produces a projective unitary representation of $G$.\\

This is the topological notion 
of central extensions that we wanted to exhibit: if $G$ is multiply connected,\i{multiply 
connected} it admits genuine 
projective representations (whose phases cannot be removed by redefinitions) due to one-dimensional unitary 
representations of its fundamental group.\footnote{Beware: a manifold being \it{multiply connected} means 
that it has a non-trivial fundamental group, and \it{not} that it has several connected components.} 
Projective representations of that type may 
equivalently be seen as exact representations of the universal cover $\widetilde{G}$ of $G$, which 
is the unique connected and simply connected group locally isomorphic to $G$.\i{universal 
cover}

\paragraph{Remark.}\i{fundamental group!of Lie group} One might be worried by the fact that only 
\it{one-dimensional} unitary representations of the fundamental group are allowed to appear in this 
construction. Indeed, if the fundamental group was non-Abelian, it would generally admit 
no non-trivial one-dimensional unitary representation. Fortunately, it turns out that the 
fundamental group of any finite-dimensional Lie group is a discrete commutative group, whose 
irreducible unitary representations are necessarily one-dimensional.

\subsubsection*{Rotations and anyons}

\begin{figure}[t]
\centering
\includegraphics[width=0.60\textwidth]{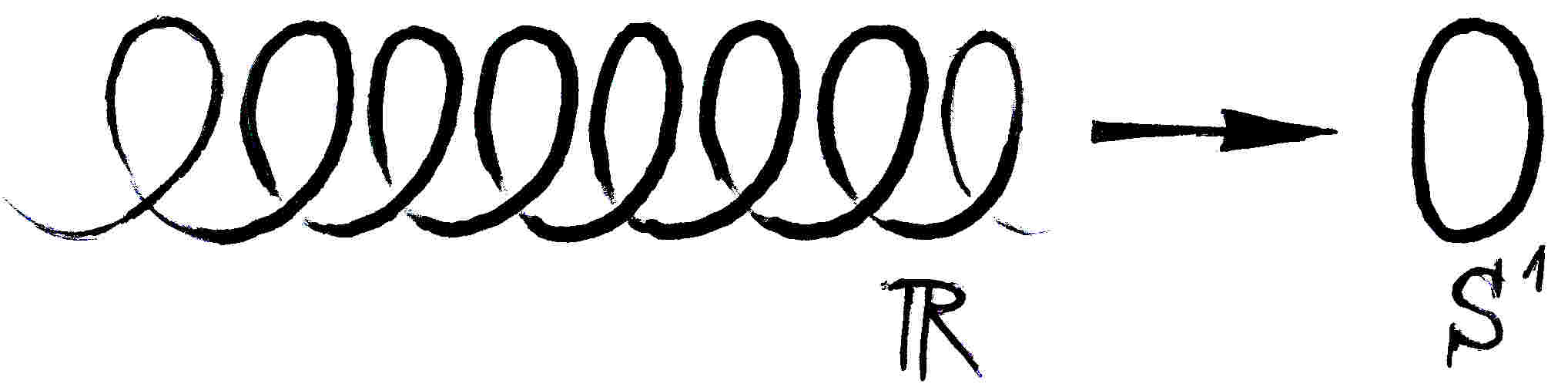}
\caption{The group $\un$ is diffeomorphic to a circle $S^1$, whose universal cover is the real line 
$\RR$.\i{universal cover!of S1@of $S^1$}\i{S1@$S^1$ (circle)!universal cover} 
The projection $\RR\rightarrow S^1\cong\RR/\ZZ$ is obtained by identifying points of $\RR$ that differ by 
some periodicity, typically $\theta\sim\theta+2\pi$. In particular, paths in $\RR$ 
which are not closed may be projected on closed paths in $S^1$. As an application we can 
picture topological projective representations: if $\cT$ is projective and if $\gamma$ is a closed path in 
the 
circle, the sequence $\cT[\gamma(t)]$ may not be a 
closed path in the space of operators.\label{figunicov}}
\end{figure}

The simplest example of topological projective representations arises with the group $\text{U}(1)$. The 
latter is diffeomorphic to a circle and has a fundamental group isomorphic to $\ZZ$ (see fig.\ 
\ref{figunicov}). Any 
exact irreducible, unitary representation of $\text{U}(1)$ takes the form\i{U1@$\un$}
\be
\cT:\text{U}(1)\rightarrow\CC^*:
\theta\mapsto e^{is\theta}
\label{tu1}
\ee
where $\theta$ is identified with $\theta+2\pi$, as a consequence of which the ``spin'' $s$ is an integer. 
For example, when $s=2$, a rotation by $\theta=\pi$ is represented by the identity. (We will 
see in section \ref{sePoTri} that the label 
$s$ actually \it{is} the spin of a particle in certain representations of the Poincar\'e groups.) But there 
is a subtlety: $\text{U}(1)$ is multiply connected and admits topological projective 
representations, which from the viewpoint of quantum mechanics are just as acceptable as exact ones. For 
example, the map (\ref{tu1}) with $s=1/2$ definitely isn't an exact representation because a 
full rotation by $2\pi$ is now represented by an inversion, 
$\cT[2\pi]=e^{i\pi}=-1$.\i{projective representation!of U1@of $\un$} Nevertheless, in 
quantum mechanics, the vectors $\Psi$ and $\cT[2\pi]\cdot\Psi$ define the same state by virtue of the 
identification (\ref{ia}), so in this sense $\cT[2\pi]$ acts as an ``almost-identity'' operator. More 
generally, formula (\ref{tu1}) is a projective representation of $\text{U}(1)$ for \it{any} real value of the 
spin $s$.\\

The example just described occurs in Nature. Indeed, fermions 
\i{fermion} provide a well-known example of projective representations, as already suggested above by the 
case $s=1/2$. By the 
spin-statistics theorem,\i{spin-statistics} all fermions have half-integer spins, and therefore transform 
according to a projective representation of the Lorentz group. The latter is multiply connected (its 
fundamental group is $\ZZ_2$), which is why it admits
projective representations in the first place. We will return to the representation theory of the Lorentz 
group (as a subgroup of Poincar\'e) 
in much greater detail in section \ref{relagroup}. In the cases where arbitrary real values of spin are 
allowed by quantum mechanics, as for example in three space-time dimensions, the particles whose spin is 
neither an integer nor a half-integer are known as \it{anyons}.\i{anyon} We will encounter this phenomenon in 
section \ref{sebmspar} when dealing 
with 
BMS$_3$ particles.

\subsection{Classifying projective representations}

Given a group $G$, suppose we wish to find all its projective unitary representations. The above 
considerations provide an algorithm that allows us, in principle, to solve that problem:\i{projective 
representation!classification}\i{classification!of projective reps}
\begin{itemize}
\item First find the universal cover $\widetilde{G}$ of $G$ to take care of topological central 
extensions.\label{pgm}
\item Then find the most general central extension $\widehat{\widetilde{G}}$ of $\widetilde{G}$ in order to 
take care of differentiable central extensions. (We will deal with the actual definition of these
extensions in the next section.)
\item Finally, consider an \it{exact} unitary representation of $\widehat{\widetilde{G}}$; any projective 
unitary representation of $G$ may be seen as a representation of that type.
\end{itemize}
Thus we now have a systematic procedure allowing us to build 
arbitrary projective unitary representations of symmetry groups in quantum mechanics. We will apply it later 
to the Virasoro algebra (section \ref{sevirep}) and the 
BMS$_3$ group (section \ref{sebmspar}), where central extensions play a crucial role.

\section{Lie algebra cohomology}
\label{s1.2}

This section is devoted to a thorough investigation of the concept of central extensions at the Lie-algebraic 
level. In 
fact, we shall describe the more general framework of Lie algebra cohomology and we will show how 
statements on algebraic central extensions can be recast in that language. The group-theoretic analogue of 
this construction is relegated to section \ref{gcoho}.

\subsection{Cohomology}
\label{licoh}

Let $\mg$ be a Lie algebra with Lie bracket $[\cdot,\cdot]$. We recall that a \it{representation} of $\mg$ in 
a vector space $\VV$ is a linear map $\sT:\mg\rightarrow\text{End}(\VV)$ such that 
$\sT[X]\circ\sT[Y]-\sT[Y]\circ\sT[X]=\sT\big[[X,Y]\big]$ for all Lie algebra elements 
$X,Y$.\footnote{Throughout this thesis the elements of a Lie algebra $\mg$ will be denoted as $X$, $Y$, etc. 
Representations of Lie algebras will be denoted by script capital letters such as $\sR$, $\sS$, 
$\sT$.}\i{representation!of Lie algebra}\i{Lie algebra!representation}

\paragraph{Definition.} Let $k$ be a non-negative integer, $\sT$ a representation of $\mg$ in $\VV$. Then a 
$\VV$-valued \it{$k$-cochain} on $\mg$ is a 
continuous, multilinear, completely antisymmetric map\i{cochain!on Lie algebra}\footnote{Cochains on Lie 
algebras will be denoted by lowercase sans serif letters such as $\sfc$, $\sfs$, etc.}
\be
\sfc:
\underbrace{\mg\times\cdots\times\mg}_{k\text{ times}}\rightarrow\VV:
(X_1,...,X_k)\mapsto\sfc(X_1,...,X_k).
\label{coch}
\ee
\vspace{.1cm}

In other words, a $\VV$-valued $k$-cochain on $\mg$ is a $k$-form on $\mg$ with values in $\VV$; note that 
$0\leq k\leq\text{dim}(\mg)$. A zero-cochain on $\mg$ is a vector in $\VV$ while a $\dim(\mg)$-cochain is a 
volume form on $\mg$. We denote the space of 
$\VV$-valued $k$-cochains on $\mg$ by $\cC^k(\mg,\VV)$ and we define the associated cochain 
complex $\cC^*(\mg,\VV)\equiv\oplus_{k=0}^{\text{dim}(\mg)}\cC^k(\mg,\VV)$. The latter is 
sometimes called the \it{Chevalley-Eilenberg complex}.

\paragraph{Definition.} The \it{Chevalley-Eilenberg differential}\i{Chevalley-Eilenberg 
differential}\i{differential!for Lie algebra} 
$\sfd:\cC^*(\mg,\VV)\rightarrow\cC^*(\mg,\VV)$ is defined by 
$\text{dim}(\mg)$ linear maps
\be
\sfd_k:\cC^k(\mg,\VV)\rightarrow\cC^{k+1}(\mg,\VV):
\sfc\mapsto \sfd_k\sfc
\nn
\ee
where $k$ runs from $0$ to $\text{dim}(\mg)-1$ and the $(k+1)$-cochain $\sfd_k\sfc$ is given by
\begin{align}
(\sfd_k\sfc)(X_1,...,X_{k+1})\equiv
&
\sum_{1\leq i<j\leq k+1}
(-1)^{i+j-1}
\sfc\big([X_i,X_j],X_1,...,\widehat{X_i},...,\widehat{X_j},...,X_{k+1}\big)\nn\\
\label{chevd}
& +
\sum_{1\leq i\leq k+1}(-1)^i\sT[X_i]\cdot\sfc\big(X_1,...,\widehat{X_i},...,X_{k+1}\big)
\end{align}
for all $X_1,...,X_{k+1}$ in $\mg$; the hat denotes omission. Note that the representation $\sT$ 
of $\mg$ in $\VV$ appears explicitly in this definition. In 
particular, when $\sT$ is trivial, formula (\ref{chevd}) simplifies since its 
last 
line disappears.

\subsubsection*{Cocycles and coboundaries}

Using the fact that $\sT$ is a representation,
one can verify that the Chevalley-Eilenberg differential (\ref{chevd}) is nilpotent:\i{nilpotent}
\be
\sfd_k\circ\sfd_{k-1}
=0
\qquad\forall\,k=0,...,\text{dim}(\mg)
\label{dd}
\ee
where it is understood that the ``extreme differentials'' are $\sfd_{-1}:0\rightarrow\VV:0\mapsto0$ 
and $\sfd_{\text{dim}\,\mg}:\cC^{\text{dim}\,\mg}(\mg,\VV)\rightarrow0:\sfc\mapsto0$. Accordingly, one adapts 
the standard terminology of differential forms to cochains on a Lie algebra: a 
\it{$k$-cocycle}\i{cocycle!on Lie algebra} is a $k$-cochain $\sfc$ such that $\sfd_k\sfc=0$; a 
\it{$k$-coboundary} is a $k$-cochain $\sfc$ of the form $\sfc=\sfd_{k-1}\sfb$, where $\sfb$ is 
some 
$(k-1)$-cochain. By virtue of property (\ref{dd}), one has $\text{Im}(\sfd_{k-1})\subseteq\text{Ker}(\sfd_k)$ 
for each $k$ (any coboundary is a cocycle). One can 
therefore define the \it{$k^{\text{th}}$ cohomology space}\i{cohomology!of Lie 
algebras} of $\mg$ with coefficients in $\VV$ as the quotient of the space of $k$-cocycles by the space of 
$k$-coboundaries:
\be
\cH^k(\mg,\VV)\equiv\text{Ker}(\sfd_k)/\text{Im}(\sfd_{k-1}).
\label{coh}
\ee
A $k$-cocycle is said to be \it{trivial}\i{trivial cocycle}\i{cocycle!trivial} if its equivalence class 
vanishes in $\cH^k$, i.e.\ if the cocycle 
is a coboundary; the cocycle is \it{non-trivial} otherwise. When $\VV=\RR$ 
with $\sT$ the trivial representation of $\mg$, we write $\cH^k(\mg,\RR)\equiv\cH^k(\mg)$.\\

Isomorphic Lie algebras have the same cohomology for any choice of the representation $\sT$. Thus,
cohomology is a way to associate invariants with Lie algebras: if two algebras have different cohomology 
spaces, then they cannot be 
isomorphic. This is analogous to, say, de Rham cohomology\i{de Rham cohomology} in differential geometry, as 
manifolds with 
different de Rham cohomologies cannot be diffeomorphic.\i{de Rham cohomology}

\subsubsection*{Low degree cohomologies}

There is a simple interpretation for the lowest cohomology spaces. For example, zero-cocycles are vectors 
$v\in\VV$ that are invariant under $\mg$ in the sense that\i{invariant vector}
\be
\sT[X]\cdot v=0
\qquad
\text{for all }X\in\mg\,,
\label{invv}
\ee
so the zeroth cohomology space of $\mg$ classifies the invariants of the representation $\sT$. Similarly, 
one-cocycles are known as \it{derivations}\i{derivation!of Lie algebra} of $\mg$ and are classified by the 
first cohomology space 
$\cH^1(\mg,\VV)$. In the particular case where $\sT$ is trivial and $\VV=\RR$, a one-cocycle is a linear map 
$\sfc:\mg\rightarrow\RR$ such that $\sfc([X,Y])=0$ for all Lie algebra elements $X,Y$. Hence the 
first real cohomology space of $\mg$ can be written as
\be
\cH^1(\mg)
\cong
\mg/[\mg,\mg]\,,
\label{hiperfecto}
\ee
which motivates the 
following definition:

\paragraph{Definition.} A Lie algebra $\mg$ is \it{perfect}\i{perfect Lie algebra}\i{Lie algebra!perfect} if 
$\mg=[\mg,\mg]$, i.e.\ 
if 
any Lie algebra element can be written as the bracket of two other elements.\\

It follows from (\ref{hiperfecto}) that $\mg$ is perfect if and only if $\cH^1(\mg)$ vanishes. We will use 
this property in section 
\ref{centalg} when defining central extensions.\\

By the definitions above, a two-cochain is 
an antisymmetric map $\sfc:\mg\times\mg\rightarrow\VV$. It is a 
coboundary if
\be
\sfc(X,Y)
\refeq{chevd}
(\sfd_1\sfk)(X,Y)=\sfk([X,Y])-\sT[X]\cdot \sfk(Y)+\sT[Y]\cdot \sfk(X)
\label{trivv}
\ee
for some one-cochain $\sfk$; and it is a cocycle if
\be
\begin{split}
& \sfc([X,Y],Z)+\sfc([Y,Z],X)+\sfc([Z,X],Y)=\\
& =\sT[X]\cdot \sfc(Y,Z)+\sT[Y]\cdot \sfc(Z,X)+\sT[Z]\cdot \sfc(X,Y).
\end{split}
\label{2coc}
\ee
As we shall see shortly, when $\sT$ is trivial, a two-cocycle defines a \it{central 
extension} of $\mg$. Thus the second cohomology of $\mg$ classifies its extensions. More generally, 
cohomology may be seen as a measure of flexibility: Lie algebras with high-dimensional cohomology groups can 
be ``deformed'' in many inequivalent ways; by contrast, Lie algebras with trivial cohomology are ``rigid'' in 
the sense that any deformation is equivalent to no deformation at all.

\paragraph{Remark.} Here we have been using the word ``deformation'' in a vague way, but there is an 
exact definition of the notion of deformations. Namely, a (true) \it{deformation}\i{deformation of Lie 
algebra}\i{Lie algebra!deformation} of a Lie 
algebra $\mg$ is 
a Lie algebra $\tilde\mg$ that coincides with $\mg$ as a vector space, but whose brackets are
\be
\tilde[X,Y\tilde]
=
[X,Y]+\sfc(X,Y)
\label{CAXY}
\ee
where $[\cdot,\cdot]$ is the bracket in $\mg$ while $\sfc$ is a $\mg$-valued two-cocycle on 
$\mg$,\footnote{It is understood that the relevant representation of $\mg$ in this case is the adjoint, 
$\sT[X]\cdot Y\equiv[X,Y]$.} such that the image of $\sfc$ belongs to its kernel. The latter condition means 
that $\sfc\big(X,\sfc(Y,Z)\big)=0$ for all Lie algebra elements $X,Y,Z$; together with the 
fact that $\sfc$ is a cocycle, this ensures that (\ref{CAXY}) is a Lie bracket.

\subsubsection*{Examples}

For finite-dimensional semi-simple Lie algebras, cohomology is trivial:

\paragraph{Whitehead's lemma.}\i{Whitehead's lemma}\label{white} Let $\mg$ be a finite-dimensional 
semi-simple Lie algebra, $\sT$ an irreducible, finite-dimensional representation of $\mg$ in a space $\VV$. 
Then
\be
\cH^k(\mg,\VV)=0
\quad
\text{for all $k>0$.}
\label{whitehead}
\ee
\vspace{.1cm}

Despite this result, examples of non-trivial cohomologies do exist in physics. For instance, 
let $\sfc$ be an arbitrary non-vanishing antisymmetric bilinear form on $\RR^2$, and view the latter as an 
Abelian Lie algebra. \label{heis} Then $\sfc$ defines a non-trivial, real-valued two-cocycle on $\RR^2$, so 
the real-valued second cohomology of $\RR^2$ is non-trivial; in fact one can prove that
\be
\cH^2(\RR^2)\cong\RR.
\label{hhei}
\ee
We shall see below that this property is related to the (three-dimensional) Heisenberg algebra, which is 
crucial for quantum mechanics. Other important examples of algebras with non-trivial cohomology spaces 
include the Galilei algebra (section \ref{galisec}), the Virasoro algebra (chapter \ref{c4}) and the $\bms$ 
algebra (chapter \ref{c6}).

\subsection{Central extensions}
\label{centalg}

\paragraph{Definition.} Let $\mg$ be a (real) Lie algebra and let $\sfc\in\cC^2(\mg,\RR)$ be a real 
two-cocycle on $\mg$. Then $\sfc$ defines a \it{central extension}\i{central extension!of Lie algebra} $\hmg$ 
of $\mg$, which is a Lie algebra whose underlying vector space
\be
\hmg=\mg\oplus\RR
\qquad\text{(as vector spaces)}
\label{hatg}
\ee
is endowed with the centrally extended Lie bracket
\be
\big[(X,\lambda),(Y,\mu)\big]
\equiv
\big([X,Y],\sfc(X,Y)\big).
\label{centb}
\ee
In particular, elements of $\hmg$ are pairs $(X,\lambda)$ where $X\in\mg$ and $\lambda\in\RR$, so that 
$\RR$ is an Abelian subalgebra of $\hmg$. The bracket (\ref{centb}) satisfies the Jacobi identity on 
account of the fact 
that $\sfc$ is a two-cocycle with respect to a trivial representation of $\mg$ (so that the 
right-hand side of eq.\ (\ref{2coc}) vanishes).\\

In (\ref{centb}) we displayed the definition of central extensions in 
intrinsic terms thanks to the two-cocycle $\sfc$. The same definition can be written in terms of 
Lie algebra generators: let 
$\{t_a|a=1,...,n\}$ be a basis of $\mg$ with brackets $[t_a,t_b]={f_{ab}}^c\,t_c$. Then a central 
extension $\hmg$ of $\mg$ is a Lie algebra generated by the basis elements $T_a\equiv(t_a,0)$ together with a 
central element $\cZ=(0,1)$, whose Lie brackets read
\be
[T_a,T_b]={f_{ab}}^c\,T_c+c_{ab}\,\cZ\,
\label{tatb}
\ee
where $c_{ab}=\sfc(t_a,t_b)$, while all brackets with $\cZ$ vanish. The cocycle 
condition on $\sfc$ then becomes the requirement
\be
{f_{ab}}^dc_{dc}+{f_{bc}}^dc_{da}+{f_{ca}}^dc_{db}=0
\nn
\ee
for the coefficients $c_{ab}$. Note that this construction can be readily generalized to multiple central 
extensions $\hmg=\mg\oplus\RR^N$, in which case there are $N$ central generators $\cZ_1,...,\cZ_N$.

\subsubsection*{Non-trivial central extensions}

When the two-cocycle $\sfc$ is trivial in the sense of Lie algebra cohomology, it takes the form 
(\ref{trivv}) in 
terms of some one-cocycle $\sfk$ and the map
\be
\mg\rightarrow\hmg:X\mapsto\big(X,\sfk(X)\big)
\label{redef}
\ee
is an injective homomorphism of Lie algebras. The central extension is then said to be 
\it{trivial}:\i{trivial central extension}\i{central extension!trivial} the 
cocycle $\sfc$ can 
be absorbed by the ``redefinition'' (\ref{redef}),\i{trivial central extension} and $\hmg$ is isomorphic 
to the direct sum $\mg\oplus\RR$ as a Lie 
algebra. By contrast, when $\sfc$ is non-trivial, it defines 
a 
non-zero element in the second cohomology space $\cH^2(\mg)$; such a two-cocycle cannot be removed by a 
mere redefinition, and the central extension is \it{non-trivial}.\\

For example, as on page 
\pageref{heis}, consider the Abelian Lie algebra $\RR^2$ and let $\sfc$ be 
a non-zero antisymmetric bilinear form on $\RR^2$. We then define the three-dimensional \it{Heisenberg 
algebra}\i{Heisenberg algebra} as the algebra $\RR^3=\RR^2\oplus\RR$ whose elements are pairs $(X,\lambda)$, 
endowed with the Lie bracket (\ref{centb}). Since $\sfc$ is non-trivial, so is the central extension. If we 
choose a basis $\{Q,P\}$ of $\RR^2$ such that 
$\sfc(Q,P)=1$ and if we call $Z$ the central element $(0,1)$, the commutation relations of the Heisenberg 
algebra take the form
\be
[Q,P]=Z.
\label{comH}
\ee
Property (\ref{hhei}) says that there is only one linearly independent central extension of $\RR^2$, 
i.e.~that Heisenberg algebras built using different (non-zero) two-cocycles $\sfc$ are mutually isomorphic. 
This can be generalized to higher dimensions: by seeing $\RR^{2n}$ as an 
Abelian Lie algebra and taking $\sfc$ an arbitrary non-zero $2n$-form on $\RR^{2n}$, the Lie algebra defined 
by 
the bracket (\ref{centb}) is the $(2n+1)$-dimensional Heisenberg algebra.

\subsubsection*{Universal central extensions}

It is important to know how many inequivalent central extensions an 
algebra may possess. This leads to the following notion:

\paragraph{Definition.} A central extension $\hmg$ of $\mg$ is \it{universal}\i{universal central 
extension}\i{central extension!universal} if, for any other central extension $\hmg'$ of $\mg$, there exists 
a unique isomorphism of 
Lie algebras $\hmg'\cong\hmg$.\\

As it turns out, any perfect Lie algebra admits a universal central extension.\i{perfect 
Lie algebra!universal central extension} By virtue of (\ref{hiperfecto}), this is to say that any algebra 
such that $\cH^1(\mg)=0$ admits a universal central extension. For example we will see in chapter 
\ref{c4} that the Virasoro algebra is the universal central extension of the Lie algebra of vector fields on 
the circle.

\newpage

\section{Group cohomology}
\label{gcoho}

This section is devoted to the group-theoretic analogue of the considerations of the previous pages. We start 
by discussing generalities on group cohomology before focussing on central extensions of groups.

\subsection{Cohomology}

Let $G$ be a Lie group, $\cT:G\rightarrow\text{GL}(\VV)$ a representation of $G$ in a vector space $\VV$.

\paragraph{Definition.} Let $k\geq0$ be an integer. A $\VV$-valued \it{$k$-cochain}\i{cochain!on group} on 
$G$ is a smooth map\footnote{Cochains on a group will be denoted by capital sans serif symbols 
such as $\sfC$, $\sfS$, etc.}
\be
\sfC:
\underbrace{G\times\cdots\times G}_{k\text{ times}}\rightarrow\VV:
(g_1,...,g_k)\mapsto\sfC(g_1,...,g_k).
\label{gcoc}
\ee
\vspace{.1cm}

Note that, in contrast to the Lie-algebraic definition (\ref{coch}), there is no restriction on $k$. The new 
ingredient in the 
group-theoretic context is the requirement that the map (\ref{gcoc}) be smooth. As in the case of Lie 
algebras, we denote by $\cC^k(G,\VV)$ the vector space of 
$\VV$-valued $k$-cochains on $G$ and we let $\cC^*(G,\VV)=\oplus_{k=0}^{+\infty}\cC^k(G,\VV)$ be the 
associated cochain complex. The space of zero-cochains is just $\VV$.

\paragraph{Definition.} The \it{differential}\i{differential!for group}\i{Chevalley-Eilenberg differential} 
$\sfd:\cC^*(G,\VV)\rightarrow\cC^*(G,\VV)$ is defined by the maps
\be
\sfd_k:\cC^k(G,\VV)\rightarrow\cC^{k+1}(G,\VV):
\sfC\mapsto \sfd_k\sfC
\nn
\ee
where $k\in\NN$ and the $(k+1)$-cochain $\sfd_k\sfC$ is given by
\be
\begin{split}
(\sfd_k\sfC)(g_1,...,g_{k+1})\equiv
& \;\cT[g_1]\cdot \sfC(g_2,...,g_{k+1})+(-1)^{k+1}\,\sfC(g_1,...,g_k)\\
& +\sum_{i=1}^k(-1)^i\,\sfC(g_1,...,g_ig_{i+1},...,g_{k+1})
\end{split}
\label{gd}
\ee
for all $g_1,...,g_{k+1}$ in $G$.\\

The differential (\ref{gd})
satisfies the key property (\ref{dd}), so the usual machinery of homological 
algebra applies: one defines a \it{$k$-cocycle}\i{cocycle!on group} as a closed $k$-cochain, that is, a 
cochain $\sfC$ such that $\sfd_k\sfC=0$. One also defines a \it{$k$-coboundary} to 
be an 
exact $k$-cochain, i.e.\ one that can be written as the differential of a $(k-1)$-cochain. As before any 
coboundary is trivially a cocycle, so one defines the \it{$k^{\text{th}}$ cohomology 
space}\i{cohomology!of groups} of $G$ with values in $\VV$ as the quotient of 
the space of $k$-cocycles by the space of $k$-coboundaries:
\be
\cH^k(G,\VV)\equiv\text{Ker}(\sfd_k)/\text{Im}(\sfd_{k-1}).
\nn
\ee
A $k$-cocycle is \it{trivial}\i{trivial cocycle}\i{cocycle!trivial} if its class in $\cH^k(G,\VV)$ vanishes; 
it is non-trivial otherwise. 
When $\VV=\RR$ with $\cT$ the trivial representation, we write $\cH^k(G,\RR)\equiv\cH^k(G)$.

\subsubsection*{Interpretation}

As in the case of Lie algebras, cohomology spaces are invariants that measure the flexibility of a group 
structure; isomorphic Lie groups have the same cohomology.\i{cohomology!interpretation} This interpretation 
is simplest to illustrate with the cohomology 
spaces of lowest degree.\\

A $\VV$-valued zero-cocycle on $G$ is a vector $v\in\VV$ such that 
$(\sfd_0v)(f)=\cT[f]\cdot v-v=0$ for any group element $f$. Accordingly, the zeroth cohomology space of $G$ 
classifies vectors 
$v\in\VV$ that are left invariant\i{invariant vector} by $G$. This is the group-theoretic analogue of 
(\ref{invv}).\\

A $\VV$-valued one-cocycle is a (smooth) map $\sfS:G\rightarrow\VV$ satisfying the property\i{one-cocycle}
\be
\sfS(fg)
=
\cT[f]\cdot\sfS(g)+\sfS(f)\qquad\forall\,f,g\in G.
\label{1coc}
\ee
Given a one-cocycle $\sfS$, one defines the associated \it{affine module}\i{affine module} as 
the space $\VV\oplus\RR$ acted upon by the following representation $\widehat\cT$ of $G$:
\be
\widehat\cT[f]\cdot(v,\lambda)
\equiv
\big(\cT[f]\cdot v+\lambda\,\sfS(f),\lambda\big).
\label{affi}
\ee
The cocycle condition (\ref{1coc}) ensures that $\widehat\cT$ is indeed a representation. In addition one can 
show that affine modules defined using different one-cocycles are equivalent if (and only if) their cocycles 
differ by a 
coboundary. Thus $\cH^1(G,\VV)$ classifies affine $G$-modules based on $\VV$. For example, in section 
\ref{seschwa} we 
will see that the Schwarzian derivative is a one-cocycle on the group of diffeomorphisms 
of the circle; this is why we denote the cocycle in (\ref{affi}) by $\sfS$. The corresponding affine module 
will be the coadjoint representation of the 
Virasoro group and the parameter $\lambda$ left invariant by (\ref{affi}) will be a Virasoro central charge. 
More generally one can think of the term $\lambda\sfS[f]$ in (\ref{affi}) as an anomaly that adds an 
inhomogeneous term to the otherwise homogeneous transformation law of $v$ under $G$.\\

Two-cocycles lead to the notion of group extensions; in particular, when $\VV=\RR$ with $\cT$ the trivial 
representation, $\cH^2(G)$ classifies central extensions of $G$. Indeed, when $\sfC$ is a real two-cocycle on 
$G$, the requirement $\sfd_2\sfC=0$ becomes the cocycle condition (\ref{cocy}); the central extension is 
trivial when $\sfC$ is a coboundary, i.e.\ if it takes the form (\ref{triv}) for some one-cochain 
$\sfK$. We will return to central extensions of groups in section \ref{cgroup}.

\subsubsection*{Relation to Lie algebra cohomology}

One may ask how group and Lie algebra cohomology are related. The following result provides a first answer:

\paragraph{Proposition.} Let $G$ be a Lie group, $\mg$ its Lie algebra. Let $\VV$ be a vector space, $\cT$ a 
smooth representation of $G$ in $\VV$, and $\sT$ the representation of $\mg$ corresponding to $\cT$ by 
differentiation. Then, for any non-negative integer $k$, there is a homomorphism\i{differentiation of 
cocycles}\i{cocycle!differentiation}
\be
\cH^k(G,\VV)\rightarrow\cH^k(\mg,\VV):[\sfC]\mapsto\left[\delta\sfC\right]
\label{HACCOK}
\ee
given by\i{infinitesimal cocycle}
\be
\delta\sfC(X_1,...,X_k)
\equiv
\frac{\der^k}{\der t_1...\der t_k}
\left.
\left[\sum_{1\leq i_1<...< i_k\leq k}
\epsilon_{i_1...i_k}\sfC\big(
e^{t_{i_1}X_{i_1}},...,e^{t_{i_k}X_{i_k}}
\big)
\right]
\right|_{t_1=0,...,t_k=0}
\nn
\ee
for all $X_1,...,X_k$ in $\mg$, with $e^X$ the exponential of $X\in\mg$ and $\epsilon_{i_1...i_k}$ 
the Levi-Civita symbol with $k$ indices (and $\epsilon_{12...k}\equiv+1$). For $k=2$ this can be rewritten as
\be
\delta\sfC(X,Y)
=
\frac{\der^2}{\der t\,\der s}
\left.\Big[
\sfC\left(e^{tX},e^{sY}\right)-\sfC\left(e^{sY},e^{tX}\right)
\Big]\right|_{t=0,\,s=0}.
\label{diffcc}
\ee
The fact that (\ref{HACCOK}) is a homomorphism ensures that, if $\delta\sfC$ is a non-trivial 
cocycle, then 
$\sfC$ itself is non-trivial. The converse is not true since the map need not be injective: a 
non-trivial cocycle $\sfC$ may well be such that $\delta\sfC$ is trivial.\\

We will use formula (\ref{diffcc}) in section \ref{sedicomo} to relate the Virasoro 
algebra to the Virasoro group. The key point here is that any 
differentiable group cocycle $\sfC$ admits an algebraic analogue $\delta\sfC$. The converse 
problem is to start from a Lie algebra cocycle, say $\sfc$, and ask whether there exists 
a 
group cocycle whose differential is $\sfc$. This is the problem of integrating Lie algebra cocycles to group 
cocycles,\i{integration of cocycles}\i{cocycle!integration} and it is generally much more complicated than 
differentiation. However, for ``sufficiently 
connected'' Lie groups, the Van Est theorem\i{Van Est theorem} states that integration is trivial because 
group and Lie algebra cohomologies coincide (see e.g.\ \cite{guieu2007algebre}). In particular, when the 
universal cover 
of a group is homotopic to a point, the cohomology of the universal cover coincides with that of the Lie 
algebra.

\subsection{Central extensions}
\label{cgroup}

Here we return in more detail to the notion of centrally extended groups, already outlined around 
(\ref{groupop}). For simplicity we deal only with simply connected groups, so as to avoid the 
topological complications of section \ref{susefamitop}. Including these subtleties would lead to a definition 
of 
central extensions somewhat more general (see e.g.\ \cite{guieu2007algebre}) than the one given here:

\paragraph{Definition.} Let $G$ be a Lie group, $\sfC$ a real two-cocycle on $G$. Then the associated 
\it{centrally extended group}\i{central extension!of group}\i{group!central extension} $\hG$ is 
topologically a product $G\times\RR$ whose elements are pairs 
$(f,\lambda)$ with $f\in G$ and $\lambda\in\RR$, endowed with a group operation (\ref{groupop}).\\

It is 
straightforward to generalize this definition to the case where $\RR$ is replaced by an arbitrary 
(additive) Abelian 
group such as $\RR^N$.

\subsubsection*{Non-trivial central extensions}

As in the Lie-algebraic case, a central extension of $G$ is \it{trivial}\i{trivial central 
extension}\i{central extension!trivial} if the 
two-cocycle $\sfC$ defining the group operation (\ref{groupop}) is a coboundary (\ref{triv}) for some 
one-cochain $\sfK$. Then the map $G\rightarrow\hG:f\mapsto\big(f,\sfK(f)\big)$
is an injective homomorphism whose Lie-algebraic analogue is (\ref{redef}), and $\hG$ is isomorphic, 
as a group, to the direct product $G\times\RR$. Thus any trivial central extension can be absorbed by a 
redefinition of the 
group, and is irrelevant as regards projective representations. By contrast, when the cohomology class of 
$\sfC$ is a non-zero vector in $\cH^2(G)$, the central extension cannot be removed by a redefinition and is 
said to be 
\it{non-trivial}.

\paragraph{Example.} Let us find the group corresponding to the $(2n+1)$-dimensional Heisenberg 
algebra.\i{Heisenberg group} Consider the Abelian additive group $G=\RR^n\times\RR^n$ (whose elements are 
pairs of column vectors $(\alpha,\beta)$) and define the \it{Heisenberg group} as
\be
\hG
\equiv
\left\{\left.
\begin{pmatrix}
1 & \alpha^t & \lambda\\
0 & \II_n & \beta\\
0 & 0 & 1
\end{pmatrix}
\right|
\alpha,\beta\in\RR^n,
\;
\lambda\in\RR
\right\}
\label{heihei}
\ee
where $\II_n$ denotes the $n\times n$ identity matrix and $\alpha^t$ is the transpose of $\alpha$. The group 
operation in $\hG$ is given by matrix 
multiplication and can be written as
\be
(\alpha,\beta,\lambda)\cdot(\alpha',\beta',\lambda')
=
\big(\alpha+\alpha',\beta+\beta',\lambda+\lambda'+\alpha^t\cdot\beta'\big)
\label{hohot}
\ee
where $\alpha^t\cdot\beta'\equiv\alpha^i\beta'^i$ is the Euclidean scalar product of $\alpha$ and $\beta'$. 
Thus the Heisenberg group is a central extension of $\RR^{2n}$ defined by the 
two-cocycle
\be
\sfC\big((\alpha,\beta),(\alpha',\beta')\big)
=
\alpha^t\cdot\beta'.
\label{hicocy}
\ee
By differentiation, one can associate with $\sfC$
a Lie algebra cocycle given by (\ref{diffcc}). For example, when $n=1$ (and writing elements of the Lie 
algebra $\RR^2$ as pairs $X=(x,y)$),
\be
\delta\sfC\big((x,y),(x',y')\big)
\stackrel{\text{(\ref{diffcc})}}{=}
\frac{\der^2}{\der t\,\der s}\left.\left(tx\cdot sy'-sx'\cdot ty\right)\right|_{t=0,\,s=0}
=
xy'-yx'.
\nn
\ee
This is a non-zero antisymmetric bilinear form on $\RR^2$, hence defining the Heisenberg algebra 
of (\ref{comH}). Note that this is an example of ``cocycle 
integration'': we have found the 
explicit group two-cocycle whose differential
defines the Heisenberg Lie algebra.

\subsubsection*{Universal central extensions}

Universal central extensions of groups can be defined exactly as for
Lie algebras. A central extension $\hG$ of $G$ is \it{universal}\i{universal central extension}\i{central 
extension!universal} if, for 
any 
other central extension $\hG'$ of $G$ by $A$, there exists a 
unique isomorphism $\hG\rightarrow\hG'$.\\

As in the algebraic case, there is a simple criterion for knowing when a group admits a universal central 
extension. A group is said to be \it{perfect}\i{perfect group}\i{group!perfect} if it coincides with the 
group of its 
commutators,\i{commutator} i.e.~if any $f\in G$ can be written as $f=ghg^{-1}h^{-1}$ for some 
$g,h\in G$. It turns out that 
any perfect group admits a universal central extension. In chapters \ref{c4} and \ref{c6} we will see that 
both $\Diff$ and $\BMS$ are perfect groups, so that their central extensions are universal.

\chapter{Induced representations}
\label{c1b}
\markboth{}{\small{\chaptername~\thechapter. Induced representations}}

In the previous chapter we learned how to deal with projective representations: given a symmetry group, we 
are 
to find its universal cover and its most general central extension. Exact representations of this central 
extension then account for all projective 
representations of the original group.
The remaining problem then is to write down explicit representations, so our goal in this chapter is to build 
Hilbert spaces of wavefunctions acted upon by a group of unitary transformations. Guided by group actions on 
homogeneous spaces, we will be led to the method of \it{induced representations}. Their basic principle is 
very simple: starting from a representation of some subgroup $H$ of a group $G$, one induces a 
representation of $G$ that acts on wavefunctions which live on the quotient space 
$G/H$.\\

Induced representations are ubiquitous in mathematics and physics:\i{ubiquitousness of induced reps}
\begin{itemize}
\item The irreducible unitary highest-weight representations of any compact, 
simple Lie group are induced from those of its maximal torus,\i{maximal torus} i.e.\ its largest Abelian 
subgroup (whose Lie algebra is the Cartan subalgebra).\i{Cartan subalgebra}
\item Highest-weight representations of $\mathfrak{sl}(2,\RR)$ and of the Virasoro algebra are induced 
from 
representations of their $\mathfrak{u}(1)$ subalgebra generated by $L_0$ (see section \ref{sevirep}).
\item All irreducible unitary representations of the Euclidean groups, the Bargmann groups, the Poincar\'e 
groups and the BMS$_3$ group are induced 
from those of their translation subgroups combined with ``little groups'' (see chapters \ref{c2bis} and 
\ref{c7}).
\end{itemize}

The plan of this chapter is as follows.
In section \ref{sewaff} we review some basics of measure theory and the ensuing construction of Hilbert 
spaces of square-integrable wavefunctions. Section \ref{seqareg} is concerned with measures on homogeneous 
spaces and introduces quasi-regular representations --- the simplest examples of 
induced representations. In section \ref{sedefirep} we display the basic formulas of induced 
representations and list some of their elementary properties. Along the way 
we define a basis of plane waves, later to be interpreted as particles with definite momentum. This basis 
is then used in section \ref{secharapou} to compute characters. Finally, section 
\ref{sysim} is devoted to systems of imprimitivity. All these notions are crucial prerequisites for 
chapter \ref{c2bis}.\\

It would be illusory to 
present a complete 
account of the rich theory of induced representations, so we refer to Barut and Raczka 
\cite{barut1986theory} or Mackey \cite{mackey1968induced} for a more thorough exposition. For some background 
on 
measure theory, see e.g.\ \cite{rudin1976principles,rudin1987real}.

\section{Wavefunctions and measures}
\label{sewaff}

Here we start with 
general considerations on measure theory before
reviewing the construction of Hilbert spaces of square-integrable wavefunctions, independently of 
group theory. We also define Radon-Nikodym derivatives and show that Hilbert spaces of wavefunctions built 
with equivalent measures are isomorphic. For the record, our approach 
will not be mathematically rigorous, and is merely intended to give a rough picture 
of the actual mathematical theory.

\subsection{Measures}

When defining a quantum-mechanical system, one of the key ingredients is a prescription 
for computing scalar products. For the spaces of wavefunctions that we wish to consider, this requires being 
able to evaluate integrals of functions on a manifold. Integration, in turn, relies on the existence of a 
measure.

\subsubsection*{Measures}

Let $\cM$ be a set. Roughly speaking, a measure is a function $\mu$ that associates a non-negative 
number 
with essentially any subset $U$ of $\cM$. That number, denoted $\mu(U)$, ``measures'' the size of 
$U$.\i{measure} Strictly speaking, not all subsets of $\cM$ can be measured: there exists a 
family of subsets of $\cM$, called ``measurable sets'', and 
only those subsets can actually be measured. The measure 
$\mu$ then is a map
\be
\mu:\left\{\text{measurable subsets of $\cM$}\right\}\rightarrow\bar\RR^+:
U\mapsto\mu(U)
\label{mimi}
\ee
where $\bar\RR^+$ denotes the set of non-negative real numbers supplemented with $+\infty$. In order to 
qualify as a measure, this map needs to satisfy certain conditions; in particular, it must be 
\it{$\sigma$-additive}:\i{sigma additivity@$\sigma$-additivity} if $U_1$, $U_2$, etc.~are 
disjoint measurable sets, then
\be
\mu\left(\bigcup_{i=1}^{+\infty}U_i\right)
=
\sum_{i=1}^{+\infty}\mu(U_i)
\qquad
\text{when }\;U_i\cap U_j=\emptyset\;\forall\;i,j.
\label{sigmad} 
\ee
In other words, the total measure of a set consisting of several disconnected components must be the sum of 
the measures of the individual components. A measure $\mu$ on $\cM$ is said to be \it{finite}\i{finite 
measure} if $\mu(\cM)$ is finite; it is 
\it{$\sigma$-finite}\i{sigma finite measure@$\sigma$-finite measure} if $\cM$ is a 
countable union of 
measurable sets with finite measure (any finite measure is trivially $\sigma$-finite).\\

For instance, the standard translation-invariant Lebesgue measure on the real 
line $\RR$ is defined so that $\mu([a,b])=b-a$ for any closed interval $[a,b]\subset\RR$; the measure takes 
the same value for open or half-open intervals. In particular, $\RR$ is a countable union of intervals of 
finite length, so the Lebesgue measure is $\sigma$-finite. This definition is readily generalized 
to $\RR^n$.\i{Rn@$\RR^n$}\\

If $\cN$ is a topological space, a function 
$\cF:\cM\rightarrow\cN$ is said to be \it{measurable} if $\cF^{-1}(V)$ 
is a measurable subset of $\cM$ for any open set $V$ in $\cN$.\i{open set} In other words, measurable 
functions are 
those that ``preserve the structure of measurable sets''. Those are the functions that we will be allowed to 
integrate later on.

\subsubsection*{Borel measures}

Throughout this chapter and the next ones, we systematically endow $\cM$ with a topology. One can take 
advantage of this structure when defining a measure:

\paragraph{Definition.} Let $\cM$ be a topological space. A \it{Borel set} in $\cM$ is a subset 
$U\subseteq\cM$ which is either an open set, 
or a closed set, or a union or an intersection of countably many open or closed sets. A \it{Borel 
measure}\i{Borel measure} on $\cM$ is a measure whose measurable sets are the Borel sets of $\cM$.\\

Thus, Borel measures are compatible with the topology of $\cM$. In particular, any 
continuous function $\cF:\cM\rightarrow\cN$ is 
Borel-measurable. From now on, all measures are understood to be Borel. When $\cM$ is a smooth manifold, the 
data of a Borel measure is 
equivalent to that of a volume form on $\cM$. For simplicity, we 
always assume that $\cM$ is a manifold.

\subsubsection*{Integrals}

Measures can be used to integrate functions.\footnote{The concrete definition of integrals relies on a 
limiting procedure where the 
integrand is approximated by a sequence of 
locally constant functions, but we will not review these details here.} Let $\mu$ be a Borel 
measure on 
$\cM$ and $U\subseteq\cM$ a Borel set. When $\VV$ is a topological vector space and 
$\cF:\cM\rightarrow\VV:q\mapsto \cF(q)$ is a measurable function, the (Lebesgue) integral\i{Lebesgue 
integral} of $\cF$ over $U$ relative to the measure $\mu$ is 
written as\footnote{We denote points of $\cM$ as $p$, $q$, etc.\ 
to suggest thinking of them as possible momenta of a particle.}
\be
\int_U \cF(q)\,d\mu(q)
\qquad\text{or}\qquad
\int_U\cF\,d\mu\,.
\nn
\ee
In these terms, the 
measure $\mu(U)$ of a Borel set $U$ is the integral of the function $\cF(q)=1$ over $U$: 
\be
\mu(U)=\int_Ud\mu(q).
\label{mimibis}
\ee
The word ``measure'' often also refers to the quantity 
$d\mu$ appearing in this expression.\\

For example, the standard translation-invariant Lebesgue measure on 
$\RR^n$ is denoted $d\mu(x)\equiv d^nx$, with the usual rules for integration.
One can generate infinitely many other measures on $\RR^n$ by multiplying the Lebesgue measure by an 
arbitrary
function: for any non-negative measurable map $\rho:\RR^n\rightarrow\RR:x\mapsto\rho(x)$, the quantity 
$d\mu(x)=\rho(x)d^nx$ is a Borel measure on $\RR^n$. Another example is provided by the 
sphere $S^2$, which admits the rotation-invariant measure $\sin\theta\,d\theta\,d\phii$ in 
terms of polar coordinates $\theta,\phii$. 
Finally, in section \ref{relagroup} we will use the Lorentz-invariant 
measure\i{measure!Lorentz-invariant}\i{Lorentz-invariant measure}
\be
d\mu(\textbf{q})
=
\frac{d^{D-1}\textbf{q}}{\sqrt{M^2+\textbf{q}^2}}
\label{lom}
\ee
where $M^2$ is a positive parameter (the mass squared)
while $\textbf{q}=(q_1,...,q_{D-1})$ is the spatial momentum in $D$ space-time dimensions.

\subsection{Hilbert spaces of wavefunctions}
\label{suseletou}

We now have the tools needed to define Hilbert spaces of square-integrable functions. For the 
sake of generality we consider wavefunctions taking values in a complex Hilbert space $\cE$ endowed 
with a scalar product
\be
(\cdot|\cdot):
\cE\times\cE\rightarrow\CC:v,w\mapsto(v|w)\,,
\nn
\ee
which we take to be linear in its second argument and antilinear in the first one. When $\cE=\CC$ we simply 
set $(v|w)=v^*w$.

\subsubsection*{Wavefunctions}

\paragraph{Definition.} Let $\cM$ be a topological space, $\mu$ a Borel measure on $\cM$, $\cE$ a complex 
Hilbert space with scalar product $(\cdot|\cdot)$. Then an $\cE$-valued \it{square-integrable 
wavefunction}\i{square integrability}\i{wavefunction} is a measurable map $\Psi:\cM\rightarrow\cE$ such that
\be
\int_{\cM}d\mu(q)\big(\Psi(q)|\Psi(q)\big)<+\infty.
\nn
\ee
We denote by $\cL^2(\cM,\mu,\cE)$ the vector space of such functions.\\

It is tempting to turn $\cL^2(\cM,\mu,\cE)$ into a Hilbert space by declaring that the scalar product of two 
wavefunctions is the integral of their product over $\cM$, but there is a problem: wavefunctions need not be 
continuous. In particular, functions that vanish everywhere on $\cM$ except at some 
countable number of points, are strictly speaking non-zero vectors in $\cL^2$ even though all their would-be 
scalar products vanish. In the language of conformal field theory, those are ``null states''. In order 
to cure this 
pathology, one introduces the following notion:

\paragraph{Definition.} Let $\mu$ be a Borel measure on $\cM$. A property is said to be true \it{almost 
everywhere}\i{almost everywhere} on $\cM$ if there exists a 
Borel set $U\subset\cM$ such that $\mu(U)=0$ and such that the property be true 
on 
each point of $\cM\backslash U$.\\

For example, when $\cF$ and $\cG$ are functions $\cM\rightarrow\cN$, we say that $\cF=\cG$ almost everywhere 
on 
$\cM$ and write $\cF\sim\cG$ if $\cF$ and $\cG$ differ only on a set of measure zero. The relation $\sim$ is 
an equivalence relation. This solves the pathology of $\cL^2$ spaces, as one can show that integrals of 
functions that coincide almost everywhere are equal. In particular, any function 
$\cF\sim 0$ is said to \it{vanish almost everywhere};\i{almost 
everywhere} such a 
function belongs to $\cL^2$ (the integral of its square vanishes) and can now be identified with the function 
that vanishes identically on $\cM$. More precisely, let us denote by $N(\cM,\mu,\cE)$ the space of 
$\cE$-valued measurable functions on $\cM$ that vanish almost everywhere; it is a subspace of $\cL^2$ and may 
be seen as the set of null states\i{null state} (hence the notation $N$) in $\cL^2$. This leads to the 
following notion:

\paragraph{Definition.} The \it{space of square-integrable wavefunctions} on $\cM$ 
with values in $\cE$ 
relative to the measure $\mu$ is the quotient of $\cL^2$ by $N$:\i{L2 space@$L^2$ space}
\be
L^2(\cM,\mu,\cE)
\equiv
\cL^2(\cM,\mu,\cE)\big/N(\cM,\mu,\cE).
\label{modout}
\ee
This space is also simply called the ($\cE$-valued) $L^2$ space on $\cM$ relative to the measure $\mu$.\\

Elements of $L^2$ are thus equivalence classes of functions $\Psi:\cM\rightarrow\cE$, two functions being 
identified if they coincide almost everywhere. With this identification, one can endow $L^2$ with a norm 
$\|\cdot\|$ defined by
\be
\|\Psi\|^2
\equiv
\int_{\cM} d\mu(q)\big(\Psi(q)|\Psi(q)\big).
\label{lnorm}
\ee
Strictly speaking we should write the left-hand side of this definition as $\|[\Psi]\|^2$, where $[\Psi]\in 
L^2$ is the class\footnote{This class has nothing to do with the ray (\ref{ray}) despite the 
identical notation.} of $\Psi\in\cL^2$. However, we will systematically abuse notation by choosing 
arbitrarily a 
representative $\Psi$ of a class $[\Psi]$, and we use the word ``wavefunction'' to refer both 
to actual functions $\Psi:\cM\rightarrow\cE$ and to the corresponding equivalence classes in $L^2$.\\

Formula (\ref{lnorm}) is a well-defined norm on $L^2$: it is independent of the 
chosen representative for the class $[\Psi]$, and it satisfies the properties 
required for a norm. In particular, a function has zero norm if it vanishes almost everywhere, i.e.~if its 
class is the zero vector in $L^2$. This is indeed the solution of the pathology we encountered in $\cL^2$ 
spaces.\\

It can be shown that the space $L^2$ is a complete normed vector space, i.e.\ a \it{Banach space},\i{Banach 
space} with respect to the norm (\ref{lnorm}). In addition the space of (equivalence classes of) smooth 
functions with compact support is dense in 
$L^2(\cM,\mu,\cE)$, so any wavefunction can be approximated with arbitrary precision by a smooth function.

\subsubsection*{Hilbert spaces of wavefunctions}

\paragraph{Definition.} Let $\mu$ be a Borel measure on $\cM$, $\cE$ a Hilbert space with scalar 
product $(\cdot|\cdot)$. Let 
$\Phi$ and $\Psi$ be two $\cE$-valued square-integrable wavefunctions on $\cM$. Then the \it{scalar 
product}\i{scalar product} of $\Phi$ and $\Psi$ is\i{wavefunction!scalar product}
\be
\langle\Phi|\Psi\rangle
\equiv
\int_{\cM}d\mu(q)\big(\Phi(q)|\Psi(q)\big),
\label{scall}
\ee
where the integrand reduces to $\Phi^*(q)\Psi(q)$ when $\cE=\CC$. The space $L^2(\cM,\mu,\cE)$ is a Hilbert 
space with respect to this scalar product.\\

With this definition we can start interpreting $L^2(\cM,\mu,\cE)$ as the space of states of some quantum 
system. In Dirac notation we would write wavefunctions as $\Psi\equiv|\Psi\rangle$, which is indeed suggested 
by the notation (\ref{scall}). The quantum state defined by such a wavefunction is a ray 
(\ref{ray})consisting of all functions $\cM\rightarrow\cE$ that are equal almost everywhere to some constant 
multiple of $\Psi$. (Again, the notation $[\cdot]$ in (\ref{ray}) does \it{not} mean the same thing as the 
class of a wavefunction in (\ref{modout})!)\\

To interpret $\cE$-valued wavefunctions, we note the isomorphism\i{tensor product}
\be
L^2(\cM,\mu,\cE)\cong L^2(\cM,\mu,\CC)\otimes\cE.
\label{zissou}
\ee
For example suppose $\cE=\CC^2$ is the 
Hilbert space of a two-state system (as will be the case, say, for the spin-$1/2$ representation of the 
Poincar\'e group in section \ref{relagroup}). In Dirac notation, we can define an 
orthonormal basis $\{|+\rangle,\;|-\rangle\}$ of $\cE$ such that a generic (normalized) state 
of $L^2(\cM,\mu,\cE)$ takes the form
\be
|\Phi\rangle
=
\frac{1}{\sqrt{2}}
\Big(
|\phi\rangle\otimes|+\rangle
+
|\psi\rangle\otimes|-\rangle
\Big)
\label{biphi}
\ee
where $|\phi\rangle$ and $|\psi\rangle$ are normalized complex-valued wavefunctions on $\cM$. If we think of 
$\cE$ as a space of spin degrees of freedom and if $\cM$ is a space of momenta, then the 
wavefunction (\ref{biphi}) describes the propagation of two spin states with generally 
different momentum distributions accounted for by $\phi$ and $\psi$. Note that the state (\ref{biphi}) is 
typically entangled\i{entanglement} with respect to the splitting (\ref{zissou}); it is unentangled if and 
only if 
$|\psi\rangle=e^{i\lambda}|\phi\rangle$ for some real number $\lambda$. The generalization of (\ref{biphi}) 
to 
higher-dimensional spaces $\cE$ is straightforward.

\paragraph{Remark.} When dealing with unitary representations of the BMS$_3$ group in part III, we will need 
to describe square-integrable wavefunctions on infinite-dimensional manifolds (see section \ref{sebmspar}). 
Until then we will not discuss this issue.

\subsection{Equivalent measures and Radon-Nikodym derivatives}
\label{seqemes}

The definition of Hilbert spaces of square-integrable wavefunctions relies on the measure 
$\mu$ used to define the 
scalar product (\ref{scall}). Naively, one might therefore expect that the Hilbert spaces 
$L^2(\cM,\mu,\cE)$ and $L^2(\cM,\nu,\cE)$ differ if the measures $\mu$ and $\nu$ do not coincide. However 
it is 
easy to show that the space $L^2(\cM,\mu,\cE)$ is 
essentially independent of the measure 
$\mu$. Here we prove this statement while introducing 
the notion of equivalent measures and their Radon-Nikodym derivative.

\paragraph{Definition.} Let $\mu$, $\nu$ be two Borel measures on a manifold $\cM$. 
We say that $\mu$ and $\nu$ are \it{equivalent}\i{equivalent measures}\i{measure!equivalence} if they have 
the same sets of zero 
measure.\\

Equivalent measures can be vastly different, yet they are still pretty much the same with regard to measure 
theory:

\paragraph{Radon-Nikodym theorem.} Let $\mu$ and $\nu$ be equivalent $\sigma$-finite measures. Then there 
exists a measurable function $\rho:\cM\rightarrow\RR^+$ such that\i{Radon-Nikodym theorem}
\be
\nu(U)=\int_U\rho(q)d\mu(q)\qquad\text{for any Borel set $U$.} 
\label{rn}
\ee
This relation is often written in infinitesimal form
\be
d\nu(q)=\rho(q)d\mu(q)
\qquad\text{or}\qquad
\rho(q)=\frac{d\nu(q)}{d\mu(q)}.
\label{rnbis}
\ee
In addition, any other function $\tilde\rho$ satisfying this property coincides with $\rho$ almost everywhere 
on 
$\cM$. The function $\rho$ is called the \it{Radon-Nikodym derivative}\i{Radon-Nikodym derivative} of 
$\nu$ with respect to 
$\mu$.\footnote{There exist infinitely many functions that all represent equally well the 
Radon-Nikodym derivative; the theorem ensures that these functions agree, except possibly on a set of zero 
measure. 
Accordingly, we call ``the'' Radon-Nikodym derivative any function that satisfies (\ref{rn}).} A proof of 
this theorem can be found in \cite{Royden}.\\

For example, we mentioned below (\ref{mimibis}) that when $d^nx$ is the Lebesgue measure on 
$\RR^n$,\i{Rn@$\RR^n$} any non-negative function $\rho$ gives rise to a new measure $\rho(x)d^nx$. The 
Radon-Nikodym 
derivative of that measure with respect to the Lebesgue measure then coincides with the function $\rho$. In 
particular, when $\rho(x)$ only vanishes on a set of Lebesgue measure zero, the measures $d^nx$ and 
$\rho(x)d^nx$ are equivalent.

\paragraph{Remark.} When $\mu$ and $\nu$ are equivalent measures, one has
$d\mu(q)/d\nu(q)\sim\left[d\nu(q)/d\mu(q)\right]^{-1}$,
i.e.\ the Radon-Nikodym of 
$\mu$ with respect to $\nu$ is (almost everywhere) the inverse of the Radon-Nikodym of $\nu$ with respect to 
$\mu$.

\subsubsection*{Isomorphic $L^2$ spaces}

The notion of equivalent measures allows us to address the question raised above, namely whether the Hilbert 
spaces $L^2(\cM,\mu,\cE)$ and $L^2(\cM,\nu,\cE)$ differ if the 
measures $\mu$ and $\nu$ differ.

\paragraph{Proposition.} Let $\mu$ and $\nu$ be equivalent Borel measures on $\cM$, $\cE$ a Hilbert space; we 
write $L^2(\cM,\mu,\cE)\equiv L^2(\mu)$ and similarly for $\nu$. 
Then there is an isometry
\be
\cU:
L^2(\mu)\rightarrow L^2(\nu):
\Psi\mapsto\cU\cdot\Psi
\qquad
\text{with}
\qquad
(\cU\cdot\Psi)(q)\equiv\sqrt{\frac{d\mu(q)}{d\nu(q)}}\,\Psi(q)
\label{isom}
\ee
so the spaces 
$L^2(\cM,\mu,\cE)$ and $L^2(\cM,\nu,\cE)$ are isomorphic as Hilbert spaces.

\begin{proof}
The map (\ref{isom}) is manifestly linear and invertible, since the measures $\mu$ and $\nu$ are equivalent 
so that the Radon-Nikodym 
derivative $\rho=d\nu/d\mu$ is strictly positive almost everywhere. It only remains to prove that $\cU$ 
preserves the scalar products (\ref{scall}); let us denote them by 
$\langle\cdot|\cdot\rangle_{\mu}$ and $\langle\cdot|\cdot\rangle_{\nu}$ in $L^2(\mu)$ and 
$L^2(\nu)$, respectively. For any two $\mu$-square-integrable wavefunctions $\Phi$ and $\Psi$, the 
definitions (\ref{rnbis}) and (\ref{isom}) readily yield $\langle\cU\cdot\Phi|\cU\cdot\Psi\rangle_{\nu}
=\langle\Phi|\Psi\rangle_{\mu}$, which proves that $\cU$ is an isometry.
\end{proof}

This proposition says that the structure of the Hilbert space $L^2(\cM,\mu,\cE)$ does not 
depend on the measure $\mu$, since any other equivalent measure would give rise to an isomorphic 
Hilbert space. A similar phenomenon will occur in section 
\ref{simind}, where induced representations built with different scalar 
products will turn out to be equivalent.

\section{Quasi-regular representations}
\label{seqareg}

In the previous pages we have seen how to build spaces of wavefunctions. Our goal now is to endow 
such Hilbert spaces with a unitary group action. The strategy will be to take the manifold 
$\cM$ (on which 
wavefunctions are defined) to be homogeneous with respect to some group 
action, then use this action to define unitary operators. We now describe this approach after recalling some 
basic properties of group actions and measures on homogeneous spaces. This will lead to the notion of 
quasi-regular representations, which provides the simplest example of induced representations.

\subsection{Quasi-invariant measures on homogeneous spaces}
\label{semehomos}

\subsubsection*{Group actions and orbits}

\paragraph{Definition.} Let $\cM$ be a manifold, $G$ a Lie group.\footnote{As before elements of $G$ are 
written as $f$, $g$, etc.\ and the identity is denoted $e$.} An \it{action}\i{group action}\i{action of a 
group} of $G$ on $\cM$ 
is 
a smooth map $G\times\cM\rightarrow\cM:(f,q)\mapsto f\cdot q$ such that $e\cdot q=q$ and 
$f\cdot(g\cdot q)=(fg)\cdot q$ for all group elements $f,g$ and any $q\in\cM$. Equivalently, an action of $G$ 
on $\cM$ is a homomorphism from $G$ to the group $\text{Diff}(\cM)$ of diffeomorphisms of $\cM$.\\

There exist many important examples of group actions in physics: the space $\RR^n$ can be seen 
as an Abelian group acting on itself by the addition of vectors; the sphere 
$S^2$ is acted upon by rotations. More generally, any group representation is a linear 
action of 
a group on a vector space; in particular the energy-momentum of a particle in Minkowski space is acted upon 
linearly by the Lorentz group.\\

Consider an action of $G$ on $\cM$, and pick a point $p\in\cM$. The \it{orbit}\i{orbit} of $p$ is the 
submanifold of $\cM$ consisting of all points that can be reached by acting on $p$ 
with $G$:
\be
\cO_p\equiv\left\{f\cdot p|f\in G\right\}.
\label{orb}
\ee
The orbit is independent 
of the choice of $p$ in the sense that, whenever $q\in\cO_p$, we have $\cO_p=\cO_q$. The 
\it{stabilizer}\i{stabilizer} of $p\in\cM$ is the subgroup of $G$ that leaves it 
invariant,
\be
G_p\equiv\left\{f\in G|f\cdot p=p\right\}.
\label{guipure}
\ee
If $q$ is another point in $\cO_p$, and if $g\in G$ is such that $g\cdot p=q$, then the stabilizer of $q$ 
is 
$g\,G_p\,g^{-1}$, which is isomorphic to $G_p$. (In particular, one often abuses terminology by 
saying ``the stabilizer of an orbit'' instead of the stabilizer of a \it{point} on the 
orbit.) The stabilizer is a (closed) subgroup of $G$ and the orbit 
(\ref{orb}) is 
diffeomorphic to the coset space\i{quotient space}
\be
\cO_p\cong G/G_p\,.
\label{hogippy}
\ee
This diffeomorphism is explicitly given by the bijection $G/G_p\rightarrow\cO_p:f\,G_p\mapsto f\cdot p$.

\subsubsection*{Homogeneous spaces}

\paragraph{Definition.} An action of a group $G$ on $\cM$ is said to be 
\it{transitive}\i{homogeneous 
space} when for any two points $p,q\in\cM$ there 
exists a group element $f$ such that $f\cdot p=q$. The space $\cM$ is then said to be a \it{homogeneous 
space} 
for this action.\\

In particular a homogeneous space coincides with the orbit of any of its points under the group 
action: $\cM=\cO_p$ for any $p\in\cM$. It follows that any homogeneous space can be written as a coset space 
(\ref{hogippy}).\\

The 
simplest example of a $G$-homogeneous space is the group $G$ itself, with the action given by left 
multiplication:
\be
g\longmapsto L_f(g)=fg.
\label{left}
\ee
The stabilizer in that case is trivial. Note that right multiplication
\be
g\longmapsto R_f(g)=gf
\label{RighT}
\ee
is not quite a group action since $R_f\circ R_g=R_{gf}$ does not coincide with $R_{fg}$. This can 
be cured by considering right multiplication by \it{inverse} elements, i.e.\ $g\mapsto R_{f^{-1}}g=gf^{-1}$. 
In the aforementioned example of $\RR^n$, seen as an 
Abelian 
group acting on itself by the addition of vectors, left and right multiplications coincide. (This is true for 
any Abelian group.) The sphere $S^2$ is 
a 
more interesting example of homogeneous space, since it is acted upon transitively by the group of rotations 
$\text{SO}(3)$ but has a non-trivial stabilizer $\text{SO}(2)$, and is therefore diffeomorphic to 
the quotient $\text{SO}(3)/\text{SO}(2)$. More generally, one has a family of 
diffeomorphisms $S^n\cong\text{SO}(n+1)/\text{SO}(n)$.
Homogeneous spaces will play a central role in representation theory, so we will encouter many more examples 
of transitive actions 
later in this thesis.

\subsubsection*{The Haar measure}

We now initiate the study of measure theory on homogeneous spaces.

\paragraph{Definition.} Let $\cM$ be a homogeneous space with respect to the action of a group $G$; 
let $\mu$ be a Borel measure on $\cM$. We say that the measure is \it{invariant}\i{invariant 
measure}\i{measure!invariant} under 
$G$ if $\mu(f\cdot U)=\mu(U)$
for all $g\in G$ and for any Borel set $U$.\\

For example, the measure $\sin\theta\,d\theta\,d\phii$ on a sphere $S^2$ is invariant under rotations, while 
the momentum measure (\ref{lom}) is invariant under Lorentz transformations. When the homogeneous space 
$\cM$ is the group manifold $G$ itself, one has the following result:

\paragraph{Haar's theorem.} Let $G$ be a finite-dimensional Lie group. Then, up to a positive 
multiplicative constant, there exists a unique 
Borel measure on $G$ invariant under left multiplication (\ref{left}), known as the \it{left Haar measure} on 
$G$.\i{Haar measure}

\begin{proof}
Any left-invariant volume form on $G$ is the pull-back by left multiplication of a volume form on the tangent 
space $T_eG$ at the identity. Since $G$ is finite-dimensional the volume form on $T_eG$ is unique up to a 
positive multiplicative constant, so the theorem follows. (See e.g.\ \cite{abraham1978foundations} for 
details.)
\end{proof}

The same theorem would hold for 
right multiplications, although the resulting \it{right} Haar measure generally differs from the 
left one. If the group $G$ is Abelian, any 
left-invariant measure is also right-invariant. For instance, the standard measure $d^nx$ 
on $\RR^n$ is the translation-invariant Haar measure for the Abelian group $\RR^n$.

\subsubsection*{Quasi-invariant measures}

Given a homogeneous space $\cM$, we wish to integrate functions over it and we ask whether 
there exists an invariant measure. It 
turns out that this is not always the case (see e.g.~\cite{barut1986theory}), so one 
introduces the following weaker notion of invariance:

\paragraph{Definition.} Let $\cM$ be a homogeneous space with respect to the action of a group $G$. A 
Borel measure $\mu$ on $\cM$ is said to be \it{quasi-invariant}\i{quasi-invariant 
measure}\i{measure!quasi-invariant} under $G$ 
if, for any group element $f$, the measure $\mu_f$ defined by
\be
\mu_f(U)\equiv\mu(f\cdot U)\qquad\text{for any Borel set }U
\label{gnugnu}
\ee
is equivalent to $\mu$.\\

Equivalent measures are related through (\ref{rnbis}) by their 
Radon-Nikodym 
derivative. Accordingly, when $\mu$ is a quasi-invariant measure on a homogeneous space $\cM$, we denote 
the Radon-Nikodym derivative\i{Radon-Nikodym derivative} 
of $\mu_f$ with respect to $\mu$ by
\be
\rho_f(q)\equiv\frac{d\mu_f(q)}{d\mu(q)}=\frac{d\mu(f\cdot q)}{d\mu(q)}
\label{rnq}
\ee
for any $f\in G$ and any $q\in\cM$. We shall refer to it as ``the'' Radon-Nikodym derivative of $\mu$ under 
the action of $G$. Note that it satisfies the 
important property
\be
\rho_{fg}(q)
=
\rho_f(g\cdot q)\rho_f(q).
\label{rnco}
\ee
The measure $\mu$ is invariant if and only if its Radon-Nikodym derivative $\rho_f$ is equal ot one (almost 
everywhere) for any group element $f$.\\

Intuitively one can think of the Radon-Nikodym derivative (\ref{rnq}) as an anomaly:\i{Radon-Nikodym 
derivative!as anomaly}\i{anomaly} since $\mu$ is defined 
on a homogeneous space, one naively expects it to be invariant under the group action. The 
Radon-Nikodym 
derivative measures the extent to which invariance is spoiled. Taking again the example of $\RR^n$, the 
Lebesgue measure $d^nx$ is invariant under translations and rotations, but \it{not} under arbitrary 
diffeomorphisms. Indeed, 
for $f:x\mapsto f(x)$ a diffeomorphism of $\RR^n$, the Lebesgue measure transforms as
\be
d\mu_f(x)=d^n[f(x)]
=
\left|\frac{\der f}{\der x}\right|d^nx,
\label{jac}
\ee
where $|\der f/\der x|$ is the Jacobian of $f$.\i{Jacobian} Thus the Radon-Nikodym derivative of a 
quasi-invariant 
measure can also be seen as a generalization of the Jacobian.\\

Since we motivated quasi-invariant measures by the observation that invariant measures do not always 
exist, one might worry that a similar problem arises for quasi-invariant measures. Fortunately one can show 
that, in contrast to invariant measures, quasi-invariant measures \it{do} always exist on any 
finite-dimensional homogeneous space (see e.g.\ \cite{barut1986theory}). We shall discuss the 
infinite-dimensional generalization of that statement in section \ref{sebmspar}. Note that the existence of 
one quasi-invariant measure $\mu$ on $\cM$ implies the existence of infinitely many of them, since 
multiplying 
$\mu$ by any positive function yields another quasi-invariant measure.

\subsection{The simplest induced representations}
\label{simind}

We are now in position to describe quasi-regular representations.
Let $\cM$ be a manifold acted upon by a group $G$, and consider a vector space of wavefunctions 
$\Psi:\cM\rightarrow\CC$. We then readily define a representation $\cT$ of $G$ in that space by writing
\be
\boxed{
\Big.
\left(\cT[f]\cdot\Psi\right)(q)
\equiv
\Psi(f^{-1}\cdot q).}
\label{qregsimple}
\ee
Each operator $\cT[f]$ is manifestly linear, and the fact that this is indeed a representation follows from 
the fact that the map $q\mapsto f\cdot q$ is a group action. The interpretation of formula (\ref{qregsimple}) 
is simple: if the wavefunction $\Psi$ is sharply centred around some point $k$ of $\cM$, then the operator 
$\cT[f]$ maps $\Psi$ on a new wavefunction, now centred around the point $f\cdot k$. In chapter \ref{c2bis} 
the space $\cM$ will consist of the allowed momenta of a particle, $\Psi$ will be the particle's 
wavefunction (in momentum space), and the map $q\mapsto f\cdot q$ will be an action by boosts or rotations.

\begin{figure}[H]
\centering
\includegraphics[width=0.70\textwidth]{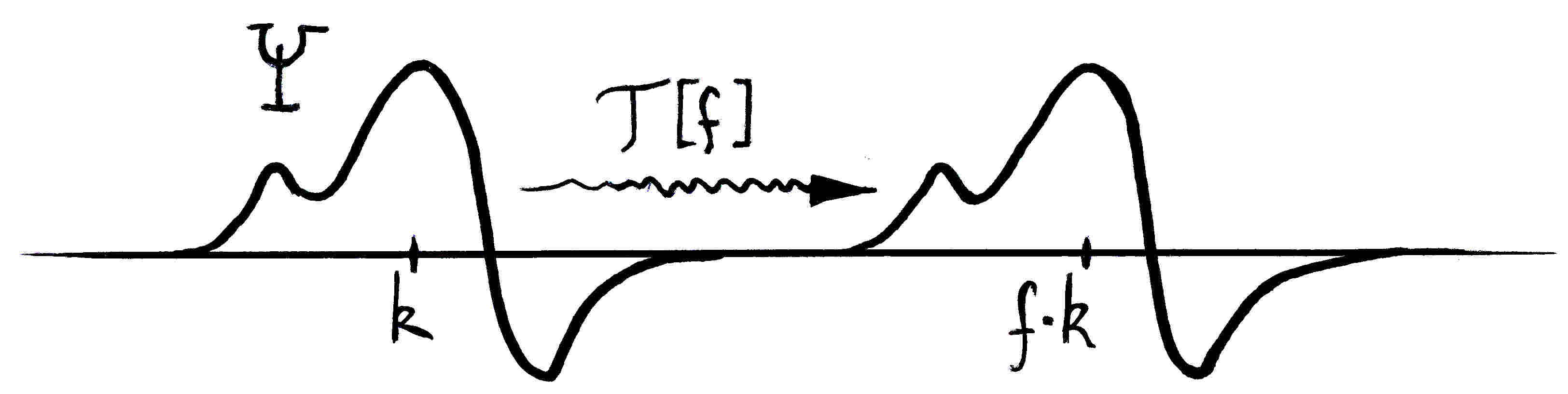}
\caption{A wavefunction $\Psi$ on $\cM=\RR$ centred around some point $k$ is acted upon by a unitary 
operator 
$\cT[f]$ that implements the transformation $k\mapsto f\cdot k$. The resulting transformed wavefunction 
$\cT[f]\cdot\Psi$ is the old one, translated by $f$.\label{movaction}}
\end{figure}

In order to interpret formula (\ref{qregsimple}) as the action of a symmetry group on a Hilbert space of 
wavefunctions, we need to make sure that each operator $\cT[f]$ is unitary. If $\cM$ is a homogeneous space 
and $\mu$ is a quasi-invariant measure on $\cM$, the scalar product of wavefunctions is (\ref{scall}) with 
$(\Phi(q)|\Psi(q))=\Phi^*(q)\Psi(q)$. Now it is easy to verify that the representation (\ref{qregsimple}) is 
generally \it{not} unitary for this scalar product:
\be
\langle\cT[f]\Phi|\cT[f]\Psi\rangle
=
\int_{\cM}d\mu(q)\Phi^*(f^{-1}\cdot q)\Psi(f^{-1}\cdot q)
\refeq{rnq}
\int_{\cM}d\mu(q)\rho_f(q)\Phi^*(q)\Psi(q).
\label{fistouille}
\ee
The far right-hand side generally does not coincide with the original scalar product (\ref{scall}) because 
it involves the Radon-Nikodym derivative (\ref{rnq}).
Thus, in order to ensure unitarity, we need to correct formula 
(\ref{qregsimple}) by a factor that compensates the non-trivial transformation law of $\mu$:

\paragraph{Definition.} Let $G$ be a Lie group acting transitively on a manifold $\cM$.
Let $\mu$ be a quasi-invariant measure on $\cM$ and let $L^2(\cM,\mu,\CC)$ be the space of
square-integrable wavefunctions on $\cM$. The \it{quasi-regular representation}\i{quasi-regular 
representation}\i{representation!quasi-regular} $\cT$ of $G$ acts on this space 
according to
\be
\left(\cT[f]\cdot\Psi\right)(q)
\equiv
\sqrt{\rho_{f^{-1}}(q)}\,\,\Psi(f^{-1}\cdot q)
\label{qreg}
\ee
for any wavefunction $\Psi$, where $\rho_f$ is the Radon-Nikodym derivative (\ref{rnq}) of $\mu$. If $\mu$ is 
invariant, the quasi-regular representation boils 
down to (\ref{qregsimple}).

\paragraph{Proposition.} The quasi-regular representation defined by (\ref{qreg}) is a unitary representation 
of $G$ in $L^2(\cM,\mu,\CC)$.

\begin{proof}
First we need to check that (\ref{qreg}) actually defines a representation, i.e.\ that $\cT[f\cdot 
g]=\cT[f]\circ\cT[g]$ for all $f,g\in G$. As in (\ref{qregsimple}), linearity of $\cT[f]$ is obvious. 
Now pick a wavefunction $\Psi\in L^2(\cM,\mu,\CC)$. At some 
point $q\in\cM$, we find
\be
(\cT[fg]\Psi)(q)
\refeq{qreg}
\left[\rho_{g^{-1}f^{-1}}(q)\right]^{1/2}
\Psi\big(g^{-1}\cdot(f^{-1}\cdot q)\big)
\nn
\ee
where we relied on the fact that $q\mapsto f\cdot q$ is a group action. Now using (\ref{rnco}) and the 
definition 
(\ref{qreg}), we can rewrite this as
\begin{align}
\label{reprep}
(\cT[fg]\Psi)(q)
& =
\left[\rho_{g^{-1}}(f^{-1}\cdot q)\rho_{f^{-1}}(q)\right]^{1/2}
\Psi\big(g^{-1}\cdot(f^{-1}\cdot q)\big)\\
\label{rekrep}
& =
\left[\rho_{f^{-1}}(q)\right]^{1/2}
\left(\cT[g]\Psi\right)(f^{-1}\cdot q)
=
\Big(\big(\cT[f]\circ\cT[g]\big)\cdot\Psi\Big)(q)\,,
\end{align}
which proves that (\ref{qreg}) is indeed a representation. To complete the proof we also have to show that 
$\cT$ is unitary for the scalar product (\ref{scall}) with $(\Phi(q)|\Psi(q))=\Phi^*(q)\Psi(q)$. Repeating 
the computation (\ref{fistouille}) we now find that the Radon-Nikodym derivative in (\ref{qreg}) yields an 
extra term in the integrand. Using (\ref{rnco}) and the fact that $\rho_e=1$, this term cancels the 
Radon-Nikodym derivative in (\ref{fistouille}) so (\ref{qreg}) is indeed unitary.
\end{proof}

\paragraph{Remark.} When the homogeneous space $\cM$ coincides with the group $G$ and is endowed with the 
invariant Haar measure, formula (\ref{qregsimple}) defines a unitary representation of $G$ known as the 
\it{regular representation}.\i{regular representation}\i{representation!regular} Quasi-regular 
representations extend this concept by 
trading the base manifold $G$ for an arbitrary homogeneous space $\cM$.

\subsubsection*{Equivalence of quasi-regular representations}

Recall that $L^2$ spaces defined with equivalent measures are isometric via the map 
(\ref{isom}). One may wonder how that statement affects quasi-regular representations: is it true that two 
representations of the form (\ref{qreg}) are equivalent if they are defined using different but equivalent 
measures?\i{equivalent measures!and representations} The answer is yes: if $\mu$ and $\nu$ are equivalent 
quasi-invariant measures on $\cM$ and if we 
denote the corresponding quasi-regular representations by $\cT_{\mu}$ and 
$\cT_{\nu}$ respectively, then the isometry (\ref{isom}) is an \it{intertwiner}\i{intertwiner}:
\be
\cU\circ\cT_{\mu}[f]=\cT_{\nu}[f]\circ\cU
\qquad
\text{for all }f\in G.
\label{inter}
\ee
Accordingly, the representations $\cT_{\mu}$ and $\cT_{\nu}$ are unitarily equivalent. This is to say that 
the quasi-regular representation (\ref{qreg}) is essentially independent of the measure $\mu$.\\

\begin{advanced}
\subsection{Radon-Nikodym is a cocycle}
\label{suseracot}
\end{advanced}

Here we show that property (\ref{rnco}) is a cohomological statement: it says that the Radon-Nikodym 
derivative is a one-cocycle with respect to the representation (\ref{qregsimple}). This is an 
anecdotal observation, so the hasty reader may go directly to section \ref{sedefirep}.

\paragraph{Proposition.} Let $\mu$ be a quasi-invariant measure on a homogeneous space $\cM$ and let 
(\ref{rnq}) be its Radon-Nikodym derivative. Then the map\i{Radon-Nikodym 
derivative!as cocycle}\i{one-cocycle!Radon-Nikodym derivative}
\be
\log\rho:G\rightarrow C^{\infty}(\cM):
f\mapsto\log(\rho_{f^{-1}})
\label{rho}
\ee
is a one-cocycle with respect to the representation (\ref{qregsimple}), with the understanding 
that $\log(\rho_{f^{-1}})$ is the function on $\cM$ mapping $q$ on $\log\big(\rho_{f^{-1}}(q)\big)$.

\begin{proof}
We need to show that the Chevalley-Eilenberg differential (\ref{gd}) of the map (\ref{rho}) vanishes. Using 
(\ref{qregsimple}) we find
\be
(\sfd\log\rho)_{fg}(q)
=
\log\rho_{g^{-1}}(f^{-1}\cdot q)+\log\rho_{f^{-1}}(q)-\log\rho_{g^{-1}f^{-1}}(q)\,,
\nn
\ee
which vanishes by virtue of property (\ref{rnco}).
\end{proof}

Let us discuss the measure-theoretic interpretation of this cohomological statement.
For 
example, suppose the map (\ref{rho}) is a trivial one-cocycle. Then
\be
\log\rho_{f^{-1}}(q)
=
(\sfd\Psi)_f(q)
\refeq{gd}
(\cT[f]\cdot\Psi)(q)-\Psi(q)
\refeq{qregsimple}
\Psi(f^{-1}\cdot q)-\Psi(q)
\nn
\ee
for some function $\Psi(q)$. Equivalently,
\be
\frac{d\mu(f\cdot q)}{d\mu(q)}
=
e^{\Psi(f\cdot q)-\Psi(q)},
\qquad\text{i.e.}\qquad
e^{-\Psi(q)}d\mu(q)
=
e^{-\Psi(f\cdot q)}d\mu(f\cdot q)\,,
\label{dmu}
\ee
which says that the quasi-invariant measure $\mu$ is actually an \it{invariant} measure in 
disguise! Indeed, the measure $\nu$ defined by $d\nu(q)=e^{-\Psi(q)}d\mu(q)$ is invariant by virtue 
of 
(\ref{dmu}). In other words, the first cohomology of $G$ with values in the space of functions on 
$\cM$ classifies the inequivalent quasi-invariant measures on $\cM$, two measures being 
equivalent if they are related to one another by a function that multiplies them. In particular, the first 
cohomology vanishes if all quasi-invariant measures on $\cM$ are equivalent to an invariant 
measure.\\

We can also rephrase this in the language of representation theory:\i{quasi-regular representation!as affine 
module}\i{affine module!and quasi-regular reps} the quasi-regular representation 
(\ref{qreg}) is a (multiplicative) affine module (\ref{affi}) on top of the original representation 
(\ref{qregsimple}). Two such modules are equivalent if the corresponding measures are equivalent.

\paragraph{Remark.} The cohomological properties of the Radon-Nikodym derivative have applications in physics:
we mentioned below (\ref{rnq}) that the Radon-Nikodym derivative may be thought of as an 
anomaly,\i{anomaly}\i{one-cocycle!anomaly} and indeed anomalies in quantum field theory are 
one-cocycles for 
the BRST\i{BRST} differential, valued in a suitable space of functions \cite{Barnich:2000zw}. Our observation 
on the 
Radon-Nikodym derivative may be seen as a baby version of that 
general statement.

\section{Defining induced representations}
\label{sedefirep}

We now extend the construction of quasi-regular representations. Suppose we have a group $G$ with some 
(closed) subgroup $H$. Given a representation $\cS$ of $H$, we wish to 
induce a corresponding representation $\cT$ of $G$. To describe this mechanism we first need to study in 
more detail the 
homogeneous manifold
\be
\cM\cong G/H\,.
\label{gihommx}
\ee
We will then define an action $\cT$ of $G$ on wavefunctions that live on $\cM$.

\subsection{Standard boosts}

In section \ref{semehomos} we introduced homogeneous spaces and observed that they can be written as coset 
spaces $G/G_p$, where $G_p$ is the stabilizer of some point $p\in\cM$. Now note that, conversely, any coset 
space $G/H$ determines a homogeneous manifold (provided $H$ is a closed subgroup of $G$). Indeed the elements 
of $G/H$ are left cosets $gH$, where $g$ spans $G$, and the action of $G$ on $G/H$ is given by left 
multiplication:\i{homogeneous space!as coset space}\i{quotient space}
$gH\mapsto f\cdot (gH)=(fg)H$.
In particular one can think of $G/H$ as a manifold $\cM$ where the coset $gH$ corresponds to the point 
$q=g\cdot p$, where $p$ is identified with the coset $eH$ at the identity. The stabilizer of $g\cdot p$ is 
$gHg^{-1}$, as noted below (\ref{guipure}). Thus from now on we describe the space $G/H$ with 
the same notation as in section \ref{semehomos}.\\

Now consider the point $p\in\cM$, identified with the identity coset $eH$ in $G/H$. Since $\cM$ is a 
homogeneous space one can map $p$ on any other point $q\in\cM$, with a group element $g\in G$ such 
that $g\cdot p=q$. Given $q$, this group element is only defined up to multiplication from the right by an 
element of $H$, since $h\cdot p=p$ for any $h\in H$.

\paragraph{Definition.} Let $G$ act transitively on $\cM\cong G/H$ and let $p\in\cM$. Then a \it{family of 
standard boosts} for $p$ on $\cM$ is a map\i{standard boost}
\be
\cM\rightarrow G:q\mapsto g_q
\qquad\text{such that}\qquad
g_q\cdot p=q.
\label{stdb}
\ee
\vspace{.1cm}

Any homogeneous manifold admits a family of standard boosts. For example, in special relativity, the map
(\ref{stdb}) would typically be an assignment of a Lorentz boost $g_q$ for each possible energy-momentum 
vector $q$ of a massive particle. If $p$ is the energy-momentum of the particle at rest, $g_q$ would map 
this momentum $p$ on the boosted momentum $q$. In fact, in the latter case this assignment can be chosen in 
such a way that $g_q$ depends continuously on $q$, so the family of standard boosts is \it{continuous}. In 
all cases of interest below, continuous families of standard boosts will exist, so from now on we always 
assume that the map (\ref{stdb}) is continuous.

\paragraph{Remark.}\label{pagesiboo} The existence of a continuous family of standard boosts is equivalent to 
that of a global section for the principal bundle $G\rightarrow G/H$,\i{principal bundle} which in turn 
amounts to saying that 
this bundle is trivial (see e.g.\ \cite{nakahara2003geometry}).\i{trivial 
bundle} In general this 
is not the case --- the typical example is the non-trivial bundle $\SO(n+1)\rightarrow S^n$, which is 
relevant to the 
Euclidean 
group in $(n+1)$ dimensions. However, all relativistic symmetry groups as well as BMS$_3$ are such that 
continuous families of standard boosts do exist,\footnote{This actually follows from the fact that typical 
momentum orbits for such groups are homotopic to a point, which then implies that the corresponding bundles 
$G\rightarrow G/H$ are trivial.} so we do not need to dwell on this subtlety any further.

\subsection{Induced representations}

\paragraph{Definition.} Let $G$ be a group acting transitively on a manifold $\cM$ endowed with 
a quasi-invariant measure $\mu$. Let $p\in\cM$ be stabilized by a closed subgroup $H$ of $G$ and let the map 
(\ref{stdb}) be a continuous family of standard boosts on $\cM$. Then the representation $\cT$ of $G$ 
\it{induced} by $\cS$\i{induced representation}\i{representation!induced} acts in the Hilbert space 
$\sH=L^2(\cM,\mu,\cE)$ according to
\be
\boxed{
\Big.
\left(\cT[f]\cdot\Psi\right)(q)
\equiv
\sqrt{\rho_{f^{-1}}(q)}\;
\cS[g_q^{-1}\,f\,g_{f^{-1}\cdot q}]\,
\Psi(f^{-1}\cdot q)}
\label{ind}
\ee
for any wavefunction $\Psi$, where $\rho$ denotes the Radon-Nikodym derivative (\ref{rnq}). It is common to 
write\i{induced representation!notation}
\be
\cT=\text{Ind}_H^G(\cS).
\label{indnot}
\ee
\vspace{.1cm}

As given here, formula (\ref{ind}) comes a bit out of the blue, so it is worth analysing its elementary 
features. First note that, if we denote the trivial representation of a group by the symbol $\textbf{1}$, 
then 
$\text{Ind}_H^G(\textbf{1})$ is the quasi-regular representation (\ref{qreg}) of $G$ on $G/H$ 
while $\text{Ind}_{\{e\}}^G(\textbf{1})$ is the regular representation. Formula 
(\ref{ind}) extends these constructions by including a non-trivial action of $H$ in an internal space $\cE$. 
Before interpreting (\ref{ind}) any further, we now verify that it is a consistent definition.

\subsubsection*{Consistency}

Up to the term involving $\cS$,
the right-hand side of (\ref{ind}) coincides with the quasi-regular representation (\ref{qreg}), so the only 
potential problem could arise from the insertion of $\cS$. But since the map (\ref{stdb}) is a family of 
standard 
boosts, we have
\be
\big(g_q^{-1}\,f\,g_{f^{-1}\cdot q}\big)\cdot p
=
g_q^{-1}\cdot\Big(f\cdot\big(g_{f^{-1}\cdot q}\cdot p\big)\Big)
=
g_q^{-1}\cdot\Big(f\cdot\big(f^{-1}\cdot q\big)\Big)
=
g_q^{-1}\cdot q
=
p\,,
\nn
\ee
so the combination $g_q^{-1}\,f\,g_{f^{-1}\cdot q}$ belongs to the stabilizer $H$ of $p$, as it 
should. Since by assumption $\Psi$ takes its values in the carrier space $\cE$ of $\cS$, we conclude that the 
right-hand side of (\ref{ind}) is well-defined. For each $f\in G$, it defines a linear operator $\cT[f]$ 
acting on $\cE$-valued functions on $\cM$.

\paragraph{Proposition.} Let $\sH=L^2(\cM,\mu,\cE)$. Then the map $\cT:G\rightarrow\text{GL}(\sH)$ defined by 
(\ref{ind}) is a unitary 
representation of $G$.

\begin{proof}
If it were not for the representation $\cS$, formula (\ref{ind}) would coincide with (\ref{qreg}); since the 
latter is a unitary 
representation of $G$, we only have to convince ourselves that this feature is not spoiled by the presence of 
$\cS$. Noting that
\be
g_q^{-1}\,fg\,g_{(fg)^{-1}\cdot q}
=
\big(g_q^{-1}\,f\,g_{f^{-1}\cdot q}\big)
\cdot
\big(g_{f^{-1}\cdot q}^{-1}\,g\,g_{(fg)^{-1}\cdot q}\big)
\label{bonafide}
\ee
and using the fact that $\cS$ is a representation of $H$, one can mimick the sequence of equations 
(\ref{reprep})-(\ref{rekrep}) for $\cT$ given by (\ref{ind}), which implies that it is indeed a 
representation. As for unitarity, it follows from the fact that $\cS$ is 
unitary: $\big(\cS[h]\Phi(q)\big|\cS[h]\Psi(q)\big)=\big(\Phi(q)\big|\Psi(q)\big)$ for all wavefunctions 
$\Phi,\Psi$, 
any point $q\in\cM$ and any $h\in H$.
\end{proof}

\subsubsection*{Interpretation}

The basic interpretation of the induced representation (\ref{ind}) is the same as for (\ref{qreg}): 
$\Psi(f^{-1}\cdot q)$ represents the fact that the wavefunction 
$\Psi$ is ``boosted'' by $f$, while the factor $\sqrt{\rho_{f^{-1}}}$ ensures unitarity. The new ingredient 
is the combination\i{Wigner rotation}
\be
\cS\left[g_q^{-1}\,f\,g_{f^{-1}\cdot q}\right]
\equiv
W_q[f].
\label{wig}
\ee
Its appearance represents the fact that, in contrast to the quasi-regular representation, wavefunctions take 
their values not in $\CC$, but in some more general ``internal'' Hilbert space $\cE$ carrying a 
representation $\cS$ of $H$.\\

When 
interpreting induced representations as particles, $H$ is typically a group of spatial rotations 
combined with space-time translations,
the space $\cE$ consists of spin degrees of freedom, and $\cS$ determines 
the value of spin. In that context the operator (\ref{wig}) is known as the \it{Wigner rotation} associated 
with $f$ at momentum $q$. The quasi-regular representation (\ref{qreg}) can thus be seen as a ``scalar'' 
induced representation, as opposed to the spinning case (\ref{ind}). Wigner rotations are trivial for scalar 
particles. Note that because $\cS$ is a representation, Wigner rotations satisfy the property 
$W_q[fg]=W_q[f]\,W_{f^{-1}\cdot q}[g]$.

\paragraph{Remark.} We mentioned on page \pageref{pagesiboo} that generic homogeneous manifolds do not admit 
continuous families of standard boosts, which invalidates the global well\--de\-fi\-ni\-te\-ness of the 
Wigner rotation 
(\ref{wig}). This problem can be cured by reformulating induced representations in terms of wavefunctions 
defined on the group manifold $G$ rather than $G/H$ (see e.g.\ \cite{barut1986theory}).\i{wavefunction!on 
group manifold}\i{induced representation!in terms of fcts on a group} In this thesis 
we systematically use the homogeneous space viewpoint (\ref{ind}), as it will suffice for all cases of 
interest below. This being said, note that the reformulation in terms of wavefunctions on $G$ is useful for 
certain applications of three-dimensional higher-spin theories \cite{David:2009xg,Gaberdiel:2010ar} due to 
the relation between induced representations and harmonic analysis\i{harmonic analysis} on homogeneous spaces 
\cite{Camporesi:1990wm,Camporesi:1995fb}.

\subsection{Properties of induced representations}
\label{senalyz}

Induced representations have a number of important properties that we now explore. We first show that the 
definition (\ref{ind}) is ``robust'' in that it depends neither on the choice of the measure $\mu$, nor on 
the choice of standard boosts (\ref{stdb}). Then we turn to the behaviour of induced representations under 
operations such as direct sums and tensor products.

\subsubsection*{Robustness}

Formula (\ref{ind}) depends not only on the inducing 
data (the group $G$, its subgroup $H$ and a spin representation $\cS$), but also on the measure $\mu$ 
on 
$\cM\cong G/H$ and on the choice of a family of standard boosts $g_q$. Naively, one expects all 
these parameters to affect the representation. We now show that this is 
not the case.\i{robustness of induced reps}\\

As far as the measure is concerned, one readily verifies that two induced representations defined with 
the same inducing data and the same standard boosts but different, though equivalent, quasi-invariant 
measures, are unitarily equivalent. The proof is essentially the same as for 
quasi-regular representations (see eq.\ (\ref{inter})), 
and the intertwiner is the map (\ref{isom}). As regards standard boosts, a similar result 
holds:

\paragraph{Proposition.} Let $\cM\cong G/H$, $\cS$ a spin representation of $H$, $\sH=L^2(\cM,\mu,\cE)$. Let 
$g:\cM\rightarrow G:q\mapsto g_q$ and $g':\cM\rightarrow G:q\mapsto g'_q$ be 
two continuous families of standard boosts and call $\cT,\cT'$ (respectively) the corresponding induced 
representations of $G$.\i{standard boost} Then there is a unitary operator
\be
\cV:
\sH\rightarrow\sH:
\Psi\mapsto\cV\cdot\Psi
\qquad\text{with}\qquad
\left(\cV\cdot\Psi\right)(q)
=
\cS\left[g_q^{-1}\cdot g'_q\right]\Psi(q)
\label{alf}
\ee
that intertwines $\cT$ and $\cT'$, which are therefore unitarily equivalent:\i{intertwiner}
\be
\cT[f]\circ\cV=\cV\circ\cT'[f]
\qquad\forall\,f\in G.
\label{equiva}
\ee

\begin{proof}
The fact that $\cV$ is a unitary operator follows from unitarity of $\cS$. Property (\ref{equiva}) then 
follows from the definitions (\ref{ind}) and (\ref{alf}).
\end{proof}

A corollary of these observations on robustness is that one may unambiguously say ``the'' representation of 
$G$ induced by the representation $\cS$ of $H$, without any reference to the measure or to the choice of 
standard boosts.

\paragraph{Remark.} The transformation (\ref{alf}) may be seen as a gauge transformation with gauge 
group $H$. Indeed the combination $g_q^{-1}g'_q\in H$ can depend on $q$ in an arbitrary way, owing 
to one's
freedom in the choice of standard boosts. It acts on wavefunctions as a momentum-dependent transformation 
$\Psi\mapsto\cV\cdot\Psi$ given by the representation $\cS$, and each such transformation maps the system 
on a unitarily equivalent one.\i{holonomy}\i{Wigner rotation!as holonomy} In fact, the differentiation of 
the operator $\cS[g_q^{-1}fg_{f^{-1}\cdot q}]$ 
defines a gauge 
field on $G/H$ valued in the Lie algebra $\mh$ of $H$, and the Wigner rotation itself may be seen as a 
holonomy (see e.g.\ 
\cite{Aravind,Williamson:2011vq}).

\subsubsection*{Operations on induced representations}

We now study the behaviour of induced representations under standard operations such as conjugation, direct 
sums and the like. The proofs are omitted and we refer to \cite{barut1986theory} for details.\\

Let $\cE$ be a Hilbert space with scalar product $(\cdot|\cdot)$. We call \it{conjugation} the map 
$C:\cE\rightarrow\cE^*_{\text{cts}}:v\mapsto(v|\cdot)$, where $\cE^*_{\text{cts}}$ denotes the space of 
continuous linear functionals\footnote{Recall that any continuous linear 
functional on a Hilbert space $\cE$ is a scalar product $(v|\cdot)$ for some fixed vector $v\in\cE$.} on 
$\cE$. Then, if $\cS$ is a unitary representation acting on $\cE$, its \it{conjugate 
representation}\i{conjugate representation}\i{representation!conjugate} is $\overline\cS\equiv 
C\circ\cS\circ C^{-1}$. In the context of induced representations, one can then show that\footnote{The symbol 
$\sim$ denotes unitary equivalence of representations.}
$\text{Ind}_H^G(\overline\cS)
\sim
\overline{\text{Ind}_H^G(\cS)}$,
i.e.\ the representation induced by the conjugate of $\cS$ is unitarily equivalent to the conjugate of the 
representation induced by $\cS$.\\

One can similarly show that induced representations behave well under direct sums and tensor 
products thanks to the unitary 
equivalences
\be
\begin{array}{rcl}
\displaystyle\text{Ind}_H^G(\cS_1\oplus\cS_2) & \sim & 
\text{Ind}_H^G(\cS_1)\oplus\text{Ind}_H^G(\cS_2)\,,\\[.3cm]
\displaystyle\text{Ind}_H^G(\cS_1\otimes\cS_2) & \sim & \text{Ind}_H^G(\cS_1)\otimes\text{Ind}_H^G(\cS_2)\,.
\end{array}
\label{indsum}
\ee
As a corollary, if $\cS$ is reducible, then $\text{Ind}_H^G(\cS)$ is reducible. 
The 
converse is not true; for instance, if $H=\textbf{1}$ is the trivial subgroup with $\cS$ the 
irreducible trivial representation, then $\cT$ is the regular representation, which is generally reducible.\\

One should also check that induction itself is a ``good'' operation on representations. This is guaranteed by 
the theorem of \it{induction in stages}:\i{induction in stages} let $H_1$ 
be a closed subgroup of $H_2$, which itself is a closed subgroup of $G$. Let $\cS$ be a unitary 
representation of $H_1$. Then one has the following unitary equivalence 
of representations:
\be
\text{Ind}_{H_1}^G(\cS)\sim\text{Ind}_{H_2}^G\big(\text{Ind}_{H_1}^{H_2}(\cS)\big).
\nn
\ee
In other words, inducing directly from $H_1$ to $G$, or from $H_1$ to $H_2$ and then to $G$, are the same 
operations.

\subsection{Plane waves}
\label{suseplaw}

For practical purposes it is convenient to rewrite formula (\ref{ind}) in a basis of \it{plane wave 
states}. In the relativistic context, they represent particles with definite momentum.

\subsubsection*{Delta functions}

Let $\cM\cong G/H$ be a homogeneous space, $\mu$ a quasi-invariant measure on $\cM$. Pick a point 
$k\in\cM$. We define the \it{Dirac distribution}\i{delta function} $\delta_k$ 
at $k$ associated with $\mu$ as the distribution such that $\langle\delta_k,\phii\rangle\equiv\phii(k)$ for 
any test function $\phii$ 
on $\cM$. Equivalently, we introduce a ``Dirac delta function'' $\delta(k,\cdot)$ such that
\be
\langle\delta_k,\phii\rangle
=
\int_{\cM}d\mu(q)\delta(k,q)\phii(q)
\equiv\phii(k).
\label{didi}
\ee
Thus the distribution $\delta_k$ acts on a test function $\phii(\cdot)$ by integrating it against 
the delta function $\delta(k,\cdot)$.\\

Note that the definition of the delta function $\delta$ relies on the measure $\mu$ since the combination 
$d\mu(q)\delta(k,q)$ is $G$-invariant by design. To make this explicit, let us denote by $\delta_{\mu}$ the 
delta function associated with $\mu$. If $\rho$ is some positive function on $\cM$ and 
$d\nu(q)=\rho(q)d\mu(q)$
is a new measure, then the delta function $\delta_{\nu}$ associated with $\nu$ differs from $\delta_{\mu}$ by 
a factor $\rho$:
\be
\delta_{\nu}(k,q)=\frac{\delta_{\mu}(k,q)}{\rho(q)}.
\label{deldel}
\ee
In particular, since $\mu$ is quasi-invariant under $G$, for any $f\in G$ we have
\be
\delta_{\mu}(f\cdot k,f\cdot q)
=
\frac{\delta_{\mu}(k,q)}{\rho_f(q)}
\label{sor}
\ee
where $\rho_f$ is the Radon-Nikodym derivative (\ref{rnq}). Thinking of the latter as a kind of 
Jacobian, eq.\ (\ref{sor}) is 
a restatement of the transformation law of the Dirac distribution under changes of 
coordinates.\i{Jacobian!and delta functions} In what follows we shall not indicate explicitly
the dependence of 
$\delta$ on the measure $\mu$.\\

The best known Dirac distribution is the one associated with the translation-invariant Lebesgue 
measure on $\RR^n$. We will encounter this delta function repeatedly so we denote it by $\delta^{(n)}$ to
distinguish it from other Dirac distributions. With that notation, the delta function 
associated with the Lorentz-invariant measure (\ref{lom}) is\i{Lorentz-invariant delta function}
\be
\delta(\textbf{k},\textbf{q})
=
\sqrt{M^2+\textbf{q}^2}\;\delta^{(D-1)}(\textbf{k}-\textbf{q}).
\label{deltakix}
\ee

\subsubsection*{Plane wave states}

\paragraph{Definition.} Let $\{e_1,...,e_N\}$ be a countable orthonormal basis\footnote{We are assuming that 
$\cE$ is a separable Hilbert space; $N$ may be infinite.} of $\cE$. For $\ell\in\{1,...,N\}$ and $k\in\cM$, 
we call \it{plane wave state} with spin $\ell$\i{plane 
wave} and momentum $k$ the wavefunction
\be
\Psi_{k,\ell}(q)
\equiv
e_{\ell}\,\delta(k,q)
\label{pwave}
\ee
where $\delta$ is the Dirac distribution associated with the measure $\mu$ on $\cM$.\\

In non-relativistic quantum mechanics, if we were describing a particle on the line $\cM=\RR$, a plane 
wave would typically be one of the states $\Psi_x=|x\rangle$ representing a particle located at the 
point $x\in\RR$ (with infinite momentum uncertainty). In the dual, momentum-space picture, a plane wave 
would be a state $\Psi_k=|k\rangle$ with definite momentum $k$ (but infinite position uncertainty). Our 
terminology is motivated by the latter viewpoint. In the Poincar\'e case a wavefunction (\ref{pwave}) will 
describe a particle with definite spin projection $\ell$ and energy-momentum 
$k$, i.e.\ a typical asymptotic state in a scattering experiment.\i{asymptotic state} With the notation of 
(\ref{biphi}), for instance, the space $\cE$ of spin degrees of freedom is two-dimensional and has a basis 
$\{|+\rangle,|-\rangle\}=\{e_1,e_2\}$ consisting of states with definite spin along the vertical axis.\\

Scalar products of plane waves can be evaluated
thanks to the definition (\ref{scall}). Using the fact that the basis of $e_k$'s is orthonormal and the 
definition (\ref{didi}) of the 
delta function, one finds
\be
\left<\Psi_{k,\ell}|\Psi_{k',\ell'}\right>
=\delta(k,k')\delta_{\ell\ell'}\,.
\label{pwavs}
\ee
This property allows us to see in which sense plane waves form a ``basis'' of the Hilbert space 
$\sH=L^2(\cM,\mu,\cE)$. Indeed, any wavefunction $\Phi:\cM\rightarrow\cE$ can be written as 
\be
\Phi(q)=\sum_{\ell=1}^N\Phi_{\ell}(q)\,e_{\ell}
\qquad\text{where}\qquad
\Phi_{\ell}(k)
=
\langle\Psi_{k,\ell}|\Phi\rangle\,.
\nn
\ee
Removing the argument $q$, this says that any wavepacket $\Phi$ is a superposition of plane waves:
\be
\Phi
=
\int_{\cM}d\mu(k)\sum_{\ell=1}^N\Phi_{\ell}(k)\Psi_{k,\ell}
=
\int_{\cM}d\mu(k)
\sum_{\ell=1}^N
\langle\Psi_{k,\ell}|\Phi\rangle\,\Psi_{k,\ell}\,.
\label{phipsi}
\ee
Note that this can be interpreted as the completeness relation\i{completeness relation}
\be
\II
=
\int_{\cM}d\mu(k)
\sum_{\ell=1}^N
\langle\Psi_{k,\ell}|\cdot\rangle\,\Psi_{k,\ell}
\label{comp}
\ee
where $\II$ is the identity operator. In the more common (but less precise) Dirac notation this would 
be a sum of 
projectors $|\Psi_{k,\ell}\rangle\langle\Psi_{k,\ell}|$. For example, for a particle on the real line, the 
Dirac form of this 
completeness relation would read $\II=\int_{\RR}dx|x\rangle\langle x|$ in position space, or 
$\II=\int_{\RR}dk|k\rangle\langle k|$ in momentum space. We will show in section \ref{sysim} that the 
existence of a family of projectors associated with $\cM$ is one of 
the key properties of induced representations.\\

From the construction of plane waves we see that induced representations are just an 
upgraded version of 
one-particle quantum mechanics, with extra freedom in the choice of the space $\cM$, the group $G$, and the 
spin states contained in $\cE$. This observation will guide us in developing our intuition of induced 
representations, especially in part III of the thesis.

\paragraph{Remark.} The quantity (\ref{pwave}) is not a square-integrable function on 
$\cM$ and therefore does not, strictly speaking, belong to the Hilbert space. The same problem arises in 
standard quantum mechanics, where the states $|x\rangle$ or $|k\rangle$ 
form a ``basis'' only in a weak sense. Intuitively one can think of plane waves (\ref{pwave}) as 
idealizations of Gaussian wavefunctions centred at $k$ in the limit where their spread 
goes to zero. A more rigorous way to include such states is to work with so-called 
\it{rigged Hilbert spaces}\i{rigged Hilbert space}\i{Hilbert space!rigged} (see e.g.\ 
\cite{Antoine1998,delaMadrid}), which are 
designed so as to include both standard square-integrable functions and 
distributions.

\subsubsection*{Boosted plane waves}

We can now rewrite formula (\ref{ind}) for induced representations in terms of 
plane waves. Choosing a plane wave (\ref{pwave}), for $f\in G$ and $q\in\cM$ we find
\be
\Psi_{k,\ell}(f^{-1}\cdot q)
=
\rho_f(k)\,\Psi_{f\cdot k,\ell}(q)
\label{labb}
\ee
where we used (\ref{sor}) and the property (\ref{rnco}).
Note what we have achieved: in the original definition (\ref{ind}) the argument of $\Psi$ changes between the 
left and the right-hand sides; here, by contrast, the 
argument will be the same, but what changes is the label specifying the momentum of the 
plane wave. Indeed, using (\ref{labb}) in formula (\ref{ind}), we find
\be
\big(\cT[f]\cdot\Psi_{k,\ell}\big)(q)
=
\sqrt{\rho_{f^{-1}}(q)}\rho_f(k)\,
\cS\left[
g_q^{-1}\,f\,g_{f^{-1}\cdot q}
\right]
\cdot
\Psi_{f\cdot k,\ell}(q).
\nn
\ee
Since the plane wave on the right-hand side contains a delta function $\delta(f\cdot k,q)$, we can 
replace all $q$'s in this expression by $f\cdot k$ and remove the argument from both sides. Using once more 
(\ref{rnco}) in $\sqrt{\rho_{f^{-1}}(q)}\rho_f(k)
=\sqrt{\rho_f(k)}$, we end up with\i{induced representation!in terms of plane waves}
\be
\cT[f]\cdot\Psi_{k,\ell}
=
\sqrt{\rho_f(k)}\;
\cS\left[
g_{f\cdot k}^{-1}\,f\,g_k
\right]
\cdot
\Psi_{f\cdot k,\ell}
\label{indsim}
\ee
This formula is the simplest rewriting of the induced representation (\ref{ind}). The only extra improvement 
we could still add is to write as 
$\left(\cS[\cdots]\right)_{\ell\ell'}$ the matrix 
element of the operator $\cS[\cdots]$ between the states $e_{\ell}$ and $e_{\ell'}$, whereupon (\ref{indsim}) 
becomes
\be
\cT[f]\cdot\Psi_{k,\ell}
=
\sqrt{\rho_f(k)}
\left(
\cS\left[
g_{f\cdot k}^{-1}\cdot f\cdot g_k
\right]
\right)_{\ell',\ell}
\cdot
\Psi_{f\cdot k,\ell'}
\label{indsimb}
\ee
with implicit summation over $\ell'$.\\

Formula (\ref{indsim}) gives a geometric picture of the states of an induced representation. Indeed, the 
label 
$k$ spans all points of $\cM$, so we can now view each point of 
$\cM$ as a quantum state. (More precisely, a point of $\cM$ is a family of $\dim(\cE)$ linearly 
independent states.) Two different points of $\cM$, say 
$k$ and $k'$, then correspond to two linearly independent states $\Psi_k$ and $\Psi_{k'}$, and a 
transformation $f$ of $\cM$ mapping $k$ on $k'=f\cdot k$ gives rise to a unitary operator $\cT[f]$ 
relating the corresponding plane waves. This is a 
\it{geometrization} of representation theory: we can ``see'' each linearly independent state of the 
representation $\cT$ as a point of 
$G/H$. This observation is at the core of the \it{orbit 
method},\i{orbit method}\i{geometric quantization}\i{representation!geometrization} which consists in 
quantizing suitable homogeneous 
manifolds 
to obtain 
unitary group representations (see chapter \ref{c3}). In three-dimensional gravity, the phase 
space of gravitational perturbations will turn out to be precisely such a 
homogeneous manifold, and its quantization will produce a Hilbert space of ``soft'' or ``boundary 
gravitons''. We will address these questions in chapter \ref{AdS3} and in part III of the thesis.

\section{Characters}
\label{secharapou}

In this section we describe the characters associated with induced representations. We start by motivating 
and defining characters in general terms, before proving the Frobenius formula. We end by discussing the 
relation between characters and fixed point theorems.

\subsection{Characters are partition functions}

Unitary representations may be seen as general models of symmetric quantum systems: any system invariant 
under 
a certain symmetry group $G$ forms a (generally reducible, generally 
projective) unitary representation of $G$. Accordingly, symmetry generators provide natural observables 
in the system, and one may ask about the properties of these observables --- typically, about their 
spectrum.\\

When a system is invariant under time translations, for instance, the corresponding symmetry generator is the 
Hamiltonian operator $H$.\i{Hamiltonian} The information about its spectrum is captured by the canonical 
\it{partition 
function}\footnote{The notation ``$Z$'' stands for the German word \it{Zustandssumme}, meaning ``sum 
over states.''}\i{partition function}
\be
Z(\beta)
=
\text{Tr}\left(e^{-\beta H}\right),
\label{canoz}
\ee
where $\beta$ is the inverse of the temperature. If the system admits extra symmetries such as, say, 
rotations, one can look for the maximal set of mutually commuting symmetry generators $Q_a$ and switch 
on their chemical potentials $\mu_a$.\footnote{Here the index $a$ runs from one to $r$, the latter being 
essentially the rank of the symmetry group.} The spectrum of these new operators, together with $H$, is 
then contained in the grand canonical partition 
function
\be
Z(\beta,\mu_1,...,\mu_r)
=
\text{Tr}\left(
\exp\bigg[-\beta\Big(H-\sum_{a=1}^r\mu_aQ_a\Big)\bigg]
\right).
\label{gc}
\ee
Now 
suppose we take $\beta$ to be purely imaginary (while keeping the $\mu_a$'s real) in this expression. Then 
the 
operator inside the trace is unitary, since it is an exponential of anti-Hermitian operators. In 
fact, it is a symmetry transformation acting in the Hilbert space
according to some unitary representation $\cT$, so we can write
\be
Z(\beta,\mu_1,...,\mu_r)
=
\text{Tr}\left(\cT[f]\right)
\nn
\ee
for some element $f$ belonging to the symmetry group $G$. This motivates the following definition:

\paragraph{Definition.} Let $\cT$ be a representation of a group $G$ in a complex vector space $\sH$. The 
\it{character} of that representation\i{character} is the map\footnote{The 
terminology of ``characters'' is due to Weber and Frobenius, and stems from the fact that irreducible 
representations of finite groups are wholly characterized by their character (see e.g.\ 
\cite[p.117]{Dieudonne} or \cite[p.783]{Gowers}).}
\be
\chi:G\rightarrow\CC:f\mapsto\chi[f]\equiv\text{Tr}\left(\cT[f]\right).
\label{char}
\ee
\vspace{.1cm}

This definition ensures that $\chi[f]$ is independent of the basis of $\sH$ used to evaluate it. As an 
application, recall 
that two group elements $f$ and $f'$ are conjugate if there exists an element 
$g\in G$ such that $f'=gfg^{-1}$, and that the conjugacy class of $f$ is\i{conjugacy class}
\be
[f]\equiv
\big\{
g\,f\,g^{-1}
\big|
g\in G\big\}.
\nn
\ee
Thus, formula (\ref{char}) ensures that
characters are \it{class functions}\i{class function} in the sense that $\chi[f]$ only depends on the 
conjugacy class of $f$, and not on $f$ itself:
\be
\chi[f]=\chi[g\,f\,g^{-1}].
\label{plainmath}
\ee

\paragraph{Remark.} The definition (\ref{char}) suggests that characters are functions on $G$. While 
this is true for finite-dimensional representations, it is \it{not} true in infinite-dimensional ones. In 
fact, characters should not be seen as functions, but rather as \it{distributions} 
\cite{sternberg1995group}. Similarly to our dealing with Dirac 
distributions as if they were ``delta 
functions'', we will not take such mathematical subtleties into account.

\subsection{The Frobenius formula}

Our derivation of the character formula for induced representations is inspired by \cite{joos1968}, although 
the formula itself appears in many textbooks on group theory; see 
e.g.~\cite{kirillov1976elements}.

\paragraph{Theorem.} The character of the induced representation $\cT=\text{Ind}_H^G(\cS)$ defined by 
(\ref{ind}) is given by the \it{Frobenius formula}\i{Frobenius formula}\i{character!for induced 
representation}\i{induced representation!character}
\be
\chi[f]
=
\boxed{
\text{Tr}\left(\cT[f]\right)
=
\int_{\cM}d\mu(k)\,\delta(k,f\cdot k)\,\chi_{\cS}[g_k^{-1}\,f\,g_k]
}
\label{frob}
\ee
where $\mu$ is a quasi-invariant measure on $\cM\cong G/H$, $\delta$ is the associated Dirac distribution, 
the $g_k$'s are standard boosts, and $\chi_{\cS}$ is the character of $\cS$.

\begin{proof}
Let $f\in G$ and let $\cT[f]$ be the associated unitary operator (\ref{ind}). We work in 
the basis of plane wave states (\ref{pwave}) so that the trace of $\cT[f]$ reads
\be
\chi[f]
=
\text{Tr}\left(\cT[f]\right)
=
\int_{\cM}d\mu(k)\sum_{\ell=1}^N\big<\Psi_{k,\ell}\big|\cT[f]\cdot\Psi_{k,\ell}\big>
\nn
\ee
where we ``sum over momenta'' thanks to the measure $\mu$ on $\cM$. Now using (\ref{indsim}) and the scalar 
products (\ref{pwavs}), we find
\begin{eqnarray}
\label{klik}
\chi[f]
& = &
\int_{\cM}d\mu(k)
\sqrt{\rho_f(k)}
\;
\sum_{\ell=1}^N
\langle\psi_{k,\ell}|\cS[g_{f\cdot k}^{-1}\cdot f\cdot g_k]
\psi_{f\cdot k,\ell}\rangle\\
& \refeq{pwavs} &
\int_{\cM}d\mu(k)
\sqrt{\rho_f(k)}\,
\delta(k,f\cdot k)
\;
\sum_{\ell=1}^N
\left(\cS[g_{f\cdot k}^{-1}\cdot f\cdot g_k]\right)_{\ell\ell}\,.\nn
\end{eqnarray}
Here the delta function $\delta(k,f\cdot k)$ allows us to trade $f\cdot k$ for $k$. In 
particular the Radon-Nikodym derivative $\rho_f(k)=d\mu(f\cdot 
k)/d\mu(k)$ reduces to unity. One then recognizes the sum $\sum_{\ell=1}^N
\left(\cS[\cdots]\right)_{\ell\ell}\equiv\chi_{\cS}[\cdots]$
as the character of $\cS$, and eq.\ (\ref{frob}) follows.
\end{proof}

The Frobenius formula (\ref{frob}) embodies the geometrization of 
representation theory mentioned at the end of section \ref{suseplaw}: the trace of an 
operator has now become an integral over a (subset of a) homogeneous space. That integral can be interpreted 
as a sum of characters of $\cS$. Before studying this formula 
further, we need to check that it satisfies the basic properties of a character.\\

First, since it is the 
character of an induced representation and since the latter is independent (up to unitary equivalence) of the 
choice of the measure $\mu$, the same should be 
true of expression (\ref{frob}). To see that this is indeed the case, recall that the 
combination $d\mu(k)\delta(k,\cdot)$ is invariant under changes of measures (as follows from the
definition (\ref{didi}) of the Dirac distribution), which then implies invariance of the character. Note in 
particular that the Radon-Nikodym derivative of the measure $\mu$ does not appear in (\ref{frob}). Secondly, 
induced representations are independent of the choice of standard boosts $g_q$; using the fact that the 
character of $\cS$ is a class function, one readily verifies that (\ref{frob}) is also independent of that 
choice. Finally, recall from (\ref{plainmath}) that characters are class functions; using (\ref{rnco}) one 
verifies that this is indeed the case with formula (\ref{frob}). Note one crucial implication of this 
fact: because of the term $\chi_{\cS}[g_k^{-1}fg_k]$ in (\ref{frob}), the character $\chi[f]$ vanishes 
if $f$ is not conjugate to an element of $H$. In other words the character of the induced representation 
$\text{Ind}_H^G(\cS)$ is supported on the points of $G$ whose conjugacy class intersects $H$.

\paragraph{Remark.} Since the character $\chi_{\cS}$ of the spin 
representation is a class function, one is naively tempted to pull the term 
$\chi_{\cS}[g_k^{-1}fg_k]$ out 
of the integral (\ref{frob}), as the notation suggests that $f$ is conjugate to $g_k^{-1}fg_k$. This is 
\it{not} true, because for generic $g,g'\in G$ one has
$\chi_{\cS}[g^{-1}fg]\neq\chi_{\cS}[g'^{-1}fg']$. As a consequence, the integral (\ref{frob}) is generally 
non-trivial.

\subsection{Characters and fixed points}

Formula (\ref{frob}) is one of the key results of this chapter. Its two most salient features are (i) the 
fact that the character of $\cT$ is completely specified by that of $\cS$ and the space $\cM$, and (ii) the 
fact that it is an integral over the points of $\cM$ that are left fixed by $f$.\i{character!and fixed 
points}\i{Frobenius formula!and fixed points}\i{localization} At first sight the 
latter 
observation is a surprise: there is no obvious reason why a sum over all states of the induced representation 
would collapse to an integral over fixed points of $f$, though in practice this is due to the scalar products 
$\langle\Psi_k|\Psi_{f\cdot k}\rangle$ in the 
trace (\ref{klik}). This collapse is an instance 
of \it{localization}:\i{localization}\i{fixed point} an integral
localizes to a small subset of points in $\cM$, so that the evaluation of (\ref{frob}) becomes child's play. 
We shall encounter this situation with the BMS$_3$ group in section 
\ref{sebmschar}.

\begin{figure}[H]
\centering
\includegraphics[width=0.20\textwidth]{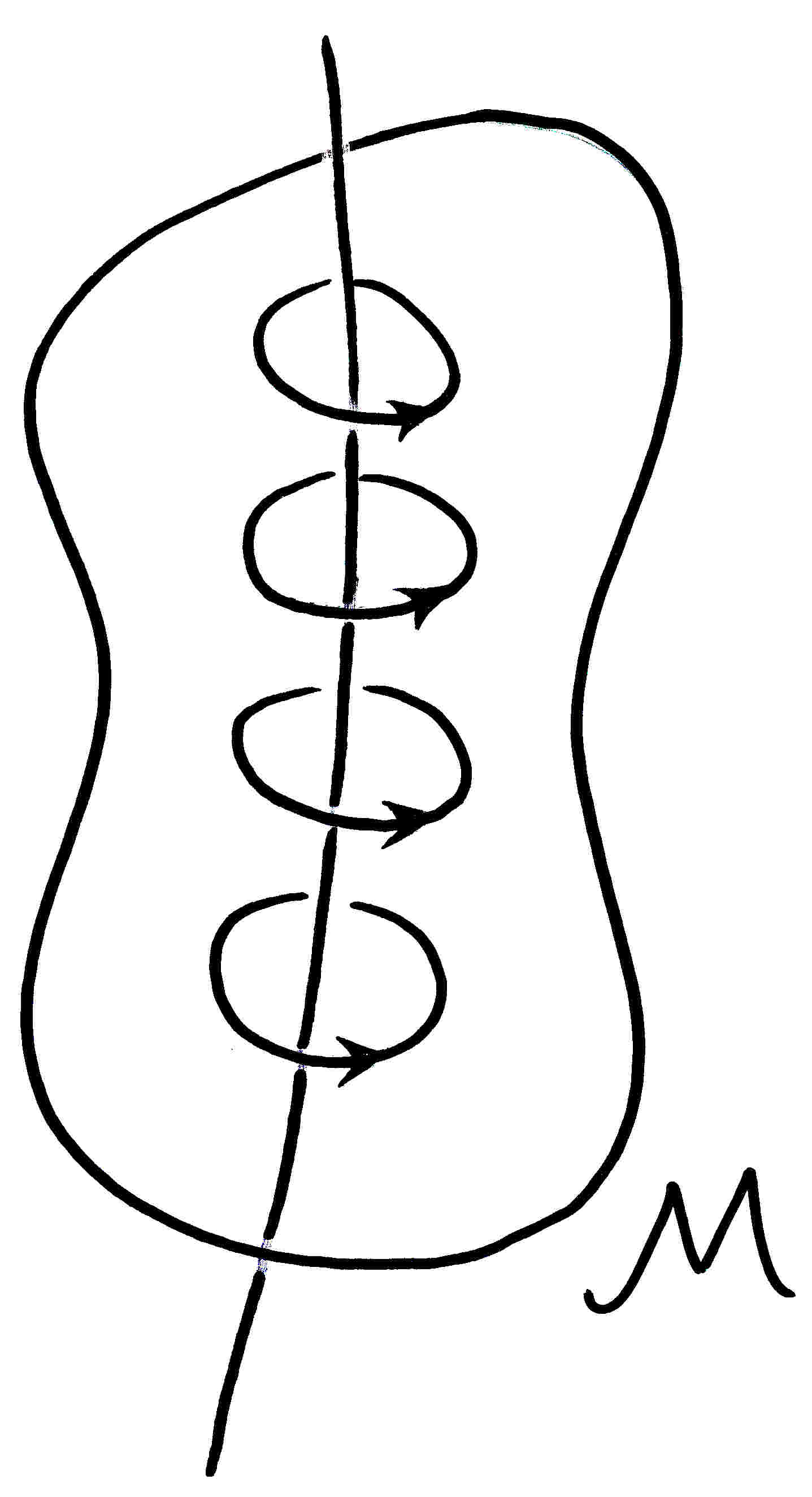}
\caption{A manifold $\cM$ acted upon by a rotation around some axis. The points that belong to the axis are 
the only ones left fixed by the rotation, and are therefore the only ones that contribute to the integral of 
formula (\ref{frob}).\label{PFix}}
\end{figure}

\paragraph{Remark.}\label{pagefix} The relation between characters of group representations and fixed point 
theorems is much deeper and more general than the superficial description given here. Indeed one can show 
\cite{AtiyahBott} (see also \cite{sternberg1995group}) that (\ref{frob}) coincides with the Lefschetz number 
of $\cT[f]$ when the latter is seen as an endomorphism acting on a space of $\cE$-valued sections on 
$\cM$.\i{Lefschetz number} In turn, the fact that $\cT[f]$ is derived by (\ref{ind}) from a diffeomorphism 
action of $f$ on $\cM$ turns out to imply that its Lefschetz number is given by the Atiyah-Bott fixed point 
theorem \cite{AtiyahBott00}.\i{Atiyah-Bott theorem}

\begin{advanced}
\section{Systems of imprimitivity}
\label{sysim}
\end{advanced}

This technical section is for advanced reading: other than for a key corollary that implies the 
exhaustivity of induced representations for semi-direct products, it is inconsequential to the remainder 
of the thesis and may be skipped in a first reading.\\

We saw in eq.~(\ref{comp}) that the identity operator $\II$ can be written as an integral of projectors 
$\Psi_k\,\langle\Psi_k|\cdot\rangle=|\Psi_k\rangle\langle\Psi_k|$. This leads to a seemingly random idea: why 
not combine these projectors into more general operators? For example, if $U$ is any Borel subset of $\cM$, 
we 
can associate with it a projection operator\i{projector}
\be
P_U
\equiv
\int_Ud\mu(k)\,
\sum_{\ell=1}^N
\langle\Psi_{k,\ell}|\cdot\rangle\,\Psi_{k,\ell}\,.
\label{proj}
\ee
In that 
language the identity operator is $\II=P_{\cM}$. As it turns out this idea is one of the key 
properties of induced 
representations, and sparked the whole development of the theory by Mackey in the fifties 
\cite{Mackey01,Mackey02,Mackey03}. In particular it leads to the \it{imprimitivity theorem}, which 
roughly states that \it{any} representation that admits a suitable 
family of projectors (\ref{proj}) is necessarily induced. An important corollary of that result is the fact 
that 
all irreducible unitary representations of semi-direct products are induced.\\

The plan of this section is the following. We first define the notion of systems of imprimitivity as suitable 
families of projection operators, and show that any induced representation admits such a family. We then 
state 
(without proof) the imprimitivity theorem, which we eventually use to define a restricted notion of 
equivalence for induced representations. The presentation is based on 
\cite{barut1986theory,mackey1968induced}, but our approach will be 
heuristic at times; we refer to \cite{kaniuth,varadarajan2007geometry} for a mathematically rigorous 
presentation.\\

\subsection{Projections and imprimitivity}

Here we describe the operators (\ref{proj}) in the framework of projection-valued measures and show that they 
form a system of imprimitivity.

\subsubsection*{Projection-valued measures}

Let us put (\ref{proj}) in a more general context. Observe that, given 
the Borel set $U$, the projector $P_U$ acts on wavefunctions 
by setting them to zero everywhere outside of $U$:
\be
\big(P_U\cdot\Psi\big)(q)
=
\left\{
\begin{array}{ll}
\Psi(q) & \text{if }q\in U,\\
0 & \text{otherwise.}
\end{array}
\right.
\label{projo}
\ee
The construction of such projectors motivates the following definition:

\paragraph{Definition.} Let $\cM$ be a manifold, $\sH$ a Hilbert 
space, $\text{End}(\sH)$ the space of linear operators in $\sH$. Then a 
\it{projection-valued measure}\i{projection-valued measure}\i{measure!projection-valued} on $\cM$ with 
respect to $\sH$ is a map
\be
P:\left\{\text{Borel subsets of $\cM$}\right\}\rightarrow\text{End}(\sH):
U\mapsto P_U
\label{mimip}
\ee
satisfying the following properties:
\begin{itemize}
\item $P_{\cM}$ is the identity operator $\II$ in $\sH$.
\item For any pair of Borel sets $U$ and $V$, we have $P_{U\cap V}=P_UP_V$; in particular each $P_U$ 
is a projector.
\item The map $P$ is $\sigma$-additive in the sense that, if $U_1$, $U_2$, etc.~are disjoint Borel 
sets,\i{sigma additivity@$\sigma$-additivity}
\be
P_{U_1\cup U_2\cup\cdots}=P_{U_1}+P_{U_2}+\cdots.
\label{sigmadd}
\ee
\end{itemize}
\vspace{.1cm}

The terminology here is inspired by measure theory: the map (\ref{mimip}) 
is an operator analogue of (\ref{mimi}) and property (\ref{sigmadd}) corresponds to (\ref{sigmad}). Thus a 
projection-valued measure measures the ``size'' of a subset $U$ not by a real number $\mu(U)$, but by the 
rank 
of a 
projector $P_U$.\\

It is easy to verify that the projectors (\ref{proj}) define a projection-valued measure on $\cM$ with 
respect 
to $\sH$. One can build such a family for any induced representation. In terms of plane waves 
(\ref{pwave}), this measure infinitesimally reads
\be
dP(k)
=
d\mu(k)\,
\sum_{\ell=1}^N
\Psi_{k,\ell}
\langle\Psi_{k,\ell}|\cdot\rangle
=
d\mu(k)\,
\sum_{\ell=1}^N
|\Psi_{k,\ell}\rangle\langle\Psi_{k,\ell}|
\equiv
d\mu(k)\II_k
\label{dproj}
\ee
at any $k\in\cM$. Here we have used both our notation and the standard Dirac one; we have also introduced an 
operator $\II_k=\sum_{\ell=1}^N
|\Psi_{k,\ell}\rangle\langle\Psi_{k,\ell}|$ such that the identity operator in $\sH$ is an integral 
$\II=\int_{\cM}d\mu(k)\II_k$. Analogously to (\ref{mimibis}), the operator $P_U$ is the 
integral of $dP$ over $U$. For a 
particle on the real line, for example, the projection-valued measure in momentum space would read 
$dP=dk\,|k\rangle\langle k|$ with $k\in\RR$.

\subsubsection*{Systems of imprimitivity}

There is one property that makes the projection-valued measure (\ref{proj}) very special. Namely, the 
transitive action of $G$ on $\cM$ gives rise to an 
action (\ref{indsim}) on wavefunctions; the latter, in turn, yields an action on the projectors 
$|\Psi_k\rangle\langle\Psi_k|$. So the fact that (\ref{proj}) acts in the space of a representation provides 
a relation between the geometry of $\cM$ and the action of $G$ on the projection-valued measure, which
motivates the following definition:

\paragraph{Definition.} Let $\cM$ be a manifold, $G$ a Lie group acting on $\cM$. Let $\cT$ be a unitary 
representation of $G$ in a Hilbert space $\sH$. Then a \it{system of 
imprimitivity}\i{imprimitivity}\i{system of imprimitivity} for $\cT$ based on $\cM$ is a projection-valued 
measure $P$ on 
$\cM$ with respect to $\sH$ such that
\be
P_{f\cdot U}=\cT[f]\circ P_U\circ \cT[f]^{-1}
\label{imp}
\ee
for all $f\in G$ and any Borel subset $U$ of $\cM$. The system is said to be \it{transitive}
if the action of $G$ on $\cM$ is transitive.\\

In this language the projectors (\ref{proj}) imply that any 
induced representation has a transitive system of imprimitivity:

\paragraph{Proposition.} Let $\cS$ be a unitary representation of a closed subgroup $H$ of $G$, 
$\cT=\text{Ind}_H^G(\cS)$ the corresponding induced representation, and $\cM\cong G/H$.\i{induced 
representation!and imprimitivity} Then 
the associated projection-valued measure (\ref{proj}) is a transitive system 
of 
imprimitivity for $\cT$ based on $\cM$. In the notation (\ref{dproj}) this is to say that
\be
dP(f\cdot k)
=
\cT[f]\circ dP(k)\circ \cT[f]^{-1}
\label{impind}
\ee
for all $k\in\cM$ and any $f\in G$. We shall refer to (\ref{proj}) as the \it{canonical system of 
imprimitivity}\i{canonical system of imprimitivity} of the induced representation $\cT$.

\begin{proof}
Transitivity is obvious, so the only subtlety is proving (\ref{imp}). Let us pick a group element $f\in 
G$ and a Borel set $U\subseteq\cM$. We start from the definition 
(\ref{proj}) and relate $\cT[f]\circ P_U\circ\cT[f]^{-1}$ to $P_{f\cdot U}$ using formula (\ref{indsim}) 
and the fact that $\cT$ is unitary:
\be
\cT[f]\circ P_U\circ\cT[f]^{-1}=
\int_Ud\mu(k)\rho_f(k)\sum_{\ell=1}^N
\cS[g_{f\cdot k}^{-1}\,f\,g_k]
\Psi_{f\cdot k,\ell}
\langle
\cS[g_{f\cdot k}^{-1}\,f\,g_k]
\Psi_{f\cdot k,\ell}|\cdot
\rangle.
\nn
\ee
Here the sum over $\ell$ allows us to cancel the two $\cS[\cdots]$'s by unitarity.
Using also $d\mu(k)\rho_f(k)=d\mu(f\cdot k)$ and renaming 
the integration variable, the right-hand side boils down to $P_{f\cdot U}$.
\end{proof}

\paragraph{Remark.} The word ``imprimitive'' means ``which is not primitive'' and was introduced by Galois 
\cite{Galois} in the context of permutation groups. The action of a group on a set shuffles 
the elements of 
this set, and the action is \it{imprimitive} if these permutations preserve some (non-trivial) partition of 
the set. In the present case the group $G$ acts on the Hilbert space $\sH$ by the induced representations 
(\ref{indsim}), and property (\ref{impind}) says that this action preserves the partition of $\sH$ 
into isomorphic subspaces $\cE_k\cong\cE$ with definite momentum $k$.

\subsection{Imprimitivity theorem}
\label{susempith}

The considerations of the previous pages open the 
door to a highly non-trivial statement, namely the fact that \it{any} representation that admits a transitive 
system of 
imprimitivity is an induced representation:

\paragraph{Imprimitivity theorem.} Let $G$ be a finite-dimensional Lie group, $H$ a 
closed subgroup of $G$. Let $\cT$ be a continuous, unitary 
representation of $G$ in some Hilbert space $\sH$ and let $P$ be a system of imprimitivity for $\cT$ 
on $\cM=G/H$. Then there exists a unitary representation $\cS$ of $H$ in some Hilbert space $\cE$ such that 
the 
pair $(\cT,P)$ is unitarily equivalent to $\big(\text{Ind}_H^G(\cS),P^{\cS}\big)$ where $P^{\cS}$ is the 
canonical 
system of imprimitivity (\ref{proj}) 
associated with 
$\text{Ind}_H^G(\cS)$. More precisely, there exists an isometry
$\cU:
L^2(G/H,\mu,\cE)
\rightarrow
\sH$,
where $\mu$ is a quasi-invariant measure on $G/H$,
that intertwines the representations $\cT$ and $\text{Ind}_H^G(\cS)$ and that satisfies
\be
\cU\circ P_U^{\cS}\circ\cU^{-1}
=
P_U
\label{intert}
\ee
for any Borel set $U\subseteq\cM$.\\

The complete proof of this theorem can be found in \cite{Orsted} and is reproduced 
in \cite{barut1986theory}. Given the representation $\cT$ and the system of 
imprimitivity $P$, the key subtlety is to construct a Hilbert space $\cE$ and a representation 
$\cS$ of $H$ in $\cE$. We will not dwell on this proof any further, but turn now to some of its applications.

\subsubsection*{Equivalent induced representations}

Here we describe a restricted notion of equivalence for induced representations, culminating with the 
observation that two induced 
representations $\cT_1,\cT_2$ are ``equivalent''  if and only if they are induced 
from equivalent representations $\cS_1,\cS_2$. The proofs rely 
crucially on the imprimitivity theorem, but we omit them; they can be found in \cite{barut1986theory}.

\paragraph{Definition.} Let $\cT_1$ and $\cT_2$ be two representations of $G$ induced by some 
representations $\cS_1,\cS_2$ (respectively) of a subgroup $H$. Let their respective carrier spaces be
$\sH_1,\sH_2$, and let $P_1,P_2$ (respectively) be their canonical systems of imprimitivity. Then a linear 
map $A:\sH_1\rightarrow\sH_2$ 
\it{intertwines}\i{intertwiner} the pairs $(\cT_1,P_1)$ and $(\cT_2,P_2)$ if
\be
A\circ\cT_1[f]=\cT_2[f]\circ A
\qquad\text{and}\qquad
A\circ (P_1)_U=(P_2)_U\circ A
\label{intini}
\ee
for any $f\in G$ and any Borel set $U$ in $G/H$.

\paragraph{Equivalence theorem.} Let $\cS_1$ and $\cS_2$ be 
unitary representations of $H$ in the Hilbert spaces $\cE_1$ and $\cE_2$ (respectively). Let $ \cT_1$ and 
$\cT_2$ be the corresponding induced representations, and $P_1,P_2$ the associated canonical 
systems of imprimitivity. Then there exists a (continuous) vector space isomorphism between the space of 
operators intertwining $\cS_1$ and $\cS_2$ and the space of intertwiners between $(\cT_1,P_1)$ and 
$(\cT_2,P_2)$.\\

As a corollary, the space of intertwiners between $(\cT_1,P_1)$ and 
$(\cT_2,P_2)$ contains an isometry if and only if the space of intertwiners between $\cS_1$ and $\cS_2$ does. 
In other words, if we declare that the pairs $(\cT_1,P_1)$ and $(\cT_2,P_2)$ are equivalent once there exists 
an isometry $A$ satisfying (\ref{intini}), then we have\i{induced 
representation!equivalence}\i{equivalent representations}
\be
(\cT_1,P_1)\sim(\cT_2,P_2)
\qquad\text{if and only if}\qquad
\cS_1\sim\cS_2.
\label{coco}
\ee
In particular, if $\cS_1$ and $\cS_2$ are equivalent (in the usual sense), then so are the induced 
representations $\cT_1$ and $\cT_2$ (regardless of their systems of imprimitivity). The converse is not true, 
since the ``if and only if'' of (\ref{coco}) also involves the systems of imprimitivity associated 
with $\cT_1$ and $\cT_2$. In other words, saying that two induced representations are equivalent \it{without} 
saying anything about their systems of imprimitivity is not sufficient to conclude that they are induced from 
the same spin representation $\cS$.\\

As mentioned below eq.~(\ref{indsum}), irreducibility of $\cS$ does not generally imply irreducibility of the 
corresponding induced representation. In the next chapter we shall state a stronger result for semi-direct 
products, but for now we display a theorem that provides a slightly weaker criterion 
for the irreducibility of induced representations.

\paragraph{Definition.} Let $\cT$ be an induced representation, $P$ the associated 
canonical system of imprimitivity. We call the pair $(\cT,P)$ 
\it{irreducible}\i{irreducible representation}\i{induced 
representation!irreducible}\i{representation!irreducible} if the space of 
operators intertwining it with itself consists of multiples of the identity.

\paragraph{Irreducibility theorem.} Let $\cT$ be induced by $\cS$ and let $P$ be its system of imprimitivity. 
Then the pair $(\cT,P)$ is irreducible if and only if $\cS$ is 
irreducible.\\

The latter theorem shows that a suitable notion of irreducibility is preserved along the induction 
process, since an irreducible $\cS$ will lead to an induced representation $\cT$ \it{and} a system of 
imprimitivity $P$ which, together, will be considered irreducible in the above sense. But the theorem does 
\it{not} say that an induced representation $\cT$ on its own is irreducible if it is induced from an 
irreducible $\cS$.

\newpage
~
\thispagestyle{empty}

\chapter{Semi-direct products}
\label{c2bis}
\markboth{}{\small{\chaptername~\thechapter. Semi-direct products}}

In this chapter we introduce semi-direct products such as the Poincar\'e group, the Galilei group and the 
Bargmann group. We 
describe their irreducible unitary representations, which are induced from representations of their 
translation subgroup combined with a so-called \it{little group}. We interpret these representations as 
\it{particles} propagating in space-time and having definite transformation properties under the 
corresponding symmetry group. This picture will be instrumental in our study of the BMS$_3$ group.\\

The plan is as follows. In section \ref{sesemi} we define semi-direct products and introduce 
the key notions of \it{momentum orbits}, little groups and particles. We also explain why irreducible unitary 
representations are always induced, and describe these representations in general terms. The remaining 
sections are devoted to applications of these considerations. In section \ref{relagroup} we describe 
relativistic particles, i.e.\ unitary representations of the Poincar\'e group, with a particular emphasis in 
section \ref{sePoTri} on the three-dimensional setting (which will be useful when dealing with BMS$_3$). 
Section \ref{galisec} is devoted to non-relativistic particles, i.e.\ unitary representations of Bargmann 
groups. Useful 
references include \cite{mackey1968induced,barut1986theory} for the general theory, 
and \cite{weinberg1995,cornwell1984group,Bekaert:2006py} for its application to 
Poincar\'e.

\section{Representations and particles}
\label{sesemi}

In short, a semi-direct product group consists of two pieces: a non-Abelian group $G$ of transformations that 
can be interpreted as ``rotations'' or ``boosts'', and another group $A$ that consists of transformations 
analogous to translations that are acted upon by rotations and boosts. This structure is denoted $G\ltimes A$ 
and is common to the Poincar\'e groups (\ref{poin}) as well as the BMS groups (\ref{BMS})-(\ref{ebms}). In 
this section we define such groups in abstract terms, define the associated notion of ``momenta'' and 
describe their irreducible unitary representations, which we interpret as particles.

\subsection{Semi-direct products}
\label{defsemi}

\paragraph{Definition.} Let $G$ and $A$ be Lie groups; we denote elements of $G$ as $f$, $g$, etc.~and 
those of $A$ as 
$\alpha$, $\beta$, etc. Let
$\sigma:G\times A\rightarrow A:(f,\alpha)\mapsto\sigma_f(\alpha)$
be a smooth action of $G$ on $A$ where each $\sigma_f$ is an automorphism of $A$. Then the \it{semi-direct 
product}\i{semi-direct product} of $G$ and $A$ with respect to $\sigma$ is the 
group denoted
\be
G\ltimes_{\sigma}A
\qquad\text{or}\qquad
G\ltimes A
\label{semid}
\ee
whose elements are pairs $(f,\alpha)$ where $f\in G$ and $\alpha\in A$, with a group operation
\be
(f,\alpha)\cdot(g,\beta)
=
\big(f\cdot g,\alpha\cdot\sigma_f(\beta)\big).
\label{semig}
\ee
\vspace{.1cm}

This definition implies for instance that the inverse of $(f,\alpha)$ is
\be
(f,\alpha)^{-1}=\left(f^{-1},[\sigma_{f^{-1}}(\alpha)]^{-1}\right).
\label{seminv}
\ee
It follows that $A$ is a normal subgroup\i{normal subgroup} of $G\ltimes A$: 
identifying 
$A$ with the set of elements $(e,\alpha)\in G\ltimes A$ (where $e$ is the identity in $G$), one finds
\be
(g,\beta)\cdot(e,\alpha)\cdot(g,\beta)^{-1}
\refeq{semig}
\left(
e,\beta\cdot\sigma_g(\alpha)\cdot\beta^{-1}
\right)
\,\in A.
\label{asas}
\ee
It is equally easy to verify that $G$ is a subgroup of $G\ltimes A$, though it is generally 
\it{not} a 
normal subgroup. Indeed, upon identifying $G$ with the subgroup of $G\ltimes A$ consisting of elements 
$(f,e_A)$ 
(where $e_A$ is the identity in $A$), we 
find
\be
(g,\beta)\cdot(f,e_A)\cdot(g,\beta)^{-1}
\refeq{semig}
\left(gfg^{-1},\beta\cdot\sigma_{gfg^{-1}}(\beta^{-1})\right).
\ee
For this to be an element of $G$, we must require that $\beta\cdot\sigma_{gfg^{-1}}(\beta^{-1})$ 
coincides with $e_A$, which is the statement that 
$\sigma_f(\alpha)=\alpha$ for any $\alpha\in A$. Thus $G$ is a normal subgroup of $A$ if and only if its 
action $\sigma$ is trivial, in which case $G\ltimes A$ is isomorphic to the direct product $G\times A$. From 
now on we always take the action $\sigma$ to be \it{non}-trivial.

\subsubsection*{Rotations and translations}

A case of great interest, both for the general theory and for our specific purposes, occurs when $A$ is a 
\it{vector group}.\i{vector group} By this we mean a vector space endowed with the Abelian group operation 
given by the addition of vectors: $\alpha\cdot\beta\equiv\alpha+\beta$. In that case the identity in 
$A$ is the vanishing vector $e_A=0$.

\paragraph{Definition.} Let $G$ be a Lie group, $A$ a vector 
space, $\sigma$ a representation of $G$ in $A$, and consider the semi-direct product $G\ltimes_{\sigma}A$ 
whose elements are pairs $(f,\alpha)$ with group operation
\be
\boxed{
\Big.
(f,\alpha)\cdot(g,\beta)
=
\big(f\cdot g,\alpha+\sigma_f(\beta)\big)\,.
}
\label{semiop}
\ee
Elements of $G$ are then called \it{rotations}\i{rotation} or \it{boosts}\i{boost} while elements of $A$ 
are \it{translations}.\i{translation}\\

Note that, since $A$ is a vector group, the inverse (\ref{seminv}) of $(f,\alpha)$ is
$(f,\alpha)^{-1}
=
\big(
f^{-1},-\sigma_{f^{-1}}(\alpha)
\big)$.
Relation (\ref{asas}) also simplifies to $(g,\beta)\cdot(e,\alpha)\cdot(g,\beta)^{-1}
=(e,\sigma_g\alpha)$.
From now on the words ``semi-direct product'' and the notation $G\ltimes A$ will always refer to a 
group (\ref{semid}) with $A$ a vector group. (This is why the second factor in
(\ref{semid}) was denoted ``$A$'' in the first place.)\\

The terminology of ``rotations'' and ``translations'' is justified by the semi-direct products commonly 
encountered in 
physics:
\begin{itemize}
\item The \it{Euclidean group}\i{Euclidean group} in $n$ space dimensions takes the form 
(\ref{semid}) where 
rotations span the group $G=\text{O}(n)$ while translations belong to $A=\RR^n$, with the action $\sigma$ of 
rotations on translations given by the vector representation of $\text{O}(n)$.
\item The \it{Poincar\'e group}\i{Poincar\'e group} in $D$ space-time dimensions takes the form 
(\ref{semid}) where rotations and 
boosts span the Lorentz group $\text{O}(D-1,1)$ while space-time translations span $A=\RR^D$ (which is 
sometimes written $\RR^{D-1,1}$); the action $\sigma$ is the 
vector representation of the Lorentz group.
\item The BMS groups (\ref{BMS})-(\ref{ebms}) all take the form 
(\ref{semid}) with $G$ a specific non-Abelian group and $A$ an Abelian vector group of so-called 
``supertranslations''. A similar structure will hold in three space-time dimensions.
\end{itemize}

Note that the definition of $G\ltimes A$ singles out the normal 
subgroup $A$, so $G$ and $A$ live on unequal footings. In particular 
the Lie algebra of $G\ltimes A$ contains a non-trivial Abelian ideal and is \it{not} semi-simple. 
This implies that, in contrast to simple 
Lie groups, the representations of $G\ltimes A$ must 
somehow distinguish the roles of $G$ and $A$ by making them act on the carrier space in radically different 
ways. We will illustrate this in the pages that follow (see e.g.\ formula (\ref{indoo})).

\subsection{Momenta}
\label{susemoma}

Suppose we wish to build unitary representations of a semi-direct product $G\ltimes_{\sigma}A$. Where 
should we start? A simple approach is to note that the restriction to $A$ of any
unitary representation of $G\ltimes A$ is a (reducible) unitary representation of $A$. So instead of 
directly looking for 
representations of $G\ltimes A$, let us consider the simpler problem of building unitary 
representations of the group of translations, $A$.\\

We denote by $A^*$ the vector space dual to $A$.\i{A*@$A^*$ (space of momenta)} It consists of linear forms 
$p:A\rightarrow\RR:\alpha\mapsto\langle p,\alpha\rangle$, which motivates the definition of a bilinear 
pairing\i{dual space}
\be
\langle\cdot,\cdot\rangle:
A^*\times A\rightarrow\RR:
(p,\alpha)\mapsto\langle p,\alpha\rangle.
\label{pair}
\ee
Since $A$ is Abelian, any one of its irreducible unitary 
representations is one-dimensional and takes the form\i{vector group!unitary 
representation}\i{representation!of vector group}\i{unitary representation!of vector group}
\be
\cR:
A\rightarrow\CC:
\alpha\mapsto e^{i\langle p,\alpha\rangle}
\label{eipa}
\ee
for some fixed element $p$ of $A^*$. Indeed, any representation $\cR$ of $A$ 
is such that
\be
\cR[\alpha+\beta]=\cR[\alpha]\,\cR[\beta]
\label{RR}
\ee
where the right-hand side is a composition of linear operators. Assuming that $A$ has a countable 
basis so that $\alpha$ and $\beta$ have components $\alpha^i$, $\beta^i$, the derivative of (\ref{RR}) with 
respect to $\beta^i$ yields
\be
\der_j\cR[\alpha]=i\big(-i\der_j\cR[0]\big)\cR[\alpha]
\label{ava}
\ee
where each $(-i\der_j\cR[0])$ is Hermitian by unitarity. Hence $\cR[\alpha]=\exp[i(-i\der_j\cR[0])\alpha^j]$, 
which can be diagonalized into a direct sum 
of 
multiplicative operators (\ref{eipa}).\\

For example, for the Euclidean group in $n$ dimensions, $\alpha=(\alpha^1,...,\alpha^n)$ is an $n$-component 
vector and 
$\langle p,\alpha\rangle=p_i\alpha^i$ where $p=(p_1,...,p_n)$ is a ``covector''. For the Poincar\'e 
group in $D$ space-time dimensions, $\alpha=(\alpha^0,...,\alpha^{D-1})$ is a $D$-vector and $\langle 
p,\alpha\rangle=p_{\mu}\alpha^{\mu}$ for some energy-momentum covector $(p_0,...,p_{D-1})$.\label{pagemom} 
When interpreting the corresponding unitary representations as ``particles'', the quantity $p$ represents the 
particle's 
\it{momentum}\i{momentum} vector.\footnote{More precisely the momentum \it{vector} is obtained by raising 
the indices of the \it{covector} $p$ thanks to some metric on $A$ left invariant by $G$, but we will keep 
referring to $p$ as the ``momentum vector''.} Accordingly, from now on the dual space $A^*$ 
will be called the \it{space of momenta}, and its elements will be denoted as $p$, $q$ or $k$. In 
the BMS$_3$ groups, translations and momenta are vectors with infinitely many components. Note that 
two irreducible representations of the form (\ref{eipa}) are equivalent if and only if their momenta 
coincide.

\paragraph{Remark.} In proving that all irreducible unitary representations of $A$ takes the
form (\ref{eipa}), we relied crucially on eq.~(\ref{RR}). The latter assumes that $\cR$ is an 
exact representation of $A$, which is not a restrictive assumption as long as there exists no 
central extension of $G\ltimes A$ that turns $A$ into a non-Abelian group. The Poincar\'e groups, the 
Bargmann 
groups and the BMS$_3$ group all satisfy this property, so one may safely assume that $A$ is Abelian even 
upon 
switching on central extensions. By contrast, the symmetry group of warped conformal field theories\i{warped 
CFT} \cite{Detournay:2012pc} is a semi-direct product whose central extension 
makes translations non-Abelian \cite{Afshar:2015wjm}.

\subsection{Orbits and little groups}
\label{suserbilly}

We now ask how irreducible, unitary representations of the Abelian group $A$ are embedded in unitary 
representations of the larger group $G\ltimes_{\sigma}A$. Let $\cT$ be a unitary representation of the 
latter; then its restriction to $A$ is, in general, reducible. It is typically a direct sum, or rather 
a 
direct integral, of irreducible representations (\ref{eipa}):\footnote{Our notation here is not 
mathematically precise; we refer to \cite{barut1986theory} for a more rigorous treatment.}\i{direct integral}
\be
\cT[(e,\alpha)]
=
\int_{\cO}d\mu(q)\,e^{i\langle q,\alpha\rangle}\,\II_q
\qquad\forall\,\alpha\in A.
\label{tint}
\ee
Here $\cO$ is a certain subset of $A^*$, $\mu$ is some measure on $\cO$, and each $\II_q$ is an identity 
operator 
acting in a suitable Hilbert space at momentum $q$. The question then is:
\be
\begin{array}{c}
\text{\it{What is the minimal set $\cO$ of momenta}}\\
\text{\it{appearing in the decomposition (\ref{tint})?}}
\end{array}
\label{question}
\ee
The answer will lead to the notion of \it{orbits}; hence the notation ``$\cO$'' in (\ref{tint}).\\

Call $\sH$ the Hilbert space of the representation $\cT$. Suppose there exists a 
subspace $\cE$ of $\sH$ where translations are represented by multiplicative operators (\ref{eipa}) 
with a certain momentum $p$:
\be
\cT[(e,\alpha)]\Big|_{\cE}=e^{i\langle p,\alpha\rangle}\,\II_{\cE}
\qquad\forall\,\alpha\in A,
\label{phase}
\ee
where $\II_{\cE}$ is the identity operator in $\cE$. We shall refer to 
this property by saying that the representation $\cT$ ``contains the momentum $p$''. Now pick some group 
element $f\in G$. By virtue (\ref{semiop}) and since $\cT$ is a 
representation, one has
\be
\cT[(e,\alpha)]\cdot\cT[(f,0)]=\cT[(f,0)]\cdot\cT[(e,\sigma_{f^{-1}}\alpha)].
\label{titisig}
\ee
One can then act with both sides of this equation on the space $\cE$; the last term on the right-hand side 
produces a multiplicative operator (\ref{phase}) with $\alpha$ replaced by $\sigma_{f^{-1}}\alpha$. This 
operator is a c-number and therefore commutes with $\cT[(f,0)]$. We conclude that, on the space 
$\cT[(f,0)]\cdot\cE\equiv\cE'$, all translations are again represented by multiplicative operators, but now 
with an additional insertion of $\sigma_{f^{-1}}$ in the phase $\langle p,\alpha\rangle$:
\be
\cT[(e,\alpha)]\Big|_{\cE'}
=
e^{i\langle p,\sigma_{f^{-1}}\alpha\rangle}\,\II_{\cE'}
\qquad\forall\,\alpha\in A.
\label{phase'}
\ee
This motivates the following definition for the action of boosts on momenta:

\paragraph{Definition.} For any momentum $p\in A^*$ and any $f\in G$, we write 
\i{rotation!action on momenta}\i{contragredient representation}\i{dual 
representation}
\be
\sigma^*_f(p)
\equiv
p\circ\sigma_{f^{-1}}\,,
\label{sstar}
\ee
i.e.\ $\langle \sigma^*_f(p),\alpha\rangle
\equiv
\langle p,\sigma_{f^{-1}}\alpha\rangle$ for all translations $\alpha$.
This defines a representation $\sigma^*$ of $G$ in the space of momenta, known as the \it{dual 
representation}\i{dual representation}\i{representation!dual} corresponding to $\sigma$. To reduce clutter, 
we will often denote it by
\be
\sigma^*_f(p)\equiv f\cdot p\,.
\label{simplinot}
\ee
\vspace{.1cm}

In terms of the dual representation (\ref{simplinot}) we can rewrite 
(\ref{phase'}) as
$\cT[(e,\alpha)]\big|_{\cE'}
=
e^{i\langle f\cdot p,\alpha\rangle}\,\II_{\cE'}$,
where $\cE'=\cT[(f,0)]\cdot\cE$. Thus, whenever the representation $\cT$ contains a momentum $p$, 
compatibility with the structure of 
$G\ltimes A$ implies that it also contains the boosted momentum $f\cdot p$, where 
$f$ is any element of $G$. This is the answer to the question (\ref{question}): if there exists a momentum 
$p$ such that (\ref{phase}) holds, 
then the representation also contains all momenta that belong to the \it{orbit}\i{orbit} 
(\ref{orb}) of $p$ under $G$,\i{orbit}\i{momentum!orbit}
\be
\boxed{\Big.\cO_p\equiv\left\{f\cdot p\,|\,f\in G\right\}.\Big.}
\label{hop}
\ee
This orbit is the minimal set of momenta needed to cook up a representation of $G\ltimes A$; we will 
see below that it is also sufficient. In fact, the whole classification of irreducible unitary 
representations 
of $G\ltimes A$ will be provided by a partition of the space of momenta into $G$-orbits. Note that this 
partition is scale-invariant in the following sense: since the action of boosts on momenta is linear, the 
orbits $\cO_p$ and $\cO_{\lambda p}$ are diffeomorphic for any real number $\lambda\neq 0$.\\

Each orbit $\cO_p$ is a homogeneous space for the action (\ref{sstar}) of $G$. Accordingly we define the 
\it{little group}\i{little group} of a momentum $p$ as the set of rotations that leave it fixed,\i{stabilizer}
\be
G_p\equiv\left\{f\in G\,|\,f\cdot p=p\right\}.
\label{gilipi}
\ee
It is the stabilizer (\ref{guipure}) for the action of $G$ on the homogeneous space 
$\cO_p$. As in (\ref{hogippy}) there is a diffeomorphism $\cO_p\cong G/G_p$.
Note that the little group of the vanishing momentum $p=0$ is the whole group $G$.\\

The notion of orbits is perhaps the one most important concept needed to understand representations of 
semi-direct products. We will encounter it repeatedly later on. Orbits hint at a 
geometrization 
of representation theory analogous to the one mentioned at the end of section \ref{sedefirep}, and 
therefore suggest 
that representations of $G\ltimes A$ are closely related to induced representations. In the next pages we 
will confirm this 
intuition by showing how to associate representations of $G\ltimes A$ with a given 
orbit.\\

Note that the action (\ref{sstar}) of $G$ on the space of momenta leaves the pairing (\ref{pair}) 
invariant in the sense that $\langle f\cdot p,\sigma_f\alpha\rangle=\langle p,\alpha\rangle$. This has an 
important implication: when $A$ is finite-dimensional it is isomorphic to its dual, so (\ref{pair}) defines a 
non-degenerate bilinear form on 
$A$ and the action $\sigma^*$ of $G$ on momenta is equivalent to $\sigma$. We shall see illustrations of this 
in the Poincar\'e groups. By contrast, 
when 
$A$ is \it{infinite}-dimensional, $\sigma^*$ may not be equivalent to $\sigma$ despite the property 
$\langle f\cdot p,\sigma_f\alpha\rangle=\langle p,\alpha\rangle$. This observation will be relevant to the 
BMS$_3$ group in part III.

\subsection{Particles}
\label{souci}

We now explain how to build irreducible unitary representations of $G\ltimes A$ starting from a 
momentum orbit $\cO_p$. Inspired by the 
Poincar\'e group, we refer to such representations as \it{particles}. We start by describing scalar particles 
and identify them with induced representations of $G\ltimes A$. This identification will then allow us to 
introduce spin.

\subsubsection*{Scalar particles}

Let $p\in A^*$ be a momentum with orbit (\ref{hop}). The latter is a homogeneous space and therefore admits a 
quasi-invariant measure $\mu$. Let then $\sH=L^2(\cO_p,\mu,\CC)$ be the Hilbert space of square-integrable 
wavefunctions in momentum space,\i{wavefunction!in momentum space}
\be
\Psi:\cO_p\rightarrow\CC:
q\mapsto\Psi(q).
\label{psiqqu}
\ee
The scalar product of wavefunctions is (\ref{scall}),
with $(\Phi(q)|\Psi(q))=\Phi^*(q)\Psi(q)$.\\

Now let us endow $\sH$ with a unitary action $\cT$ of $G\ltimes A$. In other words, if 
$\Psi\in\sH$ 
is a wavefunction, we wish to define the object
\be
\cT[(f,\alpha)]\cdot\Psi
\label{tfa}
\ee
where $(f,\alpha)$ belongs to $G\ltimes A$ and where $\cT[(f,\alpha)]$ is some unitary operator. Linearity 
implies that the result should be proportional to $\Psi$, so
\be
\big(\cT[(f,\alpha)]\cdot\Psi\big)(q)
=
\left(
\text{some number}
\right)
\times
\Psi(\text{some point on }\cO_p)
\nn
\ee
where the unknown quantities may depend on $q$, $f$ and $\alpha$. Note that the quantity multiplying 
$\Psi(\cdots)$ on the right-hand side must be a \it{number}, as opposed to an operator, because $\Psi$ takes 
its values in $\CC$ (this will change upon adding spin). Now recall that the reason for introducing orbits in 
the first place was to represent translations by multiplicative operators 
(\ref{eipa}). Accordingly the translation $\alpha$ in (\ref{tfa}) should produce a momentum-dependent phase 
factor:
\be
\big(
\cT[(f,\alpha)]\cdot\Psi
\big)(q)
=
e^{i\langle q,\alpha\rangle}
\times
\Psi(\text{some point on }\cO_p)\,.
\nn
\ee
Finally, since $\Psi$ is a wavefunction in 
momentum space, its argument on the right-hand side should represent the fact that a boost $f$ maps a 
particle with momentum $k$ 
on a particle with momentum $f\cdot k$. This is exactly the situation encountered in the quasi-regular 
representation (\ref{qregsimple}) so we can borrow that construction:\i{particle!scalar}\i{induced 
representation!of semi-direct product}\i{representation!of semi-direct product}
\be
\big(\cT[(f,\alpha)]\cdot\Psi\big)(q)
=
e^{i\langle q,\alpha\rangle}\,
\Psi(f^{-1}\cdot q)\,.
\label{qregpart}
\ee
In particular, the intuition depicted in fig.\ \ref{movaction} remains valid.\\

Formula (\ref{qregpart}) defines a representation $\cT$ of $G\ltimes A$, as can be verified by following 
the same steps as for the quasi-regular representation (\ref{qregsimple}). It is also irreducible by virtue 
of the fact that the orbit $\cO_p$ is a homogeneous space. Finally, it is unitary if the measure $\mu$ in 
(\ref{scall}) is invariant under $G$. If the 
measure has a non-trivial Radon-Nikodym derivative (\ref{rnq}), the 
representation (\ref{qregpart}) can be made unitary by 
inserting a compensating term in front of the exponential, as in (\ref{qreg}):\i{particle!scalar}\i{scalar 
particle}
\be
\big(\cT[(f,\alpha)]\cdot\Psi\big)(q)
=
\sqrt{\rho_{f^{-1}}(q)}\;\,e^{i\langle q,\alpha\rangle}\,\Psi(f^{-1}\cdot q).
\label{indoo}
\ee
We call this representation a \it{scalar particle} with momentum orbit $\cO_p$. Note how 
translations and rotations have radically different roles: translations multiply 
wavefunctions by momentum-dependent phase factors, while boosts move them around on the orbit by 
changing their argument. In particular, pure translations act as
\be
\cT[(e,\alpha)]\cdot\Psi(q)
=
e^{i\langle q,\alpha\rangle}\,\Psi(q).
\label{trabi}
\ee
Thinking of wavefunctions as sections of a complex line bundle over $\cO_p$ with fibres $\cE_q\cong\CC$, 
formula (\ref{trabi}) can be rewritten symbolically as
\be
\cT[(e,\alpha)]
=
\int_{\cO_p}d\mu(q)\,e^{i\langle q,\alpha\rangle}\,\II_q
\label{tinta}
\ee
where $\II_q$ is the identity operator in the fibre at $q$. This is precisely the anticipated expression
(\ref{tint}).

\begin{figure}[H]
\centering
\includegraphics[width=0.60\textwidth]{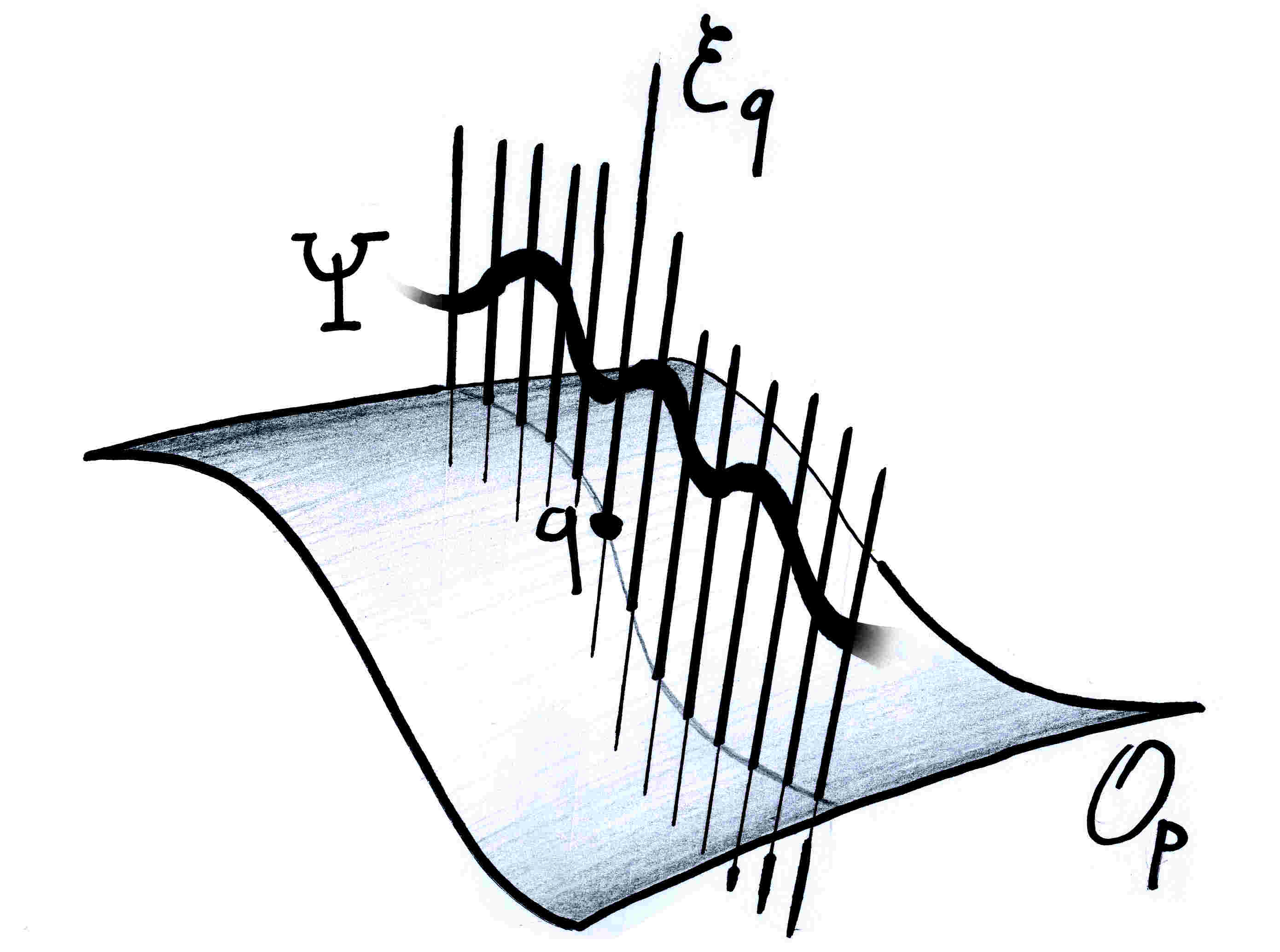}
\caption{A momentum orbit $\cO_p$ crossed by one-dimensional fibres isomorphic to $\CC$.\i{line bundle} (For 
simplicity the fibres are depicted as if they were real rather than complex.) The fibre at $q\in\cO_p$ is 
denoted $\cE_q$ and the disjoint union of such fibres is a (complex) line bundle over $\cO_p$. A wavefunction 
$\Psi$ is a section of that bundle. Translations (\ref{tinta}) act by complex multiplication $z\mapsto 
e^{i\langle q,\alpha\rangle}z$ in each fibre $\cE_q$.\label{figoFIBB}}
\end{figure}

\subsubsection*{Particles are induced representations}

Formula (\ref{indoo}) is almost identical to the quasi-regular representation (\ref{qreg}), and more 
generally to the induced representation (\ref{ind}). To investigate this relation, let $G_p$ be the 
little group (\ref{gilipi}) of $p$ and consider the subgroup $G_p\ltimes A$ of $G\ltimes A$. 
Define a map\i{particle!as induced representation}
\be
\cS:G_p\ltimes A\rightarrow\CC:
(f,\alpha)\mapsto\cS[(f,\alpha)]\equiv e^{i\langle p,\alpha\rangle},
\label{scapin}
\ee
which is a one-dimensional representation of $G_p\ltimes A$. Indeed, for all $(f,\alpha)$ and 
$(g,\beta)$ belonging to $G_p\ltimes A$, $\cS$ preserves the group structure in the sense that
\begin{align}
\cS[(f,\alpha)]\cdot\cS[(g,\beta)]
& \refeq{scapin}
e^{i\langle p,\alpha\rangle+i\langle p,\beta\rangle}
\stackrel{f\in G_p}{=}
e^{i\langle p,\alpha\rangle+i\langle f^{-1}\cdot p,\beta\rangle}\nn\\
& \refeq{sstar}
e^{i\langle p,(\alpha+\sigma_f\beta)\rangle}
\refeq{scapin}
\cS[(fg,\alpha+\sigma_f\beta)]
\refeq{semiop}
\cS[(f,\alpha)\cdot(g,\beta)].
\nn
\end{align}
Furthermore, (\ref{scapin}) is unitary so we can use it to induce 
a unitary representation
\be
\cT=\text{Ind}_{G_p\ltimes A}^{G\ltimes A}(\cS)
\label{tinder}
\ee
of $G\ltimes A$. Using the general formula (\ref{ind}) and the diffeomorphisms $\cO_p\cong 
G/G_p\cong(G\ltimes A)/(G_p\ltimes A)$,
we see that the induced representation (\ref{tinder}) acts on wavefunctions exactly in the way displayed in 
eq.\ (\ref{indoo}). Note that the little group $G_p$ is represented trivially in the ``spin'' representation 
(\ref{scapin}). This 
is why we say that the particle (\ref{indoo}) is \it{scalar}: its states are essentially unaffected by the 
rotations that span $G_p$. The picture (\ref{tinder}) suggests a simple generalization of this 
behaviour, as we now explain.

\subsubsection*{Spinning particles}

To generalize (\ref{indoo}), let $\cR$ be an irreducible, unitary representation of $G_p$ in some space $\cE$ 
and consider 
the spin representation
\be
\cS:G_p\ltimes A\rightarrow\text{GL}(\cE):
(f,\alpha)\mapsto e^{i\langle p,\alpha\rangle}\,\cR[f]\,.
\label{rspin}
\ee
This reduces to (\ref{scapin}) when $\cR$ is trivial, and the corresponding induced representation of 
$G\ltimes A$ is
\be
\cT
=
\text{Ind}_{G_p\ltimes A}^{G\ltimes A}(\cS)
=
\text{Ind}_{G_p\ltimes A}^{G\ltimes A}\big(e^{i\langle p,\cdot\rangle}\,\cR\big)\,.
\label{texas}
\ee
Its action on wavefunctions is analogous to 
(\ref{ind}) and generalizes (\ref{indoo}):\i{spinning particle}\i{particle!spinning}\i{induced 
representation!of semi-direct product}
\be
\boxed{
\left(\cT[(f,\alpha)]\cdot\Psi\right)(q)
=
\sqrt{\rho_{f^{-1}}(q)}\;\,e^{i\langle q,\alpha\rangle}\,
\cR[g_q^{-1}\,f\,g_{f^{-1}\cdot q}]\cdot
\Psi(f^{-1}\cdot q)\,,
}
\label{spipa}
\ee
where the map $g:\cO_p\rightarrow G:q\mapsto g_q$ is a continuous family of standard boosts (\ref{stdb}). In 
contrast to (\ref{indoo}), $\Psi$ now takes its values in $\cE$ rather than $\CC$.\\

We 
call the representation (\ref{spipa}) a \it{spinning particle}\i{particle!spinning} with spin $\cR$ and 
momenta belonging to $\cO_p$. It is an irreducible unitary representation of $G\ltimes A$ acting on the 
Hilbert space $\sH=L^2(\cO_p,\mu,\cE)$. The operator\i{Wigner rotation}
\be
\cR[g_q^{-1}\, f\, g_{f^{-1}\cdot q}]
\equiv
W_q[f]
\label{winota}
\ee
is the \it{Wigner rotation}\i{Wigner rotation} (\ref{wig}) associated with $f$ and $q$. It is the 
transformation that corresponds to $f$ in the space of 
internal degrees of freedom $\cE$ at $q$ and it entangles momentum and spin degrees 
of freedom. The decomposition (\ref{tinta}) 
still holds in the spinning case, with $\II_q$ the identity operator in the fibre $\cE_q\cong\cE$ at $q$.\\

From this point on, the whole machinery of induced representations applies to unitary 
representations (\ref{spipa}) of $G\ltimes A$. In particular they are independent of the choice of the 
quasi-invariant measure $\mu$ and of the family of standard boosts $g_q$. 
The plane waves (\ref{pwave}) provide a basis of the Hilbert space and 
represent one-particle states with definite momentum and definite spin. They
transform under $G\ltimes A$ according to\i{plane wave}\i{particle!with 
definite momentum}\i{induced representation!in terms of plane waves}
\be
\cT[(f,\alpha)]\cdot\Psi_{k,\ell}
=
\sqrt{\rho_f(k)}\;\,
e^{i\langle f\cdot k,\alpha\rangle}
\cR\left[
g_{f\cdot k}^{-1}\cdot f\cdot g_k
\right]
\cdot
\Psi_{f\cdot k,\ell}\,.
\label{klakra}
\ee
This is just formula (\ref{indsim}) applied to (\ref{texas}); it is the plane wave analogue of 
(\ref{spipa}). Using this, one can go on and evaluate characters along the lines that led to the Frobenius 
formula (\ref{frob}). One finds\i{Frobenius formula!for semi-direct products}\i{character!for semi-direct 
product}\i{semi-direct product!character}
\be
\chi[(f,\alpha)]
=
\Tr\big(\cT[(f,\alpha)]\big)
=
\int_{\cO_p}d\mu(k)\,\delta(k,f\cdot k)\,
e^{i\langle k,\alpha\rangle}
\chi_{\cR}[g_k^{-1}fg_k]\,,
\label{fropo}
\ee
where $\chi_{\cR}$ is the character of the representation $\cR$ of $G_p$. As before one can check that this 
formula defines a class function and that $\chi[(f,\alpha)]$ vanishes when $f$ is not conjugate to an element 
of the little group. The delta function localizes the integral to the momenta that are left fixed by the 
action of $f$ on $\cO_p$.

\subsection{Exhaustivity theorem}
\label{susexh}

Eq.\ (\ref{spipa}) is an irreducible unitary representations of 
$G\ltimes A$. As it turns out, \it{all} irreducible representations of $G\ltimes A$ take this form for some 
momentum orbit $\cO_p$ and some spin $\cR$. We refer to this property as the \it{exhaustivity 
theorem}\i{exhaustivity theorem}\i{induced 
representation!exhaustive} for induced representations.\\

This theorem has enormous practical value: it provides the classification of all irreducible unitary 
representations of a semi-direct product $G\ltimes_{\sigma}A$ when $A$ is a vector group. This classification 
can be performed thanks to the following algorithm:\label{algo}\i{representation!of semi-direct 
product}\i{semi-direct product!unitary representations}\i{classification!of 
particles}\i{particle!classification}
\begin{enumerate}
\item Consider the space of momenta, $A^*$. For each $p\in A^*$, determine the orbit $\cO_p$ given by 
(\ref{hop}). This foliates $A^*$ into disjoint momentum orbits, and each point of $A^*$ belongs to 
exactly one orbit.
\item We call \it{set of orbit representatives} a set of momenta\i{orbit!representative} that exhaust 
all orbits in a non-redundant way, in the sense that (i) each orbit contains one of the representatives, and 
(ii) different representatives belong to different orbits. Find a set of orbit representatives, compute the 
little group of each representative, and find standard boosts connecting each
representative 
to the points of its orbit.
\item For each representative $p$ with little group $G_p$, classify all irreducible unitary 
representations of $G_p$. Given such a representation $\cR$, the associated induced representation of 
$G\ltimes A$ is (\ref{spipa}).
\end{enumerate}
We will illustrate 
this classification for the Poincar\'e groups in section \ref{relagroup} and for the Bargmann groups in 
section \ref{galisec}, and of course for BMS$_3$ in chapter \ref{c7}.\\

The proof of the exhaustivity theorem is essentially an upgraded version of our arguments in section 
\ref{suserbilly} and relies on two crucial ingredients: the first is the commutativity of the vector group 
$A$, and the second is the imprimitivity theorem of section \ref{sysim}. Thanks 
to commutativity, any unitary representation of $A$ can be written as a direct integral (\ref{tint}) of 
irreducible representations specified by certain momenta $q\in A^*$. (This is known as the \textsc{snag} 
theorem.\i{SNAG theorem}) This implies that any unitary representation $\cT$ of 
$G\ltimes A$ is 
imprimitive. Indeed, relation (\ref{titisig}) can be rewritten as
\be
\cT[(f,0)]\cdot\cT[(e,\alpha)]\cdot\cT[(f,0)]^{-1}
=
\cT[(e,\sigma_f\alpha)]
\nn
\ee
whereupon the direct integral representation (\ref{tint}) yields
\be
\cT[(f,0)]\cdot
d\mu(q)\II_q
\cdot\cT[(f,0)]^{-1}
=
d\mu(f\cdot q)\II_{f\cdot q}\,,
\label{s129q}
\ee
which is precisely the statement (\ref{impind}) that the projection-valued measure $d\mu(q)\II_q$ is a system 
of imprimitivity for $\cT$ on $A^*$. The imprimitivity theorem then implies that the representation $\cT$ is 
induced. The last step of the proof consists in showing that, if $\cT$ is irreducible, then the measure $\mu$ 
in (\ref{s129q}) localizes to a single momentum orbit. We refer to \cite{barut1986theory} for details.

\paragraph{Remark.} The exhaustivity theorem relies on an extra technical assumption that we haven't 
mentioned so far. Namely, one says that $G\ltimes A$ is \it{regular}\i{regular semi-direct 
product}\i{semi-direct product!regular} if the 
space of momenta $A^*$ and the action (\ref{sstar}) of $G$ are such that $A^*$ contains a countable family of 
Borel sets, each a union of momentum orbits, such that each orbit is the limit of a decreasing sequence of 
such sets. As it turns out regularity is necessary for the measure $\mu$ in (\ref{s129q}) to be localized on 
a momentum orbit. All semi-direct products treated in part I of this thesis are regular. As for the BMS$_3$ 
group of part III, the issue of regularity will be discussed briefly in section \ref{sebmspar}.

\section{Poincar\'e particles}
\label{relagroup}

In this section and the next ones we study examples of semi-direct products to illustrate induced 
representations. Here we deal with the Poincar\'e group --- the isometry group of Minkowski space --- whose 
representations describe relativistic particles. Following the algorithm of page \pageref{algo} we will find 
that these particles are classified by two parameters: their mass and their spin. In view of treating the 
BMS$_3$ group in part III, we relegate the detailed description of relativistic particles in three dimensions 
to section \ref{sePoTri}.\\

The plan is the following. First we define the Poincar\'e group as a semi-direct product of 
the Lorentz group with the group of space-time translations. Then we turn to the classification of its 
momentum orbits and describe the corresponding particles. We also compute their characters and end with the 
observation that Lorentz transformations generally entangle momentum and spin degrees of freedom.\\

The classification of relativistic particles was first performed by Wigner \cite{Wigner:1939cj}, and their 
relation to wave equations was worked out in \cite{Bargmann}. These results are among the foundations of 
quantum mechanics and field theory; see e.g.\ 
\cite{barut1986theory,weinberg1995,cornwell1984group,Bekaert:2006py}.

\subsection{Poincar\'e groups}
\label{susePodef}

\subsubsection*{Lorentz transformations}

We consider the vector space $\RR^D$; its elements are column vectors $\alpha$, $\beta$, etc.~with 
components $\alpha^{\mu}$, $\beta^{\mu}$ where $\mu=0,1,...,D-1$. Here $\RR^D$ is to be interpreted as a 
space-time manifold with dimension $D\geq2$. We endow $\RR^D$ with a non-degenerate 
bilinear form given by the \it{Minkowski metric},\i{Minkowski space}
\be
(\alpha,\beta)
\equiv
\eta_{\mu\nu}\alpha^{\mu}\beta^{\nu},
\qquad
(\eta_{\mu\nu})
=
\bmm
-1 & 0 & 0 & \cdots & 0 \\
0 & 1 & 0 & \cdots & 0 \\
0 & 0 & 1 & \cdots & 0 \\
\vdots & \vdots & \vdots & \ddots & \vdots \\
0 & 0 & 0 & \cdots & 1
\emm.
\label{minky}
\ee
We write 
$(\alpha,\alpha)\equiv\alpha^2$ for any $\alpha$. The sign of $\alpha^2$ determines whether $\alpha$ 
is\i{time-like vector}\i{space-like vector}\i{light-like vector}\i{null vector} 
time-like, null or space-like, corresponding respectively to $\alpha^2<0$, $\alpha^2=0$ or $\alpha^2>0$.

\paragraph{Definition.} The \it{Lorentz group}\i{Lorentz group} $\text{O}(D-1,1)$ in $D$ dimensions is the 
group of 
linear transformations $\RR^D\rightarrow\RR^D:\alpha\mapsto f\cdot\alpha$ that preserve (\ref{minky}) in the 
sense that
\be
(f\cdot\alpha,f\cdot\beta)=(\alpha,\beta).
\label{inbil}
\ee
It 
consists of $D\times D$ matrices $f=({f^{\mu}}_{\nu})$ such that
\be
f^t\cdot\eta\cdot f=\eta,
\qquad\text{i.e.}\qquad
{f^{\lambda}}_{\mu}\eta_{\lambda\rho}{f^{\rho}}_{\nu}
=
\eta_{\mu\nu}
\label{lel}
\ee
where the dot denotes matrix multiplication. In particular, 
the Minkowskian norm is left invariant by Lorentz transformations.

\subsubsection*{Topology of Lorentz groups}

\begin{figure}[h]
\centering
\includegraphics[width=0.70\textwidth]{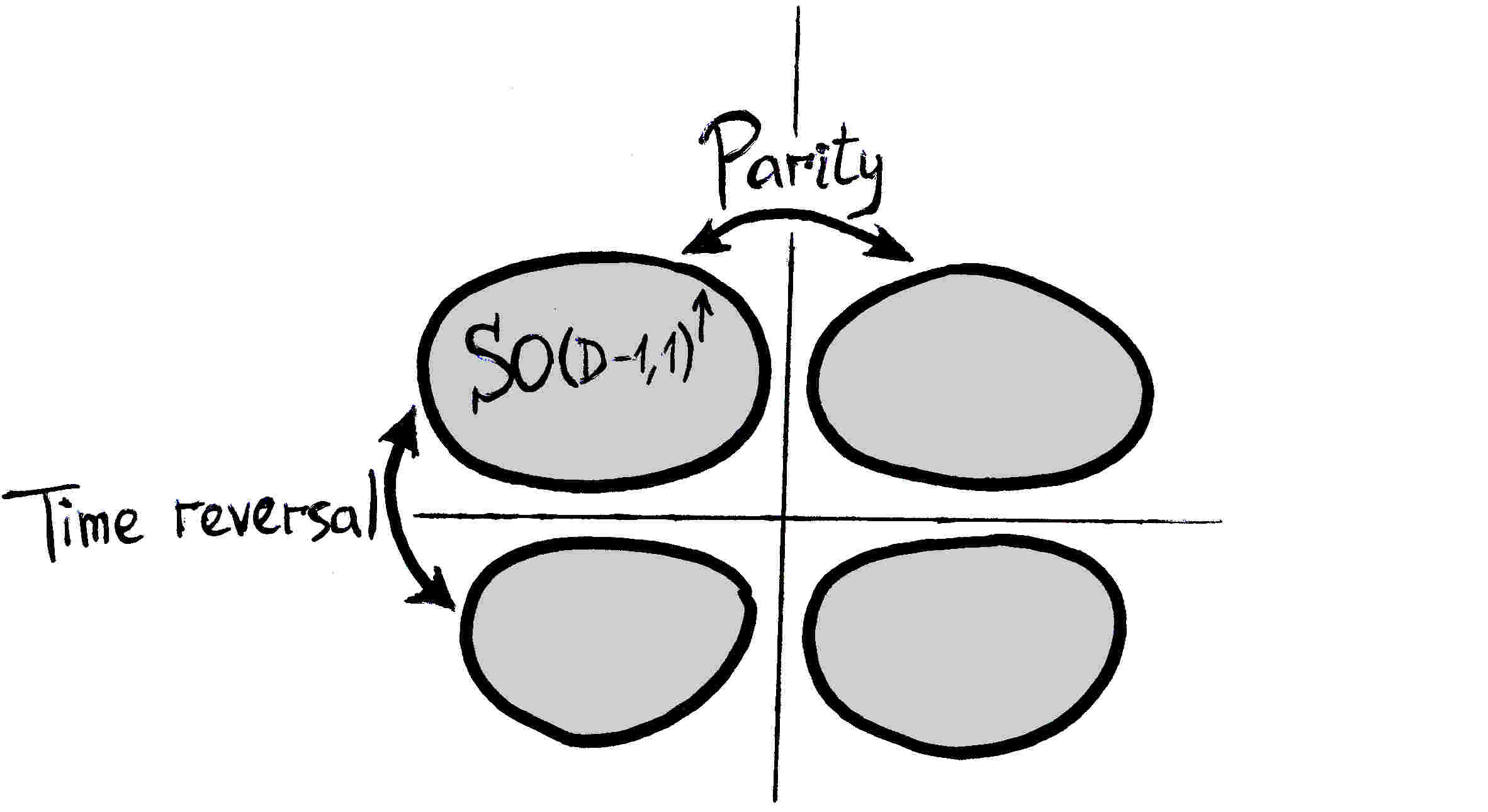}
\caption{The four connected components of the Lorentz group. The upper left component is the proper 
orthochronous Lorentz group $\text{SO}(D-1,1)^{\uparrow}$. It can be mapped on the other components using 
parity and time reversal. In particular the proper Lorentz group is generated by 
$\text{SO}(D-1,1)^{\uparrow}$ together with time reversal, while the orthochronous Lorentz group is generated 
by $\text{SO}(D-1,1)^{\uparrow}$ together with parity.\label{figoLor}}
\end{figure}

The Lorentz group $\text{O}(D-1,1)$ is disconnected. Indeed, any Lorentz matrix $f$ has determinant 
$\det(f)=\pm1$. This cuts the group in two pieces consisting of matrices with positive and negative 
determinant, corresponding to transformations that 
preserve or 
break (respectively) the orientation of the spatial coordinates. The subgroup of 
$\text{O}(D-1,1)$ consisting of Lorentz matrices with positive determinant is the \it{proper} Lorentz 
group, $\SO(D-1,1)$.\i{proper Lorentz group} Any improper Lorentz 
matrix is the product of a proper Lorentz transformation with parity. In addition, one can show that any 
Lorentz 
matrix $f$ satisfies $|{f^0}_0|\geq 1$, which again cuts the 
Lorentz group in two pieces: matrices with positive or negative ${f^0}_0$, corresponding to 
transformations that preserve or invert (respectively) the orientation of the arrow of time. The subgroup 
consisting of Lorentz 
transformations with positive ${f^0}_0$ is the \it{orthochronous} Lorentz group 
$\text{O}(D-1,1)^{\uparrow}$.\i{orthochronous Lorentz group} 
Any 
Lorentz matrix that reverts the arrow of time is the product of an orthochronous Lorentz 
matrix with time reversal. The situation is depicted in fig.\ \ref{figoLor}.\\

In this section we focus on the connected Lorentz group, i.e.\ the proper 
or\-tho\-chro\-nous 
Lorentz group $\SO(D-1,1)^{\uparrow}$.\i{connected Lorentz group}\i{Lorentz 
group!topology}\i{SO21@$\text{SO}(2,1)^{\uparrow}$} The latter 
satisfies an 
important property known as \it{standard decomposition}:\i{standard decomposition theorem} any proper, 
orthochronous Lorentz transformation is a product 
$f=R_1\cdot\Lambda\cdot R_2$, where $R_1$ and $R_2$ are spatial rotations and $\Lambda$ is a pure boost 
\cite{HenneauxGroupe}. In what follows we often refer to $\SO(D-1,1)^{\uparrow}$ simply as ``the 
Lorentz group''.\\

The Lorentz group is not simply connected: in space-time dimension $D\geq 4$, its fundamental group is 
isomorphic to $\ZZ_2$. The universal cover of the connected Lorentz group\i{universal cover!of 
Lorentz group} is then called the \it{spin 
group},\i{spin group} so that
\be
\SO(D-1,1)^{\uparrow}\cong\text{Spin}(D-1,1)/\ZZ_2
\label{sospink}
\ee
where the $\ZZ_2$ subgroup of $\text{Spin}(D-1,1)$ consists of the identity matrix and its opposite. In four 
dimensions, $\text{Spin}(3,1)=\text{SL}(2,\CC)$. In $D=3$ dimensions the situation is a bit different; we 
will return to it in the next section. In any case the Lorentz group is always multiply connected, and 
therefore 
admits topological projective representations; this will be important for representations of Poincar\'e.

\subsubsection*{Poincar\'e groups}

\paragraph{Definition.} The \it{Poincar\'e group}\i{Poincar\'e group} or \it{inhomogeneous Lorentz 
group} in $D$ space-time dimensions is the semi-direct product
\be
\text{IO}(D-1,1)
\equiv
\text{O}(D-1,1)\ltimes\RR^D
\label{piggbix}
\ee
whose elements are pairs $(f,\alpha)$ where $f$ is a Lorentz transformation, $\alpha$ a space-time 
translation. The group operation is
$(f,\alpha)\cdot (g,\beta)
=
(f\cdot g,\alpha+f\cdot\beta)$
where the dots on the right-hand side denote matrix multiplication and the action of matrices on column 
vectors. The \it{connected Poincar\'e group} is the largest connected subgroup of 
(\ref{piggbix}),\i{connected Poincar\'e group}\i{Poincar\'e group!topology}
\be
\text{ISO}(D-1,1)^{\uparrow}
\equiv
\SO(D-1,1)^{\uparrow}\ltimes\RR^D\,,
\label{pig}
\ee
and its universal cover is\i{universal cover!of Poincar\'e group}
\be
\text{Spin}(D-1,1)^{\uparrow}\ltimes\RR^D
\label{ppxa}
\ee
where spin transformations act on $\RR^D$ according to the composition of the homomorphism given by 
(\ref{sospink}) with the vector representation of the Lorentz group.\\

The Poincar\'e group turns out to have no 
algebraic central extensions, so its only non-trivial projective transformations are of topological origin. 
The Poincar\'e Lie algebra is generated by $D(D-1)/2$ Lorentz generators and $D$ translation generators; we 
will not display their brackets here.

\subsection{Orbits and little groups}
\label{susepor}

From now on we focus on the connected Poincar\'e group (\ref{pig}), to which we refer simply as ``the 
Poincar\'e group''.

\subsubsection*{Momenta and orbits}

The space of Poincar\'e momenta is $(\RR^D)^*=\RR^D$; its 
elements are 
$D$-dimensional covectors $p=(p_0,p_1,...,p_{D-1})$, where $p_0$ is to be interpreted as the energy of a 
relativistic particle, while $\bbp=(p_1,...,p_{D-1})$ is its spatial 
momentum.\footnote{Strictly speaking the energy of the particle is $p^0=-p_0$, but this detail will not 
affect our discussion so we neglect it for simplicity.}\i{energy-momentum}\i{Poincar\'e 
momentum}\i{momentum!for Poincar\'e group}\i{relativistic momentum} Given a 
momentum $p$ and a space-time translation 
$\alpha$, the pairing (\ref{pair}) is $\langle p,\alpha\rangle
=p_{\mu}\alpha^{\mu}$.\\

The Minkowski metric (\ref{minky}) provides a Lorentz-invariant pairing between translation vectors and can 
be used to define an isomorphism
\be
\cI:\RR^D\rightarrow(\RR^D)^*:\alpha\mapsto(\alpha,\cdot)
\label{ii}
\ee
where the components of $(\alpha,\cdot)$ are those of $\alpha$ lowered with the Minkowski metric. Using 
$\cI$, one verifies that the action $\sigma^*$ of Lorentz transformations on momenta is equivalent to 
their action $\sigma$ on translations:
\be
\sigma^*_f=
\cI
\circ
\sigma_f
\circ
\cI^{-1}.
\label{sequiv}
\ee
As a consequence, momentum orbits coincide with orbits of translations under Lorentz transformations, and 
consist of momenta $q$ with constant Minkowskian norm squared $q^2$. We thus conclude 
that\i{orbit!for Poincar\'e}\i{momentum!orbit}
\be
\begin{array}{c}
\text{\it{the orbits of momenta of relativistic particles are connected hyperboloids}}\\
\text{\it{specified by an equation of the form $q_0^2-\bbq^2=\text{const.}$ in $\RR^D$.}}
\end{array}
\nn
\ee
The only exception to this rule is the trivial orbit of the vanishing momentum $p=0$, which contains only 
one point. The word ``connected'' appears here because we are dealing with the connected Poincar\'e group 
(\ref{pig}). By contrast the momentum orbits of (\ref{piggbix}) are generally disconnected.\\

\begin{figure}[h]
\centering
\begin{minipage}{0.45\textwidth}
\centering
\includegraphics[width=\linewidth]{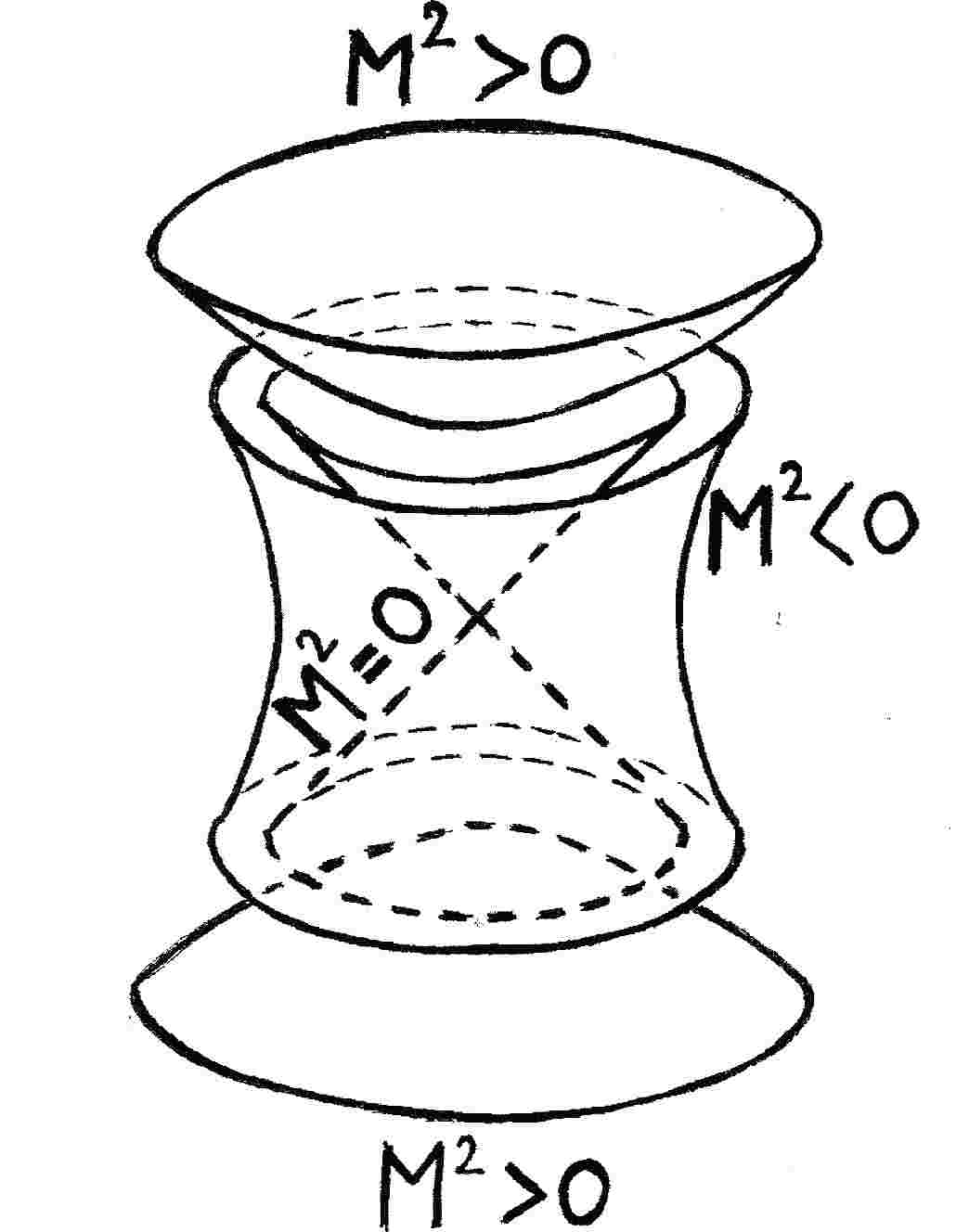}\par
(a)
\end{minipage}\hfill
\begin{minipage}{0.1\textwidth}
\centering
\includegraphics[width=\linewidth]{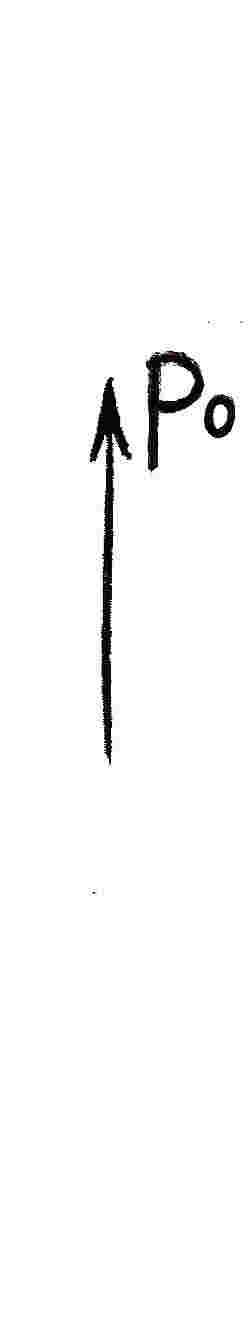}\par
~
\end{minipage}\hfill
\begin{minipage}{0.45\textwidth}
\centering
\includegraphics[width=\linewidth]{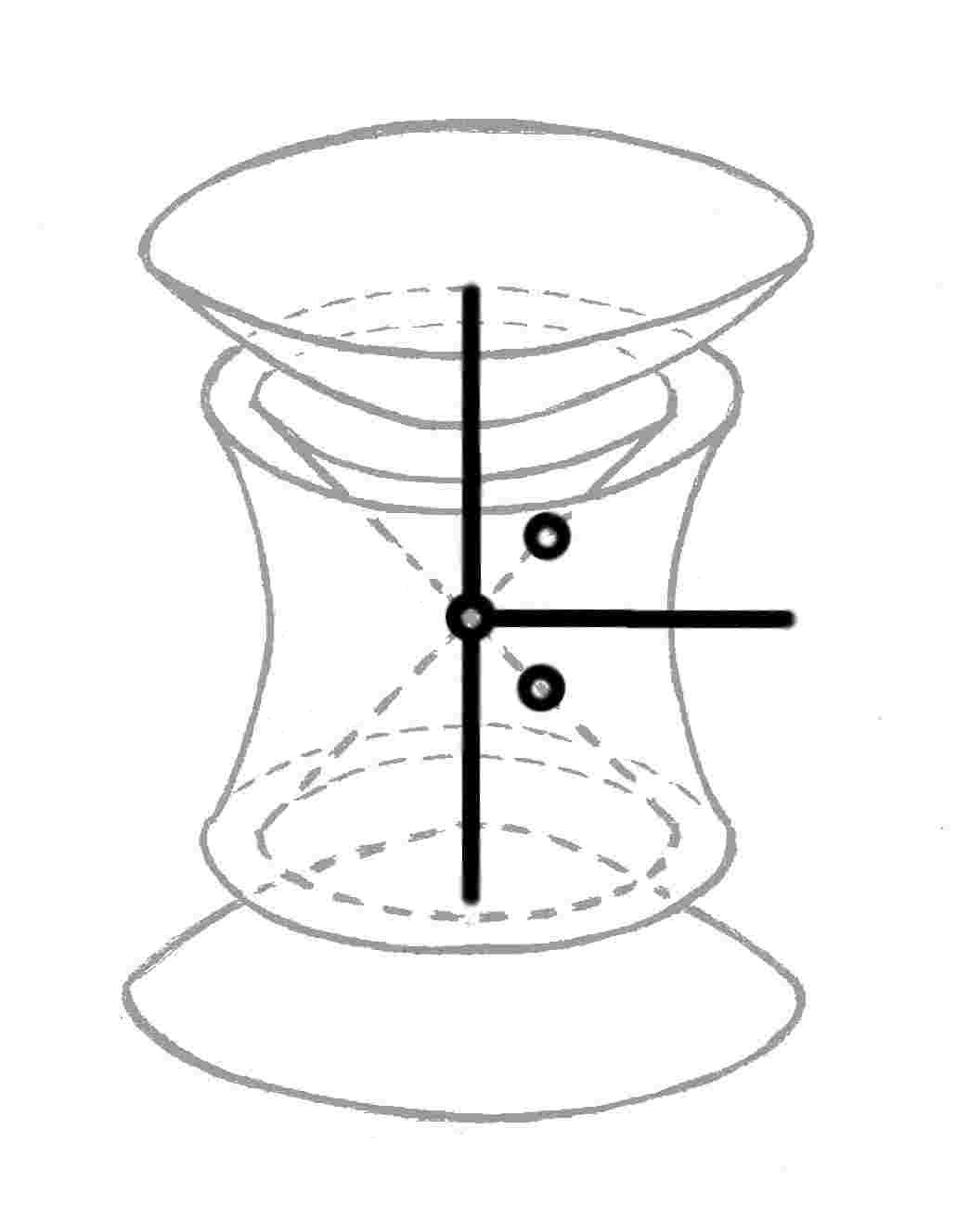}\par
(b)
\end{minipage}
\caption{On the left, fig.\ (a) represents a few momentum orbits of the Poincar\'e group in three dimensions, 
embedded in $\RR^3$ with the vertical axis corresponding to $p_0$ and the two horizontal axes (not 
represented 
in the figure) corresponding to spatial components of momentum. Orbits can be massive, massless or tachyonic 
depending on whether $M^2$ is positive, vanishing or negative, respectively. The cross in the middle is the 
trivial orbit of $p=0$, consisting of a single point. On the right, fig.\ (b) is a schematic representation 
of momentum orbits: each point of the diagram corresponds to an orbit representative, where massive orbits 
are represented by a vertical line, tachyonic ones by a horizontal line, and discrete orbits (the two 
massless ones and the trivial one) by dots. This schematic representation will be useful in parts II and III 
for the interpretation of BMS$_3$ supermomentum orbits.\label{figDepIdu}}
\end{figure}

The connected Poincar\'e group has six distinct families of momentum orbits, which we now describe. Further 
details 
can be found e.g.\ in \cite{cornwell1984group,Bekaert:2006py}.
\begin{itemize}
\item Let $p=0$ be the vanishing momentum. Its orbit $\cO_0=\{0\}$ contains a single point. Its 
little group is the whole Lorentz group.
\item Let $p$ be a timelike momentum with positive energy, $p_0>0$. Its orbit 
$\cO_p$ is \it{massive with positive energy} and consists of momenta $q$ satisfying\i{massive particle}
\be
q_0^2-\bbq^2=M^2>0,
\qquad q_0>0,
\label{qsq}
\ee
where we have introduced the \it{mass squared}\i{mass squared} $M^2\equiv-p^2$. We can choose 
as orbit representative the rest frame momentum\i{rest frame}
\be
p=(M,0,...,0),
\qquad
M>0.
\label{pm}
\ee
The little group of (\ref{pm}) is the group of spatial rotations\i{little group!for relativistic particle}
\be
G_p
=
\SO(D-1)
\label{piligoma}
\ee
consisting of proper Lorentz transformations that leave the time coordinate fixed. In particular, 
the orbit is diffeomorphic to the quotient
\be
\cO_p\cong\SO(D-1,1)^{\uparrow}/\SO(D-1)\cong\RR^{D-1}
\label{opim}
\ee
and its points can be labelled by the spatial components of momentum (since the zeroth component is then 
determined by eq.~(\ref{qsq})). Massive 
orbits 
with different masses are disjoint.
\item Let $p$ be a time-like momentum with negative energy, $p_0<0$. Its orbit is \it{massive 
with negative energy} and consists of momenta $q$ satisfying (\ref{qsq}) with $q_0<0$. A typical orbit 
representative is (\ref{pm}) with $M<0$, and orbits with different masses are disjoint. The 
little group is a again $\SO(D-1)$.
\item Let $p$ be a null momentum with positive energy, $p_0>0$. Its orbit 
$\cO_p$ is \it{massless with positive energy}.\i{massless particle} It consists 
of momenta $q$ 
satisfying 
(\ref{qsq}) with $M^2=0$. A typical orbit representative is
\be
p=(E,E,0,...,0)
\label{pe}
\ee
where the energy $E$ is positive; different values of $E$ yield the same orbit. 
Note that there is no rest frame for massless particles. The little 
group of (\ref{pe}) is isomorphic to the Euclidean group
\be
G_p
\cong
\SO(D-2)\ltimes\RR^{D-2}
=
\text{ISO}(D-2).
\label{piliso}
\ee
In particular, the orbit is 
diffeomorphic to the quotient
\be
\cO_p\cong\SO(D-1,1)^{\uparrow}/\text{ISO}(D-2)\cong\RR\times S^{D-2}
\label{opis}
\ee
and its points can be labelled by the spatial components of momentum (since the zeroth component is 
then determined by $q^2=0$).
\item Let $p$ be a null energy-momentum vector with negative energy. Its orbit is \it{massless with negative 
energy} and consists of null momenta $q$ with $q_0<0$. A typical orbit representative is 
(\ref{pe}) with 
negative $E$. The little 
group is (\ref{piliso}) and the orbit can be represented as a quotient (\ref{opis}).
\item Let $p$ be a space-like momentum. Its orbit is \it{tachyonic} and consists 
of momenta $q$ satisfying 
(\ref{qsq}) with $M^2<0$.\i{tachyon} A typical orbit representative is
\be
p=\big(0,0,...,0,\sqrt{-M^2}\big).
\label{pr}
\ee
The little group is the lower-dimensional Lorentz 
group $\SO(D-2,1)^{\uparrow}$ consisting of transformations that leave the spatial coordinate $x^{D-1}$ 
fixed. In particular the orbit is diffeomorphic to the quotient
\be
\cO\cong\SO(D-1,1)^{\uparrow}/\SO(D-2,1)^{\uparrow}\cong\RR\times S^{D-2}.
\label{tachyorbi}
\ee
Tachyonic orbits with different negative values of $M^2$ are 
disjoint. Note that rotations always allow us to map $p$ on $-p$, which is why any tachyonic orbit 
representative can be written as (\ref{pr}).
\end{itemize}
This enumeration exhausts all Poincar\'e momentum orbits. Among the six families of orbits, three contain 
only 
one orbit: the trivial orbit and the two massless orbits. The remaining 
three 
families all contain infinitely many orbits labelled by a non-vanishing mass squared, corresponding to 
massive particles and tachyons. These orbits and their representatives are 
schematically depicted in fig.\ \ref{figDepIdu}.\\

To complete the description of orbits we now display standard boosts for massive particles (the other cases 
are less important for our purposes so we skip them). We take as orbit representative the momentum (\ref{pm}) 
of a particle at rest, and look for a family of boosts $g_q$ such that $g_q\cdot p=(\sqrt{M^2+\bbq^2},\bbq)$ 
that depend continuously on $\bbq$. One readily verifies that the matrices \cite{jackson}\i{boost}\i{standard 
boost!for Poincar\'e}\i{pure boost}
\be
g_q
=
\bmm
\sqrt{1+\bbq^2/M^2} & q_j/M \\
q_i/M & \delta_{ij}+\frac{q_iq_j}{\bbq^2}\left(\sqrt{1+\bbq^2/M^2}\,-1\right)
\emm
\label{gqgq}
\ee
satisfy these requirements. Here $i,j=1,...,D-1$ are spatial indices. Each such matrix is a boost in the 
direction $\bbq/|\bbq|$ with rapidity $\text{arccosh}[\sqrt{1+\bbq^2/M^2}]$.

\paragraph{Remark.} The little groups displayed in (\ref{piligoma}) and (\ref{piliso}) hold for the connected 
Poincar\'e group (\ref{pig}). If we replace the latter by its universal cover (\ref{ppxa}), then the little 
groups are replaced by their double covers (assuming that $D\geq 4$). In particular the little group of 
massive particles becomes $\text{Spin}(D-1)$ while that of massless particles becomes 
$\text{Spin}(D-2)\ltimes\RR^{D-2}$, with the convention that $\text{Spin}(2)$ is the double cover of 
$\text{SO}(2)$. Note that $\text{Spin}(3)=\text{SU}(2)$.

\subsection{Particles}
\label{susePaTho}

According to the exhaustivity theorem of section \ref{susexh}, the momentum orbits in fig.\ \ref{figDepIdu} 
roughly classify relativistic particles. The states of each particle are wavefunctions on its momentum 
orbit, valued in a spin representation of the little group and transforming under Poincar\'e transformations 
according to formula (\ref{spipa}).\i{induced representation!of Poincar\'e group} Provided we know all 
irreducible unitary representations of all little 
groups, we have effectively classifed all irreducible unitary 
representations of the Poincar\'e group.

\subsubsection*{Vacuum}

Vacuum representations of Poincar\'e are those whose orbit $\cO_0=\{0\}$ is trivial and is left invariant 
by the whole Lorentz group.\i{Poincar\'e vacuum} In that case a spin representation is a (projective) 
irreducible unitary 
representation $\cR$ of $\SO(D-1,1)^{\uparrow}$. The latter is simple but non-compact, so its only 
finite-dimensional irreducible unitary transformation is the trivial one; the corresponding induced 
representation of Poincar\'e is trivial as well. All other irreducible unitary 
representations of the Lorentz group are infinite-dimensional; the corresponding induced representations of 
Poincar\'e are such that translations act trivially, while Lorentz transformations act non-trivially on an 
infinite-dimensional Hilbert space $\cE$ of spin-like degrees of freedom. These representations can be 
interpreted as ``vacua with spin''\i{spinning vacuum}\i{vacuum with spin} but are generally discarded as 
unphysical.

\subsubsection*{Massive particles}

The momenta of a massive particle with mass $M$ span an orbit (\ref{qsq}) with little group 
(\ref{piligoma}).\i{massive particle} 
The spin representation $\cR$ then is a finite-dimensional, irreducible, generally projective unitary 
representation of $\SO(D-1)$ 
specified by some highest weight $\lambda$.\i{spin} For example, when $D=4$, $\cR$ is a highest-weight 
representation of $\SO(3)=\text{SU}(2)/\ZZ_2$ with spin $s\geq0$; the latter is either an integer or a 
half-integer. The carrier space of $\cR$ has dimension $2s+1$ and is generated by states 
$|-s\rangle,|-s+1\rangle,...,|s-1\rangle,|s\rangle$ with definite spin projection along a prescribed axis. In 
that case the highest weight $\lambda$ coincides with $s$. The higher-dimensional case is analogous except 
that the number of coefficients specifying $\lambda$ is the rank $\lfloor(D-1)/2\rfloor$ of $\SO(D-1)$. We 
will illustrate this point in section \ref{seRROT} when dealing with partition functions of higher-spin 
fields in Minkowski space.\\

Given a spin representation, the remainder of the construction is straightforward: formula (\ref{lom}) yields 
a Lorentz-invariant measure that can be used to define scalar products (\ref{scall}) of wavefunctions, and 
the Poincar\'e representation acts according to (\ref{spipa}) with the Radon-Nikodym derivative set to 
$\rho_f=1$ thanks to the choice of measure. For example, when $D=4$ and $s=1/2$, any state takes the form 
(\ref{biphi}).

\subsubsection*{Massless particles}

The description of massless particles is analogous to that of massive ones, up to the key difference that the 
massless little group is the Euclidean group (\ref{piliso}).\i{massless particle} It is a semi-direct product 
(\ref{semid}) with 
an Abelian normal subgroup, so the exhaustivity theorem ensures that its irreducible unitary representations 
are induced and classified by momentum-like orbits of their own. From the Poincar\'e viewpoint each induced 
representation of (\ref{piliso}) is a spin representation for a massless particle.\\

In that context one makes the distinction between two types of massless particles: 
particles with \it{discrete spin}\i{discrete spin} are those given by spin representations of (\ref{piliso}) 
with vanishing Euclidean momentum. These are Euclidean analogues of the ``spinning vacua'' described earlier, 
except that they are finite-dimensional. They amount to making the action of $\RR^{D-2}$ in (\ref{piliso}) 
trivial, and coincide with (projective) irreducible unitary representations of $\SO(D-2)$. Thus massless 
particles with 
discrete spin have a finite-dimensional space of spin degrees of freedom. By contrast, massless particles 
with \it{continuous} or \it{infinite spin}\i{continuous spin}\i{infinite spin} are those whose spin 
representations of (\ref{piliso}) have non-trivial Euclidean momentum. The space of spin degrees of freedom 
is infinite-dimensional in that case, since it consists of wavefunctions on a Euclidean momentum orbit 
$\SO(D-2)/\SO(D-3)\cong S^{D-3}$. Particles with continuous spin are generally discarded on the grounds that 
they are unphysical, although they have recently been described in a field-theoretic 
framework \cite{Schuster:2013pxj,Schuster:2013vpr,Schuster:2013pta}.

\subsubsection*{Tachyons}

Tachyons are particles moving faster than light.\i{tachyon} Their little group is $\SO(D-2,1)^{\uparrow}$. It 
is simple 
and non-compact, so tachyons either have no spin at all, or have continuous spin. They are generally 
considered as unphysical.

\subsection{Massive characters}
\label{supaka}

Having completed the enumeration of relativistic particles, we now evaluate characters of massive irreducible 
unitary representations of the Poincar\'e group. Massless characters are relegated to section \ref{susemaka}. 
These computations are important for our purposes, as we will rely on them in 
chapter \ref{c8}. To our 
knowledge, Poincar\'e characters were first studied in \cite{joos1968,nghiem1969} before reappearing more 
recently in 
\cite{Oblak:2015sea,Garbarz:2015lua,Campoleoni:2015qrh}.

\subsubsection*{Setting the stage}

By virtue of the Frobenius formula (\ref{fropo}), the character of an induced representation vanishes when
evaluated on a transformation $f$ that does 
not belong to the little group. Since the character 
is a class function, only the conjugacy class of $f$ matters for 
the final result. Accordingly, for a massive particle in $D$ dimensions we let $f$ be a rotation\i{rotation}
\be
f
=
\bmm
1 & 0 & 0 & \cdots & 0 & 0 & 0\\
0 & \cos\theta_1 & -\sin\theta_1 & \cdots & 0 & 0 & 0\\
0 & \sin\theta_1 & \cos\theta_1 & \cdots & 0 & 0 & 0\\
\vdots & \vdots & \vdots & \ddots & \vdots & \vdots & \vdots\\
0 & 0 & 0 & \cdots & \cos\theta_r & -\sin\theta_r & 0\\
0 & 0 & 0 & \cdots & \sin\theta_r & \cos\theta_r & 0\\
0 & 0 & 0 & \cdots & 0 & 0 & 1
\emm
\label{fodd}
\ee
written here for even $D$ with $r=\lfloor (D-1)/2\rfloor$; if $D$ is odd we erase the last row and the last 
column. We assume for simplicity that all angles $\theta_1,...,\theta_r$ are non-zero.\\

Now let $\mu$ be a quasi-invariant measure on a momentum orbit with mass $M$, and let $\delta$ be the 
corresponding delta function. Given an arbitrary space-time translation $\alpha$, our goal is to 
evaluate the character of $(f,\alpha)$ using eq.\ (\ref{fropo}). To do this we treat separately odd and even 
space-time dimensions.

\subsubsection*{Odd dimensions}

For odd $D$ we erase the last row and column of (\ref{fodd}). Then
the Frobenius formula (\ref{fropo}) localizes to the unique rotation-invariant point of 
the momentum orbit, namely the momentum at rest $p=(M,0,0,...,0)$. This allows us to simplify (\ref{fropo}) 
by setting $k=p$ in the exponential and the little group character, and pulling them out of the 
integral. Denoting by $\lambda$ the spin of the particle (it is a highest weight for $\SO(D-1)$), 
we find
\be
\chi[(f,\alpha)]
=
e^{iM\alpha^0}\chi^{(D-1)}_{\lambda}[f]
\int_{\cO_p}d\mu(k)\,\delta(k,f\cdot k)
\label{Xixi}
\ee
where the replacement of $k$ by $p$ has projected the translation $\alpha$ on its time component $\alpha^0$. 
The little group character $\chi^{(D-1)}_{\lambda}[f]$ is some function of the angles $\theta_1,...,\theta_r$ 
that we 
do not need to write down at this stage (in practice it follows from the Weyl character formula and is 
displayed in eq.\ (\ref{JESUS}) below). To obtain (\ref{Xixi}) it only remains to 
evaluate the integral of 
the delta function. As coordinates on the orbit we choose the spatial components of momentum, in terms of 
which the Lorentz-invariant measure on $\cO_p$ is (\ref{lom}) and the corresponding delta function is 
(\ref{deltakix}). We thus get
\be
d\mu(k)\,\delta(k,q)
=
\frac{d^{D-1}\bbk}{\sqrt{M^2+\bbk^2}}\;\sqrt{M^2+\bbk^2}\,\delta^{(D-1)}(\bbk-\bbq)
=
d^{D-1}\bbk\;\delta^{(D-1)}(\bbk-\bbq)\,,
\label{Cancel}
\ee
where the multiplicative factors of the measure and its delta function cancel out. Note that
the 
same cancellation would have taken place for \it{any} measure $\mu$ proportional to $d^{D-1}\bbk$, in 
accordance with the fact that induced representations are insensitive to the choice of 
measure. Applied to (\ref{Xixi}), the cancellation (\ref{Cancel}) allows us to write
\be
\chi[(f,\alpha)]
=
e^{iM\alpha^0}
\chi^{(D-1)}_{\lambda}[f]
\int_{\RR^{D-1}}d^{D-1}\bbk\;\delta^{(D-1)}(\bbk,f\cdot\bbk)
\label{Chachap}
\ee
where $f\cdot\bbk$ denotes the action of the spatial submatrix of (\ref{fodd}) on $\bbk$. The integral can be 
written as
\be
\int_{\RR^D}d^{D-1}\bbk\;\delta^{(D-1)}\big((\II-f)\cdot\bbk\big)
=
\frac{1}{\det(\II-f)}
\label{dif}
\ee
where $\II$ is the $(D-1)$-dimensional identity matrix. In terms of angles $\theta_i$ we find
\be
\det(\II-f)
=
\prod_{j=1}^r
\begin{vmatrix}
1-\cos\theta_j & \sin\theta_j \\ -\sin\theta_j & 1-\cos\theta_j
\end{vmatrix}
=
\prod_{j=1}^r4\sin^2\theta_j
=
\prod_{j=1}^r|1-e^{i\theta_j}|^2.
\label{hidif}
\ee
Plugging this into (\ref{dif}), the character (\ref{Chachap}) finally becomes\i{Poincar\'e 
character}\i{character!for Poincar\'e}
\be
\chi[(f,\alpha)]
=
e^{iM\alpha^0}
\chi^{(D-1)}_{\lambda}[f]
\prod_{j=1}^r\frac{1}{|1-e^{i\theta_j}|^2}\,.
\label{Chamouss}
\ee
Note that for a Euclidean time translation $\alpha^0=i\beta$, this quantity may be seen as the partition 
function of a relativistic particle in a rotating frame (albeit with purely imaginary angular velocity).

\paragraph{Remark.} The localization effect (\ref{dif}) is a restatement of the Atiyah-Bott fixed point 
theorem.\i{Atiyah-Bott theorem}\i{fixed point} In that context the term
\be
\frac{1}{\det(\II-f)}
\label{detif}
\ee
is the Lefschetz number\i{Lefschetz number} of the operator $\cT[(f,\alpha)]$. If $f$ was a number 
$e^{-\beta\omega}$, (\ref{detif}) would coincide with the partition function of a harmonic oscillator with 
frequency $\omega$ at temperature $1/\beta$.

\subsubsection*{Even dimensions}

For even $D$ the rotation $f$ is exactly given by (\ref{fodd}). Then
the situation is more complicated because the integral (\ref{fropo}) localizes to a line rather 
than a 
point, as in fig.\ \ref{PFix}. To make things simple we take $\alpha=(\alpha^0,0,...,0)$ to be a pure time 
translation. Formula (\ref{Xixi}) is then replaced by
\be
\chi[(f,\alpha)]
=
\chi^{(D-1)}_{\lambda}[f]\int_{\RR^{D-1}}d^{D-1}\bbk\;
e^{i\alpha^0\sqrt{M^2+\bbk^2}}\;\delta^{(D-1)}(\bbk-f\cdot\bbk)
\label{Xypo}
\ee
where we have already implemented the simplification (\ref{Cancel}). The $\SO(D-1)$ character 
$\chi^{(D-1)}_{\lambda}$ has been pulled out of the integral because, for $D$ even, boosts along the 
direction 
$k_{D-1}$ commute with rotations (\ref{fodd}). It remains once more to integrate the delta function in 
(\ref{Xypo}). As far as the first $D-2$ components of $\bbk$ are concerned, the computation is the same as in 
the odd-dimensional case and results in a factor (\ref{dif}) given by (\ref{hidif}). But the last component 
of $\bbk$ is untouched by (\ref{fodd}), so (\ref{Xypo}) becomes
\be
\chi[(f,\alpha)]
=
\chi^{(D-1)}_{\lambda}[f]
\prod_{j=1}^r\frac{1}{|1-e^{i\theta_j}|^2}
\int_{-\infty}^{+\infty}dk\,
e^{i\alpha^0\sqrt{M^2+k^2}}\,\delta^{(1)}(k-k)\,,
\label{Xapo}
\ee
where $k\equiv k_{D-1}$.
Here the last term is an infrared-divergent factor
\be
\delta(k-k)
=
\frac{1}{2\pi}\int_{-\infty}^{+\infty}dz
\equiv
\frac{L}{2\pi}
\label{tikk}
\ee
where the length scale $L$ is an infrared regulator. The integral in (\ref{Xapo}) then gives
\be
\int_{-\infty}^{+\infty}dk\,e^{i\alpha^0\sqrt{M^2+k^2}}
=
2M\,K_1(-iM\alpha^0)
\nn
\ee
where $K_1$ is the first modified Bessel function of the second kind. In conclusion we 
get\i{character!for Poincar\'e}
\be
\chi[(f,\alpha)]
=
\frac{ML}{\pi}\,K_1(-iM\alpha^0)\,\chi^{(D-1)}_{\lambda}[f]
\prod_{j=1}^r\frac{1}{|1-e^{i\theta_j}|^2}\,,
\label{Xufy}
\ee
whose Wick-rotated version can now be seen as the rotating partition function of a particle trapped in a box 
of height $L$.

\subsubsection*{Time translations}

All characters written above diverge when one of the angles $\theta_j$ goes to zero. These 
divergences are infrared since they are due to delta functions evaluated at zero in momentum space, and can 
be regularized as in (\ref{tikk}). A case of particular interest is the character of a pure time translation, 
whose Wick rotation is a canonical partition function (\ref{canoz}). Using 
once 
more the Frobenius formula (\ref{fropo}) and the cancellation (\ref{Cancel}), and letting 
$\alpha=(\alpha^0,0,...,0)$ be a pure time translation, 
we find
\be
\chi[(e,\alpha)]
=
N
\int_{\RR^{D-1}}d^{D-1}\bbk\,e^{i\alpha^0\sqrt{M^2+\bbk^2}}\,\delta^{(D-1)}(0)
\label{chata}
\ee
where $N\equiv\dim(\cE)$ is the number of spin degrees of freedom of the particle.
The infrared-divergent delta 
function can be seen as the spatial volume of the system,
\be
\delta^{(D-1)}(0)
=
\frac{1}{(2\pi)^{D-1}}\int_{\RR^{D-1}}d^{D-1}\bbx
=
\frac{V}{(2\pi)^{D-1}}\,.
\nn
\ee
Using spherical coordinates we can then rewrite (\ref{chata}) as
\begin{align}
\chi[(e,\alpha)]
& =
\frac{NV}{(2\pi)^{D-1}}
\frac{2\pi^{(D-1)/2}}{\Gamma((D-1)/2)}
\int_0^{+\infty}k^{D-2}\,dk\,
e^{i\alpha^0\sqrt{M^2+k^2}}\nn\\
\label{ChaTa}
& =
\frac{2NV}{(2\pi)^{D-1}}
\left(\frac{2\pi M}{-i\alpha^0}\right)^{(D-2)/2}\,
MK_{D/2}(-iM\alpha^0)\,,
\end{align}
where $K_{D/2}$ denotes once more a modified Bessel function of the second kind. This is the character of a 
pure time translation in a massive Poincar\'e representation. For $\alpha^0=i\beta$ purely imaginary, it
becomes the canonical partition function of a 
massive relativistic particle,\i{partition function!for massive particle}\i{massive particle!partition 
function}
\be
\text{Tr}\left(e^{-\beta H}\right)_{\text{massive particle}}
=
\frac{2NV}{(2\pi)^{D-1}}
\left(\frac{2\pi M}{\beta}\right)^{(D-2)/2}\,
M\,K_{D/2}(\beta M).
\label{AHA}
\ee

\subsection{Massless characters}
\label{susemaka}

Characters of massless Poincar\'e representations
can be evaluated along the same lines as massive ones, but 
there are subtleties due to the little group (\ref{piliso}). The latter admits both finite- and 
infinite-dimensional irreducible unitary representations, corresponding to massless particles with discrete 
or 
continuous spin, respectively. Here we focus on the discrete case.
As in the 
massive case we treat separately even and odd dimensions, this time starting with the former. For simplicity 
we take $\alpha$ to be a pure time translation.

\subsubsection*{Even dimensions}

For even $D$ the Lorentz transformation (\ref{fodd}) belongs to the little group of a massless particle since 
it leaves invariant the 
momentum vector $(E,0,...,0,E)$. The character computation then is the same as in the even-dimensional 
massive case; formula (\ref{Xapo}) still holds with $M=0$ and $\chi^{(D-1)}$ replaced by the 
character $\chi^{(D-2)}$ of a 
representation of $\SO(D-2)$ instead of $\SO(D-1)$. Note that for even $D$ these two groups have the same 
rank $r=\lfloor(D-1)/2\rfloor$, so there is no restriction on the values of the angles 
$\theta_1,...,\theta_r$ (this will change for odd $D$). Using the regulator (\ref{tikk}) and the fact that
\be
\int_{-\infty}^{+\infty}dk\,e^{i|k|(\alpha^0+i\varepsilon)}
=
-\frac{2}{i\alpha^0}\,,
\nn
\ee
one finds the character\i{character!for Poincar\'e}\i{massless 
particle!partition function}
\be
\chi[(f,\alpha)]
=
\frac{iL}{\pi\alpha^0}\,
\chi^{(D-2)}_{\lambda}[f]\prod_{j=1}^r\frac{1}{|1-e^{i\theta_j}|^2}\,.
\label{FACH}
\ee
Up to the replacement of $D-1$ by $D-2$, this is 
the limit $M\rightarrow0$ of the massive character 
(\ref{Xufy}).

\subsubsection*{Odd dimensions}

For odd $D$ the transformation (\ref{fodd}) (with the last row and column suppressed) is no longer an element 
of the little group of $(E,0,...,0,E)$ so its character vanishes if $\theta_r\neq0$. This is consistent with 
the fact that 
$\SO(D-2)$ has lower rank than $\SO(D-1)$ when $D$ is odd. Accordingly we now take $\theta_r=0$ in 
(\ref{fodd}), being understood that the last row and column are suppressed. From there on the character 
computation is identical to the cases treated above, except that the infrared divergence of the integral 
becomes worse and requires two regulators $L,L'$:
\be
\int_{\RR^2}dkdq\,e^{i\alpha^0\sqrt{k^2+q^2}}\;\delta^{(1)}(k-k)\delta^{(1)}(q-q)
=
-\frac{LL'}{2\pi(\alpha^0)^2}\,.
\label{Ira}
\ee
Massless characters in odd dimension $D$ thus read
\be
\chi[(f,\alpha)]
=
\chi^{(D-2)}_{\lambda}[f]\left(-\frac{LL'}{2\pi(\alpha^0)^2}\right)
\prod_{j=1}^{r-1}\frac{1}{|1-e^{i\theta_j}|^2}
\label{ChD}
\ee
where it is understood that $\theta_r=0$ in (\ref{fodd}) and $\chi^{(D-2)}_{\lambda}$ is a character of 
$\SO(D-2)$.
Note that this expression is \it{not} the massless limit of (\ref{Chamouss}) because in general 
$\theta_r\neq0$ in the latter formula. However, upon setting $\theta_r=0$ in (\ref{Chachap}) and regulating 
the resulting double infrared divergence as in (\ref{Ira}), the limit $M\rightarrow 0$ does produce an 
expression of the form (\ref{ChD}), albeit with a reducible representation of $\SO(D-2)$. We shall return to 
this in section \ref{seRROT}. Characters of time 
translations can be treated as in the massive case and coincide, up to spin multiplicity, with the massless 
limit of (\ref{ChaTa}).

\begin{advanced}
\subsection{Wigner rotations and entanglement}
\label{susenta}
\end{advanced}

We now analyse the Wigner rotation (\ref{winota}) and show that, for generic spinning particles, it entangles 
momentum and spin degrees of freedom. This phenomenon was first investigated 
in \cite{Peres} (see also \cite{palge2013} and the related considerations in \cite{Alsing2002,Ahn}).\\

\subsubsection*{Wigner rotations}

Consider a particle with mass $M$ and spin representation $\cR$. We wish to understand the action of the 
Wigner rotation (\ref{winota}) for an arbitrary momentum $q$ belonging to its orbit, and for a 
boost\i{boost}\i{pure boost}
\be
f
=
\bmm
\cosh\gamma & -\sinh\gamma & 0 & \cdots & 0 \\
-\sinh\gamma & \cosh\gamma & 0 & \cdots & 0 \\
0 & 0 & 1 & \cdots & 0 \\
\vdots & \vdots & \vdots & \ddots & \vdots \\
0 & 0 & 0 & \cdots & 1
\emm
\label{laboost}
\ee
with rapidity $\gamma$ in the direction $x^1$. Since the $g_q$'s are standard boosts (\ref{gqgq}), 
the 
combination $g_q^{-1}\,f\,g_{f^{-1}\cdot q}$ is a sequence of three pure boosts in a plane, so we may safely 
take $D=3$ without affecting the outcome of the computation. The momentum $q$ then reads
\be
q=
\bmm
\sqrt{M^2+Q^2}\, \\
Q\cos\phii\,\\
Q\sin\phii\,
\emm
\label{quco}
\ee
for some angle $\phii$ and some positive number $Q$. After a mildly cumbersome but straightforward 
computation, one finds a Wigner rotation matrix\i{Wigner rotation}
\be
g_q^{-1}\,f\,g_{f^{-1}\cdot q}
=
\bmm
1 & 0 & 0 \\
0 & \cos\theta & -\sin\theta \\
0 & \sin\theta & \cos\theta
\emm
\label{RotA}
\ee
whose entries are given by
\be
\begin{split}
\sin\theta
&=
-\frac{M\sin(\phii)\sinh(\gamma)}
{\sqrt{Q^2\sin^2(\phii)+(Q\cos(\phii)\cosh(\gamma)+\sqrt{M^2+Q^2}\sinh(\gamma))^2}}\,,\\
\cos\theta
&=
\frac{Q\cosh(\gamma)+\sqrt{M^2+Q^2}\cos(\phii)\sinh(\gamma)}
{\sqrt{Q^2\sin^2(\phii)+(Q\cos(\phii)\cosh(\gamma)+\sqrt{M^2+Q^2}\sinh(\gamma))^2}}\,.
\end{split}
\label{wangle}
\ee
This is a pure rotation, as it should. It represents the fact that a boost acting on a particle with non-zero 
momentum is seen, from the rest frame of the particle, as a boost combined with a rotation (\ref{RotA}) 
rather than a pure boost.
The rotation only affects spin degrees of 
freedom; scalar particles are insensitive to it.\\

The Wigner rotation (\ref{wangle}) is responsible for the phenomenon 
of \it{Thomas precession} \cite{thomas} (see also \cite{jackson}, section 11.8).\i{Thomas precession} The 
latter is visible in atomic physics, where the spin of an electron orbiting around a nucleus undergoes a slow 
precession due to the fact that the electron's acceleration is a sequence of boosts directed towards the 
nucleus.

\subsubsection*{Momentum/spin entanglement}

In (\ref{zissou}) we saw that a space of $\cE$-valued wavefunctions on $\cO_p$ is a tensor product of $\cE$ 
with the scalar space 
$L^2(\cO_p,\mu,\CC)$. For a relativistic particle, the former consists of spin degrees of 
freedom while the latter accounts for momenta (or positions after
Fourier transformation). For example, any state of a massive particle with spin $1/2$ takes the 
form (\ref{biphi}) and describes the separate propagation of the two spin states $|+\rangle$ 
and $|-\rangle$. (For simplicity we use the Dirac notation until the end of this section.)\\

Hilbert space factorizations such as (\ref{zissou}) are seldom preserved by 
unitary maps. Indeed, if $\sH=A\otimes B$ and $|\Psi\rangle\in\sH$ is a state with unit norm, the reduced 
density matrix associated with $|\Psi\rangle$ and 
acting in 
$B$ is
$\rho\equiv\Tr_A|\Psi\rangle\langle\Psi|$.\i{reduced density matrix}\i{density matrix} When $U$ is a unitary 
operator in $\sH$, it is generally 
\it{not} true that the reduced density 
matrix of $U\cdot|\Psi\rangle$ is unitarily equivalent to $\rho$. In particular, $U$ does not preserve the 
degree of entanglement between $A$ and $B$. Accordingly one may ask \cite{Peres} whether Poincar\'e 
representations spoil the splitting 
(\ref{zissou}). To answer this, consider for definiteness a massive spin $1/2$ particle in four dimensions. 
We start from a normalized $\cE$-valued wavefunction
\be
\Psi(q)
=
\psi(q)|+\rangle,
\qquad\text{i.e.}\qquad
|\Psi\rangle
=
|\psi\rangle
\otimes
|+\rangle
\label{pures}
\ee
where $\psi$ is some complex-valued wavefunction while $|+\rangle$ is one of the two members of an 
orthonormal basis $|+\rangle,|-\rangle$ of $\cE$. This state represents a particle with spin up (say along 
the $x^3$ axis) propagating with a momentum probability distribution $d\mu(q)|\psi(q)|^2$. (For definiteness 
we take the measure $\mu$ to be the Lorentz-invariant expression (\ref{lom}).) The corresponding reduced 
density matrix obtained by tracing over spin degrees of freedom acts on the scalar Hilbert space 
$L^2(\cO_p,\mu,\CC)$ and reads
\be
\rho
=
\langle+|\Big(|\psi\rangle|+\rangle\langle\psi|\langle+|\Big)|+\rangle
+
\langle-|\Big(|\psi\rangle|+\rangle\langle\psi|\langle+|\Big)|-\rangle
=
|\psi\rangle\langle\psi|\,,
\nn
\ee
which is a pure state. Now let us act on (\ref{pures}) with a Lorentz transformation $f$. According to 
(\ref{spipa}), and writing $U\equiv\cT[(f,0)]$, the resulting wavefunction is\i{Wigner rotation!producing 
entanglement}\i{entanglement!due to Wigner rotations}
\be
\left(U\cdot\Psi\right)(q)
=
W_q[f]\cdot\Psi(f^{-1}\cdot q)
\refeq{pures}
\psi(f^{-1}\cdot q)\,W_q[f]|+\rangle
\label{tastate}
\ee
where $W_q[f]$ is the Wigner rotation (\ref{winota}). Denoting 
$\phi(q)\equiv\psi(f^{-1}\cdot q)$ we now find that the entries of the reduced density matrix 
of (\ref{tastate}) are
\be
\tilde\rho(q,q')
=
\big(\phi(q)\chi_+(q)\big)
\big(\phi(q')\chi_+(q')\big)^*
+
\big(\phi(q)\chi_-(q)\big)
\big(\phi(q')\chi_-(q')\big)^*
\label{unpure}
\ee
where we have defined $\chi_{\pm}(q)\equiv\langle\pm|W_q[f]|+\rangle$. In general expression 
(\ref{unpure}) is \it{not} equal to a product $\tilde\psi(q)\tilde\psi^*(q')$ (for some complex wavefunction 
$\tilde\psi$), so the state (\ref{unpure}) is not pure! In particular the boosted state 
(\ref{tastate}) is generally entangled with respect to the splitting (\ref{zissou}), even though the 
original state (\ref{pures}) was not. The reason for this is that the Wigner rotation (\ref{RotA}) generally 
has 
non-vanishing $+-$ entries.
Note that the functions $\chi_{\pm}$ satisfy $|\chi_+|^2+|\chi_-|^2=1$ 
by virtue of the fact that Wigner rotations are unitary, so formula (\ref{unpure}) indeed defines 
a density matrix.\\

These arguments can be generalized to any unitary representation of a semi-direct product (\ref{semid}). The 
only exceptions arise (i) if the spin representation $\cR$ is one-dimensional so that 
$\cE=\CC$ and the 
tensor product (\ref{zissou}) is trivial, or (ii) if $f$ is such that $W_q[f]$ does not depend on $q$. In 
both 
situations the splitting (\ref{zissou}) is 
robust against symmetry transformations. An
example of momentum-independent Wigner rotations will be provided by the Bargmann group below. Thus 
the entanglement of spin and momentum due to Wigner rotations is a purely relativistic effect.

\section{Poincar\'e particles in three dimensions}
\label{sePoTri}

Here we apply the considerations of the previous section to the Poincar\'e group in $D=3$ space-time 
dimensions. This exercise will be a helpful guide for the description of BMS$_3$ particles in part III. 
To our knowledge, representations of Poincar\'e in three dimensions have previously been studied in 
\cite{Binegar:1981gv,Grigore:1993nm}.

\subsection{Poincar\'e group in three dimensions}
\label{suseP3}

\subsubsection*{Prelude: the group SL$(2,\RR)$}

Many properties of the Poincar\'e group in three dimensions rely on the group $\SL$, so we start by 
describing the latter. $\text{SL}(2,\RR)$\i{SL2R@$\SL$} is the group of linear 
transformations of the plane $\RR^2$ that preserve volume and orientation. It consists of real $2\times 2$ 
matrices with unit determinant:
\be
\begin{pmatrix}
a & b \\ c & d
\end{pmatrix},
\quad
ad-bc=1.
\label{slm}
\ee
The centre of $\SL$ consists of the identity matrix and its opposite, thus spanning a 
group $\ZZ_2$. Furthermore:

\paragraph{Lemma.} The group $\SL$ is connected, but not simply connected. It is homotopic to a circle and 
its fundamental group is 
isomorphic to the group of integers $\ZZ$:\i{SL2R@$\SL$!topology}
\be
\pi_1(\SL)\cong\ZZ.
\label{wiktopal}
\ee

\begin{proof}
Since the determinant of (\ref{slm}) is non-zero, the vectors $(a,b)$ and 
$(c,d)$ in $\RR^2$ are linearly 
independent. We can thus find linear combinations of these vectors that span an orthonormal 
basis of $\RR^2$. In other words there exists a real matrix
\be
\bar K=\bmm \bar\alpha & 0 \\ \bar\beta & \bar\gamma \emm
\nn
\ee
such that, for any $\SL$ matrix $S$ of the form (\ref{slm}), the product
\be
\cO\equiv\bar KS
=
\bmm \bar\alpha a & \bar\alpha b \\ \bar\beta a+\bar\gamma c & \bar\beta b+\bar\beta d\emm
\nn
\ee
belongs to the orthogonal group $\text{O}(2)$. We can make $\bar\alpha$ positive by 
setting $\bar\alpha^{-1}=\sqrt{a^2+b^2}$ and we can set $\bar\gamma=1/\bar\alpha$ so that 
$\cO\in\text{SO}(2)$. Any matrix 
$S\in\SL$ can therefore be decomposed uniquely as\i{SL2R@$\SL$!Iwasawa decomposition}\i{Iwasawa decomposition}
\be
S=
\bar K^{-1}\cO\equiv K\cO,
\qquad\text{with}\;
\cO\in\text{SO}(2)\;\text{ and }\;K=\bmm x & 0 \\ y & 1/x \emm
\label{iwa}
\ee
for some $y\in\RR$ and $x\in\RR$ strictly positive.\footnote{This is a rewriting of the Iwasawa 
decomposition.} This shows that 
$\SL$ is connected and homotopic to its maximal compact subgroup consisting of rotations
\be
\bmm
\cos\theta & -\sin\theta \\ \sin\theta & \cos\theta
\emm,
\qquad\theta\in\RR.
\label{slrot}
\ee
In particular, the fundamental group of $\SL$ is isomorphic to $\ZZ$.
\end{proof}

\subsubsection*{Lorentz transformations in three dimensions}

The definitions of section \ref{susePodef} remain valid in three dimensions. In particular the Lorentz 
group $\text{O}(2,1)$ still has four connected components as in fig.\ \ref{figoLor}, and it is still multiply 
connected. However, in contrast to the higher-dimensional case, the Lorentz group is now homotopic to a 
circle and therefore has a fundamental group isomorphic to $\ZZ$. This is a consequence of the following 
result:

\paragraph{Proposition.} There is an isomorphism\i{PSL2R@$\PSL$}\i{Lorentz group!in 3D}\i{SL2R@$\SL$!and 
SO21@and $\text{SO}(2,1)^{\uparrow}$}\i{SO21@$\text{SO}(2,1)^{\uparrow}$}
\be
\SO(2,1)^{\uparrow}\cong\SL/\ZZ_2\equiv\text{PSL}(2,\RR)
\label{isoso}
\ee
where the $\ZZ_2$ subgroup of $\SL$ consists of the identity matrix and its opposite. In particular, 
the fundamental group of the connected Lorentz group in three dimensions is isomorphic to $\ZZ$.

\begin{proof}
Our goal is to build a homomorphism
\be
\phi:\SL\rightarrow\text{O}(2,1):f\mapsto\phi[f]
\label{phiii}
\ee
and then use the property
\be
\text{Im}(\phi)\cong\SL/\text{Ker}(\phi).
\label{wkb}
\ee
Let $A$ be the Lie algebra $\sl$. Each matrix $\alpha\in\sl$ can be written as a linear 
combination\i{sl2R@$\sl$}
\be
\alpha=\alpha^{\mu}t_{\mu}
\label{amoutimou}
\ee
where the $\alpha^{\mu}$'s are real coefficients and the matrices
\be
t_0\equiv\demi\bmm 0 & 1 \\ -1 & 0 \emm,
\qquad
t_1\equiv\demi\bmm 0 & 1 \\ 1 & 0 \emm,
\qquad
t_2\equiv\demi\bmm 1 & 0 \\ 0 & -1 \emm
\label{tmus}
\ee
form a basis of $\sl$. With these 
conventions,
\be
\det(\alpha)=-\frac{1}{4}\eta_{\mu\nu}\alpha^{\mu}\alpha^{\nu}=-\frac{1}{4}\alpha^2.
\label{detX}
\ee
Now $\SL$ naturally acts on $A$ according to the adjoint representation,
\be
A\rightarrow A:\alpha\mapsto f\alpha f^{-1}.
\label{SXS-1}
\ee
This action preserves the determinant since $\det(f)=1$, so according to (\ref{detX}) it may be seen (for 
each $f$) as a Lorentz transformation. This motivates the definition of a homomorphism (\ref{phiii}) given by
\be
f\,t_{\mu}\,f^{-1}=t_{\nu}\,{\phi[f]^{\nu}}_{\mu}
\qquad\forall\,\mu=0,1,2.
\label{fS}
\ee
The entries of $\phi[f]$ are quadratic combinations 
of those of $f$, so $\phi$ is a continuous map. Since $\SL$ is connected, the image $\text{Im}(\phi)$ is 
contained in the connected Lorentz group 
$\SO(2,1)^{\uparrow}$. In fact one has $\text{Im}(\phi)
=
\SO(2,1)^{\uparrow}$,
which follows from the standard decomposition theorem for Lorentz transformations (see 
e.g.\ \cite{HenneauxGroupe,Oblak:2015qia}). The kernel of $\phi$ coincides with the centre of $\SL$, i.e.\ 
$\text{Ker}(\phi)=\{\II,-\II\}$. The isomorphism (\ref{isoso}) follows upon using (\ref{wkb}).
\end{proof}

\paragraph{Remark.} For future reference note that the homomorphism (\ref{fS}) explicitly reads
\be
\phi\left[\bmm a & b \\ c & d \emm\right]
=
\bmm
\demi(a^2+b^2+c^2+d^2) & \demi(a^2-b^2+c^2-d^2) & -ab-cd\\
\demi(a^2+b^2-c^2-d^2) & \demi(a^2-b^2-c^2+d^2) & -ab+cd\\
-ac-bd & bd-ac & ad+bc
\emm,
\label{exoskeleton}
\ee
where the argument of $\phi$ is an $\SL$ matrix. This will be useful in section \ref{sebmsboco}.

\subsubsection*{Poincar\'e group in three dimensions}

The Poincar\'e group for $D=3$ is defined as in (\ref{piggbix}) and its connected subgroup is (\ref{pig}). 
Owing to the isomorphism (\ref{isoso}), its double cover can be written as\i{Poincar\'e group!in 
3D}
\be
\SL\ltimes\RR^3
\label{SiRat}
\ee
where the action of $\SL$ on $\RR^3$ is given by (\ref{SXS-1}). The latter is in fact the adjoint 
representation so we can also rewrite (\ref{SiRat}) as
\be
\text{double cover of ISO}(2,1)^{\uparrow}
\,
=
\,
\SL\ltimes_{\Ad}\sl_{\text{Ab}}
\label{pisel}
\ee
where $\sl_{\text{Ab}}$ is the Lie algebra of $\SL$, seen as an Abelian vector group. This observation will 
turn out to be crucial in part III of this thesis. We stress that (\ref{pisel}) is \it{not} the universal 
cover of the Poincar\'e group in three dimensions, since $\SL$ is homotopic to a circle. This implies that 
(\ref{pisel}) admits topological projective representations (which are equivalent to exact representations of 
its universal cover). There are no algebraic central extensions.

\subsection{Particles in three dimensions}
\label{susePP}

Here we describe projective irreducible unitary representations of the connected Poin\-ca\-r\'e group in 
three 
dimensions and point out a few differences with respect to the higher-dimensional case described in section 
\ref{relagroup}.

\subsubsection*{Orbits and little groups}

The classification of Poincar\'e momentum orbits in three dimensions is the same as in section 
\ref{susepor} and is summarized in fig.\ \ref{figDepIdu}. Considering the double cover (\ref{pisel}) for 
definiteness, the little groups are as follows:\i{little group!for 
relativistic particle}
\begin{table}[H]
\centering
\begin{tabular}{|c|l|}
\hline
Orbit & Little group\\
\hline
Trivial & $\SL$\\
Massive & $\un\cong\SO(2)$\\
Massless & $\RR\times\ZZ_2$\\
Tachyonic & $\RR\times\ZZ_2$\\
\hline
\end{tabular}
\caption{Orbits and little groups for Poincar\'e in three dimensions.}
\label{tabOr}
\end{table}
Let us prove that these are the correct little groups. We shall use the fact that the action of Lorentz 
transformations on momenta is equivalent to its action on translations, which in turn is equivalent to the 
adjoint representation of $\SL$ according to the definition (\ref{pisel}). From that point of view a momentum 
$(p_0,p_1,p_2)$ is represented by a matrix $\eta^{\mu\nu}p_{\mu}t_{\nu}$ where $\eta_{\mu\nu}$ is the 
Minkowski metric in $D=3$ dimensions and the $t_{\mu}$'s are given by (\ref{tmus}). Explicitly the matrix is
\be
p=
\demi\bmm p_2 & -p_0+p_1 \\ p_0+p_1 & -p_2\emm.
\label{PaMatt}
\ee
In that language the little group of $p$ is the set of $\SL$ matrices that commute with (\ref{PaMatt}). It 
immediately follows that the little group of $p=0$ is $\SL$. For massive orbits we move to a rest frame where 
$p_0=M$ and $p_1=p_2=0$; the only matrices leaving $p$ fixed then are rotations (\ref{slrot}). For tachyons 
we take $p_0=p_1=0$, $p_2\neq 0$ and find that the little group consists of matrices of the form
\be
\pm\bmm e^x & 0 \\ 0 & e^{-x}\emm,
\qquad x\in\RR,
\label{TADAT}
\ee
spanning a group $\RR\times\ZZ_2$. Finally, for massless particles we take $p_0=-p_1\neq 0$ and 
$p_2=0$; the resulting little group $\RR\times\ZZ_2$ is spanned by matrices of the type
\be
\pm\bmm1 & x \\ 0 & 1\emm,
\qquad x\in\RR.
\label{TADAM}
\ee
This reproduces all little groups listed in table \ref{tabOr}.\\

Note that the little groups listed in table \ref{tabOr} are sensitive to the cover chosen in 
(\ref{pisel}). Had we chosen the standard connected Poincar\'e group $\SO(2,1)^{\uparrow}\ltimes\RR^3$, the 
little groups for massless particles and tachyons would be quotiented by $\ZZ_2$ and would reduce to $\RR$. 
For the universal cover of the Poincar\'e group, the little groups would instead get decompactified, so 
e.g.\ $\un$ 
would be replaced by $\RR$. This has 
important implications for the spin of relativistic particles in three dimensions.

\subsubsection*{Massive particles}

The properties of massive particles in three dimensions are the same as in section \ref{susePaTho}. In 
particular their momentum orbits take the form\i{massive particle!in 3D}
\be
\cO_p\cong\SO(2,1)^{\uparrow}/\un\cong\SL/S^1.
\label{Mopping}
\ee
The only subtlety is that the group of spatial rotations now is $\un\cong\SO(2)$, so the spin of a massive 
particle is a one-dimensional irreducible unitary representation of the form (\ref{tu1}) labelled by some 
number $s$. If the double cover (\ref{pisel}) was the universal cover of the Poincar\'e group, that number 
would be restricted to integer or half-integer values. However the fact that (\ref{pisel}) is homotopic to a 
circle implies that $s$ may take any real value. Thus massive particles in three dimensions can be 
\it{anyons}\i{anyon} \cite{Leinaas:1977fm,Wilczek:1982wy}. The same phenomenon will occur with massive 
BMS$_3$ particles.

\paragraph{Remark.} Wigner rotations do occur in three dimensions, but they do not lead to momentum/spin 
entanglement when the space of spin degrees of freedom is one-dimensional.

\subsubsection*{Massless particles}

The spin properties of massless particles in three dimensions are also somewhat peculiar compared to those of 
their higher-dimensional cousins.\i{massless particle!in 3D} Their little group 
$\RR\times\ZZ_2$ can be seen as a Euclidean group in one 
dimension, where $\ZZ_2$ plays the role of rotations while $\RR$ is spanned by Euclidean translations. If the 
latter is represented non-trivially, one obtains an analogue of ``continuous spin'' particles in three 
dimensions, although in the present case the space of spin degrees of freedom is actually finite-dimensional. 
By contrast, when $\RR$ is represented trivially, the spin representation boils down to an irreducible 
unitary representation of $\ZZ_2$. The latter has exactly two irreducible unitary representations (the 
trivial one and the fundamental one), so we conclude that ``discrete spin'' massless particles in three 
dimensions can only be distinguished by their statistics (bosonic or fermionic); they have no genuine 
spin. This is consistent with the fact that massless field theories in three 
dimensions either have no local degrees of freedom at all (such as in gravity or Chern-Simons theory), or 
have only scalar or Weyl fermion degrees of freedom.

\subsection{Characters}

For future reference, we now list characters of irreducible unitary representations of the Poincar\'e group 
in three dimensions. The results of sections \ref{supaka} and \ref{susemaka} apply, so
the character of a rotation by $\theta$ combined with an arbitrary translation $\alpha$ in a Poincar\'e 
representation with mass $M$ and spin $s$ is given by formula (\ref{Chamouss}),\i{Poincar\'e 
character!in 3D}\i{character!for Poincar\'e}
\be
\chi[(\text{rot}_{\theta},\alpha)]
=
e^{iM\alpha^0+is\theta}\frac{1}{|1-e^{i\theta}|^2}
=
e^{iM\alpha^0+is\theta}\frac{1}{4\sin^2(\theta/2)}\,,
\label{ChRot}
\ee
where we have replaced the little group character by $\chi_{\lambda}[f]=e^{is\theta}$. In part III we shall 
encounter the BMS$_3$ generalization of this expression. Similarly the character (\ref{AHA}) of Euclidean 
time translations becomes\i{partition function!for massive particle}
\be
\text{Tr}\left(e^{-\beta H}\right)_{\text{massive particle}}
=
\frac{V}{2\pi\beta^2}(1+\beta M)e^{-\beta M}.
\label{TaBou}
\ee
Characters of massless particles with discrete spin are given by formula (\ref{ChD}) with $D=3$, $r=1$ and 
$\chi_{\lambda}=\pm1$.

\begin{advanced}
\section{Galilean particles}
\label{galisec}
\end{advanced}

In this section we classify irreducible unitary representations of the Bargmann groups, i.e.\ 
\it{non-relativistic} or 
\it{Galilean 
particles}.\i{non-relativistic particle}\i{Galilean particle} This example will be useful as a comparison to 
the relativistic case, and will also involve a 
dimensionful
central charge that makes it similar to the centrally extended BMS$_3$ group of part III. 
This being said, the material exposed in this section is not crucial for our later considerations, so it 
may be skipped in a first reading.
The plan is similar 
to that of section \ref{relagroup}: after defining Bargmann groups, we classify their orbits and litte 
groups, describe non-relativistic particles and compute their characters. We refer to 
\cite{deazcarraga,Bose} for further 
reading on the Bargmann groups and to \cite{Inonu} for their representations.\\

\subsection{Bargmann groups}

\subsubsection*{Galilei groups}

\paragraph{Definition.} The \it{Galilei group} in $D$ space-time dimensions is a nested semi-direct 
product\i{Galilei group}
\be
\left(\text{O}(D-1)\ltimes\RR^{D-1}\right)
\ltimes
\left(\RR^{D-1}\times\RR\right)
\label{gali}
\ee
whose elements are quadruples $(f,\bbv,\bbalpha,t)$ where $f\in\text{O}(D-1)$ is a rotation, $\bbv$ 
is a boost belonging to the first $\RR^{D-1}$, 
$\bbalpha$ 
is a spatial translation belonging to the second $\RR^{D-1}$, and $t\in\RR$ is a time translation. The group 
operation is
\be
(f,\bbv,\bbalpha,s)\cdot(g,\bbw,\bbbeta,t)
=
\big(f\cdot g,
\bbv+f\cdot\bbw,
\bbalpha+f\cdot\bbbeta+\bbv t,
s+t\big)
\label{galop}
\ee
where the dots on the right-hand side denote either matrix multiplication, or the action of a matrix on a 
column vector. The largest connected subgroup of (\ref{gali}) is obtained upon replacing $\text{O}(D-1)$ by 
$\SO(D-1)$; its universal cover is obtained by replacing $\SO(D-1)$ by its universal cover, 
$\text{Spin}(D-1)$.\\

The intricate structure (\ref{gali}) translates 
the fact that space and time live on different footings in Galilean relativity. Thus the analogue of a 
Lorentz transformation now is a pair $(f,\bbv)$, while space-time translations are pairs $(\bbalpha,t)$. 
Boosts 
and rotations span a group $\SO(D-1)\ltimes\RR^{D-1}$ while space-time translations span an Abelian group 
$\RR^D$. In particular each boost is a velocity vector $\bbv$ acted upon by rotations according to the matrix 
representation of $\text{O}(D-1)$. Since time is absolute in Galilean relativity,\i{absolute time} the last 
entry on 
the right-hand side of (\ref{galop}) is a sum $s+t$ without influence of boosts. The term $\bbv t$ of the 
third 
entry is a time-dependent translation at velocity $\bbv$. Finally, there is an Abelian subgroup 
$\RR^{2D}$ consisting of pairs
\be
(e,\bbv,\bbalpha,0)
\label{bota}
\ee
where $e$ is the identity in $\text{O}(D-1)$.\\

The Lie algebra of the Galilei group is generated by 
$(D-1)(D-2)/2$ rotation generators, $(D-1)$ boost generators, $(D-1)$ spatial translation generators, and one 
generator of time translations. We will not display their Lie brackets here.

\subsubsection*{Bargmann groups}

The Galilei group turns out to admit a 
non-trivial algebraic central extension:

\paragraph{Definition.} The \it{Bargmann group}\i{Bargmann group} in $D$ space-time dimensions is a centrally 
extended semi-direct product
\be
\text{Bargmann}(D)
\equiv
\left(\text{O}(D-1)\ltimes\RR^{D-1}\right)
\ltimes
\left(\RR^{D-1}\times\RR\right)\times\RR\,,
\label{galix}
\ee
whose elements are $5$-tuples $(f,\bbv,\bbalpha,t,\lambda)$ where $(f,\bbv,\bbalpha,t)$ belongs to the 
Galilei group 
(\ref{gali}) while $\lambda$ is a real number. The group operation is
\be
(f,\bbv,\bbalpha,s,\lambda)\cdot(g,\bbw,\bbbeta,t,\mu)
=\Big(
(f,\bbv,\bbalpha,s)\cdot(g,\bbw,\bbbeta,t),
\lambda+\mu+\bbv\cdot f\cdot\bbbeta+\demi\bbv^2t
\Big)
\label{galopex}
\ee
where the first entry on the right-hand side is given by (\ref{galop}) while $\bbv\cdot\bbbeta\equiv 
v^i\beta^i$ is 
the Euclidean scalar product of $\bbv$ and $\bbbeta$; in particular, 
$\bbv^2\equiv v^iv^i$.\\

This central extension says that the Abelian subgroup of boosts and translations (\ref{bota}) gets extended 
into a Heisenberg 
group (\ref{hohot}):\i{Heisenberg group}
\be
(e,\bbv,\bbalpha,0,\lambda)\cdot(e,\bbw,\bbbeta,0,\mu)
=
\big(e,\bbv+\bbw,\bbalpha+\bbbeta,0,\lambda+\mu+\bbv\cdot\bbbeta\big).
\nn
\ee
In other words, in quantum mechanics, spatial translations and boosts do not commute. Note that, even 
in the Bargmann group, the normal subgroup of (centrally extended) space-time translations
\be
(e,0,\bbalpha,t,\lambda)
\label{extran}
\ee
remains Abelian. Hence the exhaustivity theorem of section \ref{susexh} applies to the Bargmann group: all
Galilean particles are induced representations.\\

For $D\geq4$ space-time dimensions, eq.\ (\ref{galopex}) is 
the only algebraic central extension of the Galilei group. But for $D=3$, the 
Galilei group admits three non-trivial differentiable central extensions \cite{Bose}, one of which is the 
one displayed in (\ref{galopex}). We will not take these extra central extensions into account. As regards 
topological central extensions, the Galilean situation is 
identical to that of the Poincar\'e group. Thus Bargmann$(3)$ has a fundamental group $\ZZ$ and admits 
infinitely many topological projective representations, while for $D\geq 4$ the fundamental group of 
Bargmann$(D)$ is $\ZZ_2$, leading either to exact representations or to representations up to a sign.

\paragraph{Remark.} The Bargmann group is a limit of the Poincar\'e group as the speed of light goes to 
infinity (see e.g.\ \cite{weinberg1995}), known more accurately as an \it{In\"on\"u-Wigner contraction} 
\cite{Inonu:1953sp}.\i{In\"on\"u-Wigner contraction} We will not describe this procedure here, although we 
will encounter a very similar one 
in part III when showing that the BMS$_3$ group is an ultrarelativistic limit of two Virasoro groups.

\subsection{Orbits and little groups}
\label{susoliga}

We now classify the orbits and little groups of the Bargmann group 
(\ref{galix}). We follow the same strategy as in section \ref{susepor}.

\subsubsection*{Generalized momenta}

The Abelian normal subgroup of (\ref{galix}) consists of centrally extended translations 
(\ref{extran}). Its dual space consists of generalized momenta\i{momentum!for Bargmann group}\i{Bargmann 
group!momentum}
\be
(\bbp,E,M)
\label{pem}
\ee
paired with translations according to\footnote{The minus signs are conventional, and included for later 
convenience.}
\be
\langle(\bbp,E,M),(\bbalpha,t,\lambda)\rangle
=
\langle \bbp,\bbalpha\rangle-Et-M\lambda
\label{patem}
\ee
where $\langle\bbp,\bbalpha\rangle\equiv p_i\alpha^i$, $i=1,...,D-1$. Accordingly, $\bbp$ is dual to spatial 
translations and represents the actual momentum of a particle; $E$ is dual to time translations and
represents the particle's energy; finally $M$ is a \it{central charge}\i{central charge!for Bargmann 
group}\i{mass!as central charge}\i{dimensionful central charge}\i{non-relativistic mass} dual to the 
central entries $\lambda$ in (\ref{extran}). Working in units 
such that $\hbar=1$, eq.\ (\ref{galopex}) 
says that $\lambda$ has dimensions $[\text{distance}]\times[\text{velocity}]$ so the pairing 
(\ref{patem}) 
implies that $M$ is a mass scale:\i{dimensional analysis}
\be
[M]
=
\frac{[\text{energy}]\times[\text{time}]}{[\text{distance}]\times[\text{velocity}]}
=
[\text{mass}].
\label{gallimm}
\ee
In fact we will see below that $M$ 
is the mass of a non-relativistic particle. Note that $M$ lives on a different footing than $\bbp$ and $E$, 
which is why the relation $E=Mc^2$ is invisible in Galilean relativity.\\

According to the structure (\ref{galix}), the group $G$ acting on translations is the 
Euclidean group spanned by rotations and boosts. Its action is given by
\be
\sigma_{(f,\bbv)}(\bbalpha,t,\lambda)
=
\Big(f\cdot\bbalpha+\bbv t,t,
\lambda+\bbv\cdot f\cdot\bbalpha+\demi\bbv^2t
\Big)
\label{fiss}
\ee
by virtue of (\ref{galopex}). The pairing (\ref{patem}) then yields
the action $\sigma^*$ of boosts 
and rotations on generalized momenta:
\begin{eqnarray}
&   & \big<\sigma^*_{(f,\bbv)}(\bbp,E,M),(\bbalpha,t,\lambda)\big>=\nn\\
& \refeq{sstar} &
\!\!\!\big<(\bbp,E,M),\sigma_{(f^{-1},-f^{-1}\cdot\bbv)}(\bbalpha,t,\lambda)\big>\nn\\
& = &
\!\!\!\Big<
(\bbp,E,M),
\Big(
f^{-1}\cdot\bbalpha-f^{-1}\cdot\bbv t,
t,
\lambda-\bbv\cdot\bbalpha+\demi\bbv^2t
\Big)
\Big>\nn\\
\label{CROSSA}
& \refeq{patem} &
\!\!\!\big<\bbp,f^{-1}\cdot\bbalpha-f^{-1}\cdot\bbv t\big>
-
Et
-
M\Big(\lambda-\bbv\cdot\bbalpha+\demi\bbv^2t\Big)\,,
\end{eqnarray}
where we have used the fact that rotations preserve Euclidean 
scalar products. We can then use the Euclidean analogue of the isomorphism (\ref{ii}) to 
identify $(\RR^{D-1})^*$ with $\RR^{D-1}$ and rewrite the pairing $\langle\bbp,\bbalpha\rangle=p_i\alpha^i$ 
as a scalar product 
$\bbp\cdot\bbalpha=p^i\alpha^i=p_i\alpha_i$,
where indices are raised and lowered thanks to the Euclidean metric.
This allows us to rewrite $\bbv\cdot\bbalpha$ as 
$\langle\bbv,\bbalpha\rangle$ in (\ref{CROSSA}),
and leads to
\be
\sigma^*_{(f,\bbv)}(\bbp,E,M)
=
\Big(
f\cdot\bbp+M\bbv,
E+\bbv\cdot f\cdot\bbp+\demi M\bbv^2,M
\Big).
\label{fpem}
\ee
One may recognize here the non-relativistic transformations laws 
of momentum and energy under rotations and boosts. The mass $M$ is left unchanged, as was to be expected for 
a central charge.

\subsubsection*{Orbits}

Let us classify orbits of generalized momenta under the transformations (\ref{fpem}). 
Since the mass $M$ is invariant, it is a constant quantity specifying each 
orbit; orbits with different masses are disjoint. In particular, the orbits differ greatly depending on 
whether $M$ vanishes or not.\\

A \it{massless non-relativistic particle} is one for which $M=0$, whereupon
(\ref{fpem}) simplifies to
\be
\sigma^*_{(f,\bbv)}(\bbp,E,0)
=
\Big(
f\cdot\bbp,
E+\bbv\cdot f\cdot\bbp,0
\Big).
\label{galiless}
\ee
This implies that the norm of the momentum $\bbp$ of a massless particle is invariant under rotations 
and 
boosts. If $\bbp=0$ the particle is static (in all reference frames) and the momentum orbit is trivial. If on 
the other hand $\bbp\neq0$, then the particle moves (in all references frames); its momentum orbit 
is\i{massless particle}
\be
\cO_{(\bbp,E,0)}
=
\left\{(f\cdot\bbp,E+\bbv\cdot f\cdot\bbp,0)\big|
f\in\SO(D-1),\;\bbv\in\RR^{D-1}\right\}\nn\\
\cong
S^{D-2}\times\RR
\label{orless}
\ee
where the sphere $S^{D-2}$ is spanned by all momenta $f\cdot\bbp$ while $\RR$ is spanned by the values of 
energy. The little group is\i{little group!for non-relativistic particle}
\be
G_{(\bbp,E,0)}
=
\SO(D-2)\ltimes\RR^{D-2}
\label{ligomess}
\ee
and consists of rotations leaving $\bbp$ invariant together with boosts that are 
orthogonal to $\bbp$. Note that this is the same little group (\ref{piliso}) as for relativistic 
massless particles.\\

A \it{massive non-relativistic particle} is such that $M\neq 0$. Let $(\bbp,E)$ be its momentum and 
energy. Then the boost $\bbv=-\bbp/M$ plugged in (\ref{fpem}) maps
$(\bbp,E,M)$ on
\be
\sigma^*_{(e,-\bbp/M)}(\bbp,E,M)
=
\Big(0,E+\frac{\bbp^2}{2M},M\Big)
\label{reza}
\ee
so any massive particle admits a rest frame.\i{rest frame}
If we call $E_0\equiv E+\bbp^2/2M$, the orbit of (\ref{reza}) under rotations and boosts is a 
parabola\i{non-relativistic particle}\i{Galilean particle}
\be
\cO_{(0,E,M)}
=
\left\{\Big(M\bbv,E_0+\frac{M\bbv^2}{2},M\Big)\bigg|
\bbv\in\RR^{D-1}\right\}
\subset\RR^{D-1}\times\RR.
\label{parabol}
\ee
As orbit representative we can take the generalized momentum in the rest frame,
\be
(0,E_0,M)
\label{galirest}
\ee
where $E_0$ is an arbitrary real number; at fixed $M$, representatives with different values of $E_0$ 
define distinct orbits. The little group is the group of rotations
\be
G_{(0,E_0,M)}
=
\SO(D-1)
\label{ligomass}
\ee
in accordance with the fact that the orbit (\ref{parabol}) is diffeomorphic to the quotient 
space $\left(\SO(D-1)\ltimes\RR^{D-1}\right)/\SO(D-1)\cong\RR^{D-1}$. Note again that this is exactly the 
same little group (\ref{piligoma}) as for relativistic massive particles. Finally, pure boosts
\be
g_{\bbq}=(e,\bbq/M)
\label{sboga}
\ee
provide a continuous family of standard boosts on the orbit (\ref{parabol}) of (\ref{galirest}).
Note that 
energy is bounded from below on the orbit if and only if $M>0$.

\subsection{Particles}
\label{sugalipette}

According to the exhaustivity theorem of section \ref{susexh}, all irreducible unitary representations of 
Bargmann groups are induced, and they are classified by momentum orbits.\i{induced representation!of 
Bargmann group} Each such representation consists of 
wavefunctions on an orbit, representing the quantum states of a non-relativistic particle.\\

For example, the spin of a massive Galilean particle is an irreducible unitary representation of 
$\SO(D-1)$.\i{spin}\i{massive particle} The space of states 
of the particle then consists of wavefunctions on the orbit (\ref{parabol}) taking values in the space of 
the spin representation. Scalar products of wavefunctions are defined as usual by (\ref{scall}), where $\mu$ 
is some measure on the orbit. For convenience one can pick the standard Lebesgue measure $d^{D-1}\bbq$, which 
is left invariant by both rotations and boosts since (\ref{fpem}) says that they act on (\ref{parabol}) as 
Euclidean transformations $\bbq\mapsto f\cdot\bbq+M\bbv$.\\

In order to write down formula (\ref{spipa}) explicitly for a non-relativistic particle, we still need to 
understand the Wigner rotation (\ref{winota}). Let us evaluate it for a pair $(f,\bbv)$ at a 
point $\bbq$ belonging to the momentum orbit. We have $(f,\bbv)\cdot\bbq=f\cdot\bbq+M\bbv$,
so the standard boost (\ref{sboga}) for the momentum $(f,\bbv)^{-1}\cdot\bbq$ is $g_{(f,\bbv)^{-1}\cdot\bbq}
=
\big(e,\frac{f^{-1}\cdot\bbq}{M}-f^{-1}\cdot\bbv\big)$. Using the group operation (\ref{galop}) we read off 
the Wigner rotation
\be
g_{\bbq}^{-1}\cdot (f,\bbv)\cdot g_{(f,\bbv)^{-1}\cdot\bbq}
=
(f,0)\,.
\label{wisim}
\ee
Surprise: the Wigner rotation is blind to boosts! In fact it is momentum-independent and simply coincides 
with the rotation $f$. Thus formula (\ref{spipa}) for the transformation law of non-relativistic one-particle 
states becomes
\be
\big(\cT[(f,\bbv,\bbalpha,t,\lambda)]\cdot\Psi\big)(q)
=
e^{-iM\lambda}
e^{i\bbq\cdot\bbalpha-i\bbq^2t/2M}\,
\cR[f]\cdot
\Psi\left((f,\bbv)^{-1}\cdot\bbq\right),
\label{spipaga}
\ee
where we have also used the fact that the measure $d^{D-1}\bbq$ is invariant to cancel its Radon-Nikodym 
derivative. This result differs from the Poincar\'e transformations of relativistic particles in two key 
respects. First, Galilean Wigner rotations (\ref{wisim}) are momentum-independent, so in contrast to 
(\ref{wangle}) they do \it{not} entangle momentum and spin. In fact, there is no Thomas precession for 
non-relativistic particles. The second difference is the presence of the mass $M$: formula (\ref{spipaga}) is 
an \it{exact} representation of the Bargmann group (\ref{galopex}), but because $M\neq 0$ it is a 
\it{projective} representation of the centreless Galilei group (\ref{gali}). This can be seen by noting that 
for a pure boost $\bbv$ and a spatial translation $\bbalpha$, eq.\ (\ref{spipaga}) gives\i{projective 
representation!of Galilei group}\i{Galilei group!projective representation}
\be
\cT[\bbv]\cdot\cT[\bbalpha]
=
e^{-iM\bbv\cdot\bbalpha}\,\cT[\bbalpha]\cdot\cT[\bbv],
\label{nocotiga}
\ee
which says that boosts and spatial translations do not commute. In part III we will encounter a similar 
phenomenon with the BMS$_3$ group, whose dimensionful central charge will coincide with 
the Planck mass.

\subsection{Characters}

We now evaluate characters of massive non-relativistic particles.\i{Bargmann character} Let $M>0$ and choose 
a spin 
$\lambda$, specifying an irreducible unitary representation of the little group $\SO(D-1)$. For 
definiteness we take the rest frame energy $E_0=0$ in (\ref{parabol}). In order 
for the 
character 
(\ref{fropo}) to be non-zero we must set $\bbv=0$. Eq.\ (\ref{wisim}) then allows us to pull 
the little group character $\chi_{\cR}=\chi^{(D-1)}_{\lambda}$ out of the momentum integral:
\be
\chi[(f,0,\bbalpha,t,\lambda)]
=
e^{-iM\lambda}
\chi^{(D-1)}_{\lambda}[f]
\int_{\RR^{D-1}}d^{D-1}\bbk\;\delta^{(D-1)}(\bbk-f\cdot\bbk)\;
e^{i\bbk\cdot\bbalpha-i\bbk^2t/2M}.
\label{chu}
\ee
For simplicity we set $\lambda=0$ from now on and neglect writing this entry. We take $f$ to be a rotation 
(\ref{fodd}) with the first row and column suppressed and all angles $\theta_1,...,\theta_r$ non-zero, 
$r=\lfloor(D-1)/2\rfloor$. If $D$ is odd we also erase the last row and column. We treat separately even and 
odd dimensions.\\

If $D$ is odd, then the only fixed point of $f$ on (\ref{parabol}) is the tip $\bbk=0$. The integral of 
(\ref{chu}) localizes and (\ref{dif}) yields
\be
\chi[(f,0,\bbalpha,t)]
=
\chi_{\lambda}[f]
\prod_{j=1}^r\frac{1}{|1-e^{i\theta_j}|^2}\,.
\label{galimacha}
\ee
Note that translations do not contribute to this result. Up to the normalization of energy, it coincides with 
the relativistic character (\ref{Chamouss}).\\

If $D$ is even, then $f$ leaves fixed the whole axis $k_{D-1}$ as in fig.\ \ref{PFix}. Integrating first 
over the rotated coordinates $k_1,...,k_{D-2}$ in (\ref{chu}) and writing $k_{D-1}\equiv k$, we find
\be
\chi[(f,0,\bbalpha,t)]
=
\chi^{(D-1)}_{\lambda}[f]
\prod_{j=1}^r\frac{1}{|1-e^{i\theta_j}|^2}
\int_{-\infty}^{+\infty}dk\,\delta(0)\,
e^{ik\alpha^{D-1}-ik^2t/2M}.
\label{chuki}
\ee
Here the term $\delta(0)=\delta(k-k)$ is an infrared divergence that we 
regularize as in (\ref{tikk}) with a length scale $L$. Denoting $\alpha^{D-1}\equiv x$, we 
are left 
with the integral
\be
\int_{-\infty}^{+\infty}dk\,
e^{ikx-ik^2t/2M}
=
\left(\frac{2\pi M}{it}\right)^{1/2}e^{iMx^2/2t}
\label{galiprop}
\ee
and thus conclude
\be
\chi[(f,0,\bbalpha,t)]
=
\frac{L}{2\pi}
\chi^{(D-1)}_{\lambda}[f]
\prod_{j=1}^r\frac{1}{|1-e^{i\theta_j}|^2}
\left(\frac{2\pi M}{it}\right)^{1/2}e^{iMx^2/2t}.
\label{Yufy}
\ee
The only dependence of this expression on $\bbalpha$ appears through the 
component $\alpha^{D-1}\equiv x$, because we picked a rotation $f$ leaving fixed the direction 
$k_{D-1}$. For a general rotation, the component of $\bbalpha$ appearing in the character would be its 
projection on the axis left fixed by $f$.\\

Two comments are in order. First note that in $D=2$ 
space-time 
dimensions, (\ref{Yufy}) boils down to the quantum propagator of a free non-relativistic particle at time $t$ 
and 
separation $x$, up to an infrared-divergent factor $L$.\i{propagator} This is because the character of a pure 
spatial translation in two 
space-time dimensions is
\be
\Tr\big(\cT[(e,x,t)]\big)
=
\int_{-\infty}^{+\infty}dk\,\delta(0)\,
e^{ikx-itk^2/2M}
=
\int_{-\infty}^{+\infty}\frac{dy}{2\pi}
\int_{-\infty}^{+\infty}dk
e^{ikx-itk^2/2M}
\nn
\ee
which we can interpret as the trace of the operator $e^{iPx-iHt}$ in the Hilbert space of a free massive 
particle on the real line:
\be
\Tr\big(\cT[(e,x,t)]\big)
=
\Tr\left(e^{iPx-iHt}\right)
=
\int_{-\infty}^{+\infty}dy\bra y+x|e^{-iHt}|y\ket.
\label{gatra}
\ee
The integrand of this expression is the propagator 
of a free non-relativistic particle evaluated between $y$ and $y+x$ at time $t$ and coincides with 
(\ref{galiprop}).\\

The second comment concerns the relation between Bargmann characters and Poincar\'e characters. For even $D$, 
(\ref{Yufy}) is the non-relativistic analogue of (\ref{Xufy}) but the functions appearing in the two results 
are different. By contrast, for odd $D$, the Bargmann character (\ref{galimacha}) coincides with its 
Poincar\'e analogue (\ref{Chamouss}). This may be seen as a consequence of the phenomenon (\ref{dif}), whose 
effect is to localize the computation of the character to the region of momentum space surrounding the 
momentum at rest, that is, the non-relativistic region. By contrast, when the localization is not complete as 
is the case 
for even $D$, the momenta in the integral (\ref{galiprop}) are arbitrarily large and relativistic effects 
become important. This produces a difference between Bargmann and Poincar\'e characters. It is particularly 
apparent for characters of Euclidean time translations, which in the non-relativistic case are given by
\be
\chi[(e,0,0,-i\beta)]
=
\frac{NV}{(2\pi)^{D-1}}\int_{\RR^{D-1}}d^{D-1}\bbk\,e^{-\beta\bbk^2/2M}
=
NV\left(\frac{M}{2\pi\beta}\right)^{(D-1)/2}
\nn
\ee
where $N$ is the dimension of the spin representation. This is the non-relativistic version of (\ref{AHA}). 
For $D=3$ (and $N=1$) it reduces to $VM/(2\pi\beta)$, which is the non-relativistic limit of (\ref{TaBou}).

\chapter{Coadjoint orbits and geometric quantization}
\label{c3}
\markboth{}{\small{\chaptername~\thechapter. Coadjoint orbits and quantization}}

In the previous chapters we have seen how representation theory leads to geometric 
objects such as orbits. The purpose of this 
chapter is to describe the opposite phenomenon: starting from a \it{coadjoint orbit} of a group $G$, we will 
obtain a representation by \it{quantizing} the orbit. This construction will further explain why orbits of 
momenta classify representations of semi-direct products. In addition it will turn out to be a tool for 
understanding gravity in parts II and III.\\

The plan is as follows. We start in section \ref{seSymPa} with basic reminders on symplectic 
manifolds with symmetries, including their momentum maps. Along the way we introduce the notion of 
coadjoint orbits, which will turn out to be crucial for the remainder of this thesis. Section \ref{geoq} is 
then devoted to the quantization of symplectic manifolds, and describes in particular the relation between 
representation theory and symplectic geometry. In section \ref{gigglema} we reformulate geometric 
quantization in terms of action principles that describe the propagation of a point particle on a group 
manifold. The two last sections of the chapter are concerned with applications of these considerations to 
semi-direct products: in section \ref{secosemi} we describe the coadjoint orbits and world line actions of 
such groups in general, while in section \ref{seReWo} we illustrate these results with the 
Poincar\'e group and the Bargmann group.\\

Our language in this chapter will be slightly different than in the previous ones, as we 
rely on differential-geometric tools that were unnecessary for our earlier considerations. Useful references 
include \cite{abraham1978foundations,khesin2008geometry} for symplectic 
geometry, \cite{woodhouse1997,kirillov2004lectures} for quantization, as well as the 
(sadly unpublished) Modave lecture notes \cite{debuyl}.\i{Modave}

\paragraph{Remark.} The presentation adopted here is self-contained, but fairly dense. We urge the reader who 
is not acquainted with differential geometry to only read sections \ref{suseLILI} and \ref{suseCA}, then go 
directly to part II of the thesis. In doing so one will miss the symplectic aspects of our later 
considerations, but the other points of our presentation should remain accessible.

\section{Symmetric phase spaces}
\label{seSymPa}

In this section we study classical systems with symmetries, that is, homogeneous symplectic manifolds. We 
start by recalling a few basic facts about Lie groups and we define their adjoint and coadjoint 
representations. We then describe in general terms Poisson and symplectic structures, and show how 
such structures arise in the case of coadjoint orbits. Finally we discuss the notion of momentum maps 
associated with the symmetries of a symplectic manifold. We use the notational conventions of chapter 
\ref{c1b}.

\subsection{Lie groups}
\label{suseLILI}

A \it{Lie group}\i{Lie group} is a group $G$ which also has a structure of smooth manifold such that 
multiplication and inversion are smooth maps. In particular the operations of left and right multiplication 
defined in (\ref{left}) and (\ref{RighT}) are diffeomorphisms. We denote by $e$ the identity in $G$, and 
generic group elements are denoted $f$, $g$, etc.

\paragraph{Definition.} A vector field $\xi$ on $G$ is \it{left-invariant}\i{left-invariant vector 
field}\i{vector field!left-invariant} if 
$(L_f)_*\xi=\xi$ for all $f\in G$, i.e.\ if $(L_f)_{*g}\xi_g=\xi_{fg}$ for all $f,g\in 
G$.\footnote{Recall that the \it{differential} of a smooth map $\cF:\cM\rightarrow\cN$ at 
$p\in\cM$ is the 
map 
$\cF_{*p}:T_p\cM\rightarrow 
T_{\cF(p)}\cN:\dot\gamma(0)\mapsto\frac{d}{dt}\big[\cF(\gamma(t))\big]\big|_{t=0}$, where $\gamma(t)$ is a 
path in $\cM$ such that $\gamma(0)=p$.}\i{differential}\\

One can verify that any left-invariant vector field is given by $\xi_g=(L_g)_{*e}X$ for some tangent vector 
$X\in T_eG$. Thus the space of left-invariant vector fields is isomorphic to the tangent space of $G$ at the 
identity. We shall denote by $\zeta_X$ the left-invariant vector field on $G$ given by 
$(\zeta_X)_g=(L_g)_{*e}X$.

\paragraph{Definition.} The \it{Lie algebra}\i{Lie algebra} of $G$ is the vector space $\mg= T_eG$ 
endowed with the Lie bracket\i{Lie bracket}
\be
[X,Y]\equiv[\zeta_X,\zeta_Y]_e
\label{A196}
\ee
where the bracket on the right-hand side is the usual Lie bracket of vector fields evaluated at the 
identity.\\

One can show that the bracket (\ref{A196}) is such that $\zeta_{[X,Y]}=[\zeta_X,\zeta_Y]$. As a corollary, 
any smooth homomorphism of Lie groups $\cF:G\rightarrow H$ is such that its differential $\cF_{*e}$ at the 
identity is a homomorphism of Lie algebras. When interpreting $G$ as a symmetry group, the elements of its 
Lie algebra are seen as ``infinitesimal'' symmetries, i.e.\ transformations near the identity. In practice 
the Lie algebra structure of $\mg$ is often displayed in terms of a basis
$\{t_a|a=1,...,\dim\mg\}$ of $\mg$ with Lie brackets\i{structure constant}
\be
[t_a,t_b]
=
f_{ab}{}^c\,t_c\,.
\label{commurel}
\ee
In that context the coefficients $f_{ab}{}^c\in\RR$ are known as the \it{structure constants}\i{structure 
constant} of $\mg$ in the basis $\{t_a\}$.

\subsubsection*{Exponential map}

\paragraph{Definition.} Let $X\in\mg$, and let $\gamma_X$ be the integral curve\footnote{An 
\it{integral curve} of 
a 
vector field $\xi$ on a manifold $\cM$ is a path $\gamma(t)$ on $\cM$ such that 
$\dot\gamma(t)=\xi_{\gamma(t)}$.}\i{integral curve} of the corresponding 
left-invariant vector field $\zeta_X$ such that $\gamma_X(0)=e$. Then the \it{exponential 
map}\i{exponential map} of $G$ is
\be
\exp:\mg\rightarrow G:X\mapsto\exp[X]\equiv\gamma_X(1).
\label{ExpGa}
\ee
One can verify that, for matrix groups, this definition reduces to the standard Taylor series 
$\sum_{n\in\NN}X^n/n!$. We often denote $\exp[X]\equiv e^X$.\\

Since the exponential map is defined by a vector flow, it automatically satisfies 
$\exp[(s+t)X]=\exp[sX]\exp[tX]$ for all $s,t\in\RR$. In particular any $X\in\mg$ determines a one-dimensional 
subgroup of $G$ consisting of elements $\exp[tX]$, $t\in\RR$. Note that left-invariant vector fields are 
complete, which ensures the existence of $\exp[tX]$ for all $t\in\RR$. Finally, for any smooth homomorphism 
$\cF:G\rightarrow H$, one can show that
\be
\cF\circ\exp_G=\exp_H\circ\cF_{*e}\,.
\label{FEXTIL}
\ee

\subsection{Adjoint and coadjoint representations}
\label{suseCA}

\paragraph{Definition.} Let $G$ be a Lie group with Lie algebra $\mg$. Then the \it{adjoint 
representation}\i{adjoint representation}\i{representation!adjoint} of $G$ is the homomorphism
\be
\text{Ad}:G\rightarrow\text{GL}(\mg):
g\mapsto\text{Ad}_g
\label{Had}
\ee
where $\text{Ad}_g$ is the linear operator that acts on $\mg$ according to
\be
\text{Ad}_g(X)
=
\frac{d}{dt}\left.\left(g\,e^{tX}\,g^{-1}\right)\right|_{t=0}.
\label{ad}
\ee
Here one may freely replace $e^{tX}$ by any path $\gamma(t)$ in $G$ such that $\gamma(0)=e$ and 
$\dot\gamma(0)=X$. For matrix groups, eq.\ (\ref{ad}) reduces to $\text{Ad}_g(X)=gXg^{-1}$.\\

One can verify that this is indeed a representation of $G$. Using (\ref{FEXTIL}), one also shows that it 
satisfies the 
identity
\be
e^{\Ad_fX}
=
f\,e^X\,f^{-1}
\label{FADO}
\ee
where $e^X$ is the exponential map of $G$.
Note that the adjoint representation of any Abelian Lie group is trivial.
Finally, the adjoint representation of the Lie 
algebra $\mg$ is defined as the 
differential of (\ref{Had}) at the identity:
\be
\ad_X(Y)\equiv
\frac{d}{dt}
\big(
\Ad_{e^{tX}}(Y)
\big)\big|_{t=0}
=
[X,Y]\,.
\label{adg}
\ee
\vspace{.1cm}

In (\ref{sstar}) we saw how to define dual representations. Let us apply this to the adjoint 
representation (\ref{Had}): we write the 
dual space of $\mg$ as $\mg^*$, which consists of linear forms $p:\mg\rightarrow\RR:X\mapsto\langle 
p,X\rangle$. When interpreting $G$ as a symmetry group, the elements of the dual of $\mg$ can be 
seen as ``momenta'', or more generally conserved vectors, associated with the symmetries. In particular the 
number $\langle p,X\rangle$ then is the Noether charge associated with the symmetry generator $X$ 
when the system has ``momentum'' $p$.

\paragraph{Definition.} Let $G$ be a Lie group with Lie algebra $\mg$. Then the \it{coadjoint 
representation}\i{coadjoint representation} of $G$ is the homomorphism
\be
\Ad^*:G\rightarrow\text{GL}(\mg^*):
f\mapsto\Ad^*_f
\label{Gad}
\ee
which is dual to the adjoint representation in the sense that
\be
\Ad^*_f(p)\equiv p\circ(\Ad_f)^{-1},
\label{cocogi}
\ee
i.e. $\left<\Ad^*_f(p),X\right>\equiv\left<p,\Ad_{f^{-1}}(X)\right>$ for all $p\in\mg^*$ and any $X\in\mg$. 
From now on we refer to elements of $\mg$ and $\mg^*$ as \it{adjoint} and \it{coadjoint 
vectors}, respectively.\i{adjoint vector}\i{coadjoint vector}\\

The coadjoint representation is a linear action of $G$ on $\mg^*$. In particular one can foliate the space 
$\mg^*$ into disjoint $G$-orbits. We call the set\i{coadjoint orbit}
\be
\cW_p\equiv\left\{\Ad^*_g(p)\big|g\in G\right\}
\nn
\ee
the \it{coadjoint orbit} of $p$. It is a homogeneous space for the coadjoint action of $G$. Note that the 
coadjoint representation of any Abelian group is trivial, so its coadjoint orbits are single points. By 
contrast, coadjoint orbits of non-Abelian groups are generally non-trivial (except if $p=0$). We will see 
in section \ref{susecoo} that the coadjoint orbits of semi-direct products contain their momentum orbits.\\

The dual of the infinitesimal adjoint representation (\ref{adg}) is the differential of (\ref{Gad}) at the 
identity, i.e.\ the coadjoint representation of the Lie algebra $\mg$:
\be
\ad^*_X(p)
\equiv
\frac{d}{dt}\left.\left(\Ad^*_{e^{tX}}(p)\right)\right|_{t=0}
\refeq{cocogi}
-p\circ\ad_X
=
-p\circ[X,\cdot]\,.
\label{pixies}
\ee

\paragraph{Remark.} The adjoint and coadjoint representations of a group $G$ are generally \it{inequivalent}. 
In fact they are equivalent if and only if $\mg$ admits a non-degenerate bilinear form (which is the case 
e.g.\ for semi-simple Lie groups).

\subsection{Poisson structures}
\label{susePOSSUM}

The \it{phase space}\i{phase space} of a system is the set of its classical states. In the previous pages we 
have reviewed some basic concepts of group theory, and our goal is to eventually apply them to phase spaces 
with symmetries. Accordingly we now investigate Poisson structures and symplectic structures in more detail.

\paragraph{Definition.} Let $\cM$ be a manifold. A \it{Poisson structure} on $\cM$ is 
an antisymmetric bilinear map\footnote{From now on, real functions on $\cM$ are denoted as $\cF$, $\cG$, 
$\cH$, etc.}\i{Poisson bracket}
\be
\{\cdot,\cdot\}:C^{\infty}(\cM)\times C^{\infty}(\cM)
\rightarrow
C^{\infty}(\cM):
\cF,\cG\mapsto\{\cF,\cG\}
\nn
\ee
which satisfies the Jacobi identity and the Leibniz rule:\i{Jacobi identity}\i{Leibniz rule}
\be
\begin{array}{rl}
\{\cF,\{\cG,\cH\}\}+\{\cG,\{\cH,\cF\}\}+\{\cH,\{\cF,\cG\}\}=0 & \;\;\text{(Jacobi)},\\[.3cm]
\{\cF,\cG\cH\}=\{\cF,\cG\}\cH+\cG\{\cF,\cH\} & \;\;\text{(Leibniz)}.
\end{array}
\nn
\ee
This map is called the \it{Poisson bracket} on 
$\cM$, and the pair $\big(\cM,\{\cdot,\cdot\}\big)$ is a \it{Poisson manifold} or a 
\it{phase space}.\\

The Poisson bracket endows the space of functions $C^{\infty}(\cM)$ with a 
structure of Lie algebra; the Leibniz identity implies in addition that the map
\be
\{\cF,\cdot\}:C^{\infty}(\cM)\rightarrow\CM:
\cG\mapsto\{\cF,\cG\}
\nn
\ee
is a derivation for any function $\cF\in\CM$.\i{derivation}\footnote{A \it{derivation} of an 
algebra $\cA$ is 
a linear map $D:\cA\rightarrow\cA:a\mapsto D(a)$ that satisfies the Leibniz rule $D(a\cdot 
b)=D(a)\cdot b+a\cdot D(b)$.} These properties 
together 
endow the space $\CM$ with the structure of a \it{Poisson algebra}.\i{Poisson algebra} Note that the 
existence of a Poisson structure sets no restrictions on the dimension of $\cM$. In 
particular, odd-dimensional manifolds admit Poisson structures, e.g.\ $\cM=\RR^3$ 
with the bracket $\{\cF,\cG\}=\der_x\cF\der_y\cG-\der_y\cF\der_x\cG$. This will change once we turn to 
symplectic structures.

\paragraph{Definition.} Let $\big(\cM,\{\cdot,\cdot\}\big)$ be a Poisson manifold; let $\cH\in\CM$. We call 
\it{Hamiltonian vector field}\i{Hamiltonian vector field}\i{vector field!Hamiltonian} associated with $\cH$ 
the 
(unique) vector field $\xi_{\cH}$ on $\cM$ such that
\be
\xi_{\cH}=-\{\cH,\cdot\}.
\label{hamive}
\ee
The existence of $\xi_{\cH}$ is ensured by the one-to-one correspondence between derivations of $\CM$ and 
vector fields on $\cM$.\\

The Hamiltonian vector field associated with a function $\cH$ is a differential operator acting on functions 
on $\cM$. Its integral curves are the paths $\gamma(t)$ in $\cM$ that satisfy 
$\dot\gamma(t)=(\xi_{\cH})_{\gamma(t)}$, which in local coordinates on $\cM$ corresponds to a set of 
$\dim(\cM)$ first-order differential equations $\dot x^i(t)=\xi_{\cH}(x(t))$. These are the 
equations of motion\i{equations of motion!on Poisson manifold} associated with the Hamiltonian 
$\cH$.\i{Hamiltonian} The 
definition (\ref{hamive}) ensures that $\{\cH,\cG\}=-\xi_{\cH}(\cG)$, which implies that the equations of 
motion can be written locally as $\dot x^i=\{x^i,\cH\}$ in terms of the Poisson bracket. In particular one 
has $\{\cH,\cG\}=0$ if and only if $\cG$ is constant along integral curves of $\xi_{\cH}$. Note also that
\be
[\xi_{\cF},\xi_{\cG}]
=
-
\xi_{\{\cF,\cG\}}\,,
\label{Xifidji}
\ee
so the Lie brackets of Hamiltonian vector fields are Hamiltonian.\\

Now consider the set of all Hamiltonian vector fields on $\cM$; at a point $p\in\cM$, they
span a subspace of the tangent space $T_p\cM$. By taking this span for all $p\in\cM$, one obtains a 
subbundle of the tangent bundle $T\cM$ (i.e.\ a distribution on $\cM$). Because brackets of 
Hamiltonian vector fields are Hamiltonian, Frobenius' theorem\i{Frobenius' theorem}\i{foliation} implies 
that Hamiltonian vector fields 
yield a foliation of $\cM$ into so-called \it{symplectic leaves}.\i{symplectic leaf} Two points
belong to the same 
leaf if they can be joined by the integral curve of a Hamiltonian vector field. In the example 
of $\RR^3$ mentioned above, symplectic leaves are planes $z=\text{const}$. This leads to the definition of 
symplectic manifolds.

\subsection{Symplectic structures}

\paragraph{Definition.} Let $\cM$ be a manifold. A \it{symplectic form}\i{symplectic form} on $\cM$ is a 
closed, non-degenerate two-form $\omega$ on $\cM$.\footnote{Closedness means $d\omega=0$, where $d$ is the 
exterior derivative. Non-degeneracy means 
that for all $p\in\cM$, any vector $v\in T_p\cM$ such that $\omega_p(v,w)=0$ for all $w\in T_p\cM$ 
necessarily vanishes.} The pair $(\cM,\omega)$ is a \it{symplectic manifold}.\i{symplectic manifold}\\

Non-degeneracy means that, in local coordinates, the 
components $\omega_{ij}$ of $\omega$ form an invertible antisymmetric matrix. This implies that all 
symplectic manifolds are even-dimensional. Note that any 
symplectic manifold admits a \it{Liouville volume form}\i{Liouville volume form}
\be
\mu
\equiv
\underbrace{\omega\wedge...\wedge\omega}_{\text{dim}(\cM)/2\text{ times}}.
\label{volom}
\ee
\vspace{.1cm}

The symplectic leaves described above are prime examples of symplectic manifolds: they are 
endowed with a symplectic form $\omega$ such that $\omega(\xi_{\cF},\xi_{\cG})\equiv\{\cF,\cG\}$; this 
condition determines $\omega$ unambiguously because symplectic leaves are, by definition, spanned by the 
integral curves of Hamiltonian vector fields. Another common example is the phase space $\cM=\RR^{2n}$ of a 
non-relativistic particle 
in $\RR^n$, with coordinates $(q^1,...,q^n,p_1,...,p_n)$ and symplectic form
\be
\omega=dq^i\wedge dp_i
\qquad
\text{(implicit sum over $i=1,...,n$)}.
\label{omeq}
\ee

\subsubsection*{Canonical symplectic form}

The symplectic structure (\ref{omeq}) has an important generalization: consider a manifold $\cQ$ describing 
the configuration 
space of a classical system (so $\dim\cQ$ is the number of Lagrange variables). 
The corresponding phase space is the \it{cotangent bundle} $T^*\cQ$,\i{cotangent bundle} which consists of 
pairs $(q,\alpha)$ 
where 
$q\in\cQ$ and $\alpha\in T_q^*\cQ$. These pairs are generally interpreted as describing a ``position'' $q$ 
and a ``momentum'' $\alpha$, but we will see below that the interpretation stemming from semi-direct 
products is different: $q$ will in fact be a momentum (with $\cQ$ a momentum orbit), while $\alpha$ will 
essentially be a position (or rather a translation vector). The symplectic form on $T^*\cQ$ is defined as 
follows. We let
\be
\pi:T^*\cQ\rightarrow \cQ:
(q,\alpha)\mapsto q
\label{ProCC}
\ee
be the natural projection and define the \it{Liouville one-form} $\theta$ on $T^*\cQ$ by\i{Liouville 
one-form}
\be
\left<
\theta_{(q,\alpha)},\cV
\right>
\equiv
\left<
\alpha
,
\pi_{*(q,\alpha)}\cV
\right>
\label{lifo}
\ee
for any vector $\cV\in T_{(q,\alpha)}T^*\cQ$. Then $\omega\equiv-d\theta$ is the \it{canonical 
symplectic form}\i{canonical symplectic form} on $T^*\cQ$. In the example (\ref{omeq}), 
$\cQ=\RR^n$.\\

Let us verify that $\omega=-d\theta$ is indeed symplectic. We choose local coordinates $(q^1,...,q^n)$ on 
some 
open set $U\subset \cQ$ and denote by $(q^i,p_j)$ the 
corresponding local coordinates on $\pi^{-1}(U)$, so that the form 
$\alpha\in T_q^*\cQ$ reads $\alpha=p_j(dq^j)_q$. Given a 
vector $\cV\in 
T_{(q,\alpha)}T^*\cQ$, one can write
\be
\cV=a^i\frac{\der}{\der q^i}+b_j\frac{\der}{\der p_j}
\qquad
\Rightarrow\qquad
\pi_{*(q,\alpha)}\cV
=
a^i\frac{\der}{\der q^i}\,.
\nn
\ee
Thus the differential of the projection (\ref{ProCC}) projects $\cV$ on its part tangent to $\cQ$.
The definition (\ref{lifo}) then implies that $\theta=p_idq^i$, so
\be
\omega=-d\theta=dq^i\wedge dp_i
\label{dabu}
\ee
is definitely a closed, non-degenerate two-form. It coincides locally with (\ref{omeq}).

\paragraph{Remark.} The \it{Darboux theorem}\i{Darboux theorem} states that any point of a symplectic 
manifold has a neighbourhood with local coordinates $(q^i,p_j)$ such that the symplectic form 
reads (\ref{dabu}). Thus any symplectic 
manifold is locally equivalent to a cotangent bundle.

\subsubsection*{Hamiltonian vector fields revisited}

Any symplectic manifold can be endowed with a Poisson structure by mimicking the symplectic leaves described 
at the end of section \ref{susePOSSUM}. This relies on a new definition of Hamiltonian vector fields:

\paragraph{Definition.} Let $(\cM,\omega)$ be a symplectic manifold, $\cF\in\CM$. The \it{Hamiltonian vector 
field} $\xi_{\cF}$ associated with $\cF$ is defined by\i{Hamiltonian vector field}\i{vector field!Hamiltonian}
\be
i_{\xi_{\cF}}\omega=\omega(\xi_{\cF},\cdot)\stackrel{!}{=}d\cF.
\label{ixi}
\ee
Conversely, a vector field $\zeta$ is \it{Hamiltonian} if there exists a function $\cF$ such that 
$\zeta=\xi_{\cF}$.\\

Hamiltonian vector fields can be used to define Poisson brackets in the same way as on symplectic leaves of 
Poisson manifolds: for any two functions $\cF$, $\cG$ we write\i{Poisson bracket}
\be
\{\cF,\cG\}\equiv\omega(\xi_\cF,\xi_\cG).
\label{pois}
\ee
In terms of this bracket the definition (\ref{ixi}) is equivalent to our earlier definition of Hamiltonian 
vector fields in (\ref{hamive}).
In local coordinates the definition (\ref{ixi}) reads
\be
\omega_{ij}\xi_{\cF}^i=\der_j\cF
\qquad\Leftrightarrow\qquad
\xi_{\cF}^i=\der_j\cF\omega^{ji}
\label{oij}
\ee
where $\omega_{ij}$ are the components of $\omega$ and $\omega^{ij}$ is the matrix inverse of $\omega_{ij}$. 
Accordingly, the bracket (\ref{pois}) can be written as $\{\cF,\cG\}=-\omega^{ij}\der_i\cF\der_j\cG$.\\

Note that symplectic manifolds only contain kinematical data: they tell us the available combinations of 
``positions'' and ``momenta'' --- those combinations 
are classical states. Classical observables\i{classical observable}\i{observable} then are 
real-valued functions on phase space. Once we declare 
that a certain observable $\cH$ is the Hamiltonian, time evolution is given locally by the equations of 
motion $\dot x=\{x,\cH\}$.

\subsubsection*{Symplectomorphisms}

\paragraph{Definition.} Let $(\cM,\omega)$ and $(\cN,\Omega)$ be symplectic manifolds. A 
\it{symplectomorphism}\i{symplectomorphism}\i{canonical transformation} (or \it{canonical transformation}) 
from $\cM$ to $\cN$ is a 
diffeomorphism $\phi:\cM\rightarrow\cN$ that preserves the symplectic structure in the sense 
that\i{pullback}\footnote{Recall that the \it{pullback} of a tensor field $T$ of rank $k$ on a manifold 
$\cN$ 
by a map $\phi:\cM\rightarrow\cN$ is defined by 
$(\phi^*T)_p(v_1,...,v_k)\equiv T_{\phi(p)}(\phi_{*p}v_1,...,\phi_{*p}v_k)$
for any $p\in\cM$ and all $v_1,...,v_k\in T_p\cM$.}
\be
\phi^*\Omega=\omega.
\label{symo}
\ee
Then $(\cM,\omega)$ and $(\cN,\Omega)$ are said to be 
\it{symplectomorphic}.\\

When $\phi:\cM\rightarrow\cN$ is a symplectomorphism, it preserves Poisson brackets in the sense that 
$\{\phi^*\cF,\phi^*\cG\}=\phi^*\{\cF,\cG\}$ for all functions $\cF,\cG$ on $\cN$, where the brackets on the 
left and on the right-hand side are those of $\cM$ and $\cN$, respectively. Note that the flow\i{flow} of any 
Hamiltonian vector field on $\cM$ defines a one-parameter family of symplectomorphisms of $\cM$.

\subsection{Kirillov-Kostant structures}

We now describe phase spaces whose structure is entirely determined by group theory. They are prototypes 
for all phase spaces with symmetries.

\subsubsection*{Kirillov-Kostant Poisson bracket}

\paragraph{Definition.} Let $G$ be a Lie group with Lie algebra $\mg$. Then the \it{Kirillov-Kostant Poisson 
bracket}\i{Kirillov-Kostant bracket} on $\mg^*$ is defined as
\be
\left\{\cF,\cG\right\}(p)
\equiv
\langle p,[\cF_{*p},\cG_{*p}]\rangle
\label{kkba}
\ee
where $\cF,\cG\in C^{\infty}(\mg^*,\RR)$ and $\cF_{*p}$ denotes the differential of $\cF$ at 
$p\in\mg^*$.\footnote{$\cF_{*p}$ is a linear map from $T_p\mg^*\cong\mg^*$ 
to 
$T_{\cF(p)}\RR\cong\RR$ and therefore belongs to $(\mg^*)^*\cong\mg$.}\\

One can associate a Hamiltonian vector field (\ref{hamive}) with any function $\cF$ on a 
Poisson manifold. In the present case one has the following result:

\paragraph{Proposition.} Let $\cF$ be a real function on 
$\mg^*$, $\xi_{\cF}$ the corresponding Hamiltonian vector field. The associated evolution 
equation is the \it{Euler-Poisson equation} on $\mg^*$,\i{Euler-Poisson equation}
\be
\dot\gamma(t)
=
(\xi_{\cF})_{\gamma(t)}
=
\ad^*_{\cF_{*\gamma(t)}}\big(\gamma(t)\big)\,.
\label{euler}
\ee

\begin{proof}
Let $\cG$ be a function on $\mg^*$ and take $p\in\mg^*$. We are going to compute $\xi_{\cF}(\cG)$ at $p$ in 
two 
different ways. First, $(\xi_{\cF})_p$ is a vector tangent to 
$\mg^*$ at $p$ and may therefore be seen as an element of $\mg^*$ (since $\mg^*$ is a vector space). 
But $(\xi_{\cF})_p(\cG)$ only depends on the differential of $\cG$ at $p$, so we may write
\be
(\xi_{\cF})_p(\cG)
=
\langle(\xi_{\cF})_p,\cG_{*p}\rangle.
\label{xihef}
\ee
Secondly, using (\ref{hamive}) and the bracket (\ref{kkba}), one has
\be
(\xi_{\cF})_p(\cG)
=
-\{\cF,\cG\}
=
-
\left<
p,[\cF_{*p},\cG_{*p}]
\right>
=
-
\left<
p,\ad_{\cF_{*p}}(\cG_{*p})
\right>
\refeq{pixies}
\big<\ad^*_{\cF_{*p}}(p),\cG_{*p}\big>\,.
\nn
\ee
Comparing this with (\ref{xihef}), eq.\ (\ref{euler}) follows.
\end{proof}

\paragraph{Corollary.} The symplectic leaves of the Kirillov-Kostant bracket are the coadjoint orbits of $G$. 
In particular, all (finite-dimensional) coadjoint orbits have even dimension.\i{coadjoint orbit!as 
symplectic leaf}

\begin{proof}
By the above proposition, the Hamiltonian vector field $\xi_{\cF}$ associated with a function $\cF$ and 
evaluated 
at a 
point $p\in\mg^*$ is
\be
(\xi_{\cF})_p
=
\ad^*_{\cF_{*p}}(p).
\label{xihad}
\ee
Now, given an adjoint vector $X\in\mg$, we can always find a real function $\cF$ on $\mg^*$
such that $\cF_{*p}=X$. Accordingly eq.~(\ref{xihad}) implies that the integral curves of all possible 
Hamiltonian vector fields going through $p$ span the coadjoint orbit of $p$.
\end{proof}

For future reference it is useful to rewrite the Kirillov-Kostant bracket in terms of a basis $\{t_a\}$ of 
$\mg$ with Lie brackets (\ref{commurel}).
Any adjoint vector can then be written as $X=X^at_a$. If $\{(t^a)^*|a=1,...,n\}$ denotes the 
corresponding dual basis of $\mg^*$, so that $\bra(t^a)^*,t_b\ket=\delta^a_b$, any coadjoint vector can be 
written as $p=p_a(t^a)^*$ with 
real components $p_a$. This defines global coordinates $\{p_a|a=1,...n\}$ on $\mg^*$, where each 
$p_a$ is a real function on $\mg^*$ that associates with a coadjoint vector $p$ its 
component $p_a$. To apply (\ref{kkba}) we note that the differential $(p_a)_*$ of $p_a$ acts 
on the basis vectors $\frac{\der}{\der p_c}$ according to
\be
(p_a)_*\Big(\frac{\der}{\der p_c}\Big)
=
\frac{\der p_a}{\der p_c}
=
\delta_a^c\,.
\label{papib}
\ee
But since 
$\mg^*$ is a vector space we can identify $T_p\mg^*$ with $\mg^*$ by declaring that $\der/\der p_c$ 
coincides with $(t^c)^*$, so in fact the differential satisfies 
$(p_a)_*((t^c)^*)=\delta^c_a$. With this identification the differential $(p_a)_*$ belongs to the dual of the 
dual, $(\mg^*)^*=\mg$, and may be seen as an adjoint vector. Property (\ref{papib}) says that this 
adjoint vector is precisely $t_a$. The Poisson bracket follows:
\be
\{p_a,p_b\}
=
f_{ab}{}^c\,p_c\,.
\label{fabipe}
\ee
In parts II and III we will see that the Poisson brackets of three-dimensional gravity coincide with 
Kirillov-Kostant brackets for suitable asymptotic symmetry groups.

\paragraph{Remark.} The Euler-Poisson equation (\ref{euler}) has numerous applications in physics, 
particularly when there exists an invertible and self-adjoint ``inertia operator'' 
$\cI:\mg\rightarrow\mg^*$.\i{inertia operator}\footnote{Here self-adjointness means that 
$\langle\cI(X),Y\rangle=\langle\cI(Y),X\rangle$ for any two adjoint vectors $X,Y$.} Indeed one can then 
consider a quadratic Hamiltonian 
function $\cF(p)=\demi\langle p,\cI^{-1}p\rangle$ and eq.\ (\ref{euler}) becomes
$\dot\gamma(t)
=
\ad^*_{\cI^{-1}\gamma(t)}\gamma(t)$.
For $G=\SO(3)$ this coincides with the equations of motion of a free rigid body; for the Virasoro group, it 
leads to the Korteweg-de Vries equation\i{Korteweg-de Vries equation} (see 
\cite{khesin2008geometry} for details).

\subsubsection*{Kirillov-Kostant symplectic form}

Since the coadjoint orbits of $G$ are symplectic leaves of the Kirillov-Kostant Poisson bracket, they have a 
symplectic structure given by (\ref{pois}):

\paragraph{Definition.} Let $G$ be a Lie group, $p\in\mg^*$ a coadjoint vector with orbit $\cW_p$. Then the 
\it{Kirillov-Kostant(-Souriau) symplectic form} at $q\in\cW_p$ is given by\i{Kirillov-Kostant symplectic form}
\be
\omega_q(\ad^*_Xq,\ad^*_Yq)
=
\langle q,[X,Y]\rangle
\label{kksym}
\ee
where $X,Y\in\mg$.\\

Here $\ad^*_Xq$ and $\ad^*_Yq$ are ``infinitesimal displacements'' of $q$ and represent 
generic tangent vectors of $\cW_p$ at $q$. One can verify that (\ref{kksym}) is 
closed and non-degenerate on $\cW_p$, so it is indeed a symplectic form. In addition it is invariant under 
the 
coadjoint action of $G$ in the sense that $(\Ad^*_f)^*(\omega)=\omega$
for all $f\in G$. Thus each coadjoint orbit of $G$ is a homogeneous space 
equipped with a $G$-invariant symplectic structure. In this sense it is a symmetric phase space. We will see 
below that, for instance, each coadjoint orbit of the Poincar\'e group coincides with the space of classical 
states of a relativistic particle with definite mass and (classical) spin.

\subsection{Momentum maps}
\label{susemom}

Noether's theorem states that any classical system with a Lie group of symmetries possesses conserved 
quantities. Here we investigate this statement in the framework of symplectic geometry. Until the end of this 
section $\cM$ is understood to be a manifold acted upon by some group $G$ according to $q\mapsto f\cdot q$.

\subsubsection*{Group actions and infinitesimal generators}

\paragraph{Definition.} Let $G\times\cM\rightarrow\cM:(f,q)\mapsto f\cdot q$ be a smooth action of a Lie 
group $G$ on a manifold $\cM$. Then the \it{infinitesimal generator}\i{infinitesimal generator} of the action 
corresponding to $X\in\mg$ is the 
vector field $\xi_X$ on $\cM$ defined by
\be
(\xi_X)_q
\equiv
\frac{d}{dt}\left.\left(e^{tX}\cdot q\right)\right|_{t=0}\,.
\label{infige}
\ee
One can show (see e.g.\ \cite{abraham1978foundations}) that this definition implies
\be
[\xi_X,\xi_Y]
=
-\xi_{[X,Y]}
\label{OPOBRA}
\ee
where the Lie bracket on the left-hand side is that of vector fields, while the bracket on the 
right is that of $\mg$, given by
(\ref{A196}).\\

For example, the representations 
(\ref{adg}) and (\ref{pixies}) are infinitesimal generators of the adjoint 
and coadjoint representations of $G$, respectively.\footnote{Property (\ref{OPOBRA}) does not contradict the 
fact that the adjoint and coadjoint representations of $\mg$ are actual representations, i.e.\ for example 
that $\ad_X\ad_Y-\ad_Y\ad_X=\ad_{[X,Y]}$. Indeed, the vector fields in (\ref{OPOBRA}) are derivations acting 
on functions on $\cM$, while $\ad_X$ and $\ad^*_X$ are generally understood as linear operators acting on 
$\mg$ and $\mg^*$, respectively.}
In this language the tangent space at $q$ of an orbit 
(\ref{orb}) consists of all vectors of the form $(\xi_X)_q$, where $X$ spans the Lie algebra $\mg$. The flow 
of $\xi_X$ is $\RR\times\cM\rightarrow\cM:(t,q)\mapsto e^{tX}\cdot q$. In what follows we study group 
actions where the manifold $\cM$ is symplectic.

\subsubsection*{Momentum maps}

Let $(\cM,\omega)$ be a symplectic manifold. An action of $G$ on $\cM$ is \it{symplectic}\i{symplectic group 
action}\i{group action!symplectic} if each map $q\mapsto f\cdot q$ is a symplectomorphism, in which case $G$ 
is a symmetry group of 
$\cM$. If $\xi_X$ denotes the infinitesimal generator (\ref{infige}) of a symplectic action, then 
$\cL_{\xi_X}\omega=0$.

\paragraph{Definition.} Let $G\times\cM\rightarrow\cM:(f,q)\mapsto f\cdot q$ be a symplectic group action. A 
\it{momentum map}\i{momentum!map} for this action is a smooth 
map
\be
\cJ:\cM\rightarrow\mg^*:p\mapsto\cJ(p)
\label{CalJ}
\ee
such that, for any $X\in\mg$,
\be
i_{\xi_X}\omega
=
d\left<\cJ(\cdot),X\right>
\label{moment}
\ee
where $\xi_X$ is the infinitesimal generator (\ref{infige}) and 
$\left<\cJ(\cdot),X\right>$ is the real function on $\cM$ that associates with $q\in\cM$ the value 
$\left<\cJ(q),X\right>$. In the sequel we write $\langle\cJ(\cdot),X\rangle\equiv\cJ_X$, to which we also 
refer as a ``momentum map''.\\

The definition (\ref{moment}) can be 
compared to that of Hamiltonian vector fields, eq.\ (\ref{ixi}), and is equivalent to the statement
\be
\xi_X
=
\xi_{\cJ_X}
=
-\{\cJ_X,\cdot\}.
\label{xixi}
\ee
Here $\xi_X$ is the infinitesimal generator (\ref{infige}), while $\xi_{\cJ_X}$ is the Hamiltonian 
vector field associated with the function $\cJ_X$ and $\{\cdot,\cdot\}$ is the Poisson bracket (\ref{pois}). 
Thus the function $\cJ_X$ generates the transformation corresponding to $X\in\mg$ in phase space, in the 
sense that for any function $\cF\in\CM$ we have $\{\cJ_X,\cF\}=-\xi_X(\cF)=-\delta_X\cF$. From this 
observation we can derive an important corollary: if $X,Y\in\mg$ and if we consider the corresponding 
functions $\cJ_X$ and $\cJ_Y$, their Poisson bracket acts on classical observables according to
\begin{eqnarray}
\big\{\{\cJ_X,\cJ_Y\},\cF\big\}
& \!\!= &
\!\!\big\{\cJ_X,\{\cJ_Y,\cF\}\big\}-\big\{\cJ_Y,\{\cJ_X,\cF\}\big\}\nn\\
\label{Karate}
& \!\!\refeq{xixi} &
\!\![\xi_X,\xi_Y](\cF)
\refeq{OPOBRA}
-\xi_{[X,Y]}(\cF)
\refeq{xixi}
\big\{\cJ_{[X,Y]},\cF\big\}
\end{eqnarray}
where in the first equality we have used the Jacobi identity. Since this property is true for any function 
$\cF$, it is tempting to remove it from both ends of the equation and conclude that the momentum map 
provides a representation of the Lie algebra $\mg$. However, this hasty argument overlooks one crucial 
possibility, namely the fact that brackets of momentum maps may include a \it{central term} that commutes 
with all functions on phase space (see \cite{abraham1978foundations} or appendix 5 of \cite{arnold}). 
Thus we conclude that:

\paragraph{Proposition.} Provided phase space is connected, the Poisson 
algebra of momentum maps is a representation of the Lie algebra $\mg$ up to a (classical) central 
extension:\i{classical central extension}\i{central extension!classical}
\be
\{\cJ_X,\cJ_Y\}
=
\cJ_{[X,Y]}+\sfc(X,Y)
\qquad
\forall\,X,Y\in\mg\,,
\label{jixiy}
\ee
for some real two-cocycle $\sfc$ on $\mg$. If the phase space has several connected 
components, there may be several different cocycles (one for each connected component).\\

This statement is equivalent to saying that momentum maps provide a projective representation of $\mg$, or 
alternatively an \it{exact} representation of a central extension of $\mg$. It is the classical analogue of 
the symmetry representation theorem of section \ref{s1.1}; it will be crucial once we use geometric 
quantization to produce unitary representations. In parts II and III we will see that the surface charges 
generating asymptotic symmetries in 
gravity provide a shining illustration of this phenomenon.

\paragraph{Remark.} Not all symplectic group actions have a momentum map, as there may be no function $\cJ_X$ 
such that (\ref{xixi}) holds. If such a function exists for each 
$X\in\mg$, then the action does admit a momentum map and is said to be \it{Hamiltonian}.\i{Hamiltonian group 
action}\i{group action!Hamiltonian} Note that, if $\cJ$ and $\cJ'$ are momentum maps for the 
same group
action, then (\ref{moment}) implies that they differ by a constant coadjoint vector (provided $\cM$ is 
connected).

\subsubsection*{Noether's theorem}

The momentum map gives the conserved quantity $\cJ(p)\in\mg^*$ associated with each classical state 
$p\in\cM$. As anticipated earlier, coadjoint vectors may thus be seen as ``conserved vectors'' for 
symmetric phase spaces. This interpretation stems from the following fundamental 
result:

\paragraph{Noether's theorem.} Let $q\mapsto f\cdot q$ be a Hamiltonian action of $G$ on $(\cM,\omega)$ with 
momentum map $\cJ$. Let also $\cH\in\CM$ be a classical observable invariant 
under $G$ in the sense that $\cH(f\cdot q)=\cH(q)$ for all $f\in G$ and all $q\in\cM$, and let $\xi_{\cH}$ be 
the associated Hamiltonian vector field (\ref{hamive}).\i{Noether's theorem} Then, for any integral curve 
$\gamma(t)$ of 
$\xi_{\cH}$ with initial condition $\gamma(0)$, one has
\be
\cJ\big(\gamma(t)\big)=\cJ\big(\gamma(0)\big)
\label{jft}
\ee
for any time $t$ belonging to the domain of the curve. In other words the $\dim\mg$ 
components of the coadjoint vector $\cJ(\gamma(t))$ are \it{conserved} when the equations of motion 
$\dot\gamma=(\xi_{\cH})_{\gamma}$ are satisfied.

\begin{proof}
The Hamiltonian $\cH$ is invariant under $G$, so $\frac{d}{dt}\cH(e^{tX}\cdot p)=0$ for any 
$p\in\cM$. Since any integral curve of $\xi_X$ takes the form $e^{tX}\cdot p$ for some initial condition $p$, 
this is to say that $\xi_X(\cH)=\xi_{\cJ_X}(\cH)=0$, so $\cH$ is constant along integral curves of 
$\cJ_X$; equivalently, $\cJ_X$ is constant along integral curves of $\xi_{\cH}$.
\end{proof}

In a translation-invariant system the momentum map associates a momentum vector with any point in phase 
space. Similarly, in a rotation-invariant system it coincides 
with angular momentum. Finally, in a two-dimensional conformal field theory, it coincides with the stress 
tensor of a given field configuration. We will illustrate these statements below. In the remainder of this 
section we build momentum maps for specific families of symplectic 
manifolds.

\subsubsection*{Momentum maps for coadjoint orbits}

Let us build a momentum map (\ref{moment}) for a coadjoint orbit $\cW_p$ of some group $G$, endowed with the 
Kirillov-Kostant symplectic form (\ref{kksym}). First we note that any path $\gamma(t)$ in $\cW_p$ can be 
written as $\gamma(t)=\Ad^*_{f(t)}p$ for some path $f(t)$ in $G$. If $\gamma(0)=q$ and 
$\dot\gamma(0)=\ad^*_Yq$ for some $Y\in\mg$, then we find
\be
\omega_q\big(\ad^*_Xq,\dot\gamma(0)\big)
=
\omega_q\big(\ad^*_Xq,\ad^*_Yq\big)
\refeq{kksym}
\bra q,[X,Y]\ket
=
\bra\ad^*_Yq,X\ket
\nn
\ee
for any $X\in\mg$. Since $\ad^*_Xq$ is the infinitesimal generator $\xi_X$ of the coadjoint action of $G$ on 
$\cW_p$, the far left-hand side of this equation coincides with $(i_{\xi_X}\omega)_q\big(\dot\gamma(0)\big)$. 
As a 
consequence the momentum map (\ref{moment}) should be such that
\be
\bra\ad^*_Yq,X\ket
=
\frac{d}{dt}\big(\bra\cJ(\gamma(t)),X\ket\big)\big|_{t=0}
=
\bra\cJ_{*q}\,\ad^*_Yq,X\ket
\nn
\ee
for all $X\in\mg$. This implies that the differential $\cJ_{*q}:T_q\cW_p\rightarrow\mg^*$ is just the 
inclusion, and determines $\cJ$ up to a constant coadjoint vector. In particular:

\paragraph{Proposition.} The inclusion of the coadjoint orbit $\cW_p$ in $\mg^*$,\i{coadjoint 
orbit!momentum map}
\be
\cJ:\cW_p\xhookrightarrow{}\mg^*:q\mapsto q,
\label{INK}
\ee
is a momentum map for the coadjoint action of $G$ on $(\cW_p,\omega)$ when $\omega$ is the Kirillov-Kostant 
symplectic form (\ref{kksym}).\\

This result implies that the action of a Lie group on its coadjoint orbits is \it{always} Hamiltonian. 
In fact one can show that any symplectic manifold endowed with a transitive Hamiltonian action of some group 
$G$ is a covering space of a coadjoint orbit of $G$. In this sense coadjoint orbits are 
``universal'' homogeneous phase spaces.\\

Note that the momentum map (\ref{INK}) automatically realizes $\mg$ 
symmetry without central extensions. Indeed, if $X,Y$ belong to $\mg$ and if $\cJ_X,\cJ_Y$ are the 
corresponding momentum maps, then at a point $p\in\mg^*$ we find
\be
\{\cJ_X,\cJ_Y\}(p)
\refeq{kkba}
\bra p,[(\cJ_X)_{*p},(\cJ_Y)_{*p}]\ket
\refeq{INK}
\bra p,[X,Y]\ket
=
\cJ_{[X,Y]}(p).
\nn
\ee
This is exactly the statement (\ref{jixiy}) with a vanishing cocycle $\sfc$. However,
one should keep in mind that 
the group $G$ itself may be centrally extended.

\subsubsection*{Momentum maps for cotangent bundles}

\paragraph{Proposition.} Consider a symplectic action of $G$ on $\cM$. Let $\omega=-d\theta$ for some 
symplectic potential $\theta$. If the group action leaves $\theta$ 
invariant,
then\i{cotangent bundle!momentum map}
\be
\cJ_X\equiv\langle\theta,\xi_X\rangle
\label{prej}
\ee
defines a momentum map (\ref{CalJ}) that satisfies (\ref{jixiy}) with a vanishing cocycle $\sfc=0$.

\begin{proof}
Since the action leaves $\theta$ invariant, one has $\cL_{\xi_X}\theta=0$ for any $X\in\mg$. Writing the Lie 
derivative as $\cL_{\xi}=d\circ i_{\xi}+i_{\xi}\circ d$ and using $\omega=-d\theta$, this is equivalent to 
$d\langle\theta,\xi_X\rangle=-i_{\xi_X}d\theta=i_{\xi_X}\omega$. One may recognize this as the definition 
(\ref{moment}) of a momentum map given by (\ref{prej}). In order to prove that (\ref{jixiy}) is satisfied 
with a vanishing cocycle $\sfc$, we evaluate the Poisson bracket
\be
\{\cJ_X,\cJ_Y\}
=
\demi\left[
\xi_Y\langle\theta,\xi_X\rangle-\xi_X\langle\theta,\xi_Y\rangle
\right]
=
\demi\omega(\xi_X,\xi_Y)-\demi\langle\theta,[\xi_X,\xi_Y]\rangle.
\nn
\ee
Here $\omega(\xi_X,\xi_Y)=\{\cJ_X,\cJ_Y\}$ by virtue of (\ref{pois}) and (\ref{xixi}), while 
eqs.\ (\ref{OPOBRA}) and (\ref{prej}) imply that 
$\langle\theta,[\xi_X,\xi_Y]\rangle=-\cJ_{[X,Y]}$. Eq.\ (\ref{jixiy}) follows with $\sfc=0$.
\end{proof}

Let us now apply this result to the cotangent bundle $T^*\cQ$ of a manifold $\cQ$. Let $\phi:\cQ\rightarrow 
\cQ$ be a 
diffeomorphism. We define the associated \it{point transformation}\i{point transformation} as
\be
\bar\phi:
T^*\cQ\rightarrow T^*\cQ:
(q,\alpha)\mapsto\left(\phi^{-1}(q),\alpha\circ\phi_{*\phi^{-1}(q)}\right).
\label{poita}
\ee
As one can verify, this definition ensures that
\be
\phi\circ\pi\circ\bar\phi=\pi
\label{fipi}
\ee
where $\pi:T^*\cQ\rightarrow \cQ$ is the canonical projection (\ref{ProCC}). Thanks to this property, one can 
show 
(see e.g.\ \cite{abraham1978foundations})
that 
point transformations are symplectomorphisms:

\paragraph{Proposition.} Consider $T^*\cQ$ with the symplectic form $\omega=-d\theta$, 
where 
$\theta$ is the canonical one-form (\ref{lifo}). Let $\phi:\cQ\rightarrow \cQ$ be a diffeomorphism and let 
$\bar\phi$ 
be the associated point transformation (\ref{poita}). Then
\be
(\bar\phi)^*\theta=\theta\,.
\label{calim}
\ee
In particular, point transformations are symmetries of $T^*\cQ$.\\

Suppose now that there is an action $q\mapsto f\cdot q$ of a Lie group $G$ on the manifold $\cQ$.
For clarity we will also write $f\cdot q\equiv\sigma^*_f(q)$, where the notation is purposely the same as in 
eq.\ (\ref{simplinot}).
Then one can define an action of $G$ on $T^*\cQ$ by
\be
f\cdot (q,\alpha)
\equiv
\big(
f\cdot q,\alpha\circ(\sigma^*_{f^{-1}})_{*f\cdot q}
\big).
\label{baphi}
\ee
Proposition (\ref{calim}) ensures that this action is symplectic and even preserves the Liouville one-form. 
Accordingly we can apply (\ref{prej}) to build its momentum map:

\paragraph{Proposition.} A momentum map for (\ref{baphi}) is provided by the prescription
\be
\langle\cJ(q,\alpha),X\rangle
\equiv
\langle\alpha,(\xi_X)_q\rangle
\label{jake}
\ee
where $\xi_X$ is the infinitesimal generator of the action $q\mapsto f\cdot q=\sigma^*_f(q)$.

\begin{proof}
Applying (\ref{prej}) to the case at hand, we find a momentum map $\cJ$ given by 
$\langle\cJ(q,\alpha),X\rangle
=
\langle
\theta_{(q,\alpha)},(\bar\xi_X)_{(q,\alpha)}
\rangle$
where $\bar\xi_X$ denotes the infinitesimal generator of (\ref{baphi}). Then
(\ref{fipi}) implies that
$\pi_{*(q,\alpha)}(\bar\xi_X)_{(q,\alpha)}
=
(\xi_X)_q$, so (\ref{jake}) follows.
\end{proof}

As an application of these results one can show for instance that the momentum map given by (\ref{jake}) for 
a translation-invariant system is just the standard momentum vector, while for a rotation-invariant system it 
yields the angular momentum. See \cite{abraham1978foundations}.

\section{Geometric quantization}
\label{geoq}

Given a symmetric phase space $(\cM,\omega)$, one would like to understand how ``quantizing'' that system 
produces unitary representation of the symmetry group. This section is devoted to an overview of that 
problem, particularly as applied to coadjoint orbits. In short, the quantum Hilbert space associated with 
$(\cM,\omega)$ will consist of ``wavefunctions'', or rather sections of suitable line bundle over $\cM$, and 
will indeed provide unitary representations provided certain quantization conditions are satisfied. 
Our plan is to start by reviewing the technology of line bundles and their connections, before explaining 
how it applies to the quantization of symplectic manifolds and discussing the realization of unitary 
group representations by geometric quantization. The presentation is condensed and superficial; 
we refer to \cite{woodhouse1997,kirillov2004lectures,debuyl} for a much more detailed account of the subject.

\subsection{Line bundles and wavefunctions}
\label{suseLibou}

The basic idea of geometric quantization is to consider wavefunctions on a symplectic manifold $\cM$. These 
wavefunctions are sections of a 
complex 
line bundle over $\cM$. Recall that a 
\it{fibre bundle}\i{fibre bundle} is a quadruple $(\cL,\cM,\cF,\pi)$ where $\pi:\cL\rightarrow\cM$ is a 
projection and $\cL$ is locally diffeomorphic to the product $\cM\times\cF$, where $\cM$ is 
known as the 
\it{base space} and $\cF$ is known as the \it{fibre}. A \it{vector bundle}\i{vector bundle} is a fibre bundle 
whose fibres 
are diffeomorphic to a vector space.

\paragraph{Definition.} A \it{complex line bundle}\i{line bundle} $\cL$ over $\cM$ is a vector bundle 
$\pi:\cL\rightarrow\cM$ whose fibres are isomorphic to $\CC$.\\

Thus a complex line bundle consists of infinitely many copies of the complex plane $\CC$, one at each point 
of $\cM$, glued together in a smooth way (see fig.\ \ref{figoFIBB} with $\cO_p$ replaced by $\cM$). The 
bundle 
locally looks 
like the direct product of $\cM$ with $\CC$. When this is true globally, i.e.\ when $\cL$ is diffeomorphic to 
$\cM\times\CC$, the line bundle is said to be \it{trivial}.\i{trivial bundle}

\paragraph{Definition.} Let $\pi:\cL\rightarrow\cM$ be a complex line bundle. Then a 
\it{section}\i{section} of $\cL$ is a map $\Psi:\cM\rightarrow\cL$ such that $\pi\circ\Psi=\text{Id}_{\cM}$. 
The space of sections is denoted $\Gamma(\cM,\cL)$.\\

When $\pi:\cL\rightarrow\cM$ is trivial, the space of sections coincides with the space of complex-valued 
functions on $\cM$. For example, when $\cM=\RR^2$ is interpreted as the phase space of a particle on a line, 
complex functions $\Psi(x,p)$ on $\RR^2$ would provide sections of the trivial line bundle $\RR^2\times\CC$.

\paragraph{Remark.} If $\cM$ is a symplectic manifold and $\cL$ is a line bundle over $\cM$, not all sections 
of $\cL$ are eligible as quantum wavefunctions because they depend on too many arguments. In the example 
$\cM=\RR^2$ just given, sections $\Psi(x,p)$ depend both on positions and on momenta, which violates the 
uncertainty principle.\i{uncertainty principle} This will be remedied much later by so-called 
\it{polarization} (see sections 
\ref{suseQuaCOT}-\ref{suseQuaG}).

\subsubsection*{Connections and curvature}

In order to define quantum operators acting on wavefunctions seen as sections of a line bundle, we need a 
prescription for computing derivatives of sections. (For instance the momentum operator typically reads 
$P=-i\der_x$.) This requires a notion of covariant differentiation:

\paragraph{Definition.} Let $\cL$ be a complex line bundle over $\cM$, $\text{Vect}_{\CC}(\cM)$
the space of complex vector fields on $\cM$. A \it{connection}\i{connection} for 
$\cL$ is a map
\be
\nabla:
\text{Vect}_{\CC}(\cM)\times\Gamma(\cM,\cL)\rightarrow\Gamma(\cM,\cL):
(\xi,\Psi)\mapsto\nabla_{\xi}\Psi
\nn
\ee
which is linear on $\Gamma(\cM,\cL)$ and $\text{Vect}_{\CC}(\cM)$, and satisfies the property
$\nabla_{\cF\,\xi}\Psi=\cF\,\nabla_{\xi}\Psi$ as well as the Leibniz rule\i{Leibniz rule}
$\nabla_{\xi}(\cF\,\Psi)=\xi(\cF)\,\Psi+\cF\,\nabla_{\xi}\Psi$
for any $\cF\in C^{\infty}(\cM,\CC)$. The linear operator\i{covariant derivative} $\nabla_X$ is 
known as the \it{covariant derivative}\i{covariant derivative} along 
$\xi$.\\

A connection defines a notion of parallel transport and allows one to connect, or identify, fibres at 
different points. The extent to which these identifications deform the fibres as one moves around on the base 
manifold is measured by the \it{curvature}\i{curvature of a connection}
\be
R:
\text{Vect}_{\CC}(\cM)\times\text{Vect}_{\CC}(\cM)\times\Gamma(\cM,\cL)\rightarrow\Gamma(\cM,\cL)
\nn
\ee
which is a two-form defined for all $\xi,\zeta\in\text{Vect}_{\CC}(\cM)$
and any section $\Psi$ by
\be
R(\xi,\zeta)\Psi
\equiv
\left(\nabla_{\xi}\nabla_{\zeta}-\nabla_{\zeta}\nabla_{\xi}-\nabla_{[\xi,\zeta]}\right)\Psi.
\label{CurV}
\ee
When the curvature vanishes the connection is said to be \it{flat}.\i{flat connection} Any 
trivial vector 
bundle admits a flat connection, but the converse is not true: there exist non-trivial bundles with flat 
connections.

\subsubsection*{Hermitian structures}

Since we eventually wish to interpret sections as wavefunctions, we need to define their scalar products.

\paragraph{Definition.} A \it{Hermitian structure}\i{Hermitian structure} on a line bundle 
$\cL\rightarrow\cM$ is a smooth map
\be
\cM\times\Gamma(\cM,\cL)\times\Gamma(\cM,\cL)\rightarrow\CC:
(q,\Phi,\Psi)\mapsto\big(\Phi(q)|\Psi(q)\big).
\label{HerM}
\ee
which is linear in $\Psi$ and antilinear in $\Phi$.\footnote{The map being ``smooth'' means that, given any 
two smooth sections $\Phi,\Psi$, the assignment $q\mapsto(\Phi(q)|\Psi(q))$ is smooth.} Provided $\cM$ is 
endowed with a measure $\mu$, the Hermitian structure can be used to define a space of square-integrable 
sections with scalar product (\ref{scall}).\\

Now let $\cL$ be a complex line bundle over $\cM$ endowed with a connection $\nabla$ and a Hermitian 
structure (\ref{HerM}). We 
say that $\nabla$ is \it{Hermitian} if it is compatible with 
the Hermitian structure 
in the sense that
\be
\xi\cdot(\Phi|\Psi)
=
(\nabla_{\xi}\Phi|\Psi)+(\Phi|\nabla_{\xi}\Psi)
\label{compacon}
\ee
where $(\Phi|\Psi)$ is the function $\cM\rightarrow\CC$ whose value at $q$ is 
$\big(\Phi(q)\big|\Psi(q)\big)$. Condition (\ref{compacon}) is the Hermitian analogue of the condition of 
metric-compatibility for connections on the tangent bundle. In the realm of quantum mechanics, property 
(\ref{compacon}) will allow us to define self-adjoint operators.

\subsection{Quantization of cotangent bundles}
\label{suseQuaCOT}

Having introduced the setup, we now return to our original problem of defining a quantum Hilbert space 
associated with a symplectic manifold $(\cM,\omega)$.
In order for this definition to qualify as a consistent quantization prescription, the Hilbert space $\sH$ 
must be endowed with an operator algebra that is somehow associated with the Poisson algebra of classical 
observables. This association must be a linear map that sends a function $\cF\in\CM$ on a linear operator 
$\hat{\cF}$ in $\sH$, in such a way that\i{correspondence principle}\i{Poisson 
bracket!and commutator}\i{commutator}
\be
\big[\hat\cF,\hat\cG\big]
=
i\hbar\,\widehat{\{\cF,\cG\}}\,.
\label{haf}
\ee
Furthermore the constant function $\cF(p)=1$ must be mapped on the identity operator, i.e.\ $\hat{1}=\II$. 
Thus, the problem is to find a quantum/classical correspondence that fulfills these criteria.\\

The solution turns out to be given by so-called \it{geometric 
quantization} and consists of two steps: prequantization and polarization. Here we describe these steps for 
the simple case of cotangent bundles endowed with the symplectic form (\ref{dabu}), so that 
$\omega=-d\theta$. More general symplectic manifolds are treated in section \ref{suseQuaG}.

\subsubsection*{Prequantization}

As a first attempt at quantization, let us consider the space of complex wavefunctions on $\cM$. 
Their scalar products are then given by (\ref{scall}) where one may choose $\mu$ to be the 
Liouville volume form
(\ref{volom}). To define a linear correspondence between classical and quantum observables, one can try to 
use 
the Hamiltonian vector fields (\ref{hamive}):
\be
\cF\stackrel{?}{\longmapsto}
\hat\cF\stackrel{?}{=}-i\hbar\,\xi_{\cF}\,.
\label{fa}
\ee
Here $\hbar$ is an arbitrary (positive) constant, to be identified with Planck's constant. Indeed, using 
(\ref{Xifidji}) one verifies that (\ref{fa}) satisfies the basic consistency requirement (\ref{haf}),
and is thus at first sight a satisfactory quantization prescription.
However, the problem with (\ref{fa}) is that the trivial observable $\cF(p)=1$ is 
mapped on the zero operator $\hat\cF=0$ instead of the identity. This inconsistency
can be 
remedied by improving (\ref{fa}) as
\be
\cF\stackrel{?}{\longmapsto}
\hat\cF\stackrel{?}{=}-i\hbar\,\xi_{\cF}+\cF\,,
\label{FiKa}
\ee
where the second term on the right-hand side multiplies wavefunctions by $\cF$. This modification ensures that 
$\cF=1$ is represented by the identity operator, but now 
relation (\ref{haf}) no longer holds.\\

We seem to be stuck: how are we to define $\hat{\cF}$ in such a way that both condition (\ref{haf}) and the 
requirement $\hat{1}=\II$ be satisfied? The way out turns out to be the further 
improvement that consists in adding to (\ref{FiKa}) the momentum map 
(\ref{prej}):\i{observable}\i{prequantization}
\be
\cF
\longmapsto
\hat\cF
=
-i\hbar\,\xi_{\cF}-\langle\theta,\xi_{\cF}\rangle+\cF,
\label{hafi}
\ee
where $\theta$ is such that $\omega=-d\theta$. Indeed, using (\ref{Xifidji}) one can verify that 
the commutators of operators
(\ref{hafi}) close according to (\ref{haf}), and furthermore the constant observable $\cF=1$ is represented, 
as it 
should, by the identity operator $\hat\cF=\II$. Thus, provided $\theta$ exists, the prescription (\ref{hafi}) 
is a consistent quantization of the algebra of classical observables on $\cM$.\\

In the present case we are assuming that $\cM=T^*\cQ$ is a cotangent 
bundle, so $\theta$ is just the Liouville one-form (\ref{lifo}) and (\ref{hafi}) is a globally well-defined 
differential operator that quantizes the classical observable $\cF$. One says that cotangent bundles are 
\it{quantizable}. For example, on $\RR^{2n}$ with the symplectic form (\ref{omeq}) the position and momentum 
operators given by prequantization are
\be
\hq^j=i\hbar\frac{\der}{\der p_j}+q^j,
\qquad
\hp_j=-i\hbar\frac{\der}{\der q^j}\,.
\nn
\ee
Note that (\ref{hafi}) may be seen as a differential operator
\be
\hat\cF
=
-i\hbar\nabla_{\xi_{\cF}}+\cF
\label{XuPik}
\ee
where $\nabla_{\xi}=\xi-\frac{i}{\hbar}\bra\theta,\xi\ket$ is a covariant derivative determined by the 
connection whose connection one-form is $\theta$. From this viewpoint the symplectic potential is seen as an 
Abelian gauge field on $\cM=T^*\cQ$, and the corresponding field strength/curvature is the symplectic form 
$\omega=-d\theta$.

\subsubsection*{Polarization}

Since the symplectic form is exact, the map (\ref{hafi}) provides a globally well-defined quantization 
prescription and our job 
here 
is almost done. But there is still a problem: the would-be wavefunctions 
$\Psi:\cM\rightarrow\CC$ depend at this stage on \it{all} coordinates of $\cM=T^*\cQ$. For example, on 
$\RR^{2n}$ we would 
have $\Psi=\Psi(q^i,p_j)$. In particular, in the current situation one could easily devise a wavefunction 
with 
arbitrarily accurate values of position and momentum, violating Heisenberg uncertainty. The purpose of 
\it{polarization}\i{polarization} is to cure this pathology by cutting in half the number of coordinates on 
which wavefunctions are allowed to depend.\\

In the case of cotangent bundles it is common to declare that polarized wavefunctions only depend on the 
coordinates of $\cQ$, and not on the transverse coordinates in each fibre $T_q^*\cQ$. On $\RR^{2n}$ this 
would correspond to saying that polarized wavefunctions $\Psi(q^i)$ do not depend on the coordinates $p_j$, 
which is generally interpreted by saying that wavefunctions are written ``in position space'' --- although we 
shall see 
below that the analogue of this polarization for semi-direct products leads instead to the ``momentum space'' 
picture of chapter \ref{c2bis}. The scalar product of wavefunctions is obtained by endowing the manifold 
$\cQ$ with a measure, resulting in a Hilbert space of polarized wavefunctions.\\

Polarization also affects quantum observables since they must preserve the polarization while still 
satisfying the commutation relations (\ref{haf}).\i{polarized observable} As a result, the space of 
quantizable classical observables 
is a subset of the full space $\CM$. For instance, in $\RR^{2n}$ with Darboux coordinates $q^i,p_j$ 
($i,j=1,...,n$), the classical observables whose quantization preserves the polarization $\der_p\Psi=0$ 
all take the form
\be
\cF(q,p)
=
p_j\cF^j(q)+\cG(q)
\label{SARS}
\ee
for some functions $\cF^j$, $\cG$. Observables which are not of this form do not preserve the polarization 
and are therefore not quantizable in this sense. One should keep in mind, however, that this does \it{not} 
mean that all quantum operators acting in the polarized Hilbert space are forced to take the form 
(\ref{SARS}). Rather, quantizable classical observables give rise to a vector space of Hermitian quantum 
operators, and the full algebra of quantum observables is generated by sums and products of these operators. 
For example, the non-relativistic Hamiltonian $\hat{p}^2$ is obtained by squaring the operator that quantizes 
the classical observable $p$, although there exists no quantizable classical observable whose quantization 
would yield the operator $\hat{p}^2$. In this way one essentially recovers standard quantum mechanics from the 
quantization of the phase space $T^*\cQ$.

\begin{advanced}
\subsection{Quantization of arbitrary symplectic manifolds}
\label{suseQuaG}
\end{advanced}

We now describe geometric quantization \it{without} assuming that the symplectic form is exact. As it turns 
out, 
relaxing that assumption leads to serious complications. Since these subtleties will have very few immediate 
effects on the remainder of our exposition, we urge the hasty reader to go directly to section 
\ref{susequantizationlol}.\\

As before, the requirement that the commutators of quantum observables satisfy (\ref{haf}) leads to the 
quantization prescription 
(\ref{hafi}), where the one-form $\theta$ is such that $\omega=-d\theta$. However, in contrast to 
cotangent bundles, there is in general no 
such one-form on $\cM$ because $\omega$ need not be exact. Thus the best one can do is to treat (\ref{hafi}) 
locally: if
$\{U_i|i\in\cI\}$ is a contractible open cover of $\cM$, the Poincar\'e lemma ensures that there 
exist one-forms $\theta_i$ such that
\be
\omega|_{U_i}=-d\theta_i
\qquad\forall\,i\in\cI
\label{omele}
\ee
since $\omega$ is closed. Then, \it{locally} on each $U_i$, one can define operators
\be
\hat\cF\big|_i
\equiv
-i\hbar\xi_{\cF}-\langle\theta_i,\xi_{\cF}\rangle+\cF
\label{haif}
\ee
that provide a linear correspondence between classical and quantum observables.
The problem then is to glue together operators defined on different open sets.\i{open set} On any non-empty 
intersection 
$U_j\cap U_k$ one has $d\theta_j=d\theta_k$ so there exists a function $\cG_{jk}$ on $U_j\cap U_k$ such that
\be
\theta_j-\theta_k=d\cG_{jk}\,.
\label{ajak}
\ee
Using (\ref{haif}) one can then show that the multiplicative operator
\be
\ell_{kj}\equiv e^{i\cG_{kj}/\hbar}
\label{lkj}
\ee
(acting on functions on $U_j\cap U_k$) is such that
\be
\hat\cF\big|_k
=
\ell_{kj}\circ\hat\cF\big|_j\circ\ell_{kj}^{-1}
\qquad
\text{on }U_j\cap U_k
\label{aka}
\ee
for any classical observable $\cF\in\CM$. This result indicates that the action of $\hat\cF$ on 
functions depends on whether one defines it on $U_j$ or on $U_k$. 
It is an ambiguity in the definition of the operator corresponding to $\cF$,
which threatens the consistency of the construction based on (\ref{haif}).
The way out is think of 
$\hat\cF$ as a differential operator acting not on functions, but on \it{sections} of a complex line 
bundle over $\cM$.\i{geometric quantization!and line bundles}
Indeed, if the line bundle is chosen properly, one may hope that its transition functions for 
some local trivialization associated with the open covering 
$\{U_i\}$ coincide with the multiplication maps (\ref{lkj}),
so that the local formula (\ref{haif}) provides globally well-defined differential operators acting on 
sections.\\

One is thus led to the problem of determining whether there exists a line bundle whose transition functions 
take the form (\ref{lkj}) for the covering $\{U_i|i\in\cI\}$, in such a way that the operator (\ref{haif}) 
can be written \it{globally} as (\ref{XuPik}) for a connection $\nabla$ whose local connection one-forms are 
the $\theta_i$'s. This can be addressed in the framework of \v{C}ech cohomology\i{Cech cohomology@\v{C}ech 
cohomology}, which 
we will not describe here. The bottom line is that such a line bundle with such a connection exists \it{if 
and only if} the cohomology class 
of $\omega/2\pi\hbar$ is integral in the cohomology space $\cH^2_{\text{de Rham}}(\cM,\RR)$, 
i.e.~if\i{symplectic form!integral}\i{Cech cohomology@\v{C}ech cohomology}\i{integral 
symplectic form}\i{quantization conditions}
\be
\left[\frac{\omega}{2\pi\hbar}\right]
\in
\cH^2_{\text{de Rham}}(\cM,\ZZ).
\label{intom}
\ee
This \it{quantization condition} is equivalent to demanding that the integral of $\omega/2\pi\hbar$ over any 
closed two-surface 
be an integer.\footnote{Here ``closed'' means ``compact without boundary''.} The only quantizable symplectic 
manifolds are those that satisfy this requirement.\\

The reason why we did not see this condition in the case of cotangent bundles is that their symplectic form 
is 
\it{globally} exact, so that its cohomology class vanishes and the requirement (\ref{intom}) is trivially 
satisfied. In fact one can show that the curvature two-form (\ref{CurV}) of the connection determined by 
(\ref{haif}) is $R=i\omega/\hbar$, consistently with the fact that the curvature of any 
line bundle is integral. In particular the connection used to define quantum operators (\ref{XuPik}) for 
cotangent bundles is flat.\\

Provided the quantization condition (\ref{intom}) is satisfied, one can endow the space of sections of the 
line 
bundle with a 
Hermitian structure and use it to define the scalar product (\ref{scall}) thanks to the Liouville volume form 
(\ref{volom}). One can show that the Hermitian structure can always be chosen in a way (\ref{compacon}) 
compatible with the connection determined by $\omega$, so that all operators
(\ref{haif}) are Hermitian.\i{prequantization}This completes the first step of 
geometric quantization, i.e.\ prequantization.\\

As in the case of cotangent bundles, the Hilbert space of sections produced by prequantization is ``too 
large'' in the sense that wavefunctions depend on too many arguments.\i{polarization} Polarization corrects 
this problem by 
``cutting in half'' the number of coordinates on which wavefunctions are allowed to depend. 
Since this procedure will not be directly visible in our later considerations, we skip its presentation and 
refer instead to \cite{woodhouse1997,debuyl} for a much more thorough discussion.\\

\paragraph{Remark.} When $\cM$ is a coadjoint orbit, there exists a simple reformulation of the integrality 
condition (\ref{intom}).\i{coadjoint orbit!integrality 
condition} Namely, if $\cW_p\cong G/G_p$ is the coadjoint orbit of $p\in\mg^*$ with stabilizer 
$G_p$, the Kirillov-Kostant symplectic form (\ref{kksym}) is integral if and only if there exists a 
character $\chi$ of $G_p$ 
whose differential (at the identity $e\in G$) satisfies
$d\chi_e=\frac{i}{\hbar}j\big|_{\mg_p}$,
with $\mg_p$ the Lie algebra of $G_p$. The textbook example of this phenomenon is provided by coadjoint 
orbits of $\SU(2)$, which are spheres embedded in $\mathfrak{su}(2)^*$: the quantization condition requires 
that the radius of such a sphere be an integer or half-integer multiple of $\hbar$, corresponding to the 
statement that highest-weight representations of $\su(2)$ have integer or half-integer ``spin''.

\subsection{Symmetries and representations}
\label{susequantizationlol}

We now combine the results of section \ref{seSymPa} with the tools of geometric quantization to address the 
following question: given a symplectic manifold $(\cM,\omega)$ acted upon by a group $G$, does quantization 
produce a unitary representation of $G$?\\

We will assume that the action of $G$ is Hamiltonian, with a momentum map (\ref{CalJ}). We also assume that 
we have chosen a certain value for Planck's constant $\hbar$ and that $\omega/2\pi\hbar$ is integral in the 
sense (\ref{intom}). Then $(\cM,\omega)$ is quantizable and prequantization can be carried out independently 
of the group action. In particular, for each adjoint vector $X\in\mg$ there is a classical observable $\cJ_X$ 
given by (\ref{moment}), and the corresponding operator (\ref{XuPik}) is
\be
\hat\cJ_X
=
-i\hbar\nabla_{\xi_X}+\cJ_X
\label{Jinx}
\ee
where we have used property (\ref{xixi}) to replace $\xi_{\cJ_X}$ by the infinitesimal generator 
(\ref{infige}). By virtue of (\ref{jixiy}) the map $X\mapsto\hat\cJ_X$ is a homomorphism, possibly up to a 
central extension. Thus the assignment (\ref{Jinx}) provides a (generally projective) representation of the 
Lie algebra $\mg$, acting on a space of sections on $\cM$.\\

The subtlety arises with polarization, since then the wavefunctions of the system satisfy extra conditions 
which may not be preserved by (\ref{Jinx}). To avoid such pathologies one has to choose a \it{$G$-invariant} 
polarization.\i{invariant polarization}\i{polarization} 
In that 
case each operator (\ref{Jinx}) is a well-defined Hermitian operator acting on polarized wavefunctions, and 
one obtains a projective, unitary representation of $\mg$. It was shown by Kostant \cite{Kostant1970} that, 
when the action of $G$ on $\cM$ is transitive, the homomorphism 
$X\mapsto\hat\cJ_X$ exponentiates to a unitary representation of the group $G$.\i{unitary representation!by 
quantization}\i{geometric quantization!and unitary reps} This is true in particular 
when 
$\cM$ is a 
coadjoint orbit \cite{kirillov2004lectures}. In addition, when $G$ 
is semi-simple, compact or solvable, the representations obtained in this way are irreducible. Thus 
geometric quantization does produce unitary representations of groups, which is the conclusion we were hoping 
to obtain.

\paragraph{Remark.} One can discuss semi-classical approximations in symplectic 
terms,\i{semi-classical limit!in symplectic geometry}
and this applies in particular to coadjoint orbits.
Indeed, the Liouville volume 
form (\ref{volom}) measures the ``size'' of portions of phase space and can be 
used to compare identical manifolds endowed with different symplectic structures. If $\omega$ and 
$\lambda\omega$ (with $\lambda>0$) are two symplectic forms on $\cM$, then large $\lambda$ assigns a larger 
measure to a given portion of $(\cM,\lambda\omega)$ than to the same portion in $(\cM,\omega)$. In 
this sense large $\lambda$ is a semi-classical regime with 
respect to $(\cM,\omega)$, with $1/\lambda$ playing the role of the coupling constant. 
In the case of
coadjoint orbits, by linearity, $\cW_p$ is diffeomorphic to $\cW_{\lambda p}$ 
for any $\lambda\neq0$, but the definition 
(\ref{kksym}) 
ensures that the symplectic form on $\cW_{\lambda p}$ is ``larger'' (for $\lambda>1$ say) than that on 
$\cW_p$. Thus, for 
$\lambda$ large enough the quantization of $\cW_{\lambda p}$ can be treated semi-classically. Note that this
intuition breaks down if the orbit is invariant under scalings, i.e.\ $\cW_{\lambda p}=\cW_p$.

\section{World lines on coadjoint orbits}
\label{gigglema}

In this section we reformulate the observations of the previous pages in terms of action principles and 
path integrals. In doing so we will develop a group-theoretic world line formalism, which will 
eventually allow us (in section \ref{secosemi}) to interpret representations of semi-direct products as 
actual quantized point particles propagating in space-time.\\

We will start with general geometric considerations explaining how to 
associate 
an action principle with any quantizable symplectic manifold. After a group-theoretic interlude on the 
Maurer-Cartan form, we will focus on coadjoint orbits and describe their world line actions as gauged 
non-linear Sigma models. Useful references are \cite{Alekseev01,Alekseev02,Aratyn:1990dj}; see also 
\cite{Salgado}.

\subsection{World lines and quantization conditions}
\label{susetit}

Our approach here is similar to \cite{charles1999}. Let $(\cM,\omega)$ be a symplectic 
manifold, $p\in\cM$. Since $\omega$ is closed, there exists a 
neighbourhood $U$ of $p$ such that $\omega|_U=-d\theta$ for some one-form $\theta$ on $U$. Now let
$\gamma:[0,1]\rightarrow U:t\mapsto\gamma(t)$ be a path in 
phase space contained in $U$. We can associate with it an action
\be
S\big|_U[\gamma]
\equiv
\int_{\gamma}\theta
=
-\int_{\gamma}d^{-1}\omega\,,
\label{susu}
\ee
where the notation $-d^{-1}\omega$ means ``whatever one-form $\theta$ such that $\omega=-d\theta$''. This is 
a 
purely kinematical Hamiltonian action associated with the symplectic form $\omega$. For example, when 
$\cM=\RR^{2n}$ with $\omega=dq^i\wedge dp_i=-d(p_idq^i)$, expression (\ref{susu}) is globally well-defined 
and reads
\be
S[q^i(t),p_j(t)]
=
\int_0^1dt\,p_j(t)\dot q^j(t)
\label{kines}
\ee
which is the standard reparameterization-invariant kinetic term of any Hamiltonian action. 
There is no term involving $p^2$ or any other combination of $q$'s and $p$'s because there is no 
Hamiltonian at this stage.\\

For a generic symplectic form $\omega$ the definition (\ref{susu}) is not enough: one needs an action 
principle that makes sense for \it{any} path in $\cM$, regardless of exactness. So let $\{U_i|i\in\cI\}$ be 
a contractible open covering of $\cM$ such that 
$\omega|_{U_i}=-d\theta_i$ for each $i\in\cI$. We can then write an action (\ref{susu}) on each $U_i$, but we 
can also attempt to define $S[\gamma]$ for \it{any} path $\gamma$ by\i{geometric action functional}\i{world 
line!on symplectic manifold}\i{action functional!on symplectic manifold}
\be
S[\gamma]
\equiv
-\int_{\gamma}d^{-1}\omega\,.
\label{geoma}
\ee
We refer to this functional as the \it{geometric action} for $(\cM,\omega)$ evaluated on the path 
$\gamma$; its definition follows from the geometry of $\cM$. In particular, when a group $G$ acts on 
$\cM$ by symplectomorphisms, the action automatically has global $G$ symmetry. In section \ref{seReWo} we 
will interpret (\ref{geoma}) as the action of a point particle in space-time.\\

The action (\ref{geoma}) can be evaluated as follows. Given a path $\gamma$, we can
cover its image by open sets $U_j$, with $j\in\cJ\subset\cI$. If only one $U_j$ suffices we can simply use 
the 
original definition (\ref{susu}) to evaluate the action. If there are two open sets,\i{open set} say $U_1$ 
and $U_2$, 
then we call $\gamma_j$ the portion of the path $\gamma$ contained in $U_j$ (for $j=1,2$) and $\gamma_{12}$ 
the portion contained in $U_1\cap U_2$. We can then define
\be
S[\gamma]
\equiv
\int_{\gamma_1}\theta_1+\int_{\gamma_2}\theta_2-\int_{\gamma_{12}}\theta_1
\label{susub}
\ee
where the last term removes the overcounting due to a double integration on $U_1\cap U_2$. There is a 
subtlety 
in this expression: we chose to write $\omega|_{U_1\cap U_2}=-d\theta_1$ in the last term, 
but we could equally well have chosen $\omega=-d\theta_2$; this would have 
given a different compensating term in (\ref{susub}), hence a different value for the action! This is a 
problem at first sight, but one may recall that the action as such need not be a single-valued 
functional on the space of paths in phase space. The truly important quantity is the complex number
\be
e^{iS[\gamma]/\hbar}
\label{eis}
\ee
which determines the path integral measure and leads to transition amplitudes in the quantum theory. Thus we 
are free to have a multivalued action as 
long as all ambiguities are integer multiples of $2\pi\hbar$. This is in effect a \it{quantization condition} 
on the parameters of the action.\\

A simple reformulation of this condition is obtained by considering a closed path $\gamma$ 
(so $\gamma(0)=\gamma(1)$) and evaluating the action along that path. Using Stokes' theorem one can write
\be
S[\gamma]
=
-\oint_{\gamma}d^{-1}\omega
=
-\int_{\Sigma_{\gamma}}\omega
\label{soint}
\ee
where $\Sigma_{\gamma}$ is a two-surface with boundary $\gamma$. As expected this is a multivalued functional 
of $\gamma$. Requiring the exponential $e^{iS[\gamma]}$ to be 
single-valued then implies that the integral of $\omega$ over any closed two-surface 
must be an integer multiple of $2\pi\hbar$\i{quantization conditions!from path 
integral},
which is the old Bohr-Sommerfeld quantization condition\i{Bohr-Sommerfeld quantization} and coincides with 
the integrality requirement 
(\ref{intom}) mentioned above.
One can also show, more generally, that this condition is 
sufficient to ensure that (\ref{eis}) is single-valued on the space of paths. Thus the quantization 
condition determined by the action functional (\ref{geoma}) coincides with the condition that follows from 
geometric 
quantization. This applies, in particular, to the coadjoint orbits of any Lie group.\\

Given an action (\ref{geoma}) that satisfies the quantization condition, one
can choose a Hamiltonian $\cH\in\CM$ 
and compute transition amplitudes using path integrals with the action functional
\be
S[\gamma]
=
-\int_{\gamma}d^{-1}\omega
-\int_0^Tdt\,\cH(\gamma(t)).
\label{geomami}
\ee
Note that this expression is no longer invariant under time reparameterizations for generic choices of the 
Hamiltonian function.

\paragraph{Remark.} The geometric actions (\ref{geoma}) associated with coadjoint orbits of centrally 
extended loop groups describe certain families of Wess-Zumino-Witten models\i{Wess-Zumino-Witten model} 
\cite{Aratyn:1990dj}. In that context the single-valuedness of (\ref{eis}) 
leads to the quantization of the Kac-Moody level \cite{Wess:1971yu,Novikov:1982ei,Witten:1983tw}.\i{Kac-Moody 
level}

\subsection{Interlude: the Maurer-Cartan form}

When the phase space $\cM$ is a coadjoint orbit $\cW_p$ of a group $G$, any path 
$\gamma$ on $\cW_p$ can be written as
\be
\gamma(t)
=
\Ad^*_{f(t)}p
\label{gamad}
\ee
for some path $f(t)$ in $G$. Geometric actions such as (\ref{geoma}) can then be seen as functionals of paths 
on a group manifold. This reformulation turns out to rely on the Maurer-Cartan form of $G$, which we now 
study.

\paragraph{Definition.} Let $G$ be a Lie group and let $L_f:G\rightarrow G:g\mapsto f\cdot g$ denote 
left multiplication by $f\in G$. Then the (left) \it{Maurer-Cartan form} on $G$ is\i{Maurer-Cartan form}
\be
\Theta_f
\equiv
(L_{f^{-1}})_{*f}\,.
\label{lemaca}
\ee
At any point $f$, the map $\Theta_f$ is the differential of left multiplication by $f^{-1}$.\\

It follows from (\ref{lemaca}) that the Maurer-Cartan form at $f$ is an isomorphism 
beween the tangent spaces $T_fG$ and $T_eG$, the latter being identified as usual with the Lie algebra of 
$G$. Thus $\Theta$ is a $\mg$-valued 
one-form on $G$ and may be seen as a section of the vector bundle $T^*G\otimes\mg$. It is also left-invariant 
in the sense that
\be
L_g^*(\Theta)
=
\Theta
\label{LefTheta}
\ee
for any group element $g$. When $G$ is a matrix group, any $f$ can be written as a matrix and the entries of 
$f$ define local 
coordinates on $G$. One can then think of $df$ as the matrix whose entries are the differentials of these 
coordinates, and the left Maurer-Cartan form can be written as
\be
\Theta_f
=
f^{-1}\cdot df\,.
\label{rimaca}
\ee
One can similarly define a right Maurer-Cartan form $(R_{f^{-1}})_{*f}$, where $R$ denotes right 
multiplication (\ref{RighT}). Its expression for a matrix group is $df\cdot f^{-1}$.

\paragraph{Proposition.} The Maurer-Cartan form (\ref{lemaca}) satisfies the \it{Maurer-Cartan 
equation}
\be
(d\Theta)(\xi,\zeta)
+
\left[
\Theta(\xi),\Theta(\zeta)
\right]
=
0
\label{macakol}
\ee
for all vector fields $\xi,\zeta$ on $G$, where $[\cdot,\cdot]$ denotes the Lie bracket 
(\ref{A196}) in $\mg$.

\begin{proof}
Recall that the exterior derivative of $\Theta$ is such that, for all vector fields $\xi,\zeta$,
\be
(d\Theta)(\xi,\zeta)
\equiv
\xi\cdot\Theta(\zeta)-\zeta\cdot\Theta(\xi)
-\Theta([\xi,\zeta])\,,
\label{extedeva}
\ee
where $[\cdot,\cdot]$ is the Lie bracket 
of vector fields. If $\xi$ and 
$\zeta$ are left-invariant, they can be written as $\xi_g=(L_g)_{*e}X$ and 
$\zeta_g=(L_g)_{*e}Y$ for some adjoint vectors $X,Y$. Then (\ref{lemaca}) implies that $\Theta(\xi)=X$ is 
constant on $\mg$, and (\ref{extedeva}) reduces to
\be
(d\Theta)(\xi,\zeta)
+\Theta([\xi,\zeta])=0.
\label{ligo}
\ee
By left-invariance we may write $\Theta([\xi,\zeta])=[\Theta(\xi),\Theta(\zeta)]$ where the bracket on the 
right-hand side now is the Lie bracket (\ref{A196}) of $\mg$. Eq.\ (\ref{ligo}) then takes the form 
(\ref{macakol}) save for the fact that $\xi$ and $\zeta$ are left-invariant. This condition can be relaxed 
upon recalling that the span of left-invariant vector fields at a point $g\in G$ is the whole tangent space 
$T_gG$.
\end{proof}

\subsubsection*{Kirillov-Kostant from Maurer-Cartan}

Thanks to (\ref{gamad}), the Maurer-Cartan form provides a convenient rewriting of the 
Kirillov-Kostant symplectic form (\ref{kksym}) in terms of vectors tangent to a group manifold. Indeed, let
\be
\pi:G\rightarrow\cW_p:g\mapsto\Ad^*_g(p)
\label{projax}
\ee
be the natural projection. We then define a two-form $\bbomega$ on $G$ by
\be
\bbomega
\equiv
\pi^*\omega
\label{bibome}
\ee
where $\omega$ is the Kirillov-Kostant symplectic form (\ref{kksym}). One may think of $\bbomega$ as the 
analogue of (\ref{kksym}) on the group $G$.

\paragraph{Lemma.} Let $g\in G$, and consider tangent vectors $v,w\in T_gG$. Then\i{Kirillov-Kostant 
symplectic form!and Maurer-Cartan}\i{Maurer-Cartan form!and 
Kirillov-Kostant}
\be
\bbomega_g(v,w)
=
\big<
p,
[\Theta_g(v),\Theta_g(w)]
\big>
\label{kksygo}
\ee
where the bracket on the right-hand side is that of $\mg$.

\begin{proof} The definition of the two-form (\ref{bibome}) explicitly reads
\be
(\pi^*\omega)_g(v,w)
=
\omega_{\pi(g)}(\pi_{*g}v,\pi_{*g}w)
\stackrel{!}{=}
\bbomega_g(v,w).
\label{compacom}
\ee
We can represent the vector 
$v$ by a path $\gamma$ in $G$ such that $\dot\gamma(0)=v$, so that
\be
\pi_{*g}(v)
=
\frac{d}{dt}
\big(
\pi(\gamma(t))
\big)\big|_{t=0}
=
\frac{d}{dt}\left.\left(
\Ad^*_{\gamma(t)}(p)
\right)\right|_{t=0}.
\label{inica}
\ee
In turn we can write $\gamma=g\cdot\gamma_0(t)$ where $\gamma_0(0)=e$ is the 
identity. Then $\dot\gamma_0(0)$ belongs to the Lie algebra $T_eG=\mg$ of $G$ and is given by
\be
\dot\gamma_0(0)
=
\frac{d}{dt}\left.
\left(
g^{-1}\cdot
\gamma(t)
\right)
\right|_{t=0}
=
\frac{d}{dt}
\Big(
L_{g^{-1}}\big(\gamma(t)\big)
\Big)
\Big|_{t=0}
=
(L_{g^{-1}})_{*g}(v)
\refeq{lemaca}
\Theta_g(v)
\label{gadozo}
\ee
where we used $v=\dot\gamma(0)$. We can now use this in (\ref{inica}) to obtain
\be
\pi_{*g}(v)
=
\frac{d}{dt}\left.\left(
\Ad^*_{g}(\Ad^*_{\gamma_0(t)}(p))
\right)\right|_{t=0}
\refeq{pixies}
\Ad^*_{g}(\ad^*_{\dot\gamma_0(0)}p)
\refeq{gadozo}
\Ad^*_{g}(\ad^*_{\Theta(v)}p).
\nn
\ee
Eq.\ (\ref{kksygo}) follows upon plugging this result (and its analogue for $w$) in (\ref{compacom}).
\end{proof}

Formula (\ref{kksygo}) is sometimes rewritten as
\be
\bbomega
=
\demi
\bra
p,[\Theta\,\wedge,\,\Theta]
\ket
\label{kksygobis}
\ee
where $[\Theta\,\wedge,\,\Theta]$ is the $\mg$-valued two-form such that 
$[\Theta\,\wedge,\,\Theta]_g(v,w)\equiv 2[\Theta_g(v),\Theta_g(w)]$ for all tangent vectors $v,w\in T_gG$.
In what follows we call $\bbomega$ the \it{symplectic form on $G$} since it is related by 
(\ref{bibome}) to the Kirillov-Kostant symplectic form (in particular $d\bbomega=0$), but one should 
keep in mind that this terminology is actually incorrect: 

\paragraph{Lemma.} The two-form $\bbomega$ is degenerate. Its kernel consists of left-invariant vector 
fields $\zeta_X$ for which $X$ belongs to the Lie algebra 
of the stabilizer of $p$.

\begin{proof}
Let $v,w\in T_gG$; there are two (unique) adjoint vectors $X,Y\in T_eG=\mg$ such that 
$v=(L_g)_{*e}X$ and $w=(L_g)_{*e}Y$, so that $\Theta_g(v)=X$ and similarly for $w$.
Formula (\ref{kksygo}) can then be rewritten as
\be
\bbomega_g(v,w)
=
\bra
p,[X,Y]
\ket
=
-\langle\ad^*_Xp,Y\rangle.
\label{vanilla}
\ee
The kernel of $\bbomega$ consists of vectors $v=(L_g)_{*e}X$ such that (\ref{vanilla}) vanishes 
for any $Y\in\mg$, which is to say that $\ad^*_Xp=0$. The latter property holds if and only if $X$ belongs to 
the Lie algebra of the stabilizer of $p$.
\end{proof}

This lemma confirms that $\bbomega$ is not a symplectic form because its components 
do not form an invertible matrix. The rank of $\bbomega$ is $\text{dim}(G)-\text{dim}(G_p)$, where $G_p$ is 
the stabilizer of $p$. This number 
coincides (as it should) with the dimension of the coadjoint orbit of $p$, which proves by the way that the 
original form 
(\ref{kksym}) on $G/G_p\cong\cW_p$ is invertible. 

\subsection{Coadjoint orbits and Sigma models}
\label{suseki}

The ``symplectic form'' (\ref{kksygobis}) is a closed two-form, and is therefore locally exact. As such it 
can be used to define a kinetic action functional analogous to (\ref{geoma}),\i{geometric action 
functional!on group}
\be
S[f(t)]
\equiv
-\int_{f(t)}d^{-1}\bbomega,
\label{sigabb}
\ee
whose argument is a path $f(t)$ in $G$. Using the Maurer-Cartan equation (\ref{macakol}) in (\ref{kksygo}), 
one can write
\be
\bbomega_f
=
-\bra p,d\Theta_f\ket
=
-d\big(\bra p,\Theta\ket\big)_f
\nn
\ee
where the exterior derivative goes through the coadjoint vector $p$ by linearity. Thus the action 
(\ref{sigabb}) becomes\i{geometric 
action functional!as non-linear Sigma model}\i{non-linear Sigma model}\i{geometric action 
functional!and Maurer-Cartan form}\i{Maurer-Cartan form!and geometric action}\i{world line!on coadjoint 
orbit}\i{sigma model}
\be
S[f(t)]
=
\int_{f(t)}
\bra
p,\Theta
\ket
=
\int_0^Tdt\,\big<p,\Theta_{f(t)}\big(\dot f(t)\big)\big>\,.
\label{actisi}
\ee
It describes the dynamics of paths 
$f(t)\in G$ and may be seen as the (kinetic piece of the) action of a non-linear Sigma model. When $G$ is a 
simple matrix group, adjoint and coadjoint vectors can be identified so that $\bra 
p,\cdot\ket=\text{Tr}[X\cdot]$ for some $X\in\mg$, and (\ref{rimaca}) allows us to recast the 
integrand of (\ref{actisi}) in the form $\text{Tr}\big[Xf^{-1}\dot f\,\big]$.\\

Note that the global $G$ symmetry of (\ref{actisi}) is manifest: if $f(t)$ is a path in $G$ and $g\in G$ is 
an arbitrary constant group element, then left-invariance of $\Theta$ readily implies $S[g\cdot 
f(t)]=S[f(t)]$. In addition (\ref{actisi}) is the integral of a one-form and is thus 
invariant under redefinitions of the time parameter $t$. As in 
(\ref{geomami}) one can include a Hamiltonian in the action, at the cost of breaking time 
reparameterization invariance.\\

A key subtlety with (\ref{actisi}) is that the group variable $f(t)$ is
the group 
element that appears in a coadjoint action $\Ad^*_{f(t)}p$, as in (\ref{gamad}). The latter coadjoint vector 
is invariant under multiplication of $f(t)$ from the 
right by any (generally time-dependent) group element $h(t)$ belonging to the stabilizer of $p$. This 
means that (\ref{actisi}) should be invariant under gauge 
transformations $f(t)\mapsto f(t)\cdot h(t)$, and therefore describes a \it{gauged} non-linear 
Sigma model. Let us check that (\ref{actisi}) does indeed admit such a 
symmetry.\i{gauged Sigma model} Using the Leibniz rule we find
\be
S[f\cdot h]
=
\int_0^T
\bra
p
,
\Theta_{f(t)h(t)}
\left(
(R_h)_{*f(t)}\dot f(t)
\right)
\ket
+
\int_0^T
\bra
p
,
\Theta_{f(t)h(t)}
\left(
(L_{f(t)})_{*h(t)}\dot h(t)
\right)
\ket
\label{fitem}
\ee
where the adjoint vector paired with $p$ in the first term can be rewritten as
\be
\Theta_{f(t)h(t)}
\left(
(R_h)_{*f(t)}\dot f(t)
\right)
=
\Ad_{h^{-1}}
\Theta_{f(t)}\dot f(t)
\nn
\ee
thanks to the definitions (\ref{lemaca}) and (\ref{ad}).
This implies that the first term of (\ref{fitem}) coincides with the original action (\ref{actisi}). As 
for the 
second term in (\ref{fitem}), we use left-invariance of $\Theta$ to rewrite it as a Sigma model action 
evaluated on a path wholly 
contained in the stabilizer $G_p$:
\be
S[h(t)]
=
\int_0^T
dt
\bra p,\Theta_{h(t)}\dot h(t)\ket.
\label{wiggaboum}
\ee
The counterpart of $h(t)$ in the coadjoint orbit of $p$ is the constant path $\Ad^*_{h(t)}p=p$, but 
in the Sigma model it carries a generally non-vanishing action (\ref{wiggaboum}). Thus 
gauge-invariance of (\ref{actisi}) may be true, but is not obvious at this stage since 
the gauge-transformed action (\ref{fitem}) differs from (\ref{actisi}) by the extra term (\ref{wiggaboum}). 
To reconcile this observation with the much desired gauge-invariance of (\ref{actisi}), we note that the 
exterior derivative of the integrand of (\ref{wiggaboum}) vanishes. Indeed, for all $v,w\in T_hG_p$ the 
Maurer-Cartan equation (\ref{macakol}) yields
\be
d
\bra p,\Theta\ket_h
(v,w)
=
-\bra
p,[\Theta_h(v),\Theta_h(w)]
\ket
\refeq{pixies}
\bra
\ad^*_{\Theta_h(v)}p,\Theta_h(w)
\ket
=0
\nn
\ee
where the last equality follows from the fact that $\Theta_h(v)$ belongs to the Lie algebra of the 
stabilizer of $p$. Thus the integrand of (\ref{wiggaboum}) is closed, and is therefore locally exact. In 
particular, for a path $h(t)$ located in a sufficiently small neighbourhood of the identity in $H$, there 
exists a function $\cF(t)$ such that
\be
\bra
p,\Theta_{h(t)}\dot h(t)
\ket
=
\dot\cF(t)
\label{locaxa}
\ee
for any $t\in[0,T]$. The integral (\ref{wiggaboum}) 
of this quantity is a boundary term,\i{geometric 
action functional!and gauge invariance} so the action functional (\ref{actisi}) is indeed 
gauge-invariant, albeit up to boundary terms that can be cancelled by requiring for instance that initial and 
final configurations be fixed.\\

We stress that this gauge symmetry is unavoidable if (\ref{actisi}) is
interpreted as the Sigma model version of the action (\ref{geoma}) on a coadjoint orbit. In 
particular the inclusion of a Hamiltonian is now 
subject to a constraint: in order to reproduce (\ref{geomami}), the Hamiltonian expressed in terms of group 
variables must be invariant under stabilizer gauge transformations.

\begin{advanced}
\subsection{Coadjoint orbits and characters of $\SL$}
\label{suseseka}
\end{advanced}

As an application of the above considerations, we now
classify the coadjoint orbits of 
$\SL$ and quantize some of them, showing along the way 
that they are equivalent to one-dimensional harmonic oscillators. As an application we evaluate $\SL$ 
characters by geometric quantization. We refer e.g.\ to \cite{Taylor:1992xt,Taylor:1993zp} 
for more details on the coadjoint orbits of $\SL$, and to \cite{Alekseev01,Aratyn:1990dj} for similar 
computations in more general cases. This section is not crucial for the remainder of the thesis and may be 
skipped in a first reading.\\

\subsubsection*{Coadjoint orbits of $\SL$}

For the basic properties of the group $\SL$ we refer to section \ref{sePoTri}.
Its Lie algebra $\sl$ consists of real, traceless $2\times 2$ matrices. Any such matrix is a 
real linear combination\i{sl2R@$\sl$} $X=x^{\mu}t_{\mu}$ of basis elements (\ref{tmus}) whose brackets 
read\i{sl2R@$\sl$!commutation relations}
\be
[t_{\mu},t_{\nu}]
=
{\epsilon_{\mu\nu}}^{\rho}\,t_{\rho}\,.
\label{silcomi}
\ee
Here $\epsilon_{\mu\nu\rho}$ is the completely antisymmetric tensor such that $\epsilon_{012}=+1$, and 
indices are raised and lowered using the Minkowski metric $\eta_{\mu\nu}=\text{diag}(-\,+\,+)$. For future 
reference we also note that, in the complex basis
\be
\ell_0\equiv-t_0\,,
\qquad
\ell_1\equiv t_2-it_1\,,
\qquad
\ell_{-1}\equiv t_2+it_1\,,
\label{eltimm}
\ee
the Lie brackets (\ref{silcomi}) take the form
\be
i[\ell_m,\ell_n]=(m-n)\ell_{m+n}
\label{LaLLA}
\ee
for $m,n=-1,0,1$. On account of the isomorphism (\ref{isoso}) this can also be seen as the Lorentz algebra in 
three dimensions.\\

The $\sl$ algebra 
has a non-degenerate bilinear form
\be
(X,Y)
\equiv
2\,\Tr(XY)
=
\eta_{\mu\nu}x^{\mu}x^{\nu}
\label{sexy}
\ee
which is left invariant by the adjoint action (\ref{SXS-1}) of $\SL$. We can then use the isomorphism 
(\ref{ii}) to intertwine the adjoint and coadjoint 
representations of $\SL$ as in (\ref{sequiv}). In particular we may identify coadjoint with adjoint 
vectors, and coadjoint orbits coincide with 
adjoint orbits under that identification.\i{SL2R@$\SL$!coadjoint orbits}\i{coadjoint orbit!of SL2R@of $\SL$} 
Those are 
exactly the momentum orbits of the Poincar\'e group in 
three dimensions, which were described in sections \ref{relagroup} and \ref{sePoTri}. This 
provides the classification of coadjoint orbits of $\SL$ and an exhaustive family of 
orbit representatives is depicted schematically in fig.\ \ref{figDepIdu}.\\

Note that the fact that coadjoint orbits of $\SL$ coincide with Poincar\'e momentum orbits in three 
dimensions follows from the structure $G\ltimes_{\Ad}\mg_{\text{Ab}}$ of the double cover (\ref{pisel}) of 
the Poincar\'e group. We will encounter a 
similar structure in the BMS$_3$ group, albeit with an infinite-dimensional group $G$.

\subsubsection*{Kirillov-Kostant symplectic form}

We can write any coadjoint vector of $\SL$ as 
$q=q_{\mu}(t^{\mu})^*$ where $(t^{\mu})^*=\eta^{\mu\nu}(t_{\nu},\cdot)$ is 
the dual basis corresponding to (\ref{tmus}). The components $q_{\mu}$ are global coordinates on $\sl^*$ and 
their Kirillov-Kostant Poisson brackets read\i{SL2R@$\SL$!Kirillov-Kostant bracket}\i{Kirillov-Kostant 
bracket!for SL2R@for $\SL$}
\be
\{p_{\mu},p_{\nu}\}
=
\epsilon_{\mu\nu}{}^{\rho}\,p_{\rho}
\label{Piper}
\ee
on account of (\ref{silcomi}) and the general result (\ref{fabipe}). Now consider a ``massive'' orbit
\be
\cW_p
=
\Big\{
q_{\mu}(t^{\mu})^*\Big|q_0=\sqrt{h^2+q_1^2+q_2^2}\,
\Big\}
\cong
\SL/\un
\label{silorbitg}
\ee
with orbit representative $p=h(t^0)^*$ and stabilizer $\un$. We denote the ``mass'' of the orbit by $h$ 
rather than $M$ because its quantization will eventually correspond to a representation of $\sl$ with highest 
weight $h$ (or more precisely $h+1/2$).
The restriction of (\ref{Piper}) to the 
orbit gives rise to the Kirillov-Kostant symplectic form (\ref{kksym}), which we now evaluate.\\

We can label the points of (\ref{silorbitg}) by their ``spatial components'' $(q_1,q_2)$. In order to write 
down (\ref{kksym}) in these coordinates, we need a dictionary between the components 
$x^{\mu}$ of $X$ and those of the corresponding vector field $\ad^*_Xq$ in terms 
of the coordinates $q_1$, $q_2$. We first evaluate $\ad^*_Xq$ for $X=x^{\mu}t_{\mu}$; using 
(\ref{silcomi}), for any adjoint vector 
$Y=y^{\mu}t_{\mu}$ we find
$\langle\ad^*_Xq,Y\rangle=-\langle 
q,x^{\mu}y^{\nu}\epsilon_{\mu\nu}{}^{\rho}\,t_{\rho}\rangle$. 
For $q$ belonging to (\ref{silorbitg}) one obtains
\be
\begin{split}
\ad^*_Xq=
& \;\Big(\!\!-q_1X^2+q_2X^1\Big)(t^0)^*
+\Big(\!\!-\sqrt{h^2+q_1^2+q_2^2}X^2-q_2X^0\Big)(t^1)^*\\
& +\Big(\sqrt{h^2+q_1^2+q_2^2}X^1+q_1X^0\Big)(t^2)^*.
\end{split}
\label{variqq}
\ee
This is an infinitesimal variation of $q$ tangent to $\cW_p$. Any such 
variation can be expressed in terms of the coordinates $(q_1,q_2)$: for an infinitesimal variation $(\delta 
q_1,\delta q_2)$ of $(q_1,q_2)$, the variation of $q_0$ on the orbit (\ref{silorbitg}) is
\be
\delta q_0
=
\frac{q_1\delta q_1+q_2\delta q_2}{\sqrt{h^2+q_1^2+q_2^2}}\,.
\label{Qdel}
\ee
The variation of $q$ produced by a vector $v=v_1\der_{q_1}+v_2\der_{q_2}$ tangent to $\cW_p$ takes the 
same form with $\delta q_i$ replaced by $v_i$. Given such a vector $v$ at $q$ one may ask what Lie algebra 
element $X$ is such that $v=\ad^*_Xq$. Owing to (\ref{variqq}) and (\ref{Qdel}) we may choose
\be
x^0=0\,,
\qquad
x^1
=
\frac{V_2}{\sqrt{h^2+q_1^2+q_2^2}}\,,
\qquad
x^2
=
\frac{-V_1}{\sqrt{h^2+q_1^2+q_2^2}}\,.
\label{idefox}
\ee
This solution to $v=\ad^*_Xq$ is not unique for a given $v$ due to the non-trivial stabilizer $\un$, but it 
is all we need for evaluating the Kirillov-Kostant symplectic form. Indeed, using (\ref{kksym}) 
and the fact that the orbit (\ref{silorbitg}) is two-dimensional, we 
find\i{SL2R@$\SL$!Kirillov-Kostant form}\i{Kirillov-Kostant symplectic form!for SL2R@for $\SL$}
\be
\omega=
\frac{dq_2\wedge dq_1}{\sqrt{h^2+q_1^2+q_2^2}}\,,
\label{symsl}
\ee
which coincides (up to sign) with the Lorentz-invariant volume form (\ref{lor}) on the mass shell $\cW_p$.
One can rewrite it in global Darboux coordinates
\be
P\equiv
\left(
\frac{2\sqrt{h^2+q_1^2+q_2^2}-2h}{q_1^2+q_2^2}
\right)^{1/2}q_1,
\qquad
Q\equiv
\left(
\frac{2\sqrt{h^2+q_1^2+q_2^2}-2h}{q_1^2+q_2^2}
\right)^{1/2}q_2
\label{dazzle}
\ee
such that (\ref{symsl}) simply becomes
\be
\omega
=
dQ\wedge dP.
\label{symsida}
\ee
Hence the Kirillov-Kostant symplectic form on the orbit (\ref{silorbitg}) is globally exact and the 
quantization condition (\ref{intom}) is trivially satisfied for any value of $h$.

\subsubsection*{Characters as path integrals}

We can now quantize the orbit $\cW_p$ with the symplectic form (\ref{symsida}). The associated line 
bundle is trivial and its sections are just complex-valued functions on $\cW_p$; polarized sections can be 
chosen to depend only on the 
coordinate $Q$. One can then
evaluate characters of suitable unitary 
representations of $\SL$ by computing traces of operators in the resulting Hilbert space, as follows.\\

The character of a representation is the trace of the exponential of a certain 
Lie algebra generator. When interpreting the latter as a Hamiltonian, the character may be 
seen as a partition function. Here we take the Hamiltonian to be the generator of rotations, corresponding to 
the basis element $t_0$ in (\ref{tmus}). As a function on phase space the Hamiltonian maps the point 
$q_{\mu}(t^{\mu})^*$ on its component $q_0$, so on the orbit (\ref{silorbitg}) we have
\be
\cH=\frac{1}{\ell}\sqrt{h^2+q_1^2+q_2^2}
\label{hamisil}
\ee
where we have included a prefactor\footnote{We denote by $c$ the speed of light in the vacuum.} $\hbar 
c/\ell\equiv 1/\ell$ to ensure that $\cH$ has dimensions of energy (we think of $h$, $q_{\mu}$ as being 
dimensionless). In Darboux coordinates $Q,P$, we find the Hamiltonian of a harmonic 
oscillator:\i{SL2R@$\SL$!and 
harmonic oscillators}\i{harmonic 
oscillator}
\be
\cH=\frac{h}{\ell}+\frac{1}{2\ell}(P^2+Q^2).
\label{harmozik}
\ee
Thus the quantization of the orbit $\cW_p$ with the Hamiltonian (\ref{hamisil}) is a quantum harmonic 
oscillator on the line!\\

This tremendous simplification allows us to evaluate characters. In principle we 
could use the path integral formalism, but the operator approach is much simpler since we know the spectrum 
of 
the Hamiltonian. Its eigenvalues are
\be
\frac{h+1/2}{\ell},\;\frac{h+3/2}{\ell},\;\frac{h+5/2}{\ell},...,\frac{h+1/2+n}{\ell},...
\nn
\ee
each with unit multiplicity. In particular the partition 
function at temperature $1/\beta$ is that of a harmonic oscillator, $e^{-\beta 
h/\ell}/(2\sinh[\beta/\ell])$. For future reference we rewrite it as follows: we call $L_0$ the operator that 
generates rotations so that $\hat\cH=\frac{1}{\ell} L_0$, and we write $e^{-\beta/\ell}\equiv q$. We also 
allow 
$\beta$ to be complex as long as its real part is positive. Then the partition function can be written 
as\i{SL2R@$\SL$!character}\i{character!for SL$(2,\RR)$}
\be
\Tr\left(q^{L_0}\right)
=
\frac{q^{h+1/2}}{1-q}\,.
\label{silchar}
\ee
In section \ref{suseHiS} we will show that this is the character of a unitary representation of the $\sl$ 
Lie algebra with highest weight $h+1/2$. In the present case one can think of the ``$1/2$'' as a quantum 
correction to the classical weight $h$.

\section{Coadjoint orbits of semi-direct products}
\label{secosemi}

We now apply the considerations of the previous sections to the semi-direct products\footnote{Here the words 
``semi-direct product'' refer to a group (\ref{semid}) with an \it{Abelian vector 
group} $A$.} described in chapter \ref{c2bis}. In particular we explain how the induced representations of 
section \ref{sesemi} emerge from geometric quantization of coadjoint orbits. The plan is as follows. We first 
work out general expressions for the adjoint representation, the Lie bracket and the coadjoint representation 
of any semi-direct product.\footnote{The sequence ``group $\leadsto$ adjoint $\leadsto$ coadjoint'' will be 
ubiquitous in this thesis.} Then we expose a general classification 
of 
coadjoint orbits, seen as fibre bundles over cotangent bundles of momentum orbits. Finally we turn to 
geometric quantization and describe the world line actions associated with coadjoint orbits. The 
considerations of this section can be found e.g.\ in \cite{Rawnsley1975}, and also in more recent works 
\cite{Barnich:2015uva,Baguis1998,Li1993}. The textbooks \cite{Guillemin,Ali2000} contain detailed 
computations and examples.

\subsection{Adjoint representation of $G\ltimes A$}

We consider a semi-direct product (\ref{semid}) with $A$ a vector group. Then the Lie algebra of $G\ltimes 
A$ is a \it{semi-direct sum}\i{semi-direct sum}
\be
\mathfrak{g}\inplus_{\Sigma}A,
\label{galila}
\ee
where $\mg$ is the Lie algebra 
of $G$ and $A$ is its own Lie algebra since it is a vector group. The symbol $\Sigma$ denotes the 
differential of the action $\sigma$ at the identity,
$\Sigma:
\mg\rightarrow\text{End}(A):
X\mapsto\Sigma_X$,
where $\Sigma_X$ is the infinitesimal generator (\ref{infige}) associated with $X$:
\be
\Sigma_X:A\rightarrow A:
\alpha\mapsto
\Sigma_X\alpha\equiv\frac{d}{dt}\left.\left(\sigma_{e^{tX}}\alpha\right)\right|_{t=0}\,.
\label{sideff}
\ee
We will denote elements of (\ref{galila}) as pairs $(X,\alpha)$ 
where $X\in\mg$ and $\alpha\in A$; in the terminology of (\ref{semiop}), $X$ is an 
infinitesimal rotation/boost while $\alpha$ is a translation.\\

The adjoint representation of $G\ltimes A$ is given by (\ref{ad}), which yields\i{semi-direct product!adjoint 
representation}\i{adjoint representation!of semi-direct product}
\begin{eqnarray}
\Ad_{(f,\alpha)}(X,\beta)
& \!\!\refeq{semiop} &
\!\!\frac{d}{dt}
\Big(
fe^{tX}f^{-1},
\alpha+t\sigma_f\beta-\sigma_{fe^{tX}f^{-1}}\alpha
\Big)
\Big|_{t=0}\nn\\
\label{adsemi}
& \!\!= &
\!\!\left(
\Ad_fX,
\sigma_f\beta-\Sigma_{\Ad_fX}\alpha
\right)
\end{eqnarray}
where the symbol ``$\Ad$'' on the right-hand side denotes the adjoint representation of $G$. (More generally, 
in case of ambiguous notations, the argument of a group action determines which group it 
refers to.) In particular, rotation generators transform according to the adjoint 
representation of $G$, while translations are subject to mixed transformations involving both the finite 
action $\sigma$ and its 
differential $\Sigma$.\\

From (\ref{adsemi}) one can read off the Lie bracket in $\mg\inplus A$ upon using (\ref{adg}):\i{semi-direct 
sum!Lie bracket}\i{Lie bracket!for semi-direct sum}
\be
\big[(X,\alpha),(Y,\beta)\big]
=
\big([X,Y],\Sigma_X\beta-\Sigma_Y\alpha\big).
\label{semibra}
\ee
The presence of $\Sigma$ on the right-hand side justifies calling $\mg\inplus A$ a \it{semi}-direct sum.
Note that, if $A$ was non-Abelian, the second entry on the right-hand side would include a 
bracket of generators of $A$.\\

The structure of the algebra (\ref{semibra}) can be made more transparent by choosing a basis. Let $t_a$ be a 
basis of $\mg$ satisfying the brackets (\ref{commurel}), and let $\alpha_i$ be a basis of $A$ (here 
$a=1,...,\dim\mg$ and $i=1,...,\dim A$). Introducing the basis elements
\be
j_a\equiv (t_a,0),
\qquad
p_i\equiv(0,\alpha_i)
\nn
\ee
that generate the semi-direct sum $\mg\inplus A$, the Lie bracket (\ref{semibra}) yields
\be
[j_a,j_b]=f_{ab}{}^c\,j_c\,,
\qquad
[j_a,p_i]=g_{ai}{}^k\, p_k\,,
\qquad
[p_i,p_j]=0
\label{jAppi}
\ee
where $g_{ai}{}^k\, p_k\equiv\Sigma_{t_a}p_i$ so that the coefficients $(g_a)_i{}^k$ are the entries of the 
matrix representing the linear operator $\Sigma_{t_a}:A\rightarrow A$ in the basis $\alpha_i$. The brackets 
(\ref{jAppi}) make the semi-direct structure manifest since the bracket $[j,p]$ gives $p$'s while the bracket 
$[p,p]$ vanishes on account of the fact that $A$ is Abelian. This structure will appear repeatedly in this 
thesis.

\subsection{Coadjoint representation of $G\ltimes A$}

The space of coadjoint vectors of $G\ltimes A$ is the dual of the semi-direct sum (\ref{galila}),
\be
\mg^*\oplus A^*.
\label{semidico}
\ee
Its elements are pairs $(j,p)$ where $j\in\mg^*$ and $p\in A^*$, paired with adjoint vectors according 
to\i{momentum}\i{angular momentum}\i{semi-direct product!coadjoint vector}\i{coadjoint vector!for 
semi-direct product}
\be
\big<(j,p),(X,\alpha)\big>
=
\langle j,X\rangle
+
\langle
p,\alpha
\rangle
\label{papa}
\ee
where the first pairing $\langle\cdot,\cdot\rangle$ on the right-hand side is that of $\mg^*$ with $\mg$ 
while 
the second one pairs $A^*$ with $A$. Note that $A^*$ is precisely 
the space of momenta (see section \ref{sesemi}), while $\mg^*$ is dual to infinitesimal rotations and may be 
seen as a space of angular momentum vectors. This is consistent with the general interpretation of 
coadjoint vectors as conserved quantities (see section \ref{suseCA}) and justifies the notation $(j,p)$.\\

The coadjoint representation of $G\ltimes A$ acts on the space (\ref{semidico}). In order 
to write it down, it is convenient to introduce the following notation:

\paragraph{Definition.} The \it{cross product}\i{cross product} of translations and momenta is the bilinear 
map
$A\times A^*\rightarrow\mg^*:
(\alpha,p)
\mapsto
\alpha\times p$
given for any $X\in\mg$ by
\be
\langle
\alpha\times p,X
\rangle
\equiv
\langle
p,\Sigma_X\alpha
\rangle\,.
\label{copo}
\ee
The notation is justified by the fact that $\times$ coincides with the vector 
product when $G\ltimes A$ is the Euclidean group in three dimensions.\\

With 
this notation the coadjoint action of $G\ltimes A$ is given by
\begin{eqnarray}
\big<
\Ad^*_{(f,\alpha)}(j,p),
(X,\beta)
\big>
& \!\!\refeq{cocogi} &
\!\!\big<
(j,p),
\Ad_{(f,\alpha)^{-1}}(X,\beta)
\big>\nn\\
\label{ziu}
& \!\!\refeq{adsemi} &
\!\!\left<
(j,p),
\left(
\Ad_{f^{-1}}X,
\sigma_{f^{-1}}\beta+\Sigma_{\Ad_{f^{-1}}X}\sigma_{f^{-1}}\alpha
\right)
\right>.
\qquad
\end{eqnarray}
We can use $\Sigma_X(\sigma_f\alpha)=
\sigma_f(\Sigma_{\Ad_{f^{-1}}X}\alpha)$
to rewrite this as
\be
\big<
\Ad^*_{(f,\alpha)}(j,p),
(X,\beta)
\big>
\refeq{papa}
\langle j,\Ad_{f^{-1}}X\rangle
+
\left<
p,\sigma_{f^{-1}}\beta+\sigma_{f^{-1}}\Sigma_X\alpha
\right>\,.
\label{kook}
\ee
In the first term of the right-hand side we recognize the coadjoint representation of $G$; the 
part of the second term involving $\beta$ is the transformation law (\ref{sstar}) of momenta; the 
last term involves the cross product 
(\ref{copo}) of $\sigma^*_fp$ with $\alpha$. Collecting all these terms and removing the argument 
$(X,\beta)$, we conclude that the coadjoint representation of $G\ltimes A$ is\i{semi-direct 
product!coadjoint representation}\i{coadjoint representation!of semi-direct product/sum}
\be
\boxed{
\Big.
\Ad^*_{(f,\alpha)}(j,p)
=
\left(
\Ad^*_fj+\alpha\times\sigma^*_fp
\,,\,
\sigma^*_fp
\right)}
\label{acoga}
\ee
where we keep the notation $\sigma^*_fp$ instead of the simpler $f\cdot p$ to avoid confusion. Note that the 
coadjoint action of the translation group $A$ affects only angular momenta, since the 
transformation of $p$ only involves $f\in G$. The translation $\alpha$ contributes a term 
$\alpha\times\sigma^*_fp$, which for trivial $f$ boils down to the cross product $\alpha\times p$; 
this contribution can be identified with a combination of \it{orbital angular momentum}\i{orbital angular 
momentum}\i{angular momentum!orbital versus spin} and the \it{centre of 
mass vector},\i{centre of mass} while the \it{spin angular 
momentum}\i{spin angular momentum} is contained in $\Ad^*_fj$. We will return to 
this interpretation below.\\

From (\ref{acoga}) we obtain the coadjoint representation (\ref{pixies}) of $\mg\inplus_{\Sigma} 
A$:\i{semi-direct sum!coadjoint representation}
\be
\ad^*_{(X,\alpha)}(j,p)
=
\big(
\ad^*_Xj+\alpha\times p
\,,\,
\Sigma^*_Xp
\big)\,,
\label{oogutak}
\ee
where $\Sigma^*_Xp\equiv-p\circ\Sigma_X$. We will use this formula below when dealing with the 
Kirillov-Kostant symplectic form.

\paragraph{Remark.} We shall 
see in chapter \ref{c6} that the space of asymptotically Minkowskian solutions of Einstein's equations in 
three dimensions
spans (a subset of) the space of the coadjoint representation of the BMS$_3$ group. Each metric will then be 
labelled by a pair $(j,p)$, where $j$ and $p$ are certain functions on the celestial circle that can be 
interpreted as the angular momentum aspect and the Bondi mass aspect, respectively.

\subsection{Coadjoint orbits}
\label{susecoo}

Let us now classify the coadjoint orbits of a semi-direct product. This may be seen as a classification of 
all classical particles, analogous to the quantum classification worked out in section \ref{sesemi}. The 
coadjoint orbit of $(j,p)$ is the set \i{coadjoint orbit!of semi-direct 
product}\i{semi-direct product!coadjoint orbit}
\be
\cW_{(j,p)}
=
\big\{
\Ad^*_{(f,\alpha)}(j,p)\big|(f,\alpha)\in G\ltimes A
\big\}
\label{cocobi}
\ee
embedded in $\mg^*\oplus A^*$, with $\Ad^*_{(f,\alpha)}(j,p)$ given by (\ref{acoga}). In order to 
classify all such orbits, we will assume that the orbits (\ref{hop}) and little groups (\ref{gilipi}) 
of 
induced representations are known. Due to the second entry of the right-hand side of
(\ref{acoga}), involving only $\sigma^*_f p$, each $\cW_{(j,p)}$ is a fibre bundle over the orbit
$\cO_p$. The fibre above $q=\sigma^*_f p$ is the set 
\be
\Big\{
\left(
\Ad^*_g\Ad^*_fj
+
\alpha\times q,
q
\right)
\Big|
g\in G_q,\,\alpha\in A
\Big\}\,.
\nn
\ee
It remains to understand the geometry of these fibres and the relation
between fibres at different points. Note that in the degenerate case $p=0$ the orbit $\cW_{(j,0)}$ 
is simply the coadjoint orbit of $j$ under $G$; in particular $\cW_{(0,0)}$ contains only one 
point. Accordingly we take $p\neq0$ until the end of this section.

\subsubsection*{Warm-up: Scalar orbits}

We start by describing scalar orbits, that is, coadjoint orbits that contain points with vanishing angular 
momentum $j=0$.\i{coadjoint orbit!scalar} The terminology is justified by the 
fact that each orbit is a homogeneous phase space invariant under $G\ltimes A$, whose quantization 
yields the Hilbert space of a particle transforming under a unitary representation of $G\ltimes 
A$. Saying that an orbit contains points with $j=0$ then means that there exists a frame where the particle's 
spin 
vanishes, i.e.\ that the particle is scalar.\\

So let us describe an orbit $\cW_{(0,p)}$. With $j=0$ the first entry of the right-hand 
side of (\ref{acoga}) reduces to
\be
\alpha\times\sigma^*_f p.
\label{rhs}
\ee
Keeping $q=\sigma^*_f p$ fixed, the set spanned by angular momenta of this form is
\be
A\times q
\equiv
\left\{\alpha\times q|\alpha\in A\right\}
\subset\mg^*
\label{awedgeq}
\ee
and coincides with the set of orbital angular momenta that can be reached by acting 
with translations on a particle with momentum $q$. The geometric interpretation of (\ref{awedgeq}) is as 
follows. Recall first that the tangent space of $\cO_p$ at $q$ can be identified with
the space of ``small displacements'' of $q$ generated by infinitesimal boosts: 
\be 
T_q\cO_p =
\big\{
\Sigma^*_X q
\big|
X\in\mg
\big\}\subset A^*.
\label{tSpace}
\ee
Here $\Sigma^*_X q=0$ if and only if $X$ belongs to the Lie algebra
$\mg_q$ of the little group $G_q$, so
(\ref{tSpace}) is isomorphic to the coset space $\mg/\mg_q$. It
follows that the cotangent space\i{cotangent space} $T_q^*\cO_p$ at $q$ is the
annihilator of $\mg_q$ in $\mg^*$,\i{annihilator}
\be
T_q^*\cO_p
=
\mg_q^0
\equiv
\big\{
j\in\mg^*
\big|
\langle j,X\rangle=0\;\;\forall\,X\in\mg_q
\big\}
\subset\mg^*\,,
\label{cotaspa}
\ee
which provides the sought-for interpretation:

\paragraph{Lemma.} The cotangent space (\ref{cotaspa}) coincides with the set (\ref{awedgeq}): 
\be
T_q^*\cO_p
=
\mg_q^0
=
A\times q.
\label{inclus}
\ee

\begin{proof}
Let $X\in\mg$ be an infinitesimal rotation leaving $q$ invariant, i.e.\ $\Sigma^*_Xq=0$. One then has
$\langle\alpha\times q,X\rangle=0$ for any translation 
$\alpha$, so $\alpha\times q$ belongs to the annihilator $\mg_q^0$. By 
(\ref{cotaspa}) this implies that the span $A\times q$ is contained in $T^*_q\cO_p$. To prove 
(\ref{inclus}) we need to show the opposite inclusion, i.e.\ that any element of the annihilator $\mg_q^0$ 
can 
be written as $\alpha\times q$ for some $\alpha\in A$. To see this, consider the linear function
\be
\tau_q:
A\rightarrow\mg_q^0:
\alpha\mapsto\alpha\times q
\label{toki}
\ee
mapping a translation on the associated orbital angular momentum. The rank of this map is 
$\text{dim}(A)-\text{dim}[\text{Ker}(\tau_q)]$, where
\be
\text{Ker}(\tau_q)
=
\big\{
\alpha\in A
\big|
\langle\Sigma^*_Xq,\alpha\rangle=0\;\forall\;[X]\in\mg/\mg_q
\big\}.
\label{quak}
\ee
The elements of this kernel are translations constrained by $\text{dim}(\mg)-\text{dim}(\mg_q)$ independent 
conditions (the subtraction of $\text{dim}(\mg_q)$ comes from the quotient by $\mg_q$).
This implies that
$\text{dim}[\text{Ker}(\tau_q)]
=
\text{dim}(A)-\text{dim}(\mg)+\text{dim}(\mg_q)$,
from which we conclude that the rank of $\tau_q$ is
\be
\text{dim}[\text{Im}(\tau_q)]
=
\text{dim}(\mg)
-
\text{dim}(\mg_q)
=
\text{dim}(\mg_q^0).
\nn
\ee
It follows that $\tau_q$ is surjective, which was to be proven.
\end{proof}

We have just shown that the span (\ref{awedgeq}) at each $q\in\cO_p$ is the cotangent space of 
$\cO_p$ at $q$. Since $j=0$, this analysis exhausts all points of $\cW_{(0,p)}$ and we conclude 
that\i{coadjoint orbit!as cotangent bundle}\i{momentum!orbit}\i{cotangent bundle!of 
momentum orbit}
\be
\begin{array}{c}
\text{\it{the scalar coadjoint orbits of $G\ltimes A$}}\\
\text{\it{are cotangent bundles of momentum orbits.}}
\end{array}
\label{scacoga}
\ee
In mathematical terms we would write $\cW_{(0,p)}
=
T^*\cO_p$.
In particular, if we have classified all momentum orbits of $G\ltimes A$, then we already know the 
classification 
of all scalar coadjoint orbits $\cW_{(0,p)}$. Note that the map (\ref{toki}) allows us to express the 
stabilizer of $(0,p)$ in a compact way: it is a semi-direct product
\be
\text{Stabilizer of }(j,p)\text{ }
=
\text{ }
G_p\ltimes\text{Ker}(\tau_p)
\label{stabisca}
\ee
where $G_p$ is the little group of $p$.

\subsubsection*{Spinning orbits}

We now turn to spinning orbits, which generally contain no point with vanishing total angular 
momentum.\i{coadjoint orbit!spinning} To begin, we pick a 
coadjoint vector $(j,p)$ and 
restrict our attention to rotations $f$ that belong to the little group $G_p$. The resulting span is
\be
\Big\{
\left(\Ad^*_f j+\alpha\times p,p\right)
\Big|
f\in G_p,\,\alpha\in A
\Big\}
\label{blah}
\ee
and is a subset of the full orbit (\ref{cocobi}). In general 
$\Ad^*_f(j)\neq j$ because the little group $G_p$ need not be included in the stabilizer of $j$ for the 
coadjoint action
of $G$. Noting that the cross product (\ref{copo}) satisfies the property $\Ad^*_f(\alpha\times p)
=
\sigma_f \alpha \times\sigma^*_f p$, and using the fact that $f$ fixes $p$, we rewrite (\ref{blah}) as
\be
\big\{
\left(
\Ad^*_f\left(j+\beta\times p\right),p
\right)
\big|
f\in G_p,\;\beta\in A
\big\}
\label{blahBis}
\ee
where $\beta$ is related to the $\alpha$ of (\ref{blah}) by $\beta=\sigma_{f^{-1}}\alpha$. In particular we 
have
\be
\text{Stabilizer of }(j,p)\text{ }
=
\text{ }
(G_j\cap G_p)
\ltimes\text{Ker}(\tau_p)
\label{stabispi}
\ee
where $G_j$ is the stabilizer of $j$ for the coadjoint action of $G$ and all the remaining notation is as 
before. This extends (\ref{stabisca}) to the case $j\neq 0$.\\

The rewriting (\ref{blahBis}) allows us to see that translations along $\beta$ can modify 
at will all 
components of $j$ that point along directions in the annihilator $\mg_p^0$. The only piece of $j$
that is left unchanged by the action of translations is its restriction to $\mg_p$,\i{classical 
spin}\i{spin!classical}
\be
j\big|_{\mathfrak{g}_p}\equiv j_p\,.
\label{jeress}
\ee
Accordingly the set (\ref{blahBis}) is diffeomorphic to a product
\be
\underbrace{\left\{\Ad^*_f j_p|f\in G_p\right\}}_{\displaystyle\cW_{j_p}}
\times
\underbrace{\left\{\alpha\times p|\alpha\in A\right\}}_{\displaystyle T_p^*\cO_p}\,,
\quad\quad
\label{blahTris}
\ee
where we recognize the cotangent space (\ref{inclus}) and
where $\cW_{j_p}\subset \mg_p^*$ denotes the coadjoint orbit of
$j_p\in\mg_p^*$ under the little group $G_p$. This is in fact our main conclusion:
when $\cW_{(j,p)}$ is seen as a fibre bundle over $\cO_p$, the fibre
at $p$ is a product (\ref{blahTris}) of the cotangent space of
$\cO_p$ at $p$ with the coadjoint orbit of the projection $j_p$ of
$j$ under the action of the little group of $p$.\\

Inspecting (\ref{blahTris}), note in particular how the little group orbit $\cW_{j_p}$ 
factorizes from the cotangent space $A\times p$ due to translations. This splitting is reminiscent of the 
representation (\ref{rspin}) of $G_p\ltimes A$, where the operators representing $f\in G_p$ and $\alpha\in A$ 
live on very different footings (and actually commute). Recall that we used this representation to induce an 
irreducible representation (\ref{texas}) of the full group $G\ltimes A$. What we see in (\ref{blahTris}) is 
the classical analogue of this little group representation; upon quantization, the sub-orbit (\ref{blahTris}) 
will precisely produce a representation of the form (\ref{rspin}), and its extension to the full orbit 
$\cW_{(j,p)}$ will correspond to the induction (\ref{texas}). In particular the projection (\ref{jeress}) is 
a classical definition of spin.\i{classical spin}\i{spin!classical} We shall return to this below.\\

The arguments that led from (\ref{blah}) to the result (\ref{blahTris}) can be run at 
any 
other point $q$ on $\cO_p$, except that the little group is $G_q$ instead of $G_p$. Thus the fibre
above any point $q=\sigma^*_f p \in\cO_p$ is a product of the cotangent space of $\cO_p$ at $q$ with the 
$G_q$-coadjoint orbit $\cW_{(\Ad^*_fj)_q}$, where $(\Ad^*_fj)_q$ denotes the restriction
of $\Ad^*_fj$ to $\mg_q$. But little groups at different points of $\cO_p$ are isomorphic: if one 
chooses standard boosts $g_q\in G$ such that $\sigma^*_{g_q}(p)=q$, then $G_q=g_q\cdot G_p\cdot g_q^{-1}$
and $\mg_q=\Ad_{g_q}\mg_p$. Therefore $\cW_{(\Ad^*_fj)_q}$ is diffeomorphic to $\cW_{j_p}$ for any 
$q=\sigma^*_f p\in\cO_p$; the relation between the fibres above $q$ and $p$ is given by the
coadjoint action of $G\ltimes A$.

\subsubsection*{Classification of coadjoint orbits}

The conclusions of the previous paragraph can be used to classify the orbits of $G\ltimes A$. We start with 
some terminology:

\paragraph{Definition.} Let $(j,p)$ be a coadjoint vector of the semi-direct product $G\ltimes_{\sigma}A$. 
The corresponding \it{bundle of little group orbits} is\i{bundle of little group orbits}
\be
\cB_{(j,p)}
\equiv
\left\{
\Big(
(\Ad^*_f j)_{\sigma^*_f p},\sigma^*_f p
\Big)
\Big|f\in G
\right\}. 
\label{bundleLGOBis}
\ee
\vspace{.1cm}

According to our earlier observations, the bundle of little group orbits associated with $(j,p)$ is really 
the same as the coadjoint orbit $\cW_{(j,p)}$, except that the cotangent spaces at each point of $\cO_p$ 
are ``neglected'' since translations do not appear in (\ref{bundleLGOBis}). Thus $\cB_{(j,p)}$ is a fibre 
bundle over $\cO_p$, the fibre $F_q$ at $q\in\cO_p$ being a coadjoint orbit of the little 
group $G_q$. The relation between fibres at different points of $\cO_p$ is given by the coadjoint action of 
$G\ltimes A$, or explicitly
\be
(k,q)\in F_q
\quad\text{iff}\quad
\exists\,f\in G\text{ such that }
k=\left(\Ad^*_f j\right)_q\text{ and }q=\sigma^*_f p.
\nn
\ee
Conversely, suppose that two elements $p\in A^*$ and $j_0\in\mg_p^*$
are given. The group $G$ can be seen as a principal $G_p$-bundle over
$\cO_p$, equipped with a natural $G_p$-action by multiplication from
the left in each fibre. In addition $G_p$ acts on the coadjoint orbit
$\cW_{j_0}$, so one can define an action of $G_p$ on
$G\times\cW_{j_0}$ by
\be
(f,k)\in G\times\cW_{j_0}
\stackrel{g\in G_p}{\longmapsto}
\left(g\cdot f,\Ad^*_g(k)\right).
\nn
\ee
The corresponding bundle of little group orbits is
defined as the associated bundle
\be
\cB_{(j_0,p)}\equiv\left(G\times\cW_{j_0}\right)/G_p.
\label{bundleLGO}
\ee
Thus one can associate a bundle of
little group orbits (\ref{bundleLGOBis}) with each coadjoint orbit of $G\ltimes A$; conversely, starting 
from any bundle
of little group orbits as defined in (\ref{bundleLGO}), one can build
a coadjoint orbit of $G\ltimes A$ by choosing any $j\in\mg^*$ such that
$j_p=j_0$ and taking the orbit $\cW_{(j,p)}$. In other words the
classification of coadjoint orbits of $G\ltimes A$ is equivalent to the
classification of bundles of little group orbits \cite{Rawnsley1975,Baguis1998}.\\

These arguments yield the complete picture of coadjoint orbits of $G\ltimes A$:\i{semi-direct 
product!coadjoint orbit}\i{coadjoint orbit!of semi-direct product}
\be
\begin{array}{c}
\text{\it{the coadjoint orbit $\cW_{(j,p)}$ is a fibre bundle over $\cO_p$, where}}\\
\text{\it{the fibre at $q\in\cO_p$ is a product of the cotangent space $T^*_q\cO_p$}}\\
\text{\it{with a coadjoint orbit of the little group $G_q$.}}
\end{array}
\label{gecoga}
\ee
Equivalently,
$\cW_{(j,p)}$ is a fibre bundle over the cotangent bundle $T^*\cO_p$, the fibre above
$(q,\alpha\times q)\in T^*\cO_p$ being a coadjoint orbit of
$G_q$. To exhaust all coadjoint orbits of $G\ltimes A$, one can proceed as
follows:\i{classification!of coadjoint orbits}\i{coadjoint orbit!classification}
\begin{enumerate}
\item Pick an element $p\in A^*$ and compute its momentum orbit $\cO_p$ under
  the action $\sigma^*$ of $G$; let $G_p$ be the corresponding little group.
\item Pick $j_p\in\mg_p^*$ and compute its coadjoint orbit under the
  action of $G_p$.
\end{enumerate}
The set of all orbits $\cO_p$ and of all coadjoint orbits of the
corresponding little groups classifies the coadjoint orbits of
$G\ltimes A$. Put differently, suppose one has classified the following
objects:
\begin{enumerate}
\item The orbits of $G$ for the action $\sigma^*$, with an exhaustive set of orbit
  representatives $p_{\lambda}\in A^*$ and corresponding little groups
  $G_{\lambda}$, with $\lambda\in\cI$ some index such that
  $\cO_{p_{\lambda}}$ and $\cO_{p_{\lambda'}}$ are disjoint
  whenever $\lambda\neq\lambda'$;
\item The coadjoint orbits of each $G_{\lambda}$, with an exhaustive set of orbit
  representatives $j_{\lambda,\mu}\in\mg_{\lambda}^*$,
  $\mu\in\cJ_{\lambda}$ being some index such that
  $\cW_{j_{\lambda,\mu}}$ and $\cW_{j_{\lambda,\mu'}}$ are
  disjoint whenever $\mu\neq\mu'$.
\end{enumerate}
Then the set
\be
\left.
\big\{
\left(
j_{\lambda,\mu},p_{\lambda}
\right)
\right|
\lambda\in\cI,\mu\in\cJ_{\lambda}
\big\}
\subset
\mg^*\oplus A^*
\label{exor}
\ee
is an exhaustive set of orbit representatives for the coadjoint orbits of $G\ltimes A$. The 
(generally continuous) indices $\lambda,\mu$ label the orbits uniquely. This algorithm is a classical 
analogue of the classification of representations described in section \ref{sesemi}, since it classifies the 
phase spaces of all ``particles'' associated with $G\ltimes A$.

\subsection{Geometric quantization and particles}
\label{suseqipa}

We now describe the quantization of coadjoint orbits of semi-direct products and argue that it yields
Hilbert spaces of one-particle states as described in chapter \ref{c2bis}.

\subsubsection*{A remark on cotangent bundles}

Before studying quantization we briefly digress on cotangent bundles and their canonical symplectic form 
$\omega=-d\theta$, where $\theta$ is the Liouville one-form (\ref{lifo}). Our goal is to rewrite the 
symplectic form on $T^*\cQ$ in a simpler way. For 
a 
sufficiently small open neighbourhood $U$ of $q\in\cQ$, the preimage
$\pi^{-1}(U)$ is diffeomorphic to the product $U\times T^*_q\cQ$. Hence the tangent 
space $T_{(q,\alpha)}T^*\cQ$ can be written as a direct sum\i{cotangent bundle!tangent space}
\be
T_{(q,\alpha)}T^*\cQ
\;\cong\;
T_q\cQ\oplus T_{\alpha}T^*_q\cQ
\;\cong\;
T_q\cQ\oplus T^*_q\cQ\,
\label{tangentDec}
\ee
which justifies writing its elements as $\cV=(v,\beta)$, where $v\in T_q\cQ$ and 
$\beta\in T^*_q\cQ$. The differential of (\ref{ProCC}) at $(q,\alpha)$ then reads
$\pi_{*(q,\alpha)}(v,\beta)=v$ 
and 
the Liouville 
one-form (\ref{lifo}) reduces to\i{Liouville one-form}
\be
\theta_{(q,\alpha)}(v,\beta)=\langle\alpha,v\rangle.
\label{LiouvilleBis}
\ee
Accordingly one finds that the canonical symplectic form $\omega=-d\theta$ is\i{canonical symplectic form}
\be
\omega_{(q,\alpha)}
\left((v,\beta),(w,\gamma)\right)
=
\langle\gamma,v\rangle
-\langle\beta,w\rangle\,
\label{LiouvilleTris}
\ee
which is just a more intrinsic rewriting of the standard $\omega=dq\wedge dp$. As we now show, this 
reformulation is useful for coadjoint orbits of semi-direct products.

\subsubsection*{Quantization}

Suppose we wish to quantize a coadjoint orbit $\cW_{(j,p)}$ of $G\ltimes A$; let 
$\omega$ be its Kirillov-Kostant symplectic form (\ref{kksym}). Since the Lie bracket in 
$\mg\inplus A$ is (\ref{semibra}), the symplectic form evaluated at the point 
$\left(\Ad^*_fj+\alpha\times q,q\right)\equiv (\kappa,q)$ in $\cW_{(j,p)}$ reads\i{semi-direct 
product!Kirillov-Kostant form}\i{Kirillov-Kostant symplectic form!for semi-direct product}
\be
\begin{split}
& \omega_{\left(\kappa,q\right)}
\big(
\ad^*_{(X,\beta)}(\kappa,q),\ad^*_{(Y,\gamma)}(\kappa,q)
\big)=\\
& =
\left<
\Ad^*_fj,[X,Y]
\right>
+\left<
\alpha\times q,[X,Y]
\right>
+
\left<
\gamma\times q,X
\right>
-
\left<\beta\times q,Y\right>
\,.
\end{split}
\label{omegaH}
\ee
In the two last terms of this expression we recognize the Liouville symplectic form (\ref{LiouvilleTris}) 
on the cotangent 
bundle $T^*\cO_p$
when $\alpha\times q$ is seen as an element of $T^*_q\cO_p$ thanks to (\ref{inclus}). On the other 
hand the first term of 
(\ref{omegaH}) looks like the natural symplectic form (\ref{kksym}) on the $G$-coadjoint orbit of $j$. 
In 
particular, when $X$ and $Y$ belong to the Lie algebra $\mg_q$ of the little group at $q$, the second term in 
(\ref{omegaH}) vanishes and the first one reduces to
\be
\bra
\Ad^*_fj,[X,Y]
\ket
=
\big<
(\Ad^*_fj)_q,[X,Y]
\big>
\nn
\ee
where we use the notation (\ref{jeress}). This is the natural symplectic form on the coadjoint orbit 
$\cW_{(\Ad^*_fj)_q}$, so if 
we see $\cW_{(j,p)}$ as a fibre bundle over $T^*\cO_p$ with typical fibre
$\cW_{j_p}$, restricting the symplectic form (\ref{omegaH}) to a fibre gives back the 
symplectic form on the little group's coadjoint orbit. This observation actually follows from a more general 
result, which states that the coadjoint 
orbits of a semi-direct product are obtained by \it{symplectic induction}\i{symplectic induction} from the 
coadjoint orbits of its 
little 
groups.\i{symplectic induction} 
Symplectic induction is the classical analogue of the method of induced representations that yields 
irreducible unitary representations of semi-direct products.
We will not dwell on the details of this construction 
and refer e.g.\ to \cite{Baguis1998,Duval:1991jn} for a much more 
thorough treatment.\\

For quantization to be 
possible, the symplectic form (\ref{omegaH}) must be integral in the sense (\ref{intom}). 
But the Liouville two-form (\ref{LiouvilleTris}) is exact, so its de 
Rham cohomology class vanishes and demanding that (\ref{omegaH}) be 
integral reduces to demanding integrality of the symplectic form on the coadjoint orbit 
of 
the little group. We conclude (see e.g.\ \cite{Li1993} for the proof):\i{semi-direct product!quantization 
conditions}

\paragraph{Theorem.} Let $G\ltimes A$ be a semi-direct product, $(j,p)$ a coadjoint vector with coadjoint 
orbit $\cW_{(j,p)}$. Then $\cW_{(j,p)}$ is prequantizable if and only if
the corresponding $G_p$-coadjoint orbit $\cW_{j_p}$ is prequantizable.\\

Provided the little group orbit $\cW_{j_p}$ is quantizable, one obtains 
a unitary representation $\cR$ of the little group $G_p$ acting on polarized sections of a line bundle 
over $\cW_{j_p}$. These sections are spin states; as in section \ref{sesemi}, 
we denote their 
Hilbert space by $\cE$. Upon declaring that the polarized sections on $T^*\cO_p$ depend only on the 
coordinates of the momentum orbit $\cO_p$, polarized sections on the whole orbit $\cW_{(j,p)}$ can be seen as 
$\cE$-valued wavefunctions in momentum space. Assuming that there exists a 
quasi-invariant measure $\mu$ on $\cO_p$, the Hilbert space $\sH$ obtained by quantizing $\cW_{(j,p)}$ 
becomes a tensor product (\ref{zissou}) of $\cE$ with the space of square-integrable functions 
$\cO_p\rightarrow\CC$. This exactly reproduces the construction of section \ref{sesemi}.

\subsubsection*{Recovering induced representations}

As the last step of quantization, we now need to understand how the group $G\ltimes A$ acts on polarized 
sections, or equivalently what differential operators represent the Lie algebra $\mg\inplus A$ on sections. 
Recall that these operators take the general form (\ref{Jinx}) where $\cJ$ is a momentum map (\ref{moment}) 
while 
$\xi_X$ is an infinitesimal generator (\ref{infige}) for the Lie algebra element $X$. In the present case $X$ 
is replaced by a pair $(X,\alpha)\in\mg\inplus A$. Furthermore, since the phase space is a coadjoint orbit, 
the momentum map is an inclusion (\ref{INK}) and the infinitesimal generator is 
$\xi_{(X,\alpha)}=\ad^*_{(X,\alpha)}$.\\

Let us describe this in more detail in the scalar case $j=0$, so that $\cW_{(0,p)}=T^*\cO_p$. Then the 
Kirillov-Kostant symplectic form coincides with the canonical symplectic form on $T^*\cO_p$ and the operator 
(\ref{Jinx}) representing a Lie algebra element $(X,\alpha)$ is
\be
\hat\cJ_{(X,\alpha)}\Big|_{(\beta\times q,q)}
=
-i\hbar\,\ad^*_{(X,\alpha)}(\beta\times q,q)+\langle q,\alpha\rangle
\nn
\ee
when evaluated at a point $(\beta\times q,q)$ belonging to $T^*\cO_p$. Polarized sections are functions 
$\Psi:\cW_{(0,p)}\rightarrow\CC:(\beta\times q,q)\mapsto\Psi(q)$ since they only depend on momenta 
$q\in\cO_p$. Upon acting on such a function the operator $\hat\cJ_{(X,\alpha)}$ yields
\be
\hat\cJ_{(X,\alpha)}\cdot\Psi(q)
=
-i\hbar\,(\Sigma^*_Xq)\cdot\Psi+\langle q,\alpha\rangle\Psi(q)
\label{AhhO}
\ee
where $\Sigma^*_Xq\in T_q\cO_p$ acts on $\Psi$ according to
\be
(\Sigma^*_Xq)\cdot\Psi
\equiv
-\frac{d}{dt}\Psi(\sigma^*_{e^{-tX}}q)\Big|_{t=0}\,.
\nn
\ee
Thus all observables $\hat\cJ_{(X,\alpha)}$ are polarized and can be quantized so as to satisfy 
(\ref{haf}).\\

Formula (\ref{AhhO}) describes the action of Hermitian operators $\hat\cJ_{(X,\alpha)}$ on 
wavefunctions $\Psi:\cO_p\rightarrow\CC$, provided the measure $\mu$ on $\cO_p$ is invariant under $G$. It 
can 
be rewritten as
\be
\Big(\hat\cJ_{(X,\alpha)}\cdot\Psi\Big)(q)
=
i\hbar\frac{d}{dt}
\Big[
e^{-i\langle q,t\alpha\rangle/\hbar}\,\Psi(e^{-tX}\cdot q)
\Big]
\Big|_{t=0}
\nn
\ee
and thus corresponds by differentiation to the \it{finite} transformation law\i{particle!by geometric 
quantization}
\be
\big(\cT[(f,\alpha)]\Psi\big)(q)
=
e^{-i\langle q,\alpha\rangle/\hbar}\,\Psi(f^{-1}\cdot q)
\label{TaPsi}
\ee
where
the map $\cT$ is a representation of $G\ltimes A$ such that
\be
\cT\big[(e^{tX},t\alpha)\big]
=
\exp\left[-\frac{it}{\hbar}\hat\cJ_{(X,\alpha)}\right].
\nn
\ee
When the measure $\mu$ on $\cO_p$ defining the scalar product of wavefunctions is invariant under $G$, 
formula (\ref{TaPsi}) is a unitary representation of $G\ltimes A$ that coincides (up to a sign due to 
different conventions) with a scalar induced representation (\ref{spipa}). We have thus 
recovered induced representations by quantization! The argument can be generalized to spinning 
representations and to quasi-invariant measures \cite{Li1993,Duval:1991jn}, although we will not prove it 
here. Thus we conclude:

\paragraph{Theorem.} Let $G\ltimes A$ be a semi-direct product, $\cW_{(j,p)}$ one of its coadjoint 
orbits.\i{unitary representation!by quantization}\i{geometric quantization!and unitary 
reps}\i{representation!by quantization} 
Then the unitary representation of $G\ltimes A$ obtained by geometric quantization of $\cW_{(j,p)}$ is an 
induced representation of the form (\ref{spipa}) with momentum orbit $\cO_p$ and spin $j_p$.

\paragraph{Remark.} This theorem says nothing about the exhaustivity of the procedure: it does not guarantee 
that \it{all} induced representations can be obtained by quantization. In 
fact it is easy to work out explicit examples where certain induced representations cannot follow from 
geometric quantization, for instance if the little 
group is not connected. In this sense geometric 
quantization is somewhat weaker than the full theory of induced representations exposed in section 
\ref{sesemi}.

\subsection{World lines}
\label{susegaborka}

Geometric actions for semi-direct products can be obtained following the 
general method described in section \ref{gigglema}. As we now show they can be interpreted as world line 
actions describing the motion of a point particle (generally with spin) in ``space-time'' $A$. We will rely 
on the Sigma model picture (\ref{actisi}).\\

We start by evaluating the left Maurer-Cartan form (\ref{lemaca}) for a semi-direct product with 
multiplication (\ref{semiop}). In order to describe a vector tangent to $G\ltimes A$ at the point 
$(f,\alpha)$, consider a path in $G\ltimes A$ given by
\be
\gamma(t)=\big(g(t),\beta(t)\big)
\label{goubila}
\ee
with $g(0)=f$, $\beta(0)=\alpha$ and $\dot\gamma(0)\equiv v$. Using the group operation (\ref{semiop}) in 
$G\ltimes A$, we then find\i{semi-direct product!Maurer-Cartan form}\i{Maurer-Cartan form!for semi-direct 
product}
\be
\Theta_{(f,\alpha)}(v)
\refeq{lemaca}
\frac{d}{dt}
\Big[
\big(
f^{-1}g(t),
\sigma_{f^{-1}}\beta(t)-\sigma_{f^{-1}}\alpha
\big)
\Big]
\Big|_{t=0}
=
\big(
\Theta_f\oplus\sigma_{f^{-1}}
\big)(v)\,,
\label{macapo}
\ee
where on the far right-hand side $\Theta$ denotes the Maurer-Cartan form on $G$. The direct sum refers to 
the fact that the tangent space $T_{(f,\alpha)}(G\ltimes A)$ is isomorphic to $T_fG\oplus A$. Using 
(\ref{macapo}) we can now write the Sigma model action (\ref{actisi}) associated with the orbit of a 
coadjoint vector $(j,p)\in\mg^*\oplus A^*$:\i{world line}
\be
S[f(t),\alpha(t)]
\refeq{papa}
\int_0^Tdt\big< j,\Theta_{f(t)}(\dot f(t))\big>
+
\int_0^Tdt\bra\sigma^*_{f(t)}p,\dot\alpha(t)\ket.
\qquad
\label{actifruit}
\ee
This can be recast in intrinsic terms as\i{semi-direct product!geometric action functional}\i{geometric 
action functional!for semi-direct product}\i{sigma model}
\be
S[f(t),\alpha(t)]
=
\int_{f(t)}\bra j,\Theta\ket
+
\int_{(f(t),\alpha(t))}\bra\sigma^*_{~}p,d\alpha\ket
\label{sispi}
\ee
where $\bra\sigma^*_{~}p,d\alpha\ket$ is the one-form on $G\ltimes A$ that gives 
$\langle\sigma^*_fp,\beta\rangle$ when evaluated at $(f,\alpha)$ and acting on a vector $(v,\beta)$. Note 
that this is just the sum of the Sigma 
model action (\ref{actisi}) on $G$ with a purely kinetic scalar action functional
\be
S_{\text{scalar}}[f(t),\alpha(t)]
=
\int_{(f(t),\alpha(t))}\bra\sigma^*_{~}p,d\alpha\ket
\label{sisca}
\ee
describing a point particle propagating in $A$ along a path $\alpha(t)$ with momentum 
$q(t)=\sigma^*_{f(t)}p$. In 
particular the group 
$A$ is now interpreted as ``space-time''.\i{semi-direct product!and 
space-time} Expression (\ref{sisca}) also has a gauge symmetry with gauge group (\ref{stabisca}), and 
it is invariant under redefinitions of the time parameter. As in (\ref{geomami}), adding a 
Hamiltonian generally spoils reparameterization symmetry. In the example of the Poincar\'e group below the 
condition $p(t)\in\cO_p$ will be a 
constraint generating time reparameterizations. Note that this condition only applies 
to momenta $q(t)\in A^*$, 
while the position of the particle, $\alpha(t)\in A$, is completely unconstrained.

\section{Relativistic world lines}
\label{seReWo}

In this section we study coadjoint orbits of Poincar\'e groups and show that the corresponding geometric 
actions describe world lines of relativistic particles. At the end we also turn to Galilean world lines and 
show that the corresponding partition functions coincide with Bargmann characters.
These topics have been studied previously in a number of 
references. The papers \cite{Cushman2006,Hudon} deal with the classification 
problem (see also \cite{CushmanMass}); the books \cite{Balachandran1983,Duval:1991xg,souriau1997} describe 
particles in terms of quantization of Poincar\'e coadjoint orbits; finally the papers 
\cite{Teitelboim:1981ua,Henneaux:1982ma,Henneaux:1987cp,Alekseev:1988tj} describe the relation between 
world line actions and propagators of relativistic fields.

\subsection{Coadjoint orbits of Poincar\'e}

The classification of coadjoint orbits of the Poincar\'e group is
an application of the 
general algorithm described in section \ref{susecoo}:\i{Poincar\'e group!coadjoint orbits} all 
of them are fibre bundles over momentum orbits, the fibre 
being a coadjoint orbit of the corresponding little group.\i{coadjoint orbit!of Poincar\'e group} Since 
momentum orbits have been classified in 
section \ref{relagroup}, the classification of coadjoint orbits is straightforward. Quantizing any 
coadjoint orbit yields an irreducible, 
unitary representation of the Poincar\'e group, i.e.\ the Hilbert space of a relativistic particle.\\

As an example consider the (double cover of the) Poincar\'e 
group in three dimensions, (\ref{pisel}). Its momentum orbits 
coincide 
with $\SL$ coadjoint orbits, and the little groups are stabilizers of $\SL$ coadjoint vectors. 
All stabilizers are one-dimensional and Abelian, except for the trivial orbit whose little 
group is $\SL$. This implies that all Poincar\'e coadjoint orbits are cotangent bundles of momentum 
orbits, except in the case $p=0$ for which $\cW_{(j,0)}$ coincides 
with the coadjoint orbit of $j$ under $\SL$. The set of coadjoint orbit representatives for Poincar\'e can be 
obtained by following the algorithm 
outlined above (\ref{exor}).

\subsection{Scalar world lines}
\label{susepact}

Let us consider a massive scalar coadjoint orbit of the Poincar\'e group in space-time dimension $D$. We wish 
to work out the corresponding Sigma model action (\ref{sisca}). We refer to 
\cite{Teitelboim:1981ua,Henneaux:1982ma,Henneaux:1987cp,Alekseev:1988tj} for a similar approach and for 
spinning generalizations.\\

The action principle describing a scalar world line is (\ref{sisca}). We choose a basis $e_{\mu}$ of $\RR^D$ 
such that each translation can be written as $\alpha=\alpha^{\mu}e_{\mu}$. The dual basis consists of momenta 
$(e^{\mu})^*$ such that $\langle p,\alpha\rangle=p_{\mu}\alpha^{\mu}$ for $p=p_{\mu}(e^{\mu})^*$. The 
argument of the action functional (\ref{sisca}) is a path in $G\ltimes A$, which we denote 
$(f(\tau),x(\tau))$ in order to distinguish the time parameter $\tau$ along the world line from
the time coordinate $t=x^0$. With the coordinates $p_{\mu}$ just described we have 
$\big(\sigma^*_{f(\tau)}p\big){}_{\mu}\equiv p_{\mu}(\tau)$ for some orbit representative $p$, and the action 
becomes\i{world line}
\be
S[p(\tau),x(\tau)]
=
\int_0^T d\tau\,
p_{\mu}(\tau)\dot x^{\mu}(\tau)
\qquad
\text{with a constraint}
\qquad
p_{\mu}(\tau)p^{\mu}(\tau)=-M^2
\;\;\forall\,\tau,
\nn
\ee
where indices are raised and lowered using the Minkowski metric. The constraint accounts for the fact that 
momenta must belong to a massive orbit. It can be incorporated in the action thanks to a Lagrange multiplier 
$N(\tau)$:\i{Lagrange multiplier}
\be
S[p(\tau),x(\tau),N(\tau)]
=
\int_0^T d\tau
\Big[
p_{\mu}(\tau)\dot x^{\mu}(\tau)
-
N(\tau)\left(p_{\mu}(\tau)p^{\mu}(\tau)+M^2\right)
\Big].
\label{pactila}
\ee
The 
equations of motion enforce the constraint\i{mass shell}\i{constraint}
\be
\phi
\equiv
p_{\mu}p^{\mu}+M^2
=
0
\label{constraint}
\ee
and describe a point particle propagating in space-time with constant momentum:
\be
\dot p_{\mu}=0,
\qquad
\dot x^{\mu}=2Np^{\mu}.
\label{pakom}
\ee
Note how the non-trivial dynamics emerges from the fact that momenta span an orbit, even though we haven't 
included 
any Hamiltonian.\\

To rewrite (\ref{pactila}) in Lagrangian form, we use the second equation of motion in 
(\ref{pakom}) to express momenta in terms of velocities:
\be
p^{\mu}=\frac{\dot x^{\mu}}{2N}\,.
\label{pigoda}
\ee
Contracting this with $p_{\mu}$ and using the mass shell constraint (\ref{constraint}) then gives
\be
-M^2=\frac{\dot x^{\mu}\dot x_{\mu}}{4N^2}\,.
\label{pinam}
\ee
Since our goal is to describe a massive particle, its trajectory must be time-like so we 
require that $\dot x^{\mu}$ remains inside the light-cone at any time $\tau$, which gives $\dot x^{\mu}\dot 
x_{\mu}<0$. This implies that (\ref{pinam}) has two real solutions $N$; we choose the 
positive one,
\be
N=\frac{\sqrt{-\dot x^{\mu}\dot x_{\mu}}}{2M}\,.
\label{posiwan}
\ee
Together with (\ref{pigoda}) this defines an invertible Legendre transformation from the space of 
positions and velocities $\{(x^{\mu},\dot x^{\mu})\}$ to the space of 
positions and constrained momenta supplemented with a Lagrange multiplier,\i{Legendre transformation}
\be
\Big\{
(x^{\mu},p_{\mu},N)
\Big|
x\in\RR^D,\;p\in\RR^D\,\text{ such that }\,p^2=-M^2,\;N>0
\Big\}.
\nn
\ee
Upon expressing $p$ and $N$ in terms of $\dot x$ thanks to this correspondence, the Hamiltonian action 
(\ref{pactila}) can be rewritten as
\be
S[x(\tau),\dot x(\tau)]
=
-M\int_0^Td\tau\,\sqrt{-\dot x^{\mu}\dot x_{\mu}}.
\label{lagui}
\ee
This is an action functional describing the dynamics of a scalar relativistic particle, with the 
Lagrangian $-M\sqrt{-\eta_{\mu\nu}\dot x^{\mu}\dot x^{\nu}}$. We 
have thus recovered the metric structure of space-time from the coadjoint orbits of its isometry group.\\

One can also run the argument in reverse and recover the Hamiltonian action from the 
Lagrangian one. In doing so one discovers that the mass shell condition is a primary constraint 
generating time reparameterizations while $N(\tau)$ is a lapse function along the world line. The 
canonical 
Hamiltonian then reads $\cH=2N(p^2+M^2)$ and vanishes on the constraint surface, as usual for generally 
covariant systems.

\paragraph{Remark.}  Starting from the Hamiltonian action (\ref{pactila}), one can evaluate the associated 
transition amplitude as a path integral. This computation was performed in 
\cite{Teitelboim:1981ua,Henneaux:1982ma,Henneaux:1987cp,Alekseev:1988tj} and the result turns out to coincide 
with the Feynman propagator of a free scalar field with mass $M$. This observation is one of the starting 
points of the \it{world line formalism}\i{world line formalism} of quantum field theory 
\cite{Feynman1951,Strassler:1992zr,Strassler:1993km}, 
where
scattering amplitudes are reformulated in 
terms 
of point particles propagating in space-time.

\begin{advanced}
\subsection{Galilean world lines}
\label{seGaWo}
\end{advanced}

Here we study coadjoint orbits of the Bargmann group (\ref{galix}) and write down world line actions for 
scalar non-relativistic particles. We also show how these actions account for Bargmann characters.\\

\subsubsection*{Scalar world lines}

The classification of coadjoint orbits of the Bargmann group follows from the general 
considerations of section \ref{susecoo}, combined with the classification of momentum orbits and little 
groups described in section \ref{galisec}.\i{Bargmann group!coadjoint orbits} In what follows we study the 
geometric action associated with one such coadjoint orbit with mass $M>0$ and spin $j=0$. The 
orbit then is a cotangent bundle $T^*\cO_p$, where $\cO_p$ is a massive momentum orbit 
(\ref{parabol}). The corresponding representation describes a scalar non-relativistic particle.\\

In order to write down the action (\ref{sisca}) we use the same trick as in (\ref{pactila}) to express the 
integrand in 
components and absorb the constraint $p(\tau)\in\cO_p$ with a Lagrange multiplier $N(\tau)$. (The time 
parameter along the world line is once again denoted as $\tau$, in order to distinguish it from the time 
coordinate $t$.) Using the pairing (\ref{patem}) the world line action reads\i{non-relativistic 
particle}\i{Galilean particle}\i{world line!non-relativistic}
\be
S\big[x(\tau),t(\tau),p(\tau),E(\tau),N(\tau)\big]
=
\int_0^Td\tau
\left[
p_i\dot x^i
-
E\dot t
-N\left(\frac{p^2}{2M}-E\right)
\right]
\label{galacticon}
\ee
where $i=1,...,D-1$. In principle we should also include a time-dependent central term $\lambda(\tau)$ 
(recall the last entry of (\ref{extran})), but one readily verifies that its contribution to the action is a 
boundary term so we neglect it from now on. This being said, note that the presence of the central 
extension is crucial in giving rise to the constraint $E\approx p^2/2M$ obtained by varying $N$. The 
equations of motion obtained by varying $E$ give $N=\dot t$, so we can once more interpret $N(\tau)$ as a 
lapse function along the world line. Plugging the solution of the equations of motion of $N$ and $E$ 
into 
(\ref{galacticon}), we get
\be
S\big[x(\tau),t(\tau),p(\tau)\big]
=
\int_0^Td\tau
\left[
p_i\dot x^i
-
\frac{p^2}{2M}\dot t
\right],
\nn
\ee
which we recognize as the action of a free non-relativistic particle 
moving in $\RR^{D-1}$, written in a reparameterization-invariant way 
(see e.g.\ chapter 4 of \cite{Henneaux:1992ig}). By expressing 
the action as an integral over the ``real time'' $t=t(\tau)$, we find\i{free 
particle}
\be
S[x(t),p(t)]
=
\int_0^Tdt
\left[
p_i\dot x^i
-
\frac{p^2}{2M}
\right]
\label{galactis}
\ee
where the dot now denotes differentiation with respect to $t$.

\subsubsection*{Path integrals and characters}

From now on we take $D=3$ for simplicity. Our goal is to plug the action (\ref{galactis}) into a path 
integral so as to recover the Bargmann character (\ref{galimacha}) for $r=1$. Note that the steps leading 
from the original Hamiltonian action (\ref{galacticon}) to the 
quadratic action (\ref{galactis}) all go through in the path integral since they amount to integrating out 
variables on which the action depends linearly.\\

We wish to evaluate the rotating partition function of a massive Galilean 
particle,\i{partition function}
\be
Z(\beta,\theta)
=
\Tr
\left(
e^{-\beta\hat H+i\theta\hat J}
\right),
\label{galized}
\ee
where $\hat H=\hat p^2/2M$ is the Hamiltonian and $\hat J$ is the angular momentum operator\i{angular 
momentum}
\be
\hat J=\hat x^1\hat p_2-\hat x^2\hat p_1.
\label{galiang}
\ee
The trace (\ref{galized}) can be interpreted as the partition function of a free 
non-relativistic particle in a frame that rotates at imaginary angular velocity $i\theta/\beta$.\i{angular 
potential} There are at 
least two equivalent ways to evaluate it. The first is to compute a time-sliced 
path integral\i{path integral}\i{time slicing}
\be
Z(\beta,\theta)
=
\int\limits_{x(\beta)=x(0)}
\!\!\!\!\!\!\cD x\cD p\,
\exp
\left[
-\int_0^{\beta}d\tau
\left(
-ip_j\dot x^j+\frac{p^2}{2M}-i\theta(x^1p_2-x^2p_1)
\right)
\right]
\label{galipath}
\ee
where $\cD x\cD p$ is the standard path integral measure of quantum mechanics. In the argument of the 
exponential we recognize the 
Euclidean section of (\ref{galactis}) supplemented by a term proportional to $\theta J$. 
Expression (\ref{galipath}) may thus be seen as the canonical partition function (\ref{canoz}) of a system 
with effective Hamiltonian $\hat H_{\text{eff}}=\hat H-\frac{i\theta}{\beta}\hat J$.
The second way is to realize that the operator $\hat J$ generates rotations in 
the plane. Thus if we introduce a basis of states $|x^1,x^2\rangle$ localized at $(x^1,x^2)$, the trace 
(\ref{galized}) is a (finite-dimensional) integral
\be
Z(\beta,\theta)
=
\int_{\RR^2}dx^1dx^2
\big<
R_{\theta}\cdot(x^1,x^2)
\big|
e^{-\beta H}
\big|
x^1,x^2
\big>
\label{galuxip}
\ee
where $R_{\theta}\cdot(x^1,x^2)$ denotes the action of a rotation by $\theta$ on the vector $(x^1,x^2)$. From 
this second viewpoint, the partition function is a trace over transition 
amplitudes between initial and final states that are rotated with respect to each other. Since transition 
amplitudes can be written as path integrals, expression (\ref{galuxip}) is a path integral in disguise and 
takes the same form as (\ref{galipath}) up to two key 
differences: (i) the term $i\theta J$ no longer appears in the exponential, and (ii) the periodicity 
condition 
on paths is $x(\beta)=R_{\theta}\cdot x(0)$ instead of $x(\beta)=x(0)$.\\

The two methods just described give identical results, but we pick the second one for simplicity. Recall that 
the propagator of a free massive particle on a plane is (in Dirac 
notation)\i{propagator}
\be
\big< x',t\big|
e^{-iHt}
\big|x,0\big>
=
\frac{M}{2\pi it}
\exp\left[
\frac{iM|x'-x|^2}{2t}
\right]
\label{galipipi}
\ee
where $|\cdot|$ is the Euclidean norm. From this we find the Euclidean propagator\i{propagator}
\be
\big<R_{\theta}\cdot x,t\big|
e^{-\beta H}
\big|x,0\big>
=
\frac{M}{2\pi\beta}
\exp\left[-
\frac{M}{2\beta}(1-\cos\theta)x^2
\right]
\label{zigama}
\ee
where $x^2\equiv|x|^2$. To obtain the partition 
function (\ref{galuxip}) we integrate (\ref{zigama}):
\be
Z(\beta,\theta)
=
\int_{\RR^2}d^2x\,
\frac{M}{2\pi\beta}\,
\exp\left[-
\frac{M}{2\beta}(1-\cos\theta)x^2
\right].
\label{gagatin}
\ee
For $\theta\neq0$ (modulo $2\pi$) this is 
just a Gaussian integral and the result is precisely a character (\ref{galimacha}) with $r=1$.\i{Bargmann 
character} We conclude 
that the space obtained by quantizing a 
massive coadjoint orbit of the Bargmann group coincides with the Hilbert space of a free, massive, 
non-relativistic particle.

\paragraph{Remark.} Having seen the computation of the trace of $e^{-\beta H+i\theta J}$ in Bargmann 
representations, one may 
wonder if the result can be analytically continued to the 
grand canonical partition function\i{angular velocity}
\be
Z(\beta,\Omega)
=
\Tr\left(
e^{-\beta(H-\Omega J)}
\right)
\label{galizedo}
\ee
where $\Omega$ is a \it{real} angular velocity, describing
the 
thermodynamics of a system in a real rotating frame. This corresponds to taking $\theta=-i\beta\Omega$ purely 
imaginary in (\ref{galized}). If we were to evaluate (\ref{galizedo}), we would be led to 
expression 
(\ref{zigama}) with $1-\cos\theta=1-\cosh(\beta\Omega)<0$, which is a serious problem: the integral 
(\ref{gagatin}) would diverge. Intuitively this divergence is 
due to the fact that free particles move all over space without any potential that prevents them from 
escaping to infinity when put in a rotating frame. This divergence is typical of 
rotating characters in flat space and can also be seen in the Poincar\'e characters of section 
\ref{relagroup}. By contrast, the partition function (\ref{galizedo}) of a 
two-dimensional harmonic oscillator\i{harmonic oscillator} is well-defined as long as the angular velocity 
$\Omega$ is smaller than 
the oscillator's natural frequency.

\renewcommand{\afterpartskip}{}
\part*{Part II\\[.3cm]
Virasoro symmetry and AdS$_3$ gravity}
\addcontentsline{toc}{part}{II\ Virasoro symmetry and AdS$_3$ gravity}
\begin{center}
\begin{minipage}{.9\textwidth}
~\\
~\\
~\\
~\\
~\\
In this part we initiate the study of infinite-dimensional symmetry groups by analysing the group of 
diffeomorphisms of the circle, whose central extension is the Virasoro group. Upon defining the latter, we 
classify its coadjoint orbits, i.e.\ orbits of CFT stress tensors under conformal 
transformations in two dimensions. As an application we 
show how Virasoro symmetry is realized in asymptotically Anti-de Sitter gravity in three dimensions and 
interpret unitary representations of the Virasoro algebra from a gravitational perspective. Note that 
Virasoro coadjoint orbits will play a key role for BMS$_3$ particles in part III, as they will coincide with 
their supermomentum orbits.
\end{minipage}
\end{center}
\newpage
~
\thispagestyle{empty}

\chapter{The Virasoro group}
\label{c4}
\markboth{}{\small{\chaptername~\thechapter. The Virasoro group}}

In the first part of this thesis we have introduced some general tools for dealing with symmetries in quantum 
mechanics. Our goal is to eventually apply these tools to the BMS$_3$ group in three dimensions. Accordingly, 
in this chapter and the two next ones we address a necessary prerequisite for these considerations by 
studying the central extension of the group of diffeomorphisms of the 
circle, i.e.\ the Virasoro group. The latter is part of the asymptotic symmetry group of many 
gravitational systems, where it essentially consists 
of conformal transformations of celestial circles. It also accounts for the symmetries 
of two-dimensional conformal field theories and thus illuminates certain aspects of holography 
in general, and AdS$_3$/CFT$_2$ in particular.\\

A word of caution is in order at the outset regarding the interpretation of the Virasoro group from a 
gravitational viewpoint. While diffeomorphisms in general relativity are generally thought of as gauge 
redundancies, the group $\Diff$ that we shall study here should by no means be understood in that sense. On 
the contrary, it should be interpreted as a \it{global} space-time symmetry group on a par with $\SL$ or the 
Poincar\'e group. In fact, in the BMS$_3$ case, $\Diff$ will be an infinite-dimensional 
extension 
of the Lorentz group in three dimensions. Accordingly this chapter and the next one may be seen as a detailed 
investigation of a group that extends Lorentz symmetry in an infinite-dimensional way.\\

Our plan for this chapter is the following. In section \ref{sedifici} we define the group $\Diff$ 
of diffeomorphisms of the circle as an infinite-dimensional Lie group, and we describe its adjoint 
representation, its Lie algebra $\Vect$, and its coadjoint representation. Section \ref{sedicomo} is devoted 
to its cohomology; in particular we introduce the Gelfand-Fuks cocycle and its integral, the 
Bott-Thurston cocycle, which respectively define the Virasoro algebra and the Virasoro group. In section 
\ref{seschwa} we study the Schwarzian derivative, which will lead to a unified picture of Virasoro 
cohomology. 
Finally, in section \ref{sevigo} we 
define the Virasoro group and work out its adjoint and coadjoint representations; the latter coincides with 
the transformation law of two-dimensional CFT stress tensors under conformal transformations.\\

Regarding references, the holy book on the Virasoro group is \cite{guieu2007algebre} 
while \cite{khesin2008geometry} is a pedagogical introduction to infinite-dimensional group theory. Some 
familiarity with two-dimensional CFT may come in handy at this stage; see 
e.g.\ \cite{Ginsparg:1988ui,DiFrancesco:1997nk,blumenhagen2009introduction}.

\section{Diffeomorphisms of the circle}
\label{sedifici}

In this section we study the elementary properties of the group $\Diff$. We first briefly mention issues 
related to infinite-dimensional Lie groups, then define $\Diff$ and show that its Lie 
algebra consists of vector fields on the circle. We also introduce densities on the circle, i.e.\ primary 
fields, display the coadjoint representation of $\Diff$, and discuss certain properties of the exponential 
map.

\subsection{Infinite-dimensional Lie groups}
\label{sefidiligg}

The diffeomorphisms of any manifold depend on an infinity of parameters and therefore span an 
infinite-dimensional group. One would like this group to be smooth in a certain sense, which leads to the 
problem of defining infinite-dimensional Lie groups and manifolds. Here we review this question in broad 
terms; we refer e.g.\ to \cite{kriegl} for a much more complete presentation.\\

In the same way that any finite-dimensional manifold looks locally like 
$\RR^n$, one would like to find the prototypical infinite-dimensional topological vector space $\VV$ such 
that infinite-dimensional manifolds be locally homeomorphic to $\VV$. As it turns out, taking $\VV$ to be a 
\it{Fr\'echet space}\i{Fr\'echet space} leads to a well-defined theory of differentiation and smoothness, 
which can then be used to define Fr\'echet manifolds. Roughly speaking, Fr\'echet spaces are vector spaces 
that generalize Banach spaces. For example 
the space 
$C^{\infty}(\cM)$ of smooth functions on a finite-dimensional manifold $\cM$ is a Fr\'echet space (but 
\it{not} a Banach space). A \it{Lie-Fr\'echet group}\i{Lie-Fr\'echet group}\i{infinite-dimensional group} 
then is a group endowed with a 
structure of Fr\'echet manifold such that multiplication and inversion are smooth. For instance the group 
$\text{Diff}(\cM)$ of diffeomorphisms of a compact finite-dimensional manifold $\cM$ is a Lie-Fr\'echet 
group. 
From now on we refer to infinite-dimensional Lie-Fr\'echet groups simply as ``infinite-dimensional 
groups''.\\

Infinite-dimensional manifolds are strikingly different from finite-dimensional ones in many respects. For 
example the notion of ``tangent vectors'' is ambiguous in infinite dimension, and the lack of 
existence/uniqueness\i{existence and uniqueness} theorems makes other seemingly obvious definitions fail, 
such as the notion of integral 
curves. We will encounter a similarly counter-intuitive phenomenon below, when explaining that the 
exponential 
that maps vector fields on diffeomorphisms is not locally surjective, in contrast with its 
finite-dimensional counterpart.\\

In the remainder of this section we deal with the group of diffeomorphisms of the circle as an 
infinite-dimensional Lie(-Fr\'echet) group. In particular we will think of its Lie algebra as its tangent 
space at the identity, identified with the space of left-invariant vector fields, from which the remaining 
definitions will follow.

\subsection{The group of diffeomorphisms of the circle}
\label{susegapif}

We consider the unit circle\i{S1@$S^1$ (circle)} $S^1=\big\{e^{i\phii}\in\CC\big|\phii\in[0,2\pi[\big\}$. Its 
fundamental 
group is isomorphic to $\ZZ$ and its universal cover is the 
real line $\RR$, with a projection\i{S1@$S^1$ (circle)!universal cover}\i{universal cover!of S1@of $S^1$}
\be
\sfp:\RR\rightarrow S^1:\phii\mapsto e^{i\phii}
\label{prosifi}
\ee
depicted in fig.\ \ref{figunicov}.
This allows us to think of $S^1$ as the quotient $\RR/2\pi\ZZ$ of the real line by the equivalence relation 
$\phii\sim\phii+2\pi$, since the kernel of $\sfp$ consists of translations of $\RR$ by integer multiples 
of $2\pi$.

\subsubsection*{Diffeomorphisms in the complex plane}

A \it{diffeomorphism of the circle}\i{diffeomorphism}\i{DiffS1@$\Diff$} is a smooth bijection 
$F:S^1\rightarrow 
S^1$ whose inverse is also smooth. We 
denote the group of all such maps by $\Diff$, with the group operation given by 
composition:\i{composition}
\be
F\cdot G\equiv F\circ G
\qquad
\forall\,F,G\in\Diff.
\label{compogo}
\ee
$\Diff$ is an infinite-dimensional Lie group that inherits its smooth structure from that of the 
Fr\'echet manifold of smooth maps $S^1\rightarrow S^1$. Given an orientation on $S^1$, diffeomorphisms may 
preserve it or break it. In 
particular the set of diffeomorphisms that preserve orientation is a subgroup of $\Diff$, denoted $\Diffp$ 
and called the group of \it{orientation-preserving diffeomorphisms of the 
circle}.\i{orientation-preserving}\i{DiffS1@$\Diff$!orientation-preserving} We will prove below that $\Diffp$ 
is connected.\\

For practical purposes it is useful to describe diffeomorphisms of the circle in terms of the $2\pi$-periodic 
coordinate $\phii$ of (\ref{prosifi}). A diffeomorphism then is a map $F:e^{i\phii}\mapsto F(e^{i\phii})$ 
where $F(e^{i\phii})$ has unit 
norm. As an example one can verify that the set of transformations of the form
\be
F(e^{i\phii})
=
\frac{Ae^{i\phii}+B}{\bar Be^{i\phii}+\bar A}\,,
\qquad
|A|^2-|B|^2=1
\label{ProJJ}
\ee
is a subgroup of $\Diffp$ isomorphic to the connected Lorentz group in three dimensions, 
$\SO(2,1)^{\uparrow}\stackrel{\text{(\ref{isoso})}}{\cong}\text{PSL}(2,\RR)$. Rigid rotations 
are given by $A=e^{i\theta/2}$ and $B=0$:\i{rigid rotation}
\be
F(e^{i\phii})
=
e^{i(\phii+\theta)}
=
e^{i\theta}e^{i\phii}.
\label{firot}
\ee
Similarly, the typical orientation-changing diffeomorphism is the parity 
transformation\i{parity}
\be
F(e^{i\phii})=e^{-i\phii}.
\label{parici}
\ee
Any parity-changing 
diffeomorphism of the circle can be written as the composition of 
(\ref{parici}) with an orientation-preserving transformation. There appears to be no analogue of 
time-reversal 
in $\Diff$. All in all, $\Diffp$ is an infinite-dimensional cousin of the connected 
Lorentz group 
in three dimensions, $\SO(2,1)^{\uparrow}$, while $\Diff$ extends the orthochronous Lorentz group 
$\text{O}(2,1)^{\uparrow}$. See also fig.\ \ref{figoDiffA} below.

\subsubsection*{Diffeomorphisms in real coordinates}

Given a 
diffeomorphism $F:S^1\rightarrow S^1$, there exists a diffeomorphism 
$f:\RR\rightarrow\RR$ of the real line such that
\be
F(e^{i\phii})
=
e^{if(\phii)},
\qquad
\text{i.e.}
\qquad
F\circ\sfp=\sfp\circ f
\label{difipro}
\ee
in terms of the projection (\ref{prosifi}). In order for $f$ to be compatible with 
the 
periodicity of $\phii$, we must require that $f(\phii+2\pi)=f(\phii)\pm2\pi$, where the plus sign 
corresponds to an orientation-preserving diffeomorphism while the minus sign corresponds to an 
orientation-changing one. In this language the rotation (\ref{firot}) corresponds to 
$f(\phii)=\phii+\theta$ while the parity transformation (\ref{parici}) is $f(\phii)=-\phii$.

\paragraph{Definition.} A smooth map $f:\RR\rightarrow\RR$ is 
\it{$2\pi\ZZ$-equivariant} 
if\i{Z equivariance@$\ZZ$-equivariance}\i{2piZ equivariance@$2\pi\ZZ$-equivariance}\i{equivariant function}
$f(\phii+2\pi)
=
f(\phii)+2\pi$.
Any such map can be written as $f(\phii)=\phii+u(\phii)$, where $u$ is 
$2\pi$-periodic.\\

In these terms, any orientation-preserving diffeomorphism $F$ of the circle is a projection 
(\ref{difipro}) of a $2\pi\ZZ$-equivariant diffeomorphism $f$ of the real line, that is, a smooth 
function $f:\RR\rightarrow\RR$ such that
\be
\boxed{
\Big.
f'(\phii)>0\,,
\qquad
f(\phii+2\pi)=f(\phii)+2\pi
}
\label{fidif}
\ee
for any $\phii\in\RR$, where prime denotes differentiation with respect to $\phii$. The group 
operation (\ref{compogo}) then becomes
\be
f\cdot g=f\circ g
\label{compo}
\ee
where $f,g$ correspond to $F,G$ according to (\ref{difipro}). From now on we always describe $\Diff$ in terms 
of diffeomorphisms of $\RR $ satisfying the properties (\ref{fidif}). By the way, this is why we have kept 
writing group elements as ``$f\,$'' throughout this thesis.\\

Note that the diffeomorphism $F$ does not determine $f$ uniquely: one can add to 
$f(\phii)$ an arbitrary 
constant multiple of $2\pi$ without affecting $F=e^{if}$. This ambiguity can be removed 
by 
requiring e.g.\ that $f(0)$ belongs to the interval $[0,2\pi[$. One says that $f$ is a 
\it{lift}\i{lift (of diffeomorphism)} of $F$, and there are 
infinitely many lifts for a given $F$. For our purposes it is only important that giving $f$ determines 
$F=e^{if}$ uniquely, so that we can consistently write all orientation-preserving diffeomorphisms of the 
circle in the form 
(\ref{fidif}).

\subsection{Topology of $\Diff$}

\paragraph{Lemma.} The group $\Diffp$ of orientation-preserving diffeomorphisms is 
con\-nec\-ted,\i{DiffS1@$\Diff$!connected components} and $\Diff$ 
has two connected components related by parity.

\begin{proof}
Let $f(\phii)$ be a diffeomorphism of $\RR$ that satisfies (\ref{fidif}), and consider the corresponding 
diffeomorphism of the circle given by (\ref{difipro}). We wish to show that there exists a continuous path 
that connects $f$ to the identity. Consider therefore the one-parameter family of functions\i{homotopy} 
\be
f_t(\phii)=(1-t)f(\phii)+t\phii\,,
\qquad
t\in[0,1]\,.
\label{fitopy}
\ee
For each $t$, $f_t$ satisfies (\ref{fidif}) and 
therefore defines a diffeomorphism of the circle. At $t=0$ it coincides with $f$ while at $t=1$ it is the 
identity. See fig.\ \ref{figoTOP}.
\end{proof}

\begin{figure}[h]
\centering
\includegraphics[width=0.70\textwidth]{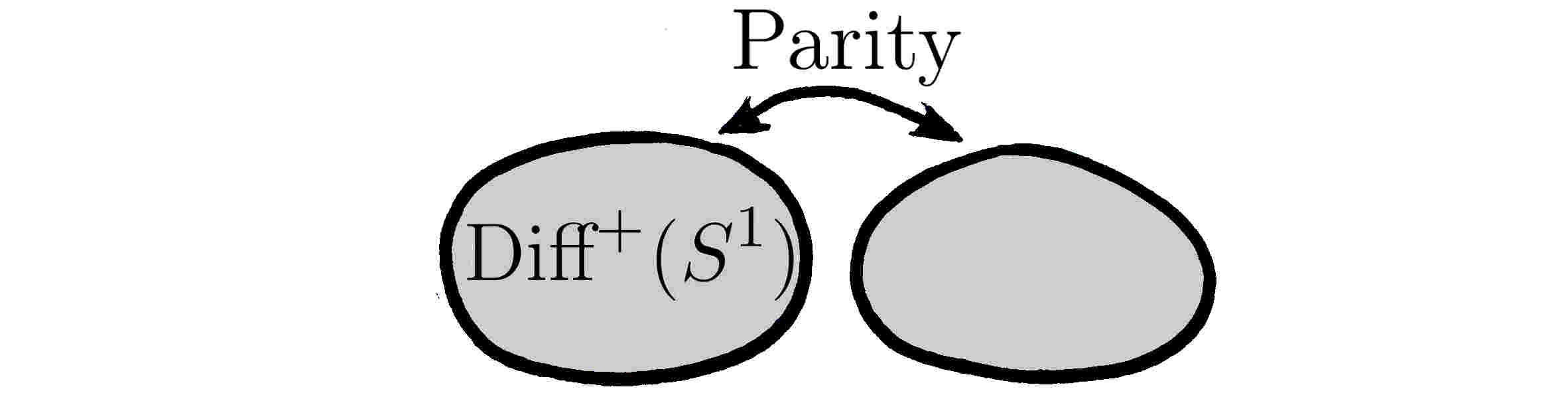}
\caption{The two connected components of $\Diff$ are related by parity. Compare with the connected 
components of the Lorentz group in fig.\ \ref{figoLor}.\label{figoDiffA}}
\end{figure}

For the purposes of representation theory it is important to know the fundamental group of $\Diffp$, as it 
determines whether $\Diffp$ has topological central extensions. In that context the key result is the 
following:

\paragraph{Lemma.} $\Diffp$ is homotopic to a circle, so its fundamental group 
is\i{DiffS1@$\Diff$!fundamental group}\i{fundamental group!of $\Diff$}\i{universal cover!of 
DiffS1@of $\Diff$}\i{DiffS1@$\Diff$!universal cover}
\be
\pi_1\left(
\Diffp
\right)
\cong\ZZ.
\label{pidifi}
\ee
Its 
universal 
cover $\Diffc$ is the group of $2\pi\ZZ$-equivariant diffeomorphisms of $\RR$, with the projection given by 
(\ref{difipro}).

\begin{proof}
We follow \cite{lurie}. The key to the proof is to realize that $\Diffp$ is homotopic to its subgroup 
$\text{Isom}^+(S^1)$ of orientation-preserving isometries of the circle (for the standard flat metric). Since 
$\text{Isom}^+(S^1)$ is a group $\un$ of rigid rotations, it will follow that $\Diffp$ has the homotopy type 
of a 
circle and therefore has a fundamental group $\ZZ$. So let us prove the homotopy equivalence 
$\Diffp\sim\text{Isom}^+(S^1)$. Call $\text{Diff}_0^+(S^1)$ the group of orientation-preserving 
diffeomorphisms of the circle leaving the point $\phii=0$ fixed. Since isometries of $S^1$ are 
rotations, there exists a decomposition
\be
\Diffp=\text{Diff}^+_0(S^1)\cdot\text{Isom}^+(S^1).
\label{tibid}
\ee
Indeed, any diffeomorphism of the circle is the 
composition of a rigid rotation with a diffeomorphism leaving $\phii=0$ fixed; both 
$\text{Diff}^+_0(S^1)$ and $\text{Isom}^+(S^1)$ are groups and their intersection only contains the 
identity. Now note that any diffeomorphism 
preserving $\phii=0$ admits a unique lift $f$ such that 
$f(0)=0$ and $f(2\pi)=2\pi$, so we can think of $\text{Diff}^+_0(S^1)$ as the set of $2\pi\ZZ$-equivariant 
diffeomorphisms of $\RR$ that fix the point $\phii=0$; this identification is one-to-one. It only remains to 
observe that the maps (\ref{fitopy}) define a homotopy whose effect at $t=1$ is to retract the whole 
$\text{Diff}_0^+(S^1)$ on the identity. As a result 
$\text{Diff}^+_0(S^1)$ is 
homotopic to a point, and so by (\ref{tibid}) $\Diffp$ is homotopic to a circle. Unwinding this circle 
gives rise to the group of
$2\pi\ZZ$-equivariant diffeomorphisms of $\RR$, which therefore span the universal cover of $\Diffp$.
\end{proof}

This lemma confirms the interpretation of $\Diffp$ as an infinite-dimensional analogue of $\PSL$, since the 
latter is also homotopic to a circle (see section \ref{sePoTri}). In particular formula (\ref{tibid}) 
is the $\Diff$ analogue of the Iwasawa decomposition (\ref{iwa}) of 
$\SL$. Since $\text{Diff}_0^+(S^1)$ has the homotopy type of a point, the group $\Diffp$ may be seen as an 
infinite-dimensional cylinder $S^1\times\RR^{\infty}$ where $S^1$ consists of rigid rotations while 
$\RR^{\infty}$ is spanned by infinite-dimensional generalizations of boosts. Note that property 
(\ref{pidifi}) implies the existence of topological projective representations of $\Diff$. Applied to $\BMS$, 
it 
will imply that the spin of massive particles is not quantized (as in the Poincar\'e group in three 
dimensions).\\

\begin{figure}[t]
\centering
\includegraphics[width=0.40\textwidth]{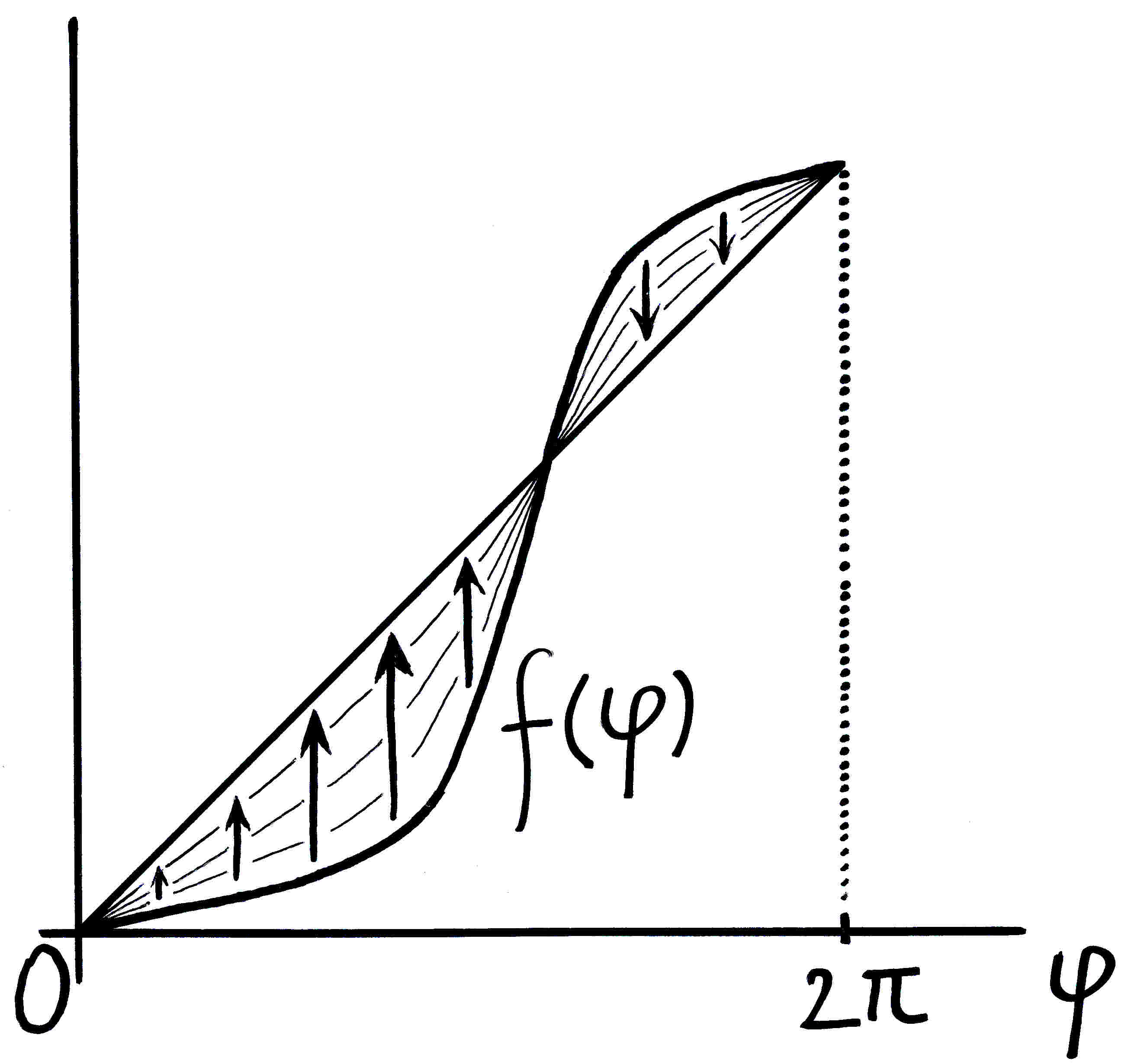}
\caption{The homotopy (\ref{fitopy}) turns a diffeomorphism $f(\phii)$ (here leaving the point $\phii=0$ 
fixed) into the identity. It implies both that the group $\Diffp$ is connected and that it is homotopic to a 
circle.\label{figoTOP}}
\end{figure}

In what follows we focus on the universal cover $\Diffc$ rather than $\Diff$ or $\Diffp$, except if 
explicitly stated otherwise. To reduce clutter we will abuse notation\label{abunot} by writing $\Diff$ for 
the universal cover, instead of the more accurate notation $\Diffc$. Accordingly, from now on elements 
of $\Diff$ are diffeomorphisms $f$, $g$, etc. of the real line satisfying the properties 
(\ref{fidif}). The inverse of $f$ will be denoted $f^{-1}$ and is such that 
$f(f^{-1}(\phii))=f^{-1}(f(\phii))=\phii$.

\subsection{Adjoint representation and vector fields}

We can now look for the Lie algebra of $\Diff$, which is identified with the 
tangent 
space at the identity and corresponds to infinitesimal diffeomorphisms. It is 
intuitively clear that this algebra is a space of functions, since a diffeomorphism close to the 
identity can be written as\i{diffeomorphism!infinitesimal}
\be
f(\phii)=\phii+\epsilon X(\phii)
\label{infidif}
\ee
where $\epsilon$ is ``small'' and $X(\phii)$ is a function on the circle. A more subtle problem is to 
determine 
the adjoint action of diffeomorphisms on this Lie algebra, and to deduce the expression of the Lie 
bracket. In order to work this out we pick a path $\gamma:\RR\rightarrow\Diff:t\mapsto\gamma_t$ such 
that $\gamma_0$ is the identity and
\be
\gamma_t(\phii)
=
\phii+tX(\phii)+\cO(t^2)
\label{gasite}
\ee
for small $t$. The adjoint representation is defined by (\ref{ad}), so we find
\be
\big(\Ad_f(X)\big)(\phii)
=
\frac{d}{dt}
\Big[
f\big(\gamma_t\left(f^{-1}(\phii)\right)\big)
\Big]\Big|_{t=0}
\refeq{gasite}
\frac{d}{dt}
\Big[
f\big(
f^{-1}(\phii)
+
tX\left(f^{-1}(\phii)\right)
\big)
\Big]\Big|_{t=0}\,.
\label{gastrique}
\ee
Since $t$ is ``small'' in this expression, we can Taylor expand
\be
f\left(
f^{-1}(\phii)
+
tX\left(f^{-1}(\phii)\right)
\right)
=
\phii
+
tX\left(f^{-1}(\phii)\right)f'\left(f^{-1}(\phii)\right)
+\cO(t^2)
\label{coti}
\ee
where we have used $f(f^{-1}(\phii))=\phii$. The derivative of the latter equation implies
\be
f'(f^{-1}(\phii))=\frac{1}{(f^{-1})'(\phii)}
\label{invide}
\ee
which can be plugged into (\ref{coti}) and thus provides the adjoint representation\i{DiffS1@$\Diff$!adjoint 
representation}\i{adjoint representation!of $\Diff$}
\be
\big(\Ad_f(X)\big)(\phii)
=
\frac{X\left(f^{-1}(\phii)\right)}{(f^{-1})'(\phii)}\,.
\label{advecis}
\ee
This formula is the transformation law of $X(\phii)$ under a diffeomorphism $\phii\mapsto f(\phii)$. It shows 
in particular that $X(\phii)$ in (\ref{infidif}) is \it{not} just a function on the circle, due to the 
derivative of $f^{-1}$ in (\ref{advecis}). Using 
(\ref{invide}), we can also rewrite it by evaluating the left-hand side at $f(\phii)$ 
rather than $\phii$:
\be
\big(\Ad_f(X)\big)(f(\phii))
=
f'(\phii)X(\phii)
\label{advec}
\ee
We recognize here the transformation law of the component $X(\phii)$ of a vector field \i{VectS1@$\Vect$}
\be
X(\phii)\frac{\der}{\der\phii}
\label{vectoci}
\ee
under a diffeomorphism $f$. We shall denote by $\Vect$
the space of smooth vector fields on $S^1$, whose elements will be written as $X$, $Y$, etc. We have just 
shown that $\Vect$ is the Lie algebra of $\Diff$; it now remains to find the Lie bracket.\\

Take once more a path $\gamma_t$ in $\Diff$ that 
satifies (\ref{gasite}). Picking a vector field $Y\in\Vect$ and a point $\phii\in[0,2\pi]$, let us evaluate
\be
\left(\ad_X(Y)\right)(\phii)
\refeq{adg}
\frac{d}{dt}
\big(
\Ad_{\gamma_t}(Y)
\big)
\big|_{t=0}(\phii)
\refeq{advecis}
\frac{d}{dt}\left.\left(
\frac{Y(\gamma_t^{-1}(\phii))}{(\gamma_t^{-1})'(\phii)}
\right)\right|_{t=0}\,.
\label{burundi}
\ee
Here (\ref{gasite}) implies $(\gamma_t^{-1})'(\phii)=1-t X'(\phii)$ as well as $Y(\gamma_t^{-1}(\phii))
=
Y(\phii)-tX(\phii)Y'(\phii)$ to first order in $t$. Plugging these expressions in (\ref{burundi}) we 
obtain $\ad_XY=-XY'+YX'$, where it is understood that both sides are evaluated at the same point $\phii$. 
We conclude that the Lie bracket of $\Vect$, seen as the Lie algebra of the 
group $\Diff$, is the \it{opposite} of the standard Lie bracket of vector fields:\i{Lie 
bracket!of vector fields}
\be
[X,Y]_{\text{Lie algebra}}
=
-[X,Y]_{\text{Vector fields}}.
\label{bakopo}
\ee
This is in fact a common phenomenon: as a consequence of (\ref{OPOBRA}),
the group 
$\text{Diff}(\cM)$ of diffeomorphisms of a (compact, finite-dimensional) manifold $\cM$ is a Lie-Fr\'echet 
group whose Lie algebra is the space $\text{Vect}(\cM)$ endowed with the 
\it{opposite} of the 
standard Lie bracket of vector fields \cite{abraham1978foundations,khesin2008geometry,guieu2007algebre}. Thus:

\paragraph{Proposition.} The Lie algebra of the group $\Diff$ is the space $\Vect$ of vector fields on the 
circle, with the Lie bracket (\ref{bakopo}) given by the opposite of the standard Lie bracket of vector 
fields.\\

In what follows we will bluntly neglect the sign subltety and 
endow $\Vect$ with the usual bracket
\be
[X,Y]=
\big(
X(\phii)Y'(\phii)-Y(\phii)X'(\phii)
\big)\frac{\der}{\der\phii}.
\label{libici}
\ee
This is a harmless abuse of conventions and may be seen as an alternative 
definition of the Lie bracket for groups of diffeomorphisms. With that abuse the Lie algebra of $\Diff$ 
becomes $\Vect$ with the \it{usual} Lie bracket of vector fields.

\subsubsection*{Witt algebra}

Since all functions on the circle can be expanded in Fourier series, any vector field is a (generally 
infinite) complex linear combination of generators
\be
\ell_m\equiv e^{im\phii}\der_{\phii}\,,
\qquad
m\in\ZZ.
\label{ellem}
\ee
The brackets (\ref{libici}) of these generators 
can be written as\i{Witt algebra}
\be
i[\ell_m,\ell_n]
=
(m-n)\ell_{m+n}\,,
\label{witt}
\ee
where one may recognize the \it{Witt algebra} of conformal field theory. It is an infinite-dimensional 
extension of the $\sl$ algebra (\ref{LaLLA}) spanned by $\ell_{-1},\ell_0,\ell_1$. The latter consists of 
vector fields $X(\phii)\der_{\phii}$ with
$X(\phii)=X_0+X_1\cos\phii+X_2\sin\phii$ for $X_{\mu}\in\RR$, and generates diffeomorphisms of the form 
(\ref{ProJJ}). In particular the constant vector field $\ell_0$ generates rigid rotations of the circle.

\subsection{Primary fields on the circle}
\label{susepifrakt}

Formula (\ref{advec}) gives the transformation law of vector fields on the circle under diffeomorphisms. 
Similarly, a one-form $\alpha(\phii)d\phii$ would transform as $\alpha\mapsto f\cdot\alpha$, 
where\i{one-form}\i{DiffS1@$\Diff$!action on one-forms}
\be
(f\cdot\alpha)(f(\phii))
=
\frac{\alpha(\phii)}{f'(\phii)}\,,
\label{tiform}
\ee
while a function $\alpha(\phii)$ would simply transform as 
$(f\cdot\alpha)(f(\phii))=\alpha(\phii)$. 
Vector fields, 
one-forms and functions can all be seen as sections of suitable vector bundles on the circle, which suggests 
that they can be generalized to sections of tensor product 
bundles such as $TS^1\otimes\cdots\otimes TS^1$ or $T^*S^1\otimes\cdots\otimes T^*S^1$.

\paragraph{Definition.} A \it{density} of weight $h\in\RR$ on the circle is an expression of the 
form\i{density on the circle}\i{weight}\i{conformal weight}\i{primary field}
\be
\alpha
=
\alpha(\phii)(d\phii)^h
\label{density}
\ee
where $\alpha(\phii)$ is a smooth function on the circle; it acts on the tangent space $T_{\phii}S^1$ 
according to 
$\bra\alpha(\phii)d\phii^h,V\der_{\phii}\ket\equiv\alpha(\phii)V^h$.\i{FhS1@$\cF_h(S^1)$ (space of 
densities)} We denote by 
$\cF_h(S^1)$ 
the vector space of densities of weight $h$.\\

When $h$ is an integer, a density of weight $h$ is a section of
\be
\underbrace{T^*S^1\otimes\cdots\otimes T^*S^1}_{|h|\text{ times}}
\;
\text{ if } h\geq 0,
\qquad\text{or}\qquad
\underbrace{TS^1\otimes\cdots\otimes TS^1}_{|h|\text{ times}}
\;
\text{ if } h<0.
\nn
\ee
In particular, a density is a vector field when $h=-1$, a one-form when $h=1$, and a function 
when $h=0$. The definition (\ref{density}) generalizes these notions to arbitrary real values 
of $h$. The notation $\alpha$ is justified by the fact that in the BMS$_3$ group, 
supertranslations will be densities with weight $-1$.\\

Expression (\ref{density}) suggests that the density $\alpha(\phii)(d\phii)^h$ is 
a coordinate-independent quantity. Indeed, under a diffeomorphism $f:\phii\mapsto f(\phii)$, it transforms as
\be
(f\cdot \alpha)(\phii)
\equiv
\left((f^{-1})'(\phii)\right)^h
\alpha(f^{-1}(\phii))
\label{tidensibis}
\ee
or equivalently as
\be
(f\cdot \alpha)(f(\phii))
\equiv
\frac{\alpha(\phii)}{(f'(\phii))^h}\,.
\label{tidensi}
\ee
This reduces to (\ref{advec}) for $h=-1$ and to (\ref{tiform}) for $h=1$. If we think of 
$f(\phii)$ as a ``conformal transformation'' of the circle, then eq.\ (\ref{tidensi}) coincides with the 
transformation law of a (chiral) primary field of weight $h$. It provides a representation of $\Diff$ in 
the space $\cF_h(S^1)$. This representation is infinite-dimensional and generally 
non-unitary because the would-be ``scalar product''
\be
\int_0^{2\pi}d\phii\,\alpha(\phii)\beta(\phii)\,,
\qquad
\alpha,\beta\in\cF_h(S^1)
\label{scaww}
\ee
is not left invariant by (\ref{tidensi}) for generic values of $h$. The only exception is the case of 
spinor fields, $h=1/2$. One can think of (\ref{tidensi}) as a 
$\Diff$ generalization of the various finite-dimensional (but non-unitary) irreducible representations of the 
Lorentz group. The number $h$ can then be thought of as a spin label, in the same way that finite-dimensional 
Lorentz representations correspond to transformation laws of relativistic fields with definite spin. (Beware: 
the word ``spin'' here does not refer to the notion of ``spin'' encountered in representations of semi-direct 
products. These two notions are related in that the Lorentz spin of a quantum field determines the Poincar\'e 
spin of the corresponding particles, but they are nevertheless different concepts.)\\

From (\ref{tidensi}) one can read off the transformation law of densities under infinitesimal 
diffeomorphisms, 
that is, under vector fields on the circle. Taking $f(\phii)=\phii+\epsilon X(\phii)$ in (\ref{tidensi}) one 
finds, to first order in $\epsilon$,
\be
(f\cdot\alpha)(\phii)
=
\alpha(\phii)-\epsilon
\big[
X(\phii)\alpha'(\phii)+h\,\alpha(\phii)X'(\phii)
\big].
\label{startingpoint}
\ee
We then define\i{infinitesimal transformation}
\be
X\cdot \alpha(\phii)
\equiv
-
\frac{(f\cdot \alpha)(\phii)-\alpha(\phii)}{\epsilon}
\label{infitransfo}
\ee
and obtain
\be
X\cdot\alpha
=
X\alpha'+h\,\alpha X'.
\label{infident}
\ee
As before, one may recognize here the infinitesimal transformation law of a primary field $\alpha$ of weight 
$h$ 
under an conformal transformation generated by $X$.

\subsection{Coadjoint representation of $\Diff$}

\subsubsection*{Dual spaces}

We mentioned around (\ref{scaww}) that the integral of the product of two densities with the same weight is 
generally
not invariant under diffeomorphisms. However, there does exist a $\Diff$-invariant 
pairing of densities. Indeed, consider the space $\cF_h(S^1)$ of densities of weight $h$. Its dual space 
consists of all linear 
forms\i{dual space}
\be
p:\cF_h(S^1)\rightarrow\RR:\alpha\mapsto\bra p,\alpha\ket.
\label{densidu}
\ee
Since $\cF_h(S^1)$ is infinite-dimensional, its dual space is pathological: the map (\ref{densidu}) need not 
be continuous and therefore does not preserve the differentiable structure of $\cF_h(S^1)$. Accordingly one 
generally restricts attention to the space of continuous 
linear forms (\ref{densidu}); the latter coincides with the space of distributions on 
the circle.\i{distribution} In addition, for concrete computations it is much more convenient to consider 
only \it{regular} 
distributions, that is, distributions that can be written in the form
\be
\bra p,\alpha\ket
=
\frac{1}{2\pi}
\int_0^{2\pi}d\phii\,p(\phii)\alpha(\phii)
\label{denpar}
\ee
where $p(\phii)$ is a smooth function on the circle. We 
will call the space of such distributions the \it{smooth} or \it{regular dual}\i{smooth dual}\i{regular 
dual} 
of $\cF_h(S^1)$. As a 
vector space, it is isomorphic to the space $C^{\infty}(S^1)$ of smooth functions on the circle. Note that 
any 
distribution can be obtained as the limit of a 
sequence of regular distributions, so in this sense we are not missing 
anything even when restricting attention to regular distributions. The regular dual of $\cF_h(S^1)$ will 
be denoted as $\cF_h(S^1)^*$. The notation in (\ref{densidu}) is justified by the fact that, in BMS$_3$, $p$ 
will be an infinite-dimensional supermomentum vector dual to supertranslations.\\

Since $\cF_h(S^1)$ carries a representation (\ref{tidensi}), it is natural to ask how 
the dual representation (\ref{sstar}) acts on the regular dual. By definition, one has $\bra f\cdot 
p,\alpha\ket=\bra 
p,f^{-1}\cdot\alpha\ket$ for any diffeomorphism $f$, any density $\alpha\in\cF_h(S^1)$ and any smooth 
distribution $p\in\cF_h(S^1)^*$. Using (\ref{tidensibis}) and the pairing (\ref{denpar}), we get
\be
\bra f\cdot p,\alpha\ket
=
\frac{1}{2\pi}
\int_0^{2\pi}d\phii\,
p(\phii)(f'(\phii))^h\alpha(f(\phii))\,.
\label{lukita}
\ee
If now we rewrite $\bra f\cdot p,\alpha\ket$ as an integral (\ref{denpar}) with the integration variable 
$\phii$ 
replaced by $f(\phii)$, the condition that $\bra f\cdot p,\alpha\ket$ matches the right-hand side of 
(\ref{lukita}) for any $\alpha$ readily provides
\be
(f\cdot p)(f(\phii))
=
\frac{p(\phii)}{(f'(\phii))^{1-h}}\,.
\nn
\ee
This is the transformation law (\ref{tidensi}) with $h$ replaced by $1-h$:

\paragraph{Proposition.} There is an isomorphism\i{FhS1@$\cF_h(S^1)$ (space of densities)!dual 
space} $\cF_h(S^1)^*\cong\cF_{1-h}(S^1)$
which is compatible with the natural action of $\Diff$ on these spaces. In addition, the pairing between 
$\cF_h(S^1)$ and $\cF_{1-h}(S^1)$ is $\Diff$-invariant in the sense that $\bra f\cdot p,f\cdot\alpha\ket
=\bra p,\alpha\ket$ for all $f\in\Diff$ and all densities $p\in\cF_{1-h}(S^1)$, $\alpha\in\cF_h(S^1)$.\\

Thus the duals of densities with weight $h$ are densities with weight 
$1-h$, and vice-versa. One can apply this to the examples encountered above:
\begin{itemize}
\item the duals of vector fields ($h=-1$) are quadratic densities ($h=2$);
\item the duals of functions ($h=0$) are one-forms ($h=1$);
\item the duals of spinor fields ($h=1/2$) are spinor fields (i.e.\ $\cF_{1/2}(S^1)$ is 
self-dual).
\end{itemize}
Note that, for all values 
of $h$ except $h=1/2$, the conformally invariant pairing (\ref{denpar}) is \it{not} a scalar product since 
its arguments are 
densities whose transformation laws under $\Diff$ differ. This should be contrasted with the 
finite-dimensional examples encountered in chapter \ref{c2bis}, where the existence of 
an invariant bilinear form led to the equivalence of $\sigma^*$ and $\sigma$. We shall see below that this 
difference is crucial for coadjoint orbits of the Virasoro group (section \ref{secovobi}) and hence for the 
supermomentum orbits of 
the BMS$_3$ group (see part III).

\subsubsection*{Coadjoint representation}

Since the adjoint representation of 
$\Diff$ is the transformation law (\ref{advec}) of vector fields, we now know that the \it{coadjoint} 
representation of $\Diff$ is the transformation law of quadratic densities:\i{DiffS1@$\Diff$!coadjoint 
representation}\i{coadjoint representation!of $\Diff$}
\be
(\Ad^*_fp)(f(\phii))
=
\frac{p(\phii)}{(f'(\phii))^2}
\label{coadif}
\ee
This can also be written infinitesimally as
\be
\ad^*_Xp
=
Xp'+2X'p.
\label{infcoadif}
\ee
In CFT terminology, vector fields are infinitesimal conformal transformations and their duals
are (quasi-)primary fields with weight $h=2$, that is, CFT stress tensors. Indeed formula (\ref{coadif}) is 
the 
transformation law of a stress tensor $p(\phii)$ if we think of the map $\phii\mapsto f(\phii)$ as a 
conformal transformation. Similarly, the duals of functions are 
primary fields with weight $h=1$, i.e.\ currents. From now on we sometimes refer to 
$\Diff$-invariance as ``conformal invariance''. Note that at this point we haven't included any central 
charge yet; this will change once we turn to the Virasoro group.

\subsection{Exponential map and vector flows}
\label{susexpat}

We mentioned above that each vector field $X(\phii)\der_{\phii}$ may be seen as an infinitesimal 
diffeomorphism; let us make this more precise. Given a vector field $X(\phii)\der_{\phii}$, its integral 
curves\i{integral curve} are paths $\phii(t)$ on the circle that satisfy the evolution 
equation
\be
\dot\phii(t)
=
X(\phii(t)).
\label{dophixit}
\ee
In particular, when $t=\epsilon$ is ``small'' one finds that $\phii(\epsilon)$ takes the form (\ref{infidif}) 
with initial condition $\phii(0)=\phii$.
Equation (\ref{dophixit}) is an ordinary differential equation in one dimension and $X(\phii)$ is smooth, so 
given an initial condition $\phii(0)$, the solution exists and is unique. We define 
the \it{flow} of $X$ as the one-parameter family of diffeomorphisms that maps a ``time'' $t$ and an initial 
condition $\phii$ on the point $\phii(t)$ obtained by solving (\ref{dophixit}) with this initial condition. 
If we call this solution $\tilde\phii(t,\phii)$, then the flow of $X$ is\i{flow}
\be
\phi_X:
\RR\times S^1\rightarrow S^1:
(t,\phii)\mapsto\tilde\phii(t,\phii).
\label{flow}
\ee
For example the flow of the constant vector field $X(\phii)=1$ is given by
$\tilde\phii(t)=\phii+t$ and consists of rigid rotations by $t$, as already anticipated above. Using the 
notion of flow, one can define an 
exponential map for $\Diff$:

\paragraph{Definition.} The \it{exponential map} of the group $\Diff$ is\i{exponential map!for 
$\Diff$}\i{DiffS1@$\Diff$!exponential map}
\be
\exp:\Vect\rightarrow\Diff:
X(\phii)\der_{\phii}\mapsto\exp[X]\equiv\phi_X(1,\cdot)
\label{difexpuk}
\ee
where $\phi_X$ is the flow (\ref{flow}) of $X$. In other words the diffeomorphism $\exp[X](\phii)$ is 
obtained by requiring that the equality
\be
\int_{\phii}^{\exp[X](\phii)}
\frac{d\phi}{X(\phi)}
=
1
\label{doufexpa}
\ee
holds for any initial condition $\phii$.\\

In any finite-dimensional Lie group, the exponential map (\ref{ExpGa}) is a local diffeomorphism, so any 
group element belonging to a suitable neighbourhood of 
the identity can be written as the exponential of an element of the Lie algebra. However, this is not so for 
groups of diffeomorphisms: one can show that the exponential map (\ref{difexpuk}) 
does not 
define 
a local chart on $\Diff$ in that it is neither locally injective, nor locally surjective. The idea of the 
proof is to build and explicit family of diffeomorphisms that are arbitrarily close to the identity but 
cannot 
be written as exponentials of vector fields. This being said, the exponential map is always well-defined on a 
Lie-Fr\'echet group, 
even when it is not locally surjective. See \cite{guieu2007algebre,khesin2008geometry} for details.

\section{Virasoro cohomology}
\label{sedicomo}

As emphasized in chapter \ref{c1}, cohomology is crucial 
for quantum-mechanical applications: it measures the possible ``deformations'' of a 
group structure (e.g.\ central 
extensions), which typically do occur in quantum mechanics. When an algebra is finite-dimensional and 
semi-simple, 
Whitehead's lemma (\ref{whitehead}) ensures that there are essentially no deformations; the same is true of 
the Poincar\'e group (\ref{pig}). By contrast, we have seen how crucial cohomology is for the 
Galilei group (\ref{galix}), since its central extension gives rise to the notion of mass.\\

With this motivation, the present section is devoted to the cohomology of $\Diff$ and its Lie algebra. These 
considerations will eventually 
lead to the definition of the Virasoro algebra, so we refer to them as ``Virasoro cohomology''. We 
will start by describing the real cohomology groups of $\Vect$ and of $\Diff$, then turn to cohomologies 
whose cochains are primary fields on the circle. The results summarized here are discussed at greater length 
in \cite{guieu2007algebre}.

\paragraph{Remark.} We use the notation and conventions of chapter \ref{c1}, and all 
cochains are required to be smooth. Lie algebra cochains are denoted by lowercase sans serif letters such as 
$\sfb$, $\sfc$, $\sfs$, etc.\ while group cochains are denoted by uppercase letters $\sfB$, $\sfC$, $\sfS$, 
etc.

\subsection{The Gelfand-Fuks cocycle}
\label{surecovec}

Here we derive the first and second real cohomology groups of $\Vect$; in 
particular we introduce the Gelfand-Fuks cocycle, which will eventually give rise to the Virasoro algebra. We 
also describe higher-degree real cohomology groups.

\subsubsection*{Cohomology in degrees one and two}

The computation of the first cohomology of $\Vect$ is immediate: since any vector field can be 
written as 
the bracket (\ref{libici}) of two other vector fields, the first cohomology group 
(\ref{hiperfecto}) of 
$\Vect$ vanishes:\i{VectS1@$\Vect$!cohomology}\i{cohomology!of VectS1@of $\Vect$}
\be
\cH^1\big(\Vect\big)=0.
\label{hone}
\ee
In other words there is no non-trivial real one-cocycle on $\Vect$. The second cohomology of $\Vect$ is far 
more interesting:

\paragraph{Theorem.} The second real cohomology space of $\Vect$ is one-dimensional. It is generated by the 
class of the \it{Gelfand-Fuks cocycle}\i{Gelfand-Fuks cocycle}
\be
\sfc(X,Y)
\equiv
-\frac{1}{24\pi}
\int_0^{2\pi}d\phii X(\phii)Y'''(\phii)
\label{gefuks}
\ee
whose expression in the basis (\ref{ellem}) is
\be
\sfc(\ell_m,\ell_n)
=
-i\frac{m^3}{12}\delta_{m+n,0}\,.
\label{gefuksmn}
\ee

\begin{proof}
Let $\sfc$ be a real two-cocycle on $\Vect$. Then $\sfd\sfc=0$ where $\sfd$ is the Chevalley-Eilenberg 
differential (\ref{chevd}) for the trivial representation $\sT$. In terms of the basis (\ref{ellem}), the 
statement $\sfd\sfc=0$ is tantamount to
\be
\sfc([\ell_m,\ell_n],\ell_p)
+
\sfc([\ell_n,\ell_p],\ell_m)
+
\sfc([\ell_p,\ell_m],\ell_n)
=
0
\label{cociden}
\ee
for all integers $m$, $n$, $p$. Taking $p=0$ and using the antisymmetry of $\sfc$ we get
\be
\sfc(\ell_0,[\ell_m,\ell_n])
=
\sfc([\ell_0,\ell_m],\ell_n)
+
\sfc(\ell_m,[\ell_0,\ell_n]).
\label{clm}
\ee
Here we can interpret the left-hand side as the differential of the one-cochain $\sfk=\sfc(\ell_0,\cdot)$, so 
the left-hand side is exact while the right-hand side is a Lie derivative\footnote{We denote by $\sfi$ the 
interior product of cochains.}\i{Lie derivative}\i{cochain!Lie derivative}
\be
\sfc([\ell_0,\ell_m],\ell_n)
+
\sfc(\ell_m,[\ell_0,\ell_n])
=
\left((\sfi_{\ell_0}\circ\sfd+\sfd\circ\sfi_{\ell_0})\cdot \sfc\right)(\ell_m,\ell_n)
=
(\cL_{\ell_0}\sfc)(\ell_m,\ell_n)
\label{lidicoc}
\ee
where we used $\sfd\sfc=0$. Since the left-hand side of (\ref{clm}) is exact, we 
conclude that Lie derivation with respect to $\ell_0$ leaves the cohomology class of $\sfc$ invariant. (In 
geometric terms $\ell_0$ generates rotations, so this says that the 
cohomology class of $\sfc$ is invariant under rotations.) This allows us to turn $\sfc$ 
into a rotation-invariant cocycle. Indeed, let $\sfb$ be a one-cochain and define 
$\tilde\sfc\equiv\sfc+\sfd\sfb$, which has the same cohomology class as $\sfc$. The 
Lie derivative of $\tilde\sfc$ with respect to $\ell_0$ now is
\be
\cL_{\ell_0}\tilde\sfc
=
\cL_{\ell_0}\sfc+\cL_{\ell_0}\sfd\sfb
\refeq{clm}
\sfd\sfk+\sfd(\sfi_{\ell_0}(\sfd\sfb))
=
\sfd\big(
\sfk+\sfd\sfb(\ell_0,\cdot)
\big).
\label{vanikou}
\ee
In order to make $\tilde\sfc$ invariant under rotations, we need 
to choose $\sfb$ such that (\ref{vanikou}) vanishes. One verifies that the definition
\be
\sfb(\ell_m)=\frac{i}{m}\sfc(\ell_0,\ell_m)
\qquad\text{for }m\neq 0
\label{betazo}
\ee
satisfies this requirement for any $\sfb(\ell_0)$. Thus, from now on we work only with the 
rotation-invariant cocycle $\tilde\sfc$ and we rename it into $\sfc$ for simplicity. 
Then we have $\sfc(\ell_0,\ell_m)=0$, and eq.\ (\ref{clm}) becomes
\be
\sfc([\ell_0,\ell_m],\ell_n)
+
\sfc(\ell_m,[\ell_0,\ell_n])
=
0
\label{cinvit}
\ee
for all integers $m,n$. The Lie brackets 
(\ref{witt}) then yield
\be
(m+n)\,\sfc(\ell_m,\ell_n)=0
\label{ourstatement}
\ee
and thus imply that $\sfc(\ell_m,\ell_n)=0$ whenever $m+n$ is non-zero. Writing 
$\sfc(\ell_m,\ell_n)=c_m\delta_{m+n,0}$ for some coefficients $c_{m}=-c_{-m}$, we are left with the 
task 
of determining the $c_m$'s with $m>0$. 
Returning to the cocycle identity (\ref{cociden}) with $p=-m-1$ and using once more the brackets 
(\ref{witt}), 
we find
\be
c_{m+1}
=
\frac{(2+m)c_m-(2m+1)c_1}{m-1}
\label{corecu}
\ee
for $m\geq2$. This shows that all $c_m$'s are determined recursively by $c_1$ and $c_2$. In particular, we 
now know that the cohomology space $\cH^2\big(\Vect\big)$ is at most two-dimensional; the choices 
$c_m=m^3$ and $c_m=m$ are indeed two linearly independent solutions of the recursion relations 
(\ref{corecu}). Now note that, if $\sfc$ is a coboundary 
$\sfc=\sfd\sfk$ for 
some one-cochain $\sfk$, then
\be
\sfc(\ell_m,\ell_n)
=
\sfd\sfk(\ell_m,\ell_n)
\refeq{chevd}
\sfk([\ell_m,\ell_n])
\refeq{witt}
-i(m-n)\sfk(\ell_{m+n})
\nn
\ee
so that $\sfc(\ell_m,\ell_{-m})=-2im\sfk(\ell_0)$ always depends linearly on $m$. Accordingly, the solution 
$c_m=m$ of the recursion relations (\ref{corecu}) yields a trivial cocycle, while $c_m=m^3$ is non-trivial. 
We conclude that, up to a coboundary, any non-trivial two-cocycle on $\Vect$ reads
\be
\sfc(\ell_m,\ell_n)
=
\cN\,m^3\delta_{m+n,0}
\label{micube}
\ee
for some normalization $\cN\neq0$. In particular, $\cH^2\big(\Vect\big)$ is one-dimensional.
\end{proof}

\subsubsection*{Higher degree cohomologies}

The real cohomology groups of $\Vect$ increase in complexity as their degree 
becomes higher. 
Since we will not need any degree higher than two, we restrict ourselves here to a qualitative 
description of the result (details can be found in \cite{guieu2007algebre}).\\

The first step is to fix the kind of cochains one wants to study. For $\Vect$ it is natural to consider
\it{local} real cochains\i{local cochain}\i{cochain!local}
\be
\sfc:\Vect^k\rightarrow\RR:(X_1,...,X_k)\mapsto\sfc(X_1,...,X_k)
\label{locochain}
\ee
that take the form of an integral over $S^1$ of some ``cochain density'' $\sC$:
\be
\sfc(X_1,...,X_k)=
\int_0^{2\pi}\!\!d\phii\;
\sC\left(
X_1(\phii),X_1'(\phii),...,X_1^{(n_1)}(\phii),
...,
X_k(\phii),...,X_k^{(n_k)}(\phii)
\right).
\nn
\ee
Here the word \it{local} is used in the same sense as in field theory. The Gelfand-Fuks cocycle 
(\ref{gefuks}) is 
of that 
form, with a density $\sC(X,Y)\propto XY'''$. With this restriction on the allowed cochains, one can 
study the resulting cohomology groups $\cH^k_{\text{loc}}\big(\Vect\big)$. The result is as follows:

\paragraph{Proposition.} The real local cohomology groups $\cH^k_{\text{loc}}\big(\Vect\big)$ are all trivial
except if $k$ is equal to $0$, $2$ or $3$, in which case the cohomology group is 
one-dimensional:\i{VectS1@$\Vect$!cohomology}\i{cohomology!of VectS1@of $\Vect$}
\be
\cH^k_{\text{loc}}\big(\Vect\big)
=
\begin{cases}
\RR & \text{if }k\in\{0,2,3\}\\
0 & \text{otherwise}.
\end{cases}
\label{HakkaL}
\ee
The generator of $\cH^0_{\text{loc}}$ is the class of any non-zero constant function on $\Vect$; that of 
$\cH^2_{\text{loc}}$ is the Gelfand-Fuks cocycle (\ref{gefuks}). Finally $\cH^3_{\text{loc}}$ is generated by 
the class of the \it{Godbillon-Vey cocycle}\i{Godbillon-Vey cocycle}
\be
\int_0^{2\pi}d\phii\;\text{det}\bmm X & Y & Z \\ X' & Y' & Z' \\ X'' & Y'' & Z'' \emm
\label{govey}
\ee
where it is understood that the integrand is evaluated at $\phii$.\\

Note that the \it{full}, generally non-local, cohomology groups of $\Vect$ do not coincide with 
(\ref{HakkaL}) because they contain classes of wedge products of the Gelfand-Fuks and Godbillon-Vey 
cocycles.\i{non-local cohomology}\i{cohomology!non-local} 
For example $\sfc\wedge\sfc$ is a non-trivial, non-local four-cocycle on $\Vect$ when $\sfc$ is the 
Gelfand-Fuks cocycle.

\paragraph{Remark.} In this work the Godbillon-Vey cocycle 
(\ref{govey}) will be unimportant. However, it does play a key role in a specific context, as it was shown in 
\cite{Barnich:2015tba} that it is responsible for the unique non-trivial 
gauge-invariant deformation of a higher-spin Chern-Simons action in three dimensions with gauge 
algebra $\Vect\inplus_{\ad^*}\Vect^*$.

\subsection{The Bott-Thurston cocycle}
\label{surecodif}

We now turn to the low-degree real cohomology groups of the universal cover $\Diffc$ of the group of 
orientation-preserving diffeomorphisms of the circle. We show in particular how one can build a non-trivial 
two-cocycle corresponding to Gelfand-Fuks by integration, and known as the \it{Bott-Thurston cocycle}. The 
latter will eventually lead to the definition of the Virasoro group. As before, we abuse notation by denoting 
the universal cover $\Diffc$ simply as $\Diff$.

\subsubsection*{Cocyclic recipes}

We start by describing a general algorithm for building two-cocycles on a group 
\cite{guieu2007algebre,ovsienko2004projective}. Let $\cM$ be an orientable manifold endowed with a volume 
form 
$\mu$. For any orientation-preserving 
diffeomorphism $f:\cM\rightarrow\cM$, we define a function $\sfT[f^{-1}]$ on $\cM$ by
\be
f^*\mu\equiv e^{\sfT[f^{-1}]}\mu.
\label{genonsen}
\ee
This function can be thought of as a modified derivative of $f$. It appears to have no standard name in the 
literature but we will use it repeatedly below in the case $\cM=S^1$, so from now on we refer to 
$\sfT[f^{-1}]$ as the \it{twisted derivative} of $f$.\i{twisted derivative}

\paragraph{Lemma.} The map $\sfT:\text{Diff}^+(\cM)\rightarrow C^{\infty}(\cM)=f\mapsto\sfT[f]$
defined by (\ref{genonsen}) is a $C^{\infty}(\cM)$-valued one-cocycle on $\text{Diff}^+(\cM)$, where the 
action of diffeomorphisms on functions is given by
\be
(f\cdot\cF)(p)=\cF(f^{-1}(p))
\label{actafi}
\ee
for 
$f\in\text{Diff}^+(\cM)$, $\cF\in C^{\infty}(\cM)$ and $p\in\cM$.

\begin{proof}
We need to show that $\sfd\sfT=0$ with the Chevalley-Eilenberg differential (\ref{gd}) and the representation 
$\cT$ 
given by 
the action (\ref{actafi}) of diffeomorphisms on functions.\footnote{The fact that the same letter denotes the 
cocycle $\sfT$ 
and the representation $\cT$ is merely a notational coincidence.} If $f,g$ are orientation-preserving 
diffeomorphisms, one readily verifies from the definition (\ref{genonsen}) that $\sfT$ satisfies the cocycle 
property (\ref{1coc}).
\end{proof}

Now let us consider another recipe, seemingly unrelated to and just as random as the previous one.
Take two vector spaces $\VV$ and $\WW$ acted upon by a group $G$ according to representations $\cS$ 
and $\cT$, respectively, and let $\Omega:\VV\times\VV\rightarrow\WW:(v,v')\mapsto\Omega(v,v')$ be an 
antisymmetric bilinear map such that
\be
\Omega\big(\cS[f]v,\cS[f]v'\big)
=
\cT[f]\,\Omega(v,v')
\label{geckobi}
\ee
for any group element $f\in G$ and all $v,v'\in\VV$. Finally, let $\sfT:G\rightarrow\VV$ be a $\VV$-valued 
one-cocycle on $G$ with respect to the representation $\cS$.

\paragraph{Lemma.} The map
\be
\sfC:G\times G\rightarrow\WW:
(f,g)\mapsto
\sfC(f,g)\equiv\Omega\big(\sfT[f],\sfT[fg]\big)
\label{bicocy}
\ee
is a $\WW$-valued two-cocycle on $G$.

\begin{proof}
We need to show that $\sfd\sfC=0$ for the Chevalley-Eilenberg differential (\ref{gd}), given that $\WW$ is 
acted upon by $G$ 
according 
to the representation $\cT$. Using the fact that $\Omega$ is bilinear and antisymmetric together with 
property (\ref{geckobi}), one readily verifies by brute force that this is indeed the case.
\end{proof}

\subsubsection*{The Bott-Thurston cocycle}

The two constructions just described can be used to define a non-trivial two-cocycle on the group $\Diff$. 
We will first use (\ref{genonsen}) to define a one-cocycle on $\Diff$, then plug it into 
(\ref{bicocy}) for a well chosen map $\Omega$ to obtain the desired two-cocycle.\\

We consider the circle $S^1$ endowed with the flat volume form $\mu=d\phii$. Under a diffeomorphism 
$\phii\mapsto f(\phii)$ we have $(f^*\mu)_{\phii}
=
d(f(\phii))
=
f'(\phii)d\phii
=
e^{\log(f'(\phii))}\mu$,
so (\ref{genonsen}) provides a $C^{\infty}(S^1)$-valued twisted derivative\i{twisted derivative}
\be
\sfT[f](\phii)
\equiv
\log\left[(f^{-1})'(\phii)\right],
\label{logif}
\ee
which is a one-cocycle.
We use square brackets to denote the argument of $\sfT$ because the latter is a functional on 
$\Diff$; then $\sfT[f](\phii)$ is the function $\sfT[f]$ evaluated at $\phii$.\\

To apply the construction (\ref{bicocy}), we also need to find a bilinear antisymmetric map 
$\Omega:C^{\infty}(S^1)\times C^{\infty}(S^1)\rightarrow\RR$ which is invariant under $\Diff$ in the sense 
that (\ref{geckobi}) holds when $\cT$ is the trivial representation while $\cS$ is the action (\ref{actafi}) 
of 
$\Diff$ on functions. A natural guess is
\be
\Omega:
C^{\infty}(S^1)\times C^{\infty}(S^1)\rightarrow\RR:
(\cF,\cG)\mapsto
\int_0^{2\pi}d\phii\,\cF(\phii)\cG'(\phii)
=
\int_{S^1}\cF d\cG\,,
\label{AMEG}
\ee
which is manifestly antisymmetric (integrate by parts) and reparameterization\--in\-va\-riant (the integrand 
is 
analogous to the $p\dot q$ of Hamiltonian actions). Applying the prescription (\ref{bicocy}) with the 
one-cocycle (\ref{logif}), we obtain the following real two-cocycle:

\paragraph{Definition.} The \it{Bott-Thurston cocycle} on $\Diff$ is \cite{Bott}\i{Bott-Thurston cocycle}
\begin{eqnarray}
\label{tibidik}
\sfC(f,g)
& \!\!\equiv &
\!\!-\frac{1}{48\pi}
\int_{S^1}\sfT[f]\,d\sfT[f\circ g]\\
\label{tibiduk}
& \!\!\refeq{logif} &
\!\!-\frac{1}{48\pi}
\int_0^{2\pi}d\phii\,
\log\left[(f^{-1})'(\phii)\right]
\left(
\log\left[((f\circ g)^{-1})'\right]
\right)'
(\phii)\,
\end{eqnarray}
where $d$ denotes the exterior derivative on the circle.\\

By construction, the Bott-Thurston cocycle satisfies the cocycle identity (\ref{cocy}),\i{cocycle condition}
\be
\sfC(f,gh)+\sfC(g,h)
=
\sfC(fg,h)+\sfC(f,g)\,.
\label{btcoid}
\ee
This will be instrumental in ensuring that $\sfC$ yields a well-defined centrally extended group.
For future applications it is useful to rewrite (\ref{tibiduk}) in a slightly simpler way, which relies on 
the following result:

\paragraph{Lemma.} If $\sfT$ is the twisted derivative (\ref{logif}), then for all $f,g\in\Diff$ one has
\be
\int_{S^1}\sfT[f]d\sfT[f\circ g]
=
\int_{S^1}\sfT[(f\circ g)^{-1}]d\sfT[g^{-1}]\,.
\label{bottlem}
\ee

\begin{proof}
We use two key properties: the first is the fact that $\sfT$ is a 
one-cocycle with respect to the action of $\Diff$ on $C^{\infty}(S^1)$, so
\be
\sfT[f\circ g]
=
\sfT[f]+\sfT[g]\circ f^{-1},
\label{fiktoci}
\ee
and the second is a property that follows from the definition (\ref{logif}) and eq.\ (\ref{invide}):
\be
\sfT[f]\circ f
=
-\sfT[f^{-1}].
\label{cifinv}
\ee
We then find that 
(\ref{tibidik}) can be rewritten as
\be
\int_{S^1}\sfT[f]\,d\sfT[f\circ g]
=
-\int_{S^1}\sfT[f^{-1}]\,
d\sfT[g]
=
\int_{S^1}\sfT[(f\circ g)^{-1}]\,d\sfT[g^{-1}]\,,
\nn
\ee
which was to be proven.
\end{proof}

Thanks to this lemma we can write the Bott-Thurston cocycle (\ref{tibiduk}) in a more convenient way, 
without $f^{-1}$'s all around the place:
\be
\sfC(f,g)
=
-\frac{1}{48\pi}
\int_{S^1}
\log(f'\circ g)\,d\log(g')
=
-\frac{1}{48\pi}
\int_{S^1}\sfT[(f\circ g)^{-1}]d\sfT[g^{-1}].
\label{btcoll}
\ee
This is the definition that we will be using from now on.\\

At this stage the Bott-Thurston cocycle seems to be coming out of the blue. However it turns out that 
(\ref{btcoll}) is, in fact, a very natural quantity. We will explain this in greater detail in section 
\ref{seschwa}, but for now we simply note the following relation:

\paragraph{Proposition.} The Bott-Thurston cocycle (\ref{btcoll}) is the integral of the Gelfand-Fuks cocycle 
(\ref{gefuks}) in the sense of formula (\ref{diffcc}):\i{Bott-Thurston cocycle!and 
Gelfand-Fuks}\i{Gelfand-Fuks cocycle!and Bott-Thurston}
\be
\sfc(X,Y)
=
-
\frac{d^2}{dt\,ds}
\Big[
\sfC\left(e^{tX},e^{sY}\right)-\sfC\left(e^{sY},e^{tX}\right)
\Big]
\Big|_{t=0,\,s=0}\,.
\label{btgefuks}
\ee
In particular, the Bott-Thurston cocycle is non-trivial.

\begin{proof}
We consider infinitesimal diffeomorphisms $f(\phii)=\phii+tX(\phii)+\cO(t^2)$ 
and $g(\phii)=\phii+sY(\phii)+\cO(s^2)$.
Then $\log\left(f'\circ g\right)=tX'$ and $\log(g')=sY'$ so that
\be
\sfC(e^{tX},e^{sY})
\refeq{btcoll}
-\frac{1}{48\pi}
\int_0^{2\pi}d\phii\,tX'(\phii)sY''(\phii)
\nn
\ee
to first order in $t,s$. Relation (\ref{btgefuks}) follows. It also follows that the Bott-Thurston cocycle is 
non-trivial, since the Gelfand-Fuks cocycle is non-trivial.
\end{proof}

\paragraph{Remark.} The bilinear map (\ref{AMEG}) is a non-trivial two-cocycle on the Abelian Lie algebra 
$C^{\infty}(S^1)$ of smooth functions on the circle. It defines a central extension of $C^{\infty}(S^1)$ that 
can be interpreted in several ways: either as an infinite-dimensional Heisenberg algebra, or as a 
$\mathfrak{u}(1)$ Kac-Moody algebra.\i{Kac-Moody algebra} This kind of central extension occurs for instance 
in the realm of 
warped conformal field theories \cite{Detournay:2012pc,Afshar:2015wjm}.

\subsection{Primary cohomology of $\Vect$}
\label{supicovec}

Here we study some of the cohomology groups of $\Vect$ in spaces of densities (i.e.\ primary fields). As in 
the real-valued case described earlier we consider the cohomology defined by \it{local} cochains, which in 
the present case take the form\i{local cochain}\i{cochain!local}
\be
\sfc[X_1,...,X_k]
=
\sC\big[X_1(\phii),X_1'(\phii),...,X_1^{(n_1)}(\phii),
...,
X_k(\phii),...,X_k^{(n_k)}(\phii)
\big](d\phii)^h
\nn
\ee
for some weight $h$. The functional $\sC$ depends on the $X_i$'s and finitely many of their derivatives, all 
evaluated at the same point $\phii$. We denote the corresponding cohomology spaces by 
$\cH^k\big(\Vect,\cF_{\lambda}(S^1)\big)$. In order to avoid technical 
considerations we state 
the results without proof and refer to \cite{guieu2007algebre} for details.

\paragraph{Theorem.} If the weight $h$ is not a non-negative integer, 
then\i{VectS1@$\Vect$!cohomology}\i{cohomology!of VectS1@of $\Vect$}
\be
\cH^k\big(\Vect,\cF_h(S^1)\big)=0
\quad
\text{for all $k\in\NN$.}
\label{HADDIP}
\ee
In particular $\cH^2\big(\Vect,\Vect\big)=0$, so there exists no non-trivial deformation of $\Vect$.\\

The result (\ref{HADDIP}) implies that the non-trivial primary cohomology of $\Vect$ is localized only on 
non-negative 
integers with weights $h\in\NN$.
Here we briefly describe the non-trivial first cohomology groups ($k=1$) for the cases $h=0,1,2$ that will be 
useful below.\\

The case $h=0$ corresponds to one-cochains taking values in the space of functions on the circle. It turns 
out that there are exactly two linearly independent, non-trivial one-cocycles in that case, namely 
$\tilde\sfc[X](\phii)=X(\phii)$ and\i{twisted derivative}
\be
\sft[X](\phii)
=
X'(\phii)\,.
\label{tutuk}
\ee
The latter may be recognized as the infinitesimal cocycle corresponding to the twisted derivative 
(\ref{logif}) by 
differentiation.\\

At weight $h=1$ we are in the realm of cochains taking values in the space $\Omega^1(S^1)$ of one-forms; 
in particular one can show that the corresponding first cohomology is one-dimensional, generated by the 
(class of the) one-cocycle
\be
\sfw[X](\phii)
=
X''(\phii)d\phii\,.
\label{wacocy}
\ee
The notation $\sfw$ is because this cocycle is relevant to certain aspects \cite{Afshar:2015wjm} of 
warped\i{warped CFT} conformal symmetry \cite{Detournay:2012pc}.\\

Finally, when $h=2$, cochains take their values in the space $\cF_2(S^1)$ of quadratic densities on the 
circle. In particular one can show that the first cohomology space is one-dimensional, generated by the 
(class of the) \it{infinitesimal Schwarzian derivative}\i{infinitesimal Schwarzian 
derivative}\i{Schwarzian derivative!infinitesimal}
\be
\sfs[X](\phii)
=
X'''(\phii)d\phii^2.
\label{insa}
\ee
One can go on and similarly classify all cohomology groups with higher weight $h$. Since we 
will not need these results here, we refrain from 
displaying them (see e.g. \cite{goncharova1972,fuks1986}). Instead, we now relate the cocycles (\ref{tutuk}), 
(\ref{wacocy}) and (\ref{insa}) to one-cocycles on $\Diff$.

\subsection{Primary cohomology of $\Diff$}
\label{supicodif}

The complete classification of density-valued cohomology groups of $\Diff$ is beyond the scope of this 
presentation, so we refer to \cite{guieu2007algebre} for a more detailed discussion. Here we simply 
note that the Lie algebra one-cocycles mentioned above can be integrated to non-trivial group 
one-cocycles:\i{DiffS1@$\Diff$!cohomology}
\begin{itemize}
\item The cocycle (\ref{tutuk}) can be 
integrated to the twisted derivative (\ref{logif}), which we used to build the Bott-Thurston 
cocycle. Indeed, for 
$f(\phii)=\phii+\epsilon X(\phii)$,\i{twisted derivative} formula (\ref{logif}) reduces to 
$\sfT[f]=-\epsilon\,\sft[X]$.
\item The cocycle (\ref{wacocy}) can be integrated to
\be
\sfW[f](\phii)
=
d\log[(f^{-1})'(\phii)]
=
d\,\sfT[f](\phii)
\label{dilogif}
\ee
where $d$ denotes the exterior derivative on the circle. As mentioned above this ``warped 
derivative'' has 
been used recently \cite{Afshar:2015wjm} to describe certain aspects of warped conformal field 
theories.\i{warped CFT} Note that in these terms the Bott-Thurston cocycle (\ref{tibidik}) 
is $\sfC(f,g)\propto\int\sfT[f]\otimes\sfW[f\circ g]$.\i{Bott-Thurston cocycle}
\end{itemize}
The one-cocycle (\ref{insa}) can similarly be related to the $\cF_2$-valued Schwarzian derivative on $\Diff$, 
although the integration is somewhat less trivial than in the two cases just described. The Schwarzian 
derivative is crucial for our upcoming considerations, so the 
whole next section is devoted to it.

\section{On the Schwarzian derivative}
\label{seschwa}

\paragraph{Definition.} Let $f\in\Diff$. Then the 
\it{Schwarzian derivative}\footnote{The name refers to H.\ Schwarz, who first introduced the object 
(\ref{swag}); it is the same 
Schwarz as in the Cauchy-Schwarz inequality.} of $f$ at $\phii$ is\i{Schwarzian derivative}
\be
\sfS[f](\phii)
\equiv
\frac{f'''(\phii)}{f'(\phii)}
-
\frac{3}{2}\left(
\frac{f''(\phii)}{f'(\phii)}
\right)^2.
\label{swag}
\ee
Many references use the notation $\{f;\phii\}$, but we will stick to $\sfS[f](\phii)$ instead.\\

In this section we investigate the many properties of the Schwarzian derivative. We will start by showing 
that it is (related to) a one-cocycle on $\Diff$ taking its values in the space $\cF_2(S^1)$ of quadratic 
densities, and that it corresponds to the Lie algebra cocycle (\ref{insa}) by differentiation. We will then 
show that it is related to the Bott-Thurston cocycle by the so-called Souriau 
construction. We will also describe the remarkable symmetry properties of the 
Schwarzian derivative under the $\PSL$ subgroup of $\Diff$, and obtain as a by-product the 
expression of Lorentz transformations in terms of diffeomorphisms of the circle. (In chapter \ref{c6} these 
transformations will turn out to be actual Lorentz transformations on the celestial circle.)

\subsection{The Schwarzian derivative is a cocycle}
\label{souriausec}

Here we show that the Schwarzian derivative is the one-cocycle corresponding to (\ref{insa}) by integration. 
Note that the relation between (\ref{swag}) and (\ref{insa}) is obvious: upon taking $f(\phii)=\phii+\epsilon 
X(\phii)$ in (\ref{swag}), one finds $\sfS[f]=\epsilon X'''$ to first 
order in $\epsilon$. The non-trivial problem is showing that the Schwarzian derivative is actually a 
cocycle:

\paragraph{Proposition.} The Schwarzian derivative (\ref{swag}) defines a one-cocycle\i{Schwarzian 
derivative!as one-cocycle}\i{one-cocycle!Schwarzian derivative}
\be
\Diff\rightarrow\cF_2(S^1):
f\mapsto \sfS[f^{-1}](\phii)d\phii^2
\nn
\ee
valued in the space of quadratic densities on the circle.

\begin{proof}
We start by noting that the definition (\ref{swag}) implies\i{Schwarzian derivative!cocycle identity}
\be
\sfS[f\circ g]
=
\Ad^*_{g^{-1}}\sfS[f]+\sfS[g]
=
\sfS[g]+(g')^2\sfS[f]\circ g\,,
\label{swapro}
\ee
where $\Ad^*$ denotes the coadjoint 
representation (\ref{coadif}) of $\Diff$. Upon defining $\tilde\sfS[f]\equiv \sfS[f^{-1}]d\phii^2$, one 
obtains a map that 
associates a quadratic density with any diffeomorphism $f$, and which satisfies
\be
\tilde\sfS[f\circ g]
=
\tilde\sfS[f]+((f^{-1})')^2\tilde\sfS[g]\circ f^{-1}
=
\tilde\sfS[f]+\Ad^*_f\tilde\sfS[g]
\label{qequ}
\ee
by virtue of (\ref{swapro}). This is precisely the 
cocycle identity (\ref{1coc}).
\end{proof}

\subsubsection*{The Souriau construction}

We now study the relation between the Schwarzian derivative and the Bott-Thurston cocycle, which follows from 
the so-called Souriau construction.

\paragraph{Definition.} Let $G$ be a Lie group with Lie algebra $\mg$, $\sfC:G\times G\rightarrow\RR$ a real 
two-cocycle on $G$. Then the \it{Souriau cocycle} associated with 
$\sfC$ is the map $G\mapsto\mg^*:f\mapsto \sfS[f^{-1}]$ defined by\i{Souriau cocycle}\i{one-cocycle!Souriau}
\be
\frac{d}{dt}
\left.\Big[
\sfC(f,e^{tX})+\sfC(fe^{tX},f^{-1})
\Big]\right|_{t=0}
\equiv
-\frac{1}{12}
\bra
\sfS[f],X
\ket
\label{souriau}
\ee
for any $f\in G$ and any adjoint vector $X\in\mg$. (The normalization is chosen so that $\sfS$ 
eventually 
coincides with the Schwarzian derivative.)

\paragraph{Proposition.} The Souriau cocycle is a 
one-cocycle on $G$ valued in the space of coadjoint vectors.

\begin{proof}
Since $\sfC$ is a real two-cocycle, it is clear that the left-hand side of (\ref{souriau}) defines a real
linear function of $X\in\mg$, that is, a coadjoint vector. The latter only depends on $f$ so we can 
certainly write it as $\sfS[f]$, which defines the map $\sfS$. The problem is to show that the map $f\mapsto 
\sfS[f^{-1}]$ is actually a one-cocycle. For this we let $X\in\mg$, pick two group elements 
$f,g\in G$, and write
\be
\bra
\sfS[(fg)^{-1}],X
\ket
\refeq{souriau}
\frac{d}{dt}\left.
\Big[
\sfC(g^{-1}f^{-1},e^{tX})
+
\sfC(g^{-1}f^{-1}e^{tX},fg)
\Big]
\right|_{t=0}.
\label{messire}
\ee
On the other hand, if $\Ad^*$ 
denotes the 
coadjoint representation of $G$, we have
\begin{align}
\label{mesar}
& \bra\Ad^*_f\sfS[g^{-1}]+\sfS[f^{-1}],X\ket=\\
& \refeq{souriau}
\frac{d}{dt}\left.
\Big[
\sfC(g^{-1},e^{t\Ad_{f^{-1}}X})+
\sfC(g^{-1}e^{t\Ad_{f^{-1}}X},g)+
\sfC(f^{-1},e^{tX})+
\sfC(f^{-1}e^{tX},f)
\Big]
\right|_{t=0}\,.\nn
\end{align}
Using the cocycle identity (\ref{btcoid}) together with property (\ref{FADO}), one can then show by brute 
force that (\ref{mesar}) coincides with the right-hand side of (\ref{messire}).
\end{proof}

In the case of the group $\Diff$, the Souriau construction yields the Schwarzian derivative from the 
Bott-Thurston cocycle. Let us check this explicitly: taking $g(\phii)=\phii+tX(\phii)$ in (\ref{btcoll}), one 
finds\i{Souriau cocycle!and Schwarzian derivative}\i{Schwarzian derivative!as Souriau 
cocycle}\i{Bott-Thurston cocycle!and Schwarzian derivative}\i{Schwarzian derivative!and Bott-Thurston}
\begin{align}
\sfC(f,e^{tX})
& =
-\frac{t}{48\pi}\int_0^{2\pi}d\phii\,
\left[
\frac{f'''}{f'}-\left(\frac{f''}{f'}\right)^2
\right]X(\phii),\nn\\
\sfC(f\circ e^{tX},f^{-1})
& =
\text{($t$-independent)}
-\frac{t}{48\pi}\int_0^{2\pi}d\phii\,
\left[
\frac{f'''}{f'}-2\left(\frac{f''}{f'}\right)^2
\right]X(\phii)\nn
\end{align}
to first order in $t$. It then follows that relation (\ref{souriau}) holds when $\sfS$ is the Schwarzian 
derivative (\ref{swag}) and $\langle\cdot,\cdot\rangle$ is the pairing (\ref{denpar}) of $\Vect$ with its 
dual. Note that by taking 
$f(\phii)=\phii+sY(\phii)$ 
with small $s$, the Schwarzian derivative reduces to $\sfS[f]=sY'''$. Upon differentiating with respect to 
$s$ 
in the right-hand side of (\ref{souriau}), we recover precisely the Gelfand-Fuks cocycle (\ref{gefuks}).

\subsubsection*{Virasoro universality}

At this point the cohomological constructions of the previous pages are starting to fit in a 
global picture of Virasoro cohomology: eq.\ (\ref{btgefuks}) relates the Bott-Thurston cocycle to the 
Gelfand-Fuks cocycle (\ref{gefuks}), while (\ref{souriau}) relates it to the Schwarzian derivative, which in 
turn is the integral of the infinitesimal cocycle (\ref{insa}). In addition the integral of the latter with a 
vector field on the circle reproduces the Gelfand-Fuks cocycle. The common feature of all these expressions 
is the 
occurrence of third derivatives such as $f'''$ or $X'''$, which will indeed play a key role in the sequel 
(and give rise to the term $m^3$ in (\ref{gefuksmn})).\\

In this sense, all these cocycles are really one and the 
same quantity, albeit expressed in very different ways. Depending on one's viewpoint, one may decide that the 
most fundamental quantity is the Gelfand-Fuks cocycle, or the Schwarzian derivative, or Bott-Thurston. Our 
point of view will be that the Bott-Thurston cocycle is the most fundamental of them all, since it 
yields the other ones by differentiation:
\be
\begin{array}{c}
\text{Bott-Thurston $\sfC$}\\
\text{\textcolor{gray}{Souriau}}\Bigg\downarrow\text{\textcolor{white}{Souriau}}\\
\text{Schwarzian derivative $\sfS$}\\
\;\text{\textcolor{gray}{~~differentiate}}\Bigg\downarrow{\textcolor{white}{differentiate}}\\
\text{infinitesimal Schwarzian $\sfs$}\\
\text{\textcolor{gray}{pair with $\Vect$}}\Bigg\downarrow\text{\textcolor{white}{pair with $\Vect$}}\\
\text{Gelfand-Fuks $\sfc$}
\end{array}
\nn
\ee

\subsection{Projective invariance of the Schwarzian}
\label{suseschwi}

There exists a deep relation between the circle, the projective line and the Schwarzian 
derivative \cite{ovsienko2004projective}, which in turn leads to powerful symmetry properties under 
the group $\SL/\ZZ_2=\PSL$. Our goal here is to explore this relation. 
Accordingly we start with a short 
detour through one-dimensional projective geometry, before recovering the Schwarzian derivative as a 
quantity that measures the extent to which diffeomorphisms deform the projective structure. Along the way we 
will
encounter the expression of Lorentz transformations in terms of diffeomorphisms of the 
circle.

\subsubsection*{The projective line}

Consider the plane $\RR^2$ and define the \it{projective line} $\RR P^1$ to be the space 
of its one-dimensional subspaces.\i{projective line}\i{RP1@$\RR P^1$ (projective line)} Equivalently $\RR 
P^1$ is the space of 
straight lines in $\RR^2$ going through the origin, i.e.\ a quotient of 
$\RR^2\backslash\{(0,0)\}$ by the equivalence relation
\be
(x,y)\sim(x',y')
\qquad
\text{if}
\qquad
\exists\,\lambda\in\RR^*\text{ such that }(x,y)=\lambda(x',y').
\nn
\ee
Denoting by $[(x,y)]$ the equivalence class of $(x,y)\in\RR^2$, the projective line is thus
\be
\RR P^1
=
\big\{
[(x,y)]
\big|(x,y)\in\RR^2\backslash\{(0,0)\}
\big\}.
\label{rpin}
\ee
In topological terms the projective line is a circle centred at the origin in $\RR^2$ with 
antipodal points identified.\i{S1@$S^1$ (circle)!as projective line} This is to say 
that $\RR P^1\cong S^1/\ZZ_2$, where $\ZZ_2$ acts on $S^1$ by 
rotations. Since any group $\ZZ_n$ acting on the circle by rotations of 
$2\pi/n$ is such that $S^1/\ZZ_n\cong S^1$, the projective line is actually diffeomorphic to a circle:
\be
\RR P^1\cong S^1.
\label{rps}
\ee
As a result, all considerations concerning the group of diffeomorphisms of $S^1$ can be recast in terms of 
projective 
geometry, and vice-versa.\\

The diffeomorphism (\ref{rps}) can be made explicit in terms of well chosen coordinates. Indeed, in 
terms of (\ref{rpin}), the projective line is a union $\RR P^1
=
\left\{
[(\zeta,1)]|\zeta\in\RR
\right\}
\cup
\left\{
[(1,0)]
\right\}$,
so the projective coordinate
\be
\zeta\equiv x/y
\label{procor}
\ee
is a local coordinate on $\RR P^1$ that misses only one point, namely the class of $(1,0)$. In 
this sense the projective line is a real line $\RR$ with an extra ``point at infinity''.\i{point at 
infinity}\\

\begin{figure}[H]
\centering
\includegraphics[width=0.40\textwidth]{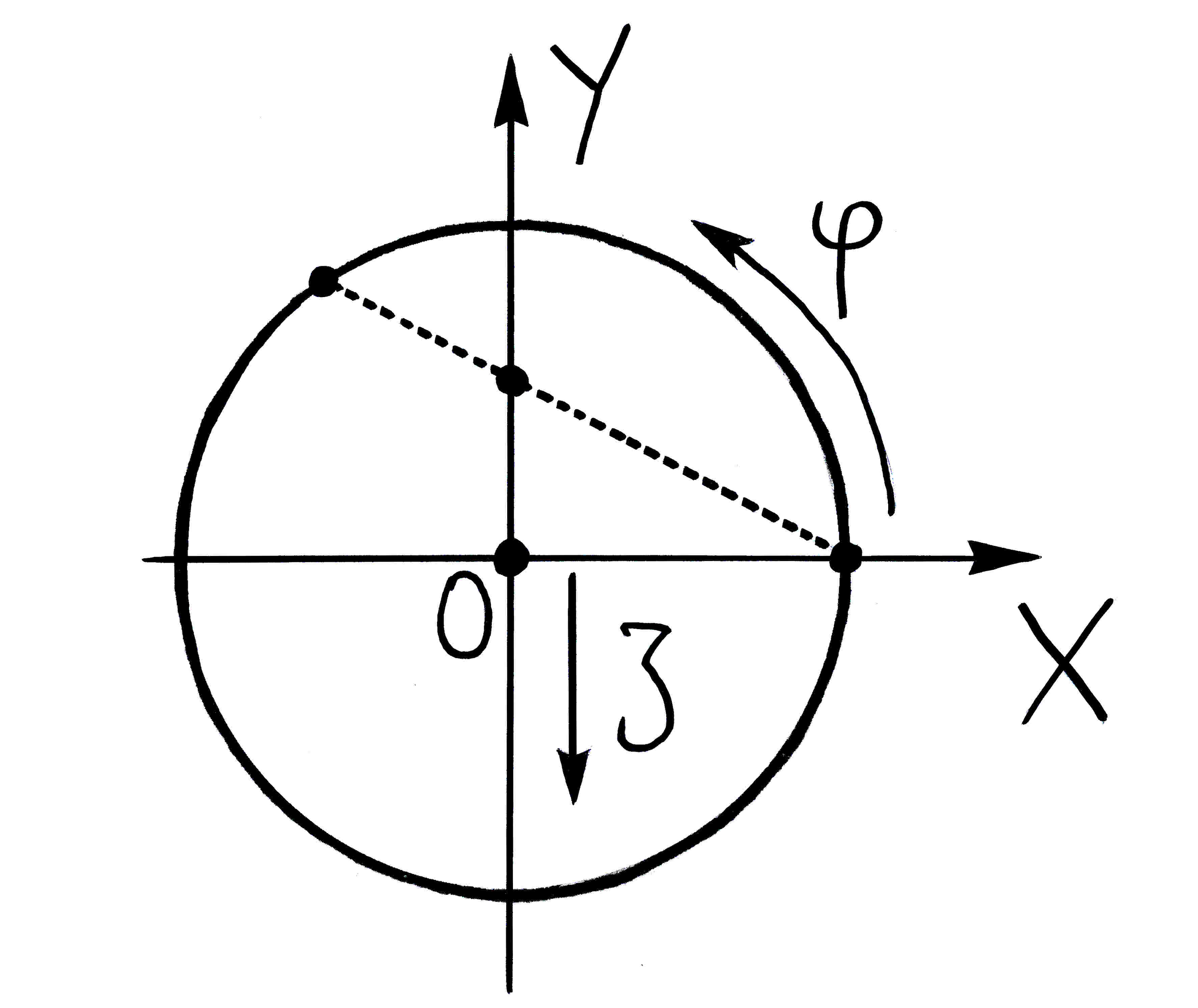}
\caption{The stereographic coordinate (\ref{zetaphi}) is obtained by projecting the points of 
a unit circle in the $(X,Y)$ plane on the $Y$ axis, along a straight line that goes through the ``east pole'' 
$(1,0)$. Upon 
writing $X=\cos\phii$ and $Y=\sin\phii$ and declaring that the projective coordinate $\zeta$ 
is \it{minus} the $Y$ coordinate of the projection, one obtains (\ref{zetaphi}). The minus 
sign is included so as to preserve the orientation of the circle in the sense that 
$d\zeta/d\phii>0$.\label{figoZeta}}
\end{figure}

This is exactly 
the same situation as
with the stereographic coordinate on a circle. For later convenience we define this coordinate in terms of an 
angular coordinate $\phii$ on the circle by\i{stereographic coordinate}\i{S1@$S^1$ (circle)!stereographic 
coordinate}
\be
\zeta
=
-\text{cot}(\phii/2)
=
\frac{e^{i\phii}+1}{ie^{i\phii}-i}
\label{zetaphi}
\ee
(see fig.\ \ref{figoZeta}).
The diffeomorphism (\ref{rps}) is then obtained by identifying this stereographic coordinate with the 
projective coordinate (\ref{procor}).

\subsubsection*{Projective transformations}

The projective line inherits a symmetry from the linear action of
$\text{GL}(2,\RR)$ on $\RR^2$. Explicitly, an invertible matrix
\be
\bmm
a & b \\ c & d
\emm
\label{abcd}
\ee
acts on the coordinate (\ref{procor}) as a \it{projective transformation}\i{projective transformation}
\be
\zeta\mapsto\frac{a\zeta+b}{c\zeta+d}.
\label{proto}
\ee
Any such transformation is independent of the determinant of (\ref{abcd}), which we can therefore 
set to one without loss of generality. Furthermore the overall sign of the matrix is irrelevant, so 
the transformations 
(\ref{proto}) span a \it{projective group}\i{projective 
group}\i{PSL2R@$\PSL$}\i{Lorentz group!as projective group}
\be
\text{PGL}(2,\RR)
\equiv
\text{GL}(2,\RR)/\RR^*
\cong
\SL/\ZZ_2
\equiv
\text{PSL}(2,\RR).
\nn
\ee
According to (\ref{isoso}) this is the connected Lorentz group in three dimensions.\\

Upon identifying the projective coordinate (\ref{procor}) with the stereographic coordinate (\ref{zetaphi}), 
the transformation (\ref{proto}) can be reformulated in terms of the angular coordinate $\phii$.
Using (\ref{proto}) and the inverse of (\ref{zetaphi}), we find that projective transformations act on 
$e^{i\phii}$ according to\i{S1@$S^1$ (circle)!projective transformation}
\be
e^{i\phii}
\mapsto
e^{if(\phii)}
=
\frac{A e^{i\phii}+B}{\bar B e^{i\phii}+\bar A}
\label{protophi}
\ee
where the complex coefficients
\be
A=\demi(a+ib-ic+d),
\qquad
B=\demi(a-ib-ic-d)
\label{unhappy}
\ee
are such that $|A|^2-|B|^2=1$. The family of transformations (\ref{protophi}) spans a subgroup $\PSL$
of $\Diff$ that we already anticipated in (\ref{ProJJ}). In section \ref{sebmsboco} we shall interpret that 
subgroup as the Lorentz group acting on null infinity. For infinitesimal parameters 
$A=1+i\epsilon$ and $B=\varepsilon$ (with $\epsilon\in\RR$ and $\varepsilon\in\CC$), formula (\ref{protophi}) 
becomes an infinitesimal diffeomorphism
\be
f(\phii)
=
\phii+2\epsilon+2\,\text{Re}(\varepsilon)\,\cos\phii-2\,\text{Re}(\varepsilon)\,\sin(\phii).
\nn
\ee
This is an $\sl$ transformation generated by the vector fields $\ell_0$, $\ell_1$ and 
$\ell_{-1}$  mentioned below (\ref{witt}). Conversely, any transformation (\ref{protophi}) belongs to the 
flow of an $\sl$ vector field.

\paragraph{Remark.} The relation between $S^1$ and $\RR P^1$ discussed here has a complex analogue $\CC 
P^1\cong S^2$, where $\CC P^1$ is the complex projective line. In this generalization 
the projective coordinate (\ref{procor}) becomes a complex coordinate $z$ and coincides with the 
stereographic coordinate (\ref{bondi}) of $S^2$. The projective transformations of $\CC P^1$ then are 
M\"obius transformations (\ref{lor}),\i{Mobius transformation@M\"obius transformation} i.e.\ Lorentz 
transformations in four dimensions.

\subsubsection*{Cross ratios and the Schwarzian derivative}

Given the projective line $\RR P^1$, one may look for projective invariants, i.e.\ quantities 
that 
are left invariant by the transformations (\ref{proto}). For example, consider 
four points on $\RR P^1$ whose projective coordinates are $\zeta_1,\zeta_2,\zeta_3,\zeta_4$ and define their 
\it{cross ratio}\i{cross ratio}
\be
[\zeta_1,\zeta_2,\zeta_3,\zeta_4]
\equiv
\frac{(\zeta_1-\zeta_3)(\zeta_2-\zeta_4)}{(\zeta_1-\zeta_2)(\zeta_3-\zeta_4)}\,.
\nn
\ee
This number is a projective invariant, as one can verify by direct computation. Now take a diffeomorphism 
$f:S^1\rightarrow S^1$, where we think of $S^1$ as a projective 
line (\ref{rps}). This diffeomorphism can be written in terms of the projective coordinate (\ref{procor}), 
giving 
rise to a map $\zeta\mapsto\sfff(\zeta)$. The 
explicit correspondence between $f$ and $\sfff$ follows from (\ref{zetaphi}) and reads
\be
\sfff\big(\!-\!\text{cot}(\phii/2)\big)
=
-\text{cot}\big(f(\phii)/2\big)\,.
\label{fufur}
\ee
In 
general, $\sfff$ is not a projective transformation (\ref{proto}) and therefore spoils the 
projective structure of $S^1$. This can be measured by taking a point with coordinate $\zeta$ on $\RR 
P^1$ together with three other nearby points, then evaluating the change in their cross-ratio under the 
action of 
$\sfff$. Let 
therefore 
$\zeta_1=\zeta+\epsilon$, $\zeta_2=\zeta+2\epsilon$ and $\zeta_3=\zeta+3\epsilon$ be three points close to 
$\zeta$. These points move under the action of $\sfff$, and one can show that
\be
\big[\sfff(\zeta),\sfff(\zeta_1),\sfff(\zeta_2),\sfff(\zeta_3)\big]
=
[\zeta,\zeta_1,\zeta_2,\zeta_3]-2\epsilon^2\, \sfS[\sfff](\zeta)+\cO(|\epsilon|^3)
\nn
\ee
where $\sfS$ is the Schwarzian derivative (\ref{swag}). Thus we have recovered the Schwarzian derivative as a 
measuring device that tells us how much the 
diffeomorphism $\sfff$ spoils the projective structure of $\RR P^1$. From this one concludes:

\paragraph{Proposition.} Let $\sfff,\sfg$ be diffeomorphisms of the projective line. Then\i{Schwarzian 
derivative!PSL2R-invariant@$\PSL$-invariant}
\be
\sfS[\sfff\circ\sfg](\zeta)=\sfS[\sfg](\zeta)
\label{iffiswa}
\ee
if and only if 
$\sfff$ is a projective transformation (\ref{proto}). In particular, $\sfS[\sfff](\zeta)=0$ if and only if 
$\sfff(\zeta)$ is a projective transformation of the form (\ref{proto}).\\

In technical terms, a one-cocycle $\sfS$ on a group $G$ with a subgroup $H$ is said to be 
\it{$H$-relative}\i{relative cocycle} if $\sfS[h]=0$ for all $h\in H$. Thus (\ref{iffiswa}) says that the 
Schwarzian derivative is a $\PSL$-relative cocycle. In 
conformal 
field theory this corresponds to the statement that the Schwarzian derivative is blind to 
M\"obius transformations (\ref{lor}).

\subsubsection*{Schwarzians on the circle}

All the above considerations can be reformulated in terms of the angular coordinate $\phii$ using 
the correspondence (\ref{zetaphi}). Here we
work out this rewriting for projective transformations (\ref{protophi}).
To begin, note that (\ref{protophi}) precisely takes the form 
of a projective transformation (\ref{proto}) in terms of the coordinate $e^{i\phii}$. Accordingly,
\be
\sfS\big[e^{if(\phii)}\big](e^{i\phii})=0.
\label{sizero}
\ee
In order to 
go from (\ref{sizero}) to $\sfS[f](\phii)$ we use the cocycle identity (\ref{swapro}) repeatedly. First 
we 
write
\be
\sfS[f](\phii)
=
\sfS[\log(e^{if(\phii)})](\phii)
\refeq{swapro}
\sfS[e^{if(\phii)}](\phii)
+\left(if'(\phii)e^{if(\phii)}\right)^2\sfS[\log](e^{if(\phii)})\,,
\label{beginning}
\ee
where
$\sfS[\log](x)
\refeq{swag}
\frac{1}{2x^2}$.
The first term on the far right-hand side of (\ref{beginning}) involves
\be
\sfS[e^{if(\phii)}](\phii)
\refeq{swapro}
\sfS[e^{i\phii}](\phii)+\left((e^{i\phii})'\right)^2\sfS[e^{if(\phii)}](e^{i\phii})
\refeq{sizero}
\sfS[e^{i\phii}](\phii)
\refeq{swag}
\demi\,,
\nn
\ee
which finally gives
\be
\sfS[f](\phii)
=
\demi\big[
1-\big(f'(\phii)\big)^2
\big]
\label{swagiff}
\ee
when $f(\phii)$ is given by (\ref{protophi}). We will put this formula to use in the next chapter.\\

Note that these observations can be generalized to infinitely many other families of diffeomorphisms 
of 
the circle. Indeed, pick a positive integer $n\in\NN^*$ and take formula (\ref{protophi}) with $\phii$ 
replaced by $n\phii$ and $f(\phii)$ replaced by $nf(\phii)$:
\be
e^{inf(\phii)}
=
\frac{A e^{in\phii}+B}{\bar B e^{in\phii}+\bar A},
\quad
|A|^2-|B|^2=1.
\label{protophin}
\ee
This defines a diffeomorphism of the 
circle, and the family of such diffeomorphisms also spans a subgroup of $\Diff$ which is locally isomorphic 
to $\SL$. The difference 
with respect to the case $n=1$ discussed above is that the corresponding Lie algebra $\sl$ is generated by 
$\ell_0$, $\ell_n$ and $\ell_{-n}$, and that the actual group spanned by such transformations is an $n$-fold 
cover of $\PSL$; we shall denote this cover by 
$\text{SL}^{(n)}(2,\RR)/\ZZ_2
\equiv
\text{PSL}^{(n)}(2,\RR)$.\i{PSL2R@$\PSL$!n-fold cover@$n$-fold cover} One can also 
verify that the Schwarzian 
derivative of the diffeomorphism $f$ defined by (\ref{protophin}) satisfies
\be
\sfS[f](\phii)
=
\frac{n^2}{2}
\left[
1-(f'(\phii))^2
\right],
\label{swagiffn}
\ee
generalizing the case $n=1$ of (\ref{swagiff}).

\paragraph{Remark.} One can show that the restriction of the Bott-Thurston cocycle (\ref{btcoll}) to the 
$\PSL$ subgroup (\ref{protophi}) coincides with the unique non-trivial two-cocycle on $\PSL$. The 
latter acts on the hyperbolic plane $\HH^2$ by isometries of the form (\ref{proto}), where 
$\zeta\in\CC$ has positive imaginary part,\i{central extension!of PSL2R@of 
$\PSL$}\i{PSL2R@$\PSL$!two-cocycle} and the 
two-cocycle associates with two such transformations 
$\sfff,\sfg$ the area of the triangle with corners $i$, $\sfff^{-1}(i)$ and 
$\sfg^{-1}\circ\sfff^{-1}(i)$. See 
\cite{guieu2007algebre} for details.

\section{The Virasoro group}
\label{sevigo}

We are now in position to 
describe the central extension of $\Diff$. This discussion is crucial for our purposes, as all symmetry 
groups of the later chapters will be variations on the basic Virasoro pattern. As announced above, our 
viewpoint is that the fundamental Virasoro structure is that of the group, from 
which the rest follows. Accordingly we start this section by reviewing general properties of centrally 
extended groups, which we then apply to the Virasoro group whose
adjoint and coadjoint representations follow by differentiation. We also define the Virasoro algebra and 
end by displaying the Kirillov-Kostant Poisson bracket on its dual.

\subsection{Centrally extended groups revisited}
\label{suserevvit}

Let $\hG$ be a central extension of some Lie group $G$, with group operation (\ref{groupop}) in terms of 
some two-cocycle $\sfC$. Here we work out its adjoint and 
coadjoint representations.

\subsubsection*{Adjoint representation}

Since the group $\hG$ consists of pairs $(f,\lambda)$ where $f\in G$ and $\lambda\in\RR$, its Lie algebra 
$\hmg$ consists of pairs $(X,\lambda)$ where $X\in\mg$. The adjoint representation of $\hG$ then follows 
from (\ref{ad}): for $X\in\mg$, $f\in G$ and $\lambda,\mu\in\RR$ we find
\be
\hAd_{(f,\mu)}(X,\lambda)
=
\frac{d}{dt}\left.\Big[
\big(
f\circ e^{tX}\circ f^{-1},t\lambda+\sfC(f,e^{tX})+\sfC(f\circ e^{tX},f^{-1})
\big)
\Big]\right|_{t=0}
\label{puggy}
\ee
where the hat in $\hAd$ stresses that this is the adjoint representation of a centrally extended 
group, as opposed to that of $G$. Note that $\mu$ acts trivially, so we can lighten the notation by 
writing $\hAd_f$ 
instead of $\hAd_{(f,\mu)}$. This follows from the fact that $\hG$ is a central 
extension of $G$ so that ``central elements'' (i.e.\ the real numbers that enter in the second slot of 
$(f,\mu)$) act trivially on everything, which is a general property of centrally extended groups.\\

It then remains to compute the two entries on the 
right-hand side (\ref{puggy}). The first entry yields the adjoint representation of $G$, while the 
second is precisely the expression (\ref{souriau}) defining the Souriau 
cocycle $\sfS$ associated with $\sfC$. We conclude that 
the adjoint representation of $\hG$ reads\i{central extension!adjoint representation}\i{adjoint 
representation!of centrally extended group}
\be
\hAd_f(X,\lambda)
=
\Big(
\Ad_fX,\lambda-\frac{1}{12}\bra \sfS[f],X\ket
\Big)
\label{advir}
\ee
where the ``$\Ad$'' on the right-hand side is the adjoint representation of $G$.

\subsubsection*{Centrally extended algebra}

From the adjoint representation of a group one can read off the Lie brackets (\ref{adg}) of its algebra. Let 
therefore $(X,\lambda)$ and $(Y,\mu)$ belong to the centrally extended Lie 
algebra $\hmg$. Using (\ref{advir}) we find
\be
\big[(X,\lambda),(Y,\mu)\big]
=
\frac{d}{dt}\left[
\Big(
\Ad_{e^{tX}}Y,\mu-\frac{1}{12}\bra \sfS[e^{tX}],Y\ket
\Big)
\right]_{t=0}.
\label{azerbaidjan}
\ee
The first entry on the right-hand side is the same as expression (\ref{adg}) in $G$; accordingly it boils 
down to the standard Lie bracket of $\mg$, which we denote as $[X,Y]$. The second entry involves the 
differential of 
the Souriau cocycle,\i{Souriau cocycle!infinitesimal}\i{infinitesimal Souriau cocycle}
\be
\sfs[X]\equiv\frac{d}{dt}\left.\sfS[e^{tX}]\right|_{t=0}\,,
\label{infisour}
\ee
paired with $Y\in\mg$. We therefore define a two-cocycle $\sfc$ on $\mg$ by
\be
\sfc(X,Y)
\equiv
-\frac{1}{12}\bra\sfs[X],Y\ket
\label{essaimon}
\ee
and the bracket of $\hmg$ takes the centrally extended form (\ref{centb}):\i{central extension!Lie 
bracket}\i{Lie bracket!for centrally extended algebra}
\be
\Big[
(X,\lambda),(Y,\rho)
\Big]
=
\big(
[X,Y],
\sfc(X,Y)
\big)
=
\Big(
[X,Y],
-\frac{1}{12}\bra\sfs[X],Y\ket
\Big).
\label{kellebel}
\ee
The fact that (\ref{essaimon})
is indeed a two-cocycle is inherited from the Souriau cocycle. Note 
that the 
central terms $\lambda,\mu$ commute with everything, as they should. In terms of Lie algebra generators the 
bracket (\ref{kellebel}) takes the general form (\ref{tatb}).

\subsubsection*{Coadjoint representation}

The Lie algebra $\hmg$ is spanned by pairs $(X,\lambda)$, so its dual consists of pairs $(p,c)$ 
where 
$p$ belongs to $\mg^*$ while $c\in\RR$ is a real number, paired with adjoint vectors 
according to
\be
\big<(p,c),(X,\lambda)\big>
=
\bra p,X\ket+c\lambda
\label{difarotux}
\ee
where the pairing $\langle\cdot,\cdot\rangle$ on the right-hand side is that of $\mg^*$ with $\mg$. The 
number $c$ is known as a 
\it{central charge}.\i{central charge} The coadjoint transformation law of $(p,c)$ follows from the 
definition 
(\ref{cocogi}). In 
particular, since central elements act trivially in the adjoint representation (\ref{puggy}), we can safely 
write $\hAd{}^*_{(f,\lambda)}\equiv\hAd{}^*_f$ for any $(f,\lambda)\in\hG$, where the hat on top of $\Ad^*$ 
indicates that we refer to a representation of the centrally extended group. If then $(X,\lambda)\in\hmg$ and 
$(p,c)\in\hmg^*$, one obtains\i{coadjoint representation!of centrally extended group}\i{central 
extension!coadjoint representation}
\begin{eqnarray}
\bra
\hAd{}^*_f(p,c),(X,\lambda)
\ket
& \!\!\refeq{advir} &
\!\!\Big<
(p,c),\Big(\Ad_{f^{-1}},\lambda-\frac{1}{12}\bra \sfS[f^{-1}],X\ket\Big)
\Big>\nn\\
& \!\!\refeq{difarotux} &
\!\!\bra p,\Ad_{f^{-1}}X\ket+c\lambda-\frac{c}{12}\bra \sfS[f^{-1}],X\ket\nn
\end{eqnarray}
where the pairing $\bra\cdot,\cdot\ket$ on the right-hand side of the last equation is the 
centreless pairing of $\mg^*$ with $\mg$. In particular the first term is simply the 
coadjoint representation of $G$. Removing the dependence on $X$, we conclude that
\be
\hAd{}^*_f(p,c)
=
\left(
\Ad^*_fp-\frac{c}{12}\sfS[f^{-1}],c
\right).
\label{covirzero}
\ee
Note that the central charge $c$ is left invariant by the coadjoint representation, as it should. 
Crucially, it also appears in the first entry and thus affects the transformation 
law of $p$.\i{affine module} In abstract terms, formula (\ref{covirzero}) is the affine $G$-module 
(\ref{affi}) associated 
with the 
Souriau cocycle.\\

The coadjoint action can be differentiated, which yields a representation (\ref{pixies}) of the 
Lie algebra $\hmg$. Using (\ref{kellebel}) with the two-cocycle (\ref{essaimon}) one finds
\be
\had{}^*_X(p,c)
=
\left(
\ad^*_Xp+\frac{c}{12}\sfs[X]
,
0
\right)
\label{azizash}
\ee
where the $\ad^*$ on the right-hand side is the coadjoint representation of $\mg$ while $\sfs$ is the 
infinitesimal Souriau cocycle (\ref{infisour}). In the remainder of this section we apply these 
considerations to the Virasoro group.

\subsection{Virasoro group}

\paragraph{Definition.} The \it{Virasoro group}\i{Virasoro group} is the universal central extension of 
$\Diff$. It is diffeomorphic to the product $\Diff\times\RR$ and its elements are pairs $(f,\lambda)$ where 
$f\in\Diff$ and $\lambda\in\RR$, with a group operation (\ref{groupop}) where $\sfC$ is the Bott-Thurston 
cocycle (\ref{btcoll}). Explicitly:\i{central extension!of DiffS1@of $\Diff$}
\be
(f,\lambda)\cdot(g,\mu)
=
\Big(f\circ g,\lambda+\mu+\sfC(f,g)\Big).
\label{vigop}
\ee
We shall denote the Virasoro group by $\hDiff$.\\

As in the previous sections we abuse notation and terminology by simply calling ``Virasoro group'' what is 
really the universal cover of the maximal connected subgroup of the Virasoro group. It should in fact be 
written as $\widehat{\widetilde{\text{Diff}}}{}^+(S^1)$, while 
we 
denote it by $\hDiff$ to reduce clutter.

\subsection{Adjoint representation and Virasoro algebra}
\label{susevira}

As a vector space, the Lie algebra of the Virasoro group is equivalent to the direct sum 
$\Vect\oplus\RR$. In particular, Virasoro adjoint vectors are pairs $(X,\lambda)$ where 
$X=X(\phii)\der_{\phii}$ is a vector field on the circle and $\lambda$ is a real number.\i{adjoint vector} 
The adjoint representation of the Virasoro 
group follows from the group operation 
(\ref{vigop}) and the 
definition (\ref{ad}). Thus the adjoint representation takes the form (\ref{advir}), where 
the adjoint representation of $\Diff$ is the transformation law (\ref{advec}) of vector fields, while $\sfS$ 
is the Schwarzian derivative (\ref{swag}). This result will be instrumental 
in our 
definition of the centrally extended BMS$_3$ group in section \ref{sebms3}.\\

The adjoint representation of a group yields the Lie brackets (\ref{adg}) of its algebra. In the present 
case this definition leads to an awkward sign
(\ref{bakopo}), which we absorb by declaring that the Lie bracket of the 
Virasoro algebra is defined by
\be
\big[(X,\lambda),(Y,\mu)\big]
\equiv
-\frac{d}{dt}\left[\hAd{}_{e^{tX}}(Y,\mu)\right]_{t=0}\,.
\label{minicort}
\ee
With this 
definition formula (\ref{azerbaidjan}) holds up to an overall minus sign on the right-hand side. Using then 
the infinitesimal Schwarzian derivative (\ref{insa}), the pairing (\ref{denpar}) allows us to recognize the 
Gelfand-Fuks cocycle (\ref{gefuks}) in $\bra\sfs[X],Y\ket$. Thus the Lie bracket of the algebra of the 
Virasoro group takes the form (\ref{kellebel}), or explicitly\i{Virasoro algebra}\i{central extension!of 
VectS1@of $\Vect$}
\be
\big[
(X,\lambda),(Y,\mu)
\big]
=
\big([X,Y],\sfc(X,Y)\big)
\label{vibraphone}
\ee
where $[X,Y]$ is the usual Lie bracket of vector fields.

\paragraph{Definition.} The \it{Virasoro algebra} is the Lie algebra 
$\hVect=\Vect\oplus\RR$ endowed with the 
Lie bracket (\ref{vibraphone}).\i{universal central extension}\i{central extension!universal} It is the 
universal central extension of $\Vect$.\footnote{Universality 
follows from the fact that the first real cohomology of $\Vect$ vanishes.}\\

In the physics literature it is customary to rewrite the Virasoro algebra in a form analogous to 
(\ref{witt}). Let us therefore define the basis elements
\be
\cL_m\equiv(\ell_m,0),
\qquad
\cZ\equiv(0,1)
\label{vimodes}
\ee
where the $\ell_m$'s are given by (\ref{ellem}). The bracket (\ref{vibraphone}) then yields 
$[\cZ,\cZ]=[\cZ,\cL_m]=0$, as well as
\be
i[\cL_m,\cL_n]
=
i[(\ell_m,0),(\ell_n,0)]
\refeq{vibraphone}
\left(i[\ell_m,\ell_n],i\sfc(\ell_m,\ell_n)\right)\,.
\nn
\ee
Using the Witt algebra (\ref{witt}) and eq.\ (\ref{gefuksmn}), we can rewrite this as\i{Virasoro 
algebra!commutation relations}
\be
i[\cL_m,\cL_n]=
(m-n)\cL_{m+n}+\frac{\cZ}{12}m^3\delta_{m+n,0}\,,
\label{vira}
\ee
which is indeed the standard expression of the Virasoro algebra 
\cite{Ginsparg:1988ui,DiFrancesco:1997nk,blumenhagen2009introduction}. In this form it can be seen as a 
central extension of the Witt algebra (\ref{witt}), with a central term involving the celebrated 
$m^3\delta_{m+n,0}$. As mentioned below (\ref{gefuksmn}), the latter is a remnant 
of the third derivative of $Y$ in the Gelfand-Fuks cocycle (\ref{gefuks}), while the $\delta_{m+n,0}$ follows 
from the integration over the circle and reflects the fact that the cocycle is invariant under rotations.

\paragraph{Remark.} The generator $\cZ$ of eq.\ (\ref{vimodes}) should rightfully be called the ``central 
charge'' of the Virasoro algebra, since it is a Lie algebra element that commutes with everything. However, 
in keeping with the standard physics terminology, we will also use the word ``central charge'' 
to refer 
to the \it{dual} of $\cZ$, which is just a real number $c$ (see the coadjoint representation 
below). This ambiguous terminology should not lead to any confusion.

\subsection{Coadjoint representation}

\subsubsection*{Coadjoint vectors}

Virasoro adjoint vectors are pairs $(X,\lambda)$ where $X$ is a vector field and 
$\lambda$ a real number. Accordingly the smooth dual of the Virasoro algebra consists of pairs $(p,c)$ where 
$p=p(\phii)d\phii^2$ is a quadratic density and $c\in\RR$ is a real number, paired with adjoint vectors 
according to the centrally extended generalization (\ref{difarotux}) of (\ref{denpar}):\i{coadjoint 
vector!for Virasoro}
\be
\big<
(p,c),(X,\lambda)
\big>
\equiv
\frac{1}{2\pi}\int_0^{2\pi}d\phii\,p(\phii)X(\phii)+c\lambda.
\label{virpar}
\ee
We refer to pairs $(p,c)$ as Virasoro coadjoint vectors; they span the space
$\hVect^*$.\\

It is worth mentioning that coadjoint vectors are crucial physical quantities in all theories enjoying 
$\Diff$ symmetry, and in particular all conformal field theories in two dimensions. Indeed the function 
$p(\phii)$ is nothing but (the chiral component of) a CFT stress tensor,\i{stress tensor} while $c$ is a CFT 
central charge.\i{central charge} 
Expression (\ref{virpar}) then coincides (up to central terms) with the Noether charge associated with a 
symmetry generator $X(\phii)\der_{\phii}$, seen as an infinitesimal conformal 
transformation. 
More precisely, in a CFT on a Lorentzian cylinder, the coordinate $\phii$ would be replaced by one of the 
two light-cone coordinates $x^{\pm}$ and $p(\phii)$ would become $p(x^+)$ or $\bar p(x^-)$. This is 
consistent with the interpretation of coadjoint vectors as Noether currents, thanks to the momentum maps 
of section \ref{seSymPa}.

\paragraph{Remark.} Our notation is somewhat non-standard in that we denote by $p(\phii)$ what would normally 
be written as 
$T(\phii)$, where $T$ stands for the stress tensor. This choice has to do with our motivations: 
we shall see in chapter \ref{c6} that Virasoro coadjoint vectors play the role of \it{supermomentum vectors} 
for the BMS$_3$ group. They will be infinite-dimensional generalizations of the Poincar\'e 
momenta $p_{\mu}$, so the notation $p(\phii)$ is introduced here to suggest thinking of coadjoint vectors as 
quantities related to energy and momentum. In fact this interpretation also holds in CFT, since a stress 
tensor is nothing but an energy-momentum density.

\subsubsection*{Coadjoint representation}

The transformation law of Virasoro coadjoint vectors follows from the definition (\ref{cocogi}). In 
particular, since central elements act trivially in the adjoint representation (\ref{puggy}), we may write 
$\hAd{}^*_{(f,\lambda)}\equiv\hAd{}^*_f$ for any $(f,\lambda)$ belonging to the Virasoro group. If then we 
let $(X,\lambda)\in\hVect$ be an adjoint vector and $(p,c)\in\hVect^*$ be a coadjoint one, formula 
(\ref{covirzero}) still holds upon letting $\sfS$ be the Schwarzian derivative. The central charge $c$ 
is left invariant by the coadjoint representation, as it should, but it also affects the transformation 
law of $p(\phii)$. Accordingly, from now on we often write the coadjoint representation of the 
Virasoro group \it{without} including a second slot for the central charge, since the latter is invariant. 
With this simplified notation formula (\ref{covirzero}) boils down to\i{Virasoro group!coadjoint 
representation}\i{coadjoint representation!of Virasoro group/algebra}
\be
\widehat{\text{Ad}}{}^*_fp
=
\Ad^*_fp-\frac{c}{12}\sfS[f^{-1}].
\label{covira}
\ee
For future reference it will be useful to rewrite this in detail, in terms of 
functions 
on the circle. Evaluating both sides of the equation at a point $\phii$ on the 
circle, we obtain
\be
\big(
\widehat{\text{Ad}}{}^*_fp
\big)(\phii)
=
\big[(f^{-1})'(\phii)\big]^2p(f^{-1}(\phii))-\frac{c}{12}\sfS[f^{-1}](\phii)
\label{covirax}
\ee
by virtue of the centreless coadjoint action (\ref{coadif}). The 
formulas are much simpler if we evaluate eq.\ (\ref{covira}) at $f(\phii)$; using the cocycle identity 
(\ref{swapro}) we find
\be
\boxed{\Bigg.
\big(\widehat{\text{Ad}}{}^*_fp\big)(f(\phii))
=
\frac{1}{(f'(\phii))^2}
\left[
p(\phii)+\frac{c}{12}\sfS[f](\phii)
\right].}
\label{covi}
\ee
This is a transparent expression of the coadjoint representation of the Virasoro group, with 
$\sfS[f](\phii)$ given by (\ref{swag}). It is the most important equation of this chapter. We will sometimes 
refer to the two terms on the right-hand side as the ``homogeneous term'' and the 
``central'' or ``inhomogeneous term'', respectively.\i{stress tensor!transformation law}\i{quasi-primary 
field} The formula can also be recognized as the transformation 
law of a CFT stress tensor $p(\phii)$ with a central charge $c$; in that context $p(\phii)$ 
is said to be a \it{quasi-primary field}\i{quasi-primary field} with weight two.
In the next chapter we will classify the orbits of this 
action, which in chapter \ref{c6} will turn out to be supermomentum orbits labelling BMS 
particles in three dimensions.\\

The differential (\ref{azizash}) of formula (\ref{covi}) is a representation 
of the 
Virasoro algebra. Taking an infinitesimal diffeomorphism $f(\phii)=\phii+\epsilon X(\phii)$, we treat the 
homogeneous term $\Ad^*_fp$ as in (\ref{startingpoint}) and find
$\Ad^*_fp=p-\epsilon(Xp'+2X'p)$
to first order in $\epsilon$ (both sides of the equation are evaluated at the same point). For the 
Schwarzian derivative we use $\sfS[f^{-1}]=-\epsilon X'''$. Defining
\be
\widehat{\text{ad}}{}^*_Xp(\phii)
\equiv
-
\frac{(\widehat{\text{Ad}}{}^*_fp)(\phii)-p(\phii)}{\epsilon}
\nn
\ee
as in (\ref{infitransfo}), we end up with the coadjoint representation of the Virasoro 
algebra:\i{Virasoro algebra!coadjoint representation}
\be
\widehat{\text{ad}}{}^*_Xp
=
Xp'+2X'p-\frac{c}{12}X''',
\label{covinf}
\ee
where both sides are evaluated at the same point.\footnote{Eq.\ (\ref{covinf}) can also be written as 
$\widehat{\text{ad}}{}^*_X(p,c)\equiv-(p,c)\circ\had{}_X$, where the infinitesimal adjoint representation of 
the Virasoro algebra is defined with a sign such that that $\had{}_X(Y,\mu)=\big([X,Y],\sfc(X,Y)\big)$ 
coincides with the bracket (\ref{vibraphone}).} This is the Virasoro version of eq.\ (\ref{azizash}). In 
the homogeneous term we recognize the primary transformation law (\ref{infident}), while the central term 
involves the infinitesimal Schwarzian (\ref{insa}).

\subsection{Kirillov-Kostant bracket}
\label{suseKKK}

In order to make contact with physics, let us describe the Kirillov-Kostant Poisson
bracket (\ref{kkba}) for the Virasoro group. In that case the bracket eats functions on $\hVect^*$, i.e.\ 
functionals $\cF[\,p(\phii),c]$. In practice, since any quadratic density $p(\phii)d\phii^2$ can be 
Fourier-expanded as\i{Fourier series}
\be
p(\phii)=\sum_{m\in\ZZ}p_me^{-im\phii},
\label{surfou}
\ee
the Fourier modes $p_m=p_{-m}^*$ define global coordinates on $\Vect^*$. Any functional $\cF[\,p(\phii),c]$ 
can 
then be 
seen as a function of the variables $p_m$ and $c$, so it suffices to know the Poisson brackets of 
these variables in order to find the Poisson brackets of functions on $\hVect^*$.\\

Now recall the basis (\ref{vimodes}) of the Virasoro algebra and let $\left\{(\cL_m)^*,\cZ^*\right\}$ denote 
the 
corresponding dual basis, such that $\langle \cL_m^*,\cL_n\rangle=\delta_{mn}$ and $\bra\cZ^*,\cZ\ket=1$. 
Using the 
pairing (\ref{virpar}) we find that, as coadjoint vectors,
\be
(\cL_m)^*
=
\big(
(\ell_m)^*,0
\big)
=
\big(
e^{-im\phii}d\phii^2,0
\big),
\qquad
\cZ^*=(0,1).
\label{starbasis}
\ee
Thus, when writing a quadratic density as a 
Fourier series (\ref{surfou}), the parameters $p_m,c$ 
are actually coordinates on $\hVect^*$ defined with respect to the basis (\ref{starbasis}):
\be
\big(p(\phii)d\phii^2,c\big)
=
\sum_{m\in\ZZ}p_m\cL_m^*+c\,\cZ^*.
\nn
\ee
Accordingly, eq.\ (\ref{fabipe}) implies that the Kirillov-Kostant Poisson bracket of these coordinates 
reproduces the Lie brackets (\ref{vira}):\i{Kirillov-Kostant bracket!for Virasoro}\i{Virasoro 
group!Kirillov-Kostant bracket}
\be
i\{p_m,p_n\}
=
(m-n)p_{m+n}+\frac{c}{12}m^3\delta_{m+n,0}\,,
\label{virapois}
\ee
while all Poisson brackets involving the central charge $c$ vanish. The key difference between 
(\ref{virapois}) and 
(\ref{vira}) is that the latter is an abstract Lie bracket, while the former is its phase space 
realization.\\

The bracket (\ref{virapois}) is well-known to physicists. Indeed, the standard way to introduce the 
Virasoro algebra in CFT textbooks is to expand the stress tensor in modes as in 
(\ref{surfou}), and then compute their Poisson brackets. Upon 
quantization, the operator $i\widehat{\{\cdot,\cdot\}}$ coincides with the commutator $[\cdot,\cdot]$ and the 
resulting quantum commutators span a Virasoro algebra (\ref{vira})-(\ref{virapois}), generally 
with a non-zero central charge $c$.\\

Note that each coordinate $p_m$ can be seen as the function on $\hVect^*$ that maps $(p,c)$ on 
$\bra(p,c),\cL_m\ket$.
As 
mentioned below (\ref{virpar}), the object $\bra(p,c),\cL_m\ket$ may be thought of as the Noether charge 
associated with the symmetry generator $\cL_m$, so the Poisson bracket (\ref{virapois}) can be interpreted as 
a 
Poisson bracket of Noether charges. We shall see in chapters \ref{AdS3} and \ref{c6} that the 
Poisson brackets of surface charges in three-dimensional gravity coincide with the Kirillov-Kostant brackets 
on the dual of suitable asymptotic symmetry algebras (albeit with definite values of the central charges).

\newpage
~
\thispagestyle{empty}

\chapter{Virasoro coadjoint orbits}
\label{c4bis}
\markboth{}{\small{\chaptername~\thechapter. Virasoro orbits}}

In this chapter we classify the coadjoint orbits of the Virasoro group. Aside from their usefulness in the 
study of conformal symmetry, they are crucial for our purposes because they will turn out to coincide with 
the supermomentum orbits that classify BMS$_3$ particles. As we shall see, despite being 
infinite-dimensional, these orbits behave very much like the finite-dimensional coadjoint orbits of 
$\SL$.\\

The plan is as follows. In section \ref{secovobi} we describe the problem and explain how it can be addressed 
in terms of two invariant quantities, namely the conjugacy class of a certain monodromy matrix and the 
winding number of a related curve taking its values in a circle. We then use this approach in section 
\ref{sevirorepss} to display explicit orbit representatives. Finally, section \ref{senepost} is devoted to 
a discussion of energy positivity in the Virasoro context.\\

Coadjoint orbits of the Virasoro group were first classified in \cite{Lazutkin,Kirillov1981} and are 
described in many later papers \cite{segal1981,Witten:1987ty,Bakas1988,GayBalmaz} and textbooks 
\cite{guieu2007algebre,khesin2008geometry}. The presentation of this chapter 
follows \cite{Balog:1997zz}.

\section{Coadjoint orbits of the Virasoro group}
\label{secovobi}

In this section we explain the methods used to classify coadjoint orbits of the Virasoro group. We start by 
describing the simple (but pathological) classification of orbits at zero central charge, before discussing 
certain basic aspects of centrally extended orbits. We then turn to the correspondence between Virasoro 
coadjoint vectors and Hill's operators, which yields two invariant quantities that can be used to classify 
the orbits. These invariants are (i) the conjugacy class of a monodromy matrix and (ii) the winding number of 
a curve on the real line whose target space is a circle.

\subsection{Centreless coadjoint orbits}
\label{susebaby}

We start our investigation with a problem that is much simpler than the full classification of coadjoint 
orbits of the Virasoro group $\hDiff$, namely the classification of orbits at vanishing central charge, $c=0$.
Those are orbits of the centreless group $\Diff=\Diffc$, whose
coadjoint representation is given by eq.\ (\ref{coadif}).\\

Let us pick a coadjoint vector $\big(p(\phii)d\phii^2,c=0\big)$ and denote its coadjoint orbit by 
$\cW_{(p,0)}$. For now, suppose for simplicity that 
$p(\phii)$ is strictly positive for all $\phii\in[0,2\pi]$. One can then verify that the 
integral\i{DiffS1@$\Diff$!coadjoint orbit}\i{coadjoint orbit!of $\Diff$}
\be
\sqrt{M}
\equiv
\frac{1}{2\pi}\int_0^{2\pi}d\phii\sqrt{p(\phii)}
\label{qupi}
\ee
is invariant under the action (\ref{coadif}) of $\Diff$ on $p$. This actually follows from the fact that $p$ 
is a 
quadratic density, so its square root is a one-form and can be integrated on the circle in a 
$\Diff$-invariant way. With this notation the diffeomorphism
\be
f(\phii)\equiv\int_0^{\phii}d\phi\sqrt{\frac{p(\phi)}{M}}
\label{voodoo}
\ee
maps $p(\phii)$ on the constant coadjoint vector $f\cdot p=M$. Thus any strictly positive 
coadjoint vector $p$ belongs to the orbit of a constant $M$ given by (\ref{qupi}), which is the 
``mass'' associated with $p(\phii)$. The stabilizer of $p(\phii)=M$ is 
the set of 
diffeomorphisms $f$ such that
$M=M/(f'(\phii))^2$.
Since we set $f'>0$ to preserve orientation, the only solution is $f'=1$ and the 
stabilizer of $p=M$ is the group $\un$ of rigid rotations $f(\phii)=\phii+\theta$. Thus the orbit of any 
strictly positive coadjoint vector is diffeomorphic to the quotient space $\Diff/S^1$\i{DiffS1S1@$\Diff/S^1$}.
The same analysis applies, up to signs, to strictly negative coadjoint vectors. Note that $\Diff/S^1$ has 
codimension one in $\Diff$.\\

Of course, coadjoint vectors may well vanish at certain points of the circle, and in particular they can 
change sign; the previous analysis must then be modified. For example, suppose $p(\phii)$ is everywhere 
non-negative, but vanishes at the point $\phii=0$. We will say that $p(\phii)$ has a ``double zero'' at 
$\phii=0$,\i{double zero} since both $p(\phii)$ and $p'(\phii)$ vanish there. Then the integral (\ref{qupi}) 
is still 
invariant on the orbit of $p$, but it is no longer true that $p(\phii)$ can be mapped on a constant because 
the corresponding would-be diffeomorphism (\ref{voodoo}) is degenerate: its derivative vanishes at 
$\phii=0$. We conclude that the orbit of $p$ is now specified by two invariant statements: first, the fact 
that the integral of $\sqrt{p}$ takes the value (\ref{qupi}), and second, the fact that $p(\phii)$ has one 
double zero. More generally, if $p(\phii)$ is everywhere non-negative but has $N$ double zeros at the points 
$\phii=\phii_1,...,\phii_N$, then the $N$ integrals
\be
\int_{\phii_i}^{\phii_{i+1}}d\phii\sqrt{p(\phii)}
\nn
\ee
(where $\phii_{N+1}\equiv\phii_1$) are invariants specifying the orbit of $p$. The orbit is labelled by the 
values of these integrals together with their ordering (which is $\Diffp$-invariant) 
and the statement that all its elements have exactly $N$ double zeros. In particular, the orbit has 
codimension $N$ in $\Diff$ since it is specified by $N$ continuous parameters.\\

A similar treatment can be applied to coadjoint vectors that change sign on the circle, i.e.\ functions 
$p(\phii)$ having \it{simple} zeros (where $p'$ does not vanish).\i{simple zero} The number of such points is 
always even 
since $p(\phii)$ is $2\pi$-periodic, so let us suppose $p(\phii)$ has $2N'$ simple zeros. Then 
the 
integral of $\sqrt{|p(\phii)|}$ between any two consecutive zeros is $\Diff$-invariant as before, so the 
orbit of $p$ is specified by the $2N'$ values of these integrals, by their ordering and by the sign of 
$p(\phii)$ on one of the intervals where it does not vanish. From this we also deduce the general 
classification of orbits for quadratic densities with a finite number of zeros: if $p(\phii)$ has $N$ double 
zeros and $2N'$ simple zeros, its orbit is specified by the values of $N+2N'$ integral invariants of the form
\be
\int_{\phii_i}^{\phii_{i+1}}d\phii\sqrt{|p(\phii)|}
\label{tinderly}
\ee
(where $\phii_i$ and $\phii_{i+1}$ are any two consecutive zeros), together with the ordering of these 
invariants, the specification of whether the points $\phii_i$ and $\phii_{i+1}$ are simple or double zeros, 
and the sign of $p$ on a given interval, say $[\phii_1,\phii_2]$. The orbit of $p$ then has codimension 
$N+2N'$ in $\Diff$; in particular there exist orbits with arbitrarily high codimension.\\

As we can see here, centreless coadjoint orbits are somewhat messy: they can be specified 
by an arbitrarily large number of parameters. Besides, we haven't even discussed the case of coadjoint 
vectors $p(\phii)$ that vanish on a whole open set\i{open set} in $S^1$ --- these have an 
infinite-dimensional little 
group and their orbits have infinite codimension in $\Diff$. In 
particular, 
coadjoint orbits can have arbitrary (even or odd) codimension in $\Diff$. This is in sharp contrast with 
finite-dimensional Lie groups, where all coadjoint orbits are even-dimensional since they are symplectic 
manifolds. 
In the case of $\Diff$, coadjoint orbits are still symplectic, but they need not satisfy ``codimension 
parity'':\i{codimension parity}\i{symplectic manifold!codimension parity} the fact that a given orbit has 
codimension $N$ does not imply that there are no orbits with 
codimension $N\pm 1$. We shall see that this pathology does not occur when the Virasoro central charge is 
non-zero, where all orbits have codimension one or three.

\subsection{Basic properties of centrally extended orbits}
\label{suseconoze}

Let us turn to coadjoint orbits of the Virasoro group at \it{non-zero} central charge $c\neq0$.
From now on we pick some non-zero value for $c$ and we stick to it; for definiteness we take $c>0$, although 
all our considerations also apply to $c<0$ after a few straightforward sign modifications. In principle our 
goal is to address the following problems:\i{classification!of coadjoint orbits}\i{Virasoro 
group!coadjoint orbit}
\begin{enumerate}
\item Classify all Virasoro coadjoint orbits with central charge $c$.
\item Find a non-redundant, exhaustive set of orbit representatives.
\item Given a coadjoint vector $p(\phii)d\phii^2$ (at central charge $c$), write down the 
diffeomorphism $f\in\Diff$ that maps it on one of the orbit representatives.
\end{enumerate}
If we manage to satisfy these criteria, we will have fully classified the orbits of the Virasoro group (at 
non-zero 
central charge).\\

While this task was relatively easy in the centreless case thanks to the integral invariants 
(\ref{tinderly}), 
it turns out to be much more complicated in the centrally extended case; the 
remainder of this chapter is devoted to orbits at non-zero central charge, where the classification will rely 
on elaborate 
techniques involving monodromy matrices. For now we simply describe the most elementary aspects of some of 
these orbits.

\subsubsection*{Stabilizers}

Suppose we are given a coadjoint vector $\big(p(\phii)d\phii^2,c\big)$. Since the central charge is 
invariant, the orbit of $(p,c)$ under the Virasoro group can be represented as\i{Virasoro orbits}\i{coadjoint 
orbit!of Virasoro group}
\be
\cW_{(p,c)}
=
\Big\{
\hAd{}^*_fp
\Big|f\in\Diff
\Big\},
\label{virorbit}
\ee
where $\hAd^*_fp$ is given by (\ref{covi}). It is an infinite-dimensional manifold, so obtaining information 
on its geometry sounds at first like an impossible task. Accordingly, instead of actually trying to picture 
the orbit as such, let us look for 
the stabilizer $G_p$ of $p$, which is a subgroup of $\Diff$ such that
\be
\cW_{(p,c)}
\cong
\hDiff/(G_p\times\RR)
\cong
\Diff/G_p\,.
\label{whoopy}
\ee
The stabilizer consists of diffeomorphisms $f(\phii)$ such 
that
\be
p(f(\phii))
=
\frac{1}{(f'(\phii))^2}
\left[
p(\phii)+\frac{c}{12}\sfS[f](\phii)
\right].
\label{stabbeq}
\ee
Given $p(\phii)$, this is a highly non-linear differential equation for 
$f(\phii)$; if we could actually solve it, we would know the stabilizer.\\

To make things simpler let us look only for the Lie algebra of the stabilizer, rather than the 
stabilizer itself. This algebra is spanned by vector fields $X$ that leave $p(\phii)$ invariant, which 
according to (\ref{covinf}) amounts to the requirement\i{stabilizer!for Virasoro coadjoint 
vector}\i{Virasoro stabilizer}
\be
Xp'+2X'p-\frac{c}{12}X'''=0\,.
\label{stabbo}
\ee
This is already a lot easier than eq.\ (\ref{stabbeq}): it is a \it{linear} third order equation for the 
function $X(\phii)$, assuming that the function $p(\phii)$ is known. A number of important consequences 
follow from this equation. The first is that, for non-zero $c$, it admits at most three linearly independent 
solutions:

\paragraph{Lemma.} The stabilizer of $p(\phii)d\phii^2$ at non-zero central charge is at most 
three-dimensional.\\

This is already a sharp difference with respect to the centreless case, where stabilizers had 
arbitrarily high dimension. If we were on a line 
rather than a circle, we would actually conclude from (\ref{stabbo}) that the stabilizer is \it{always} 
three-dimensional; but the requirement of periodicity restricts 
the space of allowed solutions $X$ for a given $p$, as we shall see momentarily.

\subsubsection*{Constant coadjoint vectors}

\begin{figure}[h]
\centering
\includegraphics[width=0.30\textwidth]{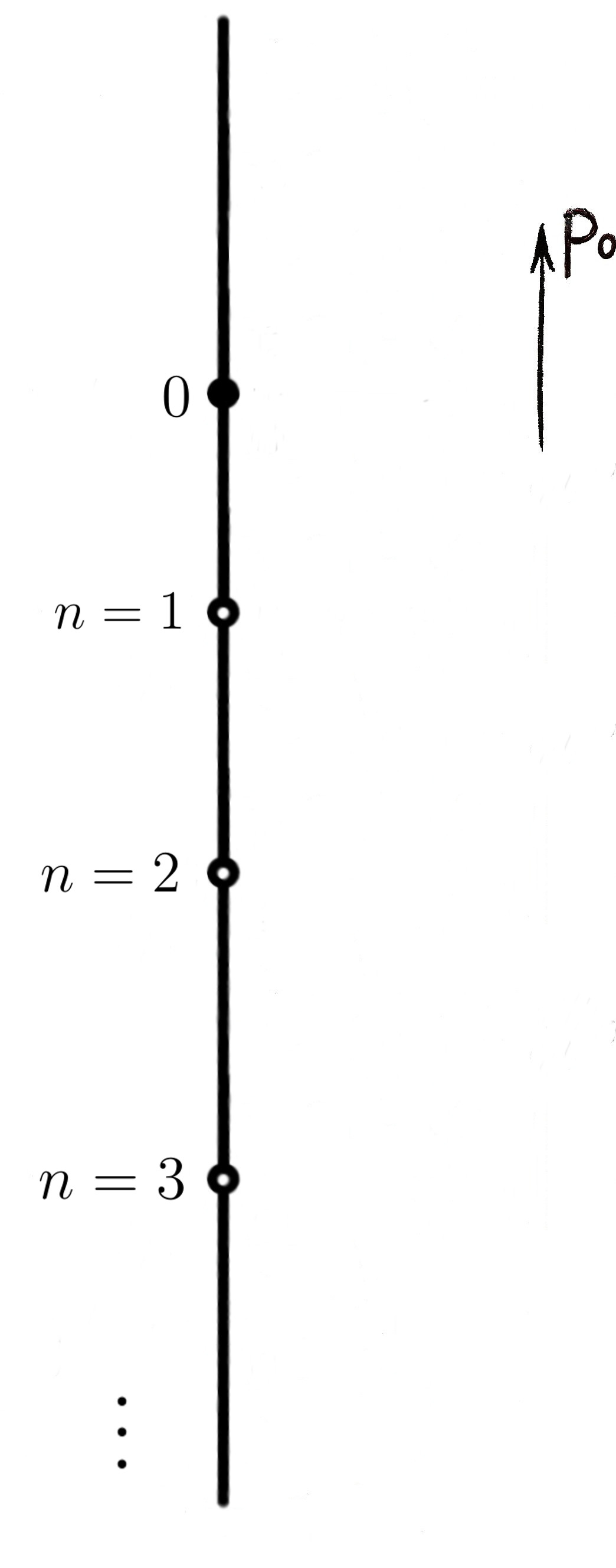}
\caption{The map of Virasoro coadjoint orbits with constant representatives. The open dots labelled by 
$n=1,2,3,...$ indicate the location of the exceptional points $-c/24$, $-4c/24$, $-9c/24$, 
etc.\label{figorbivir}}
\end{figure}

It is worth exploring the solutions of (\ref{stabbo}) in the simple case where $p(\phii)=p_0$ is a 
constant.\i{p0@$p_0$ (constant supermomentum)} 
The equation then reduces to
\be
X'''-\frac{24p_0}{c}X'=0\,,
\nn
\ee
whose general solution is a sum of exponentials
\be
X(\phii)
=
A+B\,e^{\sqrt{\frac{24p_0}{c}}\,\phii}+C\,e^{-\sqrt{\frac{24p_0}{c}}\,\phii}
\label{listab}
\ee
where $A$ is real while $B$ and $C$ are generally complex coefficients, being understood that 
$\sqrt{24p_0/c}$ 
is purely imaginary when $p_0<0$. For generic values of $p_0$,\i{generic constant} the 
only $2\pi$-periodic 
solution of this type is a constant $X(\phii)=\text{const}$. In that case the stabilizer is 
one-dimensional, and consists of rigid rotations of the circle.
But there also exist exceptional values of $p_0$ whose stabilizer is larger, namely\i{exceptional 
constant}
\be
p_0=-\frac{n^2c}{24}
\label{pn}
\ee
where $n\in\NN^*$ is a positive integer. At such values the exponentials in (\ref{listab}) are $e^{\pm 
in\phii}$ and 
the corresponding vector field $X$ is automatically $2\pi$-periodic (and real upon setting $C=B^*$). Thus, 
for exceptional constants (\ref{pn}), the stabilizer is three-dimensional. Its Lie 
algebra is isomorphic to $\sl$; we will see below that the stabilizer itself is an $n$-fold cover of 
$\PSL$. In particular, orbits of generic constants $p_0$ are radically different from orbits of exceptional 
constants (\ref{pn}). The situation is depicted in fig.\ \ref{figorbivir}.\\

If one thinks of $\Diff$ as a group of conformal transformations and identifies 
$p(\phii)$ with the stress tensor of a CFT, one can 
recognize in (\ref{pn}) with $n=1$ the vacuum value of a stress tensor on the cylinder:\i{vacuum coadjoint 
vector}\i{Virasoro vacuum}
\be
p_{\text{vac}}=-\frac{c}{24}\,.
\label{pivacu}
\ee
We shall see below that this interpretation is indeed correct, as the coadjoint orbit of $p_{\text{vac}}$ 
turns out to be the lowest-lying orbit with energy bounded from below and has its
energy minimum at $p_{\text{vac}}$.
By contrast, the points (\ref{pn}) with $n\geq2$ belong to orbits with energy 
unbounded from below.

\subsection{Hill's equation and monodromy}
\label{sehill}

At this point we have seen the most 
basic features of Virasoro orbits at $c\neq0$; we now describe the first step of the complete 
classification by explaining the relation between orbits and monodromy matrices for solutions of Hill's 
equations. To lighten the notation, from now on we write 
$\hAd{}^*_fp\equiv f\cdot p$.\\

Until the next section it will be convenient to think of the coordinate 
$\phii$ as spanning the real line $\RR$, without identification $\phii\sim\phii+2\pi$ (this will be 
justified below). Accordingly we now reinstate the notation $\Diffc$ for the universal cover of 
the group of orientation-preserving diffeomorphisms of the circle and we think of it as a subgroup of 
$\text{Diff}^+(\RR)$. Functions on the circle then are $2\pi$-periodic functions on $\RR$. We also sometimes 
use the words ``conformally invariant'' or ``conformally equivalent'' to refer to objects 
that are $\Diffc$-invariant or $\Diffc$-equivalent, respectively.

\subsubsection*{Virasoro symmetry of Hill's equation}

The key idea of the classification is the following: given a coadjoint vector 
$(p,c)$, we can associate with it a differential operator
\be
\Delta_{(p,c)}
\equiv
-\frac{c}{6}\frac{\der^2}{\der\phii^2}+p(\phii)
\label{hillo}
\ee
where $\phii$ is a coordinate on the real line, $p(\phii)$ is $2\pi$-periodic, and the operator 
$\Delta_{(p,c)}$ acts on suitable densities on the real line. The normalization in front of $\der_{\phii}^2$ 
is chosen so as to ensure that 
the operator has good transformation properties under $\Diffc$, as we shall see below. Note that the 
crucial term $\der_{\phii}^2$ disappears if $c=0$, which is why the considerations that follow 
apply exclusively to orbits with non-zero central charge.

\paragraph{Definition.} Let $(p,c)$ be a Virasoro coadjoint vector with $c\neq0$. The associated \it{Hill's 
equation} is the second-order, linear differential equation\i{Hill's equation}
\be
-\frac{c}{6}\psi''(\phii)+p(\phii)\psi(\phii)=0
\label{hill}
\ee
for the real-valued function $\psi(\phii)$ on the real line. With the notation (\ref{hillo}) this is 
just the statement 
$\Delta_{(p,c)}\cdot\psi=0$.\\

Hill's equation may be seen as a non-relativistic Schr\"odinger equation on the real line for a 
``wavefunction'' 
$\psi(\phii)$ with a 
periodic ``potential energy'' $p(\phii)$, up to the fact that 
$\psi$ is real and need not be square-integrable.\i{Sturm-Liouville operator}\footnote{Note that Hill's 
operator (\ref{hillo}) coincides 
with a Sturm-Liouville operator with periodic potential.} Thus we can 
associate an equation (\ref{hill}) with each coadjoint vector $(p,c)$, and vice-versa.\\

The transformation law of $(p,c)$ under $\Diffc$ determines that of Hill's 
operator 
(\ref{hillo}). Using (\ref{covira}) and the centreless transformation law (\ref{coadif}), we find
\be
\Delta_{(f\cdot p,c)}(\phii)
=
-\frac{c}{6}\der_{\phii}^2+
p(f^{-1}(\phii))\big((f^{-1})'(\phii)\big)^2-\frac{c}{12}\sfS[f^{-1}](\phii)\,,
\label{hilfim}
\ee
which is indeed very different from the original operator (\ref{hillo}). The key point, however, is that the 
associated Hill's equation (\ref{hill}) can be made conformally invariant by choosing a suitable 
transformation 
law for $\psi(\phii)$:

\paragraph{Lemma.} If $\psi(\phii)$ is a density with weight $-1/2$ on the real line, then Hill's 
equation (\ref{hill}) is invariant under $\Diffc$.

\begin{proof}
To simplify formulas, let us act on Hill's operator with a diffeomorphism $f^{-1}$ rather than $f$ so that 
$\big(f^{-1}\cdot\psi\big)(\phii)=\psi(f(\phii))(f'(\phii))^{-1/2}$. Then $f^{-1}$ maps the left-hand side of 
Hill's equation (\ref{hill}) on
\be
\begin{split}
-\frac{c}{6}\der_{\phii}^2\big[\psi(f(\phii))(f'(\phii))^{-1/2}\big]
+p(f(\phii))\psi(f(\phii))(f'(\phii))^{3/2}\qquad\\
\qquad-\frac{c}{12}\sfS[f](\phii)\,(f'(\phii))^{-1/2}\psi(f(\phii))
\end{split}
\label{thisguy}
\ee
where the term with a second derivative can be written as
\be
\der_{\phii}^2\big[\psi(f(\phii))(f'(\phii))^{-1/2}\big]
=
\psi''(f(\phii))(f'(\phii))^{3/2}
-\demi\psi(f(\phii))(f'(\phii))^{-1/2}\sfS[f](\phii)\,.
\nn
\ee
Here the term involving the Schwarzian derivative cancels that of (\ref{thisguy}); the latter expression can 
therefore be 
rewritten as
\be
\big(f'(\phii)\big)^{3/2}
\left[
-\frac{c}{6}\psi''(\phii)+p(\phii)\psi(\phii)
\right]\,.
\nn
\ee
Provided $\psi$ solves (\ref{hill}), this vanishes.
\end{proof}

As a consequence of this lemma, the map that associates Hill's equations with Virasoro coadjoint vectors 
$(p,c)$ is $\Diffc$-invariant. Thus, Hill's equation is an invariant associated with each 
coadjoint orbit of the Virasoro group, and classifying Virasoro orbits is equivalent to 
classifying all $\Diffc$-inequivalent Hill's equations.

\subsubsection*{Monodromy}

Hill's equation (\ref{hill}) is a second-order linear differential equation, so its solutions span a 
two-dimensional vector space. Let $\psi_1$ and $\psi_2$ be two linearly independent solutions. We define 
their \it{Wronskian} as\i{Wronskian}
\be
W\equiv\text{det}\bmm \psi_1 & \psi_2 \\ \psi_1' & \psi_2' \emm
=
\psi_1\psi_2'-\psi_2\psi_1'\,.
\label{Wikk}
\ee
The Wronskian is constant on the real line ($W'=0$) by virtue of Hill's equation. Furthermore $W$ does not 
vanish since $\psi_1$ and $\psi_2$ are linearly independent. (Conversely, if the Wronskian does not vanish, 
then the solutions $\psi_1,\psi_2$ are linearly independent.) Thus we can always choose
\be
W[\psi_1,\psi_2]=-1\,.
\label{wroko}
\ee
We will refer to this equality as the ``Wronskian condition'' and to the solutions that satisfy it as being 
``normalized''. Note that the Wronskian (\ref{Wikk}) is invariant under $\Diffc$ when the 
$\psi_i$'s transform as densities of weight $-1/2$, regardless of them solving Hill's equation:
\be
W[f\cdot\psi_1,f\cdot\psi_2](\phii)=W[\psi_1,\psi_2](f^{-1}(\phii))\,.
\nn
\ee
In particular, 
for solutions of Hill's equation, the Wronskian is constant:
\be
W[f\cdot\psi_1,f\cdot\psi_2]=W[\psi_1,\psi_2]
\qquad
\text{when $\psi_1,\psi_2$ solve (\ref{hill}).}
\label{wokinvar}
\ee
\vspace{.1cm}

Hill's equation is a differential equation on the real line 
$\phii\in\RR$ with a $2\pi$-periodic potential $p(\phii)$. Its solutions need not be periodic, 
but they do satisfy certain constraints due to the periodicity of $p(\phii)$:

\paragraph{Lemma.} Let $p(\phii)$ be $2\pi$-periodic and let $\psi_1,\psi_2$ be linearly independent 
solutions of Hill's equation (\ref{hill}). Then there exists a \it{monodromy matrix}\i{monodromy} 
$\sfM\in\SL$ such that, for any $\phii\in\RR$,\i{Hill's equation!monodromy}
\be
\bmm \psi_1(\phii+2\pi) \\ \psi_2(\phii+2\pi) \emm
=
\sfM\cdot
\bmm \psi_1(\phii) \\ \psi_2(\phii) \emm.
\label{demon}
\ee

\begin{proof}
Let $\psi_1,\psi_2$ be two linearly independent solutions of (\ref{hill}), and define 
$\tilde\psi_i(\phii)\equiv\psi_i(\phii+2\pi)$ for $i=1,2$. Then the Wronskian associated with 
$\tilde\psi_{1,2}$ takes the same value as that of $\psi_{1,2}$; furthermore, the functions $\tilde\psi_i$ 
solve the same Hill's 
equation as the functions $\psi_i$ since $p(\phii)$ is $2\pi$-periodic. This implies that there exists a 
real matrix $\sfM$ such that (\ref{demon}) holds
for any $\phii\in\RR$. Since the $\psi_i$'s and the $\tilde\psi_i$'s have the same Wronskian, $\sfM$ must 
have 
unit determinant.
\end{proof}

Thus we can associate a monodromy matrix with any Virasoro coadjoint vector and any pair of (normalized) 
solutions of 
the corresponding Hill's equation. From now 
on we use the notation\i{solution vector}
\be
\Psi\equiv\bmm \psi_1 \\ \psi_2 \emm
\label{pizizi}
\ee
for the ``solution vector'' associated with the basis of solutions $\psi_1,\psi_2$. Relation (\ref{demon}) 
then becomes $\Psi(\phii+2\pi)=\sfM\cdot\Psi(\phii)$. If we were to choose another normalized basis of 
solutions, say $(\phi_1,\phi_2)=\Phi^t$, there would be a linear relation $\Phi=S\cdot\Psi$ between 
solution vectors, for some matrix $S\in\SL$. Accordingly the monodromy matrix $\sfM_{\Phi}$ associated with 
$\Phi$ would 
be 
related to the monodromy $\sfM_{\Psi}$ of $\Psi$ by
$\sfM_{\Phi}=S\sfM_{\Psi}S^{-1}$.
Thus the monodromy matrix changes by conjugation in $\SL$ under changes of bases of normalized solutions. In 
particular, the \it{conjugacy class}\i{conjugacy class} of $\sfM$,
\be
[\sfM]\equiv
\big\{S\sfM S^{-1}\big|S\in\SL\big\}\,,
\nn
\ee
is invariant under changes of bases. It depends only on the function $p(\phii)$, and not on 
the choice of solutions $\Psi$.\\

We have shown above that Hill's equation is invariant under $\Diffc$ in the sense that, if $\psi$ solves 
the equation with a potential $p(\phii)$, then $f\cdot\psi$ solves the same equation with a potential $f\cdot 
p$. In addition we have seen in (\ref{wokinvar}) that this transformation preserves the Wronskian 
condition, so that normalized solutions remain normalized under $\Diffc$. Accordingly, if $\Psi$ is a 
normalized solution vector for the potential $p$, then $f\cdot\Psi$ is 
a normalized solution vector for $f\cdot p$. And now comes the key argument: since 
both Hill's equation and the action of $\Diffc$ on $\Psi$ are linear, the monodromy matrix of $f\cdot\Psi$ 
coincides with that of $\Psi$. We therefore conclude:

\paragraph{Theorem.} Let $c\neq0$ and denote by $[\sfM]_{(p,c)}$ the conjugacy class of any monodromy matrix 
$\sfM$
associated with the Hill's equation (\ref{hill}) specified by $p(\phii)$ and $c$. Then there is a 
well-defined map
\be
\left\{\text{Virasoro orbits at central charge $c$}\right\}
\rightarrow
\left\{\text{Conjugacy classes of $\SL$}\right\}
\label{mappy}
\ee
that associates with a coadjoint orbit $\cW_{(p,c)}$ the equivalence class $[\sfM]_{(p,c)}$ of the 
corresponding monodromy matrix. In particular, Virasoro coadjoint vectors with the same central charge 
but non-conjugate monodromy matrices do not belong to the same orbit.\\

This result illustrates the power of Hill's operators. It provides 
a rough classification of Virasoro orbits by allowing us to distinguish orbits with non-conjugate 
monodromies and may be seen as an infinite-dimensional analogue of the classification of coadjoint orbits of 
$\SL$ according to the value of the ``mass squared''. In particular the trace $\text{Tr}(\sfM)$ is 
a conformally invariant quantity. However, the classification is not precise in that two 
orbits whose monodromy matrices are 
conjugate may well be different: the map (\ref{mappy}) need not be injective (and we shall see below 
that it is not). To 
make further progress we need to investigate Hill's equation in more detail.\\

For future reference, note the following: thanks to the fact that the conjugacy class of the monodromy matrix 
is independent of the 
choice of a solution vector $\Psi$ for Hill's equation associated with $(p,c)$, one can write its trace as a 
Wilson loop\i{monodromy!as Wilson loop}\i{Wilson loop}
\be
\text{Tr}(\sfM)
=
\text{Tr}\left(
P\,\exp\left[\int_0^{2\pi}d\phii\bmm 0 & 1 \\ 6p(\phii)/c & 0 \emm\right]
\right)
\label{wilson}
\ee
where $P$ denotes path ordering. This quantity is conformally invariant, so one can replace 
$(p,c)$ by any coadjoint vector $(q,c)$ belonging to its orbit without affecting the value of (\ref{wilson}). 
In particular, if $(p,c)$ belongs to the orbit of a constant coadjoint vector $(p_0,c)$ with positive $p_0$, 
the trace reads
\be
\text{Tr}(\sfM)
=
2\cosh\bigg(2\pi\sqrt{\frac{6p_0}{c}}\;\bigg).
\label{traCos}
\ee
The same formula holds for negative $p_0$, with $\cosh(ix)=\cos(x)$. We will put it to use in section 
\ref{sebmspar} when defining the mass of BMS$_3$ particles.

\subsubsection*{Hill's equation and stabilizers}

The stabilizer of a coadjoint vector $(p,c)$ consists of diffeomorphisms that satisfy (\ref{stabbeq}). Let us 
see how this information is related to Hill's equation (\ref{hill}). First note
that, if $\psi_1$ and $\psi_2$ are linearly independent solutions of Hill's equation, 
then the combinations
\be
\psi_1^2,
\quad
\psi_1\psi_2,
\quad
\psi_2^2
\label{hillprod}
\ee
all solve the stabilizer equation (\ref{stabbo}). These products are generally not $2\pi$-periodic and 
therefore do not represent vector fields on the circle, but one can show that there always exist either one 
or 
three $2\pi$-periodic 
linear combinations of these products.\i{Virasoro stabilizer} This confirms our earlier observation that the 
stabilizer of all 
orbits is either one- or three-dimensional. Note that, being $-1/2$-densities on the circle, the products 
(\ref{hillprod}) were bound to be densities of weight $-1$, i.e.\ vector fields.\\

Let us now see how the stabilizer $G_p$ of $(p,c)$ is described in the Hill language. If $f\in G_p$ and if 
$\Psi$ is
a normalized solution vector of Hill's equation associated 
with $(p,c)$,
the action of $G_p$ on $\Psi$ is such that 
$f\cdot\Psi$ provides another normalized solution vector for the same equation. Accordingly there exists some 
(constant) $\SL$ matrix $A_f$ such that
\be
f\cdot\Psi=A_f^{-1}\Psi.
\label{ficidop}
\ee
In addition we have seen that the action of $\Diff$ leaves the monodromy matrix invariant, so the monodromy 
of $f\cdot\Psi$ coincides with the monodromy $\sfM$ of $\Psi$. Combining this statement with (\ref{ficidop}) 
we 
conclude that $A_f^{-1}\sfM A_f=\sfM$,
which is to say that $A_f$ belongs to the stabilizer $G_{\sfM}$ of $\sfM$ with respect to conjugation. In 
addition the inversion $A_f^{-1}$ in (\ref{ficidop}) ensures that $A_{fg}=A_fA_g$, so we conclude:

\paragraph{Lemma.} Let $(p,c)$ be a Virasoro coadjoint vector with $c\neq0$, $\Psi$ a normalized solution 
vector of the associated Hill's equation. Let $G_p$ be the stabilizer of $p$ for the coadjoint action 
(\ref{covi}) and let $G_{\sfM}$ be the stabilizer of $\sfM$ for conjugation. Then the map
\be
\cA:G_p\rightarrow G_{\sfM}:f\mapsto\cA(f)\equiv A_f
\label{camorphi}
\ee
defined by (\ref{ficidop}) is a homomorphism.\\

This map relates the stabilizer of $p$ to that of the corresponding 
monodromy matrix.\i{Hill's equation!inequivalent solutions} In particular it allows us to classify the 
conformally inequivalent solutions of Hill's 
equation at fixed $(p,c)$. 
Indeed, the set of normalized solution vectors of 
Hill's equation at $p$ with fixed monodromy $\sfM$ is in one-to-one correspondence with the elements of 
$G_{\sfM}$, so the set of orbits of the stabilizer $G_p$ in that set of solutions is a quotient
\be
G_{\sfM}/\text{Im}(\cA)
\label{gima}
\ee
where $\text{Im}(\cA)$ is the image of (\ref{camorphi}). Two solution vectors are conformally equivalent if 
and only if they belong to 
the same orbit under $G_p$, i.e.\ if they define the same point in (\ref{gima}).

\paragraph{Remark.} The fact that the products of half-densities (\ref{hillprod}) solving Hill's 
equation produce integer densities solving the stabilizer equation (\ref{stabbo}) is reminiscent of the fact 
that the ``square'' of two Killing spinors is a Killing vector. This correspondence is exactly 
realized in three-dimensional gravity: eq.\ 
(\ref{stabbo}) turns out to coincide with the Killing equation expressed in terms of a suitable component 
$X$ of a vector field on space-time, while Hill's equation (\ref{hill}) corresponds to the Killing spinor 
equation for a suitable spinor component (see e.g.\ eq.\ (16) in \cite{Barnich:2014cwa}).

\subsection{Winding number}

The conjugacy class of monodromy matrices provides a continuous parameter that roughly classifies Virasoro 
orbits. We now describe a second invariant quantity which, combined with monodromies, will provide a precise 
classification of orbits. This second invariant turns out to be the discrete winding number of a path 
in the circle.\\

Let $\psi_1$ and $\psi_2$ be normalized solutions of Hill's equation 
(\ref{hill}). They have non-zero weight under 
$\Diffc$, but their ratio
\be
\eta(\phii)\equiv\frac{\psi_1(\phii)}{\psi_2(\phii)}
\label{etaphi}
\ee
transforms under $\Diffc$ as a function (i.e.\ a zero-weight density). It blows up at the 
zeros of $\psi_2$, so it is more convenient to think of it as a curve
\be
\eta:\RR\rightarrow\RR P^1:\phii\mapsto\eta(\phii)
\label{etapath}
\ee
whose expression is (\ref{etaphi}) in terms of the projective coordinate (\ref{procor}). The points where 
$\eta$ diverges are then mapped by $\eta$ on the ``point at infinity'' in $\RR 
P^1$. Since $\RR P^1$ is diffeomorphic to the circle (\ref{rps}), we can also think 
of $\eta$ as a path in $S^1$ whose expression in stereographic coordinates is (\ref{etaphi}).

\subsubsection*{Coadjoint vectors from projective curves}

As in (\ref{pizizi}) we denote the basis of solutions $\psi_{1,2}$ by $\Psi$. Then the quasi-periodicity 
(\ref{demon}) of $\Psi$ implies a similar ``projective'' monodromy for $\eta(\phii)$,\i{projective 
monodromy}\i{monodromy!projective}
\be
\eta(\phii+2\pi)
=
\frac{a\,\eta(\phii)+b}{c\,\eta(\phii)+d}
\label{monopro}
\ee
where $a,b,c,d$ are the entries of the monodromy matrix $\sfM$. If we let $\Phi=A\Psi$ be 
another normalized basis of solutions with $A\in\SL$, the curve $\tilde\eta=\phi_1/\phi_2$ 
corresponding to $\Phi$ by (\ref{etaphi}) is related to $\eta$ by a projective transformation of the form 
(\ref{proto}). In particular eq.\ (\ref{iffiswa}) implies that the Schwarzian derivative of $\eta$ 
with respect to $\phii$ is left unchanged by such a transformation. Thus the Schwarzian 
derivative of $\eta$ is invariant under changes of (normalized) bases of solutions of Hill's equation, which 
is consistent with the following observation:

\paragraph{Lemma.} Let $\psi_1$ and $\psi_2$ be normalized solutions of Hill's equation (\ref{hill}) and 
$\eta\equiv\psi_1/\psi_2$. Then
the function $p(\phii)$ is specified by the solutions of its Hill's 
equation:
\be
\sfS[\eta](\phii)
=
-\frac{12}{c}\,p(\phii)\,.
\label{setaphi}
\ee

\begin{proof}
By virtue of the Wronskian condition (\ref{wroko}),
\be
\eta'
=
\frac{1}{(\psi_2)^2}\,.
\label{etaprime}
\ee
It then follows from the definition (\ref{swag}) that $\sfS[\eta](\phii)
=-2\frac{\psi_2''(\phii)}{\psi_2(\phii)}$,
which coincides with the right-hand side of (\ref{setaphi}) upon using Hill's equation (\ref{hill}).
\end{proof}

This lemma says that the correspondence between Virasoro coadjoint vectors and solutions of Hill's equation 
goes bothways: Hill's equation specifies certain solutions, which in turn uniquely determine the 
periodic potential $p(\phii)$ via (\ref{setaphi}). In particular the coadjoint transformation law 
(\ref{covi}) of $p$ can 
be rewritten in terms of the scalar transformation law of (\ref{etaphi}) plugged into (\ref{setaphi}).

\subsubsection*{Winding numbers}

One can think of $\eta(\phii)$ as a path in the 
circle with a ``time parameter'' $\phii$. Eq.\ (\ref{etaprime}) then says that $\eta'(\phii)>0$, so 
$\eta(\phii)$ always spins around 
the circle in the same direction. We therefore introduce the following terminology:

\paragraph{Definition.} The \it{winding number} $n\in\NN$ of 
$\eta(\phii)$\i{winding number} is the number of laps around the circle performed by $\eta$ in a ``time 
interval'' 
of length $2\pi$.\\

We will illustrate the computation of the winding number in the next section, when describing explicit 
Virasoro orbit representatives. For now note that $\eta(\phii)$ transforms under $\Diffc$ as a function, so 
its winding number is conformally invariant:

\paragraph{Proposition.} Let $c\neq0$ and let $n_{(p,c)}\in\NN$ be the winding number 
of the curve $\eta$ associated with the Hill's equation (\ref{hill}) specified by $p(\phii)$ and 
$c$. Then there is a well-defined map
\be
\left\{\text{Virasoro orbits at central charge $c$}\right\}
\rightarrow
\NN:
\cW_{(p,c)}\mapsto n_{(p,c)}\,.
\label{mappo}
\ee
In particular, Virasoro coadjoint vectors with the same central charge but different winding numbers do 
not belong to the same orbit.\\

This supplements our previous observation (\ref{mappy}) that the conjugacy classes of 
monodromy 
matrices yield a rough classification of Virasoro coadjoint orbits. In fact, these two invariants together 
provide the complete classification of Virasoro orbits.
\i{Virasoro orbits!classification} Indeed one can show that the map that associates a pair $([\sfM],n)$ 
with each Virasoro coadjoint orbit is injective, provided $[\sfM]$ is the conjugacy 
class of the monodromy matrix and $n$ is the winding number. Note however that the map is not surjective, as 
some pairs $([\sfM],n)$ do not belong to its image. We now verify this by brute force by describing orbit 
representatives.

\section{Virasoro orbit representatives}
\label{sevirorepss}

Virasoro coadjoint orbits are classified by two parameters, one of them continuous (the conjugacy class of 
the monodromy $\sfM$), 
the other discrete (the winding number $n$). In this section we display explicit orbit representatives for 
all admissible pairs $([\sfM],n)$, after a brief review of conjugacy classes in $\SL$. We end with a 
picture of orbits that extends fig.\ \ref{figorbivir}. 
As before, we assume that the central charge $c$ is positive.

\subsection{Prelude: Conjugacy classes of $\SL$}
\label{susecoclas}

In order 
to classify the conjugacy classes of $\SL$, we note that the trace of an $\SL$ matrix is 
invariant under conjugation; matrices 
with different traces cannot be conjugate. This motivates the following terminology: 
\i{SL2R@$\SL$!conjugacy classes}\i{elliptic matrix}\i{parabolic matrix}\i{hyperbolic 
matrix}
\be
\text{
An $\SL$ matrix $\sfM$ is }
\begin{cases}
\text{\it{elliptic}} & \text{if }|\Tr(\sfM)|<2;\\
\text{\it{parabolic}} & \text{if }|\Tr(\sfM)|=2;\\
\text{\it{hyperbolic}} & \text{if }|\Tr(\sfM)|>2.
\end{cases}
\nn
\ee
Each conjugacy class of $\SL$ is contained in one of these three families, but 
each family contains several conjugacy classes. The elliptic and hyperbolic families contain 
infinitely many conjugacy classes since they depend on a continuous parameter (the trace of $\sfM$). Note 
that 
the trace of $\sfM$ determine the properties of its eigenvalues:
\begin{center}
\begin{tabular}{rcl}
$\sfM$ is elliptic & $\leftrightarrow$ & distinct complex eigenvalues;\\
$\sfM$ is parabolic & $\leftrightarrow$ & degenerate real eigenvalue $\pm 1$;\\
$\sfM$ is hyperbolic & $\leftrightarrow$ & distinct real eigenvalues.
\end{tabular}
\end{center}
We now determine the conjugacy classes contained in each family. The computations are very similar to those 
of section \ref{sePoTri} where we determined the orbits of momenta for the Poincar\'e group in three 
dimensions.

\paragraph{Lemma (elliptic family).} Let $\sfM$ be elliptic.\i{conjugacy 
class!elliptic/hyperbolic/parabolic} 
Then 
it is conjugate to a unique rotation matrix 
\be
\bmm \cos(2\pi\omega) & \sin(2\pi\omega) \\ -\sin(2\pi\omega) & \cos(2\pi\omega) \emm
\label{monodel}
\ee
where $\omega$ belongs to the set $\,]0,1/2[\,\cup\,]1/2,1[\,$. The stabilizer of (\ref{monodel}) is the 
$\un$ rotation subgroup (\ref{slrot}) of $\SL$.

\begin{proof}
In the elliptic family, the eigenvalues of $\sfM$ are complex conjugates of one another with non-zero 
imaginary part. Since
$\det(\sfM)=1$, they can be written as $e^{\pm 2\pi i\omega}$ where $\omega$ belongs to the 
open interval $]0,1[$ without loss of generality, but differs from $1/2$. Let $v\in\CC^2$ be 
an eigenvector of $\sfM$ such that $\sfM\cdot v=e^{2\pi i\omega} v$. This vector is complex and linearly 
independent of its complex conjugate $\bar v$; the latter is an 
eigenvector of $\sfM$ with eigenvalue $e^{-2\pi i\omega}$. Then $v+\bar v$ and $i(v-\bar v)$ are linearly 
independent real vectors; we can choose the norm of $v$ in such a way that the (real) matrix $S$ expressing 
$\sfM$ in the basis $\{v+\bar v,i(v-\bar v)\}$ has unit determinant. Then $S\sfM S^{-1}$ takes the form 
(\ref{monodel}). The stabilizer consists of all matrices that commute with (\ref{monodel}) and is readily 
seen to consist of rotations.
\end{proof}

\paragraph{Lemma (parabolic family).} Let $\sfM$ be parabolic. Then it is conjugate to 
exactly one of the 
following six matrices:
\be
\pm
\bmm
1 & 0 \\ 0 & 1
\emm,
\qquad
\pm
\bmm
1 & 1 \\ 0 & 1
\emm,
\qquad
\pm
\bmm
1 & -1 \\ 0 & 1
\emm.
\label{paracon}
\ee
The stabilizer of the first two matrices is the whole group $\SL$, while the stabilizer $\RR\times\ZZ_2$ of 
the four remaining 
ones consists of triangular matrices (\ref{TADAM}).

\begin{proof}
When $\sfM$ is parabolic, its eigenvalues are either both $1$ or both $-1$. Let $\lambda$ be the eigenvalue 
of 
$\sfM$ and let $v\in\RR^2$ be a (real) eigenvector of $\sfM$. Let $v'$ be another vector such that $\{v,v'\}$ 
is a 
basis of $\RR^2$, and choose the normalization of $v$ and $v'$ in such a way that the matrix $S$ expressing 
$\sfM$ in this basis has unit determinant. Then
\be
S\sfM S^{-1}
=
\bmm
\lambda & x \\ 0 & \lambda
\emm
\label{lax}
\ee
where $\lambda=\pm1$ and $x$ is an arbitrary real number. For $x=0$ we find the first two matrices in the 
list (\ref{paracon}), each of which is alone in its conjugacy class. For non-zero $x$, note that
\be
\bmm
y & 0 \\ 0 & 1/y
\emm
\bmm
1 & \pm 1 \\ 0 & 1
\emm
\bmm
1/y & 0 \\ 0 & y
\emm
=
\bmm
1 & \pm y^2 \\ 0 & 1
\emm\,,
\label{AYAYA}
\ee
so for $\lambda=+1$, $\sfM$ is conjugate to the second matrix in (\ref{paracon}) if $x>0$ and to the third 
one if $x<0$, in both cases with an overall plus sign. The situation is similar when $\lambda=-1$, but with 
the 
minus sign. The proof ends with the observation that all matrices in (\ref{paracon}) belong to 
disjoint conjugacy classes. The stabilizer is obtained by direct computation.
\end{proof}

\paragraph{Lemma (hyperbolic family).} Let $\sfM$ be hyperbolic. Then it is conjugate 
to a unique matrix of the form
\be
\pm
\bmm
e^{2\pi\omega} & 0 \\ 0 & e^{-2\pi\omega}
\emm
\label{xox}
\ee
where $\omega$ is a strictly positive real number. Its stabilizer is the group 
$\RR\times\ZZ_2$ consisting 
of matrices (\ref{TADAT}) of the same form as (\ref{xox}) but without restriction on $\omega\in\RR$. 

\begin{proof}
Since $\sfM$ is hyperbolic, it has two distinct real eigenvalues $\lambda$ and $1/\lambda$, where 
$\lambda\in\RR^*$. Let $v$ and $v'$ be two eigenvectors of $\sfM$ for these eigenvalues; we can 
normalize them so that the matrix $S$ expressing $\sfM$ in the basis $\{v,v'\}$ has unit determinant. Then 
$S\sfM S^{-1}$ takes the form (\ref{xox}) with $e^{2\pi\omega}=\lambda$ or $e^{2\pi\omega}=1/\lambda$. The 
ordering of eigenvalues can be 
changed thanks to the $\SL$ matrix $\bmm 0 & 1 \\ -1 & 0 \emm$, so we are free to pick $\omega>0$, and this 
specifies uniquely the conjugacy class of the matrix $\sfM$. Finding the stabilizer is straightforward.
\end{proof}

From now on we say that a Virasoro orbit is elliptic, parabolic or hyperbolic if the associated monodromy 
matrix is of one of those three 
types, respectively. In addition we will distinguish parabolic orbits associated with $\pm\II$ from 
parabolic orbits
associated 
with the four other matrices in (\ref{paracon})\i{degenerate parabolic matrix} by referring to the 
former as ``degenerate'' and to the 
latter as ``non-degenerate''.

\subsection{Elliptic orbits}
\label{suselliptic}

Here we initiate the classification of Virasoro coadjoint orbits, by studying those whose monodromy is 
elliptic. Parabolic and hyperbolic orbits will be investigated in sections \ref{DEGANNA} to \ref{NODEGANNA}.

\subsubsection*{Finding orbit representatives}

Let $c>0$ and suppose that $q(\phii)d\phii^2$ is a quadratic density such that the monodromy of Hill's 
equation (\ref{hill}) is 
elliptic.\i{elliptic Virasoro orbit} (We denote the quadratic density by $q$ rather than $p$, because the 
latter will 
eventually be the ``representative'' of the orbit of $q$.) Then we can choose a 
solution vector $\Psi$ whose monodromy matrix takes the form (\ref{monodel}) for some angle $2\pi\omega$ 
which is not an integer multiple of $\pi$. The function
\be
X_q(\phii)\equiv\psi_1^2(\phii)+\psi_2^2(\phii)
\label{expiation}
\ee
is strictly positive and $2\pi$-periodic; it is a vector field on the circle, 
since it is a quadratic combination of $-1/2$-densities such as (\ref{hillprod}). In fact, it belongs to 
the Lie algebra of the stabilizer of $q$ since it solves equation (\ref{stabbo}). In addition it is 
invariant under the action of the stabilizer of $\sfM$ and is therefore a well-defined functional of 
$q(\phii)$, which justifies the notation $X_q$. Conversely, $X_q$ determines $q(\phii)$ since 
Hill's equation implies
\be
q
=
\frac{c}{6}\,\frac{\psi_1''\psi_1+\psi_2''\psi_2}{X_q}
\refeq{expiation}
\frac{c}{6}
\left[
\demi\frac{X_q''}{X_q}
-\frac{1}{4}\left(\frac{X_q'}{X_q}\right)^2
-\frac{1}{X_q^2}
\right].
\label{piggix}
\ee
We would have obtained the same formula upon using eq.\ (\ref{setaphi}) with $\eta=\psi_1/\psi_2$. 
Our goal now is to build a diffeomorphism $g_q\in\Diffc$ such that $q$ is obtained by acting with $g_q$ on
a suitable orbit representative $p$. In the language of 
induced representations, the maps $g_q$ will be ``standard boosts'' on the orbit of $p$.\\

Let us define the negative number
\be
p_0
\equiv
-\frac{c}{6}
\left[\int_0^{2\pi}\frac{d\phii}{X_q(\phii)}\right]^2,
\label{piziro}
\ee
where the notation ``$p_0$'' will be justified
below. This number is well-defined since $X_q(\phii)$ never vanishes, and it is invariant under $\Diffc$ 
since $X_q$ 
is a vector field. We can then define a diffeomorphism $f\in\Diffc$ by
\be
f(\phii)
\equiv
\frac{2\pi}{\sqrt{6|p_0|/c}}
\int_0^{\phii}\frac{d\phi}{X_q(\phi)}.
\label{bagaku}
\ee
This quantity is the inverse of the sought-for standard boost since eq.\ (\ref{piggix}) can be written as
\be
q(\phii)
=
p_0(f'(\phii))^2-\frac{c}{12}\sfS[f](\phii)\,,
\label{holdme}
\ee
which we recognize as the coadjoint action (\ref{covirax}) of
\be
g_q\equiv f^{-1}
\label{vistaboost}
\ee
on the constant coadjoint vector $p(\phii)=p_0<0$. In conclusion:

\paragraph{Proposition.} Let $(q,c)$ with $c>0$ be a Virasoro coadjoint vector with elliptic monodromy. Then 
it belongs to the orbit of a constant coadjoint vector $(p,c)$ with $p(\phii)=p_0$, where the value of $p_0$ 
is determined by $q(\phii)$ according to (\ref{piziro}) with $X_q$ given by (\ref{expiation}) in terms 
of normalized solutions of the Hill's equation of $(q,c)$. In addition, the diffeomorphism $g_q$ 
defined as the inverse of (\ref{bagaku}) is a \it{standard boost} for the orbit of $p$ in the sense 
that\i{standard boost}
\be
g_q\cdot p=q
\label{gikidota}
\ee
where the dot denotes the coadjoint action (\ref{covi}).

\subsubsection*{Monodromy and winding number}

Let us now see how the parameter (\ref{piziro}) is related to the monodromy matrix. At 
$p=p_0$, Hill's equation (\ref{hill}) reads
\be
-\frac{c}{6}\psi''-|p_0|\psi=0
\label{hillcon}
\ee
where we write $p_0=-|p_0|$ to emphasize that this is a harmonic oscillator equation with frequency
\be
\omega=\sqrt{6|p_0|/c}\,.
\label{hofrik}
\ee
A basis of solutions satisfying 
the Wronskian condition (\ref{wroko}) is provided by
\be
\psi_1(\phii)
=
\frac{1}{\sqrt{\omega}}\sin(\omega\phii),
\qquad
\psi_2(\phii)
=
\frac{1}{\sqrt{\omega}}\cos(\omega\phii).
\label{ellisol}
\ee
The corresponding monodromy matrix $\sfM$ is readily seen to take the 
form (\ref{monodel}) with $\omega$ given by (\ref{hofrik}) in terms of $p_0/c$. The fact that the monodromy 
matrix is elliptic implies that $\omega$ is 
\it{not} an integer multiple of $1/2$, which is equivalent to saying that
\be
p_0\neq-\frac{n^2c}{24}\,.
\label{picnico}
\ee
In the language of section \ref{suseconoze}, the constant orbit representative $p_0$ must be 
\it{generic}\i{generic constant}\i{p0@$p_0$ (constant supermomentum)!generic} in order for its orbit to be 
elliptic. By contrast, the 
exceptional orbit 
representatives 
(\ref{pn}) will turn out to have degenerate parabolic monodromy (see below).\\

Thus different values of $p_0$ generally define disjoint orbits since their monodromy matrices 
(\ref{monodel}) are not conjugate. However, at 
this stage we cannot tell whether
\be
p_0\qquad\text{and}\qquad-\Big(\sqrt{|p_0|}+\sqrt{\tfrac{c}{6}}\,N\Big)^2
\label{whichp}
\ee
belong to different orbits when $N\in\NN$ since 
their monodromy matrices coincide (their angles differ by $2\pi N$). This issue is settled by the 
winding number (\ref{mappo}): the curve 
(\ref{etaphi}) associated with the solutions (\ref{ellisol}) is
\be
\eta(\phii)
=
\tan(\omega\phii)\,,
\label{etOmm}
\ee
which can be seen as a path on a circle written in terms of a stereographic coordinate 
$\eta=\tan(\theta/2)$, where the coordinate $\theta\in\RR$ is identified as $\theta\sim\theta+2\pi$. In terms 
of $\theta$ the path (\ref{etOmm}) is a rotation around the circle at constant velocity, 
$\theta(\phii)=2\omega\phii$. The number of laps performed by this path around the circle when $\phii$ goes 
from zero to $2\pi$ is the winding number\footnote{We denote the 
winding number by $n_p$ instead of $n_{(p_0,c)}$ to reduce clutter.}\i{winding number!for elliptic Virasoro 
orbit}
\be
n_{p}
=
\lfloor2\omega\rfloor
\refeq{hofrik}
\left\lfloor\sqrt{\frac{24|p_0|}{c}}\right\rfloor
\label{elliwind}
\ee
where $\lfloor\cdot\rfloor$ denotes the integer part.
Thus the winding number associated with $p_0<0$ takes a definite value in each interval 
$\,]-\frac{(n+1)^2c}{24},-\frac{n^2c}{24}[\,$, and jumps by one unit every time $p_0$ takes one of the 
exceptional values (\ref{pn}). For instance $n_p=0$ when $p_0$ belongs to $]-c/24,0[\,$, while $n_p=1$ when 
$p_0\in\,]-c/6,-c/24[\,$, and so on.\\

In conclusion, the orbits of two generic constants $p_0$ and $\tilde p_0$ are disjoint if 
and only if these constants differ. We have thus recovered the lower part $(p_0<0)$ of fig.\ 
\ref{figorbivir}. 
As a bonus we can now assign a 
monodromy matrix determined by (\ref{hofrik}), and a winding number (\ref{elliwind}), with each point on that 
part. In particular the integers $n$ written on the left of the $p_0$ axis can be interpreted as winding 
numbers for constants $p_0$ located between $-(n+1)^2c/24$ and $-n^2c/24$.

\subsubsection*{Stabilizers}

To conclude the description of orbits of generic constants $p_0<0$, it remains to find their 
stabilizer.
As anticipated in (\ref{listab}), one shows 
that the stabilizer of $p_0$ is the group $\un$ of rigid rotations $f(\phii)=\phii+\theta$ (or more 
precisely its universal cover $\RR$ when dealing with $\Diffc$). The coadjoint orbit of $(p_0,c)$ can thus be 
written as
\be
\cW_{(p_0,c)}
\cong
\Diffc/\RR
\cong
\Diffp/S^1,
\label{ellorbidif}
\ee
which may be seen as an infinite-dimensional generalization of the orbit 
$\SL/S^1$ of $\SL$. The latter coincides with the momentum orbit (\ref{Mopping}) of a massive Poincar\'e 
particle in three dimensions; in the same way, we shall see in section \ref{sebmspar} that (\ref{ellorbidif}) 
is the supermomentum orbit of a massive BMS$_3$ particle.

\paragraph{Remark.} The stabilizer $\un$ coincides with the stabilizer of 
the monodromy matrix (\ref{monodel}), so the quotient (\ref{gima}) consists of a single point. This implies 
that all conformally inequivalent normalized solutions of the Hill equation associated with $(p_0,c)$ can be 
obtained 
by acting with rotations on the solution (\ref{ellisol}).

\subsection{Degenerate parabolic orbits}
\label{DEGANNA}

\subsubsection*{Orbit representatives}

We now turn our attention to coadjoint vectors $(q,c)$ whose monodromy matrix is of the ``degenerate'' 
parabolic type (\ref{paracon}), i.e.\ coincides with $\pm\II$.\i{parabolic Virasoro orbit!degenerate} We 
proceed as in the elliptic case. In particular the monodromy matrix still ensures that (\ref{expiation}) is a 
positive, $2\pi$-periodic vector field belonging to the Lie algebra of the stabilizer 
of $q(\phii)$. The negative number (\ref{piziro}) is still well-defined and $\Diffc$-invariant, 
and formula (\ref{bagaku}) provides a diffeomorphism of $S^1$ such that eq.\ (\ref{holdme}) holds. Then 
(\ref{vistaboost}) is a standard boost that maps the constant coadjoint vector $p_0$ on $q(\phii)$; in 
particular the proposition surrounding (\ref{gikidota}) still holds up to the replacement of the word 
``elliptic'' by ``degenerate parabolic''.\\

As in the elliptic case we can choose constant coadjoint vectors as orbit representatives. The 
corresponding Hill's equation then reads (\ref{hillcon}) and admits the normalized solutions (\ref{ellisol}), 
but the monodromy matrix is $\pm\II$. 
\i{parabolic Virasoro orbit!and exceptional constants} Such a monodromy matrix $\sfM$ only occurs when $p_0$ 
takes the 
exceptional 
form (\ref{pn}) for some strictly positive integer $n$, in which case\i{exceptional 
constant}\i{p0@$p_0$ (constant supermomentum)!exceptional}
\be
p_0=-\frac{n^2c}{24}
\qquad
\text{and}
\qquad
\sfM=(-1)^n\II.
\label{monodegepa}
\ee
By contrast, elliptic orbits \it{never} contain an exceptional constant; this 
is a sharp difference between elliptic and degenerate parabolic orbits.\\

The monodromy matrix (\ref{monodegepa}) implies that two exceptional constants specified by integers $n,n'$ 
can belong to the same orbit only if $n$ and $n'$ have the same parity; but at this stage we cannot tell if 
two orbits with the same parity are disjoint. As in the elliptic case we can address this 
question by studying the winding number of the curve (\ref{etaphi}) associated with the solutions 
(\ref{ellisol}). One can verify that the winding number coincides with the number $n$ specified by 
$p_0=-n^2c/24$, which implies that any two orbits of exceptional constants specified by different values of 
$n>0$ 
are disjoint. In conclusion, we have now recovered the dots in the lower part of fig.\ 
\ref{figorbivir}, and the values of $n$ displayed there coincide with winding numbers. In particular the 
orbit 
at $n=1$ will be called the
\it{vacuum orbit} from now on;\i{vacuum orbit!for Virasoro} in the context of Riemann surfaces, it is known 
as 
\it{universal Teichm\"uller space} \cite{guieu2007algebre,Nag}.\i{universal Teichm\"uller space} Note that 
the orbit of $p_0=0$ does \it{not} 
have degenerate parabolic 
monodromy, and so has not yet been accounted for by our classification of Hill's equations.

\subsubsection*{Stabilizers}

We now study the stabilizers of orbits of exceptional constants $p_0=-n^2c/24$. We 
saw below (\ref{listab}) that the stabilizer is three-dimensional for such values, and is 
generated by the vector fields
\be
\frac{\der}{\der\phii}\,,
\qquad
\sin(n\phii)\frac{\der}{\der\phii}\,,
\qquad
\cos(n\phii)\frac{\der}{\der\phii}\,.
\label{seltusta}
\ee
The Lie algebra of the stabilizer is therefore isomorphic to $\sl$, but different values of $n$ define 
non-conjugate embeddings of $\sl$ in $\Vect$. In fact one can verify using (\ref{swagiffn}) that
the finite 
diffeomorphisms that span the stabilizer of $p_0=-n^2c/24$ (and that reduce to (\ref{seltusta}) close to the 
identity) are projective 
transformations (\ref{protophin}) spanning a group $\text{PSL}^{(n)}(2,\RR)$ (the $n$-fold cover of 
$\PSL$). In conclusion:\i{PSL2R@$\PSL$!n-fold cover@$n$-fold cover}

\paragraph{Lemma.} The stabilizer of $p_0=-n^2c/24$ for the coadjoint action of $\Diffc$ (resp.\ $\Diffp$) is 
the group $\widetilde{\text{PSL}}{}^{(n)}(2,\RR)$ (resp.\ $\text{PSL}^{(n)}(2,\RR)$) spanned by 
diffeomorphisms $f(\phii)$ given by (\ref{protophin}), where $\widetilde{\text{PSL}}{}^{(n)}(2,\RR)$ is the 
universal cover of the $n$-fold cover of 
$\text{PSL}(2,\RR)=\SL/\ZZ_2$. The coadjoint orbit of $(p_0,c)$ can be written as
\be
\cW_{\big(-\frac{n^2c}{24},c\big)}
\cong
\Diffc/\widetilde{\text{PSL}}{}^{(n)}(2,\RR)
\cong
\Diffp/\text{PSL}^{(n)}(2,\RR)
\label{factipuce}
\ee
The Lie algebra of the stabilizer is generated by the vector fields 
(\ref{seltusta}).\\

In section \ref{sebmscodj} we will interpret the orbit of $p_0=-c/24$ as the set of gravitational 
perturbations 
around Minkowski space. In that context the little group $\PSL$ will be seen as the Lorentz 
group 
in three dimensions. The remaining exceptional 
values $p_0=-n^2c/24$ (with $n\geq2$) will be interpreted as conical excesses where one turn around the 
origin of space spans an angle $2\pi n$.

\paragraph{Remark.} An important difference between elliptic and degenerate pa\-ra\-bo\-lic orbits is that, 
in 
the latter case, the stabilizer of the monodromy matrix (\ref{monodegepa}) is the whole group 
$\SL$, which does not leave the combination 
(\ref{expiation}) invariant. Nevertheless, the integral 
(\ref{piziro}) is still independent of the choice of the normalized solution vector $\Psi$ because, in that 
specific case, any $\SL$ transformation $\Psi\mapsto S\Psi$ is equivalent to the 
action of a diffeomorphism of the circle belonging to the stabilizer of $p_0$; since the integral 
(\ref{piziro}) is invariant under diffeomorphisms, it follows that it is also invariant under $\Psi\mapsto 
S\Psi$ for any $S\in\SL$.

\subsection{Hyperbolic orbits without winding}

Consider a Virasoro coadjoint vector $(q,c)$ whose monodromy matrix is of the hyperbolic type 
(\ref{xox}) with some $\omega>0$. We shall 
see that hyperbolic orbits differ greatly 
depending on the winding number of the curve (\ref{etaphi}), so we focus here on the case of zero winding; 
the non-zero case will be treated in 
section \ref{sewynG}.

\subsubsection*{Finding orbit representatives}

Let $\psi_1$ and $\psi_2$ be normalized solutions of Hill's equation associated with $(q,c)$ and let 
$\eta=\psi_1/\psi_2$. Since the 
winding number of $\eta$ is zero, we can choose our solution vector such that $\psi_2$ has 
no zeros 
on the real line. Then $\eta(\phii)$ 
is smooth and, by virtue of (\ref{xox}), we have
\be
\eta(\phii+2\pi)
=
e^{4\pi\omega}\eta(\phii)
\label{monodeta}
\ee
so $\eta(\phii)$ is monotonically increasing on $\RR$ (since $\omega>0$). As in the case of elliptic 
orbits, our goal is to find a ``standard boost'' $g_q$ whose inverse $g_q^{-1}\equiv f$ 
will map $q(\phii)$ on a suitably chosen orbit representative. To do so we define
\be
f(\phii)
\equiv
\frac{1}{2\omega}\log(\eta(\phii))
\refeq{etaphi}
\frac{1}{2\omega}\log\left(\frac{\psi_1(\phii)}{\psi_2(\phii)}\right)
\label{hyperstabo}
\ee
which belongs to $\Diffc$ by virtue of (\ref{monodeta}). Now if we set
\be
p_0\equiv\frac{c\,\omega^2}{6}\,,
\label{pizirobis}
\ee
we can use (\ref{setaphi}) and the cocycle identity 
(\ref{swapro}) to write
\be
q(\phii)
=
p_0(f'(\phii))^2-\frac{c}{12}\sfS[f]\,.
\label{tight}
\ee
As in (\ref{holdme}) we 
recognize the coadjoint action of $g_q=f^{-1}$ on the constant coadjoint vector $p(\phii)=p_0>0$, 
and thus conclude:\i{hyperbolic Virasoro orbit}

\paragraph{Proposition.} Let $(q,c)$ with $c>0$ be a Virasoro coadjoint vector with
hyperbolic monodromy and zero winding number. Then it belongs to the orbit of a constant coadjoint
vector $(p_0,c)$, where $p_0>0$ is determined by the monodromy matrix according to (\ref{pizirobis}). In 
addition the diffeomorphism $g_q$
defined as the inverse of (\ref{hyperstabo}) is a standard boost for the orbit of $p$ in the sense 
(\ref{gikidota}).\\

Note that the definition (\ref{pizirobis}) coincides with eq.\ (\ref{hofrik}) for $p_0>0$. 
Roughly speaking, ``hyperbolic orbits are an analytic continuation of elliptic orbits to imaginary values of 
the monodromy parameter $\omega$''. This is analogous to the fact that tachyonic momentum orbits may be seen 
as 
massive orbits with imaginary mass.

\subsubsection*{Stabilizers}

At $p=p_0$, 
Hill's equation (\ref{hill}) reads $-\frac{c}{6}\psi''+|p_0|\psi=0$
where we stress that the sign of the potential term is opposite to the one in (\ref{hillcon}). 
A 
basis of solutions satisfying the Wronskian condition (\ref{wroko}) is provided by
\be
\psi_1^{\pm}(\phii)=\pm\frac{1}{\sqrt{2\omega}}\,e^{\omega\phii},
\qquad
\psi_2^{\pm}(\phii)=\pm\frac{1}{\sqrt{2\omega}}\,e^{-\omega\phii}
\label{solehyp}
\ee
where $\omega>0$ is given by (\ref{hofrik}). The corresponding monodromy matrix is (\ref{xox}).\\

We have 
seen in (\ref{listab}) that the stabilizer is
one-dimensional for $p_0>0$, and that it consists of rotations of the circle. Thus the stabilizer of $p_0$ is 
a group $\un$ of rigid rotations (or more precisely its universal cover $\RR$ when dealing with $\Diffc$). In 
particular the orbit of $(p_0,c)$ for $p_0>0$ and $c>0$ is diffeomorphic to
\be
\cW_{(p_0,c)}\cong\Diffc/\RR\cong\Diffp/S^1.
\label{hyperorbidif}
\ee
As in (\ref{ellorbidif}) this orbit 
may be seen as an infinite-dimensional generalization of $\SL$ orbits of the type $\SL/S^1$. However, the 
orbit differs from those of negative $p_0$'s in that the two choices of 
signs in (\ref{solehyp}) are conformally inequivalent. Indeed, the stabilizer $G_{\sfM}$ of 
the 
matrix (\ref{xox}) under conjugation is isomorphic to $\RR\times\ZZ_2$ while the universal cover of the 
little group of $p_0$ is just $\RR$. Accordingly the quotient (\ref{gima}) contains two points, indicating 
that there are two inequivalent normalized families of solutions to Hill's equation at $(p_0,c)$; these two 
families are labelled by the sign $\pm$ in (\ref{solehyp}).\\

In terms of fig.\ \ref{figorbivir}, we have now completed our understanding of almost the whole real line 
$p_0\in\RR$, since we now know that the orbits that pass through $p_0>0$ are of hyperbolic type without 
winding. The only remaining mystery is the orbit of $p_0=0$, and of course 
all the orbits that do not contain constant representatives.

\subsection{Hyperbolic orbits with winding}
\label{sewynG}

\subsubsection*{Building orbit representatives}

We now consider a Virasoro coadjoint vector $(q,c)$ with hyperbolic monodromy (\ref{xox}) but strictly 
positive winding number $n>0$.\i{hyperbolic Virasoro orbit} The classification 
of orbits of such vectors is more 
involved than in the previously encountered cases, so we proceed in a ``backwards'' fashion. Namely, 
suppose we are given a pair of smooth real functions $\psi_1$, $\psi_2$ on $\RR$, chosen in such a way that 
they satisfy the Wronskian condition (\ref{wroko}). Then it is automatically true that the function 
$p(\phii)$ defined by
\be
p
\equiv
\frac{c}{6}\,\frac{\psi_1''}{\psi_1}
=
\frac{c}{6}\,\frac{\psi_2''}{\psi_2}
\label{defiqip}
\ee
is smooth for any constant $c>0$. If in addition there exists a monodromy matrix 
$\sfM$ such that (\ref{demon}) holds, then $p(\phii)$ is $2\pi$-periodic and $\psi_1,\psi_2$ are solutions of 
the corresponding Hill's equation. This procedure provides a way to build 
Virasoro coadjoint vectors out of functions $\psi_i$; in particular, in order to prove that there exist 
Virasoro orbits with hyperbolic monodromy and non-zero winding number, it suffices to find two normalized 
functions $\psi_i$ satisfying these criteria, and the identification of the corresponding Virasoro coadjoint 
vectors will follow.\\

Thus, let $\omega>0$ be a strictly positive real number and let $n>0$ be a positive integer. Let us define 
the positive function
\be
F_{n,\omega}(\phii)
\equiv
\cos^2(n\phii/2)
+\left(
\sin(n\phii/2)+\frac{2\omega}{n}\cos(n\phii/2)
\right)^2
\label{fnomega}
\ee
as well as
\begin{align}
\label{mojito}
\psi_1(\phii)
& \equiv
\frac{e^{\omega\phii}}{\sqrt{F_{n,\omega}(\phii)}}\sqrt{\frac{2}{n}}
\left(
\sin(n\phii/2)
+\frac{\omega}{n}\cos(n\phii/2)
\right),\\
\label{pinacolada}
\psi_2(\phii)
& \equiv
\frac{e^{-\omega\phii}}{\sqrt{F_{n,\omega}(\phii)}}\sqrt{\frac{2}{n}}
\cos(n\phii/2).
\end{align}
Since $F_{n,\omega}$ is 
strictly positive, the $\psi_i$'s are smooth functions. They satisfy the Wronskian condition (\ref{wroko}) 
and their monodromy 
matrix is (\ref{xox}). Their ratio is
\be
\eta(\phii)
=
e^{2\omega\phii}\tan(n\phii/2)+\frac{\omega}{n}
\nn
\ee
and describes a path on the circle with varying velocity and winding number $n$. It follows that the 
function 
$p(\phii)$ defined by (\ref{defiqip}) is a Virasoro coadjoint vector with hyperbolic monodromy (\ref{xox}) 
and winding number $n>0$. It is explicitly given by\i{tachyonic 
BMS$_3$ particle}\i{supermomentum!tachyonic}
\be
p(\phii)
=
\frac{c\,\omega^2}{6}
+\frac{c}{12}\,\frac{n^2+4\omega^2}{F_{n,\omega}(\phii)}
-\frac{c}{8}\,\frac{n^2}{F_{n,\omega}^2(\phii)}
\label{pitachy}
\ee
in terms of the function (\ref{fnomega}). We have thus built explicit orbit representatives with hyperbolic 
monodromy and non-zero winding number.\\

It is 
worth spending some time to interpret formula (\ref{pitachy}). Let us take $\omega$ small and expand $p$ 
around 
$\omega=0$. To first order in $\omega$, we get
\be
p(\phii)
=
-\frac{n^2c}{24}+\omega\frac{nc}{3}\sin(n\phii)+\cO(\omega^2)\,.
\label{pitachydef}
\ee
The leading term in $p$ is an exceptional constant $-n^2c/24$, so we can think of (\ref{pitachy}) as a 
deformation of that constant.\i{exceptional constant!deformation} The term of order one in $\omega$ in 
(\ref{pitachydef}) is 
proportional to 
$\sin(n\phii)$, which is one of the elements of the Lie algebra of the stabilizer of $-n^2c/24$. This
ensures that the deformation does not belong to the orbit of $-n^2c/24$. Indeed, all deformations that 
\it{do} belong to that orbit take the form
\be
\had^*_X\left(-\frac{n^2c}{24}\right)
\refeq{covinf}
-\frac{c}{12}\left(
n^2X'+X'''\right)
\nn
\ee
for some vector field $X$, where the term $n^2X'+X'''$ annihilates the contribution of the modes 
$\sin(n\phii)$ or $\cos(n\phii)$.\\

\begin{figure}[t]
\centering
\includegraphics[width=0.30\textwidth]{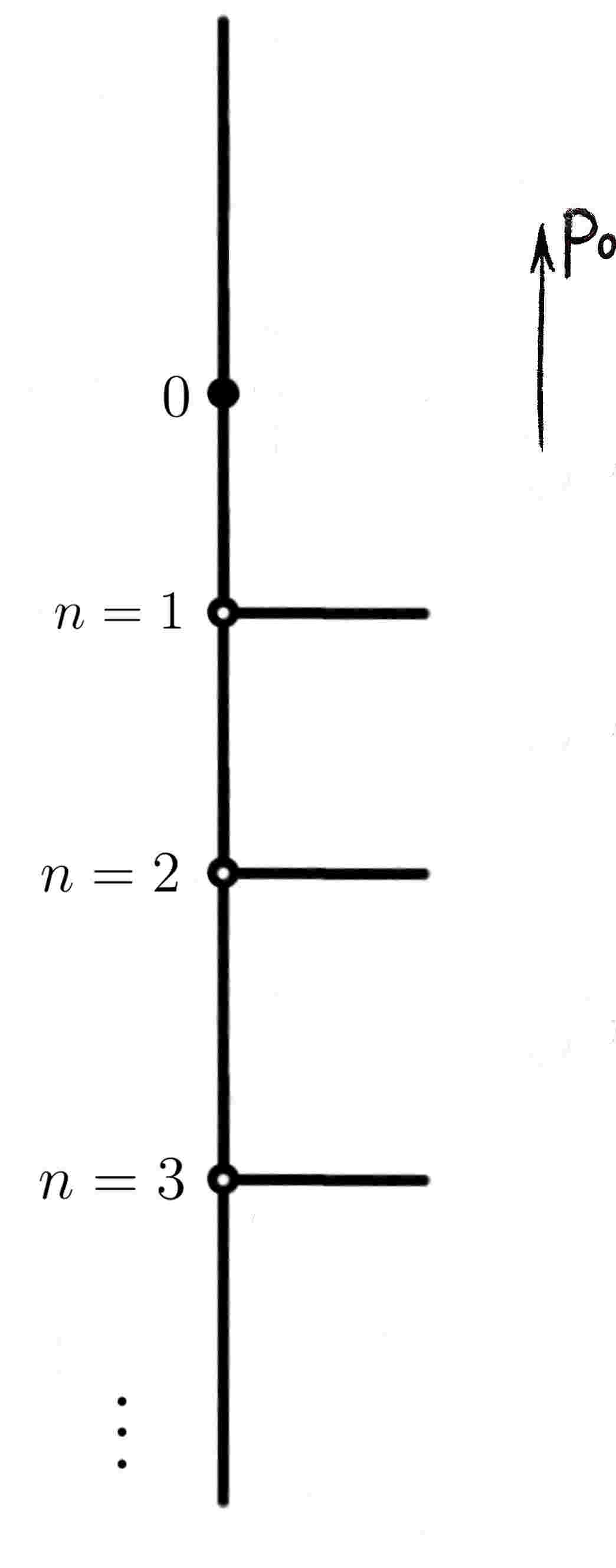}
\caption{A partial map of Virasoro orbits, including orbits of constant coadjoint vectors together with 
tachyonic orbits. Compare to fig.\ \ref{figorbivir}.\label{figorbiviralmost}}
\end{figure}

In section \ref{sebmspar} we will interpret (\ref{pitachy}) as the supermomentum of a 
BMS$_3$ tachyon with imaginary mass proportional to $\omega^2$. Accordingly, from now on we refer to 
hyperbolic Virasoro orbits with non-zero monodromy as \it{tachyonic 
orbits}. They are our first example of orbits that do not admit any constant 
representative, so
they are \it{not} accounted for by fig.\ \ref{figorbivir}. In order to include them in our ``map of coadjoint 
orbits'', we think of them as orbits of deformations 
(\ref{pitachydef}) of exceptional constants. With this viewpoint and the ``tachyonic'' terminology, it is 
natural to identify this kind of deformation with the horizontal line in fig.\ 
\ref{figDepIdu}{\textcolor{blue}{b}} that represents 
tachyonic orbits of Poincar\'e. Accordingly we represent tachyonic 
Virasoro orbits by a horizontal line to the right of the point 
labelled ``$n$'' in fig.\ \ref{figorbivir}. With this convention our schematic representation of Virasoro 
orbits becomes the one displayed in fig.\ \ref{figorbiviralmost}. It remains to 
understand which orbit contains the point $p_0=0$, and to find the 
remaining orbits that have no constant representative. Before doing so, we address a few 
minor points regarding tachyonic orbits:
\begin{itemize}
\item The construction that led from (\ref{fnomega}) to (\ref{pitachy}) did produce Virasoro coadjoint 
vectors with 
the desired monodromy and winding number, but it is not clear at this stage that \it{any} coadjoint vector 
satisfying these properties can be mapped on (\ref{pitachy}). However, this turns out to be the case; in this 
sense, the orbit representatives (\ref{pitachy}) exhaust all orbits with hyperbolic monodromy and 
non-zero winding. See \cite{Balog:1997zz} for the proof.
\item The Lie 
algebra of the stabilizer of (\ref{pitachy}) is spanned by the periodic linear combinations of the functions 
(\ref{hillprod}). 
As it turns out, the only periodic combination in this case is the product $\psi_1\psi_2$. The latter has 
$2n$ simple zeros inside $[0,2\pi[$ and generates a 
non-compact group $\RR$. In addition the function (\ref{pitachy}) is 
periodic with period $2\pi/n$, so the stabilizer must contain a group $\ZZ_n$ consisting of rotations by 
integer multiples of $2\pi/n$. In fact, one can show (see \cite{Balog:1997zz}) that the stabilizer of $p$ in 
$\Diffp$ is isomorphic to a product $\RR\times\ZZ_n$, while its stabilizer in the universal cover $\Diffc$ 
is $\RR\times T_{2\pi/n}$ where $T_{2\pi/n}$ is the group of translations of the real line by integer 
multiples of $2\pi/n$. We conclude that the orbit of (\ref{pitachy}) is 
diffeomorphic to
\be
\cW_{(p,c)}
\cong
\Diffc/(\RR\times T_{2\pi/n})
\cong
\Diffp/(\RR\times\ZZ_n).
\label{tachyorbitiv}
\ee
\end{itemize}

\subsection{Non-degenerate parabolic orbits}
\label{NODEGANNA}

Here we include the last missing pieces of our description of Virasoro orbits.
When the monodromy matrix is non-degenerate parabolic, it is conjugate to one 
of the four last elements in the list (\ref{paracon}). As in the hyperbolic case we discuss zero and non-zero 
windings separately.\i{parabolic Virasoro orbit!non-degenerate}

\subsubsection*{Zero winding}

At zero winding we proceed as in the elliptic and $n=0$ hyperbolic cases, i.e.\ we look for
standard boosts. Let therefore $(q,c)$ be a Virasoro coadjoint vector such that a normalized 
solution 
vector $\Psi=(\psi_1\;\psi_2)^t$ associated with the corresponding Hill's equation has non-degenerate 
pa\-ra\-bo\-lic monodromy and zero winding number. The monodromy matrices in (\ref{paracon}) imply that
\be
\psi_1(\phii+2\pi)=\pm\big(\psi_1(\phii)+\varepsilon\psi_2(\phii)\big),
\qquad
\psi_2(\phii+2\pi)=\pm\psi_2(\phii)
\label{wannabe}
\ee
where $\varepsilon$ is a priori $+1$ or $-1$. The corresponding curve (\ref{etaphi}) satisfies
\be
\eta(\phii+2\pi)
=
\eta(\phii)+\varepsilon
\label{etaphichoice}
\ee
and (\ref{etaprime}) implies
that
$\varepsilon$ must actually be equal to $+1$. The opposite sign corresponds to changing the orientation in 
the space of solutions of Hill's equation, so with our choice of orientation for $\psi_1,\psi_2$, only the 
value $\varepsilon=+1$ gives rise to an admissible monodromy matrix. Then the 
function
\be
f(\phii)\equiv2\pi\,\eta(\phii)
\label{fifikapu}
\ee
is a $2\pi\ZZ$-equivariant diffeomorphism of the real line, and property (\ref{setaphi}) implies that
\be
q(\phii)=-\frac{c}{12}\sfS[f](\phii)\,.
\nn
\ee
As in eqs.\ (\ref{holdme}) and (\ref{tight}), we recognize the coadjoint action of $g_q\equiv f^{-1}$:

\paragraph{Proposition.} Let $(q,c)$ with $c>0$ be a Virasoro coadjoint vector with non\--de\-ge\-ne\-rate 
parabolic 
monodromy and vanishing winding number.\i{parabolic Virasoro orbit} Then it belongs to the orbit of $(0,c)$ 
and the inverse of 
the diffeomorphism (\ref{fifikapu}) is a standard boost in the sense of eq.\ (\ref{gikidota}).\\

Thus we have finally found the orbit of $p_0=0\,$! It was the only point of fig.\ \ref{figorbivir} that was 
still eluding us. We now know that its orbit has parabolic type. The corresponding stabilizer is the group 
$\un$ of rigid rotations (as for 
all positive or generic constants $p_0$), and there are two conformally inequivalent solutions of Hill's 
equation at $p_0=0$, namely
$\psi_1^{\pm}=\pm\phii$, $\psi_2^{\pm}(\phii)=\pm1$.
The orbit can be represented as a quotient space
\be
\cW_{(0,c)}
\cong
\Diffc/\RR
\cong
\Diffp/S^1
\nn
\ee
and is diffeomorphic to the orbits (\ref{ellorbidif})-(\ref{hyperorbidif}) of generic or positive 
constants.

\subsubsection*{Non-zero winding}

At non-zero winding our strategy will be similar to that used in the hyperbolic case with winding: we rely on 
the fact that formula (\ref{defiqip}) always defines a $2\pi$-periodic function 
$p(\phii)$ when $\psi_1$ and $\psi_2$ satisfy the Wronskian condition and admit a well-defined monodromy, 
which allows us to build orbit representatives.\\

Thus, pick a number $\varepsilon\in\{\pm1\}$ and let $n\in\NN^*$ be a non-zero winding number. Let us define 
the positive function
\be
H_{n,\varepsilon}(\phii)
\equiv
1+\frac{\varepsilon}{2\pi}\sin^2(n\phii/2)
\label{hnomega}
\ee
as well as
\begin{align}
\label{suliste}
\psi_1(\phii)
& \equiv
\frac{1}{\sqrt{H_{n,\varepsilon}(\phii)}}
\left(
\frac{\varepsilon\phii}{2\pi}\sin(n\phii/2)-\frac{2}{n}\cos(n\phii/2)
\right),
\\
\label{soliste}
\psi_2(\phii)
& \equiv
\frac{1}{\sqrt{H_{n,\varepsilon}(\phii)}}\sin(n\phii/2)\,.
\end{align}
Since the function 
$H_{n,\varepsilon}$ is strictly positive, the $\psi_i$'s are smooth
functions. They satisfy the Wronskian condition (\ref{wroko}) and their monodromy
matrix is one of the four matrices on the right in the list (\ref{paracon}), with the off-diagonal entry 
coinciding with $\varepsilon$ and the overall $\pm1=(-1)^n$. The curve (\ref{etaphi}) corresponding to this 
basis of solutions is
\be
\eta(\phii)
=
\frac{\varepsilon\phii}{2\pi}-\frac{2}{n}\text{cot}(n\phii/2)
\nn
\ee
and has winding number $n$. This is all as in the hyperbolic 
case below eq.\ (\ref{pinacolada}). It follows that 
the function $p(\phii)$ defined by (\ref{defiqip}) is a Virasoro coadjoint vector with non-degenerate 
parabolic monodromy and winding number $n>0$, explicitly given by\i{supermomentum!massless}
\be
p(\phii)
=
\frac{c}{12}\,\frac{n^2}{H_{n,\varepsilon}(\phii)}
-\frac{c}{8}\,\frac{n^2(1+\varepsilon/2\pi)}{H_{n,\varepsilon}^2(\phii)}\,.
\label{pikachu}
\ee
\vspace{.1cm}

As in the hyperbolic case, one can think of (\ref{pikachu}) as 
a deformation of a suitable constant. However, in contrast ot (\ref{pitachy}), expression 
(\ref{pikachu}) seemingly contains no continuous parameter that one could tune to ``small'' values since 
$\varepsilon$ 
is only allowed to take the values $\pm1$. In order to solve this problem, recall from (\ref{AYAYA}) that the 
matrices
\be
\bmm 1 & \varepsilon \\ 0 & 1 \emm
\qquad\text{and}\qquad
\bmm 1 & \lambda\varepsilon \\ 0 & 1 \emm
\label{labacon}
\ee
are conjugate in $\SL$ for any positive real number $\lambda$. Accordingly we could just as well have chosen 
the 
representatives of non-degenerate parabolic conjugacy classes to involve an arbitrary positive parameter 
$\varepsilon$; the limit $\varepsilon\rightarrow0$ then may be taken since it does not affect the conjugacy 
class 
of the monodromy matrix.
The corresponding coadjoint vector is (\ref{pikachu}) and its expansion to first order in $\varepsilon$ 
reads\i{parabolic Virasoro orbit!by deformation of a constant}\i{exceptional constant!deformation}
\be
p(\phii)
=
-\frac{n^2c}{24}\left(
1+\frac{\varepsilon}{2\pi}(1+2\cos\phii)
\right)
+\cO(\varepsilon^2).
\label{pikachudef}
\ee
As in (\ref{pitachydef}), the leading term is an exceptional constant (\ref{pn}) and we can think of 
(\ref{pikachudef}) as a deformation thereof. The deformation is designed so that it does not belong 
to the orbit of $-n^2c/24$. When dealing with
BMS$_3$ supermomentum orbits in section \ref{sebmspar}, we will interpret (\ref{pikachu}) as the 
supermomentum of a massless BMS$_3$ 
particle. Accordingly, from now on we refer to 
non-degenerate parabolic orbits with non-zero winding as \it{massless orbits}.\i{Virasoro 
orbits!massless} Note that the statement 
that 
the matrices (\ref{labacon}) are conjugate is tantamount to saying that massless orbits are scale-invariant.\\

To conclude our analysis we state (without proof) a few features of massless orbits:
\begin{itemize}
\item One can show that the orbit representatives (\ref{pikachu}) are exhaustive in that 
any coadjoint vector belonging to a massless orbit can be brought in that form by a suitable diffeomorphism. 
See appendix C of \cite{Balog:1997zz}.
\item The Lie algebra of the stabilizer of (\ref{pikachu}) is generated by the vector field $X=\psi_2^2$, 
which has $n$ double zeros. In fact, as in the hyperbolic case, the stabilizer is isomorphic to 
$\RR\times\ZZ_n$, but the generator of the $\RR$ part of that group is 
\it{not} the same as in the hyperbolic case. The orbit is diffeomorphic to a quotient of 
$\Diffp$ by this stabilizer, or equivalently a quotient of $\Diffc$ by $\RR\times T_{2\pi/n}$ where 
$T_{2\pi/n}$ is the same discrete translation group as in (\ref{tachyorbitiv}).
\item Up to $\Diffc$ transformations, the solution (\ref{suliste})-(\ref{soliste}) is the unique solution of 
Hill's equation with non-degenerate parabolic monodromy
\be
(-1)^n\bmm 1 & \varepsilon \\ 0 & 1\emm
\label{MaH}
\ee
and winding number $n$.
\end{itemize}

\subsection{Summary: a map of Virasoro orbits}
\label{susesumav}

The above analysis exhausts all coadjoint orbits of the Virasoro group. Since these orbits will play a key 
role in the remainder of this thesis, we now briefly summarize the salient features of the 
classification.\\

\begin{figure}[h]
\centering
\includegraphics[width=0.30\textwidth]{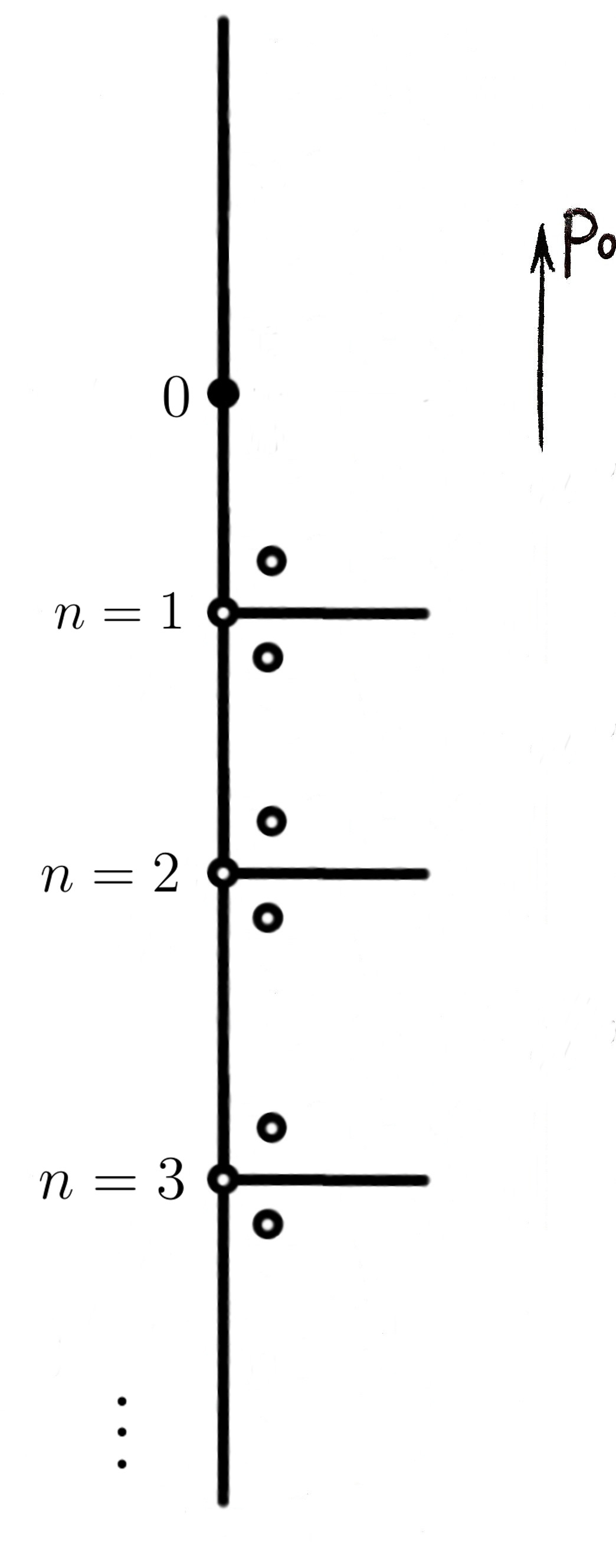}
\caption{The map of Virasoro coadjoint orbits at positive central charge. 
Note the similarity with fig.\ \ref{figDepIdu}{\textcolor{blue}{b}}. Roughly speaking, the map consists of an 
infinity of copies of 
Poincar\'e momentum orbits glued together and labelled by the winding 
number $n$. Locally (near a node $n$), the two pictures look identical. This 
is not surprising given that Poincar\'e momentum orbits in three dimensions coincide with $\SL$ coadjoint 
orbits, which in turn are classified similarly to the conjugacy 
classes of $\SL$ that were instrumental for Virasoro coadjoint orbits. This hints that 
there exists a relation between Virasoro and Poincar\'e symmetry; we shall see in part III
that this relation is embodied by the BMS$_3$ group.\i{map of Virasoro 
orbits}\i{Virasoro orbits!map}\label{vifig}}
\end{figure}

The schematic drawings of figs.\ \ref{figorbivir} and \ref{figorbiviralmost} represent Virasoro 
orbits. The only orbits which are not accounted for by these pictures are massless ones; in order to include 
them we use the 
same 
trick as in fig.\ \ref{figDepIdu}{\textcolor{blue}{b}}, where massless orbits are represented by two dots near 
the origin (one with 
positive energy, the other with negative energy). We will use the same notation here, except that such a pair 
of massless orbits occurs for all positive integers $n\in\NN^*$. With this convention, 
fig.\ \ref{figorbiviralmost} turns into the complete map of Virasoro coadjoint orbits displayed in fig.\ 
\ref{vifig}.\\

Each point in that map represents an orbit representative; different 
points correspond to different representatives and define disjoint orbits. All orbits 
are now accounted for since the orbit representatives are exhaustive. The vertical line represents orbits 
that 
contain a constant orbit representative:\i{Virasoro orbits!with constant representative}
\begin{itemize}
\item For generic $p_0<0$ the orbit has elliptic monodromy determined by eq.\ (\ref{hofrik}). Its winding 
number is given by (\ref{elliwind}), so the points of fig.\ \ref{vifig} located between $n$ and 
$n+1$ have winding number $n$ (while points such that $-c/24<p_0<0$ have zero winding 
number).
\item For exceptional values $p_0=-n^2c/24$ with $n\in\NN^*$, the orbit has degenerate parabolic monodromy 
determined by (\ref{monodegepa}). Its winding number is $n$. In particular, the orbit at $n=1$ is the 
\it{vacuum orbit}.
\item For $p_0>0$, the orbit has hyperbolic monodromy with zero winding, and the conjugacy class of the 
monodromy matrix is determined by (\ref{hofrik}).
\item The orbit of $p_0=0$ has non-degenerate parabolic monodromy with zero winding.
\end{itemize}
On the other hand, the points of fig.\ \ref{vifig} that do \it{not} 
belong to the vertical axis represent orbits that do \it{not} contain any constant representative:
\begin{itemize}
\item Each horizontal line starting at a point labelled by $n$ represents a family of tachyonic orbits with 
winding number $n$. The orbit representatives are given by (\ref{pitachy}) and involve a continuous 
parameter $\omega>0$ that determines the corresponding monodromy matrix (\ref{xox}).
\item Each pair of dots surrounding a tachyonic line at $n$ represents the two massless orbits with 
winding 
number $n$. The orbit representatives are given by (\ref{pikachu}) and involve a discrete parameter 
$\varepsilon=\pm1$ that determines the corresponding monodromy matrix (\ref{MaH}).
\end{itemize}

Focussing for definiteness on the multiply connected group $\Diffp$, the stabilizers of Virasoro orbits are 
as 
follows:\i{Virasoro orbits!stabilizer}
\begin{table}[H]
\centering
\begin{tabular}{|c|l|}
\hline
Orbit & Stabilizer\\
\hline
Vacuum-like $p_0=-n^2c/24$, $n\geq1$ & $\text{PSL}^{(n)}(2,\RR)$\\
Elliptic & $\un$\\
Hyperbolic, zero winding & $\un$\\
Non-degenerate parabolic, zero winding & $\un$\\
Massless, winding $n\geq 1$ & $\RR\times\ZZ_n$\\
Tachyonic, winding $n\geq 1$ & $\RR\times\ZZ_n$\\
\hline
\end{tabular}
\caption{Virasoro coadjoint orbits and their stabilizers.}
\label{tabVir}
\end{table}
In the universal cover of the Virasoro group the first four entries of the right column would be replaced by 
their universal covers, while the two last ones would be replaced by $\RR\times T_{2\pi/n}$. This should be 
compared with (and is very similar to) the list of Poincar\'e little groups in table \ref{tabOr}. Note that, 
at $n=1$, the Virasoro 
stabilizers are quotients by $\ZZ_2$ of their Poincar\'e counterparts. This is because table \ref{tabOr} 
lists the little groups given by the \it{double cover} (\ref{pisel}) of the Poincar\'e group.

\paragraph{Remark.} Fig.\ \ref{vifig} may be misleading since it suggests that all Virasoro 
orbits of constant coadjoint vectors are of a similar type, which is clearly not the case since orbits 
of constants $p_0>0$ are hyperbolic while those of 
(generic) constants $p_0<0$ are elliptic. In this sense, the map of orbits would have been more accurate if 
we had represented the orbits of $p_0>0$ by a \it{horizontal} line to suggest that they have the same type of 
monodromy as the tachyonic orbits; see e.g.\ fig.\ 1 of
\cite{Balog:1997zz}. Our convention in fig.\ \ref{vifig} is motivated instead by the fact that the value of 
$p_0$ essentially measures energy (see below), so that higher points in fig.\ \ref{vifig} 
have higher energy.

\section{Energy positivity}
\label{senepost}

In this section we investigate the boundedness properties of an energy functional on Virasoro 
orbits. This question is motivated both by its use in two-dimensional conformal field theory, and by 
its applications in three-dimensional gravity. We start by defining the Virasoro energy functional, before 
showing that the Schwarzian derivative satisfies an ``average lemma'' which will 
play a key role for this functional's boundedness. We then show that the only orbits 
with energy 
bounded from below are either orbits of constants $p_0\geq-c/24$, or the massless orbit at winding $n=1$ and 
monodromy $\varepsilon=-1$.
To reduce clutter we return to our earlier abusive notation by writing as $\Diff$ the 
universal cover of the group of orientation-preserving diffeomorphisms of the circle.
Relevant references include \cite{Balog:1997zz,guieu2007algebre} as usual.

\subsection{Energy functional}

The group $\Diff$ can be interpreted as (part of) the symmetry group of a two-dimensional conformal field 
theory. In that context the quadratic density 
$p(\phii)d\phii^2$ is (a component of) the stress tensor of the theory, and its zero-mode\i{Virasoro 
energy}\i{energy functional}\i{Virasoro energy}\i{Bondi mass (aspect)!as Virasoro energy functional}
\be
E[p]
\equiv
\frac{1}{2\pi}\int_0^{2\pi}d\phii\,p(\phii)
\label{enep}
\ee
is the associated energy. We shall refer to this quantity as the \it{Virasoro energy 
functional} evaluated at $p$. If the theory admits a configuration whose stress 
tensor is $p(\phii)$, then consistency with 
conformal symmetry requires that it also admits configurations with stress tensor $f\cdot p$, where 
$f\in\Diff$ and the dot denotes the coadjoint action (\ref{covi}) for some definite 
value of the central charge. The energy functional varies under conformal transformations, since
\be
E[f\cdot p]
=
\frac{1}{2\pi}\int_0^{2\pi}
\frac{d\phii}{f'(\phii)}
\left[
p(\phii)+\frac{c}{12}\sfS[f](\phii)
\right]
\label{enefp}
\ee
generally differs from (\ref{enep}).\\

Now consider a CFT with central charge $c>0$ and let $\bbOmega$ be the space of its stress tensors 
$p(\phii)$; in general $\bbOmega$ is a certain subset of the space $\cF_2(S^1)$ 
of
quadratic densities. Since any quantum system with a well-defined vacuum is expected to have energy 
bounded from below, the map
\be
\bbOmega\rightarrow\RR:p\mapsto E[p]
\label{bimogan}
\ee
should
be bounded from below. In addition, consistency with conformal symmetry implies that $\bbOmega$ is a union of 
Virasoro coadjoint orbits. One is thus led to the following question:
\be
\begin{array}{c}
\text{\it{Which of the Virasoro coadjoint orbits of fig.\ \ref{vifig} have}}\\
\text{\it{energy bounded from below under conformal transformations?}}
\end{array}
\label{enequest}
\ee
In the sequel we will refer to orbits with energy bounded from below as orbits ``with positive 
energy'', although their energy (\ref{enep}) may actually be negative for some field configurations 
$p(\phii)$.\\

Note that all orbits have energy unbounded from \it{above}. Indeed the term involving the Schwarzian 
derivative in (\ref{enefp}) can be written as
\be
\frac{c}{24\pi}\int_0^{2\pi}
\frac{d\phii}{f'(\phii)}\sfS[f](\phii)
=
-\frac{c}{24\pi}\int_0^{2\pi}d\phii\,\sfS[f^{-1}](\phii)
\label{enefps}
\ee
where we have renamed the integration variable from $\phii$ to $f^{-1}(\phii)$, then used (\ref{invide}) and 
the 
cocycle identity (\ref{swapro}). Since the Schwarzian derivative can be written as
\be
\sfS[f](\phii)
=
\left(\frac{f''}{f'}\right)'-\demi\left(\frac{f''}{f'}\right)^2,
\label{depeas}
\ee
we can also recast (\ref{enefps}) in the form\i{Virasoro 
energy!unbounded from above}
\be
\frac{c}{48\pi}\int_0^{2\pi}d\phii\,\left(\frac{(f^{-1})''}{(f^{-1})'}\right)^2.
\nn
\ee
This can be made arbitrarily large for suitable choices of $f$, which proves that the energy functional $E$ 
is unbounded from above on any Virasoro orbit.

\subsection{The average lemma}

As a first step towards the answer of the question (\ref{enequest}), we focus on the piece of eq.\ 
(\ref{enefp}) that involves the Schwarzian derivative. The result that we shall describe was first derived in 
\cite{Schwartz} and was based on projective geometry (see also 
\cite{Schwartz1997,Tabachnikov}). The elementary proof given here 
is borrowed from \cite{Balog:1997zz}.

\paragraph{Average lemma.} Let $f\in\Diffc$ and let $\sfS[f](\phii)$ be its Schwarzian derivative 
(\ref{swag}) 
at $\phii$. Then the average of the Schwarzian derivative satisfies the inequality\i{average 
lemma}\i{Schwarzian derivative!average lemma}
\be
\int_0^{2\pi}d\phii\,\sfS[f](\phii)\;
\leq\;
\int_0^{2\pi}d\phii\,\demi\left(1-(f'(\phii))^2\right),
\label{avlem}
\ee
with equality if and only if $f(\phii)$ is a projective transformation of the form 
(\ref{protophi}).

\begin{proof} We consider the functional
\be
I[f]
\equiv
-
\int_0^{2\pi}d\phii\left[\demi(f'(\phii))^2+\sfS[f](\phii)\right].
\label{fictoe}
\ee
Our goal is to show that this quantity is bounded from below and that its minimum value is $-\pi$. By 
(\ref{depeas}), it only depends on 
$f'$ and $f''$. A convenient way to express 
this dependence is to define
\be
Y(\phii)\equiv f'(f^{-1}(\phii))
\refeq{invide}
\frac{1}{(f^{-1})'(\phii)}\,.
\label{defyi}
\ee
Since $f$ is a $2\pi\ZZ$-equivariant, orientation-preserving diffeomorphism, $Y(\phii)$ is strictly positive 
and $2\pi$-periodic. In terms of $Y$ we can rewrite (\ref{fictoe}) as
\be
I[Y]
=
\demi
\int_0^{2\pi}d\phii
\left[
\frac{(Y'(\phii))^2}{Y(\phii)}
-Y(\phii)
\right]
\label{fourteen}
\ee
where the integrand is well-defined since $Y>0$.
Let us denote the minimum and maximum of $Y(\phii)$ by
\be
m\equiv\min_{\phii\in[0,2\pi]}Y(\phii)\,,
\qquad
M\equiv\max_{\phii\in[0,2\pi]}Y(\phii)\,.
\label{minimax}
\ee
With this notation the function
\be
m+M-Y(\phii)-\frac{mM}{Y(\phii)}
=
\frac{1}{Y(\phii)}
\left[
\left(\frac{M-m}{2}\right)^2
-\left(Y(\phii)-\frac{M+m}{2}\right)^2
\right]
\nn
\ee
is non-negative and vanishes only 
at the points where $Y$ reaches its minimum or its maximum. Now consider the obvious inequality
\be
\left(
\frac{|Y'|}{\sqrt{Y}}
-
\sqrt{m+M-Y-\frac{mM}{Y}}
\right)^2
\geq0.
\label{mobit}
\ee
Integrating this over the circle and using (\ref{fourteen}), we obtain
\be
I[Y]
\geq
-\pi(m+M-mM)
+
\int_0^{2\pi}d\phii
\frac{|Y'|}{Y}
\sqrt{\Big(\frac{M-m}{2}\Big)^2
-\Big(Y-\frac{M+m}{2}\Big)^2}.
\label{iyer}
\ee
If there was no absolute value in the integrand on the right-hand side, we could just change the integration 
variable from $\phii$ to $Y$ using $d\phii\,Y'(\phii)=dY$; the absolute value prevents us from doing this 
globally, but we can do it locally between two consecutive extrema of the function $Y(\phii)$ (since the sign 
of $Y'$ is constant in such an interval). We can then express the right-hand side of (\ref{iyer}) in terms of 
the primitive function of the integrand,
\be
\cF(Y)
=
\int_m^Y\frac{dz}{z}
\sqrt{\left(\frac{M-m}{2}\right)^2
-\left(z-\frac{M+m}{2}\right)^2}
\equiv
\int_m^Ydz\,\cG(z),
\label{brave}
\ee
where we have introduced the function $\cG(z)$ to reduce clutter below. To see the use of this, 
consider a function $Y(\phii)$ of the following shape (the general case follows straightforwardly):

\begin{figure}[H]
\centering
\includegraphics[width=0.50\textwidth]{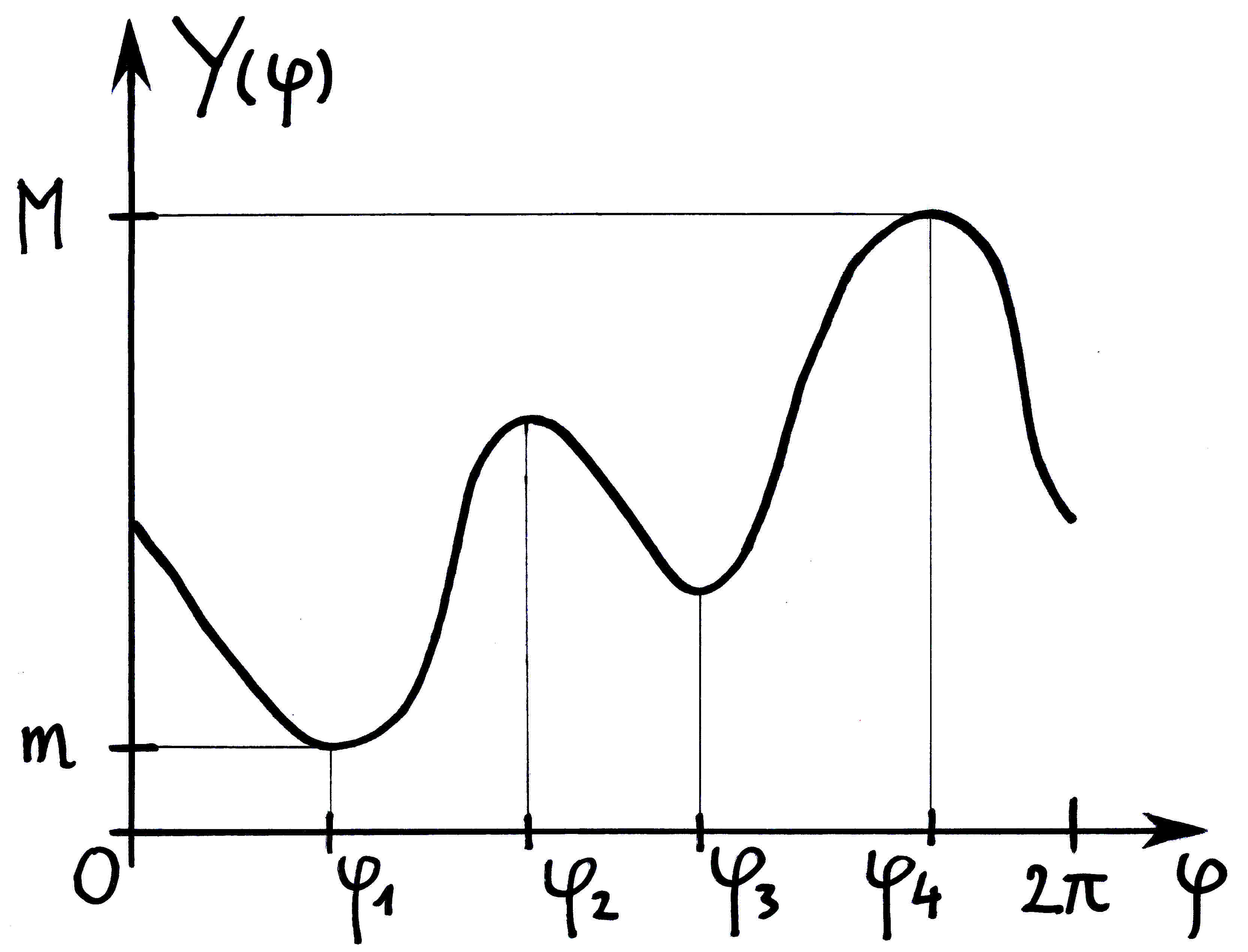}
\caption{The function $Y(\phii)$ is $2\pi$-periodic and strictly positive. Here we choose it with four local 
extrema, the global minimum being $Y(\phii_1)=m$ and the global maximum $Y(\phii_4)=M$.}
\end{figure}

This function has two local minima at $\phii_1$ and $\phii_3$ and two local maxima at $\phii_2$ and $\phii_4$ 
(the numbers of local minima and maxima coincide since $Y(\phii)$ is smooth and $2\pi$-periodic). Then 
the integral in (\ref{iyer}) can be written as
\begin{eqnarray}
&   &
\int_0^{2\pi}d\phii
\frac{|Y'|}{Y}
\sqrt{\left(\frac{M-m}{2}\right)^2
-\left(Y-\frac{M+m}{2}\right)^2}=\nn\\
& = &
\!\!\int_{Y_1}^{Y_2}dY\,\cG(Y)
-\int_{Y_2}^{Y_3}dY\,\cG(Y)
+\int_{Y_3}^{Y_4}dY\,\cG(Y)
-\int_{Y_4}^{Y_1}dY\,\cG(Y)\nn\\
& \refeq{brave} &
\!\!2\left[\cF(Y_2)+\cF(Y_4)-\cF(Y_1)-\cF(Y_3)\right]\nn
\end{eqnarray}
with the shorthand notation $Y(\phii_i)\equiv Y_i$. The same computations would work for 
arbitrarily many minima and maxima of $Y(\phii)$, with the same result: the integral is twice the sum of 
$\cF$'s evaluated at the maxima minus twice the sum of $\cF$'s evaluated at the minima.
Thus the inequality (\ref{iyer}) can be written as
\begin{eqnarray}
I[Y]
& \!\!\!\stackrel{\text{(\ref{minimax})}}{\geq} &
\!\!\!-\pi(m+M-mM)+2[\cF(M)-\cF(m)]+2[\cF(Y_2)-\cF(Y_3)]\nn\\
\label{indspu}
& \!\!\!\geq &
\!\!\!-\pi(m+M-mM)+2\cF(M)
\end{eqnarray}
where we also used the fact that $\cF(m)=0$ by virtue of the definition (\ref{brave}). Now it 
turns out that $\cF(M)=\frac{\pi}{2}\left(\sqrt{M}-\sqrt{m}\right)^2$, which allows us to rewrite 
(\ref{indspu}) as
\be
I[Y]
\geq
-\pi(m+M-mM)+\pi\left(\sqrt{M}-\sqrt{m}\right)^2
\geq
-\pi.
\label{ingf}
\ee
We conclude that $I[Y]$ is bounded from below by the value $-\pi$, 
which is exactly the inequality (\ref{avlem}).
It only remains to find the conditions under which (\ref{avlem}) becomes an equality. For this to be the 
case, the inequalities (\ref{mobit}), (\ref{indspu}) and (\ref{ingf}) must all be saturated; this occurs when 
$Y(\phii)$ satisfies the following three conditions:
\begin{itemize}
\item In order to saturate (\ref{mobit}), it satisfies the differential equation
\be
Y'^2=(m+M)Y-Y^2-mM.
\label{ignomo}
\ee
\item In order to saturate (\ref{indspu}), $Y(\phii)$ has only one minimum and one maximum, where it takes 
the values $m$ and $M$, respectively.
\item In order to saturate the second inequality of (\ref{ingf}), $M=1/m$.
\end{itemize}
To solve (\ref{ignomo}) we use (\ref{defyi}) and rewrite the equation in terms of $f^{-1}$. Using $M=1/m$ the 
derivative of (\ref{ignomo}) becomes
\be
Y'\left(1-((f^{-1})')^2-2\,\sfS[f^{-1}]\right)=0,
\label{sgfr}
\ee
which is equivalent to (\ref{swagiff}). We have shown below (\ref{seltusta}) that the only 
$f$'s satisfying this property are those that belong to the group of projective transformations 
(\ref{protophi}), which concludes the proof.
\end{proof}

\subsection{Orbits with constant representatives}

The average lemma allows us to investigate the boundedness properties of the energy 
functional (\ref{enep}) on Virasoro orbits. For now we limit ourselves to orbits that admit a 
constant representative.

\paragraph{Proposition.} The vacuum orbit, containing the point $p_{\text{vac}}=-c/24$, has energy bounded 
from below: 
\i{Virasoro energy!on vacuum orbit}\i{vacuum orbit!energy}\i{energy functional!on vacuum 
Virasoro orbit}
\be
E[f\cdot p_{\text{vac}}]\geq E[p_{\text{vac}}]=-\frac{c}{24}.
\label{enevac}
\ee
The minimum of energy is located at $p_{\text{vac}}$.

\begin{proof}
We consider formula (\ref{enefp}) with $p(\phii)=p_{\text{vac}}=-c/24$. Renaming the integration variable 
from $\phii$ to $f^{-1}(\phii)$ and using eqs.\ (\ref{invide}) and (\ref{swapro}), we find
\be
E[f\cdot p_{\text{vac}}]
=
\frac{c}{24\pi}
\int_0^{2\pi}
d\phii
\left[
-\demi\big((f^{-1})'(\phii)\big)^2
-\sfS[f^{-1}](\phii)
\right]
\nn
\ee
which we recognize as the functional (\ref{fictoe}) evaluated at $f^{-1}$. The average lemma (\ref{avlem}) 
then implies that $E[f\cdot p_{\text{vac}}]\geq-c/24$, with equality if and only if $f$ is a 
projective transformation (\ref{protophi}). Our earlier result (\ref{factipuce}) ensures 
that such transformations precisely span the stabilizer of $p_{\text{vac}}$, so the minimum of energy is 
reached at $p_{\text{vac}}$.
\end{proof}

Let us turn to other orbits containing a 
constant representative $p(\phii)=p_0$. The key will be to rewrite their energy functional as the 
vacuum energy functional, plus another term. Starting from formula 
(\ref{enefp}) we obtain
\be
E[f\cdot p_0]
=
\frac{p_0+c/24}{2\pi}
\int_0^{2\pi}
\frac{d\phii}{f'(\phii)}
+E[f\cdot p_{\text{vac}}]
\label{microcosm}
\ee
where the integral of $1/f'$ can be rewritten as
\be
\frac{1}{2\pi}\int_0^{2\pi}\frac{d\phii}{f'(\phii)}
\refeq{invide}
1+
\frac{1}{2\pi}\int_0^{2\pi}
d\phii\,
\big[(f^{-1})'(\phii)-1\big]^2
\nn
\ee
as follows from $\int_0^{2\pi}d\phii(f^{-1})'(\phii)=2\pi$. Plugging this into 
(\ref{microcosm}) and using (\ref{enevac}), we obtain
\be
E[f\cdot p_0]
\geq
p_0+
\frac{p_0+c/24}{2\pi}
\int_0^{2\pi}
d\phii\,\big[(f^{-1})'(\phii)-1\big]^2.
\nn
\ee
The 
right-hand side here is the sum of $p_0$ and an integral whose integrand is manifestly 
non-negative. This implies the following result:\i{Virasoro orbits!with positive energy}

\paragraph{Proposition.} If $p_0\geq-c/24$, then the orbit of $(p_0,c)$ has energy bounded from below, with 
the energy minimum located at $p_0$:\i{energy functional!on elliptic Virasoro orbit}
\be
p_0\geq-\frac{c}{24}
\qquad\Rightarrow\qquad
E[f\cdot p_0]\geq E[p_0]=p_0\,.
\nn
\ee

\begin{figure}[H]
\centering
\includegraphics[width=0.50\textwidth]{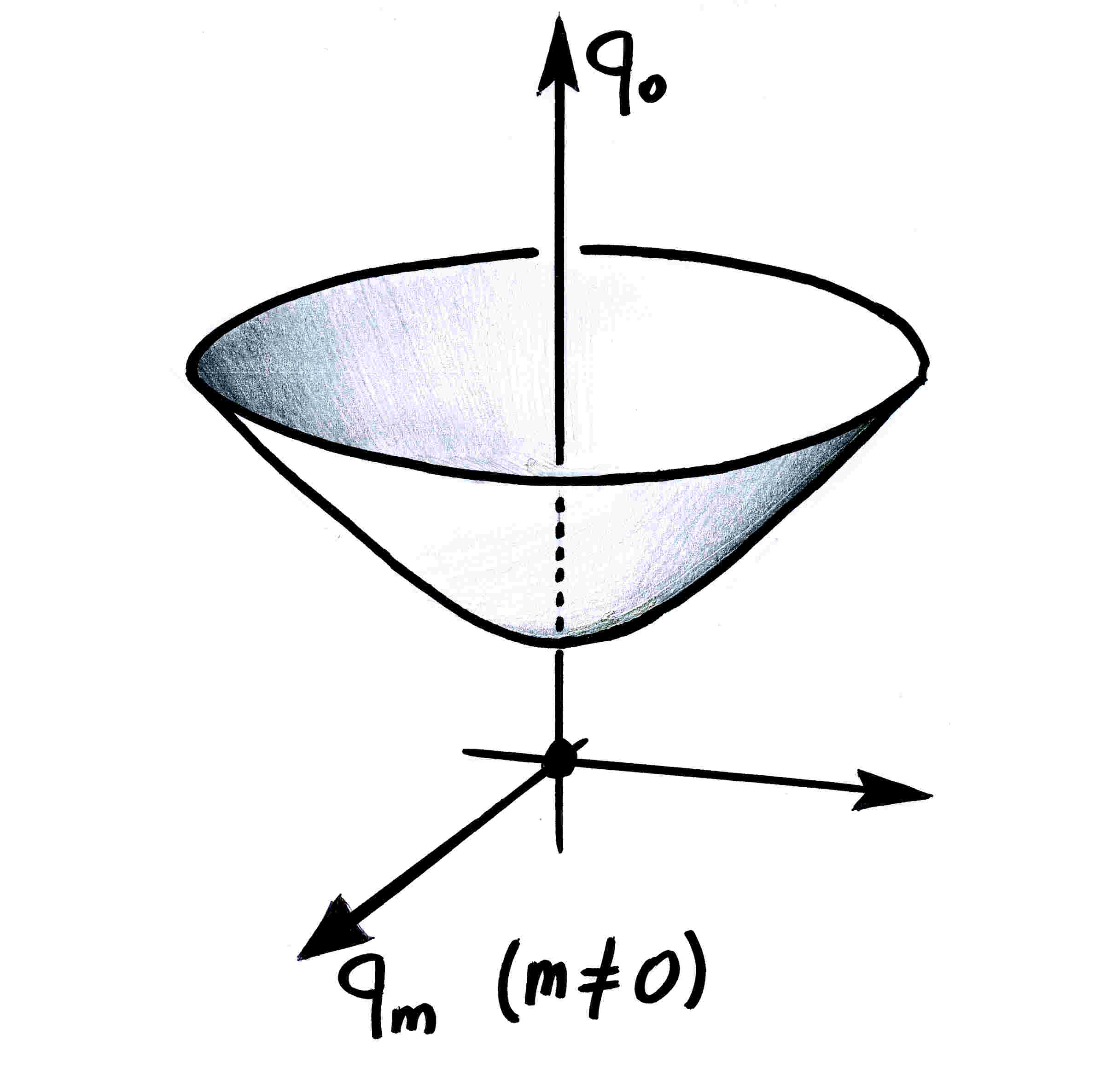}
\caption{Schematic representation of the Virasoro orbit of a constant $p_0$ located above the vacuum 
value $-c/24$, here understood to be the origin of the coordinate system. The coordinates $q_m$, 
$m\in\ZZ$ are the Fourier modes of coadjoint vectors $q(\phii)$; in particular the zero-mode $q_0=E[q]$ is 
their 
energy, which is bounded from below on the orbit. Compare to the massive 
Poincar\'e orbit with 
positive energy in fig.\ \ref{figDepIdu}{\textcolor{blue}{a}}.\label{treza}}
\end{figure}

Now what happens when $p_0$ is \it{lower} than $-c/24$? In that case energy is \it{unbounded}, as can be 
shown by finding a family of diffeomorphisms 
that lower the energy indefinitely. Indeed, consider the matrix
\be
\bmm
\cosh(\gamma/2) & \sinh(\gamma/2) \\ \sinh(\gamma/2) & \cosh(\gamma/2)
\emm
\in\SL
\label{booma}
\ee
where $\gamma\in\RR$ (the normalization is chosen for later convenience). The corresponding projective 
transformation (\ref{protophi}) is\i{boost!as projective transformation}\i{projective 
transformation!boost}\i{pure boost}
\be
e^{if(\phii)}
=
\frac{e^{i\phii}\cosh(\gamma/2)+\sinh(\gamma/2)}{-e^{i\phii}\sinh(\gamma/2)+\cosh(\gamma/2)}\,,
\label{Booss}
\ee
and one verifies that
\be
\frac{1}{f'(\phii)}
=
|e^{i\phii}\cosh(\gamma/2)+\sinh(\gamma/2)|^2
=
\cosh\gamma+\sinh\gamma\,\cos\phii\,.
\label{fagobo}
\ee
The Schwarzian derivative of $f$ is given by (\ref{swagiff}), so we find that
\be
E[f\cdot p_0]
\refeq{enefp}
\frac{p_0+c/24}{2\pi}\int_0^{2\pi}\frac{d\phii}{f'(\phii)}-\frac{c}{24}
\nn
\ee
where we have used the fact that the integral of $f'$ over $S^1$ is normalized to $2\pi$. The integral of 
(\ref{fagobo}) then yields
\be
E[f\cdot p_0]
=
(p_0+c/24)\cosh\gamma-\frac{c}{24},
\label{esoboom}
\ee
and this can become arbitrarily negative when $p_0<-c/24$. In conclusion:
\be
\begin{array}{c}
\text{\it{The coadjoint orbit $\cW_{(p_0,c)}$ of a constant $p_0$}}\\
\text{\it{has energy bounded from below if and only if $p_0\geq-c/24$.}}
\end{array}
\label{konkotab}
\ee
Thus, when $p_0<-c/24$, fig.\ \ref{treza} is no longer valid because the energy functional can 
reach arbitrarily low values in certain directions. The orbit then looks like an 
infinite-dimensional saddle instead of the hyperboloid represented in fig.\ \ref{treza}.\\

Note that the matrix (\ref{booma}) can be interpreted as the $\SL$ group element that represents a Lorentz 
boost with rapidity $\gamma$ in three dimensions\footnote{Rapidity is related to velocity $v$ by 
$\gamma=\text{arctanh}(v)$.} thanks to the isomorphism (\ref{isoso}), which also explains 
our choice of normalization. In that context, formula (\ref{esoboom}) is the transformation law of the energy 
of a particle with mass $p_0+c/24$ under Lorentz boosts. We will return to this interpretation in part III.

\subsection{Orbits without constant representatives}

We now describe the boundedness properties of the energy functional on Virasoro coadjoint 
orbits that do \it{not} admit a constant representative. As it turns out there is only one orbit with 
energy bounded from below, while all other ones have unbounded energy.\\

Consider the non-degenerate parabolic orbit with winding number $n=1$ and monodromy $\varepsilon=-1$; a 
typical orbit representative is 
given by (\ref{pikachu}). One can then prove the following result:\i{Virasoro energy!on parabolic orbits}

\paragraph{Proposition.} The energy functional on the massless orbit specified by $n=1$ and $\varepsilon=-1$ 
is 
bounded from below by $-c/24$.\i{energy functional!on parabolic Virasoro orbit}\i{parabolic 
Virasoro orbit!energy functional} There exist infinitely many points on 
the orbit whose energy is arbitrarily 
close to that value, but there is no orbit representative that realizes this value of energy.\\

We will not prove this proposition here and refer instead to \cite{Balog:1997zz}. Roughly speaking, the 
proof follows from a construction very similar to the one used in the proof of the average lemma 
(\ref{avlem}), except that it crucially relies on the parameters $n=1$, 
$\varepsilon=-1$. The proof of the fact that 
the infimum of energy is never reached on the orbit follows from the construction of a one-parameter family 
of points belonging to the orbit in such a way that they converge to the constant $-c/24$ without ever quite 
reaching it.

\begin{figure}[H]
\centering
\includegraphics[width=0.50\textwidth]{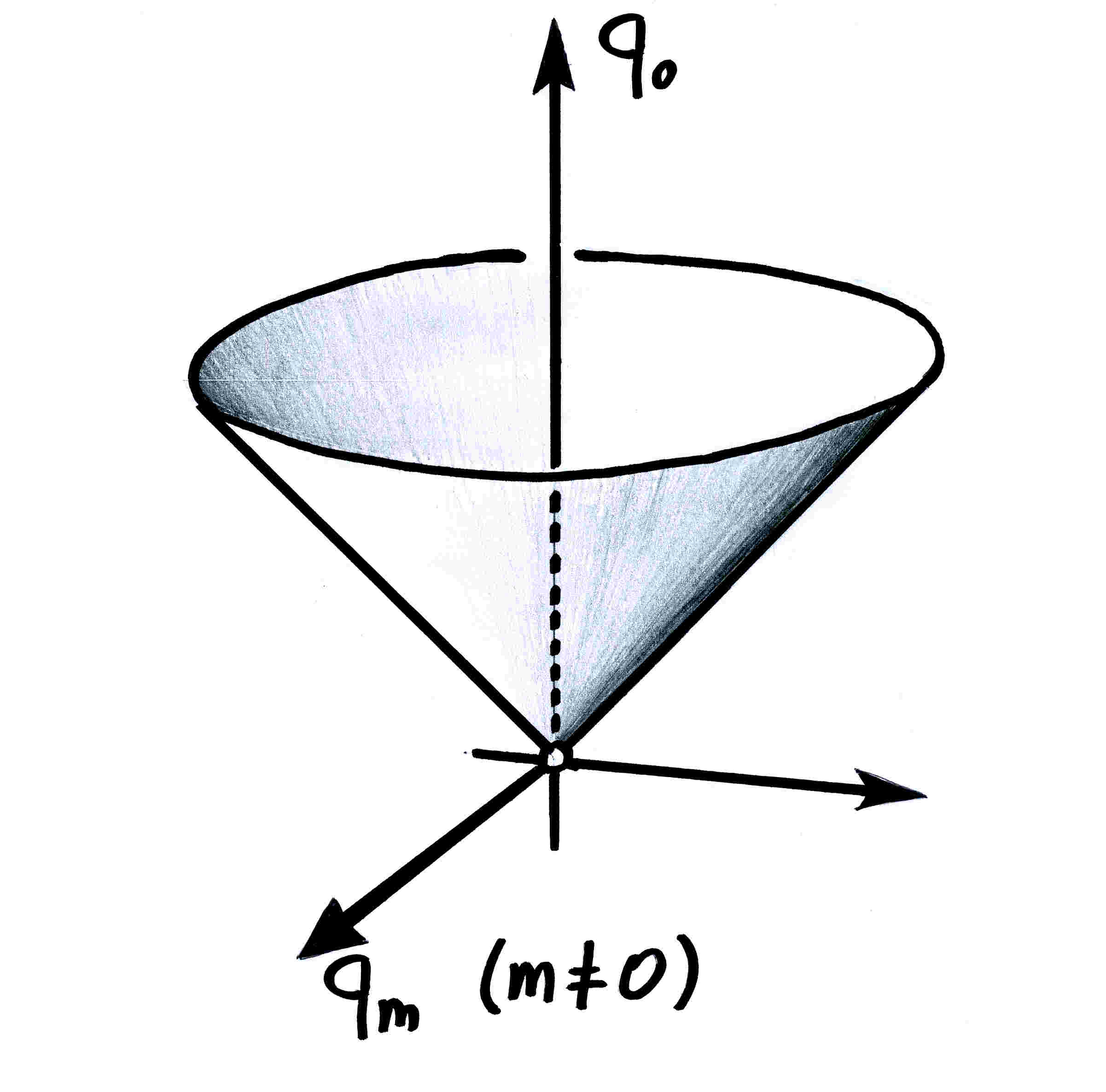}
\caption{Schematic representation of the Virasoro orbit with non-degenerate parabolic monodromy 
$\varepsilon=-1$ and winding number $n=1$. The origin of the coordinate system is the vacuum coadjoint 
vector, $q_m=-(c/24)\delta_{m0}$, which does not belong to the orbit. Energy is bounded from below on the 
orbit but its infimum is never quite reached, in 
contrast to fig.\ \ref{treza}. Compare to the massless Poincar\'e orbit with 
positive energy in fig.\ \ref{figDepIdu}{\textcolor{blue}{a}}.}
\end{figure}

One might think that the other orbits without constant representatives behave in a similar way, i.e.\ that 
they also have energy bounded from below. However, for any such orbit, it is possible to 
build a one-parameter family of orbit elements whose energy can be arbitrarily low, similarly to constant 
representatives $p_0<-c/24$. We refer again to \cite{Balog:1997zz} for explicit 
constructions. Thus one concludes that 
\i{Virasoro orbits!with positive energy}\i{energy functional!unbounded}
\be
\begin{array}{c}
\text{\it{all tachyonic or massless Virasoro orbits have \emph{unbounded} energy,}}\\
\text{\it{except the one 
with non-degenerate parabolic monodromy}}\\
\text{\it{$\varepsilon=-1$ and winding $n=1$.}}
\end{array}
\label{konkotob}
\ee

\subsection{Summary: a new map of Virasoro orbits}

\begin{figure}[h]
\centering
\includegraphics[width=0.30\textwidth]{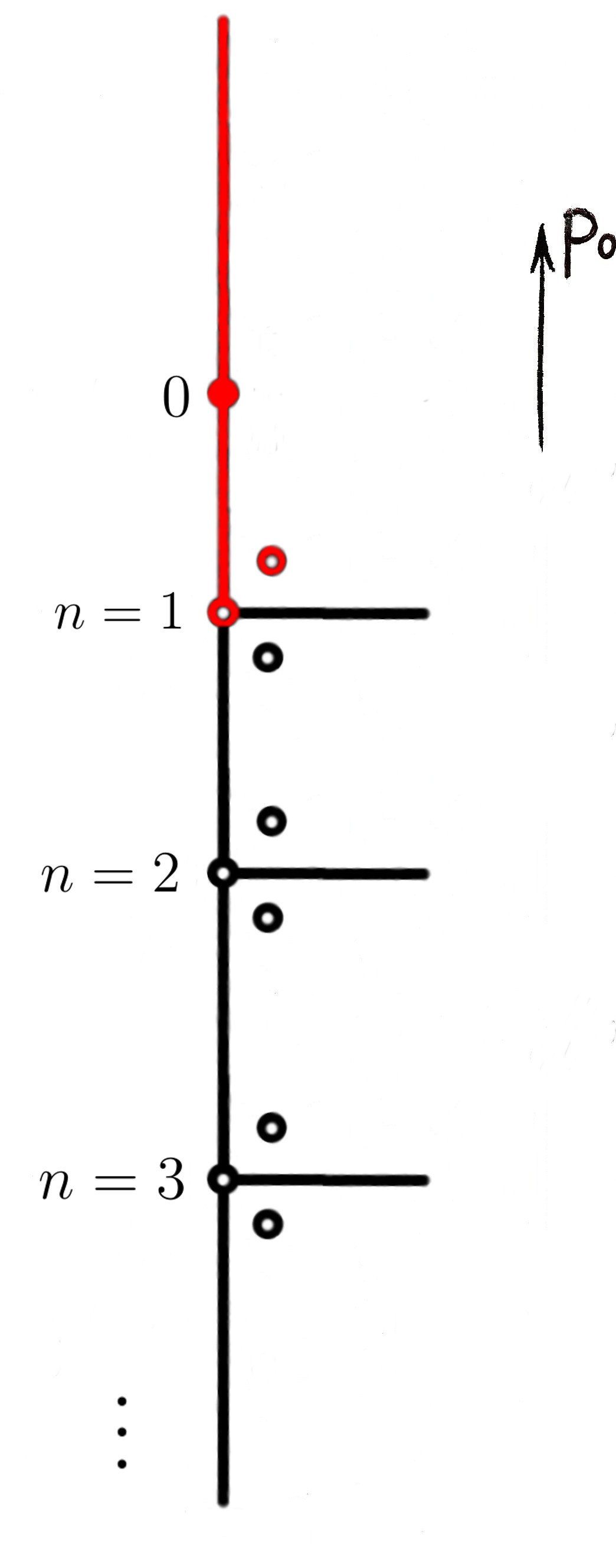}
\caption{The map of Virasoro coadjoint orbits at positive central charge.\i{Virasoro orbits!map}\i{map of 
Virasoro orbits} Orbits with energy 
bounded from below are coloured in red. Those are orbits of constants $p_0\geq-c/24$, plus 
the unique massless orbit with monodromy (\ref{MaH}) such that $\varepsilon=-1$ and winding number 
$n=1$. All other orbits have energy unbounded from below.\label{vifigibis}}
\end{figure}

The considerations of the last few pages allow us to include more information in the map of Virasoro orbits 
of fig.\ \ref{vifig}; see fig.\ \ref{vifigibis}. Its two striking features are the occurrence of 
a single orbit without constant representatives 
and positive energy, and the fact that the lowest-lying orbit with positive energy is that of 
$p_{\text{vac}}=-c/24$. This observation justifies referring to the latter orbit as the ``vacuum orbit'' and 
to 
$p_{\text{vac}}$ as the ``vacuum stress tensor''. Note that the exact same situation occurs with relativistic 
particles, as the only ones with energy bounded from below are either massive (with non-negative mass) or 
massless.

\newpage
~

\newpage
~
\thispagestyle{empty}

\chapter{Symmetries of gravity in AdS$_3$}
\label{AdS3}
\markboth{}{\small{\chaptername~\thechapter. Symmetries of gravity in AdS$_3$}}

In this chapter we explore a physical model where the Virasoro group plays 
a key role, namely three-dimensional gravity on Anti-de Sitter (AdS) backgrounds and its putative dual 
two-dimensional
conformal field theory (CFT). These considerations will be a basis and a guide for our 
study of asymptotically flat space-times in part III.\\

The plan is the following. Section \ref{seGeTri} is a prelude where we recall a few basic facts about 
(three-dimensional) gravity, in particular regarding the notion of 
asymptotic symmetries. Section \ref{sebohad} is then devoted to three-dimensional space-times whose metric 
approaches that of Anti-de Sitter space at spatial infinity; this includes Brown-Henneaux boundary conditions 
and their asymptotic symmetries, which will turn out to consist of two copies of the Virasoro group. In 
section \ref{sebitu} we describe the phase space of AdS$_3$ gravity as a hyperplane at fixed central charges 
in the space of the coadjoint representation of two Virasoro groups. Finally, in section \ref{sevirep} we 
describe 
unitary highest-weight representations of the Virasoro algebra and relate them to the quantization of the 
AdS$_3$ phase space.

\paragraph{Bibliographical remarks.} This chapter is based on several combined references. Perhaps the most 
important one is the original paper by Brown and Henneaux \cite{Brown:1986nw}, which triggered the 
development 
of the field as a whole. In that paper the authors relied on the methods of 
\cite{Geroch:1972up,Geroch1977,Ashtekar1980,ABBOTT1982259} to build surface charges associated with 
asymptotic symmetries, but our approach will be led by their Lagrangian (or 
``covariant'') reformulation \cite{Barnich:2001jy,Barnich:2003xg,Barnich:2007bf}. In 
particular, our presentation of Brown-Henneaux boundary conditions and of the associated asymptotic Killing 
vector fields follows \cite{Barnich:2010eb}. The 
general solution of the equations of motion first appeared in \cite{Banados:1998gg,Skenderis:1999nb}. It 
contains in particular the BTZ black hole, which was discovered and studied in 
\cite{Banados:1992wn,Banados:1992gq}. Finally, the 
group-theoretic approach to the gravitational phase space first appeared in 
\cite{Garbarz:2014kaa,Barnich:2014zoa,Barnich:2015uva}, 
which is also where 
the AdS$_3$ positive energy theorem was derived. (See also 
\cite{Nakatsu:1999wt,NavarroSalas:1999sr,Maloney:2007ud} for earlier related considerations.)

\section{Generalities on three-dimensional gravity}
\label{seGeTri}

Here we recall a few basic facts about classical general relativity in three dimensions. We start by 
explaining that three-dimensional Einstein gravity has no local degrees of freedom, then turn to a 
discussion of boundary conditions and the ensuing boundary terms that one adds to the action in order to make 
the variational principle well-defined. This finally leads to the concept of asymptotic symmetries and the 
important observation that the Poisson brackets of surface charges that generate these symmetries generally 
contain central extensions.

\subsection{Einstein gravity in three dimensions}

We consider an orientable three-dimensional space-time manifold $\cM$ endowed with coordinates $x^{\mu}$ 
($\mu=0,1,2$) on which we put a metric $g_{\mu\nu}$ with signature
$(-\,+\,+)$. The 
equations of 
motion are determined by the \it{Einstein-Hilbert action}, 
\i{Einstein-Hilbert action}\i{action functional!for gravity}
\be
S_{\text{EH}}[g_{\mu\nu},\Phi]
=
\frac{1}{16\pi G}\int_{\cM}d^3x\sqrt{-g}\left(R-2\Lambda\right)
\,.
\label{s174}
\ee
Here $G$ is the Newton constant in three dimensions, $R$ is 
the Ricci scalar associated with $g_{\mu\nu}$ and $\Lambda\in\RR$ is a cosmological constant. In three 
dimensions, and using units such that $c=\hbar=1$, Newton's constant $G$\i{Newton constant} is a length 
scale. 
Equivalently $1/G$ is an energy scale that coincides with the Planck mass.\i{Planck mass}\\

Upon varying the action (\ref{s174}) and neglecting all boundary terms (which we shall talk about later), one 
obtains the vacuum Einstein's equations with a cosmological constant:\i{equations of 
motion!for gravitational field}
\be
R_{\mu\nu}-\demi Rg_{\mu\nu}+\Lambda g_{\mu\nu}
=
0\,,
\qquad
\text{i.e.}
\qquad
R_{\mu\nu}=2\Lambda g_{\mu\nu}\,.
\label{s175}
\ee
What is special about three-dimensional manifolds is that their Ricci curvature wholly 
determines their 
Riemann tensor independently of the equations of motion:\i{Riemann tensor}
\be
R_{\lambda\mu\nu\rho}
=
g_{\lambda\nu}R_{\mu\rho}
-g_{\lambda\rho}R_{\mu\nu}
-g_{\mu\nu}R_{\lambda\rho}
+g_{\mu\rho}R_{\lambda\nu}
-\demi R(g_{\lambda\nu}g_{\mu\rho}-g_{\lambda\rho}g_{\mu\nu})\,.
\label{ss177}
\ee
Then 
the Einstein equations (\ref{s175}) imply that, at each point of space-time, the on-shell Riemann 
tensor is that of a maximally symmetric manifold with curvature determined by the cosmological 
constant:\i{maximally symmetric}
\be
R_{\lambda\mu\nu\rho}
=
\Lambda
(g_{\lambda\nu}g_{\mu\rho}
-g_{\lambda\rho}g_{\mu\nu})\,.
\label{t177}
\ee
In other words, any solution of Einstein's equations in three dimensions is locally isometric 
to three-dimensional de Sitter, Minkowski or Anti-de Sitter space depending on whether $\Lambda$ is positive, 
vanishing or negative respectively. This is strikingly different from higher-dimensional general relativity 
and relies on the relation (\ref{ss177}) expressing Riemann in terms of Ricci, valid only in two and 
three dimensions.\footnote{In two dimensions one has in addition $R_{\mu\nu}=\demi R g_{\mu\nu}$.} In 
technical terms it is the statement that\i{no gravitons in 3D}\i{degrees of 
freedom}\i{3D gravity}\i{gravity in 3D}
\be
\begin{array}{c}
\text{\it{there are no local degrees of freedom in three-dimensional Einstein gravity.}}
\end{array}
\nn
\ee
It follows in particular that there are no gravitational waves, hence no gravitons. Equivalently, all 
configurations of the metric are locally gauge-equivalent to empty space.
Importantly, this is \it{not} to say that the only solution of three-dimensional gravity is empty space. For 
example, any quotient of Minkowski space by some 
discrete group solves Einstein's equations, but is not globally isometric to 
Minkowski. Thus global aspects are 
essential: even though all
solutions of Einstein's equations are locally isometric, they are generally \it{not} globally isometric 
and therefore represent 
physically distinct field configurations. In this sense the absence of local degrees of freedom in 
three-dimensional gravity does not prevent the overall absence of degrees of freedom: it only means that the 
actual, physical degrees of freedom of the theory cannot be captured by a local analysis, but 
require instead a \it{global} one, taking into account topological properties of the space-time manifold. 
Field theories of this type, having no local degrees of freedom but still globally non-trivial, are called 
\it{topological field theories}.\i{topological field theory}\\

Note that the absence of local degrees of freedom is confirmed by the Hamiltonian 
formalism \cite{Teitelboim:1972vw}: picking a time direction in $\cM$, one can split the metric field into a 
lapse $N$, a shift $N^i$ and a spatial metric $g_{ij}$ with conjugate momenta $\pi^{ij}$, the indices 
$i,j\in\{1,2\}$ labelling spatial directions.\i{no gravitons in 3D!Hamiltonian counting}\i{counting 
degrees of freedom} The lapse and shift play the role of Lagrange multipliers 
enforcing the 
constraints that generate reparameterizations of time and spatial diffeomorphisms, respectively. One thus 
obtains 
three dynamical Lagrange variables $g_{ij}$ with three conjugate momenta $\pi^{ij}$, subject to three 
first-class constraints. These constraints can be solved by choosing three gauge-fixing conditions (this is 
the statement that ``first-class constraints count twice''), which reduces the number of physical degrees 
of freedom of three-dimensional Einstein gravity to
$\demi(3\times 2-3-3)=0$,
as expected. 

\paragraph{Remark.} Since three-dimensional Einstein gravity has no local degrees of freedom, it is an 
unrealistic model of the world (where gravitational waves do exist \cite{Abbott:2016blz}).\i{gravitational 
wave} This 
motivates the construction of alternative theories of three-dimensional gravity that do contain local 
degrees of freedom,\i{topologically massive gravity}\i{new massive gravity} such as topologically massive 
gravity \cite{Deser:1981wh} or new massive gravity 
\cite{Bergshoeff:2009hq}. In this thesis we shall be concerned only with 
Einstein gravity, although many of our considerations also apply to such modified theories.

\subsection{Boundary conditions and boundary terms}
\label{susebocote}

A field theory, as a Hamiltonian system, is defined by (i) its field content and Poisson brackets, and 
(ii) boundary conditions on fields and momenta.\i{fall-off conditions} The second 
point 
is crucial for gauge theories such as 
gravity. Here we explain certain generalities on boundary conditions, leaving specific definitions in 
three-dimensional gravity for later. In general terms, given a set of fields living on a manifold $\cM$, one 
chooses coordinates $(r,x)$ on $\cM$ 
and calls ``infinity''\i{infinity} the region where $r$ goes to infinity while all other coordinates are kept 
finite. One then specifies certain fall-off conditions for fields and their derivatives on that region, 
typically of the form
\be
\Phi(r,x)=\cO(r^{\#})\quad
\text{as }\;r\rightarrow+\infty
\nn
\ee
where $\Phi$ is some field and the coefficient $\#$ depends on the choice of fall-off conditions. In writing 
this it is understood that $\der_r\Phi$ is of order $\cO(r^{\#-1})$ at infinity.\footnote{This is not 
a trivial requirement; for instance the function $\sin(r^{42})/r$ is of order $\cO(1/r)$ as 
$r\rightarrow+\infty$ but its derivative is not of order $\cO(1/r^2)$.}\\

The influence of fall-offs is visible at the level of the action principle. Indeed, it is 
understood 
that the action of the theory should be plugged in an 
exponential $e^{iS}$, which is then to be integrated over field configurations in a path integral so as to 
produce quantum-mechanical transition amplitudes. In the classical limit, the leading contribution to the 
path integral should be due to on-shell field configurations;\i{differentiable action} but for this to be 
true the integrand must be 
differentiable, which is to say that the functional derivative $\delta S/\delta\Phi(x)$ is a local quantity. 
This, in turn, is only true provided the variation of the action contains no boundary terms. For instance, 
the 
variation of the Einstein-Hilbert 
action (\ref{s174})
is given by\i{Einstein-Hilbert action!variation}\i{action functional!differentiable}\i{variation of action 
functional}\i{boundary term}
\be
\begin{split}
\delta S_{\text{EH}}=
&
\;\frac{1}{16\pi G}
\int_{\cM}d^3x\sqrt{-g}
\Big(R_{\mu\nu}-\demi Rg_{\mu\nu}+\Lambda g_{\mu\nu}\Big)\delta g^{\mu\nu}\\
&
+
\frac{1}{16\pi G}
\int_{\cM}d^3x\,
\der_{\alpha}
\big(
\sqrt{-g}
g^{\mu\nu}\delta\Gamma^{\alpha}_{\mu\nu}
-
\sqrt{-g}
g^{\mu\alpha}\delta\Gamma^{\lambda}_{\lambda\mu}
\big)\,.
\end{split}
\label{s182}
\ee
The first term of this expression is the integral of the variation of the metric multiplying the vacuum 
Einstein equations, as expected. The second term is the integral of a total divergence and is therefore 
equal, by Stokes' theorem, to the flux of a vector field through the boundary $\der\cM$ of 
$\cM$.\i{boundary}\i{Stokes' theorem} Depending on one's choice of fall-off conditions for the metric, 
this boundary term may or may not vanish. If it does vanish, then the pure bulk action (\ref{s174}) can be 
legally plugged into a path integral. If it does not, then (\ref{s174}) is not differentiable 
and cannot be inserted as such in a path integral, which is to say that the semi-classical limit of a path 
integral 
involving only the action (\ref{s174}) is not given by on-shell field configurations. Accordingly, in 
order for the theory to have a well-defined semi-classical limit given by the equations of motion 
(\ref{s175}), one is generally forced to modify the pure bulk action (\ref{s174}) as
\be
S[g_{\mu\nu}]
=
S_{\text{EH}}[g_{\mu\nu}]
+
\int_{\der\cM}d^2x\,\cL(g_{\mu\nu},\der g_{\mu\nu},...)\,.
\label{s183}
\ee
Here $\cL$ is a certain Lagrangian density on the boundary of $\cM$, chosen so as to cancel the possibly 
non-vanishing boundary terms coming 
from the variation (\ref{s182}). Provided one can find a suitable $\cL$, the variation of the improved action 
(\ref{s183}) only involves the first term of (\ref{s182}) and the theory is classically consistent.\\

This explains, in terms of the action, how boundary conditions affect the definition 
of the theory. A few remarks are in order:
\begin{itemize}
\item We have been sloppy in our discussion of the notion of ``boundary''. Indeed we claimed that 
the fields of our theory live on a manifold $\cM$ and called $\der\cM$ its boundary, which we identified with 
the region $r\rightarrow+\infty$ in terms of some radial coordinate $r$. But typical space-time manifolds 
(such as $\RR^3$) actually have no boundary in the strict sense, so we should have been more 
precise:\i{conformal compactification} when we 
say that the region $r\rightarrow+\infty$ is the boundary $\der\cM$ of $\cM$, we really mean that we complete 
$\cM$ into a larger manifold, say $\overline{\cM}$, which now has a boundary, and in terms of the original 
coordinate $r$ that boundary is located at $r=+\infty$. This completed manifold $\overline{\cM}$ is 
known 
as a \it{conformal compactification} of $\cM$ \cite{Penrose1974}.
\item Aside from fall-off conditions, there is a second reason for adding boundary terms to the 
Einstein-Hilbert action. Namely, the Ricci scalar contains second-order derivatives of the metric, so in 
order to insert legally the gravity action in a path integral when the metric satisfies Dirichlet boundary 
conditions, boundary terms must be added to the 
Einstein-Hilbert action to cancel these second-order terms.\i{Gibbons-Hawking term} This is 
the 
origin of the Gibbons-Hawking-York boundary term \cite{York1972,Gibbons1977}.
\item The discussion of boundary terms clarifies in which sense a theory having no local degrees of freedom 
can still have non-trivial topological degrees of freedom: even though the bulk dynamics is 
trivial, that of the boundary is highly non-trivial!\i{boundary degree of freedom}\i{topological field 
theory!and holography}\i{holography!and topological field theories} In particular topological field 
theories, such as 
three-dimensional gravity, are the simplest examples of holographic 
systems since all their physical degrees of freedom live on the boundary of space-time. In higher space-time 
dimensions, the discussion of boundary 
terms remains the same but it is complicated by the presence of local, bulk degrees of freedom.
\item Three-dimensional Einstein gravity can be reformulated as a Chern\--Si\-mons theory whose 
gauge 
group is determined by the sign of the cosmological constant \cite{Achucarro:1987vz,Witten:1988hc} (see 
also \cite{blagojevic2001,Gomez:2013sfb,Donnay:2016iyk}).\i{Chern-Simons theory} This allows one to rewrite 
the Einstein-Hilbert 
action (plus boundary terms) as a purely two-dimensional action 
describing a field theory on the boundary of space-time, as follows from the relation 
between 
Chern-Simons theory, Wess-Zumino-Witten models and Liouville theory.\i{3D gravity!dimensional 
reduction}\i{dimensional reduction}\i{Liouville theory} It is often referred to as 
``dimensional reduction'', and was first worked out in \cite{Coussaert:1995zp} for Brown-Henneaux 
boundary 
conditions, while flat boundary conditions were studied in \cite{Barnich:2013yka}.
\end{itemize}

\subsection{Asymptotic symmetries}

Having justified the necessity of boundary terms for field theories, we now turn to gauge theories and 
explain qualitatively how one can find their global symmetries. These symmetries turn out to depend in a 
crucial way on the choice of fall-off conditions. We will first argue that the conserved charges associated 
with rigid global symmetries are strikingly different from those of gauge theories, then 
describe the 
ensuing notion of asymptotic symmetries. We conclude with the observation that the canonical generators of 
these symmetries generally satisfy a centrally extended Poisson algebra.

\subsubsection*{The problem of gauge symmetries}

Suppose we are given some gauge-invariant field theory living on a manifold $\cM$, with some bulk action 
$S[\Phi]$. The system has gauge redundancies, 
i.e.\ gauge symmetries, and one expects that there exist corresponding conserved quantities. 
The question is: how to build such conserved charges? To answer this we follow \cite{Barnich:2001jy}.\\

A naive guess is to simply apply the Noether procedure. For a field theory which is left 
invariant by certain symmetry transformations generated by some parameters $\epsilon^a$, $a=1,...,N$, with 
field and space-time transformations of the general form\i{Noether's theorem}
\be
x\mapsto x+\delta_{\epsilon}x,
\qquad
\Phi\mapsto\Phi+\delta_{\epsilon}\Phi\,,
\nn
\ee
the $N$ Noether 
currents $j_a^{\mu}$ can be obtained by ``gauging'' the symmetry, that is, replacing the rigid 
parameters $\epsilon^a$ by arbitrary functions $\epsilon^a(x)$ on space-time. The variation of the action 
then takes the form
\be
\delta S=-\int_{\cM}d^Dx\;j_a^{\mu}\,\der_{\mu}\epsilon^a
\label{obikatol}
\ee
from which one can read off the definition of the currents $j_a^{\mu}$. Their conservation follows from 
the fact that $\delta S\approx0$ on-shell, and the corresponding conserved Noether charges are the fluxes of 
these 
currents through a space-like slice $\Sigma$ of space-time:
\be
Q_a=\int_{\Sigma}(d^{D-1}x)_{\mu}j_a^{\mu}\,,
\label{togiadik}
\ee
where $(d^{D-1}x)_{\mu}$ is proportional to 
$\epsilon_{\mu\alpha_1...\alpha_{D-1}}dx^{\alpha_1}...dx^{\alpha_{D-1}}$. Equivalently, 
$(d^{D-1}x)_{\mu}\propto d^{D-1}x\cdot n_{\mu}$ where $n^{\mu}$ is the future-pointing time-like unit vector 
field orthogonal to $\Sigma$, and indices are moved thanks to the space-time metric.

\begin{figure}[H]
\centering
\includegraphics[width=0.40\textwidth]{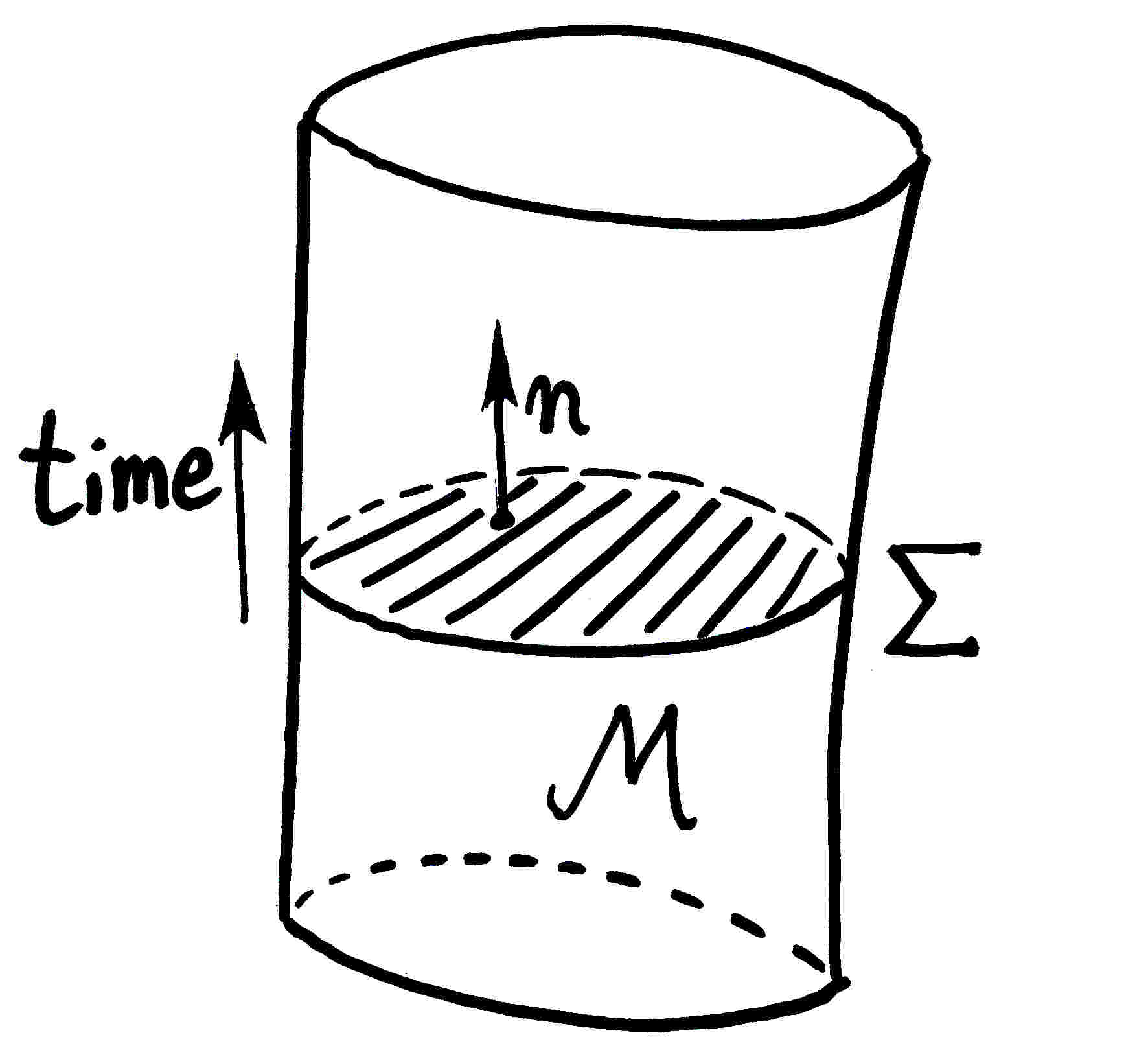}
\caption{A space-time manifold $\cM$ with an embedded space-like slice $\Sigma$ and future-pointing time-like 
normal vector $n$.}
\end{figure}

The problem with gauge symmetries now becomes apparent. Indeed, in that case the symmetry parameters 
$\epsilon^a$ are \it{already} gauged, which is to say that the right-hand side of (\ref{obikatol}) vanishes. 
This in turn implies that the Noether currents associated with gauge transformations all vanish! In 
particular, there seems to be no way of defining conserved charges of the form (\ref{togiadik}) for a gauge 
symmetry; this problem is the key difference between \it{gauge} symmetries and rigid symmetries.\\

The solution is provided by the 
following observation: the Noether current defined by (\ref{obikatol}) is not unique, as one can add 
to it the divergence of a two-form without affecting the left-hand side. In other words 
eq.\ (\ref{obikatol}) does not specify the Noether current $j_a^{\mu}$ uniquely, since the modified current
$\tilde j_a^{\mu}=j_a^{\mu}+\der_{\nu}k_a^{\mu\nu}$, where
$k_a^{\mu\nu}=-k_a^{\nu\mu}$,
satisfies the same property provided the antisymmetric tensor $k$ falls off fast enough at infinity. The 
corresponding Noether charge (\ref{togiadik}) is left unaffected by this modification provided the integral 
of 
$k$ on the boundary of $\Sigma$ vanishes; if that integral does \it{not} vanish, however, the charge receives 
an additional surface contribution of the form\i{surface charge}
\be
Q_{\text{surface}}
=
\int_{\der\Sigma}(d^{D-2}x)_{\mu\nu}\,k^{\mu\nu}
\label{qusurf}
\ee
where $(d^{D-2}x)_{\mu\nu}$ is 
proportional to $\epsilon_{\mu\nu\alpha_1...\alpha_{D-2}}dx^{\alpha_1}...dx^{\alpha_{D-2}}$. As we have just 
argued, the would-be Noether charges of a 
gauge theory can \it{only} receive surface contributions such as (\ref{qusurf}) since the corresponding 
Noether current vanishes up to the divergence of a two-form.\\

At first sight this means that the situation is even worse than expected, since the Noether 
charges of gauge theories are apparently ill-defined: there is no a priori way to associate a 
$k^{\mu\nu}$ with a given symmetry generator, so the surface integral (\ref{qusurf}) can
take any value. But in fact, this also suggests a solution to the problem: instead of trying to build a 
conserved current $j^{\mu}$, one can associate, with a gauge symmetry, a $(D-2)$-form $k^{\mu\nu}$ and define 
the corresponding charge by (\ref{qusurf}). If $k^{\mu\nu}$ is conserved 
on-shell in the sense that\i{superpotential}\i{conserved superpotential}\i{strength tensor}
$\nabla_{\mu}k^{\mu\nu}\approx 0$,
then the corresponding charge (\ref{qusurf}) is conserved by time evolution. In that context, the field 
$k^{\mu\nu}$ is called a \it{superpotential} and its integral (\ref{qusurf}) over the boundary of 
$\Sigma$ is 
known as the associated \it{surface charge}.\footnote{The term ``superpotential'' here has nothing 
to do 
with supersymmetry.} For example, in electrodynamics,\i{electric charge} the 
superpotential coincides with 
the strength tensor 
$F^{\mu\nu}$ and the corresponding surface charge is the flux of the electric field at infinity, that is, 
the total electric charge. Its conservation follows from the fact that $\der_{\mu}F^{\mu\nu}$ vanishes 
on-shell by virtue of Maxwell's equations.\\

Thus the computation of conserved 
charges for gauge symmetries boils down to the problem of associating a conserved superpotential with a given 
gauge transformation, and understanding to what extent that superpotential is unique.

\subsubsection*{Asymptotic symmetries}

While the definition (\ref{obikatol}) of the Noether current associated with a global symmetry transformation 
is straightforward, that of the superpotential associated with a gauge transformation is much 
more involved; see e.g.\ \cite{Barnich:2001jy,Barnich:2007bf,Compere:2007az}. Here we simply summarize the 
main ideas so as to apply 
them later to the specific case of three-dimensional gravity. The contruction consists of several steps:
\begin{enumerate}
\item Define the theory by choosing a bulk action, imposing certain fall-off conditions on the field 
content, and possibly adding a boundary term to the bulk action such that the full action is differentiable.
\item Find, among all possible gauge transformations, those that preserve the fall-off conditions. 
\i{allowed gauge transformation}\i{forbidden gauge transformation}\i{gauge transformation} Such gauge 
transformations are said to be \it{allowed}, as opposed to the gauge transformations that spoil the fall-off 
conditions and are therefore ``forbidden''. Allowed gauge transformations should 
then be thought of as the symmetries (global or gauge) of the theory.
\item Associate, with each allowed gauge transformation, a conserved superpotential $k^{\mu\nu}$; the latter 
depends linearly on the gauge parameters, while its dependence on the field content depends on 
the model under study. We will not write down that dependence explicitly here and refer to 
\cite{Barnich:2001jy,Compere:2007az} for details.
\item For each superpotential $k^{\mu\nu}$,\i{surface charge}\i{trivial gauge transformation} define a 
surface charge $Q$ by (\ref{qusurf}). 
If all surface 
charges associated with allowed gauge transformations are finite, then the boundary conditions 
are consistent. The allowed gauge transformations whose surface 
charges vanish are said to be \it{trivial},\i{trivial gauge transformation} while those whose surface 
charges do not vanish are 
\it{non-trivial}.
\end{enumerate}

\begin{figure}[h]
\centering
\includegraphics[width=0.60\textwidth]{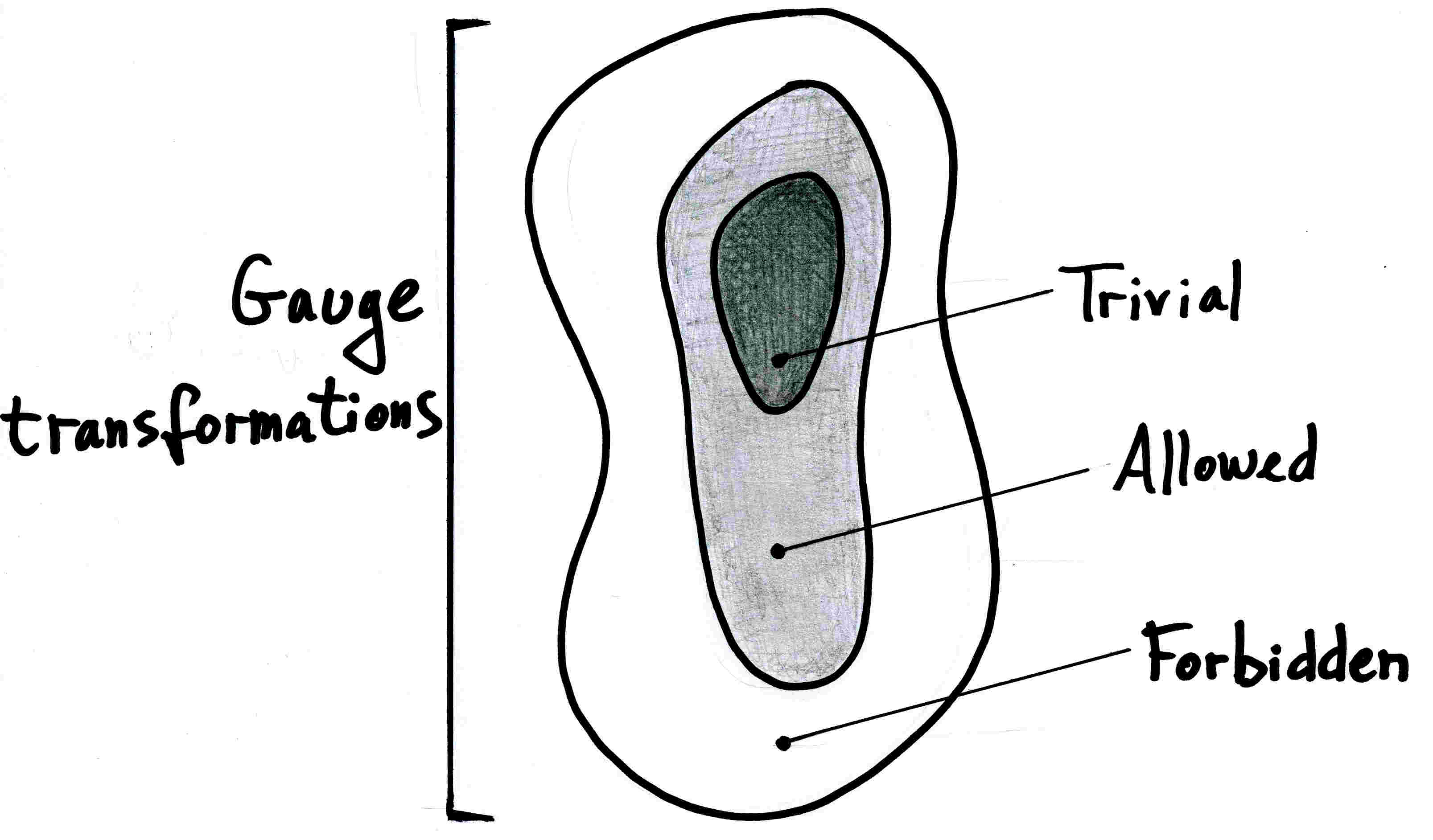}
\caption{Gauge transformations fall in three classes: forbidden transformations are those that do not 
preserve the fall-off conditions of the theory; allowed transformations are those that do, although they
generally change the state of the system; trivial transformations are those that preserve the fall-off 
conditions \it{and} leave the state of the system unchanged. In this sense trivial gauge transformations are 
actual gauge redundancies, and the global symmetry group of the system is the quotient of the 
group of allowed transformations by its subgroup of trivial gauge transformations.\label{figodisti}}
\end{figure}

This construction provides a distinction between three families of gauge transformations --- forbidden, 
allowed and trivial --- and is illustrated in fig.\ \ref{figodisti}. It is not just a matter of terminology; 
different classes of gauge transformations truly represent physically distinct notions of symmetries:
\begin{itemize}
\item Trivial gauge transformations are genuine (allowed) gauge transformations, that is, 
redundancies in the 
description of the theory.
\item Non-trivial gauge transformations are global symmetries that map a 
field configuration on a physically different one.\i{non-trivial gauge transformation} They fall off at 
infinity much slower than trivial gauge 
transformations and change the state of the system when acting on it. For example, in electrodynamics, 
non-trivial gauge transformations at spatial infinity take the form $\delta 
A_{\mu}(x)=\der_{\mu}\epsilon(x)$ with $\epsilon(x)=\text{const}$.\footnote{In practice, for constant 
$\epsilon$ this gives $\delta A_{\mu}=0$, but for fields with non-zero electric charge the transformation 
given by constant $\epsilon$'s is non-trivial.} This corresponds to a global $\un$ symmetry 
and the associated charge is the electric charge.
\item Forbidden gauge transformations are neither gauge transformations, nor even global symmetries: 
they are literally excluded from the theory since they do not leave its phase space invariant.
\end{itemize}

Note that infinitesimal gauge transformations are always endowed with a Lie bra\-cket. 
Accordingly, they span a Lie algebra. The notions introduced above then lead to the following terminology:

\paragraph{Definition.} The \it{asymptotic symmetry algebra}\i{asymptotic symmetry} of a theory 
is the quotient of the algebra 
of allowed gauge transformations by its ideal consisting of trivial transformations.\\

In the context of gravity, gauge transformations are diffeomorphisms of the space-time manifold, 
generated by certain vector fields. Allowed gauge transformations are generated by so-called 
\it{asymptotic Killing vector fields}.\i{asymptotic Killing vector} Their Lie bracket is the standard Lie 
bracket of vector fields and the 
asymptotic 
symmetry algebra coincides with the global symmetry algebra of the putative dual theory. In section 
\ref{sebohad} we will illustrate these notions in the case of AdS$_3$ gravity with 
Brown-Henneaux boundary conditions, while section \ref{sebmsboco} will be devoted to their asymptotically 
flat analogue.

\subsubsection*{Central extensions in the surface charge algebra}

The surface charges associated with asymptotic symmetries are designed in such a way that 
they implement asymptotic symmetry transformations on the fields of the theory. Explicitly, if we call $\xi$ 
some infinitesimal gauge parameter generating an allowed non-trivial gauge transformation and if we denote 
the associated surface charge by $Q[\xi]$, then the Poisson bracket of this charge with any field $\Phi$ 
takes the form\i{surface charge!as symmetry generator}\i{symmetry generator}
\be
\left\{
Q[\xi],\Phi
\right\}
=
-\delta_{\xi}\Phi
\label{dekuphi}
\ee
where the right-hand side is (minus) the variation of $\Phi$ under the transformation generated by $\xi$. 
This 
is a restatement of eq.\ (\ref{xixi}), where we noted 
that Poisson brackets with momentum maps generate symmetry transformations.\\

Since Poisson brackets satisfy the Jacobi identity, eq.\ 
(\ref{Karate}) still holds: for any two infinitesimal gauge transformations $\xi,\zeta$ and any field 
configuration $\Phi$, we have
\be
\big\{
\left\{Q[\xi],Q[\zeta]\right\}
,
\Phi
\big\}
=
\left\{
Q\big[[\xi,\zeta]\big],\Phi
\right\}\,.
\label{pojacobic}
\ee
It is tempting to remove 
the Poisson brackets from both sides of this equality and conclude that surface charges provide an 
exact representation of the asymptotic symmetry algebra. However, this 
naive removal would overlook the crucial point (\ref{jixiy})
that surface charges generally close according to a (classical) \it{central extension} of the algebra of 
asymptotic 
symmetry generators:\i{classical central extension}\i{central extension!classical}
\be
\left\{Q[\xi],Q[\zeta]\right\}
=
Q\big[[\xi,\zeta]\big]
+
\sfc(\xi,\zeta)\,.
\label{cezatro}
\ee
Here $\sfc(\xi,\zeta)$ is a real-valued two-cocycle that acts trivially on 
any field and is therefore invisible in eq.\ (\ref{pojacobic}). The point of the seminal paper 
\cite{Brown:1986nw} was to show that such non-trivial central extensions do arise in asymptotic symmetries of 
gravitational systems.

\section{Brown-Henneaux metrics in AdS$_3$}
\label{sebohad}

In this section we analyse Brown-Henneaux boundary conditions for Einstein gravity in AdS$_3$. After 
recalling some elementary geometric aspects of three-dimensional Anti-de Sitter space, we introduce 
Brown-Henneaux fall-offs and work out the corresponding asymptotic Killing vector fields. We also display the 
general solution of Einstein's equations satisfying these boundary conditions and use it to derive the 
algebra of surface charges associated with asymptotic symmetries, resulting in a direct sum of 
two Virasoro algebras with non-zero central charges. We end by describing an important family of 
Brown-Henneaux metrics that includes BTZ black holes.

\subsection{Geometry of AdS$_3$}

\subsubsection*{Anti-de Sitter space in three dimensions}

Consider the space $\RR^4=\RR^{2,2}$ endowed with coordinates $(x,y,u,v)$ and the metric
\be
ds^2=dx^2+dy^2-du^2-dv^2.
\label{s199}
\ee
Then three-dimensional \it{Anti-de Sitter} space (or simply AdS$_3$) is the submanifold 
of $\RR^{2,2}$ 
given by\i{AdS$_3$}
\be
\text{AdS}_3
\equiv
\big\{
(x,y,u,v)\in\RR^{2,2}
\big|
u^2+v^2=\ell^2+x^2+y^2
\big\}
\label{ss199}
\ee
for some parameter $\ell^2>0$, equipped with the induced metric of $\RR^{2,2}$. The parameter $\ell$ is 
called the \it{AdS radius}.\i{AdS radius} The manifold (\ref{ss199}) is diffeomorphic to a product 
$S^1\times\RR^2$ where the circle is time-like; in particular it contains closed time-like curves.\i{closed 
time-like curve} Its 
isometry group is $\text{O}(2,2)$\i{O22@$\text{O}(2,2)$} and acts transitively according to 
$x^{\mu}\mapsto\Lambda^{\mu}{}_{\nu}x^{\nu}$, where $x^{\mu}$ denotes the coordinates $(x,y,u,v)$ and 
$\Lambda$ is a $4\times 4$ matrix that preserves the ``Minkowski metric'' (\ref{s199}). 
The stabilizer for this action is isomorphic to $\text{O}(2,1)$, so there is a 
diffeomorphism\i{SO22@$\text{SO}(2,2)$}
\be
\text{AdS}_3
\cong
\text{O}(2,2)/\text{O}(2,1)
\cong
\text{SO}(2,2)/\text{SO}(2,1).
\nn
\ee

\begin{figure}[h]
\centering
\includegraphics[width=0.40\textwidth]{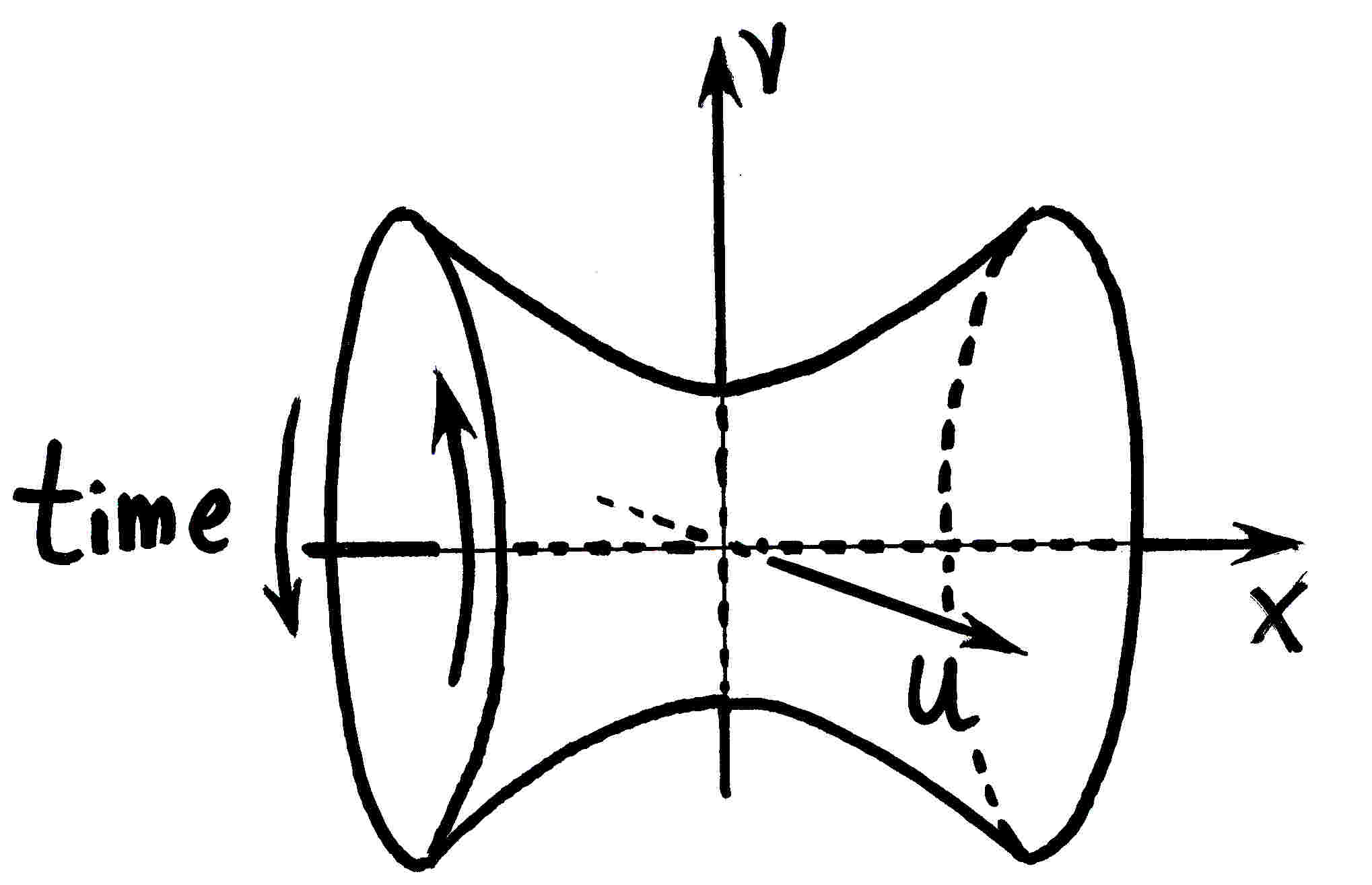}
\caption{Two-dimensional Anti-de Sitter space-time embedded in $\RR^3$ as the submanifold 
$u^2+v^2=x^2+\ell^2$ in terms of coordinates $u,v,x$ such that the mock-Minkowski metric of $\RR^3$ reads 
$-du^2-dv^2+dx^2$. Circles at constant $x$ are closed time-like curves in AdS$_2$. The 
spatial boundary of AdS$_2$ consists of two circles at $|x|\rightarrow+\infty$. For 
AdS$_3$, the boundary is a time-like torus $S^1\times S^1$.}
\end{figure}

\begin{figure}[t]
\centering
\includegraphics[width=0.20\textwidth]{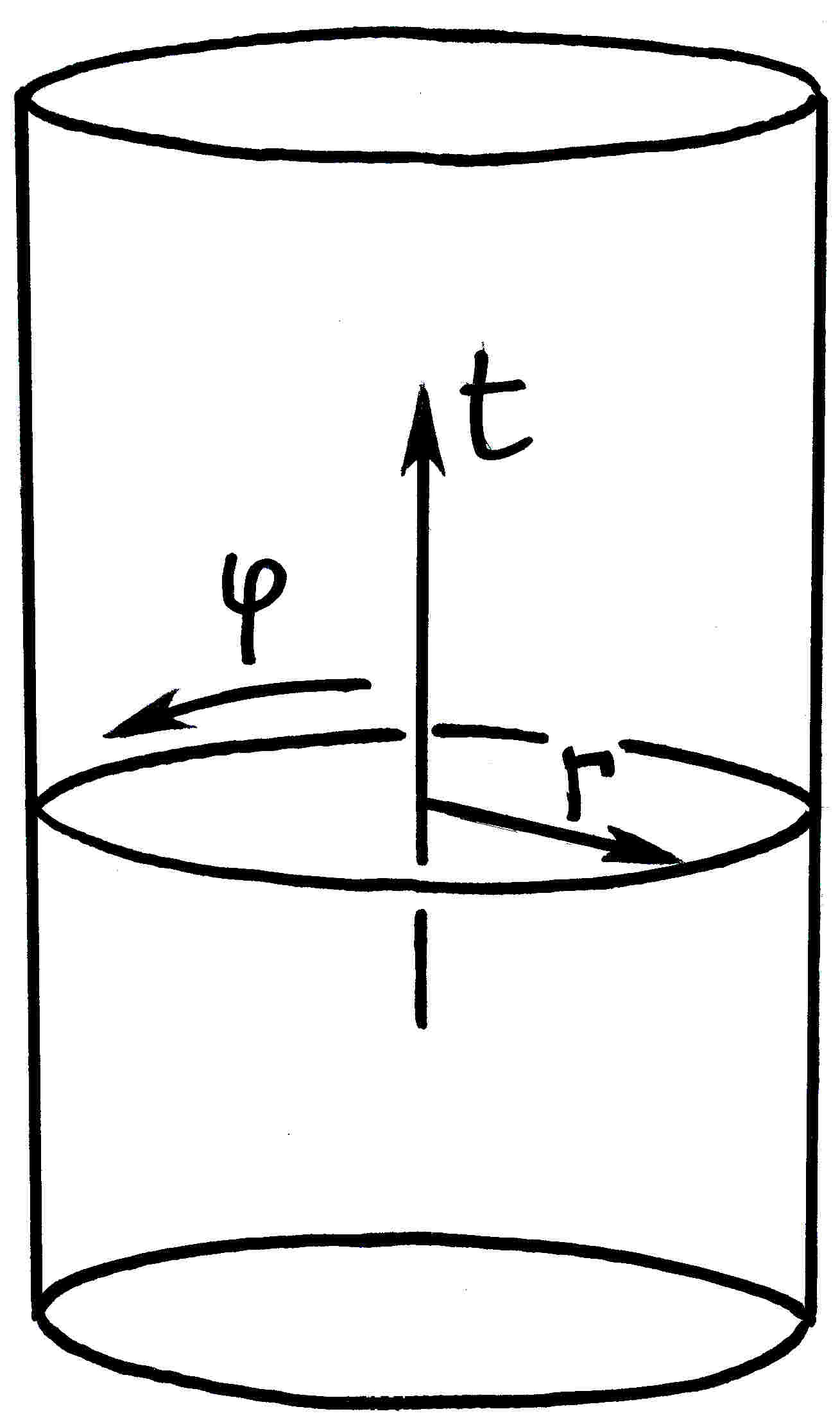}
\caption{The universal cover of three-dimensional Anti-de Sitter space-time, 
diffeomorphic to $\RR^3$. It is equivalent to the interior of a solid cylinder, which may be seen 
as the 
Penrose diagram of $\widetilde{\text{AdS}}{}_3$.\i{Penrose diagram} The 
time coordinate $t$ is directed along the axis of the 
cylinder while $r$ is 
a radial coordinate, and $\phii$ is a $2\pi$-periodic coordinate on the circle. The spatial boundary 
$r\rightarrow+\infty$ is a two-dimensional time-like cylinder spanned by the coordinates $(\phii,t)$, or 
equivalently by the light cone coordinates $x^{\pm}$.\label{FIGAX}}
\end{figure}

In practice, physical models of space-time are manifolds without closed time-like 
curves. It is therefore customary to unwind 
AdS$_3$ into its universal cover, 
$\widetilde{\text{AdS}}{}_3$, 
which is diffeomorphic to $\RR^3$ as a manifold. (Of course the metric on 
$\widetilde{\text{AdS}}_3\cong\RR^3$ is not flat!) To describe $\widetilde{\text{AdS}}_3$, we introduce new 
coordinates $(r,\phii,t)$ given on (\ref{ss199}) by\i{cylindrical coordinates}\i{coordinates!cylindrical}
\begin{align}
r & = \sqrt{x^2+y^2}=\sqrt{u^2+v^2-\ell^2},\nn\\
\label{ss200}
\phii & = \arctan(y/x),\\
t & = \ell\,\text{arctanh}(v/u).\nn
\end{align}
On AdS$_3$ the coordinate $t\in\RR$ is subject to the identification $t\sim t+2\pi\ell$, while on the 
universal cover 
$\widetilde{\text{AdS}}_3$ it takes all real values, without identification; see fig.\ \ref{FIGAX}. In terms 
of these coordinates the 
AdS$_3$ metric induced by (\ref{s199}) is
\be
ds^2
=
-(1+r^2/\ell^2)dt^2
+\frac{dr^2}{1+r^2/\ell^2}
+r^2d\phii^2\,.
\label{s200}
\ee
From now on we always refer to the universal cover $\RR^3$ of (\ref{ss199}) simply as AdS$_3$, without tilde. 
With this notation the coordinates $t\in\RR$, $r\in[0,+\infty[\,$, $\phii\in\RR$ with $\phii\sim\phii+2\pi$,
are global coordinates on AdS$_3$. In general-relativistic terms, $\text{AdS}_3$ is the (universal cover of 
the) maximally 
symmetric solution\i{maximally symmetric} of Einstein's vacuum equations in three dimensions with a negative 
cosmological constant\i{cosmological constant} $\Lambda=-1/\ell^2$.
Note at the outset that gravitation on an AdS$_3$ background is determined by two independent length scales 
$\ell$ and $G$. In particular the dimensionless coupling constant of the theory is $G/\ell$, so that the 
semi-classical regime corresponds to $\ell/G\rightarrow+\infty$.

\subsubsection*{Killing vectors}

The Killing vectors that generate isometries of AdS$_3$ can be found thanks to the embedding 
(\ref{ss199}), where ``Lorentz'' transformations are generated by the six independent vector fields 
\i{AdS$_3$!Killing vectors}\i{Killing vector}
\begin{align*}
\xi_1 & = u\der_v-v\der_u, & \xi_2 & = x\der_y-y\der_x, & \xi_3 & = u\der_y+y\der_u,\\
\xi_4 & = v\der_x+x\der_v, & \xi_5 & = u\der_x+x\der_u, & \xi_6 & = v\der_y+y\der_v.
\end{align*}
The combinations of signs appearing here are due to the metric $(+\,+\,-\,-)$ in (\ref{s199}). Upon defining
\begin{align}
\ell_0 & \equiv \tfrac{1}{2}(\xi_1+\xi_2),
& \bar\ell_0 & \equiv \tfrac{1}{2}(\xi_1-\xi_2),\nn\\[.2cm]
\label{s199b}
\ell_1 & \equiv \tfrac{1}{2}(\xi_3+\xi_4-i\xi_5+i\xi_6),
& \bar\ell_1 & \equiv \tfrac{1}{2}(-\xi_3+\xi_4-i\xi_5-i\xi_6),\\[.2cm]
\ell_{-1} & \equiv \tfrac{1}{2}(\xi_3+\xi_4+i\xi_5-i\xi_6),
& \bar\ell_{-1} & \equiv \tfrac{1}{2}(-\xi_3+\xi_4+i\xi_5+i\xi_6),\nn
\end{align}
one finds the following Lie brackets for $m,n=-1,0,1$:\i{isometry 
algebra}\i{so22@$\mathfrak{so}(2,2)$}\i{sl2R@$\sl$}
\be
i[\ell_m,\ell_n]=(m-n)\ell_{m+n}\,,
\qquad
i[\bar\ell_m,\bar\ell_n]=(m-n)\bar\ell_{m+n}\,,
\qquad
i[\ell_m,\bar\ell_n]=0\,.
\label{exibita}
\ee
This exhibits the isomorphism $\mathfrak{so}(2,2)\cong\sl\oplus\sl$,\footnote{Strictly 
speaking we have displayed this isomorphism here for the complexification of $\mathfrak{so}(2,2)$, but it 
also holds for real Lie algebras.} 
upon identifying the Lie brackets (\ref{LaLLA}).
Note that the generator of time translations is 
$\der_t=\frac{1}{\ell}(\ell_0+\bar\ell_0)$ while the generator of rotations is 
$\der_{\phii}=\ell_0-\bar\ell_0$.

\subsubsection*{Spatial infinity}

The region $r\rightarrow+\infty$ is a cylinder spanned by coordinates $(\phii,t)$ at 
space-like infinity. It is the spatial boundary $\der\cM$ of AdS$_3$.
In that region the metric (\ref{s200}) is\i{infinity}\i{spatial 
infinity}\i{AdS$_3$!boundary}
\be
ds^2\sim
\frac{\ell^2}{r^2}dr^2-r^2\left(
\frac{dt^2}{\ell^2}-d\phii^2
\right)
=
\frac{\ell^2}{r^2}dr^2
-r^2dx^+dx^-
\label{ss201b}
\ee
where we have introduced the light-cone coordinates\i{light-cone coordinates}\i{coordinates!light-cone}
\be
x^{\pm}\equiv
\frac{t}{\ell}\pm\phii\,.
\label{licordin}
\ee
For large $r$ the Killing vector fields (\ref{s199b}) are asymptotic to
\be
\ell_m\sim e^{imx^+}\der_+-\demi ime^{imx^+}r\der_r\,,
\qquad
\bar\ell_m\sim e^{imx^-}\der_--\demi ime^{imx^-}r\der_r
\nn
\ee
where $m=-1,0,1$. They generate global conformal transformations of the cylinder at infinity, including time 
translations $\ell_0+\bar\ell_0=\ell\der_t$ and rotations $\ell_0-\bar\ell_0=\der_{\phii}$. These 
expressions have the general form
\be
\xi\sim X(x^+)\der_+-\demi\der_+ X(x^+)r\der_r\,,
\qquad
\bar\xi\sim\bar X(x^-)\der_--\demi\der_-\bar X(x^-)r\der_r
\label{s201b}
\ee
where the functions $X$ and $\bar X$ are $2\pi$-periodic. Brown-Henneaux boundary conditions will be such 
that vector fields of the form (\ref{s201b}) are asymptotic symmetry 
generators for \it{arbitrary} functions $X,\bar X$.

\subsection{Brown-Henneaux boundary conditions}
\label{sebroh}

We now wish to define a family of metrics on $\RR^3$ that are ``asymptotically Anti-de Sitter'' in 
the sense that they take the form of a pure AdS$_3$ metric (\ref{s200}) at infinity. As a starting point we 
ask what is the minimum amount of metrics that we wish to include. A natural choice is to take pure AdS$_3$ 
together with conical deficits, which are obtained by cutting out a wedge out of the middle of AdS$_3$ and 
identifying its two sides. Concretely, consider the manifold 
described by coordinates $r\in[0,+\infty[\,$, $\phii\in\RR$, $t\in\RR$ subject to the 
identifications\i{AdS$_3$!conical deficit}\i{conical deficit}
\be
(r,\phii,t)
\sim
\big(r,\phii+4\pi\omega,t-2\pi A\big)
\label{s201}
\ee
for some $A\in\RR$ and $\omega>0$. (The normalization of $\omega$ is chosen for later convenience.) 
For $\omega=1/2$ and $A=0$ this reduces to the identifications that define pure AdS$_3$. For $0<\omega<1/2$ 
it is 
a 
conical deficit; for $\omega>1/2$ it is a conical excess. Since this is a global (topological) 
identification, 
the resulting pseu\-do\--Rie\-man\-nian manifold still solves Einstein's vacuum equations everywhere, except 
at the 
origin. In fact, the metric (\ref{s200}) with identifications (\ref{s201}) is the solution of Einstein's 
equations coupled to the stress tensor of a point mass at the origin.\i{point mass} Using the change of 
coordinates
\be
t'\equiv t+\frac{A}{2\omega}\phii\,,
\qquad
r'\equiv r\,,
\qquad
\phii'\equiv \frac{\phii}{2\omega}\,,
\label{cacott}
\ee
the space-time metric can be rewritten as
\be
ds^2
=
-\left(
1+\frac{r'^2}{\ell^2}
\right)(dt'-Ad\phii')^2
+
\frac{dr'^2}{1+r'^2/\ell^2}
+
4\omega^2r'^2d\phii'^2
\label{s202}
\ee
where now there are no identifications on $t'$, while $\phii'$ is $2\pi$-periodic. The term 
$A\,dt'd\phii'$ suggests that $A$ is proportional to angular momentum, as will indeed be the case below. Note 
that the integral curves of 
$\der_{\phii'}$ contain closed time-like curves unless\i{closed time-like curve}
\be
|A|\leq2\omega\ell
\qquad\text{and}\qquad
r'^2\geq\frac{A^2\ell^2}{4\omega^2\ell^2-A^2}\,.
\label{stringent}
\ee
Thus, the space-time manifold has no pathologies only in the region where $r'$ is large enough (and in 
particular in the asymptotic region $r'\rightarrow+\infty$), and provided the parameter $A$ is not too large 
compared to $\omega\ell$. Accordingly, from now on we refer to the solutions (\ref{s202}) with 
$0<\omega<1/2$ and $|A|=2\omega\ell$ as \it{extreme conical deficits}.\i{extreme conical deficit}\\

In order to find boundary conditions that genuinely describe AdS$_3$ space-times, one would like the 
asymptotic symmetry algebra to at least include $\mathfrak{so}(2,2)$. If in addition the phase space is to 
contain conical deficits (\ref{s202}), one is led to act with $\mathfrak{so}(2,2)$ 
transformations on such 
conical deficit metrics so that, if $\xi$ is an AdS$_3$ Killing vector and $g_{\mu\nu}$ is the metric of 
a conical deficit, the fall-off conditions are satisfied by the infinitesimally 
transformed metric\i{Lie derivative}
\be
g_{\mu\nu}+\cL_{\xi}g_{\mu\nu}\,.
\label{s205}
\ee
Here $\cL_{\xi}g_{\mu\nu}$ generally does \it{not} vanish because $\xi$ may not be a Killing vector for the 
conical deficit. In terms of cylindrical coordinates $(r,\phii,t)$, one thus obtains metrics that satisfy 
the fall-off conditions\i{Brown-Henneaux fall-offs} \cite{Brown:1986nw}
\be
(g_{\mu\nu})
=
\bmm
g_{rr} & g_{r\phii} & g_{rt}\\
g_{\phii r} & g_{\phii\phii} & g_{\phii t}\\
g_{tr} & g_{t\phii} & g_{tt}
\emm
\sim
\bmm
\frac{\ell^2}{r^2}+\cO(r^{-4}) & \cO(r^{-3}) & \cO(r^{-3})\\
\cO(r^{-3}) & r^2+\cO(1) & \cO(1)\\
\cO(r^{-3}) & \cO(1) & -\frac{r^2}{\ell^2}+\cO(1)
\emm.
\label{suppize}
\ee
In practice, we will impose an extra gauge-fixing condition that simplifies the computation of asymptotic 
symmetries. Namely, it turns out that the mixed components $g_{r\phii}$ and $g_{rt}$ can 
always be set to zero (identically) by applying a trivial differomorphism --- one whose 
surface charges all vanish. The subleading corrections to 
$g_{rr}=\frac{\ell^2}{r^2}+\cO(r^{-4})$ can similarly be set to zero. We refer to this gauge choice as 
the
\it{Fefferman-Graham gauge}. It leads to the following definition 
\cite{Barnich:2010eb}:

\paragraph{Definition.} Let $\cM$ be a three-dimensional manifold with a pseu\-do-\-Rie\-man\-nian metric 
$ds^2$. 
Suppose there exist local coordinates $(r,x^a)$ on $\cM$ (with $a=0,1$), defined for $r$ larger than some 
lower limit, such that the region $r\rightarrow+\infty$ is a time-like cylinder at spatial 
infinity where
the asymptotic behaviour of the metric is\i{fall-off conditions!Brown-Henneaux}
\be
ds^2
\stackrel{r\rightarrow+\infty}{\sim}
\frac{\ell^2}{r^2}dr^2
+
\big(r^2\eta_{ab}+\cO(1)\big)dx^adx^b
\label{s207}
\ee
with $\eta_{ab}dx^adx^b$ the two-dimensional Minkowski metric on the cylinder. Then we say that $(\cM,ds^2)$ 
is 
\it{asymptotically Anti-de Sitter} in the sense of Brown-Henneaux (in the Fefferman-Graham gauge), with a 
cosmological constant $\Lambda=-1/\ell^2$.\\

From now on, when dealing with AdS$_3$ gravity, we always restrict our attention to metrics satisfying 
the Brown-Henneaux boundary conditions (\ref{s207}). For practical purposes we will mostly describe the 
time-like cylinder in terms of light-cone coordinates $x^{\pm}$, in which case the label $a$ in (\ref{s207}) 
takes the values $\pm$ and the Minkowski metric on the cylinder is $\eta_{ab}dx^adx^b=-dx^+dx^-$. Note that 
asymptotically AdS$_3$ space-times need not be (and generally are not) globally diffeomorphic to AdS$_3$; in 
particular
there may be singularities in the bulk, as the definition (\ref{s207}) only requires $r$ to be larger than 
some lower limiting value. In the following pages we establish the main properties of this family of metrics, 
including their asymptotic symmetry algebra.

\paragraph{Remark.} The fact that one is allowed to choose the Fefferman-Graham gauge without losing any 
information is a general property of locally asymptotically Anti-de Sitter 
space-times \cite{Skenderis:2009nt}.\i{Fefferman-Graham expansion}\i{ambient construction}\i{conformal 
compactification} It is related to 
the Fefferman-Graham expansion of AdS metrics 
and the ambient construction of conformal structures \cite{Fefferman:2007rka}, where conformal manifolds are 
built as boundaries, or celestial spheres,\i{celestial sphere} of higher-dimensional bulk manifolds.

\subsection{Asymptotic Killing vector fields}

The asymptotic Killing vector fields associated with Brown-Henneaux boundary conditions are vector fields 
that generate diffeomorphisms which preserve the fall-offs (\ref{s207}). If $g_{\mu\nu}$ is a Brown-Henneaux 
metric and if $\xi$ is such a vector field, this is to say that
\be
\cL_{\xi}g_{rr}
=
\cL_{\xi}g_{r\pm}
=
0,
\qquad
\cL_{\xi}g_{ab}=\cO(1)
\quad(a,b=\pm)
\label{3.12}
\ee
in terms of light-cone coordinates (\ref{licordin}). Here 
the first condition follows from the fact that the components $g_{rr}=\ell^2/r^2$ and 
$g_{r\pm}=0$ are fixed, while the components $g_{ab}$ are allowed to fluctuate by terms of order 
$r^0$ at infinity.

\paragraph{Lemma.} Let $g_{\mu\nu}$ be a metric that is asymptotically AdS$_3$ in the sense (\ref{s207}) and 
let $\xi$ be a 
vector field that satisfies the properties (\ref{3.12}). Then\i{Brown-Henneaux fall-offs!asymptotic Killing 
vector}\i{asymptotic Killing vector!for AdS$_3$}
\be
\xi
=
X(x^+)\der_++\bar X(x^-)\der_--\demi\big(\der_+X(x^+)+\der_-\bar X(x^-)\big)r\der_r
+
\text{(subleading)}
\label{s208}
\ee
where $X(x^+)$ and $\bar X(x^-)$ are two arbitrary (smooth) $2\pi$-periodic functions while the subleading 
terms take the form
\be
\begin{split}
&  
-\frac{\ell^2}{2}
\der_a(\der_+X+\der_-\bar X)
\int_r^{+\infty}
\frac{dr'}{r'}g^{ab}(r',x^{\pm})
\der_b=\\
& =
\frac{\ell^2}{2r^2}
\big[
\der_-(\der_+X+\der_-\bar X)\der_+
+
\der_+(\der_+X+\der_-\bar X)\der_-
\big]
+
\cO(r^{-4})\,.
\end{split}
\label{s208sub}
\ee
These formulas associate an asymptotic Killing vector $\xi$ with an asymptotically AdS$_3$ metric 
$g_{\mu\nu}$ and a vector field $X(x^+)\der_++\bar X(x^-)\der_-$ on the cylinder; the 
dependence of $\xi$ on the latter is linear.

\begin{proof}
Let $g_{\mu\nu}$ be an asymptotically AdS$_3$ metric (\ref{s207}). We first note that the requirement 
$\cL_{\xi}g_{rr}=0$ imposes $\der_r\xi^r=\xi^r/r$, whose solution is
\be
\xi^r(r,x^{\pm})=r\cF(x^{\pm})
\label{3.15}
\ee
for some function $\cF$ on the cylinder. On the other hand the condition $\cL_{\xi}g_{r\pm}=0$ 
yields $\der_r\xi^c=-g^{ca}\frac{\ell^2}{r}\der_a\cF$,
which is solved by
\be
\xi^a
=
X^a(x^{\pm})
+
\ell^2\der_b\cF(x^{\pm})\int_r^{+\infty}\frac{dr'}{r'}g^{ab}(r',x^{\pm})
\label{3.16}
\ee
where $X^a\der_a$ is an arbitrary vector field on the cylinder. Note that the integral over $r'$ converges 
since $g_{ab}(r,x^{\pm})=r^2\eta_{ab}+\cO(1)$ by virtue of (\ref{s207}), so that the inverse is 
$g^{ab}=\frac{\eta^{ab}}{r^2}+\cO(r^{-4})$.
Plugging this in the integral of (\ref{3.16}) we find explicitly
\be
\xi^a
=
X^a(x^{\pm})
+
\frac{\ell^2}{2r^2}
\eta^{ab}\der_b\cF(x^{\pm})
+\cO(r^{-4})\,.
\label{omicron}
\ee
In light-cone coordinates (\ref{licordin}), the two-dimensional Minkowski metric reads\i{Minkowski 
space!light-cone coordinates}\i{light-cone coordinates!Minkoswki metric}
\be
(\eta_{ab})
=
\bmm
\eta_{++} & \eta_{+-}\\
\eta_{-+} & \eta_{--}
\emm
=
\bmm
0 & -1/2 \\ -1/2 & 0
\emm
\qquad\text{and}\qquad
(\eta^{ab})
=
\bmm
0 & -2 \\ -2 & 0
\emm
\nn
\ee
so that (\ref{omicron}) becomes
\be
\xi^{\pm}
=
X^{\pm}
-
\frac{\ell^2}{r^2}
\der_{\mp}\cF
+\cO(r^{-4})\,.
\label{qertf}
\ee
Finally one finds $\cL_{\xi}g_{ab}
=
r^2\left(2\cF\eta_{ab}+\cL_X\eta_{ab}\right)+\cO(1)$
where $\cL_X$ denotes the Lie derivative on the cylinder with respect to the vector field $X^a\der_a$. 
The requirement that this expression be of order one yields the conformal Killing equation for 
$X$, $\cL_X\eta_{ab}
=
-2\cF\eta_{ab}$.
Contracting this with $\eta^{ab}$ one finds $\cF=-\demi(\der_+X^++\der_-X^-)$ and the remaining constraints 
set $\der_-X^+=\der_+X^-$, which implies $X^+=X(x^+)$ and $X^-=\bar X(x^-)$.
Formula (\ref{s208}) follows, while the subleading terms (\ref{s208sub}) are produced by (\ref{3.16}).
\end{proof}

Note that the asymptotic Killing vector (\ref{s208}) takes the anticipated form (\ref{s201b}) and 
thus provides the generalization we were hoping to find. We will denote by $\xi_{(X,\bar X)}$ the asymptotic 
Killing vector field 
determined by the functions $X(x^+)$ and $\bar X(x^-)$. One can decompose these functions in 
Fourier modes and define the vector fields
\be
\ell_m
\equiv
\xi_{(e^{imx^+},0)}\,,
\qquad
\bar\ell_m
\equiv
\xi_{(0,e^{imx^-})}\,,
\label{ellemoha}
\ee
whose Lie brackets take the form (\ref{exibita}) up to subleading corrections, with indices 
$m,n$ ranging over all integer values. Thus, asymptotically, the finite-dimensional isometry algebra 
$\mathfrak{so}(2,2)$ of AdS$_3$ is enhanced to two commuting copies of the infinite-dimensional Witt algebra 
(\ref{witt}). In fact we can already anticipate the result:

\paragraph{Theorem.} The asymptotic symmetry group of AdS$_3$ gravity with Brown-Henneaux boundary conditions 
is a direct product $\Diffc\times\Diffc$ whose elements are diffeomorphisms\i{Brown-Henneaux 
fall-offs!asymptotic symmetry group}\i{asymptotic symmetry!for 
AdS$_3$}\i{conformal 
transformation}\i{Witt algebra}
\be
(x^+,x^-)\mapsto\big(f(x^+),\bar f(x^-)\big)
\label{xipiM}
\ee
acting as conformal transformations on the cylinder at spatial infinity.\\

At this stage, we have not yet proven this claim since we do not know whether all asymptotic 
Killing vector fields (\ref{s208}) have non-vanishing surface charges on the phase space; this will be done
in the following pages. Also note that we are being slightly sloppy in (\ref{xipiM}), since the 
diffeomorphisms generated by (\ref{s208}) affect the radial coordinate. Hence formula (\ref{xipiM}) 
only holds up to $1/r$ corrections; it is accompanied by transformations of the radial 
coordinate that we do not bother writing down, but that do preserve the limit $r\rightarrow+\infty$ in that 
they map $r$ on a positive $\cO(1)$ multiple of itself.

\paragraph{Remark.} In our description of asymptotic symmetries we mentioned that the algebra of 
vector 
fields (\ref{ellemoha}) is a direct sum of two Witt algebras \it{up to subleading corrections} which we did 
not take into account.\i{modified Lie bracket} This is because these corrections are unimportant: 
starting from the standard 
Lie bracket of vector fields, one can define a ``modified bracket'' that coincides with the standard one at 
infinity but ensures that the asymptotic symmetry algebra is satisfied everywhere in the bulk;\i{asymptotic 
symmetry!in the bulk} see e.g.\ \cite{Barnich:2001jy,Barnich:2010xq}.

\subsection{On-shell Brown-Henneaux metrics}
\label{sepaspav}

In order for the equations of motion to provide a true extremum of the action functional, the latter must be 
differentiable in the space of fields subject to the chosen boundary conditions. In the case of 
Brown-Henneaux 
fall-offs, one can show that the improved action\i{differentiable action!for AdS$_3$ gravity}\i{action 
functional!for gravity}\i{improved action}\i{boundary term}\i{extrinsic 
curvature}\i{Gibbons-Hawking term}
\be
S[g_{\mu\nu}]
\equiv
S_{\text{EH}}[g_{\mu\nu}]
-\frac{1}{8\pi G}\int_{\der\cM}d^2x\sqrt{-\text{det}(g_{ab})}
\Big(
K+\frac{1}{\ell}
\Big)
\nn
\ee
is differentiable in the space of metrics satisfying Brown-Henneaux boundary conditions, where 
$S_{\text{EH}}$ is the Einstein-Hilbert action (\ref{s174}) while $K$ is the 
trace of the extrinsic curvature at the boundary \cite{York,Gibbons1977} (see also 
\cite{Brown1993,Henningson:1998ey,Balasubramanian:1999re,deHaro:2000wj}).\\

With this improved action it makes sense to solve Einstein's equations in the space of metrics (\ref{s207}).
We will not review this computation here and refer to \cite{Banados:1998gg,Skenderis:1999nb,Barnich:2010eb} 
for 
details. The 
bottom line is that the general solution of the equations of motion with Brown-Henneaux boundary conditions 
in the Fefferman-Graham gauge reads\i{Brown-Henneaux metric}\i{Fefferman-Graham expansion}\i{Brown-Henneaux 
fall-offs!on-shell metric}\i{Banados metric@Ba\~{n}ados metric}\i{on-shell metric}
\be
ds^2
=
\frac{\ell^2}{r^2}dr^2
-
\Big(rdx^+-\frac{4G\ell}{r}\bar p(x^-)dx^-\Big)
\Big(rdx^--\frac{4G\ell}{r}p(x^+)dx^+\Big)
\label{s209}
\ee
where $p(x^+)$ and $\bar p(x^-)$ are arbitrary, $2\pi$-periodic functions of their arguments. The factors of 
$4G\ell$ are introduced for later convenience. We will study this space of solutions in greater detail below. 
For 
now, we only note that it is endowed with
a well-defined action of asymptotic symmetry transformations. Indeed,
we define the variation of $p$ and $\bar p$ under the action 
of an asymptotic Killing vector (\ref{s208}) by
\be
\cL_{\xi_{(X,\bar X)}}ds^2
\equiv
4G\ell\;\delta_Xp(x^+)\,(dx^+)^2
+4G\ell\;\delta_{\bar X}\bar p(x^-)\,(dx^-)^2
+\text{(subleading)},
\nn
\ee
and this variation preserves the structure of the solution (\ref{s209}).
In particular, observe that $\xi_{(X,\bar X)}$ is an exact Killing vector if the variations $\delta_Xp$ and 
$\delta_{\bar X}\bar p$ 
vanish.
Using (\ref{s208}) one finds
\be
\delta_Xp=X\der_+p+2p\der_+X-\frac{c}{12}\der_+^3X\,,
\qquad
\delta_{\bar X}\bar p=\bar X\der_-\bar p+2\bar p\der_-\bar X-\frac{\bar c}{12}\der_-^3\bar X
\label{s210}
\ee
where $c=\bar c$ is the \it{Brown-Henneaux central charge}\i{Brown-Henneaux central charge}\i{central 
charge!Brown-Henneaux}\i{classical central extension!in AdS$_3$}
\be
\boxed{
c=\bar c=\frac{3\ell}{2G}\,.
}
\label{ss210}
\ee
The transformations (\ref{s210}) are exactly those of the components of a CFT stress tensor under conformal 
transformations;\i{Brown-Henneaux metric!as CFT stress tensor} they coincide with the 
coadjoint representation (\ref{covinf}) of the 
Virasoro algebra when seeing $p(x^+)$ and $\bar p(x^-)$ as Virasoro coadjoint vectors. In that context the 
condition for $\xi_{(X,\bar X)}$ to be an exact Killing vector is equivalent to the statement that $(X,\bar 
X)$ belongs to the stabilizer of $(p,\bar p)$.
We refrain 
from interpreting these results any further for now; we will return to them in section \ref{sebitu}. Note 
that at this 
stage there is actually no reason to call 
(\ref{ss210}) a central charge: even though it does appear in (\ref{s210}) exactly as the inhomogeneous term 
of the coadjoint representation (\ref{covinf}), the specific value (\ref{ss210}) is irrelevant since 
changing the normalization of $p$ or $\bar p$ would change the value of $c$ and $\bar c$. The importance 
of the parameter (\ref{ss210}) will become apparent only from the algebra of surface charges.

\subsection{Surface charges and Virasoro algebra}
\label{sesuchal}

\subsubsection*{Surface charges}

Take an asymptotic Killing vector field (\ref{s208}) specified by the functions $X(x^+),\bar X(x^-)$, and 
choose an on-shell metric (\ref{s209}) specified by $p(x^+),\bar p(x^-)$. We wish to evaluate the 
surface charge associated with the symmetry transformation generated by $\xi_{(X,\bar X)}$ on the background 
specified by $p,\bar p$. As 
explained around eq.\ (\ref{qusurf}), this charge depends linearly on the components of $\xi_{(X,\bar 
X)}$. In addition we need to choose a ``background'' solution for which all surface charges 
vanish, which we take to be the degenerate conical deficit at $p=\bar p=0$,\i{degenerate conical 
deficit}
\be
\bar g
=
\frac{\ell^2}{r^2}dr^2-r^2dx^+dx^-.
\label{ss212}
\ee
With this normalization 
one can show that 
the conserved superpotentials corresponding to Brown-Henneaux asymptotic symmetries are such that
the surface charge (\ref{qusurf}) associated with the vector field 
$\xi_{(X,\bar X)}$ on the solution $(p,\bar p)$ is\i{surface 
charge!in AdS$_3$}\i{Brown-Henneaux fall-offs!surface charge}
\be
Q_{(X,\bar X)}[p,\bar p]
=
\frac{1}{2\pi}\int_0^{2\pi}d\phii\big[p(x^+)X(x^+)+\bar p(x^-)\bar X(x^-)\big]
\label{s212}
\ee
where $\phii=\demi(x^+-x^-)$. (See e.g.\ \cite{Barnich:2010eb} for an 
explicit computation.)\\

This charge can be interpreted in two ways: first,\i{surface charge!as dual 
Noether charge} as the Noether charge associated 
with a conformal transformation $(X,\bar X)$ in a two-dimensional CFT on the cylinder with stress tensor 
$(p,\bar p)$; second, as the pairing (\ref{virpar}) between the direct sum of two Virasoro algebras and its 
dual. This is consistent with the fact that the transformation law (\ref{s210}) coincides with the coadjoint 
representation of Virasoro. In particular, the charge associated with time translations corresponds to the 
asymptotic Killing vector $\der_t=(\der_++\der_-)/\ell$; it is the ADM mass\i{ADM mass} of the system, or 
equivalently the Hamiltonian\i{Brown-Henneaux fall-offs!Hamiltonian}\i{ADM mass}\i{Hamiltonian!in 
AdS$_3$}\i{mass!in AdS$_3$}
\be
M[p,\bar p]
=
\frac{1}{2\pi\ell}\int_0^{2\pi}d\phii\left[p(x^+)+\bar p(x^-)\right]
\label{tt212}
\ee
and it coincides (up to a factor $1/\ell$) with the sum of two Virasoro energy functionals (\ref{enep}). 
Similarly the charge 
associated with rotations, generated by the asymptotic Killing vector $\der_{\phii}=\der_+-\der_-$, is the 
angular momentum\i{angular momentum!in AdS$_3$}\i{angular 
momentum}\i{Brown-Henneaux fall-offs!angular momentum}
\be
J=\frac{1}{2\pi}\int_0^{2\pi}d\phii\left[p(x^+)-\bar p(x^-)\right]
\label{square212}
\ee
and coincides with the difference of two Virasoro energy functionals. With this normalization, pure AdS$_3$ 
(\ref{s209b}) has mass $M=-\frac{1}{8G}$; all its other surface charges vanish.

\subsubsection*{Surface charge algebra}

We now compute the 
algebra satisfied by the surface charges (\ref{s212}) under Poisson brackets. Recall that these brackets are 
such that they generate symmetry transformations according to (\ref{dekuphi}). We can apply this property 
here: if we let $(p,\bar p)$ be an on-shell metric 
(\ref{s209}), then the bracket of two charges $Q_{(X,0)}[p,\bar p]$ and $Q_{(Y,0)}[p,\bar p]$ is\i{surface 
charge!algebra}
\begin{eqnarray}
\left\{
Q_{(X,0)}[p,\bar p],Q_{(Y,0)}[p,\bar p]
\right\}
& \!\!\refeq{s210} &
\!\!-\frac{1}{2\pi}\int_0^{2\pi}d\phii\,
\left(
X\der_+p+2p\der_+X-\frac{c}{12}\der_+^3X
\right)Y(x^+)\nn\\
\label{ss213}
& \!\!= &
\!\!Q_{([X,Y],0)}[p,\bar p]\,+\,c\;\sfc(X,Y)\,.
\end{eqnarray}
In the last line we have introduced the bracket $[X,Y]$ defined as the usual Lie bracket of vector fields on 
the line, 
while $\sfc(X,Y)$ is the Gelfand-Fuks cocycle (\ref{gefuks}) expressed in the coordinate $x^+$. This is a 
Virasoro algebra (\ref{vibraphone}), with a classical central extension! The same computation would hold in 
the barred 
(antichiral) sector, while chiral and antichiral charges commute. Thus we conclude:

\paragraph{Theorem.} The algebra of surface charges associated with asymptotic symmetries of AdS$_3$ 
space-times in the sense of Brown-Henneaux is the direct sum of two Virasoro algebras with central charges 
(\ref{ss210}).\\

The Poisson bracket algebra (\ref{ss213}) can also be rewritten in terms of more conventional Virasoro 
generators. If we define the charges
\be
\cL_m\equiv Q_{(e^{imx^+},0)}[p,\bar p]\,,
\qquad
\bar\cL_m\equiv Q_{(0,e^{imx^-})}[p,\bar p]\,,
\label{ViCha}
\ee
their Poisson brackets close according to two copies of the Virasoro algebra (\ref{virapois}), up to the 
renaming $p\to\cL$. The central charges take the definite value $c=\bar c=3\ell/2G$. In particular, the 
normalization of the homogeneous term 
of the bracket fixes the normalization of the Brown-Henneaux central charge, confirming the fact that it is 
an unambiguous 
parameter specifying the phase space. In this language the mass (\ref{tt212}) and 
the angular momentum (\ref{square212}) are
\be
M=\frac{1}{\ell}(\cL_0+\bar\cL_0)\,,
\qquad
J=\cL_0-\bar\cL_0\,,
\nn
\ee
as in a two-dimensional conformal field theory.
In particular, pure AdS$_3$ has $\cL_0=\bar\cL_0=-c/24$, as does a CFT vacuum on the cylinder. Note that the 
Brown-Henneaux central charge is essentially the Planck mass measured in units of the inverse of the AdS$_3$ 
radius. Equivalently, it is the inverse of the coupling constant of the system, so the semi-classical limit 
corresponds to $c\rightarrow+\infty$.

\paragraph{Remark.} Brown-Henneaux boundary conditions are the ``standard'' boun\-da\-ry conditions for 
gravity 
on AdS$_3$ but other fall-off conditions exist as well, both in pure Einstein gravity and in modified 
theories of gravity. For instance, in the Einstein case,\i{free boundary conditions} free boundary 
conditions \cite{Troessaert:2013fma} 
extend those of Brown-Henneaux by allowing the conformal factor of the metric on the boundary to fluctuate, 
resulting in an even larger asymptotic symmetry algebra. Many other families of boundary conditions exist, 
such as the chiral 
boundary conditions of \cite{Compere:2013bya} or the AdS$_3$ boundary conditions\i{topologically massive 
gravity}\i{new massive gravity} of topologically massive 
gravity \cite{Henneaux:2009pw,Henneaux:2010fy,Skenderis:2009nt} and new massive gravity 
\cite{Liu:2009kc,Cunliff:2013en}, but we will have very little to say about these 
alternative possibilities.

\subsection{Zero-mode solutions}
\label{susezeromodes}

In order to interpret the metrics (\ref{s209}), let us study zero-mode solutions, where $p(x^+)=p_0$ and 
$\bar p(x^-)=\bar p_0$ are constants. In that case the only non-zero surface charges (\ref{s212}) are the 
Virasoro zero-modes $\cL_0=p_0$ 
and $\bar\cL_0=\bar p_0$.\i{Brown-Henneaux 
metric!zero-mode}\i{zero-mode metric}\\

At $p_0=\bar p_0=-c/24\refeq{ss210}-\ell/16G$ the space-time metric is that of pure AdS$_3$,
\be
ds^2_{\text{AdS}}
=
\frac{\ell^2}{r^2}dr^2-
\Big(rdx^++\frac{\ell^2}{4r}dx^-\Big)
\Big(rdx^-+\frac{\ell^2}{4r}dx^+\Big).
\label{s209b}
\ee
To verify that this is indeed pure AdS$_3$, note that the change of coordinates
\be
r=\frac{\ell}{2}e^{\text{arcsinh}(\bar r/\ell)}
\label{s210b}
\ee
brings this metric into the manifest AdS$_3$ form (\ref{s200}) (up to the bar on the coordinate $\bar r$) by 
virtue of the identity
\be
\frac{dr}{r}
=
\frac{d\bar r}{\sqrt{\ell^2+\bar r^2}}\,.
\nn
\ee
The 
angular momentum vanishes while the ADM mass of the solution is
\be
M_{\text{vac}}
=
\frac{1}{\ell}(\cL_0+\bar\cL_0)
=
-\frac{c}{12\ell}
=
-\frac{1}{8G}\,.
\label{s227}
\ee
This is the energy of the vacuum state of a two-dimensional CFT on the cylinder.\\

Recall that Brown-Henneaux boundary conditions were designed so as to include conical deficits. One can show 
that the zero-mode solution specified by\i{Brown-Henneaux metric!for conical deficit}\i{conical deficit}
\be
\cL_0
=
p_0
=
-\frac{\ell}{16G}\left(2\omega-\frac{A}{\ell}\right)^2,
\qquad
\bar\cL_0
=
\bar p_0
=
-\frac{\ell}{16G}\left(2\omega+\frac{A}{\ell}\right)^2
\label{ss209q}
\ee
is precisely a conical deficit (\ref{s202}) written in Fefferman-Graham coordinates provided 
$|A|/\ell<2\omega<1$. In terms of Virasoro charges (\ref{ViCha}), conical deficits have 
$-\frac{c}{24}<\cL_0,\bar\cL_0\leq 0$. The angular momentum is $J=p_0-\bar p_0=\omega A/2G$,
while the ADM mass is
\be
M=\frac{p_0+\bar p_0}{\ell}
=
-\frac{1}{8G}\left(4\omega^2+\frac{A^2}{\ell^2}\right).
\nn
\ee
Extreme conical deficits are solutions of this type for which either $p_0$ or $\bar p_0$ vanishes, or 
equivalently for which $|A|=2\omega\ell$. Conical excesses are solutions for which 
$p_0,\bar 
p_0$ are of the form (\ref{ss209q}) with $|A|\leq2\omega$ but $\omega>1/2$, and the line separating deficits 
from excesses is a section of parabola
\be
\ell M=-\frac{\ell}{8G}-\frac{2G}{\ell}J^2,
\qquad
|J|\leq\ell/4G
\label{s228}
\ee
whose endpoints are tangent to the lines $\ell M=|J|$. The solution at $p_0=\bar 
p_0=0$ is the degenerate conical deficit (\ref{ss212}) that we used to normalize charges. Note that conical 
excesses with an angle of $2\pi n$ around the origin correspond to $\omega=n/2$; for fixed 
$n$, the set of such excesses is again a section of parabola in the $(J,\ell M)$ plane specified by
\be
\ell M=-\frac{\ell}{8G}n^2-\frac{2G}{\ell}\frac{J^2}{n^2}\,,
\nn
\ee
which generalizes (\ref{s228}). Note that, for vanishing angular momentum ($A=0$), eq.\ (\ref{ss209q}) 
yields $p_0=\bar p_0=-c\,\omega^2/6$ in terms of the Brown-Henneaux central charge. This is precisely the 
relation (\ref{hofrik}) between constant elliptic Virasoro coadjoint vectors and their monodromy matrix.\\

When $p_0$ and $\bar p_0$ are positive constants, the metric (\ref{s209}) turns out to be that of a \it{BTZ 
black hole}\i{BTZ black hole} with mass $M=(p_0+\bar p_0)/\ell$ and angular momentum 
$J=p_0-\bar p_0$ 
written in Fefferman-Graham coordinates \cite{Banados:1998gg}:\i{Brown-Henneaux metric!for BTZ black 
hole}\i{black hole}
\be
ds^2_{\text{BTZ}}
=
\frac{\ell^2}{r^2}dr^2
-
\Big(
rdx^+-\frac{2G\ell}{r}(\ell M-J)dx^-
\Big)
\Big(
rdx^--\frac{2G\ell}{r}(\ell M+J)dx^+
\Big).
\label{s215}
\ee
In that context the requirement $p_0,\bar p_0\geq0$ is interpreted as a cosmic censorship condition 
$|J|\leq\ell M$, which is saturated by extremal black holes.\i{cosmic censorship} Beyond that barrier, all 
zero-mode metrics for 
which $|J|>\ell |M|$ contain closed time-like curves\i{closed time-like curve} at arbitrarily large 
$r$.\\

\begin{figure}[t]
\centering
\includegraphics[width=0.60\textwidth]{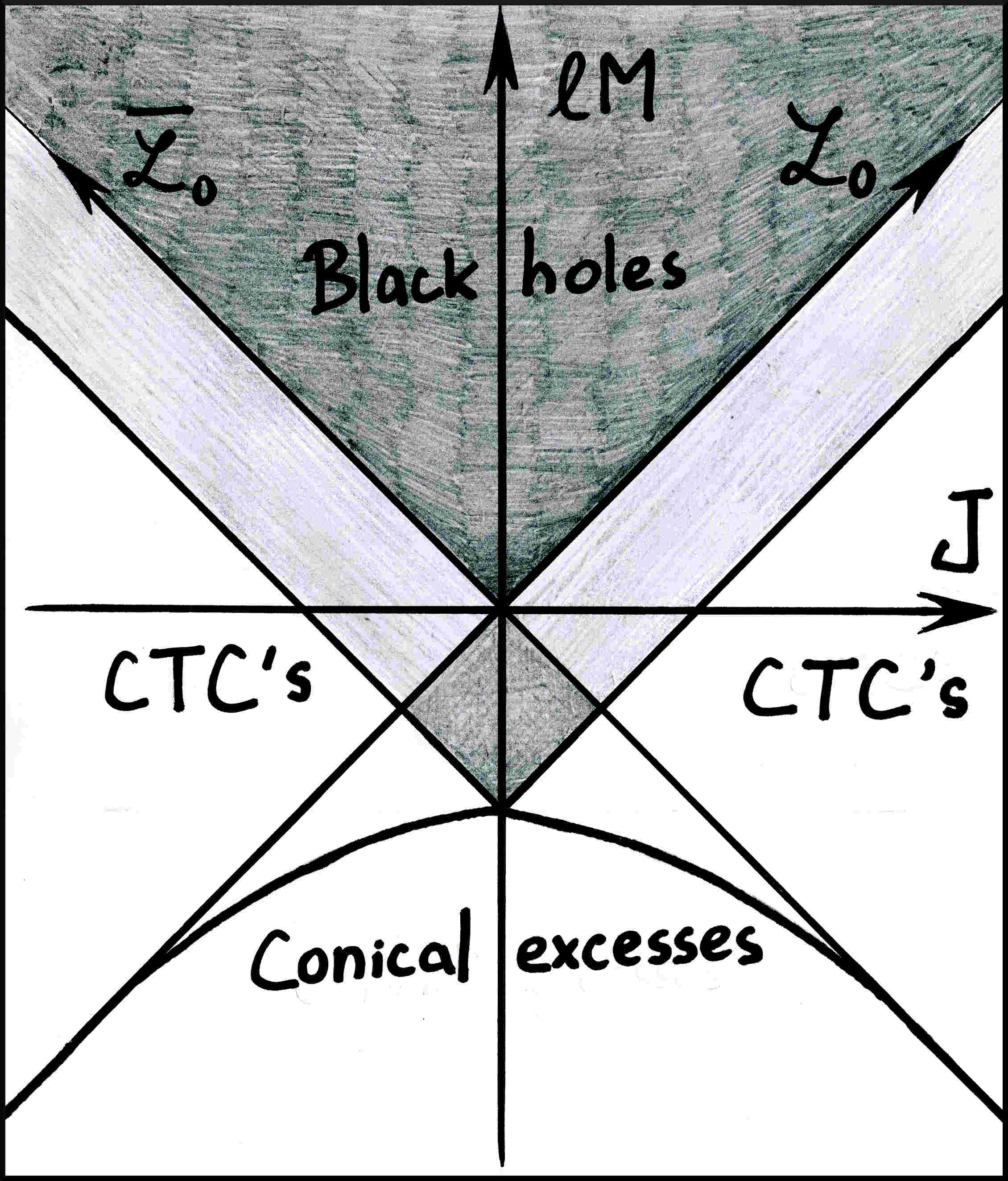}
\caption{The zero-mode solutions of AdS$_3$ gravity with Brown-Henneaux boundary conditions. The origin of 
the coordinate system $(J,\ell M)$ is the degenerate conical deficit (\ref{ss212}); the AdS$_3$ metric is 
located below, on the $\ell M$ axis, at the lower tip of the shaded square. BTZ black holes are located 
in the wedge $|J|\leq\ell M$. Conical deficits and excesses are located in the lower wedge $|J|\leq-\ell M$, 
respectively above and below the 
parabola (\ref{s228}). All metrics such that 
$|J|>\ell|M|$ contain closed time-like curves at arbitrarily large radius. Anticipating the results of 
section \ref{susePosAd}, we have shaded the 
solutions whose orbit has energy bounded from below under Brown-Henneaux transformations; those are all BTZ 
black holes, the AdS$_3$ metric, and all conical deficits such that $p_0,\bar p_0\geq-c/24$. Certain 
solutions with energy bounded from below are pathological in that they contain closed time-like 
curves at infinity --- those are 
the two 
diagonal strips surrounding the BTZ wedge.\label{figs229}}
\end{figure}

The lightest 
BTZ black hole at $M=J=0$ is the degenerate conical deficit (\ref{ss212}). This is strikingly different from 
four-dimensional black holes: in the latter case, the lightest black hole is typically empty space, whereas 
in three dimensions the lightest black hole is separated from AdS$_3$ by a classical mass gap.\i{mass gap} 
The metrics 
that fill this gap are conical deficits, i.e.\ metrics of point particles, so one can loosely say that a 
particle\i{point particle} turns into a black hole when its mass is higher than the threshold $c/24\ell$, 
which is essentially 
the Planck mass. We will encounter a similar phenomenon in flat space, though in that case black holes will 
be replaced by cosmological space-times.

\paragraph{Remark.} Since three-dimensional gravity has no local degrees of freedom, all solutions of 
Einstein's equations in three dimensions are locally isometric to AdS$_3$ and can therefore be realized as 
quotients of AdS$_3$.\i{quotient of AdS$_3$}\i{quotient space!of AdS$_3$} In 
particular, the BTZ metric (\ref{s215}) has no curvature singularity at $r=0$, 
where, as everywhere else, it is locally isometric to AdS$_3$. So how can it be a black hole?\i{BTZ black 
hole!as quotient of AdS$_3$} The answer to 
this question was clarified in \cite{Banados:1992gq}, where it was noted that the point $r=0$ is a 
singularity 
in the causal sense even though it is a regular point in the metric sense. Note that black holes obtained as 
regular identifications of AdS also exist in higher dimensions \cite{Aminneborg:1996iz,Banados:1998dc}.

\section{The phase space of AdS$_3$ gravity}
\label{sebitu}

From a Hamiltonian perspective, a phase space is a manifold consisting of ``positions and momenta'' 
endowed with a Poisson structure, and time evolution is generated by Poisson brackets with a Hamiltonian 
function $\cH$. This time evolution is in fact the one-parameter group of diffeomorphisms generated by a 
Hamiltonian vector field (\ref{hamive}); it follows that phase space 
trajectories corresponding to different initial conditions never cross, so one is free to think of phase 
space as the set of possible initial conditions of the equations of motion. In other words, one can identify 
the phase 
space of a system with the space of solutions of its equations of motion 
\cite{Crnkovic:1986ex}.\i{covariant 
phase space} This 
reformulation is at the core of the covariant approach to Hamiltonian mechanics, which is sometimes stressed 
by referring to the phase space as being \it{covariant}.\i{covariant phase space}\i{phase space!covariant} 
(In contrast to the standard Hamiltonian approach, 
the covariant 
one treats space and time coordinates on 
an equal footing.)\\

According to this viewpoint, the space of solutions (\ref{s209}) is really the phase space of AdS$_3$ gravity 
with Brown-Henneaux boundary conditions.\i{Brown-Henneaux fall-offs!phase space} The 
purpose of this section is to analyse some of its properties and 
to relate them with holography. Thus, we interpret points of phase space as 
CFT stress tensors, describe and interpret their transformation law under Brown-Henneaux 
transformations, and 
derive a positive energy theorem for AdS$_3$ gravity. Quantization is relegated to section 
\ref{sevirep}.

\subsection{AdS$_3$ metrics as CFT$_2$ stress tensors}
\label{susedaciff}

According to (\ref{s210}), the functions $p,\bar p$ that specify an on-shell metric transform under 
asymptotic symmetry 
transformations as the components of the stress tensor of a two-dimensional CFT with central charges 
(\ref{ss210}). The corresponding surface charges (\ref{s212}) generate two Virasoro algebras (\ref{ss213}). 
Accordingly, from now on we interpret Brown-Henneaux asymptotic symmetries as the global conformal 
symmetries of a two-dimensional CFT ``dual'' to AdS$_3$ gravity.\i{holography!in AdS$_3$} One can think of 
that theory as living on 
the cylindrical boundary of AdS$_3$. Its central 
charges are (\ref{ss210}) and the components of its stress tensor should be operators whose 
one-point functions are the functions $p(x^+)$ and $\bar p(x^-)$ appearing in the metric (\ref{s209}), which
is in fact a general feature of the AdS/CFT\i{AdS/CFT} 
correspondence \cite{Maldacena:1997re,Witten:1998qj,Balasubramanian:1999re}. 
Thus\i{Brown-Henneaux metric!as CFT stress tensor} the covariant phase space 
of AdS$_3$ 
gravity coincides with the space of CFT stress tensors on the cylinder at fixed central charges. The finite 
transformation laws of 
these stress tensors under conformal transformations are given by the coadjoint representation of the 
Virasoro group, eq.\ (\ref{covi}).\\

The Poisson structure on the phase space of AdS$_3$ gravity is determined by the requirement (\ref{dekuphi}) 
ensuring that surface charges generate the correct transformation laws when acting on the fields of the 
theory. For Brown-Henneaux boundary conditions this leads to the Poisson brackets of charges (\ref{ss213}), 
which coincides with the 
Kirillov-Kostant Poisson bracket (\ref{virapois}). Hence we conclude:\i{Kirillov-Kostant bracket!applied to 
AdS$_3$}

\paragraph{Theorem.} The covariant phase space of AdS$_3$ gravity with Brown\--Hen\-neaux boundary conditions 
is 
a hyperplane at fixed central charges (\ref{ss210}) embedded in the space of the coadjoint representation of 
the direct product of two Virasoro groups,\i{coadjoint representation!in gravity}\i{AdS/CFT!and coadjoint 
representation}\i{covariant phase space!as coadjoint representation}\i{phase space!as coadjoint 
representation} and endowed with the corresponding Kirillov-Kostant Poisson 
structure.\\

A loose way to interpret this theorem is to say that AdS$_3$ gravity \it{is} group theory: the whole phase 
space of the system is determined by the structure of the Virasoro group, save for the fact that the 
value of the central charge is fixed by the coupling constant. We will encounter a similar 
phenomenon in the next chapter when dealing with asymptotically flat gravity. This being said, the occurrence 
of a Virasoro coadjoint representation should not come as too big a surprise. Indeed, it is always true that 
the charges associated with certain symmetries transform under the coadjoint representation of the symmetry 
group (since these charges are nothing but momentum maps). Accordingly, the surface charges of AdS$_3$ 
gravity were bound to involve the coadjoint representation of the Virasoro group. The only surprise is that 
Virasoro coadjoint vectors exactly coincide with the functions specifying the metric, instead of being some 
complicated non-linear combinations of the entries of the metric and their derivatives. In particular, note 
that the set of on-shell Brown-Henneaux metrics (\ref{s209}) is a vector space.

\paragraph{Remark.} It is not strictly true that the whole phase space of AdS$_3$ gravity coincides 
with the dual of two Virasoro algebras. Indeed, this conclusion is entirely based on the 
asymptotic solutions (\ref{s209}), but completely overlooks the fact that some of these solutions cannot be 
extended arbitrarily far into the bulk.
This subtlety leads to (finitely many) additional directions in the complete phase space of the theory, as 
discussed in \cite{Kim:2015qoa}. We will ignore this detail since it plays a 
minor role for our purposes.

\subsection{Boundary gravitons and Virasoro orbits}
\label{susebovio}

The covariant phase space (\ref{s209}) is spanned by pairs of functions $\big(p(x^+),\bar p(x^-)\big)$. The 
zero-mode 
solutions were described in section \ref{susezeromodes}, but a generic par $(p,\bar p)$ is definitely 
\it{not} a zero-mode since $p$ and $\bar p$ may have some non-trivial profile on the circle.
If we pick one such solution at random, we can generate infinitely 
many other ones by acting on it with asymptotic symmetry transformations. The resulting manifold is 
the product of two Virasoro coadjoint orbits at central charges 
(\ref{ss210}),\i{Brown-Henneaux metric!orbit under Virasoro}
\be
\cW_{(p,c)}\times\cW_{(\bar p,\bar c)}\,.
\label{s230}
\ee
If we think of the asymptotic symmetry group as a generalization of the space-time isometry group 
$\text{O}(2,2)$, and of $(p,\bar p)$ as an infinite-dimensional generalization of 
space-time momentum, then the orbit (\ref{s230}) is an infinite-dimensional generalization of the 
standard orbits of momenta under, say, Lorentz transformations. In particular the metrics spanning the orbit 
(\ref{s230}) should be seen as boosts of the metric $(p,\bar p)$.\\

This is a good point to introduce a terminology which has come to be more or less standard 
\cite{Garbarz:2014kaa,Troessaert:2015gra,Compere:2015knw}: a metric $(p,\bar p)$ obtained by 
acting on pure AdS$_3$ with a certain asymptotic symmetry transformation\i{boundary graviton} is known
as a (classical) \it{boundary graviton}. This nomenclature is then extended to any 
metric obtained 
from a zero-mode solution by an asymptotic symmetry transformation. The name is justified by the 
fact that three-dimensional gravity has no local (bulk) degrees of freedom, but does have non-trivial 
topological (boundary) degrees of freedom visible in the arbitrariness of the pair $(p,\bar p)$ that 
specifies a solution of the equations of motion.\\

If our goal is to classify all solutions (\ref{s209}), then orbits provides a natural 
organizing criterion: rather than classifying the solutions, we can classify 
their orbits under asymptotic symmetries. Since we know the classification of Virasoro coadjoint orbits, we 
may claim to control the full covariant phase space of AdS$_3$ gravity. In particular, the 
classification 
of zero-mode solutions in fig.\ \ref{figs229} is a first step towards the full classification: 
each point in the plane $(J,\ell M)$ defines the orbit of the corresponding zero-mode solution,
and different points define distinct orbits. However, as we 
have seen in section \ref{sevirorepss}, not all orbits have constant representatives: there exist infinitely 
many conformally inequivalent on-shell 
metrics that cannot be brought into zero-mode solutions by asymptotic symmetry 
transformations. Thus the complete classification of AdS$_3$ metrics is essentially a product 
of two copies of fig.\ \ref{vifig}, where zero-mode solutions are those where both $p$ and $\bar p$ belong to 
the vertical axis of the figure.\i{phase space!foliation}\i{symplectic 
leaf!in AdS$_3$}\i{Brown-Henneaux 
fall-offs!symplectic leaves} This classification foliates the covariant phase 
space of AdS$_3$ gravity into disjoint orbits of the asymptotic symmetry group.

\begin{figure}[H]
\centering
\includegraphics[width=0.30\textwidth]{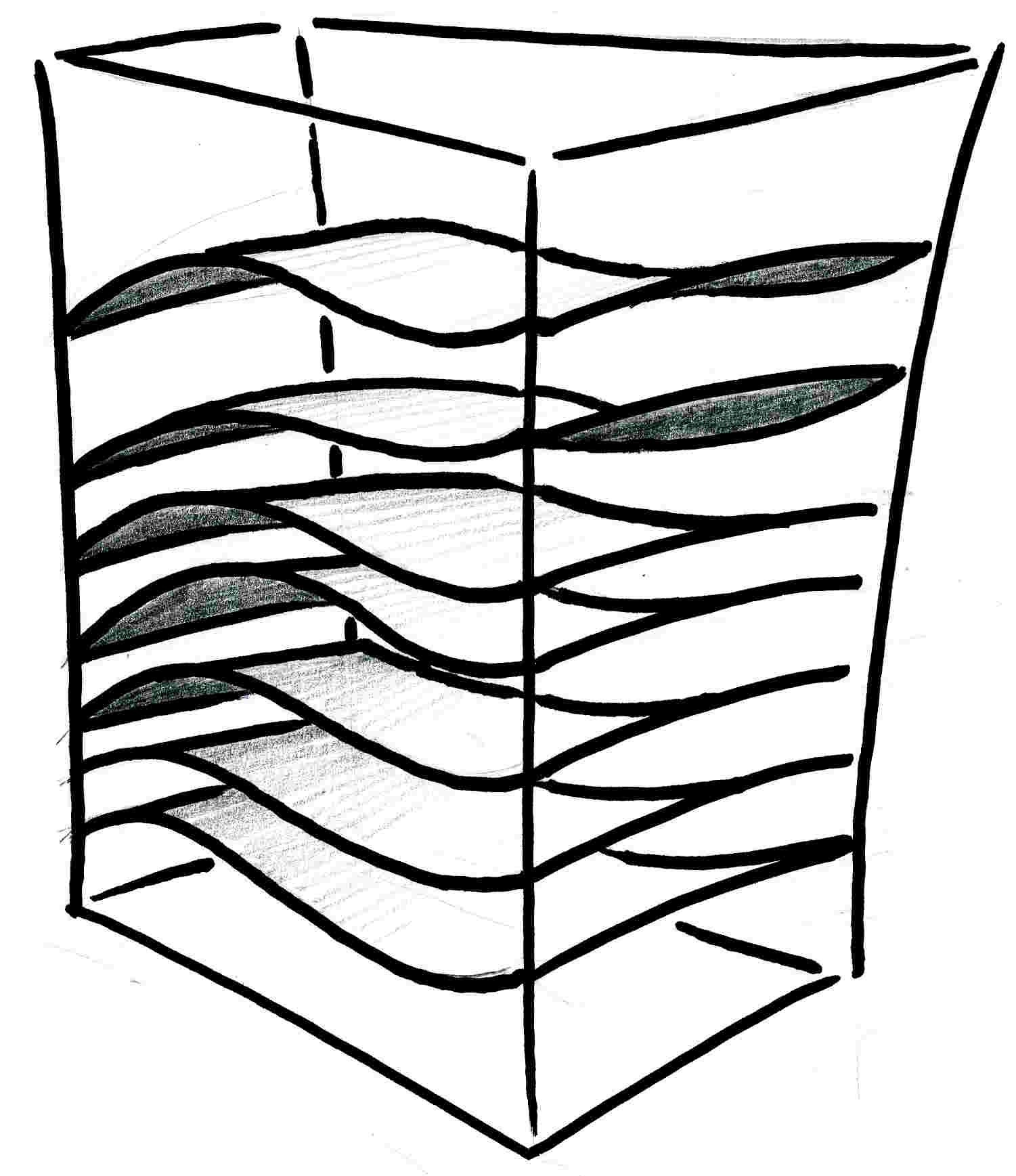}
\caption{A schematic representation of the 
AdS$_3$ phase space foliated into orbits of the asymptotic symmetry group. Solutions belonging to the same 
symplectic
leaf are related to one another by asymptotic symmetry transformations, i.e.\ ``boosts'', but there are no 
boosts that connect different leaves.\label{figoLeaf}}
\end{figure}

\paragraph{Remark.} The relation between AdS$_3$ gravity and orbits of the Virasoro group has recently been 
the object of renewed interest, as it was realized that a similar structure arises in many other contexts. 
To the author's knowledge, the first explicit mention of that relation appears in 
\cite{Martinec:1998wm,NavarroSalas:1999sr,Nakatsu:1999wt}; it is also hidden between the lines 
in \cite{Witten:2007kt,Maloney:2007ud}. The relation was later studied 
in \cite{Garbarz:2014kaa,Barnich:2014zoa} due to its implications for positive energy theorems, 
while \cite{Sheikh-Jabbari:2016unm} (see also \cite{Seraj:2016cym}) is devoted to the geometric properties of 
metrics corresponding to non-constant pairs $(p,\bar p)$.

\subsection{Positive energy theorems}
\label{susePosAd}

As in section \ref{senepost} one may ask which solutions of AdS$_3$ gravity belong to Virasoro orbits on 
which the energy functional (\ref{tt212}) is bounded from below. These solutions can then be considered as 
``physical'', in contrast to the pathological solutions whose energy can be made arbitrarily low by suitable 
asymptotic symmetry transformations. Since the transformation law of the components $(p,\bar p)$ is the 
coadjoint representation of the Virasoro group, the results of section \ref{senepost} are directly 
applicable to the problem at hand.\i{Brown-Henneaux metric!positive energy} Thus 
the only solutions $(p,\bar p)$ with energy bounded from below are 
those in which both $p$ and $\bar p$ belong to one of the orbits highlighted in red in fig.\ \ref{vifigibis}. 
More explicitly, zero-mode solutions $(p_0,\bar p_0)$ belong to orbits with energy bounded from below if and 
only if both $p_0$ and $\bar p_0$ are larger than (or equal to) the vacuum value $-c/24$. For solutions that 
do not admit a rest frame, either $p$ or $\bar p$ 
(or both) must belong to the unique massless orbit with energy bounded from below.\\

These arguments can be interpreted as a positive energy theorem for AdS$_3$ 
gravity \cite{Garbarz:2014kaa,Barnich:2014zoa}. They imply in particular that, in the diagram of zero-mode 
solutions of fig.\ \ref{figs229}, all BTZ black holes belong to orbits with energy bounded from below, while 
all conical excesses belong to orbits with unbounded energy. The pure AdS$_3$ metric also belongs to an orbit 
with energy bounded from below, while the only conical deficits whose energy is bounded from below under 
asymptotic symmetries are those located in the square $-c/24\leq p_0,\bar p_0\leq 0$. The absolute minimum of 
energy among all solutions with energy bounded from below is realized by AdS$_3$ space-time.

\paragraph{Remark.} Positive energy theorems in general relativity have a long history; in short, the 
problem is to show that energy is bounded from below in a suitably defined phase space of metrics. 
\i{positive energy theorem} This problem is 
classically addressed in four-dimensional asymptotically flat space-times, where positivity of energy was 
first proved in \cite{Schon:1981vd}. A supersymmetry-based proof can also be found in \cite{Witten:1981mf}, 
while
the case of the Bondi mass was settled shortly thereafter by
various authors --- see e.g.~\cite{Wald:1984rg} for a list of references. Note that a relation between 
positive energy theorems \cite{Brill:1968ca} and Virasoro orbits was already suggested in footnote 8 of 
\cite{Witten:1987ty}, albeit in a very different context. In section \ref{sebmscodj} we will 
encounter a positive-energy theorem in three-dimensional asymptotically flat space-times.

\section{Quantization and Virasoro representations}
\label{sevirep}

Recall from section \ref{geoq} that the quantization of 
coadjoint orbits yields unitary group representations. Assuming that this applies to the Virasoro group, 
unitary representations can be interpreted as quantized orbits of Brown-Henneaux metrics under asymptotic 
symmetry 
transformations. Accordingly, we now investigate the relation between unitary representations of 
the Virasoro algebra and AdS$_3$ quantum gravity. We start with an overview of $\sl$ and Virasoro 
highest-weight representations, which we interpret as particles dressed with 
gravitational 
degrees of freedom in AdS$_3$. We then conclude with the observation that Virasoro characters coincide with 
quantum gravity partition functions in AdS$_3$.

\paragraph{Remark.} This section is our first encounter with representations of Lie \it{algebras} in this 
thesis, so our language and notations will be somewhat different from those 
of part I. The link between 
the language of part I and Lie algebra representations will be established through 
induced modules, in section \ref{sebomodule}.

\subsection{Highest-weight representations of $\sl$}
\label{suseHiS}

Here we describe highest-weight unitary representations of the Lie algebra $\sl$, which will be useful guides 
for studying Virasoro representations. From a space-time perspective, $\sl$ is half of the isometry 
algebra of AdS$_3$, so tensor products of highest-weight representations of $\sl$ are particles propagating 
in AdS$_3$. Their Minkowskian analogue are the Poincar\'e representations of section 
\ref{sePoTri}.

\subsubsection*{Highest weights and descendants}

The Lie algebra $\sl$ of $\SL$ consists of traceless 
real $2\times 2$ matrices; its basis can be chosen as in (\ref{tmus}), but for the purpose 
of describing unitary representations it is convenient to use the complex basis\i{sl2R@$\sl$}
\be
L_0\equiv-it_0\,,
\qquad
L_1\equiv t_1+it_2\,,
\qquad
L_{-1}\equiv -t_1+it_2\,.
\label{ellemok}
\ee
Equivalently, $L_m=i\ell_m$ in terms of the basis (\ref{eltimm}) and the brackets (\ref{LaLLA}) become
\be
[L_m,L_n]=(m-n)L_{m+n}
\label{wittibi}
\ee
with $m,n=-1,0,1$.
In any unitary representation $\sT$ of 
$\sl$, the real generators $t_{\mu}$ are represented by 
anti-Hermitian operators acting in a suitable Hilbert space. In terms of the generators (\ref{ellemok}), this 
is to say that the Hermiticity conditions\i{sl2R@$\sl$!Hermiticity conditions}\i{Hermiticity 
conditions!for sl2R@for $\sl$}
\be
\sT[L_m]^{\dagger}=\sT[L_{-m}]
\label{hermiconw}
\ee
hold in a unitary representation. From now on we will abuse notation and neglect writing the representation 
$\sT$, so that $\sT[L_m]\equiv L_m$. This abuse is common in physics, so it should not lead to 
any misunderstanding. Until the end of this chapter we also use the Dirac notation instead of the less 
standard notation of part I.\\

Since the group $\SL$ is simple but non-compact, all its non-trivial unitary representations are 
infinite-dimensional. Fortunately, the complexification of $\sl$ coincides with that of $\mathfrak{su}(2)$, 
and we definitely know how to build unitary highest-weight representations of the latter. Let us therefore 
use the same approach for $\sl$: we start from a (normalized) \it{highest-weight state} $|h\rangle$ belonging 
to the Hilbert space of the representation, such that 
\i{sl2R@$\sl$!highest-weight representation}\i{highest-weight representation!of sl2R@of $\sl$}
\be
L_0|h\rangle=h|h\rangle,
\qquad
L_1|h\rangle=0\,,
\label{hiwesta}
\ee
where $h$ is the \it{highest weight}.
The conditions (\ref{hermiconw}) imply that 
the operator representing $L_0$ is Hermitian, so $h$ must be real. The interpretation of (\ref{hiwesta}) is 
that $|h\rangle$ has 
energy $h$ if we think of $L_0$ as the Hamiltonian, while $L_1$ is an annihilation operator.\footnote{The 
terminology is 
somewhat backwards, since the highest weight $h$ is actually the \it{lowest} eigenvalue of $L_0$ in the space 
of the representation; this terminological clash is standard.}\\

In order to produce a representation we must also act with operators $L_{-1}$ on the highest-weight state. 
This leads to \it{descendant states} of the form\i{sl2R@$\sl$!descendant state}\i{descendant state!for $\sl$}
\be
(L_{-1})^N|h\rangle,
\label{s72}
\ee
where the non-negative integer $N$ is the \it{level}\i{level (of descendant state)} 
of the descendant. Each descendant state is 
an eigenstate of $L_0$ with eigenvalue $h+N$, so $L_{-1}$ is analogous to a creation operator. We then 
declare 
that the carrier space $\sH$ of the representation is spanned by all linear combinations 
of descendant states. Since we want $\sH$ to be a Hilbert space, descendant states with 
different levels must be orthogonal because their eigenvalues under $L_0$ differ. Furthermore, all descendant 
states must have non-negative norm 
squared. Using the Hermiticity conditions (\ref{hermiconw}), this amounts to the 
requirement
\be
0\leq\left\|
(L_{-1})^N|h\rangle
\right\|^2
=
\big<h\big|
[(L_{-1})^N]^{\dagger}(L_{-1})^N
\big|h\big>
\refeq{hermiconw}
\big< h\big|
(L_1)^N(L_{-1})^N
\big|h\big>\,.
\nn
\ee
To ensure that this condition holds, we evaluate scalar products of descendant states. 
Thanks to the commutation relations (\ref{wittibi}) one finds\i{highest-weight representation!of sl2R@of 
$\sl$}
\be
\big< h\big|
(L_1)^N(L_{-1})^N
\big|h\big>
=
N!\,\prod_{k=0}^{N-1}(2h+k)
\label{normaD}
\ee
where we have used $\langle h|h\rangle=1$. Thus, all descendant states have strictly positive norm squared if 
and only if 
$h>0$. If $h=0$, then the representation is trivial.\\

Note that here the only restriction imposed on $h$ by unitarity is $h\geq0$. 
However, if we were to integrate a highest-weight representation of $\sl$ into an \it{exact} unitary 
representation of the group $\SL$,\i{projective representation!of SL2R@of $\SL$}\i{SL2R@$\SL$!projective 
representation} 
then $h$ would have to be an integer in order to ensure that a rotation 
by $2\pi$ is represented by the identity. On the other hand, \it{projective} representations of $\SL$ allow 
$h$ to be an arbitrary positive real number since the fundamental group of $\SL$ is isomorphic to $\ZZ$.

\subsubsection*{Representations by quantization}

The representations just described can be
identified with representations obtained by quantizing suitable coadjoint orbits of $\SL$. Indeed, note that 
the Lie algebra (\ref{wittibi}) admits a quadratic Casimir operator 
\i{sl2R@$\sl$!Casimir operator}\i{Casimir operator!for $\sl$}
\be
\cC
=
L_0^2-\demi(L_1L_{-1}+L_{-1}L_1)
\refeq{ellemok}
-t_0^2+t_1^2+t_2^2\,.
\label{silcasimir}
\ee
This operator is proportional to the identity in any irreducible representation of $\sl$. In the 
highest-weight representation (\ref{s72})
it takes the value
\be
\cC
=
h(h-1),
\label{silcash}
\ee
which proves by the way that the representation is irreducible. The fact that 
(\ref{silcasimir}) takes a constant value is reminiscent of the ``mass shell''
condition defining the coadjoint orbit (\ref{silorbitg}), and indeed the 
highest-weight representation just displayed is the quantization of such an orbit for $h>0$. The only 
subtlety is that the value of the Casimir (\ref{silcash}) is not quite $h^2$, but $h(h-1)$; the two numbers 
coincide in the semi-classical limit $h\rightarrow+\infty$, and the $h-1$ of (\ref{silcash}) may be seen as a 
quantum correction of the classical result.\\

This identification is confirmed by the computation of $\SL$ characters. Indeed, the counting argument of the 
end of section \ref{gigglema} can now be applied to the 
space $\sH$ spanned by the highest-weight state $|h\rangle$ and 
its descendants. As a result one finds that $\text{Tr}(q^{L_0})$ is precisely given by formula 
(\ref{silchar}) 
up to the replacement of 
$h+1/2$ by $h$. From a space-time perspective, the product of two such characters is the character of an 
irreducible representation of the isometry algebra $\mathfrak{so}(2,2)$ of AdS$_3$. If the two 
representations have weights $h$ and $\bar h$ say, the product of their characters can be interpreted as 
the partition function of a particle with mass $(h+\bar h)/\ell$ and spin $h-\bar h$ 
propagating in AdS$_3$.

\subsection{Virasoro modules}
\label{susebasics}

We now generalize the representation theory of $\sl$ to the Virasoro algebra.
Before doing so, a word of caution is in order: in the case of $\SL$, we were able to interpret 
highest-weight representations as quantized coadjoint orbits of the type (\ref{silorbitg});
this identification was supported by the matching of the Casimir operator (\ref{silcash}) with the definition 
of the orbit. In the case of the Virasoro algebra the situation is complicated by the 
fact that all coadjoint orbits (at non-zero central charge) are infinite-dimensional; in addition, the only 
Casimir operators of the Virasoro algebra\i{Casimir operator!for Virasoro}\i{Virasoro 
algebra!Casimir operators} are functions of its central charge \cite{feigin1983}. Accordingly, 
we start with a few comments regarding Virasoro geometric quantization, before turning to the 
construction of its highest-weight representations and the evaluation of the associated characters.

\subsubsection*{Semi-classical regime}

If one believes in the orbit method, geometric quantization applied to the coadjoint orbits of the 
Virasoro algebra should produce unitary Virasoro representations.\i{Virasoro 
orbits!quantization}\i{highest-weight representation!by quantization}\i{geometric quantization!and 
unitary reps} This 
viewpoint was adopted in 
\cite{Witten:1987ty,Taylor:1992xt,Taylor:1993zp}, with the conclusion that the quantization of orbits with 
positive energy and constant representatives indeed provides highest-weight representations in the 
large $c$ limit. By contrast, the limit of small $c$ is much more elusive, and at present it is 
not known if the discrete series of Virasoro representations at $c\leq1$ can be obtained by geometric 
quantization (see e.g.\ \cite{La:1990tv}). From a gravitational point of view, the Virasoro central charge 
(\ref{ss210}) is the AdS radius in units of the Planck length, so large $c$ corresponds to the 
semi-classical regime.\i{semi-classical limit}\i{large central charge} This is confirmed by symplectic 
geometry: the 
Kirillov-Kostant symplectic form 
(\ref{kksym}) evaluated at a constant Virasoro coadjoint vector $(p_0,c)$ is\i{Virasoro 
orbits!Kirillov-Kostant form}\i{Kirillov-Kostant symplectic form!for Virasoro}
\be
\omega_{(p_0,c)}\big((\xi_m)_{p_0},(\xi_n)_{p_0}\big)
\refeq{virpar}
-im\left(2p_0+\frac{c}{12}m^2\right)\delta_{m+n,0}
\label{kkfovira}
\ee
where $\xi_m=\ad^*_{\cL_m}$ is the vector field on $\cW_{(p_0,c)}$ that 
generates the coadjoint action of the Virasoro generator $\cL_m$ given by (\ref{vimodes}). The occurrence of 
$c$ confirms that 
the regime of large $c$ is semi-classical in the sense that a large volume is assigned to 
any portion of phase space. Conversely, small $c$ corresponds to the non-perturbative 
regime, where quantum corrections may alter classical results in a radical way.\\

This heuristic argument is consistent with the fact that geometric quantization is relatively well 
established at large $c$, but 
poorly understood at small $c$. Since applications to three-dimensional gravity rely on the semi-classical 
limit anyway, from now on we restrict ourselves to the regime of large $c$. This assumption turns out to 
greatly simplify representation theory, and allows us to think of highest-weight Virasoro representations as 
quantizations of Virasoro orbits with constant representatives.\\

This being said, to our knowledge there is as yet no strict mathematical proof of the fact 
that geometric quantization 
of Virasoro orbits produces highest-weight representations, despite numerous attempts in the literature (see 
e.g.\ \cite{AiraultMal,AMT}). Our viewpoint here will be pragmatic, and we shall assume that 
the representations obtained by quantizing such orbits are indeed highest-weight representations. This 
assumption will be supported, among other observations, by the fact that Virasoro characters match suitable 
gravitational partition functions (see section \ref{susedads}).

\subsubsection*{Highest-weight representations}

The basis of the Virasoro algebra given by (\ref{vimodes}) is such that, in any unitary representation, the 
operators representing the generators $\cL_m+\cL_{-m}$, $i(\cL_m-\cL_{-m})$ and $\cZ$ are 
\it{anti}-Hermitian. A more standard basis is given by
\be
L_m\equiv i\cL_m+i\frac{\cZ}{12}\delta_{m,0},
\qquad
Z\equiv i\cZ,
\label{operatul}
\ee
where the constant shift in $L_0$ ensures that the vacuum state has zero eigenvalue under $L_0$. According 
to this definition the operators 
representing $L_m$ and $Z$ in a unitary representation satisfy the Hermiticity conditions 
\i{Virasoro algebra!Hermiticity conditions}\i{Hermiticity conditions!for Virasoro}
\be
L_m^{\dagger}=L_{-m},
\qquad
Z^{\dagger}=Z
\label{hemingway}
\ee
where we abuse notation by denoting the basis element $L_m$ and the operator that 
represents it with the same symbol. In any irreducible representation the Hermitian central operator $Z$ 
is proportional to the identity with a coefficient $c\in\RR$, so we may write the commutation relations of 
the 
operators representing the generators (\ref{operatul}) as
\be
[L_m,L_n]
=
(m-n)L_{m+n}+\frac{c}{12}m(m^2-1)\delta_{m+n,0}\,.
\label{serge}
\ee
Here the existence of the $\sl$ subalgebra (\ref{wittibi}) is manifest, since the contribution of the central 
extension 
vanishes for $m=-1,0,1$.\\

Highest-weight representations of the Virasoro algebra (\ref{serge}) seem to have first appeared 
in \cite{FeiginFuchs01,FeiginFuchs02,FeiginFuchs03}, and were then further studied in 
\cite{segal1981,Kac1982,goddard1986}. They are built in direct analogy to the highest-weight representations 
of $\sl$. In accordance with 
geometric quantization, the parameters that specify these representations coincide with those that determine 
the corresponding coadjoint orbits. In the present case, taking the orbit of a constant coadjoint 
vector $(p_0,c)$, one defines a real 
number $h$ by
\be
p_0=h-\frac{c}{24}\,.
\label{pizohh}
\ee
This ensures that $h=0$ for the vacuum configuration, while $h\geq 0$ for orbits with energy bounded 
from below. With this notation, the representation obtained by quantizing the orbit of $(p_0,c)$ 
is obtained as follows.\\

To begin, in analogy with (\ref{hiwesta}), one defines the 
\it{highest-weight state} of the representation to be a normalized state $|h\rangle$ such that\footnote{The 
actual value of the quantum weight $h$ may differ from the classical parameter defined in (\ref{pizohh}) by 
corrections of order $\cO(1/c)$, so from now on it is understood that $h$ refers to the \it{quantum} value. 
This subtlety will have very little effect on our discussion.}
\be
L_0|h\rangle=h|h\rangle,
\qquad
L_m|h\rangle=0
\quad\text{for }m>0.
\label{KATYS}
\ee
The state $|h\rangle$ is also called a \it{primary state}.\i{primary state} Its definition ensures 
that it has energy 
$h$ under the Hamiltonian $L_0$, while the operators $L_m$ with $m>0$ are annihilation 
operators. In analogy with (\ref{s72}) one also defines \it{descendant states}\i{Virasoro 
algebra!descendant state}\i{descendant state}\i{Virasoro algebra!highest-weight 
rep}\i{highest-weight representation!of Virasoro}
\be
L_{-k_1}...L_{-k_n}|h\rangle,
\qquad
1\leq k_1\leq k_2\leq\cdots\leq k_n\,.
\label{deska}
\ee
Thus one can interpret the operators $L_{-m}$ with $m>0$ as creation operators. We will discuss the 
gravitational interpretation of this representation in section \ref{susedads}.\\

Using the commutation relations (\ref{serge}) of the Virasoro algebra, one verifies that each 
descendant (\ref{deska}) is an eigenstate of $L_0$ with eigenvalue\i{level (of 
descendant state)}
\be
h+\sum_{i=1}^nk_i\equiv h+N
\nn
\ee
where the non-negative integer $N$ is the \it{level} of the descendant. One then declares that 
the space $\sH$ of the Virasoro representation is the \it{Verma module}\i{Verma module} spanned by all linear 
combinations of 
descendant states. As in the case of $\sl$, descendants with different levels have 
different eigenvalues under $L_0$. According to (\ref{hemingway}) the latter must be Hermitian if the 
representation is to be unitary, so scalar products of descendants with different 
levels vanish. However, in contrast to $\sl$, there are in general many different 
descendant states with the same level. More precisely, at large central charge $c$, the number of different 
descendants at level $N$ is the number $p(N)$ of partitions of $N$ in distinct positive integers\i{partition 
of 
integers}\i{pN@$p(N)$ (partition of integers)} (e.g.\ $p(0)=p(1)=1$, 
$p(2)=2$, $p(3)=3$, $p(4)=5$, etc.).\\

Note that the representation whose carrier space is spanned by the descendant states (\ref{deska}) 
is an induced representation of the Virasoro algebra, i.e.\ an \it{induced module}.\i{induced 
module!for Virasoro algebra} Indeed, the conditions 
(\ref{KATYS}) define a one-dimensional representation of the subalgebra generated by $L_0$ and the $L_m$'s 
with $m>0$, and the prescription (\ref{deska}) is the algebraic analogue of the statement that wavefunctions 
live on a quotient space $G/H$ (recall section \ref{sedefirep}). By the way, a similar interpretation holds 
for the highest-weight representations of $\sl$ displayed in section \ref{suseHiS}. We will return to this 
observation in section \ref{sebomodule}.

\subsubsection*{Unitarity for Virasoro representations}

We now ask whether the vector space spanned by $|h\rangle$ and its descendants is a 
Hilbert space, given that Hermitian conjugation is defined by (\ref{hemingway}). Working only with low-level 
descendant states, one can easily derive some basic necessary conditions for unitarity. For instance, 
the only descendant at level one is $L_{-1}|h\rangle$ and its norm squared is
$\bra h|L_1L_{-1}|h\ket
=
2h$,
so a necessary condition for unitarity is $h\geq 0$. Similarly, at 
level $N$ there is a state 
$L_{-N}|h\rangle$ with norm squared
\be
\bra h|L_NL_{-N}|h\ket
\refeq{serge}
2Nh+\frac{c}{12}N(N^2-1)
\nn
\ee
whose positivity for large $N$ requires $c\geq 0$. Thus, the only values of $c$ and 
$h$ that give rise to unitary representations are positive. In terms of coadjoint 
orbits of the Virasoro group, this is to say that only orbits with positive energy can produce unitary 
representations under quantization. In order to go further one generally relies on the so-called \it{Gram 
matrix}\i{Gram matrix} of the module, whose entries are the scalar 
products of descendants. Demanding unitarity then boils down to the requirement that the Gram matrix be 
positive-definite. One can show that 
this condition is always verified by Virasoro
highest-weight 
representations at large $c$. Since this is a standard result in two-dimensional conformal 
field theory (see e.g.\ 
\cite{DiFrancesco:1997nk}), we simply state it here without proof:

\paragraph{Proposition.} Highest-weight representations of the Virasoro algebra with $c>1$ and $h>0$ are 
unitary\i{highest-weight representation!unitarity} and 
irreducible in the sense that all descendant states have strictly positive norm squared.\\

Note that, by contrast, unitary highest-weight representations at central charge $c\leq1$ have a very 
intricate 
structure due to null states,\i{null state}\i{highest-weight representation!at c1@at $c\leq1$} i.e.\ 
states with vanishing norm that are modded out of the Hilbert space as in 
the definition (\ref{modout}) of $L^2$ spaces. In that case, not all descendant states 
(\ref{deska}) are linearly 
independent, since some of them are effectively set to zero --- in this sense the Verma module is reducible. 
We will 
not take such subtleties into account here because we are interested only in the limit 
of large central charge, where null states are absent.

\paragraph{Remark.} It was recently shown \cite{Salmasian:2014wwa} that all irreducible unitary 
representations of 
the Virasoro group with a spectrum of $L_0$ bounded from below are highest-weight representations of the type 
described 
here.\i{highest-weight representation!exhaustivity}\i{classification!of Virasoro reps}\i{exhaustivity 
theorem} In this sense, highest-weight representations exhaust all 
unitary representations of the Virasoro algebra.

\subsubsection*{Vacuum representation}

In $\sl$, the representation 
with highest weight $h=0$ is trivial since all descendant states are null by virtue of eq.\ (\ref{normaD}). 
We now describe the \it{Virasoro} representation obtained by setting $h=0$, to which we refer as the 
\it{vacuum representation}.\i{Virasoro vacuum}\i{vacuum module!for Virasoro algebra}\i{highest-weight 
representation!vacuum} It is obtained by quantizing the vacuum Virasoro orbit, 
containing the point $p_{\text{vac}}=-c/24$.\\

The highest weight state $|0\rangle$ of that representation satisfies the 
properties\i{vacuum module!for Virasoro algebra}
\be
L_0|0\rangle
=
L_{-1}|0\rangle
=
L_m|0\rangle
=
0
\quad\text{for all }m>0,
\label{s82}
\ee
which ensures that the vacuum state $|0\rangle$ is invariant under the $\sl$ subalgebra generated 
by $L_{-1}$, $L_0$ and $L_1$; it is the quantum counterpart of the statement that the stabilizer of the 
vacuum orbit is the group $\PSL$. Crucially, the 
vacuum is \it{not} invariant under the higher-mode generators $L_{-2}$, $L_{-3}$ etc.\ so that the 
representation whose carrier space is spanned by all descendant states\i{Virasoro vacuum!descendant state}
\be
L_{-k_1}...L_{-k_n}|0\rangle,
\qquad
2\leq k_1\leq\cdots\leq k_n
\label{ss82}
\ee
is non-trivial. This representation is unitary for all $c>0$, and it is free of null states (i.e.\ 
irreducible) whenever $c>1$. It is in many ways 
analogous to the standard highest-weight representation generated by the descendant states (\ref{deska}), but 
the condition $L_1|0\rangle=0$ makes it slightly smaller than generic highest-weight representations.\\

It may seem puzzling that the vacuum state is \it{not} left invariant by 
all Virasoro 
generators but only by a subset thereof as in (\ref{s82}). The reason is that, at non-zero central charge, it 
is impossible to define a non-zero state that is annihilated by all Virasoro generators. Indeed, if there was 
such a state $|\tilde 0\rangle$, then we would have
\be
0
=
\langle\tilde 0|L_n L_{-n}|\tilde 0\rangle
\refeq{serge}
\langle\tilde 0|
\left(
2nL_0+\frac{c}{12}n(n^2-1)
\right)
|\tilde 0\rangle
=
\frac{c}{12}n(n^2-1)\bra\tilde 0|\tilde 0\ket,
\nn
\ee
which is a contradiction when $c\neq 0$ and $n\neq-1,0,1$. Hence the vacuum state of a Virasoro-invariant 
theory is always $\sl$-invariant but \it{never} Virasoro-invariant.
One should appreciate the counter-intuitive nature of this phenomenon: it means that even the absolute 
simplest Virasoro-invariant quantum theory is described by an infinite-dimensional Hilbert space of vacuum 
descendants. In section \ref{sebmspar} we will encounter a similar non-trivial vacuum 
representation for the BMS$_3$ group.

\subsection{Virasoro characters}
\label{suseVIKKA}

The relation between AdS$_3$ quantum gravity and Virasoro representations is most easily expressed in terms 
of partition functions, so as a preliminary we now evaluate characters of 
highest-weight representations of the Virasoro algebra at $c>1$.\\

Let $\sH$ be the Hilbert space of an irreducible, unitary 
representations of the Virasoro algebra with central charge $c$ and highest-weight $h$ (if $h=0$ we take the 
vacuum representation). We define the \it{character} of the representation as\i{highest-weight 
representation!character}\i{Virasoro algebra!character}\i{character!for Virasoro}
\be
\chi(\tau)
\equiv
\text{Tr}_{\sH}\left[q^{L_0-c/24}\right],
\qquad
q\equiv e^{2\pi i\tau}
\label{s84}
\ee
where the notation is almost the same as in eq.\ (\ref{silchar}). The parameter $\tau$ is a 
complex number with 
positive imaginary part so that, if $L_0$ is interpreted as the Hamiltonian, then $\text{Im}(\tau)$ is 
the inverse temperature and the character itself is a canonical partition 
function. The normalization factor $-c/24$ is conventional.

\subsubsection*{Generic highest-weight representations}

Let $h>0$ be a highest weight and $c>1$ a large central charge. Consider the Verma module spanned 
by all descendant states (\ref{deska}). Then there are no null states and distinct descendants are 
linearly independent, so the spectrum of $L_0$ consists of all values $h+N$, with multiplicity $p(N)$. 
Accordingly, the character (\ref{s84}) reads
\be
\chi(\tau)
=
\sum_{N=0}p(N)q^{h+N-c/24}
=
q^{h-c/24}\sum_{N=0}^{+\infty}p(N)q^N.
\label{ss84}
\ee
We now rewrite this in a more convenient way thanks to the following result:

\paragraph{Lemma.} The series (\ref{ss84}) can be rewritten as an infinite product
\be
\sum_{N=0}^{+\infty}p(N)q^N
=
\prod_{n=1}^{+\infty}\frac{1}{1-q^n}\,.
\label{s85}
\ee

\begin{proof}
We follow \cite{apostol1998}, to which we refer for a careful treatment of the convergence issues that will 
not be addressed here. To prove (\ref{s85}) we consider its right-hand side and expand each individual term 
of the infinite product as a geometric series:
\begin{align}
\prod_{n=1}^{+\infty}\frac{1}{1-q^n}
& =
\prod_{n=1}^{+\infty}
\sum_{k=0}^{+\infty}q^{nk}
=
\left(1+q+q^2+q^3+\cdots\right)
\left(1+q^2+q^4+\cdots\right)\cdots\nn\\
& =
1+\sum_{N=1}^{+\infty}\sum_{n=1}^N
\underbrace{
\sum_{\substack{k_1,k_2,...,k_n\\ k_1+k_2+\cdots+k_n=N\\ 1\leq k_1\leq k_2\leq\cdots\leq k_n}}
}_{p(N)}
\underbrace{
q^{k_1+k_2+\cdots+k_n}
}_{q^N}
=
\sum_{N=0}^{+\infty}p(N)q^N\,.
\nn
\end{align}
This concludes the argument.
\end{proof}

Thus the character (\ref{ss84}) can be rewritten as\i{highest-weight representation!character}\i{Virasoro 
character}
\be
\chi(\tau)
=
q^{h-c/24}\prod_{n=1}^{+\infty}\frac{1}{1-q^n}
=
\frac{q^{h-(c-1)/24}}{\eta(\tau)}
\label{s86}
\ee
where in the second equality we have introduced the \it{Dedekind eta function}\i{Dedekind eta function}
\be
\eta(\tau)
\equiv
q^{1/24}\prod_{n=1}^{+\infty}(1-q^n).
\label{ss86}
\ee
The result (\ref{s86}) can be seen as an infinite product of $\sl$ characters (\ref{silchar}) with 
parameters $n\tau$. Equivalently, since $\sl$ representations coincide with harmonic oscillators whose 
partition functions were written in (\ref{silchar}), one can think of a Virasoro representation as an 
infinite 
collection of harmonic oscillators. This suggests that Virasoro characters can be interpreted as quantum 
field theory partition functions, which will be confirmed in section \ref{susedads} below. Note also the 
presence of 
the ubiquitous factor $1-q^n$ in the denominator.

\paragraph{Remark.} Our derivation of (\ref{s86}) relied on the fact that the eigenvalue $h+N$ of 
$L_0$ has degeneracy $p(N)$. This is only true provided there are no null 
states, i.e.\ provided $c>1$. By contrast, for $c\leq 1$, null states generally do exist and are modded out 
of the Hilbert space of the representation. This leads to a smaller degeneracy of eigenvalues of $L_0$, hence 
to a character that is strikingly different from (\ref{s86}). We will not display characters at $c\leq 1$ 
here; see e.g. \cite{Wassermann02,Wassermann01} for explicit 
formulas.

\subsubsection*{Vacuum representation}

The character of the vacuum Virasoro representation can be evaluated in the same way as for generic 
highest-weight representations. The only subtlety is that the vacuum is $\sl$-invariant, leading 
to a reduced number of descendant states (\ref{ss82}). Explicitly, let $\Delta(N)$ denote the degeneracy of 
the eigenvalue $N$ in the space spanned by the vacuum descendants. For $N\geq2$, this degeneracy is the 
number of partitions of $N$ in positive integers which are strictly greater than one (thus $\Delta(0)=1$ by 
convention but $\Delta(1)=0$, $\Delta(2)=\Delta(3)=\Delta(4)=\Delta(5)=1$, $\Delta(6)=2$, etc.). Then the 
vacuum character is 
\be
\chi(\tau)
=
\sum_{N=0}^{+\infty}\Delta(N)q^{N-c/24}
=
q^{-c/24}\sum_{N=0}^{+\infty}\Delta(N)q^N.
\label{s87}
\ee
In order to relate $\Delta(N)$ to the usual partition $p(N)$, we note that
\be
\Delta(N)
=
p(N)
-
\bmm
\text{number of partitions of $N$}\\
\text{containing at least one ``$1$''}
\emm
=
p(N)-p(N-1)
\nn
\ee
which allows us to rewrite the character (\ref{s87}) as\i{Virasoro vacuum!character}\i{character!for 
Virasoro}\i{vacuum character!for Virasoro}
\be
\chi_{\text{vac},c}(\tau)
=
q^{-c/24}\sum_{N=0}^{+\infty}p(N)q^N(1-q)
\refeq{s85}
q^{-c/24}\prod_{n=2}^{+\infty}\frac{1}{1-q^n}.
\label{s88}
\ee
Note how the product in the denominator of (\ref{s88}) is truncated (no term $n=1$) owing to the $\sl$ 
symmetry of the vacuum state. In particular the vacuum character (\ref{s88}) is \it{not} just the limit 
$h\rightarrow0$ of the generic character (\ref{s86}).

\subsection{Dressed particles and quantization}
\label{susedads}

Since AdS$_3$ gravity has Virasoro symmetry, its quantization is expected to produce unitary representations 
of the direct sum of two Virasoro algebras. We now investigate to what extent this is the case. For 
notational simplicity we denote the Virasoro algebra by $\vir$ (instead of $\hVect$), so that the asymptotic 
symmetry algebra of AdS$_3$ gravity with Brown-Henneaux boundary conditions is $\vir\oplus\vir$.\\

First let us make the proposal more precise: the orbit of a metric $(p,\bar p)$ is a coadjoint orbit 
(\ref{s230}) of the product of two Virasoro groups with central charges (\ref{ss210}). For simplicity let us 
assume that the metric is a zero-mode and that its energy is bounded from below under asymptotic symmetry 
transformations, so $p(x^+)=p_0\geq -c/24$ and $\bar p(x^-)=\bar p_0\geq -\bar c/24$. The non-zero modes 
belonging to the orbit (\ref{s230}) may be seen as classical analogues of the descendant states 
(\ref{deska}).\i{boundary graviton!quantization}\i{geometric quantization!and boundary gravitons} Upon 
defining $h\equiv p_0+c/4$ and $\bar h\equiv\bar p_0+\bar c/24$, one expects that 
the geometric quantization of the orbit (\ref{s230}) produces the tensor product of two highest-weight 
representations of the Virasoro algebra labelled by $(h,c)$ and $(\bar h,\bar c)$.\footnote{As before, the 
quantum values of $(h,\bar h)$ may differ from their classical counterparts by $1/c$ corrections.} The same 
would be true by 
quantizing the orbit of AdS$_3$, except that the result would be the tensor product of two vacuum 
representations.\\

It is worth comparing these representations to those of $\mathfrak{so}(2,2)\cong\sl\oplus\sl$, the isometry 
algebra of AdS$_3$. Since the latter is a subalgebra of $\vir\oplus\vir$, any Virasoro 
representation with highest weights $(h,\bar h)$ contains many $\sooh$ subrepresentations with weights 
increasing 
from $(h,\bar h)$ to infinity. Thus a Virasoro representation is an $\sooh$ representation dressed with 
(infinitely many) extra directions in the Hilbert space obtained by acting with the Virasoro generators 
$L_{-2}$, $L_{-3}$, etc.\\

Now recall that a particle propagating in AdS$_3$ is an irreducible unitary representation of 
$\sooh$,\i{particle!in AdS$_3$} while asymptotic symmetries 
generalize isometries by including 
gravitational fluctuations. Accordingly one is led to interpret Virasoro representations as particles in 
AdS$_3$ dressed with some extra gravitational degrees of freedom accounted for by the modes $L_n$ that do not 
appear in the isometry algebra. These are quantum analogues of the classical ``boundary 
gravitons'' described in section \ref{susebovio}. Thus a Virasoro representation is a particle in AdS$_3$ 
dressed with boundary gravitons.\i{dressed particle}\i{particle!dressed with gravitons}\i{boundary 
graviton!dressing} In particular the vacuum representation of 
$\vir\oplus\vir$ is identified with the Hilbert space of quantum boundary gravitons around pure AdS$_3$.\\

As a verification of the fact that Virasoro symmetry is realized in AdS$_3$ quantum gravity, one may wonder 
whether the quantum partition function of gravity reproduces a (combination of) Virasoro character(s). This 
computation was carried out in \cite{Giombi:2008vd}, where the authors evaluated the one-loop partition 
function of gravity on AdS$_3$ at finite temperature $1/\beta$ and angular potential $\theta$ (both taken 
to be real). Upon combining these numbers into a modular parameter\i{modular parameter}
\be
\tau\equiv\frac{1}{2\pi}\Big(\theta+i\frac{\beta}{\ell}\,\Big),
\label{TaKatak}
\ee
it was found that the gravitational one-loop partition function reads\i{partition function!for 3D 
gravity}\i{boundary graviton!partition function}\i{Virasoro vacuum!character}\i{character!for Virasoro}\i{3D 
gravity!partition function}
\be
Z_{\text{grav}}(\beta,\theta)
=
\text{Tr}\left(q^{L_0-c/24}\bar q^{\bar L_0-c/24}\right)
=
|q|^{c/12}\prod_{n=2}^{+\infty}\frac{1}{|1-q^n|^2}
\label{provaco}
\ee
where $q\equiv e^{2\pi i\tau}$ and $\bar q$ is its complex conjugate.
This is precisely the character of the tensor product of two Virasoro vacuum representations, which confirms 
that quantized boundary gravitons around AdS$_3$ span an irreducible highest-weight representation of 
$\vir\oplus\vir$. 
The same computation can be performed for orbifolds of AdS$_3$ obtained by imposing identifications of the 
form $\phii\sim\phii+2\pi/N$ in terms of cylindrical coordinates, where $N\in\NN^*$. The corresponding metric 
is that of a conical deficit labelled by the Virasoro coadjoint vectors $p_0=\bar p_0=-c/(24N^2)$. 
For $N\geq2$ the resulting one-loop partition function is found to be
\be
Z_{\text{grav,N}}(\beta,\theta)
=
|q|^{2h}\prod_{n=1}^{+\infty}\frac{1}{|1-q^n|^2}
\label{Vanouche}
\ee
where $h=\frac{c}{24}(1-1/N^2)$. This is again the character of the tensor product of two highest-weight 
representations of the Virasoro algebra with weights $h=\bar h$, which confirms the 
interpretation of Virasoro modules as particles dressed with boundary gravitons.\\

In chapter \ref{c7} we will develop a similar interpretation for BMS$_3$ particles, which will then be 
confirmed in chapter \ref{c8} by the matching of BMS$_3$ characters with one-loop partition functions for 
asymptotically flat gravity and higher-spin theories.

\paragraph{Remark.} In \cite{Maloney:2007ud} it was conjectured that the one-loop partition function 
(\ref{provaco}) is \it{exact}\i{one-loop exactness} because it is the only expression compatible with 
Virasoro symmetry. Higher 
loop corrections would then renormalize the Brown-Henneaux central charge but would leave the $q$-dependent 
one-loop determinant unaffected. This conjecture was used to 
evaluate a Farey tail\i{Farey tail} sum representing the 
putative full, non-perturbative, partition function of AdS$_3$ gravity (see also \cite{Keller:2014xba}). To 
our knowledge 
the one-loop exactness of (\ref{provaco}) is still an unproven statement.

\subsubsection*{A note on the Fabri-Picasso theorem}

Quantizing the orbit of the AdS$_3$ metric under Brown-Henneaux transformations yields the vacuum 
representation of $\vir\oplus\vir$, whose highest-weight state $|0\rangle$ is annihilated by all $\sooh$ 
generators. But the Virasoro generators $L_m,\bar L_m$ with $m\leq-2$ do \it{not} leave the vacuum invariant. 
Thus Virasoro symmetry is spontaneously broken in AdS$_3$ gravity, and the descendant states\i{broken 
symmetry}\i{Goldstone boson}
\be
L_{-k_1}...L_{-k_n}\bar L_{-\ell_1}...\bar L_{-\ell_m}|0\rangle
\label{Ledes}
\ee
can be loosely interpreted as Goldstone bosons obtained by acting on the vacuum with broken symmetry 
generators. This interpretation has come to be standard in the realm of asymptotic symmetries; in four 
dimensions it leads to
the identification of soft graviton states with Goldstone bosons for spontaneously 
broken BMS symmetry \cite{Strominger:2013jfa}.\\

This being said, one should keep in mind that the comparison to Goldstone bosons should be handled with care. 
Indeed, spontaneously broken \it{internal} symmetries are always such that states obtained by acting 
with broken symmetry generators on the vacuum do \it{not} belong to the Hilbert space. This statement is the 
Fabri-Picasso theorem\i{Fabri-Picasso theorem} (see e.g.\ \cite{aitchison}), and it follows from the fact 
that the norm of the state 
$Q|0\rangle$ has an infrared (volume) divergence whenever $Q$ generates a broken global internal symmetry. If 
the Fabri-Picasso theorem was to hold for asymptotic symmetries, then the descendant states (\ref{Ledes}) 
would make no sense. Fortunately the situation of asymptotic symmetries is different because the charges that 
generate them are \it{surface} charges (\ref{qusurf}) rather than Noether charges. As a result, when $Q$ is a 
broken asymptotic symmetry generator, the norm squared of $Q|0\rangle$ is an integral over a compact 
manifold, and is therefore finite. In this sense spontaneously broken \it{asymptotic} symmetries behave in a 
way radically different from spontaneously broken \it{internal} symmetries.

\newpage
~
\thispagestyle{empty}

\renewcommand{\afterpartskip}{}
\part*{Part III\\[.3cm]
BMS$_3$ symmetry \\ and gravity in flat space}
\addcontentsline{toc}{part}{III BMS$_3$ symmetry and gravity in flat space}
\begin{center}
\begin{minipage}{.9\textwidth}
~\\
~\\
~\\
~\\
~\\
This part contains the original contributions of the thesis and is devoted to Bondi-Metzner-Sachs (BMS) 
symmetry in three dimensions. It starts with an introductory chapter where the definition of the BMS$_3$ 
group is motivated by asymptotic symmetry considerations. We then move on to the quantization of BMS$_3$ 
symmetry and show that irreducible unitary representations of the BMS$_3$ group, i.e.\ \it{BMS$_3$ 
particles}, are classified by supermomentum orbits that coincide with coadjoint orbits of the Virasoro group. 
We also evaluate the associated characters and show that they coincide with one-loop partition 
functions of the gravitational field at finite temperature and angular potential. Finally we extend this 
matching to higher-spin theories and supergravity in three dimensions.
\end{minipage}
\end{center}
\newpage
~
\thispagestyle{empty}

\chapter{Classical BMS$_3$ symmetry}
\label{c6}
\markboth{}{\small{\chaptername~\thechapter. Classical BMS$_3$ symmetry}}

The Bondi-Metzner-Sachs (BMS) group is an infinite-dimensional symmetry group of asymptotically flat gravity 
at null 
infinity, that extends Poincar\'e symmetry. It was originally discovered in four space-time dimensions in the 
seminal work of Bondi, 
Van der Burg, Metzner 
\cite{Bondi:1960jsa,Bondi:1962px} and Sachs \cite{Sachs:1962zza,Sachs:1962wk}. In this chapter we introduce 
BMS symmetry in three dimensions \cite{Ashtekar:1996cd} and describe its classical aspects, i.e.\ those that 
do not rely on its 
realization as the quantum symmetry group of a Hilbert space. We will show in particular that the phase 
space of asymptotically flat gravity coincides with (a hyperplane in) the coadjoint representation of the 
centrally extended BMS$_3$ group.\\

The structure is as follows. In section \ref{sebmsboco} we show how BMS$_3$ symmetry emerges from an 
asymptotic symmetry analysis. Section \ref{sebms3} is devoted to the abstract mathematical 
definition of the BMS$_3$ group and its central extension, including their adjoint and coadjoint 
representations. In section \ref{sebmscodj} we describe the phase space of three-dimensional asymptotically 
flat gravity embedded in the space of the coadjoint representation of BMS$_3$. Finally, in section 
\ref{sebmsmod} we show how BMS$_3$ symmetry can be seen as a flat limit of Virasoro symmetry.\\

This chapter is mostly based on \cite{Barnich:2014zoa,Barnich:2014kra,Barnich:2015uva}, although the first 
section follows the earlier references \cite{Barnich:2006av,Barnich:2010eb}. As usual, more specialized 
references will be cited in due time.

\section{BMS metrics in three dimensions}
\label{sebmsboco}

The purpose of this section 
is to explain how the BMS$_3$ group (and its central extension) emerges as an asymptotic symmetry of 
three-dimensional Minkowskian space-times at null infinity. In particular we describe the embedding of 
Poincar\'e transformations and the action of BMS$_3$ on the covariant phase space of the system, and observe 
the appearance of a classical central extension. We refer to section \ref{seGeTri} for some background on 
three-dimensional gravity and asymptotic symmetries in general.

\subsection{Three-dimensional Minkowski space}

\it{Minkowski space} in three dimensions is the manifold $\RR^3$ endowed with a metric whose expression in 
inertial coordinates $(t,x,y)=(x^0,x^1,x^2)$ is\i{Minkowski space}\i{inertial coordinates}
\be
ds^2=-dt^2+dx^2+dy^2
=
\eta_{\mu\nu}dx^{\mu}dx^{\nu}
\label{encomi}
\ee
where $(\eta_{\mu\nu})$ is the Minkowski metric (\ref{minky}) in $D=3$ dimensions. In general-relativistic 
terms Minkowski space-time is the maximally symmetric\i{maximally symmetric} solution of 
Einstein's equations with vanishing 
cosmological constant.\\

The isometry 
group of Minkowski space is the Poincar\'e group (\ref{piggbix}) with $D=3$: 
$\text{IO}(2,1)=\text{O}(2,1)\ltimes\RR^3$. Its elements are pairs $(f,\alpha)$ acting transitively on 
$\RR^3$,\i{Poincar\'e group}
\be
x^{\mu}\mapsto f^{\mu}{}_{\nu}x^{\nu}+\alpha^{\mu}\,,
\label{pokkary}
\ee
where $(f^{\mu}{}_{\nu})$ is a Lorentz transformation while $\alpha^{\mu}$ is a space-time translation. The 
stabilizer of the origin $x^{\mu}=0$ is the Lorentz group, confirming the obvious diffeomorphism\i{Minkowski 
space!as homogeneous space}
$\RR^3\cong
\text{IO}(2,1)/\text{O}(2,1)$.\\

While inertial coordinates are the most common in Minkowski space, a different set 
of coordinates will be more convenient for the description of BMS$_3$ symmetry. Namely, as in (\ref{bondi}), 
we define \it{retarded Bondi coordinates} $(r,\phii,u)$ by\i{Bondi coordinates}\i{retarded 
time}\i{coordinates!Bondi}
\be
r\equiv\sqrt{x^2+y^2}\,,
\qquad
e^{i\phii}
\equiv
\frac{x+iy}{r}\,,
\qquad
u\equiv t-r\,,
\label{bondibis}
\ee
whose range is $r\in[0,+\infty[\,$, $u\in\RR$, and $\phii\in\RR$ with the identification 
$\phii\sim\phii+2\pi$.
In that context the coordinate $u$ is known as \it{retarded time}.
We will also refer to the coordinates $(r,\phii,t)$ as \it{cylindrical coordinates};\i{cylindrical 
coordinates} 
they are analogous to (\ref{ss200}) in AdS$_3$. In terms of cylindrical and Bondi coordinates, 
the Minkowski metric (\ref{encomi}) reads
\be
ds^2
=
-dt^2+dr^2+r^2d\phii^2
=
-du^2-2dudr+r^2d\phii^2
\label{bominkow}
\ee
which is the three-dimensional analogue of (\ref{minkB}). Bondi coordinates are represented on the Penrose 
diagram of 
Minkowski space in fig.\ \ref{BondiC}. Note that parity\i{parity} acts as 
$\phii\mapsto-\phii$.

\subsubsection*{Killing vectors}

The Killing vector fields that generate Poincar\'e transformations (\ref{pokkary}) are simplest to write down 
in 
inertial coordinates, where they have the general form\i{Killing vector}
\be
\xi(x)
=
\big(
\alpha^{\rho}
+
X^{\mu}x^{\nu}\epsilon_{\mu\nu}{}^{\rho}
\big)\der_{\rho}\,.
\label{mikill}
\ee
Here $\alpha^{\mu}$ and $X^{\mu}$ are two arbitrary, constant vectors generating translations and Lorentz 
transformations, respectively, while $\epsilon_{\mu\nu\rho}$ is the completely antisymmetric tensor such 
that $\epsilon_{012}=1$ (indices are raised and lowered with the Minkowski metric). In particular, the 
component $\alpha^0$ is responsible for time translations, while $\alpha^1$ and $\alpha^2$ generate 
translations in the directions $x=x^1$ and $y=x^2$, respectively. The component $X^0$ is 
responsible for spatial rotations while $X^1$ and $X^2$ give rise to boosts in the directions $x^1$ and 
$x^2$, respectively.\\

For later comparison with asympotic symmetries it is convenient to rewrite the Killing vectors 
(\ref{mikill}) in Bondi coordinates (\ref{bondibis}). For pure translations we find
\be
\xi_{\text{Translation}}
=
\alpha(\phii)\der_u-\frac{\alpha'(\phii)}{r}\der_{\phii}+\alpha''(\phii)\der_r
\label{kitra}
\ee
where the function $\alpha(\phii)$ is related to the translation vector $\alpha^{\mu}$ by
\be
\alpha(\phii)=\alpha^0-\alpha^1\cos\phii-\alpha^2\sin\phii\,.
\label{afahafa}
\ee
For pure Lorentz 
transformations we similarly obtain
\be
\xi_{\text{Lorentz}}
=
\big(X(\phii)-\frac{u}{r}X''(\phii)\big)\der_{\phii}
+
uX'(\phii)\der_u
-
\big(rX'(\phii)-uX'''(\phii)\big)\der_r
\label{kilor}
\ee
where the function $X(\phii)$ is related to the boost vector $X^{\mu}$ by
\be
X(\phii)
=
X^0-X^1\cos\phii-X^2\sin\phii\,.
\label{xafahafa}
\ee
Note 
that both (\ref{kitra}) and (\ref{kilor}) depend on functions on the circle; already at this stage it is 
tempting to speculate that there exist boundary conditions such that asymptotic symmetry generators take that 
form with \it{arbitrary} functions $(X,\alpha)$ on the circle. The BMS boundary conditions below will do just 
that.\footnote{We are cheating in (\ref{kilor}), since for now there is no way 
to 
distinguish $X'''(\phii)$ from $-X'(\phii)$; the justification for this combination of derivatives 
will be provided by asymptotic symmetries.}\\

The structure of the algebra spanned by the vector fields (\ref{kitra}) and (\ref{kilor}) can be made more 
transparent by a suitable choice of basis. Thus we define the 
complexified Poincar\'e generators\i{Poincar\'e algebra}\i{Lie algebra!of Poincar\'e in 
3D}
\be
j_m\equiv
\xi_{\text{Lorentz}}\Big|_{X(\phii)=e^{im\phii}}\,,
\qquad
p_m\equiv
\xi_{\text{Translation}}\Big|_{\alpha(\phii)=e^{im\phii}}
\nn
\ee
where $m,n=-1,0,1$. The resulting Lie brackets read\i{isometry algebra}
\be
i[j_m,j_n]=(m-n)j_{m+n}\,,
\qquad
i[j_m,p_n]=(m-n)p_{m+n}\,,
\qquad
i[p_m,p_n]=0\,,
\label{yebem}
\ee
with $m,n=-1,0,1$. The Lie algebra of the BMS$_3$ group will extend these brackets by allowing arbitrary 
integer values of $m,n$, in the same way that the Witt algebra (\ref{witt}) extends $\sl$. Note that the 
structure $G\ltimes\mg$ of the Poincar\'e group (\ref{pisel}) is manifest in these relations, as the bracket 
of $j$'s with $p$'s takes exactly the same form as the bracket of $j$'s with themselves.

\subsection{Poincar\'e symmetry at null infinity}
\label{suseponull}

\subsubsection*{Null infinities and celestial circles}

In preparation for the asymptotic analysis to come, note that in Bondi coordinates the region 
$r\rightarrow+\infty$ at finite $\phii$ 
and $u$ is a cylinder spanned by coordinates $(\phii,u)$ at future null infinity.\i{null infinity} It is the 
upper cone on the boundary of the Penrose diagram of fig.\ \ref{BondiC}. This is due to our choice of 
coordinates: instead of (\ref{bondibis}) we could have defined \it{advanced} Bondi coordinates, with 
advanced time given by $v=t+r$ instead of $u=t-r$.\i{advanced Bondi coordinates} As a result 
we would have found that the 
region $r\rightarrow+\infty$ is \it{past} null infinity, but up to this difference the whole 
construction would have been the same. In this thesis we use retarded Bondi coordinates throughout, but 
it is always understood that a parallel construction exists in terms of advanced Bondi coordinates. In 
particular the 
region $r\rightarrow+\infty$ will always be \it{future} null infinity, denoted $\sI^+$.\i{I+@$\sI^+$ 
(future null infinity)}\\

Future null infinity is the region of space-time where all light rays 
eventually 
escape; similarly past null infinity is the origin of all incoming light rays. In optical terms, if we were 
living in a three-dimensional space-time, the region that we would see around us would be a circle on our 
past light-cone. As the distance from us to the circle increases, the latter approaches past null infinity. A 
similar (time-reversed) interpretation holds for future null infinity, and justifies the following 
terminology:

\paragraph{Definition.} The \it{future celestial circle}\i{celestial circle} at retarded time $u$ associated 
with the Bondi 
coordinates $(r,\phii,u)$ is the circle spanned by the coordinate $\phii$ on future null infinity, and at 
fixed time $u$.\i{celestial circle} Similarly the \it{past celestial circle} at advanced time $v$ is the 
circle at fixed time $v$ 
on past null infinity.\\

This definition is illustrated in fig.\ \ref{penrose}. From now on the words ``celestial circle'' always 
refer 
to a \it{future} celestial circle. As we shall see, BMS$_3$ symmetry 
will reformulate and generalize the action of Poincar\'e transformations on 
celestial circles, and more generally on null infinity.

\subsubsection*{Poincar\'e transformations on $\sI^+$}

Since we have rewritten Minkowski Killing vectors in Bondi coordinates, it is worth asking whether one can 
write \it{finite} Poincar\'e diffeomorphisms (\ref{pokkary})
(as opposed to infinitesimal vector fields) in Bondi 
coordinates. The answer is obviously yes, but the result is not particularly 
illuminating because the linear nature of the transformations is hidden when writing them in terms of 
$(r,\phii,u)$. Fortunately, 
Bondi coordinates are designed 
so that things simplify at null infinity; in particular it turns out that Poincar\'e transformations 
preserve the limit $r\rightarrow+\infty$ in the sense that (i) they map $r$ on a positive multiple of itself 
and (ii) they affect $\phii$ and $u$ but leave them finite. Accordingly Poincar\'e transformations are 
well-defined at null infinity and one may ask how they act on the coordinates $(\phii,u)$ spanning $\sI^+$. 
The 
procedure for finding this action is explained in greater detail in \cite{Oblak:2015qia}.\\

For definiteness we focus on the 
connected Poincar\'e group (\ref{pig}). We can use the isomorphism (\ref{isoso}) to describe Lorentz 
transformations in terms of $\SL$ matrices, the correspondence being given explicitly by (\ref{exoskeleton}). 
One then finds that a pure space-time translation $x^{\mu}\mapsto x^{\mu}+\alpha^{\mu}$ acts on null infinity 
according to\i{translation}
\be
(u,\phii)\mapsto\big(u+\alpha(\phii),\phii\big)
\qquad
\text{(translation)}
\label{utrann}
\ee
where the function $\alpha(\phii)$ is related to the components $\alpha^{\mu}$ by (\ref{afahafa}). In 
particular, pure time translations act on Bondi coordinates as $u\mapsto u+\alpha^0$, without affecting the 
other coordinates. Similarly, a Lorentz transformation specified by an $\SL$ matrix (\ref{abcd})
acts on $\sI^+$ according to\i{Lorentz group!at null infinity}
$(u,\phii)\mapsto\big(f'(\phii)u,f(\phii)\big)$
where $f(\phii)$ is a projective transformation (\ref{protophi}) of the celestial circle, with parameters 
$A,B$ given by (\ref{unhappy}). For instance, 
spatial rotations act as $\phii\mapsto\phii+\theta$, leaving all other coordinates untouched.
This is analogous to the four-dimensional 
situation described in eqs.\ (\ref{lor})-(\ref{supert}). Upon performing simultaneously a translation 
$\alpha$ and a Lorentz transformation $f$, the transformation of $(u,\phii)$ reads\i{Poincar\'e 
group!at null infinity}\i{BMS$_3$ group!at null infinity}\i{null infinity!Poincar\'e transformations}
\be
(u,\phii)\mapsto
\Big(f'(\phii)u+\alpha(f(\phii))\,,\,f(\phii)\Big)
\label{umanif}
\ee
where $f$ takes the form (\ref{protophi}) while $\alpha$ is given by (\ref{afahafa}). As we shall see below, 
the 
BMS$_3$ group acts on $\sI^+$ in the same way, except that $f(\phii)$ will be an 
arbitrary diffeomorphism of the circle and that $\alpha(\phii)$ will be an arbitrary function on the 
circle. Analogous results hold at past null infinity.\\

Note that in (\ref{umanif}) we are abusing notation slightly. Indeed, a Poin\-ca\-r\'e 
transformation is a diffeomorphism of the whole space-time (not just null infinity) and acts on all three 
coordinates $r,\phii,u$. In particular the transformation law (\ref{umanif}) only holds up to corrections of 
order $1/r$. These corrections vanish in the limit $r\rightarrow+\infty$ and leave out only the leading piece 
displayed in (\ref{umanif}), but they matter for the extension of Poincar\'e (or 
BMS) transformations from the boundary into the 
bulk.

\subsection{BMS$_3$ fall-offs and asymptotic symmetries}

We now wish to define a family of metrics on $\RR^3$ that are ``asymptotically flat'' at future null infinity 
in the sense that they approach the Minkowski metric (\ref{bominkow}) near the boundary of 
space-time. As in the AdS$_3$ case above, a good starting point is to ask what is the minimum amount of 
metrics that one wants to include; clearly, pure Minkowski space should be there, but in addition one may 
include conical deficits.\i{conical deficit} These are defined by cutting out a wedge of angular opening 
$2\pi(1-2\omega)$ 
out of the middle of space and quotienting Minkowski space-time with identifications of the 
type (\ref{s201}) in terms of cylindrical coordinates. The change of coordinates (\ref{cacott}) then 
turns the metric of (quotiented) Minkowski space into
\be
ds^2
=
(dt'-Ad\phii')^2
+
dr'^2
+
4\,\omega^2r'^2d\phii'^2,
\label{hakkaba}
\ee
which is the flat limit ($\ell\rightarrow+\infty$) of eq.\ (\ref{s202}). In these terms there are no 
identifications on $t'$, and $\phii'$ is $2\pi$-periodic. As in the AdS$_3$ case the cross-term $Adt'd\phii'$ 
suggests that $A$ is proportional to angular momentum, as will indeed be the case below. In contrast to 
AdS$_3$, however, the region 
of large $r'$ is always free of closed time-like curves since the condition for the integral 
curves of $\der_{\phii'}$ to be space-like now simply yields $r'^2>A^2/(4\omega^2)$,
without condition on the ratio of $A$ and $\omega$. This is a flat limit of the 
more stringent conditions (\ref{stringent}) encountered in AdS$_3$. We refer for instance to 
\cite{Deser:1983tn,Deser:1989cf} for a more 
thorough study of conical deficits.\\

Now suppose we wish to find boundary conditions that include
such conical deficits. If we want the asymptotic symmetry group to contain the Poincar\'e group, we are 
forced to include in the phase space all metrics obtained by performing rotations, translations and boosts of 
conical deficits. This is the same argument as in section \ref{sebohad}, where we derived 
Brown-Henneaux boundary conditions. It leads to a class of metrics with prescribed asymptotic 
behaviour at null infinity, analogous to eq.\ (\ref{suppize}) in the AdS$_3$ case. Some of the subleading 
components of the metric can then be set to zero identically as a gauge choice, which leads 
to the following definition:

\paragraph{Definition.} Let $\cM$ be a three-dimensional manifold with a pseu\-do-\-Rie\-man\-nian metric 
$ds^2$. Suppose there exist local Bondi coordinates $(r,\phii,u)$ on $\cM$, defined for $r$ larger than some 
lower limit, such that the region $r\rightarrow+\infty$ be a two-dimensional cylinder at future null 
infinity and such that the asymptotic behaviour of the metric be\i{BMS$_3$ fall-offs}\i{fall-off 
conditions!BMS$_3$}\i{BMS gauge}
\be
ds^2
\stackrel{r\rightarrow+\infty}{\sim}
\cO(1)du^2
-\big(2+\cO(1/r)\big)dudr
+r^2d\phii^2+\cO(1)dud\phii\,.
\label{asyf}
\ee
Then we say that $(\cM,ds^2)$ is \it{asymptotically flat} at future null infinity in the BMS gauge. A 
parallel construction exists at past null infinity.\\

The BMS gauge condition is the flat space analogue of the Fefferman-Graham gauge used in (\ref{s207}). We 
stress that it is truly a \it{gauge} condition in the sense of asymptotic symmetries: the diffeomorphism used 
to bring a metric from a general asymptotically flat form into the BMS gauge is trivial, as it 
does not affect the surface charges of the metric. By contrast, 
the diffeomorphisms that change the physical state of the system are generated by non-zero surface charges 
and span the asymptotic symmetry group of the system, which will turn out to be the BMS$_3$ group. From now 
on, when dealing with asymptotically flat gravity, we always restrict our
attention to metrics satisfying the BMS boundary conditions (\ref{asyf}). Note that asymptotically flat 
space-times need not be (and
generally are not) globally diffeomorphic to Minkowski space; the definition (\ref{asyf}) only requires $r$ 
to 
be larger than some
lower limiting value. Note also that there is no restriction on the sign of the fluctuating components in the 
metric (\ref{asyf}); in particular the term of order $r^0$ multiplying $du^2$ may be positive.

\subsubsection*{Asymptotic Killing vectors}

The asymptotic Killing vector fields associated with flat boundary conditions (in BMS gauge) 
are vector fields that generate diffeomorphisms which preserve the fall-off conditions (\ref{asyf}). This is 
to say that, if $g_{\mu\nu}$ is an asymptotically flat metric, its Lie derivative under such a vector field 
$\xi$ must satisfy
\be
\label{killingAsptBis}
\cL_{\xi}g_{rr}
=
\cL_{\xi}g_{r\phii}
=
\cL_{\xi}g_{\phii\phii}
=0
\ee
together with
\be
\cL_{\xi}g_{uu}=\cO(1),
\qquad
\cL_{\xi}g_{u\phii}=\cO(1),
\qquad
\cL_{\xi}g_{ur}=\cO(1/r)
\label{indico}
\ee
in terms of retarded Bondi coordinates. Here (\ref{killingAsptBis}) follows
from the fact that the components $g_{rr}=g_{r\phii}=0$ and $g_{\phii\phii}=r^2$ are fixed in the BMS gauge 
(\ref{asyf}); by contrast the components $g_{uu}$, $g_{u\phii}$ and $g_{ur}$ are allowed to fluctuate by 
terms of order $r^0$, $r^0$ and $r^{-1}$ respectively.

\paragraph{Lemma.} Let $g_{\mu\nu}$ be an asymptotically flat metric in the sense (\ref{asyf}) and let $\xi$ 
be a vector field that satisfies (\ref{killingAsptBis}) and (\ref{indico}). Then\i{BMS$_3$ 
fall-offs!asymptotic Killing vector}\i{asymptotic Killing vector!for flat space}
\be
\xi
=
X(\phii)\der_{\phii}+\big(\alpha(\phii)+uX'(\phii)\big)\der_u-rX'(\phii)\der_r
+
\text{(subleading)}
\label{fasyki}
\ee
where $X(\phii)$ and $\alpha(\phii)$ are two arbitrary (smooth) $2\pi$-periodic functions, while 
the subleading terms take the form
\be
\begin{split}
&
\left[(\alpha'+uX'')\int_r^{+\infty}\frac{dr'}{r'^2}g_{ur}\right]\der_{\phii}\\
&
+
\left[
\der_{\phii}\left((\alpha'+uX'')\int_r^{+\infty}\frac{dr'}{r'^2}g_{ur}\right)
+
\frac{1}{r^2}(\alpha'+uX'')g_{u\phii}
\right]\der_r=\\
& =
\;\frac{1}{r}(\alpha'+uX'')\der_{\phii}+\frac{1}{r}(\alpha''+uX''')\der_r+\cO(r^{-2})\,.
\end{split}
\label{subculture}
\ee
These formulas uniquely associate an asymptotic Killing vector
field $\xi$ with an asymptotically flat metric $g_{\mu\nu}$ and two functions 
$\big(X(\phii),\alpha(\phii)\big)$ on 
the celestial circle; the dependence of $\xi$ on these functions is linear.

\begin{proof}
Let $g_{\mu\nu}$ be an asymptotically flat metric (\ref{asyf}). First note that the condition 
$\cL_{\xi}g_{rr}=0$ yields $\der_r\xi^u=0$, so $\xi^u$ is $r$-independent. On the other hand the 
condition $\cL_{\xi}g_{r\phii}=0$ gives a differential equation
$\der_r\xi^{\phii}
=
-\frac{1}{r^2}g_{ru}\der_{\phii}\xi^u$,
which is solved by
\be
\xi^{\phii}
=
X(u,\phii)
+\der_{\phii}\xi^u\int_r^{+\infty}\frac{dr'}{r'^2}g_{r'u}
\label{xiphi}
\ee
where $X(u,\phii)$ is an arbitrary function on the cylinder at null infinity. The integral over 
$r'$ converges since $g_{ru}=-1+\cO(1/r)$ by virtue of (\ref{asyf}), so that
$\xi^{\phii}
=
X(u,\phii)
+\frac{1}{r}\der_{\phii}\xi^u
+\cO(1/r^2)$.
At this point we introduce a function $\alpha(u,\phii)$ defined by
\be
\xi^u
=
\alpha(u,\phii)+uX'(u,\phii)
\label{xuxuk}
\ee
(prime denotes partial differentiation with respect to $\phii$), which is allowed by virtue of the fact that 
$\xi^u$ is 
$r$-independent. In these terms the condition $\cL_{\xi}g_{\phii\phii}=0$ gives
\be
\xi^r
=
-r\der_{\phii}\xi^{\phii}-\frac{1}{r}g_{u\phii}(\alpha'+uX'')
\label{elbelga}
\ee
where $\xi^{\phii}$ is given by (\ref{xiphi}). Since we now know that the most general solution 
$\xi$ of (\ref{killingAsptBis}) is determined by two functions $X(u,\phii)$ and $\alpha(u,\phii)$ on null 
infinity, we can use the remaining conditions (\ref{indico}) to constrain these functions. Using 
first $\cL_{\xi}g_{ur}=\cO(1/r)$, we find
\be
\der_u\xi^u
=
X',
\label{oboi}
\ee
which upon rewriting $\xi^u$ as (\ref{xuxuk}) says that the combination $\der_u\alpha+u\der_uX'$ vanishes. 
The requirement $\cL_{\xi}g_{u\phii}=\cO(1)$ then yields $\der_uX=0$, which is to say that 
$X(u,\phii)=X(\phii)$ only depends on the coordinate $\phii$ on the celestial circle. Plugging this back into 
(\ref{oboi}) then yields $\der_u\alpha=0$ as well. Formula (\ref{fasyki}) follows, while the subleading terms 
(\ref{subculture}) are produced by (\ref{xiphi}) and (\ref{elbelga}).
\end{proof}

Note that the asymptotic Killing vectors (\ref{fasyki}) precisely take the anticipated form 
(\ref{kitra})-(\ref{kilor}) and generalize Poincar\'e transformations in an 
infinite-dimensional way. In particular the asymptotic symmetry group contains all space-time translations 
(corresponding to $\alpha(\phii)$ of the form (\ref{afahafa})) and all Lorentz transformations 
(corresponding to $X(\phii)$ of the form (\ref{xafahafa})). We shall denote by $\xi_{(X,\alpha)}$ the 
asymptotic Killing vector determined by the functions $X(\phii)$ 
and $\alpha(\phii)$. One verifies that the Lie brackets of such vector fields read\i{BMS$_3$ 
fall-offs!asymptotic symmetry algebra}\i{bms3 algebra@$\bms$ algebra}
\be
\left[\xi_{(X,\alpha)},\xi_{(Y,\beta)}\right]
=
\xi_{([X,Y],[X,\beta]-[Y,\alpha])}
+\text{(subleading)}
\label{dingit}
\ee
where the brackets in the subscript on the right-hand side are understood to be standard Lie brackets on the 
circle, e.g.\ $[X,\alpha]\equiv X\alpha'-\alpha X'$. The subleading terms can be neglected 
because 
they will turn out not to contribute to the surface charges; alternatively, as in the AdS$_3$ case, they can 
be absorbed by 
a redefinition of the Lie bracket such that the algebra is 
realized everywhere in the bulk \cite{Barnich:2001jy,Barnich:2010xq}.\\

The structure of the algebra 
(\ref{dingit}) can be made more transparent by decomposing the functions $\big(X(\phii),\alpha(\phii)\big)$ 
in 
Fourier modes and defining the vector
fields
\be
j_m\equiv\xi_{(e^{im\phii},0)},
\qquad
p_m\equiv\xi_{(0,e^{im\phii})}.
\label{yipika}
\ee
As one can verify, formula (\ref{dingit}) implies that their Lie brackets take the form 
(\ref{yebem}) with \it{arbitrary integer labels} $m,n$, up to subleading 
corrections.\\

Thus we now know that the asymptotic symmetries of three-dimensional Min\-kow\-skian 
space-times span an algebra that contains the Witt algebra (extending the Lorentz algebra) and an 
infinite-dimensional Abelian algebra (extending space-time 
translations).\i{superrotation}\i{supertranslation} 
The corresponding 
asymptotic 
symmetry transformations\i{asymptotic symmetry!for flat space} are referred to as \it{superrotations} and 
\it{supertranslations}, 
respectively;\footnote{The prefix ``super'' has nothing to do with supersymmetry, but stresses the fact 
that special-relativistic quantities are extended in an infinite-dimensional way.} they span an 
infinite-dimensional algebra known as 
the \it{BMS algebra in three dimensions}, 
that we shall denote as $\bms$. 
Finite BMS$_3$ transformations act on null infinity according to formula (\ref{umanif}), where $f(\phii)$ is 
an arbitrary diffeomorphism of the celestial circle while $\alpha(\phii)$ is an arbitrary function on the 
circle.\\

We refrain from analysing the group-theoretic aspects of these symmetries at this point --- this will be the 
subject of all later sections in this chapter. Instead, we now keep going in our study of asymptotically flat 
gravity; in particular we actually still have to confirm that superrotations and supertranslations 
are indeed non-trivial asymptotic symmetries, i.e.\ that the associated surface charges do not vanish.

\paragraph{Remark.} One should keep in mind that Bondi coordinates are 
\it{global},\i{Bondi 
coordinates!globality} since the 
definition (\ref{bondibis}) covers all points of Minkoswki space-time. Thus the fact that Bondi 
coordinates allow one to describe either only future or only past null infinity (and not both) does not mean 
that they cover only ``half'' of the space-time. A similar comment applies to BMS symmetry, whose definition 
in terms of space-time relies on a choice of coordinates that favours future over past null infinity (or 
vice-versa). Despite this asymmetry, it was recently realized that (for 
well-behaved 
asymptotically flat space-times \cite{Christodoulou1993}) the two definitions of BMS can be related by an 
``antipodal identification'', which leads to the application of BMS symmetry to scattering phenomena 
\cite{Strominger:2013lka,Strominger:2013jfa,He:2014laa,Cachazo:2014fwa,Kapec:2014opa,He:2014cra,Lysov:2014csa,
Strominger:2014pwa,Kapec:2014zla,Pasterski:2015tva,Kapec:2015vwa,Pasterski:2015zua,He:2015zea,Kapec:2015ena,
Strominger:2015bla, Dumitrescu:2015fej}.
A related question (as yet unsolved) is whether BMS symmetry can be defined at spatial infinity 
\cite{Henneaux2016}.

\subsection{On-shell BMS$_3$ metrics}
\label{suseshebe}

In order for the equations of motion to provide an extremum of the action functional, the latter must be 
differentiable in the space of fields satisfying certain fall-off conditions. In the case of 
asymptotically flat three-dimensional gravity, it was shown in \cite{Barnich:2013yka}, using the Chern-Simons 
formalism, that there exists a well-defined variational principle. The same conclusion was obtained more 
recently in \cite{Hartong:2015usd} in the metric formalism, with the observation that the pure 
Einstein-Hilbert action (\ref{s174}), without any extra boundary term, is differentiable in the space of 
asymptotically flat metrics.\\

Accordingly, it makes sense to ask about the general solution of Einstein's vacuum equations in the BMS 
gauge. It was shown in \cite{Barnich:2010eb} that this solution reads\i{BMS$_3$ fall-offs!on-shell 
metric}\i{on-shell metric}\i{BMS$_3$ metric}
\be
ds^2
=
8G\,p(\phii)du^2
-2dudr
+8G\big(j(\phii)+up'(\phii)\big)dud\phii
+r^2d\phii^2
\label{piment}
\ee
where $p(\phii)$ and $j(\phii)$ are arbitrary, $2\pi$-periodic functions of $\phii$. Upon evaluating 
surface
charges we will see that $p(\phii)$ and $j(\phii)$ are densities of energy and angular momentum at null 
infinity, respectively.
As in the 
earlier AdS$_3$ case (\ref{s209}), the normalization factors involving Newton's constant $G$ are included 
for later convenience.\\

The transformation law of 
the solution (\ref{piment}) under the action of asymptotic Killing vectors follows from the definition
\be
\cL_{\xi_{(X,\alpha)}}ds^2
\equiv
8G\,\delta_{(X,\alpha)}p(\phii)\,du^2
+8G\left(
\delta_{(X,\alpha)}j(\phii)+u\,\delta_{(X,\alpha)}p'(\phii)
\right)dud\phii
\label{LiDeR}
\ee
where the functions $X(\phii)$ and $\alpha(\phii)$ determine the vector field
(\ref{fasyki}). Evaluating the Lie derivative (\ref{LiDeR}) one finds
\begin{align}
\label{deltajii}
\delta_{(X,\alpha)}j
& =
Xj'+2X'j
+\alpha p'+2\alpha'p-\frac{c_2}{12}\alpha''',\\
\label{deltapii}
\delta_{(X,\alpha)}p
& =
Xp'+2X'p-\frac{c_2}{12}X'''
\end{align}
where $c_2$ is a dimensionful central charge proportional to the Planck mass 
\cite{Barnich:2006av}:\i{Planck mass}\i{c2@$c_2$ (BMS$_3$ central charge)}\i{BMS$_3$ central 
charge}\i{central charge!for BMS$_3$}\i{dimensionful central 
charge}\i{classical central extension!in flat space}
\be
\boxed{
\Big.
c_2=\frac{3}{G}\,.}
\label{god}
\ee
In this language the 
asymptotic vector field $\xi_{(X,\alpha)}$ is an \it{exact} Killing vector field for the 
metric $(j,p)$ if both variations (\ref{deltajii})-(\ref{deltapii}) vanish. The subscript 
``2'' in (\ref{god}) will be justified below.\\

The transformation law of $p$ in (\ref{deltapii}) coincides with that of a CFT stress tensor under a 
conformal transformation generated by $X$; it is the coadjoint representation 
(\ref{covinf}) of the Virasoro algebra. The transformation (\ref{deltajii}) of $j$ is 
somewhat more involved. We refrain from interpreting these results for now, as we will return to them in 
much 
greater detail in the upcoming sections. Note that at this stage all normalizations are arbitrary, 
and in particular the central charge (\ref{god}) would take another value if we chose to change the 
normalization 
of $p$.

\subsection{Surface charges and BMS$_3$ algebra}

\subsubsection*{Surface charges}

Take an asymptotic Killing vector field (\ref{fasyki}) specified by the functions 
$\big(X(\phii),\alpha(\phii)\big)$, 
and choose an on-shell metric (\ref{piment}) specified by $\big(j(\phii),p(\phii)\big)$. We wish to 
evaluate the 
surface charge associated with the symmetry transformation generated by $\xi_{(X,\alpha)}$ on the background 
specified by $(j,p)$. This charge depends linearly on the components of $\xi_{(X,\alpha)}$, as 
explained around eq.\ (\ref{qusurf}). In addition we must choose a normalization, that is, a 
``background'' solution for which we declare that all surface charges vanish. Here we take it to be the null 
orbifold at $j=p=0$,\i{degenerate conical 
deficit}\i{null orbifold}\i{orbifold}
\be
\bar g
=
-2dudr
+r^2d\phii^2.
\label{nullorbif}
\ee
With this normalization 
one can show that the surface charge (\ref{qusurf}) associated with the vector field 
$\xi_{(X,\alpha)}$ on the solution $(j,p)$ is\i{BMS$_3$ surface charge}\i{BMS$_3$ fall-offs!surface 
charge}\i{surface charge!in flat space} \cite{Barnich:2010eb}
\be
Q_{(X,\alpha)}[j,p]
=
\frac{1}{2\pi}\int_0^{2\pi}d\phii\big[j(\phii)X(\phii)+p(\phii)\alpha(\phii)\big].
\label{bokka}
\ee
It can be interpreted as the pairing 
of the $\bms$ algebra, consisting of pairs $(X,\alpha)$, with its dual consisting of pairs $(j,p)$. In 
particular, even though we haven't defined the BMS$_3$ group at this stage, we already know that the space of 
solutions (\ref{piment}) belongs to its coadjoint representation. The charge 
associated 
with time translations correponds to the 
asymptotic Killing vector $\der_u$; it is the Hamiltonian of the system,\i{BMS$_3$ 
fall-offs!Hamiltonian}\i{Hamiltonian!in flat space}\i{mass!in flat space}
\be
M
=
\cP_0
=
\frac{1}{2\pi}\int_0^{2\pi}d\phii\, p(\phii)\,,
\label{pooh}
\ee
and it allows us to interpret $p(\phii)$ as the energy density carried by the gravitational field at (future) 
null infinity. Thus $p(\phii)$ is the \it{Bondi mass aspect} associated with 
the metric 
(\ref{piment}) and its zero-mode (\ref{pooh}) is the Bondi mass.\i{Bondi mass (aspect)} More generally the 
charges associated with supertranslations ($X=0$) take the 
form\i{surface charge!for supertranslation}
\be
Q_{(0,\alpha)}[j,p]
=
\frac{1}{2\pi}\int_0^{2\pi}d\phii\,p(\phii)\alpha(\phii)\,.
\label{supertramp}
\ee
In the same way, the charge 
associated with rotations corresponds to the asymptotic Killing vector $\der_{\phii}$; it is the 
angular momentum\i{BMS$_3$ fall-offs!angular momentum}\i{angular momentum!in flat space}
\be
J
=
\cJ_0
=
\frac{1}{2\pi}\int_0^{2\pi}d\phii\, j(\phii)\,.
\label{jooh}
\ee
We can interpret $j(\phii)$ as the density of angular momentum carried by the gravitational 
field at null infinity;\i{angular momentum aspect} it is the \it{angular momentum aspect} associated with 
the metric (\ref{piment}). More generally 
all superrotation charges take the form\i{surface charge!for superrotation}
\be
Q_{(X,0)}[j,p]
=
\frac{1}{2\pi}\int_0^{2\pi}d\phii\,j(\phii)X(\phii)
\nn
\ee
and generalize centre of mass charges. With this normalization Minkowski space (\ref{bominkow}) has energy 
$M=-1/8G$ and all its other surface charges vanish.

\subsubsection*{Surface charge algebra}

We now compute the Poisson brackets of surface charges for asymptotically flat 
space-times. Recall that these brackets generate symmetry transformations 
(\ref{dekuphi}), on account of the fact that conserved charges are momentum maps (\ref{xixi}). We can apply 
this property here 
to deduce the Poisson brackets of charges: if we let $(j,p)$ be an on-shell metric 
(\ref{piment}), then the bracket of charges is
\be
\begin{split}
& 
\left\{
Q_{(X,\alpha)}[j,p],Q_{(Y,\beta)}[j,p]
\right\}=\qquad\qquad\qquad\\
& \refeq{bokka}
-\frac{1}{2\pi}\int_0^{2\pi}d\phii
\left[
\delta_{(X,\alpha)}j(\phii) Y(\phii)
+
\delta_{(X,\alpha)}p(\phii)\beta(\phii)
\right].
\end{split}
\label{antikro}
\ee
Using the infinitesimal transformation laws (\ref{deltajii})-(\ref{deltapii}) and integrating by parts one 
can then show that\i{surface charge!algebra}\i{bms3 algebra@$\bms$ 
algebra!centrally extended}\i{classical central extension!in flat space}\i{BMS$_3$ fall-offs!surface charge 
algebra}
\be
\left\{
Q_{(X,\alpha)}[j,p],Q_{(Y,\beta)}[j,p]
\right\}
=
Q_{([X,Y],[X,\beta]-[Y,\alpha])}[j,p]
+
c_2\left[\sfc(X,\beta)-\sfc(Y,\alpha)\right],
\label{bobrak}
\ee
where as in (\ref{dingit}) we denote by $[X,Y]\equiv XY'-YX'$ the standard Lie 
bracket of vector fields on the circle, while $\sfc(X,Y)$ is the Gelfand-Fuks cocycle (\ref{gefuks}). 
Thus, 
the surface charges of asymptotically flat space-times close under the Poisson bracket according to a central 
extension of the BMS$_3$ Lie algebra displayed in (\ref{dingit}). Furthermore the central extension is 
remarkably similar to that of the Virasoro algebra (\ref{vibraphone}). Again, we refrain from interpreting 
this result any further at this point, since we haven't truly defined the BMS$_3$ group yet. For future 
reference we simply note that the Poisson brackets (\ref{bobrak}) can be rewritten in terms of a discrete set 
of generators analogous to (\ref{yipika}). Namely, let us define the charges
\be
\cJ_m\equiv Q_{(e^{im\phii},0)}[j,p],
\qquad
\cP_m\equiv Q_{(0,e^{im\phii})}[j,p]
\nn
\ee
for all $m\in\ZZ$,
generalizing the Hamiltonian (\ref{pooh}) and angular momentum (\ref{jooh}). Then the bra\-cket 
(\ref{bobrak}) 
yields the algebra
\begin{align}
i\{\cJ_m,\cJ_n\} & = (m-n)\cJ_{m+n}\,,\nn\\
\label{surall}
i\{\cJ_m,\cP_n\} & = (m-n)\cP_{m+n}+\frac{c_2}{12}m^3\delta_{m+n,0}\,,\\
i\{\cP_m,\cP_n\} & = 0\,.\nn
\end{align}
This is an infinite-dimensional central extension of (\ref{yebem}), with $m,n\in\ZZ$.\\

Note that the central extension proportional to $c_2$ in (\ref{surall})
pairs superrotation generators $\cJ_m$ with supertranslation generators $\cP_m$. By 
contrast the Witt algebra spanned by 
superrotations receives no central extension. This is why we wrote the central charge (\ref{god}) with an 
index ``$2$'': the notation $c_1$ will be kept for the central charge pairing superrotation generators with 
themselves. Despite many similarities,
we stress that
$c_2$ is \it{not} a 
Virasoro central charge;\i{c2@$c_2$ (BMS$_3$ central charge)!not a Virasoro central charge} in particular it 
is a dimensionful 
quantity. This is consistent with the fact that 
the value of $c_2$ varies when changing the normalization of the charges $\cP_m$: if we 
were to define $\tilde\cP_m\equiv\lambda\cP_m$ with some non-zero real number $\lambda$, the Poisson brackets 
of $\cJ$'s and $\tilde\cP$'s would take the form (\ref{surall}) with the central charge $c_2$ replaced by 
$\lambda c_2$. Nevertheless, the value displayed in (\ref{god}) is canonical in the sense that it is the one 
provided by the normalization of the Hamiltonian (\ref{pooh}), which in turn is the surface charge associated 
with the vector field $\der_u$ in terms of Bondi coordinates.\i{BMS$_3$ central charge!as mass 
scale}\i{c2@$c_2$ (BMS$_3$ central charge)!as mass scale} In 
essence the central charge $c_2$ is 
analogous to that of the Bargmann group (\ref{galix}), which as we saw in (\ref{gallimm}) is also a mass 
scale. This is radically different from 
the Virasoro algebra, where the value of the central charge $c$ in (\ref{virapois}) is unambiguously fixed by 
the condition that the homogeneous structure constants take the form
$(m-n)$.

\paragraph{Remark.} The BMS boundary conditions given here are the flat analogue of 
Brown-Henneaux boundary conditions. In this sense they are the ``standard'' fall-offs for 
three-dimensional asymptotically flat gravity. However, it is likely that other consistent boundary 
conditions exist in Einstein gravity --- for instance adapting to flat space the free AdS$_3$ boundary 
conditions of \cite{Troessaert:2013fma}. In addition one can devise BMS-like boundary conditions for other 
theories of gravity, such as\i{topologically massive gravity} topologically massive gravity 
\cite{Bagchi:2012yk,Grumiller:2015xaa},\i{bigravity} bigravity \cite{Gonzalez:2012nv}, conformal gravity 
\cite{Afshar:2013bla} or new massive gravity\i{new massive gravity} 
\cite{Barnich:2015dvt,Troessaert:2015syk}. In particular, in parity-breaking theories such as 
{\textsc{TMG}}, one typically finds that the Virasoro algebra spanned by superrotations $\cJ_m$ 
develops a 
non-zero central charge $c_1$. Aside from this comment we will have very little to say about 
these alternative possibilities.

\subsection{Zero-mode solutions}

We focus here on zero-mode metrics, with constant $(j,p)=(j_0,p_0)$ in eq.\ 
(\ref{piment}). The only non-vanishing surface charges for such metrics are the Bondi mass (\ref{pooh}) and 
the 
angular momentum (\ref{jooh}), which coincide with $p_0$ and $j_0$ respectively.\i{zero-mode 
metric}\i{BMS$_3$ metric!zero-mode}\\

At $j_0=0$, $p_0=-c_2/24\refeq{god}-1/8G$, the metric is that of pure Minkowski space-time 
(\ref{bominkow}). Solutions 
having $p_0=-c_2/24$ but non-zero $j_0$ corresponding to ``spinning Minkowski space-time''. Note that, while 
the 
normalization of $p_0$ and $c_2$ is arbitrary, the relation
\be
p_{\text{vac}}=-\frac{c_2}{24}
\nn
\ee
is a normalization-independent statement.\footnote{Indeed, changing the normalization of $p$ would also 
change the value of the central charge that ensures that the bracket $\{\cJ,\cP\}$ takes the canonical form 
in 
eq.\ (\ref{surall}).} It suggests that Minkowski space plays the 
role of a classical vacuum for a putative dual theory; we will return to this later.\\

Solutions having $0>p_0>-c_2/24$ are conical deficits for all values of $j_0$, with a deficit angle 
$2\pi(1-2\omega)$ given by (\ref{hofrik}).\i{conical deficit} In 
particular, solutions with $p_0=0$ are 
degenerate conical 
deficits, and the solution $p_0=j_0=0$ is the null orbifold (\ref{nullorbif})\i{null orbifold} that we used 
to normalize 
charges. Solutions having $p_0<-c_2/24$ are conical excesses with an excess angle $2\pi(2\omega-1)$ given 
again by (\ref{hofrik}). For 
$p_0=-c_2n^2/24$ 
the excess angle is $2\pi(n-1)$.\\

Zero-mode solutions with positive $p$ turn out to describe flat space cosmologies, sometimes also called 
shifted boost orbifolds \cite{Cornalba:2002fi,Cornalba:2003kd}. They represent a $(2+1)$-dimensional universe 
that undergoes a big crunch followed by a big bang, where the transition between the contracting and 
expanding phases is smooth only if $j\neq0$.\i{flat space cosmology} When $j=0$ these solutions can be 
thought of 
as a compactification of the three-dimensional Milne universe.\i{Milne universe} They can also be seen as 
limits of the 
interior region of BTZ black holes as the AdS$_3$ radius goes to infinity. The lightest flat space cosmology 
has $p_0=0$ and is separated from Minkowski space-time $p_{\text{vac}}=-c_2/24$ by a classical mass gap; the 
latter is filled by conical deficits. This is very similar to the mass gap separating BTZ black holes from 
AdS$_3$.\\

\begin{figure}[t]
\centering
\includegraphics[width=0.60\textwidth]{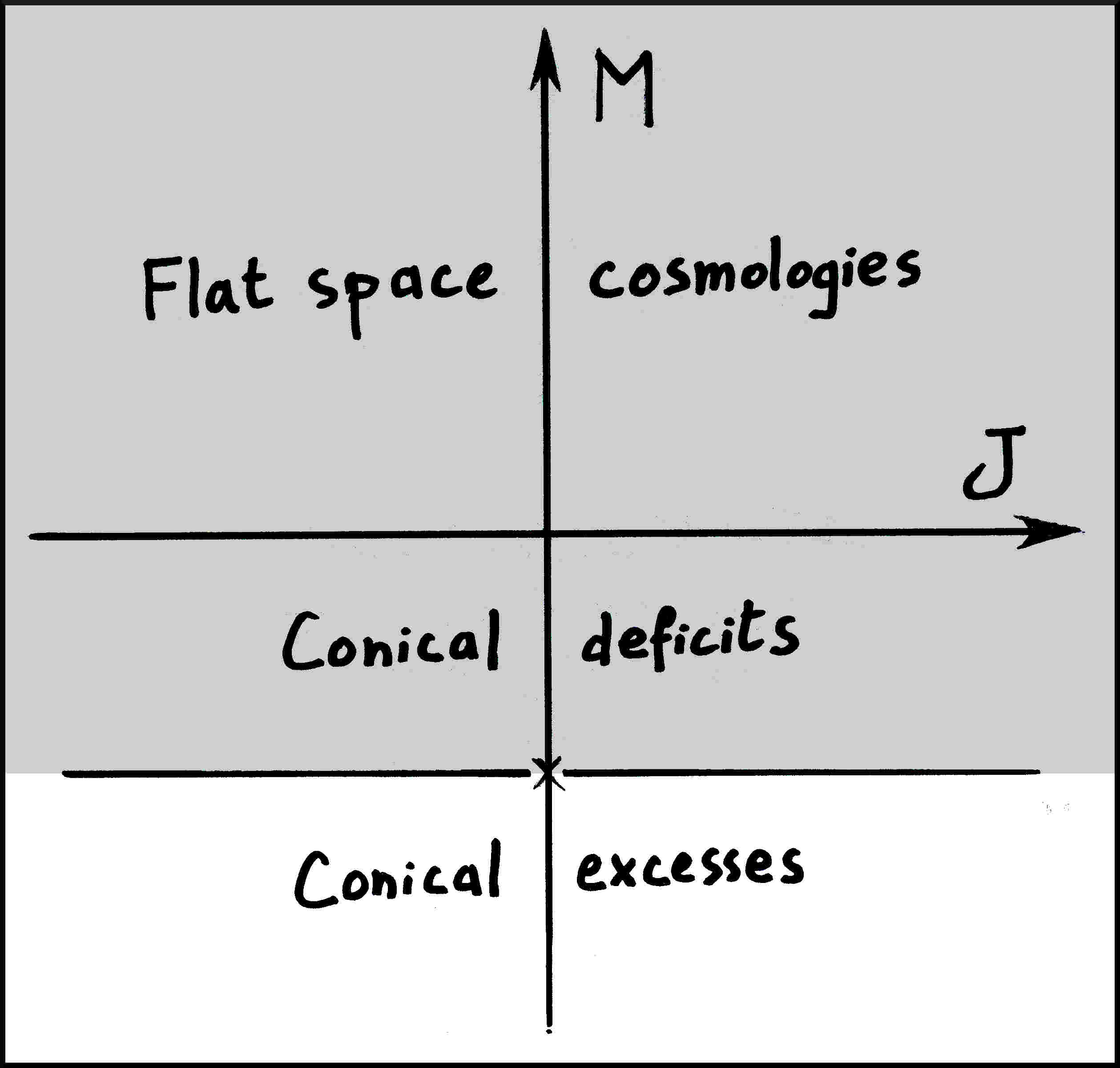}
\caption{The zero-mode solutions of asymptotically flat gravity with BMS$_3$ fall-offs. The origin 
of the coordinate system $(J,M)$ is the null orbifold (\ref{nullorbif}); the Minkowski metric is located 
below, on the $M$ axis, right between conical deficits and conical excesses. Flat space cosmologies are 
located in the region $M>0$. Conical deficits are such that $-c_2/24<M<0$ while excesses have $M<-c_2/24$. 
Anticipating section \ref{susePOUPOU}, we have shaded the solutions whose orbit has 
energy bounded 
from below under BMS$_3$ transformations; those are all flat space cosmologies and all conical excesses, plus 
Minkowski space. Note that this figure is a flat limit of fig.\ \ref{figs229}, as the 
slope of the curve $\ell M=J$ in the plane $(J,M)$ goes to zero when $\ell\rightarrow+\infty$. 
\label{figazerofla}}
\end{figure}

Note that, in contrast to AdS$_3$,  no 
cosmic censorship\i{cosmic censorship} is needed to ensure the absence of closed time-like curves at infinity 
(although closed 
time-like curves generally do exist in the bulk). In fact, the whole 
classification of flat zero-mode metrics may be seen as a limit $\ell\rightarrow+\infty$ of that of zero-mode 
metrics in AdS$_3$. The family of flat zero-mode solutions is plotted in 
fig.\ \ref{figazerofla}.

\section{The BMS$_3$ group}
\label{sebms3}

This section is devoted to a detailed description of the BMS$_3$ group and its central extension. This will 
rely on a level of abstraction that may seem offputting at first sight, but one should keep in mind 
that BMS$_3$ is an extension of Poincar\'e symmetry so that almost all statements on BMS have an analogue in 
special relativity. We urge the reader to adopt this point of view whenever there is a risk of getting lost 
in 
mathematical formulas. In particular, our notation will be consistent with the analogies between Poincar\'e 
and BMS$_3$:
\begin{table}[H]
\centering
\begin{tabular}{|c|c|c|}
\hline
Notation & Poincar\'e & BMS$_3$\\
\hline
$f$ & finite Lorentz tsf. & finite superrotation\\
$X$ & infinitesimal Lorentz tsf. & infinitesimal superrotation\\
$\alpha$ & translation & supertranslation\\
$j_m,\,\cJ_m,\,J_m$ & Lorentz generator & superrotation generator\\
$p_m,\,\cP_m,\,P_m$ & translation generator & supertranslation generator\\
$j$ & relativistic angular momentum & angular supermomentum\\
$p$ & energy-momentum & supermomentum\\
$\cZ_1,\,c_1$ & $/$ & superrotational central charge\\
$\cZ_2,\,c_2$ & $/$ & supertranslational central charge\\
\hline
\end{tabular}
\caption{Analogies between Poincar\'e and BMS$_3$.}
\end{table}

The plan of this section is as follows. Motivated by the structure of the Poincar\'e group (\ref{pisel}), we 
start by defining a notion of ``exceptional semi-direct products'' (generally centrally extended) and work 
out their adjoint and coadjoint representations. We then use asymptotic symmetries to motivate the definition 
of the BMS$_3$ group and its central extension, which turn out to be exceptional semi-direct products based 
on the Virasoro group. Finally, we write down the adjoint representation, the Lie algebra and the coadjoint 
representation of the (centrally extended) BMS$_3$ group. Throughout the section, these structures are 
compared to their Poincar\'e counterparts and to three-dimensional asymptotically flat gravity. Note that the 
material presented here relies heavily on chapters \ref{c1}, \ref{c2bis} and \ref{c4}.

\subsection{Exceptional semi-direct products}
\label{susexepto}

Here we study a general family of semi-direct products, whose structure turns out to be common to the 
Poincar\'e group (in three dimensions) and the BMS$_3$ group. We start by describing this structure and its 
central extension, then display the corresponding adjoint and coadjoint representations.

\subsubsection*{Defining exceptional semi-direct products}

\paragraph{Definition.} Let $G$ be a Lie group with Lie algebra $\mg$. The associated \it{exceptional 
semi-direct product} is the group\i{exceptional semi-direct product/sum}\i{semi-direct product!exceptional}
\be
G\ltimes_{\Ad}\mg_{\text{Ab}}
\equiv
G\ltimes\mg
\label{exCeS}
\ee
where $\mg_{\text{Ab}}$ denotes the Lie algebra of $G$ seen as an Abelian vector group acted upon by $G$ 
according to the adjoint representation. Its group operation is given by (\ref{semiop}) with the action 
$\sigma$ replaced by the adjoint.\\

As usual we denote elements of (\ref{exCeS}) as pairs $(f,\alpha)$ where $f\in G$ is a ``rotation'' while 
$\alpha\in A$ is a ``translation''. For instance the (double cover of the) Poincar\'e group (\ref{pisel}) 
takes the exceptional 
form with $G=\SL$. It is straightforward to obtain central extensions of this structure: if $\hG$ is a 
central extension of $G$ with group 
operation (\ref{groupop}) in terms of some two-cocycle $\sfC$ and if $\hmg$ is its Lie 
algebra, one can consider the exceptional semi-direct product
\be
\hG\ltimes_{\hAd}\hmg_{\text{Ab}}
\label{gigg}
\ee
where $\hAd$ denotes the adjoint representation (\ref{advir}) of $\hG$. Its elements are 
quadruples
\be
(f,\lambda;\alpha,\mu)
\label{falafel}
\ee
where $\lambda,\mu$ are real numbers, being understood that the pair 
$(f,\lambda)$ belongs to $\hG$ while $(\alpha,\mu)$ belongs to $\hmg_{\text{Ab}}$. The notation emphasizes 
the 
fact that 
``centrally extended rotations'' $(f,\lambda)$ play a role radically different from ``centrally extended 
translations'' $(\alpha,\mu)$. In fact the notation 
$\big((f,\lambda),(\alpha,\mu)\big)$ would be more accurate, but to reduce clutter we stick to 
(\ref{falafel}).\\

The group operation in (\ref{gigg}) is that of an exceptional semi-direct product based on the centrally 
extended group $\hG$. Explicitly, using the centrally extended adjoint representation (\ref{advir}), we have
\be
\begin{split}
& (f,\lambda;\alpha,\mu)\cdot(g,\rho;\beta,\nu)=\\
& \refeq{advir}
\Big(
f\cdot g,\lambda+\rho+\sfC(f,g)
\,;\,
\alpha+\Ad_f\beta,
\mu+\nu-\frac{1}{12}\bra\sfS[f],\beta\ket
\Big)
\end{split}
\label{bimop}
\ee
where $\sfC$ is the two-cocycle that defines $\hG$, $\sfS$ is the associated Souriau one-cocycle 
(\ref{souriau}), 
and 
the pairing $\bra\cdot,\cdot\ket$ is that of $\mg^*$ with $\mg$.
This is exactly the structure that we will find in the centrally extended BMS$_3$ group below, but for 
now we
first investigate the adjoint and coadjoint representations of (\ref{gigg}) in general terms.

\subsubsection*{Adjoint representation and Lie algebra}

Consider a centrally extended exceptional semi-direct product $\hG\ltimes\hmg$. Owing to the general form 
(\ref{galila}), its Lie algebra is a 
semi-direct sum\i{semi-direct sum!exceptional}
\be
\hmg\inplus_{\had}\hmg_{\text{Ab}}
\label{lalalak}
\ee
where $\had$ is the adjoint representation of $\hmg$, i.e.\ the Lie bracket (\ref{kellebel}). The elements of 
this algebra are quadruples $(X,\lambda;\alpha,\mu)$ where $(X,\lambda)$ 
belongs to $\hmg$ while $(\alpha,\mu)$ belongs to $\hmg_{\text{Ab}}$.\\

The adjoint representation of the group (\ref{gigg}) follows from formula (\ref{adsemi}). Starting 
for simplicity with the centreless group (\ref{exCeS}), it is given by\i{adjoint representation!of semi-direct 
product}
\be
\Ad_{(f,\alpha)}(X,\beta)
=
\big(
\Ad_fX,\Ad_f\beta-\ad_{\Ad_fX}\alpha
\big)
=
\Big(
\Ad_fX,\Ad_f\beta-[\Ad_fX,\alpha]
\Big)
\label{fodagott}
\ee
where the ``$\Ad$'' on the right denotes the adjoint representation of $G$ alone.\footnote{In case of 
identical notations, the subscript indicates which group we are referring to.} In the second 
equality we abuse notation by writing a bracket between $\Ad_fX\in\mg$ and $\alpha\in\mg_{\text{Ab}}$, being 
understood that we use the Lie bracket of $\mg$ and interpret the result 
as an element of $\mg_{\text{Ab}}$. The Lie bracket of the centreless Lie algebra 
$\mg\inplus_{\ad}\mg_{\text{Ab}}$ follows:\i{Lie bracket!for semi-direct sum}
\be
\big[(X,\alpha),(Y,\beta)\big]
=
\big([X,Y],\ad_X\beta-\ad_Y\alpha\big)
=
\big([X,Y],[X,\beta]-[Y,\alpha]\big)\,,
\label{fobavitt}
\ee
in accordance with the general formula (\ref{semibra}). Note that this is precisely the form of the Lie 
bracket (\ref{dingit}) of BMS$_3$ asymptotic Killing vectors.\\

The centrally extended adjoint representation corresponding to (\ref{fodagott}) can be ob\-tained in a 
similar fashion. 
Using 
(\ref{advir}) and (\ref{kellebel}) we find explicitly
\be
\begin{split}
& \hAd_{(f,\alpha)}(X,\lambda;\beta,\mu)=\qquad\qquad\qquad\qquad\qquad\\
&=
\Big(
\Ad_fX,\lambda-\frac{1}{12}\bra \sfS[f],X\ket;
\Ad_f\beta-[\Ad_fX,\alpha],
\mu-\frac{1}{12}\bra \sfS[f],\beta\ket+\frac{1}{12}\bra\sfs[\Ad_fX],\alpha\ket
\Big)
\end{split}
\label{haddasim}
\ee
where $\Ad$ on the right-hand side denotes the adjoint representation of $G$ and $[\cdot,\cdot]$ is the Lie 
bracket of $\mg$. On the left-hand side we have neglected central terms in the subscript of the adjoint 
representation, since they act trivially.\\

From the adjoint representation one can read off, by differentiation, the Lie bracket of the 
centrally extended algebra (\ref{lalalak}). One 
expects the contribution of central terms to include a cocycle $\sfc$ given by (\ref{essaimon}), and indeed 
one finds
\be
\big[
(X,\lambda;\alpha,\mu),(Y,\rho;\beta,\nu)
\big]
=
\Big(
[X,Y],\sfc(X,Y);
[X,\beta]-[Y,\alpha],
\sfc(X,\beta)-\sfc(Y,\alpha)
\Big)
\label{wisig}
\ee
where we abuse notation as in (\ref{fobavitt}). Already note that the last entry precisely takes the form of 
the central extension in the Poisson bracket (\ref{bobrak}) of flat surface charges.\\

The appearance of the same cocycle $\sfc$ in both central 
entries of (\ref{wisig}) is due to the exceptional semi-direct product structure of (\ref{gigg}). 
It implies that, when written in terms of generators, the brackets of rotations with translations take the 
same form as the brackets of rotations with themselves, 
including central terms. Explicitly, suppose we are given a basis of $\hmg\inplus\hmg_{\text{Ab}}$ consisting 
of non-central generators
\be
\cJ_a\equiv(j_a,0;0,0)\,,
\qquad
\cP_a\equiv(0,0;p_a,0)
\nn
\ee
where the $j_a$'s and $p_a$'s respectively generate $\mg$ and $\mg_{\text{Ab}}$,
together with two central elements
\be
\cZ_1\equiv(0,1;0,0)\,,
\qquad
\cZ_2\equiv(0,0;0,1)\,.
\label{zaza}
\ee
Suppose also that the Lie brackets of $\cJ_a$'s take the form (\ref{tatb}) with some structure constants 
$f_{ab}{}^c$ and some central coefficients $c_{ab}$, and let us choose the basis elements $\cP_a$ such that 
their bracket with $\cJ_a$'s takes the same form as the bracket of $\cJ_a$'s with themselves. 
This is allowed by the exceptional semi-direct product structure.
Then the bracket (\ref{wisig}) 
implies 
that the commutation relations of $\hmg\inplus\hmg_{\text{Ab}}$ are
\begin{align}
{}[\cJ_a,\cJ_b] & = {f_{ab}}^c\,\cJ_c+c_{ab}\,\cZ_1\,,\nn\\
\label{jappah}
{}[\cJ_a,\cP_b] & = {f_{ab}}^c\,\cP_c+c_{ab}\,\cZ_2\,,\\
{}[\cP_a,\cP_b] & = 0\,.\nn
\end{align}
The fact that $\cJ$'s act on $\cP$'s according to the adjoint representation is now manifest since the 
structure 
constants of the two first lines are identical. Note in particular that the 
central generator $\cZ_1$ pairs 
rotations with themselves, while 
$\cZ_2$ pairs rotations with translations. The centrally extended BMS$_3$ algebra (\ref{surall}) illustrates 
this phenomenon, as does the Poincar\'e algebra (\ref{yebem}), albeit without central extension.\\

Note that the definition of (\ref{gigg}) rules out all central extensions in the bracket 
$[\cP,\cP]$ of (\ref{jappah}), and indeed we will show in section \ref{suseRecca} that such central 
extensions never take place in the centrally extended BMS$_3$ algebra. However, for other semi-direct 
product groups, such 
extensions 
may occur; an example is the symmetry group of warped conformal field theories\i{warped CFT} 
\cite{Detournay:2012pc}, $\Diff\ltimes C^{\infty}(S^1)$.

\subsubsection*{Coadjoint representation}

The space of coadjoint vectors dual to the algebra (\ref{lalalak}) is a direct sum $\hmg^*\oplus\hmg^*$, or 
more accurately $\hmg^*\oplus\hmg^*_{\text{Ab}}$. Following the notation of section \ref{secosemi}, its 
elements are quadruples\i{coadjoint vector!for semi-direct product}
\be
(j,c_1;p,c_2)
\label{heckoda}
\ee
where $(j,c_1)$ is a centrally extended angular momentum dual to $\hmg$, while $(p,c_2)$ is a centrally 
extended momentum dual to $\hmg_{\text{Ab}}$. The real numbers $c_1,c_2$ are central charges; the first 
pairs rotation generators with themselves, while the second pairs rotations with translations. The pairing of
(\ref{heckoda}) with $\hmg\inplus\hmg_{\text{Ab}}$ is
\be
\big<
(j,c_1;p,c_2),(X,\lambda;\alpha,\mu)
\big>
=
\bra j,X\ket+\bra p,\alpha\ket+c_1\lambda+c_2\mu\,,
\label{JAPAN}
\ee
where the two pairings $\langle\cdot,\cdot\rangle$ on the right-hand side are those of $\mg^*$ with $\mg$ 
and $\mg^*_{\text{Ab}}$ with $\mg_{\text{Ab}}$, respectively. 
This is a centrally extended 
generalization of (\ref{papa}).\\

Recall that the coadjoint representation of a semi-direct product involves a cross product (\ref{copo}). For 
the centreless exceptional semi-direct product (\ref{exCeS}), we have\i{cross product}
\be
\bra\alpha\times p,X\ket
\refeq{copo}
\bra p,\ad_X\alpha\ket
=
-\bra p,\ad_{\alpha}X\ket
\refeq{pixies}
\bra\ad^*_\alpha p,X\ket
\nn
\ee
where $\ad$ and $\ad^*$ denote the adjoint and coadjoint representations of $\mg$, respectively. In other 
words,
\be
\alpha\times p=\ad^*_{\alpha}\,p
\label{execross}
\ee
where we abuse notation slightly by acting with an element of $\mg_{\text{Ab}}$ on an element of 
$\mg^*_{\text{Ab}}$. Using (\ref{acoga}), it readily follows that the coadjoint representation of the 
centreless semi-direct product (\ref{exCeS}) is given by
\be
\Ad^*_{(f,\alpha)}(j,p)
=
\left(
\Ad^*_fj,+\ad^*_{\alpha}\Ad^*_f p,\Ad^*_fp
\right)
\label{opokitu}
\ee
where the $\Ad^*$ on the right-hand side is the coadjoint representation of $G$. For example, when $G=\SL$, 
this formula is the transformation law of relativistic angular momentum $j$ and energy-momentum $p$ under 
Poincar\'e transformations in three dimensions. From (\ref{opokitu}) we also find that the coadjoint 
representation of the Lie algebra $\mg\inplus\mg_{\text{Ab}}$ is
\be
\ad^*_{(X,\alpha)}(j,p)
=
\big(
\ad^*_Xj+\ad^*_{\alpha}p,\ad^*_Xp
\big)
\label{apokitu}
\ee
in accordance with eq.\ (\ref{oogutak}).\\

The centrally extended generalization of these considerations is straightforward, if mildly technical. Using 
eq.\ 
(\ref{covirzero}) for the coadjoint action of $\hG$, formula (\ref{opokitu}) yields the coadjoint 
representation of $\hG\ltimes\hmg_{\text{Ab}}$:
\be
\begin{split}
& \hAd^*_{(f,\alpha)}(j,c_1;p,c_2)=\qquad\qquad\qquad\\
& =
\left(
\Ad^*_fj-\frac{c_1}{12}\sfS[f^{-1}]
+
\ad^*_{\alpha}\left[\Ad^*_fp-\frac{c_2}{12}\sfS[f^{-1}]\right]
+
\frac{c_2}{12}\sfs[\alpha],
c_1;
\Ad^*_fp-\frac{c_2}{12}\sfS[f^{-1}],
c_2
\right)\,.
\end{split}
\label{digotaz}
\ee
Here it is understood that all $\Ad^*$'s and $\ad^*$'s on 
the right-hand side are centre\it{less} --- they are the coadjoint representations of $G$ and $\mg$, 
respectively.\\

Formula (\ref{digotaz}) looks a bit scary but it is crucial for our purposes, so let us briefly point out 
two of its important features. First, the central charges $c_1,c_2$ are left invariant by the 
action of the group, as 
expected. Second, note that the transformation law 
of momentum is
\be
f\cdot p=\Ad^*_fp-\frac{c_2}{12}\sfS[f^{-1}]\,,
\label{fipp}
\ee
where the $\Ad^*$ on the right-hand side is that of $G$ (not $\hG$). This formula says that $p$ is 
invariant 
under translations (since it is unaffected by $\alpha$) and that its transformation law is blind to the 
central charge $c_1$, but \it{not} to $c_2$. In fact, eq.\ (\ref{fipp}) is the coadjoint 
representation (\ref{covirzero}) of the centrally extended group $\hG$ at central 
charge $c_2$. As a corollary we can already conclude that the orbits of momenta labelling unitary 
representations of (\ref{gigg}) are coadjoint orbits of the group $\hG$ at fixed central charge $c_2$; there 
is no need to master the much more complicated transformation law of angular momentum in 
(\ref{digotaz}) in order to classify such representations. This will have key consequences 
for the BMS$_3$ group below.

\paragraph{Remark.} Property (\ref{execross}) explains why we refer to the map (\ref{copo}) as a \it{cross 
product}.\i{cross product} Indeed, the (double cover of the) Euclidean group in three dimensions is an 
exceptional semi-direct 
product 
$\text{SU}(2)\ltimes_{\text{\Ad}}\mathfrak{su}(2)_{\text{Ab}}$. Since the coadjoint representation of 
$\text{SU}(2)$ is 
equivalent to the adjoint, one may identify vectors with covectors and the cross product 
(\ref{execross}) for the Euclidean group can be rewritten as $\alpha\times 
p=\ad_{\alpha}p=[\alpha,p]$. Here the Lie bracket is that of $\mathfrak{su}(2)$, so in components one 
has $(\alpha\times p)_i=\epsilon_{ijk}\alpha^jp^k$, which is the standard definition of the cross 
product in mechanics.

\subsection{Defining BMS$_3$}
\label{suseHaB}

Now that we are acquainted with exceptional semi-direct products, let us show how this structure occurs in 
three-dimensional BMS symmetry.

\subsubsection*{Centreless BMS$_3$ group}

Our first task is to move backwards from the centreless BMS$_3$ algebra (\ref{dingit}) to the corresponding 
group. The algebra consists of pairs $\big(X(\phii),\alpha(\phii)\big)$, where 
$X(\phii)\der_{\phii}$ is a vector field on the circle while $\alpha(\phii)$ is a priori just a function on 
the celestial circle. These two quantities were referred to above as infinitesimal superrotations and 
supertranslations, respectively. Together, they generate finite transformations 
(\ref{umanif}) of the cylinder at null infinity, where $f(\phii)$ is a diffeomorphism of the circle. Thus we 
already know that the BMS$_3$ group consists of pairs $(f,\alpha)$, where $f$ is a diffeomorphism of the 
circle while $\alpha$ is a function. It only remains to work out the group operation; the latter is given by 
the composition of two transformations (\ref{umanif}):
\begin{align}
(u,\phii)
& \stackrel{(g,\beta)}{\longmapsto}
\big(g'(\phii)u+\beta(g(\phii)),g(\phii)\big)\nn\\
& \stackrel{(f,\alpha)}{\longmapsto}
\Big(f'(g(\phii))\big[g'(\phii)u+\beta(g(\phii))\big]+\alpha\big(f(g(\phii))\big),f(g(\phii))\Big)\,.\nn
\end{align}
Here the last result on the right-hand side can be rewritten as
\be
\Big((f\circ g)'(\phii)u+[\alpha+\Ad_f\beta]\Big|_{(f\circ g)(\phii)},(f\circ g)(\phii)\Big)
\label{exyour}
\ee
where $\Ad_f\beta$\i{adjoint representation!of $\Diff$}\i{DiffS1@$\Diff$!adjoint representation} denotes the 
adjoint representation (\ref{advecis}) of $\Diff$ acting on $\beta$, that is, 
the transformation law of a vector field $\beta(\phii)\der_{\phii}$ under $f(\phii)$:
\be
(\Ad_f\beta)\big|_{f(\phii)}=f'(\phii)\beta(\phii).
\label{ADABA}
\ee
Expression 
(\ref{exyour}) indicates three things:
\begin{enumerate}
\item The group operation of superrotations is given by composition 
(\ref{compo}); hence finite (as opposed to infinitesimal) 
superrotations\i{superrotation}\i{DiffS1@$\Diff$!superrotation} span a group $\Diff$.\footnote{As in 
chapter \ref{c4} we describe diffeomorphisms of the circle by their 
lifts belonging to the universal cover $\Diffc$, which we abusively denote simply as $\Diff$.}
\item If it wasn't for superrotations, the group operation of supertranslations would just 
be addition, $\alpha\cdot\beta\equiv\alpha+\beta$. Thus supertranslations span an Abelian additive group 
whose elements are certain functions on the circle.
\item The action of superrotations on supertranslations is that of diffeomorphisms on vector fields, 
i.e.\ it is the adjoint representation (\ref{ADABA}) of $\Diff$. In particular, supertranslations, which so 
far we thought of as functions $\alpha(\phii)$ on the circle, should better be seen as vector fields 
$\alpha(\phii)\der_{\phii}$. The only subtlety is that these vector fields do \it{not} generate 
diffeomorphisms of celestial circles, but rather angle-dependent translations (\ref{utrann}) of retarded time 
$u$. Equivalently, each supertranslation is a density $\alpha=\alpha(\phii)(d\phii)^{-1}$ on the circle.
\end{enumerate}
These observations motivate the following definition:

\paragraph{Definition.} The centreless \it{BMS group} in three dimensions is the exceptional semi-direct 
product\i{BMS$_3$ group}
\be
\boxed{\Big.\text{BMS}_3
\equiv
\Diff\ltimes_{\Ad}\Vect_{\text{Ab}}}
\label{defbms}
\ee
where $\Diff$ is the group of diffeomorphisms of the circle while $\Vect_{\text{Ab}}$ is its Lie algebra, 
seen 
as an Abelian vector group acted upon by $\Diff$ according to the adjoint representation. Its elements are 
pairs $(f,\alpha)$; its group operation follows from the general definition (\ref{semiop}) and is given by
\be
(f,\alpha)\cdot(g,\beta)
=
\big(
f\circ g,\alpha+\Ad_f\beta
\big)
\label{guixec}
\ee
where $\Ad$ is the action (\ref{ADABA}) of $\Diff$ on vector fields. With this 
definition the action (\ref{umanif}) of BMS$_3$ on null infinity reproduces the group operation 
(\ref{guixec}).\\

The BMS$_3$ group is 
infinite-dimensional and has the announced form (\ref{exCeS}), with $G=\Diff$. Since we saw in section 
\ref{sedifici} that $\PSL$ is 
a subgroup of $\Diff$, the Poincar\'e group is obviously a subgroup of BMS$_3$.
We therefore introduce officially the following terminology:

\paragraph{Definition.} In the BMS$_3$ group (\ref{defbms}), elements of $\Diff$ are known as 
\it{superrotations}\i{superrotation} while elements of $\Vect_{\text{Ab}}$ are called 
\it{supertranslations}.\i{supertranslation}

\paragraph{Remark.} The name ``superrotation'' has come to be standard, but the geometric interpretation of 
$\Diff$ makes the terminology ``superboosts'' somewhat more appropriate.\i{superboost} Indeed, recall from 
section 
\ref{sedifici} that the group $\Diff$ is homotopic to a circle, so that the only 
superrotations spanning a compact group are those conjugate to rigid rotations 
$f(\phii)=\phii+\theta$. The other one-parameter subgroups of $\Diff$ are all non-compact and 
should be 
seen as boost groups.

\subsubsection*{Universal cover of BMS$_3$}

As in section \ref{sedifici} we should be careful about what we mean by $\Diff$. Strictly speaking, 
$\Diff$ consists of all diffeomorphisms of the circle with the composition law (\ref{compogo}); its connected 
subgroup $\Diffp$ consists of orientation-preserving diffeomorphisms. Since the group of supertranslations is 
a vector space, it is also connected and we define the \it{connected} 
BMS$_3$ group as\i{connected BMS$_3$ group}\i{BMS$_3$ group!connected}
\be
\text{BMS}_3^+
\equiv
\Diffp\ltimes_{\Ad}\Vect_{\text{Ab}}\,.
\label{BaConn}
\ee
If we think of the group of superrotations as an extension of the Lorentz group in three 
dimensions, then $\Diff$ corresponds to the disconnected orthochronous Lorentz group 
$\text{O}(2,1)^{\uparrow}$ while $\Diffp$ corresponds to the connected (orthochronous \it{and} proper) 
Lorentz 
group $\text{SO}(2,1)^{\uparrow}$. It appears that no $\Diff$ transformation corresponds to 
time reversal (which sounds reasonable since BMS symmetry is defined separately at future and past null 
infinity).\\

The group $\Diffp$ of orientation-preserving superrotations is homotopic to a 
circle, so it admits topological projective transformations that 
can be dealt with by trading it for its 
universal cover, $\Diffc$. Since the vector group of supertranslations is homotopic to a point, the BMS$_3$ 
group has the homotopy type of a circle.

\paragraph{Definition.} The \it{universal cover of the BMS group} in three dimensions is the 
exceptional semi-direct 
product\i{universal cover!of BMS$_3$ group}\i{BMS$_3$ group!universal cover}
\be
\widetilde{\text{BMS}}{}_3^+
\equiv
\Diffc\ltimes_{\Ad}\Vect_{\text{Ab}}
\label{debmscov}
\ee
where $\Diffc$ is the universal cover of the connected group $\Diffp$ and consists of $2\pi\ZZ$-equivariant 
superrotations (\ref{fidif}).\\

In particular, exact representations of $\widetilde{\text{BMS}}{}_3^+$ generally correspond to projective 
representations of BMS$_3^+$. The groups BMS$_3$, BMS$_3^+$ 
and $\widetilde{\text{BMS}}{}_3^+$ are well-defined infinite-dimensional Lie-Fr\'echet 
groups.\i{Lie-Fr\'echet group}\i{infinite-dimensional group} In what 
follows, motivated by quantum-mechanical applications, we always focus (implicitly) on the universal 
cover. 
Accordingly we abuse notation and denote the universal cover simply 
by BMS$_3$, neglecting the superscript ``$+$'' and the tilde.

\subsubsection*{Centrally extended BMS$_3$ group}

In order to define the central extension of BMS$_3$, we apply the prescription 
(\ref{gigg}) for centrally extended exceptional semi-direct products to the case $G=\Diff$:

\paragraph{Definition.} The \it{centrally extended BMS group} in three dimensions\i{BMS$_3$ 
group!central extension}\i{central extension!of BMS$_3$ group} is the 
exceptional semi-direct product
\be
\boxed{\Big.
\hBMS
\equiv
\hDiff\ltimes_{\hAd}\hVect}
\label{hibiscus}
\ee
where $\hDiff$ is the (universal cover of the) Virasoro group.\\

Since this thesis is concerned with the group $\hBMS$, let us make its definition a bit more explicit before 
going further. The elements of $\hBMS$ are quadruples $\big(f,\lambda;\alpha,\mu\big)$ 
where $f$ is a superrotation, $\alpha$ a supertranslation, while $\lambda,\mu$ are real numbers, extending 
Poincar\'e transformations as before.
In $\hBMS$, centrally extended superrotations $(f,\lambda)$ span a Virasoro group while 
extended 
supertranslations\i{supertranslation} $(\alpha,\mu)$ span an infinite-dimensional Abelian group acted upon by 
superrotations 
according to the Virasoro adjoint representation. Explicitly, the group operation in 
$\hBMS$ 
takes the form (\ref{bimop}) where $f\cdot g=f\circ g$, while
$\sfC$ is the Bott-Thurston cocycle (\ref{btcoll}) and $\sfS$ is the 
Schwarzian 
derivative 
(\ref{swag}). The pairing $\bra\cdot,\cdot\ket$ is that of $\Vect$ 
with its dual, given by (\ref{denpar}).\\

The centreless BMS$_3$ group (\ref{defbms}) is perfect, in the same way as $\Diff$; this implies that it 
admits a universal central extension. As it turns out, this is precisely achieved by $\hBMS$ (see section 
\ref{suseRecca} for the proof):

\paragraph{Theorem.} The centrally extended BMS$_3$ group (\ref{hibiscus}) is the 
universal central extension of the centreless BMS$_3$ group (\ref{defbms}).

\subsection{Adjoint representation and $\mathfrak{bms}_3$ algebra}

\subsubsection*{Lie algebra}

Since BMS$_3$ is an exceptional semi-direct product, its centreless Lie algebra takes the form 
$\mg\inplus\mg_{\text{Ab}}$, where $\mg$ is the Lie algebra of $\Diff$:\i{bms3 algebra@$\bms$ algebra}
\be
\bms
=
\Vect\inplus_{\ad}\Vect_{\text{Ab}}\,.
\label{qabomit}
\ee
Its elements are pairs $(X,\alpha)$ where $X=X(\phii)\der_{\phii}$ is an infinitesimal superrotation and 
$\alpha=\alpha(\phii)(d\phii)^{-1}$ an infinitesimal supertranslation. These functions determine 
the components of vector fields (\ref{fasyki}) generating asymptotic symmetries, so that elements of $\bms$ 
can be seen as infinitesimal BMS$_3$ 
transformations. In particular the Poincar\'e subalgebra of $\bms$ consists of 
pairs $(X,\alpha)$ whose only non-vanishing Fourier modes are the 
three 
lowest ones, as in (\ref{afahafa})-(\ref{xafahafa}). The centrally extended generalization (\ref{lalalak}) 
of this definition is immediate:

\paragraph{Definition.} The Lie algebra of $\hBMS$ is an exceptional semi-direct sum
\be
\hbms\equiv\hVect\inplus_{\had}\hVect{}_{\text{Ab}}\,.
\label{haBOP}
\ee
Its elements are quadruples $(X,\lambda;\alpha,\mu)$ where $X=X(\phii)\der_{\phii}$ is an infinitesimal 
superrotation, $\alpha=\alpha(\phii)(d\phii)^{-1}$ an infinitesimal supertranslation, while $\lambda,\mu$ are 
real numbers.

\subsubsection*{Adjoint representation}

The adjoint representation of the centreless BMS$_3$ group is given by formula (\ref{fodagott}), where the 
adjoint
action of $\Diff$ is the transformation law of vector fields (\ref{advec}). An important subtlety is that the 
Lie bracket appearing on the right-hand side is that of the Lie algebra of $\Diff$ and is therefore the 
\it{opposite} (\ref{bakopo}) of the standard bracket of vector fields. Accordingly, in terms of the usual 
Lie bracket of vector fields on the circle one would write 
the adjoint representation of $\BMS$ as
\be
\Ad_{(f,\alpha)}(X,\beta)
=
\big(
\Ad_fX,\Ad_f\beta+[\Ad_fX,\alpha]
\big)\,,
\label{bmsad}
\ee
with a plus sign instead of a minus sign in the second entry of (\ref{fodagott}). The centrally extended 
generalization of that expression is provided by eq.\ (\ref{haddasim}), where $\sfS$ is the 
Schwarzian derivative (\ref{swag}), $\sfs$ is its infinitesimal version (\ref{insa}), and
$\bra\cdot,\cdot\ket$ is the standard pairing (\ref{denpar}). Again, when writing the adjoint representation 
in terms of the standard Lie bracket of vector fields, the sign in front of the bracket of the third entry of 
(\ref{haddasim}) is a plus instead of a minus. Since we will not explicitly need the adjoint representation 
of the $\hBMS$ group, we do not display it here.

\subsubsection*{Lie brackets}

From the adjoint representation one can read off the Lie bracket of the $\hbms$ algebra. In order to 
absorb the minus sign of (\ref{bakopo}) we \it{define} the bracket to be\i{bms3 algebra@$\bms$ 
algebra!centrally extended}\i{central extension!of $\bms$ algebra}
\be
\big[(X,\lambda;\alpha,\mu),(Y,\rho;\beta,\nu)\big]
\equiv
-\frac{d}{dt}\left.
\hAd{}_{(e^{tX},t\alpha)}(Y,\rho;\beta,\nu)
\right|_{t=0}\,.
\nn
\ee
With this definition the Lie 
bracket in $\hbms$ 
takes the form (\ref{wisig}) where the brackets on the right-hand side are standard Lie brackets of vector 
fields while $\sfc$ is the Gelfand-Fuks cocycle (\ref{gefuks}). This is consistent with the algebra of 
surface charges (\ref{bobrak}).\\

The Lie algebra structure can be made more apparent by writing the bracket (\ref{wisig}) in a suitable basis. 
As in (\ref{yipika}) we define the complex superrotation and supertranslation generators of the centreless 
$\bms$ algebra,
\be
j_m\equiv\big(e^{im\phii}\der_{\phii},0\big)\,,
\qquad
p_m\equiv\big(0,e^{im\phii}(d\phii)^{-1}\big)\,,
\label{jim}
\ee
where the index $m$ runs over all integers. Their brackets take the form (\ref{yebem}). The 
corresponding basis of the centrally extended $\hbms$ algebra is
\be
\begin{split}
\cJ_m &\equiv \big(j_m,0;0,0\big)\refeq{jim}\big(e^{im\phii}\der_{\phii},0;0,0\big)\,,\\
\qquad
\cP_m &\equiv \big(0,0;p_m,0\big)\refeq{jim}\big(0,0;e^{im\phii}(d\phii)^{-1},0\big)\,,
\end{split}
\label{JimmA}
\ee
together with two central elements (\ref{zaza}), i.e.\ $\cZ_1=(0,1;0,0)$ and $\cZ_2=(0,0;0,1)$. In these 
terms 
the centrally extended bracket (\ref{wisig}) yields
\begin{align}
i[\cJ_m,\cJ_n] & = (m-n)\cJ_{m+n}+\frac{\cZ_1}{12}m^3\delta_{m+n,0}\,,\nn\\
\label{bammex}
i[\cJ_m,\cP_n] & = (m-n)\cP_{m+n}+\frac{\cZ_2}{12}m^3\delta_{m+n,0}\,,\\
i[\cP_m,\cP_n] & = 0\,.\nn
\end{align}
Up to central terms this is of the same form as the asymptotic symmetry algebra (\ref{yebem}), and it is 
consistent with the general form 
(\ref{jappah}) for centrally extended exceptional semi-direct products. In the first line we see that 
superrotations close according to a Virasoro algebra (\ref{vira}) with central generator $\cZ_1$, while the 
second 
line shows that brackets of superrotations with supertranslations take the Virasoro form with a different 
central element $\cZ_2$. The algebra of surface charges (\ref{surall}) takes that form, with 
definite values $c_1=0$, $c_2=3/G$ for the central generators $\cZ_1$, $\cZ_2$.

\paragraph{Remark.} The canonical Poincar\'e subgroup of BMS$_3$ is the one spanned by superrotations 
(\ref{protophi}) and supertranslations (\ref{afahafa}), or equivalently the one generated by basis elements 
$j_m,p_m$ with $m=-1,0,1$. But in fact, BMS$_3$ admits infinitely many other Poincar\'e 
subgroups:\i{BMS$_3$ group!Poincar\'e subgroups}\i{Poincar\'e group!embedded in BMS$_3$} each 
of them has a Lie algebra spanned by $j_n,j_0,j_{-n}$ and $p_n,p_0,p_{-n}$, 
consisting of 
superrotations of the form (\ref{protophin}) and supertranslations
\be
\alpha(\phii)
=
\alpha^0-\alpha^1\cos(n\phii)-\alpha^2\sin(n\phii)
\nn
\ee
whose only non-vanishing Fourier modes are the zero-mode and the $n^{\text{th}}$ modes.

\subsection{Coadjoint representation}
\label{susecobomis}

\subsubsection*{Angular and linear supermomentum}

The coadjoint vectors of $\hBMS$ are quadruples $\big(j,c_1;p,c_2\big)$ where 
$j=j(\phii)d\phii^2$ and $p=p(\phii)d\phii^2$ are quadratic densities on the circle, respectively dual to 
infinitesimal superrotations and supertranslations. The coefficients $c_1$ and $c_2$ are central charges. The 
pairing of $(j,c_1;p,c_2)$ with the Lie algebra $\hbms$ is given by formula (\ref{JAPAN}), or 
explicitly\i{BMS$_3$ group!coadjoint vector}\i{coadjoint vector!for BMS$_3$}
\be
\big<
(j,c_1;p,c_2),(X,\lambda,\alpha,\mu)
\big>
=
\frac{1}{2\pi}\int_0^{2\pi}d\phii\big[j(\phii)X(\phii)+p(\phii)\alpha(\phii)\big]
+c_1\lambda+c_2\mu\,.
\nn
\ee
The right-hand side of this expression coincides (up to central terms) with the surface charge (\ref{bokka}).
Inspired by the terminology of 
superrotations and supertranslations, we introduce the following nomenclature:

\paragraph{Definition.} Let $(j,p)$ be a coadjoint vector for the BMS$_3$ group. Then $p=p(\phii)d\phii^2$ is 
called a \it{supermomentum}\i{supermomentum} while $j=j(\phii)d\phii^2$ is an \it{angular 
supermomentum}\i{angular supermomentum}\i{supermomentum!angular}.\\

The embedding of the Poincar\'e algebra in $\bms$ suggests an interpretation for the lowest 
Fourier modes of
\be
j(\phii)=\sum_{m\in\ZZ}j_me^{-im\phii}
\qquad\text{and}\qquad
p(\phii)=\sum_{m\in\ZZ}p_me^{-im\phii}.
\label{jipifou}
\ee
Indeed, $p_0$ is dual to time translations and should be interpreted as the energy associated with 
$p(\phii)$; similarly, the components $p_1$ and $p_{-1}=p_1^*$ are complex linear combinations of the spatial 
components of momentum. As for $j_0$, it is the angular momentum associated with $j(\phii)$, while $j_1$ and 
$j_{-1}$ are centre of mass charges. More 
generally, the function $p(\phii)$ should be seen as an energy density on the circle --- essentially a stress 
tensor --- 
while $j(\phii)$ is an angular momentum density on the circle. In particular, it is natural to give 
dimensions of energy to the function $p(\phii)$ and the central charge $c_2$, while the function $j(\phii)$ 
and the central charge $c_1$ are dimensionless. This 
interpretation is confirmed by the surface charges (\ref{bokka}), since $p(\phii)$ is a
Bondi mass aspect while $j(\phii)$ is an angular momentum 
aspect; furthermore the central charge (\ref{god}) is indeed a mass scale.

\paragraph{Remark.} To our knowledge the terminology of ``supermomentum'' for duals of supertranslations 
dates back to \cite{McCarthy489}, and has subsequently been used throughout the BMS literature (see 
e.g.\ \cite{Ashtekar:1981bq,Moreschi:1988pc,Iyer:1988fm,Helfer1995,Flanagan:2015pxa}). The terminology of 
``angular supermomentum'', on the other hand, seems to have first appeared in \cite{Garecki00,Garecki1999} in 
relation to the problem of angular momentum, but apparently without direct relation to BMS symmetry. It was 
later independently introduced in \cite{Barnich:2015uva} in the BMS context. In \cite{Flanagan:2015pxa}, 
angular supermomentum is referred to as ``super centre of mass''.\i{super centre of mass}

\subsubsection*{Coadjoint representation}

As in the 
Virasoro case, one should 
think of the pair $(j,p)$ as the stress tensor of a BMS$_3$-invariant theory;\i{BMS$_3$ stress 
tensor}\i{stress tensor!for BMS$_3$-invariant theories}\i{BMS$_3$ metric!as stress tensor} its 
transformations under 
BMS$_3$ then coincide with the coadjoint representation, given for centrally extended exceptional semi-direct 
products by formula (\ref{digotaz}). In that expression, the central charges are invariant (as they 
should) while the transformation law of supermomentum coincides with the coadjoint representation 
(\ref{covi}) of the Virasoro group at central charge $c_2$:\i{coadjoint representation!of BMS$_3$}
\be
\big(f\cdot p\big)\big(f(\phii)\big)
=
\frac{1}{\big(f'(\phii)\big)^2}
\left[
p(\phii)+\frac{c_2}{12}\sfS[f](\phii)
\right],
\label{ATOM}
\ee
where $\sfS$ denotes the Schwarzian derivative (\ref{swag}). We stress once more that supermomentum is left 
invariant by supertranslations, as it should.\\

The transformation law of angular supermomentum is a bit more involved and translates the fact that $j$ is 
sensitive both to superrotations and to supertranslations, as it should since angular momentum and centre of 
mass charges are always defined with respect to an arbitrarily chosen origin. We refrain from describing this 
transformation law any further at this point, as we shall return to it in section \ref{susePaCCo} when 
showing that the phase space of metrics (\ref{piment}) is a hyperplane at central charges $c_1=0$, $c_2=3/G$ 
embedded in the coadjoint representation of $\hBMS$.

\subsubsection*{Kirillov-Kostant bracket}

A prerequisite for showing that the asymptotically flat phase space is a coadjoint representation is to 
understand the Kirillov-Kostant Poisson bracket of the asymptotic symmetry group. Let us do this here for 
$\hBMS$; we proceed as in section \ref{sevigo}. 
Thus let $\big\{\cJ_m^*,\cP_m^*,\cZ_1^*,\cZ_2^*\big\}$ be the dual basis corresponding to 
(\ref{JimmA}) and (\ref{zaza}). Writing any coadjoint vector as
\be
\big(
j(\phii)d\phii^2,c_1;p(\phii)d\phii^2,c_2
\big)
=
\sum_{m\in\ZZ}\big(j_m\cJ_m^*+p_m\cP_m^*\big)
+c_1\cZ_1^*+c_2\cZ_2^*\,,
\nn
\ee
the components $\{j_m,p_m,c_1,c_2\}$ are global coordinates on the dual space $\hbms{}^*$. Their Poisson 
brackets (\ref{fabipe}) take the form\i{Kirillov-Kostant bracket!for BMS$_3$}\i{BMS$_3$ 
group!Kirillov-Kostant bracket}
\begin{align}
i\{j_m,j_n\} & = (m-n)j_{m+n}+\frac{c_1}{12}m^3\delta_{m+n,0}\,,\nn\\
\label{bimipois}
i\{j_m,p_n\} & = (m-n)p_{m+n}+\frac{c_2}{12}m^3\delta_{m+n,0}\,,\\
i\{p_m,p_n\} & = 0\,.\nn
\end{align}
As is obvious here, $c_1$ is a genuine Virasoro 
central charge for superrotations, while $c_2$ is the (generally dimensionful) central charge pairing 
superrotations with supertranslations. The surface charges of asymptotically flat gravity 
satisfy the exact same algebra (\ref{surall}), with the values of central charges $c_1=0$, $c_2=3/G$. We 
will return to this in section \ref{sebmscodj}.

\begin{advanced}
\subsection{Some cohomology}
\label{suseRecca}
\end{advanced}

To conclude our abstract description of BMS$_3$ symmetry, we now show that the centrally extended group 
(\ref{hibiscus}) is in fact the universal central extension of 
the BMS$_3$ group (\ref{defbms}). (In both cases $\Diff$ is understood to denote the universal cover of the 
group of orientation-preserving diffeomorphisms of the circle.) We use the notation of 
section \ref{s1.2}. Since the proof is very similar to the construction of the Gelfand-Fuks cocycle 
(\ref{gefuks}), this section may be skipped in a first reading.\\

The $\bms$ algebra is perfect: it is equal to its Lie bracket with itself.\i{perfect Lie 
algebra}\i{bms3 algebra@$\bms$ algebra!perfect}\i{bms3 algebra@$\bms$ algebra!cohomology} This can be 
seen, for instance, by 
noting that the right-hand sides of the brackets (\ref{yebem}) span all possible $\bms$ generators. 
Accordingly it follows from (\ref{hiperfecto}) that the first real cohomology of $\bms$ vanishes: $\cH^1(\bms)
=
0$. Since the same is true of the centreless BMS$_3$ group, its central extension is universal, and it only
remains to establish the second cohomology of $\bms$.\i{cohomology!of bms3@of $\bms$}

\paragraph{Theorem.} The second real cohomology space of $\bms$ is two-dimensional. It is 
generated by the classes of the two-cocycles
\be
\sfc_1\big((X,\alpha),(Y,\beta)\big)
=
\sfc(X,Y)
\quad\text{and}\quad
\sfc_2\big((X,\alpha),(Y,\beta)\big)
=
\sfc(X,\beta)-\sfc(Y,\alpha)
\label{bemuks}
\ee
where $\sfc$ is the Gelfand-Fuks cocycle (\ref{gefuks}). Their expression in the basis 
(\ref{jim}) is
\be
\sfc_1(j_m,j_n)
=
\sfc_2(j_m,p_n)
=
-i\frac{m^3}{12}\delta_{m+n,0}\,,
\label{gifabom}
\ee
while their other components vanish. As a consequence, the Lie algebra (\ref{haBOP}) is the universal central 
extension of (\ref{qabomit}), 
and the group (\ref{hibiscus}) is the universal central extension of (\ref{defbms}).

\begin{proof}
The fact that the cocycle 
$\sfc_1$ is the only non-trivial cocycle pairing superrotation 
generators 
with themselves follows from the fact that infinitesimal superrotations span a Witt subalgebra of 
$\bms$. The considerations of section \ref{sedicomo} then carry over directly to $\bms$. Now let us ask 
whether there exists a two-cocycle $\sfc$ such 
that $\sfc(p_m,p_n)\neq0$.
The cocycle identity (\ref{2coc}) with trivial $\sT$ implies
\be
\sfc(j_0,[p_m,p_n])
+
\tilde \sfc(p_m,[p_n,j_0])
+
\tilde \sfc(p_n,[j_0,p_m])
=
(m+n) \tilde \sfc(p_m,p_n)
\stackrel{!}{=}
0\,,
\nn
\ee
where we used the Lie brackets (\ref{yebem}). This yields
$\tilde \sfc(p_m,p_n)
=
\tilde c_m\delta_{m+n,0}$
where the coefficients $\tilde c_m=-\tilde c_{-m}$ are to be determined. We now 
attempt to find a recursion relation for these coefficients; using the cocycle identity
\be
\sfc(p_{-1},[j_{-m+1},p_m])
+
\sfc(j_{-m+1},[p_m,p_{-1}])
+
\sfc(p_m,[p_{-1},j_{-m+1}])
=
0\,,
\nn
\ee
the $\bms$ algebra (\ref{yebem}) implies $(2m-1)\tilde c_1+(m-2)\tilde c_m
=0$.
Since this must be true for all integer values of $m$ we conclude that $\tilde c_1=0$, which 
in turn implies $\tilde c_m=0$ for all $m\in\ZZ$. Thus, any two-cocycle on the $\bms$ algebra has vanishing 
components $\tilde \sfc(p_m,p_n)=0$. Finally, suppose that $\sfc$ is a two-cocycle on the $\bms$ 
algebra and let us ask whether one can have $\sfc(j_m,p_n)\neq0$.
As in (\ref{clm}) we start by ensuring that the cocycle $\sfc$ is rotation-invariant by adding to it a 
suitable coboundary. Consider therefore the cocycle relation
\be
\sfc(j_0,[j_m,p_n])
=
\sfc([j_0,j_m],p_n)+\sfc(j_m,[j_0,p_n])\,.
\nn
\ee
The left-hand side can be interpreted as the differential of the one-cochain $\sfk=\sfc(j_0,\cdot)$ 
evaluated at $(j_m,p_n)$, while the right-hand side is the Lie derivative of $\sfc$ with respect to 
$j_0$. Since the left-hand side is exact we know that the cohomology class of 
$\sfc$ is left invariant by rotations; in particular we can add to $\sfc$ the differential 
$\sfd\sfb$ 
of the one-cochain
\be
\sfb(j_m)\equiv0\,,
\quad
\sfb(p_m)\equiv\frac{i}{m}\sfc(j_0,p_m)\,,
\nn
\ee
which is such that
\be
\cL_{j_0}(\sfc+\sfd\sfb)(j_m,p_n)=0\,.
\label{dhst}
\ee
From now on we simply write $\sfc$ to denote $\sfc+\sfd\sfb$.
Then, analogously to (\ref{cinvit}), eq.\ (\ref{dhst}) implies
$(m+n)\sfc(j_m,p_n)=0$
by virtue of the brackets (\ref{yebem}). In particular we can now write
$\sfc(j_m,p_n)
=
c_m\delta_{m+n,0}$
and it only remains to find the coefficients $c_m$. For this we derive a recursion relation using the cocycle 
identity
\be
\sfc(p_1,[j_{-m-1},j_m])
+
\sfc(j_{-m-1},[j_m,p_1])
+
\sfc(j_m,[p_1,j_{-m-1}])
=
0\,,
\nn
\ee
which implies
\be
(2m+1)c_{-1}
+(m-1)c_{-m-1}
+(m+2)c_m
=
0
\label{ted}
\ee
by virtue of the $\bms$ algebra (\ref{yebem}). In particular we have 
$c_0=0$ and $c_1=-c_{-1}$, which then gives $c_m=-c_{-m}$ and the recursion relation 
(\ref{ted}) can be rewritten as
\be
c_{m+1}
=
\frac{(m+2)c_m-(2m+1)c_1}{m-1}\,.
\nn
\ee
This is the same relation as in the Virasoro case, eq.\ (\ref{corecu}). In particular it is solved by 
$c_m=m$ and $c_m=m^3$, the former being a coboundary. The result (\ref{gifabom}) follows.
\end{proof}

\section{The BMS$_3$ phase space}
\label{sebmscodj}

As explained in section \ref{sebitu}, the space of solutions of a Hamiltonian system 
coincides with its phase space.  Accordingly the 
on-shell metrics (\ref{piment}) span the phase space of asymptotically flat gravity in three 
dimensions. Here we show that this space is a hyperplane at fixed central charges $c_1=0$, $c_2=3/G$ 
embedded in the coadjoint representation of the $\hBMS$ group. We also discuss this result from 
a holographic perspective and derive a positive energy theorem.

\subsection{Phase space as a coadjoint representation}
\label{susePaCCo}

The space of on-shell metrics (\ref{piment}) is spanned by pairs $(j,p)$ transforming under BMS$_3$ according 
to (\ref{deltajii})-(\ref{deltapii}). We now show that these formulas coincide with the coadjoint 
representation of the $\hbms$ algebra at central charges $c_1=0$, $c_2=3/G$. This is trivially true for the 
transformation law of $p(\phii)$ since (\ref{deltapii}) coincides with the coadjoint 
representation (\ref{covinf}) of the Virasoro algebra, which in turn is the infinitesimal version of the 
transformation law (\ref{ATOM}). As pointed out in section \ref{susecobomis}, the case 
of the angular momentum aspect is more intricate since its coadjoint transformation law is the first entry on 
the right-hand side of (\ref{digotaz}).\i{coadjoint representation!in gravity}\i{phase space!as coadjoint 
representation}\i{BMS$_3$ fall-offs!phase space} When applied to BMS$_3$, the 
latter formula must be modified slightly 
to match our conventions for $\Diff$. Namely, due to the minus sign of the Lie bracket (\ref{bakopo}), the 
$\ad^*$ of eq.\ (\ref{digotaz}) should be replaced by $-\ad^*$. Taking this subtlety into account, the 
transformation law of angular supermomentum is
\be
(f,\alpha)\cdot j
=
\Ad^*_fj
-\frac{c_1}{12}\sfS[f^{-1}]
-\ad^*_{\alpha}\left[\Ad^*_fp-\frac{c_2}{12}\sfS[f^{-1}]\right]
+\frac{c_2}{12}\sfs[\alpha]\,.
\label{SHOCK}
\ee
Here $\Ad^*$ denotes the coadjoint representation (\ref{coadif}) of $\Diff$, $\sfS$ is the Schwarzian 
derivative (\ref{swag}), $\sfs$ is its infinitesimal cousin (\ref{insa}), and $\ad^*$ is the infinitesimal 
coadjoint representation (\ref{infcoadif}) so that $\ad^*_{\alpha}\,p=\alpha p'+2\alpha'p$. In order to 
relate formula (\ref{SHOCK}) to the transformation law of the angular momentum aspect, we take an 
infinitesimal superrotation $f(\phii)=\phii+\epsilon X(\phii)$, an infinitesimal supertranslation 
$\epsilon\,\alpha(\phii)$, and define the variation of $j$ by
\be
\delta_{(X,\alpha)}j\equiv
-\frac{(f,\epsilon\,\alpha)\cdot j-j}{\epsilon}\,.
\nn
\ee
As a result we obtain
\be
\delta_{(X,\alpha)}j
=
Xj'+2X'j-\frac{c_1}{12}X'''+\alpha p'+2\alpha' p-\frac{c_2}{12}\alpha''',
\nn
\ee
which exactly coincides with (\ref{deltajii}) when $c_1=0$, as expected.\\

Using the fact that the Poisson algebra of surface charges (\ref{surall}) coincides with the Kirillov-Kostant 
bracket (\ref{bimipois}), we conclude that the (covariant) phase space of three-dimensional 
asymptotically flat gravity with BMS boundary conditions is a hyperplane $c_1=0$, $c_2=3/G$ embedded in the 
space of the coadjoint representation of the $\hBMS$ group and endowed with its Kirillov-Kostant Poisson 
bracket. This observation is the flat space analogue of the statement that the subleading components of an 
AdS space-time metric contain one-point functions of the dual CFT stress tensor.\i{dual stress 
tensor}\i{stress tensor!dual} As in AdS, this observation 
should not come as a surprise. Indeed, the coadjoint representation of BMS$_3$ was bound to appear in the 
transformation law of the momentum map of the system, and it just so happens that this map is determined by 
the entries of the metric (\ref{piment}). The truly surprising aspect of this observation is the fact that it 
is the entries of the metric, and not some non-linear combinations thereof, that determine the momentum map. 
In particular, as in AdS$_3$, the set of solutions (\ref{piment}) is a vector space.\\

In view of these results, one may ask whether the subleading components of asymptotically flat metrics can be 
interpreted as the components of the stress tensor of some dual theory, similarly to AdS/CFT. The notion of 
``dual theory'' appears to be elusive in the asymptotically flat case, essentially because
the metric becomes degenerate at null infinity, but the question can be answered regardless of this 
complication. Indeed, whatever the dual theory is, it \it{must} 
be such that its stress tensor transforms under the coadjoint representation of the BMS$_3$ group 
(generally with some non-zero central charges), by virtue of the very nature of momentum maps. Accordingly, 
the stress tensor $T$ of any BMS$_3$-invariant  theory is necessarily such that $T_{uu}=p(\phii)$ is a 
supermomentum generating supertranslations, while $T_{u\phii}=j(\phii)$ is an angular supermomentum 
generating superrotations.\\

This being said, it would be reassuring to have explicit field-theoretic illustrations of the fact that 
$(j,p)$ actually \it{is} the stress tensor of some two-dimensional field theory.\i{BMS$_3$ stress 
tensor}\i{dual stress tensor} Such an illustration is 
provided by \cite{Barnich:2013yka} (see also \cite{Barnich:2013jla}), 
where a 
two-dimensional field theory invariant under BMS$_3$ was obtained thanks to the ``dimensional reduction'' of 
three-dimensional gravity through the Chern-Simons formalism. As expected, the stress tensor of that theory 
is a pair $(j,p)$ that coincides with the functions specifying the metric (\ref{piment}), and whose BMS$_3$ 
transformations exactly take the form of the coadjoint representation (\ref{digotaz}) with central charges 
$c_1=0$, $c_2=3/G$ \cite{Barnich:2012rz}. The higher-spin \cite{Gonzalez:2014tba} and supersymmetric 
\cite{Barnich:2014cwa,Barnich:2015sca} generalizations of these considerations confirm this statement, so 
known examples of BMS$_3$-invariant field theories do support our claim that the 
functions $(j,p)$ coincide with the components of a ``dual'' stress tensor.

\subsection{Boundary gravitons and BMS$_3$ orbits}

If one picks a metric (\ref{piment}) at random, the pair $\big(j(\phii),p(\phii)\big)$ is most likely to 
consist of functions that are not constant on the circle. This is actually implied by BMS$_3$ symmetry: if we 
let $(j,p)$ be any seed solution (with $j,p$ constant or not), 
the set of metrics obtained from it by asymptotic symmetry transformations spans an infinite-dimensional 
coadjoint orbit of the $\hBMS$ group at central charges $c_1=0$, $c_2=3/G$,\i{BMS$_3$ metric!orbit under 
BMS$_3$}\i{symplectic leaf!in flat space}
\be
\cW_{(j,c_1;p,c_2)}\,.
\label{wijjipi}
\ee
The metrics belonging to this orbit are
infinite-dimensional analogues of Poincar\'e transforms of the state of a particle with momentum $p$ and 
angular momentum $j$. As in section \ref{sebitu} one may refer to the orbit (\ref{wijjipi}) as a 
space of classical ``boundary gravitons'' around the background $(j,p)$.\i{boundary graviton}\\

The fact that the phase space of flat gravity coincides with (a hyperplane in) the 
coadjoint representation of $\hBMS$ allows us to use the orbits (\ref{wijjipi}) as an organizing principle. 
As in fig.\ \ref{figoLeaf}, the space 
of solutions is foliated into disjoint $\hBMS$ orbits, each of which is a symplectic manifold. Since the 
classification of coadjoint orbits of 
$\hBMS$ 
follows from the results 
of section \ref{secosemi}, we may claim to control the full covariant phase space of asymptotically flat 
gravity. In particular the classification of zero-mode solutions in fig.\ \ref{figazerofla} is a first step 
towards the full classification: each point in the plane $(J,M)$ determines an orbit (\ref{wijjipi}), and 
different 
points define disjoint orbits. Since not all orbits have constant representatives, fig.\ \ref{figazerofla} 
is an incomplete representation of the full phase space of the 
system. The complete picture would involve the BMS$_3$ analogue of fig.\ \ref{vifig}. Note that the relation 
between metrics and $\hBMS$ orbits hints that the quantization of 
asymptotically flat gravity produces unitary representations of BMS$_3$. We will investigate this proposal in 
chapters \ref{c7} and \ref{c8}.

\subsection{Positive energy theorem}
\label{susePOUPOU}

Positive energy theorems in general relativity are commonly formulated in asymptotically flat space-times, so 
we are now in position to address the three-dimensional version of that problem. The question that we wish to 
ask is the following: which asymptotically flat metrics (\ref{piment}) have energy bounded from below 
under BMS$_3$ transformations?\i{positive energy theorem}\\

The answer follows from the fact that asymptotically flat metrics transform under BMS$_3$ according to the 
coadjoint representation (\ref{digotaz}). For our purposes the key property of that formula is the 
fact that the transformation law of $p$ is blind to supertranslations. In this sense the positive energy 
theorem in 
three-dimensional flat space is even simpler than in AdS$_3$:

\paragraph{Positive energy theorem.} The asymptotically flat metric $(j,p)$ has energy boun\-ded from below 
under BMS$_3$ transformations\i{BMS$_3$ metric!positive energy} if and only if $p$ belongs to a Virasoro 
coadjoint orbit (at central 
charge $c_2=3/G$) with energy bounded from below. This is to say that either $p$ is superrotation-equivalent 
to a constant $p_0\geq-c_2/24$, or $p$ belongs to the unique massless Virasoro orbit with bounded energy.\\

As a corollary, we now know that all conical deficits and all flat space cosmologies have energy bounded from 
below under BMS$_3$ transformations. By contrast, all conical excesses have energy unbounded from below.

\section{Flat limits}
\label{sebmsmod}

There are many similarities between the asymptotic symmetries of three\--di\-men\-sio\-nal Anti-de Sitter and 
flat space-times. Intuitively, this is because the limit 
$\ell\rightarrow+\infty$ (i.e.\ $\Lambda\rightarrow0$) of AdS$_3$ is just Minkowski space. It is 
tempting to ask if the phenomenon can be formulated in a mathematically precise way such that the conclusions 
of section \ref{sebmsboco} follow from those of section \ref{sebohad} by a suitably defined flat 
limit. This question was addressed in \cite{Barnich:2012aw}, and the answer is yes. In short, upon 
reformulating 
Brown-Henneaux boundary conditions in Bondi-like coordinates at null (rather than spatial) infinity,\i{null 
infinity!for AdS$_3$}\i{AdS$_3$!null infinity} the 
Minkowskian asymptotic Killing vectors 
(\ref{fasyki}), the on-shell metrics (\ref{piment}) 
and the surface charges (\ref{bokka}) are flat limits of their AdS$_3$ counterparts displayed in eqs.\ 
(\ref{s208}), (\ref{s209}) and (\ref{s212}) 
respectively. In particular, BMS$_3$ symmetry may be seen as a limit of two-dimensional conformal symmetry.\\

In this section we explore this flat limit from the point of view of group theory, starting from the 
definition of the AdS$_3$ asymptotic symmetry group as a set of conformal transformations of a time-like 
cylinder. We describe the limit at the level of groups, then at the level of Lie algebras, and finally at the 
level 
of the coadjoint representation. We end by pointing out a different contraction that produces the Galilean 
conformal 
algebra in two dimensions.
Considerations related to flat limits of unitary representations are relegated to section \ref{sebomodule}.

\subsection{From $\Diff$ to BMS$_3$}

In Anti-de Sitter space, spatial infinity coincides with null infinity. This observation allows one to 
reformulate Brown-Henneaux boundary conditions (originally defined at spatial infinity) in terms of 
Bondi-like coordinates $(r,\phii,u)$ at null infinity \cite{Barnich:2012aw}.\i{Brown-Henneaux fall-offs!at 
null infinity} The conclusion of this 
reformulation is that AdS$_3$ results take the same form as in the 
standard Fefferman-Graham gauge, up to the replacement of the time coordinate $t$ by a retarded time 
coordinate $u$. In particular one can introduce light-cone coordinates
\be
x^{\pm}\equiv\frac{u}{\ell}\pm\phii
\label{xippim}
\ee
in terms of which the asymptotic symmetry group acts on the cylinder at (null) infinity according to 
conformal 
transformations (\ref{xipiM}). Our goal here is to start from these transformations and 
rediscover the BMS$_3$ transformations (\ref{umanif}).\\

The way to go is to expand everything in powers of a ``small'' parameter $\epsilon=1/\ell$.\i{flat 
limit}\i{large l limit@large $\ell$ limit} In practice 
$\ell$ 
is dimensionful so it makes no sense to think of it as being 
``large''; a more precise statement would be that the dimensionless Brown-Henneaux central charge, 
proportional to $\ell/G$, must go to infinity. Despite this subtlety we will keep referring to $\ell$ as a 
``large'' parameter, keeping in mind that there exists a more precise formulation of the procedure.\\

In order to distinguish Virasoro elements from those of BMS$_3$, we denote elements of the group 
$\Diff\times\Diff$ as pairs $(\cF,\bar\cF)$ where $\cF$ and $\bar\cF$ are lifts of orientation-preserving 
diffeomorphisms of the circle satisfying the conditions (\ref{fidif}). Let then $(\cF,\bar\cF)$ be a 
conformal transformation of the cylinder with coordinates (\ref{xippim}). 
In the large $\ell$ limit the transformation of the angular coordinate $\phii$ becomes
\be
\phii
\mapsto
\demi\big(\cF(x^+)-\bar\cF(x^-)\big)
\stackrel{\ell\rightarrow+\infty}{\rightarrow}
\demi\big(\cF(\phii)-\bar\cF(-\phii)\big),
\label{fokazer}
\ee
where the combination of $\cF$'s on the far right-hand side was obtained by Taylor-expanding functions around 
$\pm\phii$ in terms of the small parameter $u/\ell$, and neglecting all terms of order $\cO(1/\ell)$.
The combination of diffeomorphisms in (\ref{fokazer}) is itself a (lift of a) diffeomorphism of the circle. 
Indeed one readily verifies that
\be
f(\phii)
\equiv
\demi(\cF(\phii)-\bar\cF(-\phii))
\label{fikazer}
\ee
satisfies the conditions (\ref{fidif}) when $\cF$ and 
$\bar\cF$ do. Let us investigate what happens with the time coordinate $u$ in the same limit. Using 
(\ref{fikazer}) we find
\be
u
\mapsto
\frac{\ell}{2}\big(\cF(x^+)+\bar\cF(x^-)\big)
\stackrel{\ell\rightarrow+\infty}{\rightarrow}
\frac{\ell}{2}\big(\cF(\phii)+\bar\cF(-\phii)\big)
+
f'(\phii)u,
\nn
\ee
where all terms $\cO(1/\ell)$ were neglected once more.
The first term on the far right-hand side is potentially divergent: typical diffeomorphisms are independent 
of $\ell$, so the first term goes to infinity in the large $\ell$ limit. Note, 
however, that the combination $\cF(\phii)+\bar\cF(-\phii)$ is $2\pi$-periodic. Thus, in order for the limit 
$\ell\rightarrow+\infty$ to work we require that there be a finite, $\ell$-independent function $\alpha$ on 
the circle such that
\be
\cF(\phii)+\bar\cF(-\phii)\equiv\frac{2}{\ell}\,\alpha(f(\phii))+\cO(1/\ell^2)
\label{zuppai}
\ee
where the argument of $\alpha$ is taken to be $f(\phii)$ for convenience. This is to say that the 
diffeomorphisms $\cF$ and $\bar\cF$ are required to depend on $\ell$ in such a way that
\be
\bar\cF(-\phii)
=
-\cF(\phii)+\cO(1/\ell).
\label{cofier}
\ee
For instance, if $\cF(\phii)=\phii+\theta$ and $\bar\cF(\phii)=\phii+\bar\theta$ are rotations, this 
condition says that $\theta-\bar\theta$ goes to zero at least as fast as $1/\ell$ in the large $\ell$ 
limit. With this choice the transformation law of $u$ reduces to
$u\mapsto f'(\phii)u+\alpha(f(\phii))$.
Including (\ref{fokazer}), we have thus reproduced the BMS$_3$ transformations (\ref{umanif}) from a flat 
limit of $\Diff\times\Diff$.\i{BMS$_3$ group!as flat limit}\i{flat limit!of Virasoro group}\i{Virasoro 
group!flat limit} In this sense the centreless BMS$_3$ group (\ref{defbms}) is a flat limit of the 
asymptotic 
symmetry group of AdS$_3$ with Brown-Henneaux boundary conditions. In particular superrotations arise in the 
form (\ref{fikazer}) while 
supertranslations (\ref{zuppai}) measure how fast $\bar\cF(\phii)$ goes to $-\cF(-\phii)$ as $\ell$ goes to 
infinity. Similar considerations would reproduce the centrally extended BMS$_3$ group (\ref{hibiscus}) as a 
contraction of the direct product of two Virasoro groups.\\

Note that the condition (\ref{cofier}) does not imply that there are 
less elements in the BMS$_3$ group than in the group $\Diff\times\Diff$. Indeed, both groups are 
infinite-dimensional Lie groups consisting of two spaces of functions on the circle and have the same 
cardinality in this sense.

\subsection{From Witt to $\bms$}
\label{suseFlalga}

The limit from $\Diff\times\Diff$ to BMS$_3$ can be reformulated in terms of Lie algebras.
Again, our notation will be slightly different from that of the previous chapters so as 
to distinguish AdS$_3$ quantities from Minkowskian quantities. Thus we consider a vector field 
$\cX(x^+)\der_++\bar\cX(x^-)\der_-$ on a two-dimensional 
cylinder and use $\der_{\pm}=\demi(\ell\der_u\pm\der_{\phii})$ to rewrite it as
\be
\frac{\ell}{2}
\big(\cX(x^+)+\bar\cX(x^-)\big)\der_u
+
\demi
\big(\cX(x^+)-\cX(x^-)\big)\der_{\phii}\,.
\label{wxcv}
\ee
In the flat limit $\ell\rightarrow+\infty$ the angular component becomes
\be
\demi\big(\cX(x^+)-\bar\cX(x^-)\big)
\stackrel{\ell\rightarrow+\infty}{\rightarrow}
\demi\big(\cX(\phii)-\bar\cX(-\phii)\big)
\equiv
X(\phii)
\label{xihuie}
\ee
where $X(\phii)$ is some function on the circle, later to be interpreted as (the component of) a 
superrotation generator. For the time component one finds
\be
\frac{\ell}{2}\big(\cX(x^+)+\bar\cX(x^-)\big)
\stackrel{\ell\rightarrow+\infty}{\rightarrow}
\frac{\ell}{2}\big(\cX(\phii)+\bar\cX(-\phii)\big)+uX'(\phii)
\equiv
\alpha(\phii)+uX'(\phii)
\label{NOVEL}
\ee
where we have once more introduced a function $\alpha$ on the circle, later to be interpreted as (a 
component of) a supertranslation generator. This time the requirement is
\be
\cX(\phii)+\bar\cX(-\phii)
=
\frac{2}{\ell}\,\alpha(\phii)
\label{ddfgh}
\ee
with a finite, $\ell$-independent $\alpha$,
and is directly analogous to the condition (\ref{zuppai}). All in all 
we find that, in the flat limit, the vector field (\ref{wxcv}) turns into\i{bms3 algebra@$\bms$ algebra!as 
flat limit}
\be
\xi_{(X,\alpha)}
\equiv
X(\phii)\der_{\phii}+\big(\alpha(\phii)+uX'(\phii)\big)\der_u
\nn
\ee
and thus coincides with the leading non-radial components of the asymptotic Killing vector field 
(\ref{fasyki}). The Lie brackets of such vector fields satisfy the centreless $\bms$ algebra; the latter is 
thus a flat limit of the direct sum of two Witt algebras. Note that from 
this perspective the fact that supertranslations have dimensions of length follows from the fact 
(\ref{ddfgh}) 
that $\alpha(\phii)$ is proportional to $\ell$.\\

The limit from Witt to $\bms$ can also be formulated in terms of 
commutation relations. Indeed, let $\ell_m=e^{imx^+}\der_+$ and 
$\bar\ell_m=e^{imx^-}\der_-$ denote the 
generators of two commuting Witt algebras (\ref{witt}). Then the correspondence (\ref{xihuie})-(\ref{ddfgh}) 
instructs us 
to define would-be superrotation and supertranslation generators\i{Witt algebra!flat limit}\i{flat limit!of 
Witt algebra}
\be
j_m\equiv\ell_m-\bar\ell_{-m}\,,
\qquad
p_m\equiv\frac{1}{\ell}(\ell_m+\bar\ell_{-m})\,.
\label{hajim}
\ee
The terminology here is consistent with the fact that, on the cylinder, $\ell_0-\bar\ell_0$ generates 
rotations while 
$\ell_0+\bar\ell_0$ generates time translations. In the basis (\ref{hajim}),
the commutation relations of the direct sum of two Witt algebras 
take the form
\be
i[j_m,j_n]=(m-n)j_{m+n}\,,
\qquad
i[j_m,p_n]=(m-n)p_{m+n}\,,
\qquad
i[p_m,p_n]=\frac{1}{\ell^2}(m-n)j_{m+n}\,.
\label{haHHa}
\ee
In the limit $\ell\rightarrow+\infty$ the last bracket vanishes and the algebra 
reduces to (\ref{yebem}), reproducing $\bms$ as expected. The same argument can be applied to the direct sum 
of two Virasoro algebras and gives rise to the centrally extended $\hbms$ algebra (see eq.\ (\ref{CiCCoh}) 
below).\\

This observation can be used to define ``flat limits'' of Lie algebras in general terms. 
Consider indeed the Lie algebra $\mg\oplus\mg$, whose generators 
we denote $t_a$ and $\bar t_a$, with identical commutation relations (\ref{commurel}) in both sectors:
\be
[t_a,t_b]
=
f_{ab}{}^c\,t_c\,,
\qquad
[\bar t_a,\bar t_b]
=
f_{ab}{}^c\,\bar t_c\,.
\label{tatabar}
\ee
Then consider the redefinitions
\be
j_a\equiv t_a+\bar t_a\,,
\qquad
p_a\equiv\frac{1}{\ell}(t_a-\bar t_a)
\label{pajakat}
\ee
where $\ell$ is some length scale that we will eventually let go to infinity.
In terms of $j$'s and $p$'s the commutation relations (\ref{tatabar}) become
\be
[j_a,j_b]=f_{ab}{}^c\,j_c\,,
\qquad
[j_a,p_b]=f_{ab}{}^c\,p_c,\,
\qquad
[p_a,p_b]=\frac{1}{\ell^2}f_{ab}{}^c\,j_c\,
\label{guiwaz}
\ee
and the limit $\ell\rightarrow+\infty$ reproduces the commutation relations 
(\ref{jappah}) of exceptional semi-direct sums (without central terms). Thus,
the flat limit of any group $G\times G$ is an exceptional semi-direct product 
$G\ltimes_{\Ad}\mg_{\text{Ab}}$. The flat limit (\ref{hajim}) giving rise to $\bms$ from two copies 
of the Witt 
algebra is a special case of that construction. Indeed, the map
\be
\ell_m\mapsto-\ell_{-m}
\label{involution}
\ee
is a Lie algebra isomorphism when the $\ell_m$'s generate a Witt algebra (\ref{witt}), so the redefinitions 
(\ref{hajim}) precisely take the form (\ref{pajakat}) with the correspondence $t_a\leftrightarrow\ell_m$ and 
$\bar 
t_a\leftrightarrow-\bar\ell_{-m}$. As it turns out, all symmetry algebras 
found so far in the realm of asymptotically flat field theories in three dimensions can be seen as flat 
limits of the type just described when compared to their AdS$_3$ counterparts.\\

The limiting procedure that turns the sum of two Witt algebras into $\bms$ is 
an example of \it{In\"on\"u-Wigner contraction} \cite{Inonu:1953sp},\i{In\"on\"u-Wigner contraction} similar 
to the relation between the
Poincar\'e group and the Galilei group. Conversely, the direct sum of two Witt algebras is a deformation of 
$\bms$. The same construction can be used to show that 
the Poincar\'e algebra is a flat limit of the AdS$_3$ isometry algebra, 
$\mathfrak{so}(2,2)\cong\sl\oplus\sl$. We will return to this in section \ref{sebomodule}.

\subsection{Stress tensors and central charges}

We now apply 
the flat limit to the coadjoint representation of two Virasoro groups, generally with non-zero central 
charges. Let therefore $T(x^+)$ and $\bar T(x^-)$ be CFT stress tensors transforming under left and 
right conformal transformations as Virasoro coadjoint vectors with central charges $c$ and $\bar c$, 
respectively. (In (\ref{s209}) we denoted these stress tensors as $p,\bar p$, but here we keep 
the 
letter $p$ for supermomenta.) They are paired with vector fields 
$\cX(x^+)\der_++\bar\cX(x^-)\der_-$ on the cylinder according to
\be
\bra(T,\bar T),(\cX,\bar\cX)\ket
=
\frac{1}{2\pi}\int_0^{2\pi}d\phii\big[
T(x^+)\cX(x^+)+\bar T(x^-)\bar\cX(x^-)
\big]
\nn
\ee
which (up to notation) is just the AdS$_3$ surface charge (\ref{s212}). As above we 
expand the functions $T$ and $\bar T$ in powers of $1/\ell$ and we define\i{BMS$_3$ 
stress tensor!as flat limit}
\be
p(\phii)\equiv\lim_{\ell\rightarrow+\infty}\frac{1}{\ell}\big(T(x^+)+\bar T(x^-)\big),
\qquad
j(\phii)+up'(\phii)\equiv\lim_{\ell\rightarrow+\infty}\big(T(x^+)-\bar T(x^-)\big).
\nn
\ee
Using (\ref{NOVEL}) and (\ref{ddfgh}) one then verifies that, in the limit $\ell\rightarrow+\infty$,
\be
T(x^+)\cX(x^+)+\bar T(x^-)\bar\cX(x^-)
=
j(\phii)X(\phii)+p(\phii)\alpha(\phii)+u(pX)'(\phii),
\nn
\ee
which coincides up to a total derivative with the integrand of the flat surface charge (\ref{bokka}). In 
other words the surface charges of flat space gravity are large $\ell$ limits of those of AdS$_3$. In 
mathematical terms this is to say that the flat limit of the coadjoint representation of the direct 
product of two Virasoro groups is the coadjoint representation of the (centrally extended) BMS$_3$ group.\\

This phenomenon also allows us to relate the central charges of the Virasoro algebra to those of $\hbms$.
(We could have done this in terms of abstract Lie algebra generators, but for comparison with 
three-dimensional gravity we do it here in terms of coadjoint vectors.)
Let us consider two Virasoro algebras with central charges $c$ and $\bar c$ that depend on $\ell$ 
as\i{central charge!flat limit}
\be
c=A\ell+B+\cO(1/\ell),
\qquad
\bar c=A\ell+\bar B+\cO(1/\ell)
\nn
\ee
where $A$, $B$ and $\bar B$ are $\ell$-independent. Then the definitions
\be
c_1\equiv\lim_{\ell\rightarrow+\infty}(c-\bar c),
\qquad
c_2\equiv\lim_{\ell\rightarrow+\infty}\frac{c+\bar c}{\ell}
\label{CiCCoh}
\ee
allows us to write the flat limit of the algebra in the $\hbms$ form (\ref{bimipois}) in terms of generators 
$(j_m,p_m)$ related to Virasoro generators $(\ell_m,\bar\ell_m)$ by (\ref{hajim}). This is the 
centrally extended analogue of the 
flat limit described in (\ref{guiwaz}). Note that for the Brown-Henneaux central charges 
(\ref{ss210}) the prescription (\ref{CiCCoh}) yields $c_1=0$ and $c_2=3/G$, which are indeed the standard 
values for asymptotically flat space-times.

\paragraph{Remark.} The fact that flat space holography\i{flat space holography}\i{holography!in flat space} 
can be studied as a flat 
limit of the AdS/CFT 
correspondence is an old idea \cite{Susskind:1998vk,Polchinski:1999ry}; see also 
\cite{Gary:2009ae,Costa:2013vza,Krishnan:2013wta}.\i{flat limit!of AdS/CFT}\i{AdS/CFT!flat limit} Here we 
have described 
its group-theoretic formulation. It 
should be noted, however, that there is no known limiting 
construction that yields BMS symmetry in \it{four} dimensions from some corresponding asymptotic symmetry in 
AdS$_4$.

\subsection{The Galilean conformal algebra}
\label{suseGaC}

The $\bms$ algebra turns out to be isomorphic to the Galilean conformal algebra in two dimensions. We now 
explain how the latter can be obtained as a non-relativistic contraction of two Witt algebras and discuss the 
extent to which Galilean conformal symmetry applies to asymptotically flat gravity in three dimensions. As in 
the earlier sections of this chapter we work only at the classical level. The quantum version of these 
considerations will be exposed in chapter \ref{sebomodule}.\\

The redefinitions (\ref{pajakat}) suggest a contraction of Witt algebras that differs from the flat limit 
(\ref{hajim}). Namely, instead of performing the involution (\ref{involution}) before taking the limit 
$\ell\rightarrow+\infty$, one can define
\be
\tilde j_m\equiv\bar\ell_m+\ell_m\,,
\qquad
\tilde p_m\equiv\frac{1}{\ell}(\bar\ell_m-\ell_m)\,.
\label{limini}
\ee
In contrast to (\ref{hajim}), this redefinition has nothing to do with the flat limit of 
AdS$_3$, but the limit $\ell\rightarrow+\infty$ still gives rise to an algebra with commutation relations 
(\ref{yebem}) upon renaming $j_m\rightarrow\tilde j_m$ and $p_m\rightarrow\tilde p_m$. The key difference is 
that now the generator of time translations is $\tilde j_0$ 
(since it coincides with $\ell_0+\bar\ell_0$) while $\tilde p_0$ generates rotations (since it 
is proportional to $\ell_0-\bar\ell_0$). More generally, with the redefinition (\ref{limini}), the generators 
of would-be supertranslations do not commute while those of would-be superrotations do commute. This is the 
opposite of the behaviour of superrotations and supertranslations in three-dimensional Einstein gravity.\\

The redefinitions (\ref{limini}) can be interpreted as a non-relativistic contraction of the 
direct sum of two Witt algebras in two dimensions, analogous to the usual In\"{o}n\"{u}-Wigner 
contraction of the Poincar\'e algebra to the Galilei 
algebra.\i{Witt algebra!non-relativistic limit} For this reason the 
algebra spanned by $\tilde j_m$'s and $\tilde p_m$'s is known 
as the \it{Galilean conformal algebra}\i{gca2@$\gca$} in two 
dimensions 
\cite{Bagchi:2009pe,Bagchi:2010eg}. It is the non-relativistic 
limit of the conformal algebra in two dimensions and, by a geometric coincidence, it is isomorphic to 
$\bms$. The Galilean conformal algebra has been extensively studied in its own 
right; see e.g.\ 
\cite{Bagchi:2009ke} for its supersymmetric extension and \cite{Bagchi:2009ca,Bagchi:2009pe} 
for its highest-weight representations. It is a fundamental tool in the non-relativistic limit 
of the AdS/CFT correspondence \cite{Bagchi:2009my}.\i{AdS/CFT!non-relativistic limit}\i{non-relativistic 
limit} In what follows we denote it
by $\mathfrak{gca}_2$.\\

At some point the isomorphism $\mathfrak{gca}_2\cong\bms$ led to the proposal that flat space 
holography (in three space-time dimensions) is described by a Galilean conformal field theory 
\cite{Bagchi:2010eg,Bagchi:2010zz}. In view of the geometric interpretation of superrotations and 
supertranslations described above, this sounds suspicious: the Galilean conformal algebra is a version of the 
$\bms$ algebra ``rotated by 90 degrees'' where the roles of the Hamiltonian and angular momentum are 
exchanged. In particular the flat limit of AdS$_3$/CFT$_2$, if it exists, should not give rise to a 
Galilean conformal field theory since the gravitational flat limit 
(\ref{hajim}) of two Witt algebras gives rise to standard $\bms$, in which $p_0$ 
generates time translations. Nevertheless, at the level of \it{classical} symmetries, there is essentially no 
distinction between $\bms$ and $\mathfrak{gca}_2$; the two are interchangeable.\i{bms3 algebra@$\bms$ 
algebra!and $\gca$ algebra} This coincidence led to many 
publications concerned with flat space holography and attempting to describe its dual 
theory as a Galilean conformal field theory; see e.g.\ 
\cite{Fareghbal:2013ifa,Banerjee:2015kcx,Fareghbal:2016hqr} and references therein. One of the goals of this 
thesis is to explain why the dual theory of asymptotically flat gravity, if it 
exists at all, cannot be a Galilean conformal field theory. The reason for this is rooted in 
the elementary observation that the correspondence $\bms\leftrightarrow\mathfrak{gca}_2$ exchanges the 
Hamiltonian and the angular momentum, but we will go much beyond that. In fact we shall see that the 
difference between $\bms$ and $\mathfrak{gca}_2$, while classically invisible, becomes 
apparent at 
the \it{quantum} level. This will rely on the induced representations developed in the next 
chapter and will be studied in much greater detail in section \ref{sebomodule}. As it turns out, the most 
striking illustration of this distinction will arise in section \ref{CAPsec:3DW} in the realm of quantum 
higher-spin theories.\\

This being said, we stress that discarding Galilean conformal field theories as putative duals for 
asymptotically flat gravity does \it{not} rule out all the conclusions of the substantial literature on flat 
space holography approached from the Galilean side. Rather, the point we wish to make is that 
those 
computations that did work in flat space while relying on $\mathfrak{gca}_2$ symmetry would have worked 
equally well in the language of BMS$_3$ symmetry. More precisely, any computation that holds for 
$\mathfrak{gca}_2$ but does not rely on its realization as a quantum symmetry algebra also holds for $\bms$, 
and therefore for asymptotically flat gravity.

\chapter{Quantum BMS$_3$ symmetry}
\label{c7}
\markboth{}{\small{\chaptername~\thechapter. Quantum BMS$_3$ symmetry}}

This chapter is devoted to irreducible unitary representations of the BMS$_3$ group, i.e.\ BMS$_3$ particles, 
which we classify and interpret. As we shall see, the classification is provided by supermomentum orbits 
that coincide with coadjoint orbits of the Virasoro group. Upon identifying supermomentum with the Bondi mass 
aspect of asymptotically flat metrics, we will be led to interpret BMS$_3$ particles as relativistic 
particles 
dressed with gravitational degrees of freedom.\\

The plan is as follows. In section \ref{sebmspar} we classify
BMS$_3$ particles according to orbits of supermomenta under superrotations. We also describe and interpret 
the resulting Hilbert spaces of wavefunctions, which we relate to the quantization of (coadjoint) orbits of 
asymptotically flat metrics under BMS$_3$. Section \ref{sebomodule} is devoted to the 
description of BMS$_3$ particles as representations of the (centrally extended) $\bms$ algebra and their 
relation to 
highest-weight representations of the Virasoro algebra; we also briefly touch upon Galilean 
representations. Finally, in section \ref{sebmschar} we evaluate characters of BMS$_3$ particles. To lighten 
the notation, from now on the words ``BMS$_3$ group'' or ``$\bms$ algebra'' implicitly refer to their 
centrally extended versions (except if stated otherwise). We also abuse notation by writing $\Diff$ to refer 
either to $\Diffp$ or to $\Diffc$, depending on the context.\\

Most of the results exposed in this chapter have been reported in the papers 
\cite{Barnich:2014kra,Barnich:2015uva,Oblak:2015sea,Campoleoni:2016vsh}. The relation between BMS$_3$ 
particles and gravitational one-loop partition functions \cite{Barnich:2015mui,Campoleoni:2015qrh} will be 
described in the next chapter. Note that the considerations that follow rely heavily on the material of 
chapter \ref{c2bis}.

\section{\ BMS$_3$ particles}
\label{sebmspar}

In high-energy physics a \it{particle} is usually defined as an irreducible unitary representation of the 
Poincar\'e group. If one takes BMS symmetry seriously, it is tempting to apply the same terminology to 
representations of BMS. Accordingly, in this section our goal is to answer the following question:
\be
\begin{array}{c}
\text{\it{Replace the word ``Poincar\'e'' by ``BMS'' in the definition of a particle.}}\\
\text{\it{What new notion of particle does one then obtain?}}\\
\text{\it{What new quantum numbers describe its degrees of freedom?}}
\end{array}
\nn
\ee
In principle this problem should be addressed in the realistic four-dimensional world. However, as mentioned 
in the introduction of this thesis, BMS symmetry in four dimensions is very poorly understood at present, so 
we will content ourselves with the more modest task of understanding irreducible unitary 
representations of BMS in three dimensions --- that is, \it{BMS$_3$ particles}.\i{BMS$_3$ particle} 
Remarkably, we will discover 
that BMS$_3$ particles are labelled by mass and spin, exactly as standard relativistic particles. As in 
section 
\ref{sevirep}, we will interpret their extra degrees of freedom as boundary 
gravitons, or equivalently soft gravitons.\\

The plan of this relatively long section is the following. We first describe the supermomentum orbits and 
little groups that 
classify BMS$_3$ particles. We shall see that these orbits are in fact coadjoint orbits of the Virasoro 
group, 
which will allow us to define massive, massless and tachyonic BMS$_3$ particles. We also 
discuss the existence of integration measures on supermomentum orbits, since such measures are required to 
define scalar products of wavefunctions. We then describe the states represented by such wavefunctions and 
interpret them as particles dressed with quantized gravitational degrees of freedom, in accordance with 
the relation between asymptotically flat metrics and the coadjoint representation of BMS$_3$. We also apply 
this 
interpretation to the vacuum representation and to spinning BMS$_3$ particles, and we conclude by discussing 
the extension of our considerations to four space-time dimensions.

\subsection{\ Orbits and little groups}
\label{susebilly}

Our goal is to understand the quantum-mechanical implementation of BMS$_3$ symmetry, at least as far as 
irreducible representations are concerned. According to section \ref{s1.1} we should leave room for 
projective 
representations; to do this we consider exact representations of the 
universal cover of the universal central extension of the connected BMS$_3$ group, that is, $\hBMS$. The 
latter was 
defined in (\ref{hibiscus}). Since 
$\hBMS$ is a semi-direct product, one expects all its irreducible unitary 
representations to be induced \`a la Wigner. These representations are classified 
by the orbits and little groups described in general terms in section 
\ref{sesemi}. Here we perform that classification.

\subsubsection*{Supermomentum orbits}

The key ingredient in the description of BMS$_3$ particles is the dual of the space of supertranslations, 
$\hVect{}_{\text{Ab}}^*$. Following the terminology of section \ref{sebms3}, its elements are 
\it{centrally extended supermomenta}\i{supermomentum}
\be
\big(p(\phii)d\phii^2,c_2\big)
\label{pikkitu}
\ee
paired with centrally extended supertranslations $(\alpha,\lambda)$ according to (\ref{virpar}) with the 
replacements $X\rightarrow\alpha$ and $c\rightarrow c_2$. As mentioned below 
(\ref{jipifou}), $p(\phii)$ has dimensions of energy; its three lowest Fourier modes form a Poincar\'e 
energy-momentum vector (in particular the 
zero-mode is the energy of $p$). More generally $p(\phii)$ is an energy density on the 
circle while the central charge $c_2$ is an energy scale. Supermomentum transforms as a Virasoro coadjoint 
vector 
(\ref{ATOM}) under 
superrotations, so the allowed supermomenta of a BMS$_3$ particle span a coadjoint orbit of the Virasoro 
group at central charge 
$c_2$. This is the first key conclusion of this section:

\paragraph{Theorem.} The orbit $\cO_p$ of a supermomentum $(p,c_2)$\i{orbit!for BMS$_3$}\i{supermomentum 
orbit}\i{classification!of BMS3 particles@of BMS$_3$ particles} under superrotations is a coadjoint orbit 
of the Virasoro group at central charge $c_2$.\\

When interpreting $p(\phii)$ as the Bondi mass aspect of an asymptotically flat metric (\ref{piment}), the 
orbit $\cO_p$ is a subset of the orbit (\ref{wijjipi}) of the metric under BMS$_3$ transformations. In that 
context the central charge $c_2$ coincides with the Planck mass (\ref{god}). Accordingly, from now on we 
restrict our attention to centrally 
extended supermomenta 
whose central charge $c_2$ is strictly positive.

\subsubsection*{Massive and massless BMS$_3$ particles}

The statement that supermomentum orbits are Virasoro coadjoint orbits is analogous to the fact that coadjoint 
orbits of $\SL$ classify the momenta of relativistic particles in three dimensions. In particular the 
map of Poincar\'e momenta in fig.\ \ref{figDepIdu}{\textcolor{blue}{b}} is embedded in 
the larger picture of fig.\ \ref{vifig}, which is now interpreted as a map of BMS$_3$ supermomenta. Thus, 
supermomentum orbits that contain a constant representative (the vertical line in fig.\ \ref{vifig}) are the 
supermomenta of BMS$_3$ particles that admit a rest frame.

\paragraph{Definition.} A \it{massive BMS$_3$ particle}\i{massive BMS$_3$ particle} is a BMS$_3$ 
particle 
whose supermomenta span a 
Virasoro coadjoint orbit that admits a generic constant representative $p_0$.\\

In this definition the word ``generic'' refers to the fact that $p_0$ should not take one of the discrete 
exceptional values $-n^2c_2/24$. Indeed the orbits containing such exceptional constants are better thought 
of 
as BMS$_3$ generalizations of the trivial representation of Poincar\'e; we will return to this 
interpretation below.\\

By contrast, supermomentum orbits 
that do \it{not} admit a constant representative describe BMS$_3$ particles 
that have no rest frame. For instance, the discrete dots that do not belong to the vertical line in 
fig.\ \ref{vifig} are BMS$_3$ generalizations of massless Poincar\'e particles, while the horizontal lines 
of fig.\ \ref{vifig} generalize tachyons.

\paragraph{Definition.} A \it{massless BMS$_3$ particle}\i{massless BMS$_3$ particle} is a BMS$_3$ 
particle 
whose supermomenta span a 
Virasoro coadjoint orbit with non-degenerate parabolic monodromy and non-zero winding number. A \it{BMS$_3$ 
tachyon}\i{BMS$_3$ tachyon}\i{tachyonic BMS$_3$ particle} is a BMS$_3$ particle 
whose supermomenta 
span a Virasoro 
coadjoint orbit with hyperbolic monodromy and non-zero winding number.\\

In these definitions the terms ``monodromy'' and ``winding number'' refer to the Virasoro invariants defined 
in 
section \ref{secovobi}. They are the BMS$_3$ generalization of 
the mass squared in the Poincar\'e group.

\subsubsection*{Little groups}

The little groups of BMS$_3$ particles coincide with the stabilizers of the corresponding 
Virasoro coadjoint 
orbits. Here, for comparison with the Poincar\'e little groups of section \ref{sePoTri}, we list the little 
groups obtained by using the central extension of the multiply connected BMS$_3$ group (\ref{BaConn}). The 
list of orbits is that of section \ref{sevirorepss} and their little groups 
are summarized in table \ref{tabVir}:\i{little group!for BMS$_3$ particle}
\begin{itemize}
\item For a massive BMS$_3$ particle, the stabilizer is the group $\un$ of spatial rotations.
\item For a vacuum-like BMS$_3$ particle whose supermomentum at rest takes the value $-n^2c_2/24$, the little 
group is an $n$-fold cover of the Lorentz group in three dimensions, $\text{PSL}^{(n)}(2,\RR)$ (with 
$n\geq 1$).
\item For a massless particle with winding number $n\geq 1$, the little group is $\RR\times\ZZ_n$.
\item For a BMS$_3$ tachyon with winding number $n\geq 1$, the little group is $\RR\times\ZZ_n$.
\end{itemize}
This list should be compared with table \ref{tabOr}. The representations of these little groups will lead to 
a notion of BMS$_3$ spin. Note that when dealing with the universal cover (\ref{debmscov}) of BMS$_3$, all 
compact directions of the above little groups get decompactified so that $\un$ is replaced by $\RR$, 
$\text{PSL}^{(n)}(2,\RR)$ is replaced by its universal cover, and $\ZZ_n$ is replaced by the group 
$T_{2\pi/n}\cong\ZZ$ of translations of $\RR$ by integer multiples of $2\pi/n$.

\subsubsection*{BMS$_3$ particles with positive energy}

It is natural to declare that physically admissible BMS$_3$ particles have
supermomentum orbits such that the energy functional (\ref{enep}) is bounded from below under superrotations. 
Finding these 
particles is the BMS$_3$ 
analogue of the question (\ref{enequest}) encountered in the Virasoro context. The solution is provided 
by the earlier results (\ref{konkotab})-(\ref{konkotob}):

\paragraph{Theorem.} A BMS$_3$ particle has energy bounded from below if and only if its supermomenta span 
one of the Virasoro orbits coloured in red in fig.\ \ref{vifigibis}.\\

Recall that Poincar\'e particles with positive energy fall in 
exactly three classes, two of which contain only one momentum orbit: massive particles, massless particles, 
and the trivial orbit. The theorem tells us that essentially the same conclusion holds for BMS$_3$ 
particles, since all supermomentum orbits with energy bounded from below belong to one of the three following 
classes:\i{BMS$_3$ particle!with positive energy}\i{positive energy theorem!for BMS$_3$ particles}
\begin{itemize}
\item the unique vacuum orbit containing the supermomentum $p_0=-c_2/24$,
\item one of the massive orbits located above the vacuum and containing a constant supermomentum 
$p_0>-c_2/24$,
\item the unique massless orbit with energy bounded from below.
\end{itemize}
From now on, when referring to BMS$_3$ particles we always implicitly refer only to particles with 
energy bounded from below (except if explicitly stated otherwise). Note that, in contrast with Virasoro 
representations, BMS$_3$ particles with unbounded energy may provide unitary representations of BMS$_3$. 
Furthermore the energy spectrum of any BMS$_3$ particle is continuous.

\subsection{\ Mass, supermomentum, central charge}
\label{susePoPhy}

In the list of physical BMS$_3$ particles, the only family with infinitely many members is the class of 
massive particles. Let us therefore describe these particles in some more detail and interpret the labels 
$(p,c_2)$ that classify them.

\subsubsection*{Defining mass}

The starting point is the observation that the vacuum supermomentum is 
$p_{\text{vac}}=-c_2/24$, 
while the supermomentum at rest of any massive BMS$_3$ particle is located above that vacuum value.

\paragraph{Definition.} Consider a massive BMS$_3$ particle with supermomentum at rest $p_0>-c_2/24$. Then 
the \it{mass} of the particle is\i{mass!of BMS$_3$ particle}\i{BMS$_3$ particle!mass}
\be
M\equiv p_0+c_2/24.
\label{bmass}
\ee
Massive BMS$_3$ particles with energy bounded from below have positive mass.\\

The definition of mass in eq.\ (\ref{bmass}) can be rewritten in a manifestly superrotation-invariant way, 
without invoking any rest frame. Indeed, recall from (\ref{traCos}) that the value of $p_0$ determines the 
trace of the monodromy matrix $\sfM$. This relation can be inverted and combined with the 
definition (\ref{bmass}), which yields\i{mass operator}\i{mass squared!for BMS$_3$}
\be
M
=
\frac{c_2}{24}
\left[
1+\Big(\frac{1}{\pi}\text{arccosh}\big[\text{Tr}(\sfM/2)\big]\Big)^2
\right]
\label{bamonod}
\ee
where we assume for definiteness that $p_0\geq0$, which is to say that $\sfM$ is hyperbolic and $M\geq 
c_2/24$. The same relation 
holds for elliptic $\sfM$, hence $M<c_2/24$, upon replacing $\text{arccosh}$ by $i\,\text{arccos}$, with the 
convention 
$\text{arccos}(1)=0$ and $\text{arccos}(-1)=\pi$.\\

Formula (\ref{bamonod}) is a superrotation-invariant definition of the mass of a BMS$_3$ particle, 
since the trace of the monodromy matrix associated with Hill's equation is Virasoro-invariant. It is a 
BMS$_3$ 
analogue of the relation
\be
M^2=E^2-\bbp^2=-p_{\mu}p^{\mu}
\nn
\ee
that determines the mass of a Poincar\'e particle from its energy-momentum $p_{\mu}$. As a 
bonus, (\ref{bamonod}) allows us to distinguish massive particles with elliptic and hyperbolic 
monodromy. This distinction is 
consistent with three-dimensional gravity, where metrics with a Bondi mass aspect $p_0<0$ are conical 
deficits 
--- i.e.\ classical particles --- while metrics with $p_0>0$ are flat space cosmologies --- Minkowskian 
analogues of BTZ black holes. Furthermore, eq.\ (\ref{bamonod}) confirms that 
massless 
BMS$_3$ particles (with positive energy) are actually massless. Indeed, the corresponding monodromy matrix is 
(\ref{MaH}) with 
winding number $n=1$. This implies that $\text{Tr}(\sfM)=-2$ for physical massless BMS$_3$ particles, 
which can be plugged into (\ref{bamonod}) and yields $M=0$ upon using 
$\text{arccosh}(-1)=i\,\text{arccos}(-1)=i\pi$.
It would have been impossible to obtain this result with the weaker definition of mass of eq.\ (\ref{bmass}), 
since massless particles have no rest frame.

\subsubsection*{Intepreting supermomentum}

In order to develop our intuition about the supermomentum vector $p(\phii)$, it is useful to 
rewrite standard Poincar\'e momenta in terms of functions on the circle. The supermomentum of a 
Poincar\'e particle with mass $M$ typically takes the form\i{supermomentum!for moving particle}
\be
p(\phii)
=
\sqrt{M^2+p_x^2+p_y^2}\,
+\,
p_x\cos\phii\,
+\,
p_y\sin\phii
-
\frac{c_2}{24}
\label{opossum}
\ee
and represents a particle moving in space with a spatial momentum $(p_x,p_y)$. Note the extra factor 
$-c_2/24$ at the end, which ensures that $M$ actually coincides with the mass of the 
particle.\\

One can then
use (\ref{ATOM}) to act with superrotations on (\ref{opossum}) and 
obtain various boosted momenta. In particular, Lorentz transformations 
take the form (\ref{protophi}) and act on the components 
$(\sqrt{M^2+p_x^2+p_y^2},p_x,p_y)$ according to 
the vector representation.\i{boost!acting on supermomentum}\i{supermomentum!transformation under boosts} 
This point can be verified by taking a supermomentum 
at 
rest, $p_0=M-c_2/24$, and acting on it with a boost (\ref{Booss}) in the direction $\phii=0$,
\be
e^{if(\phii)}
=
\frac{\cosh(\gamma/2)e^{i\phii}+\sinh(\gamma/2)}{\sinh(\gamma/2)e^{i\phii}+\cosh(\gamma/2)}
\label{saretyk}
\ee
where $\gamma$ is the rapidity (in terms of standard velocity, $\gamma=\text{arctanh}(v)$). 
Since this superrotation is of the projective form (\ref{protophi}), its 
Schwarzian derivative satisfies (\ref{swagiff}) and the corresponding transformation (\ref{ATOM}) of the 
supermomentum $p$ can be rewritten as
\be
\big(f\cdot p\big)(f(\phii))+\frac{c_2}{24}
=
\frac{1}{(f'(\phii))^2}\left[p(\phii)+\frac{c_2}{24}\right].
\label{MINAS}
\ee
This says that the combination 
$p+c_2/24$ transforms under Lorentz transformations as a centreless coadjoint vector of $\Diff$. In 
particular, the supermomentum of a massive particle at rest transforms according to (\ref{MINAS}) 
with $p(\phii)=M-c_2/24$. As a result, since $f'(\phii)$ is given by (\ref{fagobo}), the energy $E[\gamma]$ 
of the boosted particle is
\be
E[\gamma]
=
\frac{M}{2\pi}\int_0^{2\pi}\frac{d\phii}{f'(\phii)}
=
M\cosh\gamma
\nn
\ee
while the boosted spatial momentum along the $x$ direction is
\be
p_x[\gamma]
=
\frac{M}{2\pi}\int_0^{2\pi}\frac{d\phii}{f'(\phii)}\cos(f(\phii))
=
M\sinh\gamma.
\nn
\ee
The boosted spatial momentum along the $y$ direction vanishes, as it should. This confirms that Lorentz 
transformations act on a supermomentum at rest exactly as in standard special relativity (albeit 
in three space-time dimensions).\\

From these considerations we can now draw a general conclusion on the supermomentum of a (massive) BMS$_3$ 
particle. Typically, the function $p(\phii)$ will have some non-trivial profile on the circle; for instance 
the momentum of a particle moving fast in the $x$ direction is represented by a function $p(\phii)$ which is 
larger than $-c_2/24$, has a bump around the point $\phii=0$, and almost vanishes in the neighbourhood of the 
opposite point $\phii=\pi$. If the particle is obtained by a pure Poincar\'e boost from a particle at rest, 
the only non-vanishing components of its supermomentum are its three lowest Fourier modes, $p_0$, $p_1$, 
$p_{-1}$ (as in eq.\ (\ref{opossum})). Upon switching on superrotations, the supermomentum of the particle 
acquires extra Fourier modes ($p_2$, $p_3$, etc.) that dress the original Poincar\'e momentum with 
additional fluctuations.\i{gravitational dressing}\i{supermomentum!gravitational dressing} In the 
upcoming pages we will interpret these extra degrees of freedom as being of 
gravitational origin.

\begin{figure}[H]
\centering
\begin{minipage}{0.5\textwidth}
\centering
\includegraphics[width=0.7\linewidth]{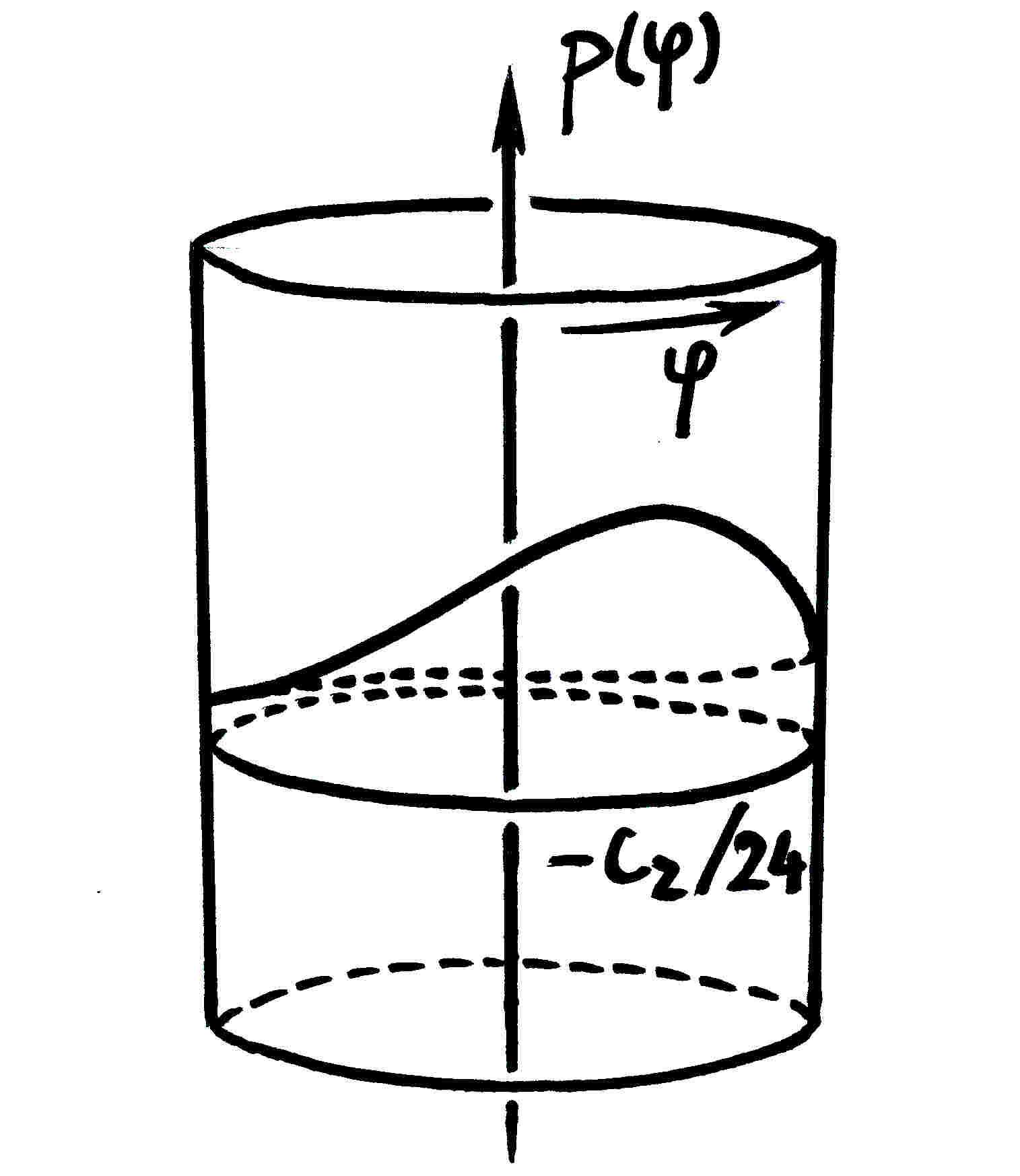}\par
(a)
\end{minipage}\hfill
\begin{minipage}{0.5\textwidth}
\centering
\includegraphics[width=0.7\linewidth]{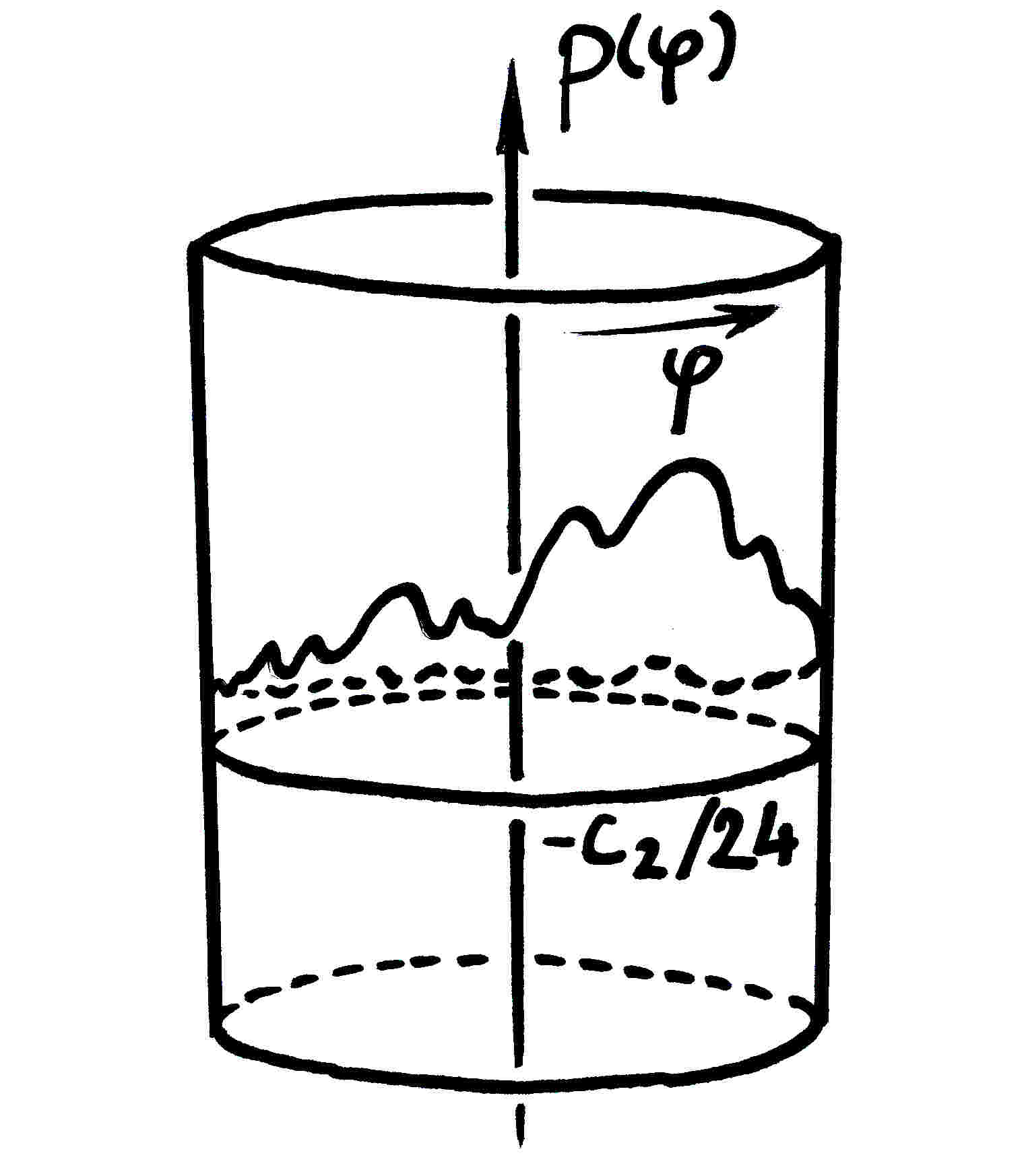}\par
(b)
\end{minipage}
\caption{Two possible supermomenta of a massive BMS$_3$ particle. In (a) the supermomentum is that of a 
boosted Poincar\'e particle, given by eq.\ (\ref{opossum}); it is a function on the circle located above the 
line $-c_2/24$ and its only non-zero Fourier modes are the three lowest ones. In (b) the function is dressed 
with extra non-vanishing Fourier modes, which results in more wiggles. These extra Fourier modes account for 
gravitational degrees of freedom that do not appear in the pure Poincar\'e case.}
\end{figure}

\subsubsection*{Interpreting the central charge}

The central charge $c_2$ is an energy scale; this is 
manifest in formula (\ref{bamonod}), where $c_2$ converts the dimensionless trace of a monodromy matrix into 
a mass $M$.\i{c2@$c_2$ (BMS$_3$ central charge)!as mass scale}\i{central charge!for BMS$_3$}\i{BMS$_3$ 
central charge}\i{Planck 
mass}\i{dimensionful central 
charge} The actual value of $c_2$ is arbitrary in principle, but in Einstein gravity it is proportional 
to the Planck mass: $c_2=3/G$. In particular, note that $c_2$ is \it{not} a 
Virasoro central charge, even though it appears in the transformation law (\ref{ATOM}) of supermomentum as if 
$p$ was a CFT stress tensor.\\

It is worth stressing the crucial importance of $c_2$ for the conclusions of the previous pages. For one 
thing, the whole classification of supermomentum orbits and the ensuing definition of massive/massless 
BMS$_3$ particles only makes sense because $c_2$ is non-zero. If $c_2$ happened 
to vanish, none of these results would hold since the corresponding supermomentum orbits would be Virasoro 
coadjoint orbits at \it{vanishing} central charge, and we saw in section \ref{secovobi} that these orbits 
are radically different from (and arguably much uglier than) their centrally extended peers. This is not to 
say that $c_2$ must be non-zero in order for the supermomentum (\ref{pikkitu}) to yield a representation of 
the 
BMS$_3$ group; in principle, representations associated with orbits having $c_2=0$ are just as acceptable as 
representations in which $c_2\neq 0$. However the application to 
gravity, and the ensuing interpretation of representations as particles, relies crucially on the fact that 
$c_2=3/G$ does \it{not} vanish. Note that the change of monodromy 
occurring at $M=c_2/24$ suggests that something radical happens with BMS$_3$ particles whose mass is higher 
than that bound. This bifurcation reflects the fact that the metric of the gravitational field surrounding 
the 
particle changes from that of a conical deficit (when $M<1/8G$) to that of a flat cosmology (when 
$M>1/8G$).

\subsection{\ Measures on superrotation orbits}
\label{susemesor}

Suppose we actually want to describe the space of states of a BMS$_3$ particle with supermomentum orbit 
$\cO_p$. If the particle is scalar, then its Hilbert space consists of complex-valued wavefunctions 
(\ref{psiqqu}) in 
supermomentum space whose scalar product (\ref{scall}) 
involves an integral over $\cO_p$ with some measure $\mu$. The latter needs 
to be quasi-invariant 
under superrotations,\footnote{We recall that the definition of quasi-invariant measures was given in section 
\ref{seqareg}.} which motivates the following question:\i{Virasoro measure}\i{Virasoro 
orbits!measure}\i{measure!on Virasoro orbit}\i{quasi-invariant measure!on Virasoro orbit}
\be
\begin{array}{c}
\text{\it{Let $\cO_p$ be a Virasoro coadjoint orbit at non-zero central charge;}}\\
\text{\it{is there a quasi-invariant Borel measure on it?}}
\end{array}
\label{quesemes}
\ee
If the answer is affirmative, then the measure is a functional one since $\cO_p$ consists of functions on the 
circle.

\subsubsection*{A conjecture}

Our viewpoint regarding the problem (\ref{quesemes}) will be pragmatical: path integral measures are used on 
a daily basis in quantum mechanics, and their efficiency in correctly predicting the values of physical 
observables is firmly established. Thus, if one
is willing to define Hilbert spaces of square-integrable functions thanks to functional
measures, their application to BMS$_3$ particles is as acceptable as in quantum physics.
In particular one may hope that Virasoro coadjoint orbits do admit quasi-invariant measures:

\paragraph{Conjecture.} Let $\cO_p$ be a Virasoro coadjoint orbit with non-zero central charge (and energy 
bounded from below). Then there exists a Borel measure $d\mu(q)$ on $\cO_p$ which is quasi-invariant under 
the action of the Virasoro group (where $q\in\cO_p$). \label{pageconje}\\

In the remainder of this thesis we will rely on this conjecture in order to define the Hilbert space of a 
BMS$_3$ particle (at least one with bounded energy). The conjecture does not say how the measure 
$d\mu(q)$ is actually defined, but this is not a problem since the results of section \ref{seqareg} imply 
that induced representations based on 
different quasi-invariant measures are unitarily equivalent. Thus, assuming that the conjecture is 
true, we 
do not really need to know anything specific about the measure.\footnote{Note that the existence of a 
quasi-invariant measure $d\mu(q)$ on $\cO_p$ implies the existence 
of infinitely many other ones, since one can always multiply the measure by a 
strictly positive smooth function $\rho(q)$ and obtain a new measure $\rho(q)d\mu(q)$.} In 
particular 
the character computation of section \ref{sebmschar}, although relying on an unknown measure, will produce an 
unambiguous 
result.\\

Aside from these basic observations, we will have very few concrete things to say about the measure. 
Nevertheless the lines that follow are devoted to a brief review of the 
literature on Virasoro measure theory, with the intent of further motivating the validity 
of the 
conjecture. The reader who is not interested in mathematical subtleties is free to go directly 
to section \ref{susebadresKA}.

\paragraph{Remark.} Recall that all irreducible unitary representations of \it{regular} semi-direct products 
are induced representations, where regularity refers to the property defined at the end of section 
\ref{sesemi}.\i{regular semi-direct product}\i{semi-direct product!regular} Accordingly, in order to claim 
that all irreducible unitary representations of BMS$_3$ are 
BMS$_3$ particles whose supermomenta span Virasoro orbits, we would have to prove that the BMS$_3$ group is a 
regular semi-direct product, which in turn relies on the existence of a measure on the space of supermomenta. 
We will not address this question here and assume instead that the standard results on 
finite-dimensional semi-direct products carry over to BMS$_3$.

\subsubsection*{Measures on Virasoro orbits}

The issue of rigorously defining path integral measures was first addressed by Wiener about a century ago, in 
the context of stochastic processes. We refer e.g.\ to the 
biographical memoir in \cite{SegalMemoir} for more references and a more 
accurate account of the development of the subject. The problem of defining a Wiener-like quasi-invariant 
measure on Virasoro coadjoint orbits is more recent, but well known. In 
the physics literature, as in our conjecture above, the question of the measure is mostly treated in a 
heuristic way motivated by quantum mechanics; see e.g.\ \cite{Alekseev1990} for such an approach. By 
contrast, there is a fair amount of mathematical literature that aims at 
solving the problem in a rigorous way, and to our knowledge no definite, widely accepted solution is known at 
present. As announced above, we do not claim to provide an answer here; rather, we shall content ourselves 
with 
a brief literature review.\\

The main motivation for defining Virasoro measures comes from representations of the Virasoro algebra and 
conformal field theory; the hope is that such measures could provide a rigorous prescription for 
the geometric quantization of Virasoro orbits. This approach to the 
problem is adopted for instance in 
\cite{AiraultMal,AiraultFrench,AMT,Airault}. In \cite{Dai} the authors tackle the issue with a 
similar 
motivation, though with different methods; in 
particular it is suggested there that a measure might be provided by an infinite-dimensional version 
of the Liouville measure (\ref{volom}) obtained by taking the Kirillov-Kostant symplectic form 
(\ref{kkfovira}) to an infinite power.\\

A somewhat different approach consists in building measures on Virasoro orbits regardless of their relation 
to conformal field theory and highest-weight representations. This approach does not simplify the problem of 
geometric quantization of Virasoro orbits, but it does have the virtue of producing the desired measure. One 
should keep in mind that the measures needed for BMS$_3$ particles generally have nothing to do with those 
needed for geometric quantization of the Virasoro group, so the fact that a measure is 
unrelated to Virasoro representations is not a problem for our purposes. As it turns out, certain results due 
to Shavgulidze \cite{ShavgulidzeTheOriginal,ShavgulidzeBis,ShavgulidzeFR,Shavgulidze} precisely point in 
that direction (see also \cite{Bogachev,Shimomura}). Indeed it was shown in \cite{ShavgulidzeTheOriginal} 
that the group of diffeomorphisms of any compact manifold can be endowed with a quasi-invariant Borel 
measure. This measure then plays a role analogous to the Haar measure (recall section \ref{seqareg}) and 
ensures that quotients of the group, such as $\Diff/S^1$, can also be endowed with a quasi-invariant 
measure. As a corollary, Virasoro orbits (all of 
which are quotients of $\Diff$) should generically admit quasi-invariant Borel measures. This is precisely 
what is needed for BMS$_3$ particles, and it is in fact our main justification for the above conjecture.

\subsection{\ States of BMS$_3$ particles}
\label{susebadresKA}

Under the assumption that there exist quasi-invariant measures on Virasoro coadjoint orbits, we have all 
the ingredients required to build explicit unitary, irreducible, projective representations of the BMS$_3$ 
group. Here we 
describe and interpret the wavefunctions that represent the quantum states of a scalar massive BMS$_3$ 
particle with mass $M>0$. Spinning particles and the vacuum representation will be described later.

\subsubsection*{The states of a BMS$_3$ particle}

The supermomenta of a massive particle span an orbit $\cO_p=\Diff/S^1$ (with implicit central charge 
$c_2>0$), which we 
assume to admit a quasi-invariant measure $\mu$. For convenience the orbit representative $p$ 
is taken to be the supermomentum $p(\phii)=p_0=M-c_2/24$ at rest. As in (\ref{psiqqu}) the particle's Hilbert 
space $\sH$ consists of complex-valued wavefunctions\i{BMS$_3$ particle!wavefunction}\i{wavefunction}
\be
\Psi:\cO_p\rightarrow\CC:q\mapsto\Psi(q)
\label{psikku}
\ee
which are square-integrable with respect to $\mu$ in the usual sense that the integral
\be
\int_{\cO_p}d\mu(q)\left|\Psi(q)\right|^2
\label{skinteg}
\ee
is finite. As usual, $\Psi$ should be thought of as a 
wavefunction in (super)momentum space representing a wavepacket that propagates with fuzzy velocity. In 
contrast to the finite-dimensional case, however, 
the map (\ref{psikku}) is really a \it{functional} since it is defined on a space of 
functions:\i{wavefunctional}
\be
\Psi:q(\phii)\mapsto\Psi[q(\phii)].
\nn
\ee
Similarly the integral (\ref{skinteg}) is actually a functional integral. Despite these complications we will 
keep using the simpler notation $\Psi(q)$ for the wavefunctions of BMS$_3$ particles. The scalar product on 
$\sH=L^2(\cO_p,\mu,\CC)$ is then defined by (\ref{scall}) with $(\Phi(q)|\Psi(q))=\Phi^*(q)\Psi(q)$.\\

The space of states of a BMS$_3$ particle carries an action of BMS$_3$ by unitary transformations. 
Since we are assuming that the particle is \it{scalar}, the action of BMS$_3$ on $\sH$ is given 
by formula (\ref{indoo}):\i{scalar particle}\i{BMS$_3$ particle!scalar}
\be
\big(\cT[(f,\alpha)]\cdot\Psi\big)(q)
=
\sqrt{\rho_{f^{-1}}(q)}\;\,e^{i\langle q,\alpha\rangle}\;\Psi(f^{-1}\cdot q)\,.
\label{tabimm}
\ee
Here $(f,\alpha)$ is an element of the BMS$_3$ group (\ref{defbms}), with $f(\phii)$ a superrotation and 
$\alpha(\phii)$ a supertranslation. The point $q\in\cO_p$ is a supermomentum vector and its pairing $\bra 
q,\alpha\ket$ with $\alpha$ is given by (\ref{denpar}). The function $\rho_f(q)$ is the (unknown) 
Radon-Nikodym derivative (\ref{rnq}) of the measure $\mu$; if by chance the measure happens to be invariant, 
then one can set $\rho_f(q)=1$. Finally, the action 
$f\cdot q$ appearing in the argument of the wavefunction on the right-hand side is the 
BMS$_3$ generalization of the action of boosts on momenta; it is given by formula (\ref{ATOM}), which is the 
coadjoint representation of the Virasoro group at central charge $c_2$.

\paragraph{Remark.} Wavefunctionals are common in quantum field theory. Indeed, the quantum state of a 
typical field theory is a wavefunctional $\Psi[\phi(\bbx)]$, where $\phi(\bbx)$ is a spatial field 
configuration. The truly striking aspect of (\ref{psikku}) is not quite the fact that it belongs to a space 
of wavefunctionals, but rather that it provides an \it{irreducible} 
representation of the symmetry group.

\subsubsection*{BMS$_3$ particles as projective representations}

The central charge $c_2\neq 0$ turns expression (\ref{tabimm}) into a 
\it{projective} representation of the centreless BMS$_3$ group. Indeed, as is manifest in (\ref{bimipois}), 
$c_2$ is reponsible 
for an extra term in the commutation relations of superrotations with supertranslations. At the 
group-theoretic level this difference is due to the extra terms of the group operation (\ref{bimop}), as 
opposed to the centreless group operation (\ref{guixec}). In the latter case we have\i{projective 
representation!of BMS$_3$}\i{BMS$_3$ particle!as projective representation}
\be
(f,0)\cdot(e,\alpha)
=
(f,\Ad_f\alpha)
=
(e,\Ad_f\alpha)\cdot(f,0)
\label{fafati}
\ee
where $e$ is the identity in $\Diff$ and $\Ad$ is the action of diffeomorphisms on vector fields on the 
circle. By contrast, in the centrally extended case (\ref{bimop}) there 
are two extra slots for central terms and the analogue of (\ref{fafati}) becomes
\begin{align}
(f,0;0,0)\cdot (e,0;\alpha,0) & = \Big(f,0;\Ad_f\alpha,-\frac{1}{12}\bra\sfS[f],\alpha\ket\Big),\nn\\
(e,0;\alpha,0)\cdot (f,0;0,0) & = \Big(f,0;\Ad_f\alpha,0\Big)\nn
\end{align}
where the two lines differ by a term proportional to $\bra\sfS[f],\alpha\ket$, with $\sfS$ the Schwarzian 
derivative (\ref{swag}). This phenomenon is analogous to the 
statement that boosts and translations do not commute in the Bargmann group (\ref{galix}). In practice it 
means that the wavefunction obtained by acting first with $(e,\alpha)$, then 
by 
$(f,0)$, differs from the one obtained by acting first with $(f,0)$, then with $(e,\Ad_f\alpha)$, by a 
constant complex phase that can be evaluated using (\ref{fipp}):
\be
\cT[(f,0)]\cdot\cT[(e,\alpha)]
=
\exp\left[-i\frac{c_2}{12}\bra \sfS[f],\alpha\ket\right]
\cT[(e,\Ad_f\alpha)]\cdot\cT[(f,0)]\,.
\label{nocoti}
\ee
This is indeed the statement (\ref{cent}) that the representation $\cT$ is projective, when seen as a 
representation of the centreless BMS$_3$ group (\ref{defbms}). It is the BMS$_3$ analogue of the Galilean 
result (\ref{nocotiga}).\\

There is an important subtlety in (\ref{nocoti}): formula (\ref{tabimm}) is a projective 
representation only if one insists on using the \it{centreless} group operation (\ref{guixec}). If instead 
one 
uses the centrally extended group (\ref{hibiscus}), projectivity is absorbed by 
the definition of the group operation (\ref{bimop}). This is a restatement of our earlier observation in 
section \ref{s1.1} that projective representations can be seen in two equivalent ways: either as 
genuine projective representations of a centreless group, or as exact (non-projective) representations of a 
centrally extended group. This same argument is the reason why massive BMS$_3$ particles can have arbitrary 
real values of spin, as we shall discuss below.

\subsubsection*{Plane waves}

As in section \ref{sedefirep} we can describe the representation (\ref{tabimm})
in terms of a basis of one-particle states with definite (super)momentum on the orbit $\cO_p$. Let therefore 
$\delta$ denote the Dirac distribution associated with the measure $\mu$ and defined by the requirement 
(\ref{didi}). In the present case $\mu$ is a functional measure, so $\delta$ is a functional 
delta distribution. For $k\in\cO_p$, we define the plane wave state with supermomentum $k$ as\i{BMS$_3$ 
particle!with definite supermomentum}\i{plane wave}\i{delta function}
\be
\Psi_k(q)\equiv\delta(k,q)
\label{pawawest}
\ee
which is now a functional analogue of eq.\ (\ref{pwave}). 
It is a typical asymptotic state in a scattering experiment.
The scalar products of plane waves are given by (\ref{pwavs}):
\be
\langle\Psi_k|\Psi_{k'}\rangle
=
\delta(k,k').
\label{PlaSca}
\ee
Strictly speaking, plane waves are not square-integrable, hence do not belong to the space of states 
of a BMS$_3$ particle. They should therefore be understood in the weaker sense that any wavepacket 
(\ref{phipsi}) can be written as an infinite sum of plane waves. With this word of caution, one may say that 
plane waves form a ``basis'' of the space of states of a BMS$_3$ particle. Their 
transformation 
law under BMS$_3$ transformations is given by eq.\ (\ref{klakra}),
\be
\cT[(f,\alpha)]\cdot\Psi_k
=
\sqrt{\rho_f(k)}\;\,
e^{i\langle f\cdot k,\alpha\rangle}\;
\Psi_{f\cdot k}\,,
\label{Tiga}
\ee
except that we have removed the spin representation $\cR$ since the particle considered here has vanishing 
spin. This formula reflects the 
fact that a wavefunction with momentum $k$ boosted by a superrotation $f$ becomes a 
wavefunction with momentum $f\cdot k$. In short, all results of chapters \ref{c1b} and \ref{c2bis} remain 
valid, up to the fact that manifolds become spaces of functions while functions become functionals.

\subsection{\ Dressed particles and quantization}
\label{susebadres}

Asymptotically flat gravity enjoys BMS$_3$ symmetry, so its quantization is expected to produce 
unitary representations of the BMS$_3$ group. More precisely, since the phase space of metrics (\ref{piment}) 
is a hyperplane $c_1=0$, $c_2=3/G$ in the 
space of the coadjoint representation of $\hBMS$, the geometric quantization of the orbit (\ref{wijjipi}) of 
a metric $(j,p)$ under asymptotic symmetry transformations is expected to produce an irreducible unitary 
representation of BMS$_3$ with supermomentum orbit $\cO_p$ and spin (\ref{jeress}) determined by the 
restriction of $j$ to 
the Lie algebra of the little group of $p$. Thus,\i{quantum gravity!and BMS$_3$ particles}\i{BMS$_3$ 
particle!as quantized metric}\i{boundary graviton!quantization}\i{geometric quantization!and boundary 
gravitons}
\be
\begin{array}{c}
\text{\it{A BMS$_3$ particle is the quantization}}\\
\text{\it{of the orbit of a metric under BMS$_3$ transformations.}}
\end{array}
\nn
\ee
In the following pages we use this observation to compare BMS$_3$ particles with standard relativistic 
particles. For simplicity we focus on a massive scalar particle whose orbit representative is taken to be the 
supermomentum $p=M-c_2/24$ at rest.

\subsubsection*{Leaking wavefunctions}

The transformation law (\ref{tabimm}) is an infinite-dimensional generalization of a scalar representation 
(with mass $M$) of the Poincar\'e group in three dimensions. Indeed, by restricting one's attention to the 
Poincar\'e subgroup of BMS$_3$, one obtains a (highly reducible) unitary representation of Poincar\'e. The 
latter contains the standard scalar irreducible representation with mass $M$, but it also contains 
an uncountable infinity of other representations with higher mass.
These extra representations 
arise because the action of Lorentz transformations on the supermomentum orbit $\cO_p\cong\Diff/S^1$ is 
not transitive; in fact, the set of Lorentz-inequivalent 
supermomenta\i{supermomentum!Lorentz-inequivalent} is an infinite-dimensional manifold
\be
\text{PSL}(2,\RR)\backslash\Diff/S^1,
\label{douquo}
\ee
which is a double quotient of $\Diff$. The quotient on the left is taken with respect to the Lorentz group 
$\text{PSL}(2,\RR)$.\\

This observation can be rephrased in a more intuitive way: it says that the 
supermomentum orbit $\cO_p=\Diff/S^1$ contains infinitely many finite-dimensional submanifolds $\SL/S^1$ 
obtained by acting on the orbit
with Lorentz transformations. Each submanifold is a standard momentum orbit (\ref{Mopping}) for massive 
Poincar\'e particles in three dimensions. Any two of those submanifolds are mutually Lorentz-inequivalent; 
the set of such 
inequivalent sub-orbits is the space (\ref{douquo}).

\begin{figure}[H]
\centering
\includegraphics[width=0.60\textwidth]{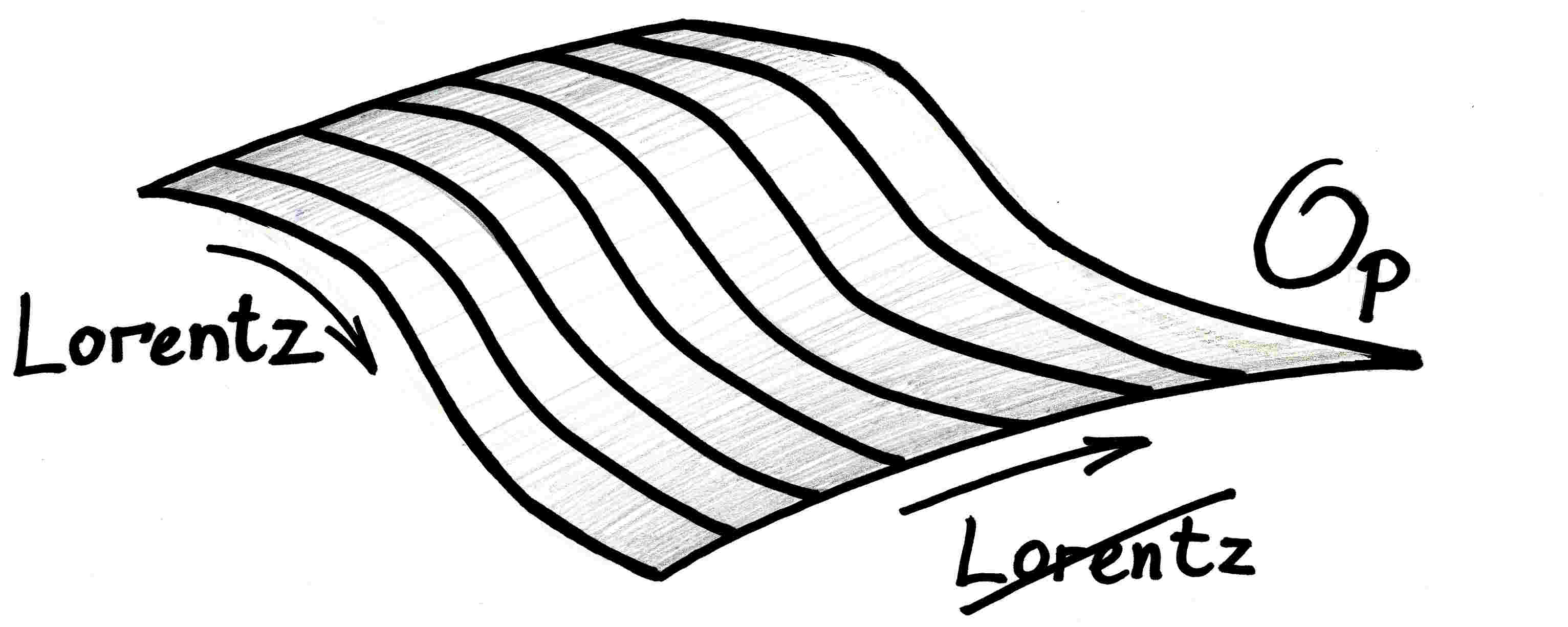}
\caption{A supermomentum orbit $\cO_p\cong\Diff/S^1$ with embedded Lorentz sub-orbits represented as curvy 
lines. Lorentz 
transformations move points along these sub-orbits. Transitions 
from one sub-orbit to another are only possible with superrotations that do not belong to the 
Lorentz subgroup of $\Diff$. One can define an equivalence relation on the supermomentum orbit by declaring 
that two points are equivalent if 
they belong to the same Lorentz sub-orbit. The quotient of $\cO_p$ by that equivalence 
relation is the double quotient (\ref{douquo}).\label{figembed}}
\end{figure}

Now, a wavefunction $\Psi(q)$ of a BMS$_3$ particle is 
never suppported on just one Lorentz sub-orbit of $\cO_p$: if it was, some components of its supermomentum 
would be sharply defined and the uncertainty principle would be violated. (By the way, we stress again 
that the plane waves (\ref{pawawest}) do \it{not} actually belong to the Hilbert space.) Rather, the 
wavefunction spreads 
over many Lorentz-inequivalent momentum orbits. In other words, the wavefunction 
$\Psi(q)$ ``leaks'' into 
the directions of supermomentum obtained by acting with superrotations that are \it{not} Lorentz 
transformations:\i{leaking wavefunction}

\begin{figure}[H]
\centering
\includegraphics[width=0.60\textwidth]{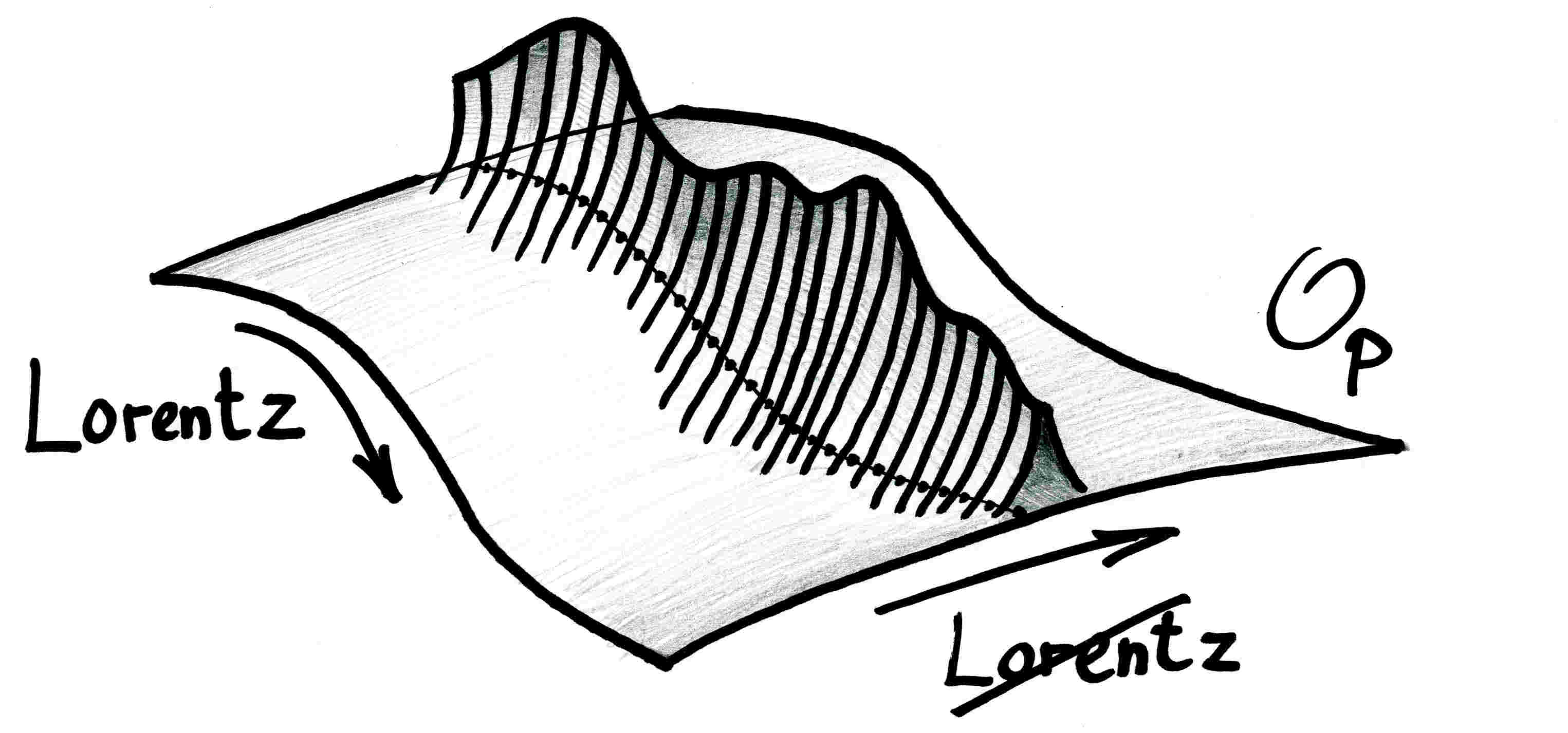}
\caption{The supermomentum orbit of fig.\ \ref{figembed}, now with a wavefunction on top. The wavefunction is 
roughly supported on a Lorentz sub-orbit, but not quite: it leaks into directions that cannot be achieved 
with 
Lorentz 
transformations.\label{figoleak}}
\end{figure}

\paragraph{Remark.} The double quotient (\ref{douquo}) is an application of the \it{induction-reduction 
theorem}\i{induction-reduction theorem} for induced representations. The latter roughly goes as follows: Let 
$G$ be a group with two (generally different) subgroups $H,H'$ and let $\cS$ be an irreducible representation 
of $H$. Then the restriction to $H'$ of the induced representation $\text{Ind}_H^G(\cS)$ is a direct integral 
of irreducible representations of $H'$ labelled by the points of the double quotient $H'\backslash G/H$. In 
(\ref{douquo}) we have applied this theorem to $G=\BMS$, $H=\un\ltimes\Vect$ and $H'=\PSL\ltimes\sl$. The 
latter is the Poincar\'e group in three dimensions while the $\un$ of $H$ is the little group for massive 
particles.

\subsubsection*{Gravitational dressing}

We have just explained that any supermomentum orbit contains infinitely many 
Poin\-ca\-r\'e\--inequivalent sub-orbits. It is natural to wonder how the extra directions of the orbit --- 
those 
that do not lie along Lorentz generators --- are to be interpreted. In other words: how should one think of 
the fact that the wavefunction of a BMS$_3$ particle leaks into directions that are forbidden by 
Poincar\'e transformations?\\

A natural guess is suggested by the very origin of the BMS$_3$ group: as we 
showed in section \ref{sebmsboco}, it is the asymptotic symmetry group of Minkowskian space-times (in three 
dimensions). Asymptotic symmetries may be thought of as generalizations of isometries that 
incorportate gravitational fluctuations. Now, the isometry group of Minkowski space-time is the Poincar\'e 
group, 
and its unitary representations are particles in the usual sense. Accordingly,\i{BMS$_3$ particle!as 
dressed particle}\i{gravitational dressing}
\be
\begin{array}{c}
\text{\it{A BMS particle is a Poincar\'e particle}}\\
\text{\it{dressed with gravitational degrees of freedom.}}
\end{array}
\label{bimbo}
\ee
\vspace{.1cm}

Let us be more precise about what we mean by ``gravitational degrees of freedom''. Classically, those are the 
classes of points on the orbit $\cO_p$ that cannot be obtained from $p$ by a Lorentz transformation; the set 
of these classes coincides with the double coset space (\ref{douquo}). Quantum-mechanically, we would 
like 
these gravitational degrees of freedom to correspond to the set of quantum states that \it{cannot} be 
obtained from the state of a particle at rest by Lorentz transformations. This can be rephrased in precise 
terms: on the Hilbert space $\sH$ of a BMS$_3$ particle we define an equivalence 
relation $\sim$ such that $\Psi\sim\Psi'$ if there exists a Poincar\'e 
transformation $(f,\alpha)$ for which $\Psi'=\cT[(f,\alpha)]\Psi$. Then the space of Poincar\'e-inequivalent 
states is the quotient
\be
\sH/\sim\,.
\label{sahasim}
\ee
The latter can also be seen as the set of quantum states obtained by acting on a state at rest with 
supertranslations and superrotations 
that do not belong to the Poincar\'e subgroup. In this 
sense it is the three-dimensional analogue of the space of \it{soft gravitons} in four dimensions, as follows 
from the recently discovered relation \cite{Strominger:2013lka,Strominger:2013jfa} between 
asymptotic symmetries and soft theorems in gauge theories. Thus,\i{soft graviton}
\be
\begin{array}{c}
\text{\it{A BMS particle is a particle dressed with soft gravitons.}}
\end{array}
\nn
\ee
Note that the terminology of ``soft gravitons'' is a bit dangerous here, since three-dimensional Einstein 
gravity has no local degrees of freedom, hence no genuine (bulk) gravitons. We already pointed out this 
subtlety in the introduction of the thesis, and our point of view remains the same: owing to the relation 
between soft gravitons and asymptotic symmetries, any theory with non-trivial asymptotic symmetries can be 
interpreted as a theory containing soft degrees of freedom, regardless of the presence of bulk 
degrees of freedom. In this sense three-dimensional gravity is a toy model for soft gravitons.

\subsection{\ The BMS$_3$ vacuum}
\label{suseother}

Having analysed massive scalar BMS$_3$ particles, we now turn to some of their cousins. Here we describe the 
vacuum BMS$_3$ representation, while spinning particles are studied in section \ref{suseSABAM}.\\

The vacuum representation of BMS$_3$ is the scalar induced representation based on the vacuum 
supermomentum orbit --- the one containing the constant $p_{\text{vac}}=-c_2/24$. The 
corresponding little group is the Lorentz 
group $\text{PSL}(2,\RR)$, so the orbit is\i{vacuum orbit!for BMS$_3$}\i{BMS$_3$ vacuum}
\be
\cO_{\text{vac}}
\cong
\Diff/\text{PSL}(2,\RR).
\label{vacorbit}
\ee
As before we assume that it admits a measure which is quasi-invariant under superrotations. The 
Hilbert space of the representation is then spanned by square-integrable wavefunctions (\ref{psikku}) on 
$\cO_{\text{vac}}$ transforming under BMS$_3$ according to (\ref{tabimm}).\\

The interesting aspect of the vacuum representation is its interpretation. Indeed, while massive particles 
exist in both the Poincar\'e group and the BMS$_3$ group, the vacuum representation is non-trivial only in 
the BMS$_3$ context. This is analogous to the observation of section \ref{sevirep} that the vacuum 
representation of the Virasoro group is non-trivial. In particular, as in Virasoro, there is no 
fully BMS$_3$-invariant definition of the vacuum at non-zero central charge; the maximal possibility is 
Poincar\'e invariance, which is indeed achieved by $p_{\text{vac}}=-c_2/24$ (with $j=0$). This reduced 
symmetry is responsible for the non-triviality of the representation and for the fact that wavefunctions of 
the vacuum representation ``leak'' into directions that cannot be reached by Poincar\'e transformations. As 
in the massive case (\ref{douquo}), the set of Lorentz-inequivalent supermomenta on the vacuum orbit is a 
double coset space
\be
\text{PSL}(2,\RR)\backslash\Diff/\text{PSL}(2,\RR)
\label{vqlui}
\ee
which parameterizes the decomposition of the vacuum BMS$_3$ representation as a direct integral of 
Poincar\'e sub-representations.\\

The presence of a non-trivial vacuum representation can be interpreted in gravitational terms. Indeed, 
according to our heuristic proposal (\ref{bimbo}), the vacuum BMS$_3$ representation consists \it{only} of 
gravitational degrees of freedom (since the corresponding Poincar\'e particle is trivial). Classically these 
degrees of freedom span the BMS$_3$ orbit of the Minkowski metric. In the terminology of section 
\ref{sebitu} those would be ``boundary gravitons'',\i{boundary graviton} or equivalently soft 
gravitons, around Minkowski space.

\subsubsection*{Dressed particles revisited}

Let us return to the interpretation of BMS$_3$ particles as dressed particles, now using the vacuum 
representation as an extra input. We start with the following observation: consider the space 
$L^2(\cM\times\cN,\mu)$ of square-integrable functions on the product space 
$\cM\times\cN$; suppose the measure $\mu$ factorizes as a product 
$\mu=\mu_{\cM}\times\mu_{\cN}$,\i{disintegration of a measure}\i{measure!disintegration} where 
$\mu_{\cM}$ is a measure on $\cM$ and $\mu_{\cN}$ is a measure on $\cN$. Then one has a tensor product 
decomposition\i{tensor product}
\be
L^2(\cM\times\cN,\mu_{\cM}\times\mu_{\cN})
\cong
L^2(\cM,\mu_{\cM})\otimes L^2(\cN,\mu_{\cN}).
\label{lituli}
\ee
Let us use this result to compare massive particles, with supermomentum orbits $\Diff/S^1$, to the vacuum 
whose orbit is $\Diff/\text{PSL}(2,\RR)$. Since both of 
these infinite-dimensional manifolds are homotopic to a point, we can 
relate them as
\be
\Diff/S^1
\cong
\big(\text{PSL}(2,\RR)/S^1\big)
\times
\big(\Diff/\text{PSL}(2,\RR)\big).
\label{difodif}
\ee
This is to say that the 
massive supermomentum orbit is a direct product $\cM\times\cN$. The first factor of the product 
is the Poincar\'e momentum orbit (\ref{Mopping}) of a massive particle in three dimensions, which suggests 
that a BMS$_3$ 
particle is equivalent to a relativistic particle ``times'' the vacuum BMS$_3$ representation. 
This can be made precise using (\ref{lituli}): if the measure $\mu$ used to define the scalar product of 
wavefunctions for a massive BMS$_3$ particle factorizes into a product on
$\Diff/\text{PSL}(2,\RR)$ and $\text{PSL}(2,\RR)/S^1$, then the Hilbert space of a massive 
BMS$_3$ particle factorizes into\i{massive BMS$_3$ particle!factorization}
\be
\sH_{\text{BMS}}
\cong
\sH_{\text{Poinc}}
\otimes
\sH_{\text{vac}}
\label{sqoei}
\ee
where $\sH_{\text{Poinc}}$ is the space of states of a massive Poincar\'e particle and 
$\sH_{\text{vac}}$ is that of the BMS$_3$ vacuum representation.\\

One can reformulate the statement (\ref{sqoei}) in the basis of plane wave states (\ref{pawawest}). Indeed, 
on 
a massive supermomentum orbit $\cO_p=\Diff/S^1$, the diffeomorphism (\ref{difodif}) allows us to write any 
supermomentum $q$ as a pair $q=(q_{\text{Poinc}},q_{\text{vac}})$ where $q_{\text{Poinc}}$ is a momentum 
vector with three components belonging to the Poincar\'e sub-orbit 
$\cO_{\text{Poinc}}=\text{PSL}(2,\RR)/S^1$, while $q_{\text{vac}}$ is a supermomentum that belongs to the 
vacuum BMS$_3$ orbit $\cO_{\text{vac}}=\Diff/\text{PSL}(2,\RR)$. Then, under 
the assumption that the measure $\mu$ on $\cO_p$ disintegrates into a product of measures on 
$\PSL/S^1$ and $\cO_{\text{vac}}$, any plane wave (\ref{pawawest}) can be written as a tensor 
product
\be
\Psi_k
=
\Psi_{k_{\text{Poinc}}}
\otimes
\Psi_{k_{\text{vac}}}
\nn
\ee
where the left and right factors of the product are plane waves on the orbits $\cO_{\text{Poinc}}$ and 
$\cO_{\text{vac}}$, respectively. A generic state of a BMS$_3$ particle is an infinite linear 
combination of such factorized plane waves. Note that none of our upcoming conclusions rely on this 
phenomenon, so in the sequel we will not necessarily assume that the measure $\mu$ disintegrates into a 
product.

\paragraph{Remark.} The decomposition (\ref{difodif}) is similar to that of the Poincar\'e $D$-momentum 
$p_{\mu}=(E,\bbp)$. In the latter case the truly covariant quantity is $p_{\mu}$ but our non-relativistic 
intuition splits it into an energy-momentum pair $(E,\bbp)$. In the same way, for BMS$_3$ particles the truly 
covariant quantity is the full supermomentum $p(\phii)$, but our intuition splits it into a pair 
$(p_{\text{Poinc}},p_{\text{vac}})$.

\subsection{\ Spinning BMS$_3$ particles}
\label{suseSABAM}

We finally turn to the spinning generalization of the BMS$_3$ representations considered above. We have 
already addressed almost all the subtleties of the construction, so we display the inclusion of spin merely 
for completeness. In short, our main conclusion will be that the spin of massive BMS$_3$ particles is 
not quantized, exactly as in the Poincar\'e group in three dimensions.\i{spin!for BMS$_3$ 
particle}\i{BMS$_3$ particle!spin} The reader who is happy to accept 
this result, or to deal only with scalar particles, may go to section \ref{susefour?}.\\

We recall from (\ref{rspin}) that spin is the label that specifies the representation of 
the little group chosen for the description of a particle. In the case of BMS$_3$ particles at 
non-zero $c_2$, the little groups were described at the end of subsection \ref{susebilly}. They are all 
either one-dimensional Abelian groups such as $\un$ or $\RR$ (possibly up to discrete factors), or $n$-fold 
covers of the Lorentz group $\text{PSL}(2,\RR)$. All these are finite-dimensional Lie groups and their 
unitary 
representations are known, so writing down generic spinning representations of BMS$_3$ is 
mostly a technical problem.\\

For definiteness, let us consider a massive BMS$_3$ particle. Its little group $\un$ consists of 
spatial 
rotations, exactly as for massive Poincar\'e particles in three dimensions. All irreducible unitary 
representations of $\un$ are of the form (\ref{tu1}), with $s$ an integer. However, the BMS$_3$ group 
(\ref{defbms}) has the same homotopy type as $\Diff$, which is homotopic to a circle. This implies that it 
admits topological projective representations of the type 
described in section \ref{s1.1}, which can be classified by considering 
exact (non-projective) representations of the universal cover (\ref{debmscov}) of BMS$_3$. In the latter 
case, the little group of massive particles gets unwrapped from $\un$ to $\RR$, whose unitary representations 
are now labelled by an arbitrary real spin $s$. Since those are the physically relevant representations, we 
conclude that the spin of a massive BMS$_3$ particle is generally an arbitrary real number. In particular, 
most massive BMS$_3$ particles are anyons.\i{anyon} This is the same conclusion as 
in the Poincar\'e group in three dimensions.\\

Now suppose we fix a certain value of mass $M$ and spin $s\in\RR$ and ask what are the states of the 
corresponding BMS$_3$ 
particle. We denote the spin $s$ representation of the little group by $\cR$; it is a one-dimensional 
representation of the form (\ref{tu1}). Thus the Hilbert space of the BMS$_3$ particle consists again of 
complex-valued wavefunctions (\ref{psikku}), but their transformation law under BMS$_3$ contains an extra 
term 
with respect to the scalar representation (\ref{tabimm}). That extra term involves the Wigner rotation 
(\ref{winota}) associated with the superrotation $f$ and the supermomentum $q$,\i{Wigner rotation}
\be
W_q[f]
=
\cR[g_q^{-1}\,f\,g_{f^{-1}\cdot q}]\,,
\label{wikipedia}
\ee
where the superrotations $g_q$ ($q\in\cO_p$) are standard boosts such that $g_q\cdot p=q$. Thus we run into 
the problem of finding standard boosts for a massive supermomentum orbit.\\

Incidentally we have already defined such standard boosts, though not in the same language.\i{standard 
boost!for BMS$_3$} Indeed, eq.\ 
(\ref{gikidota}) is precisely the definition of standard boosts for elliptic Virasoro coadjoint orbits. 
These boosts are built as follows:
\begin{enumerate}
\item Take the supermomentum $q(\phii)$ with central charge $c_2$; write down the associated Hill's equation 
(\ref{hill}) with the replacement $(p,c)\rightarrow(q,c_2)$.
\item Find two linearly independent solutions $\psi_1,\psi_2$ of Hill's equation satisfying the Wronskian 
condition (\ref{wroko}).
\item Define a vector field $X_q(\phii)$ by (\ref{expiation}).
\item Define a diffeomorphism $f$ of $S^1$ by (\ref{bagaku}), with $p_0=M-c_2/24$
\item The standard boost associated with $q$ is $g_q=f^{-1}$.
\end{enumerate}
This procedure is somewhat convoluted, but it does provide a family of standard 
boosts on the orbit of a massive supermomentum, as desired. We refrain here from actually computing 
these boosts.\\

Equipped with standard boosts one can write down the transformation law of massive BMS$_3$ particles with 
non-zero spin, given by eq.\ (\ref{spipa}):\i{induced representation}
\be
\big(\cT[(f,\alpha)]\cdot\Psi\big)(q)
=
\sqrt{\rho_{f^{-1}}(q)}\;\,e^{i\langle q,\alpha\rangle}\,
W_q[f]
\cdot
\Psi(f^{-1}\cdot q)\,,
\nn
\ee
where the notation is the same as in (\ref{tabimm}) up to the insertion of the Wigner rotation 
(\ref{wikipedia}). This can also be rewritten in terms of plane waves (\ref{pawawest}) as
\be
\cT[(f,\alpha)]\cdot\Psi_k
=
\sqrt{\rho_f(k)}\;\,
e^{i\langle f\cdot k,\alpha\rangle}\,
\cR\left[
g_{f\cdot k}^{-1}\,f\,g_k
\right]
\cdot
\Psi_{f\cdot k}
\nn
\ee
(recall eq.\ (\ref{klakra})). The interpretation of all these formulas is the same as before, up to the extra 
Wigner rotation. In particular, a spinning BMS$_3$ particle is a spinning Poincar\'e 
particle dressed with soft gravitons.

\subsection{\ BMS particles in four dimensions?}
\label{susefour?}

Having described BMS particles in three dimensions, we now ask to what extent our observations apply to the
realistic four-dimensional case. To answer this we first briefly review some previous 
literature on BMS in four dimensions and its representations, before exposing our viewpoint on 
the matter in light of the more recent developments relating BMS symmetry to soft theorems.\footnote{To our 
knowledge there is, at present, no detailed review on BMS 
symmetry. Accordingly the literature review provided here cannot fail to be biased by the author's ignorance; 
we apologize in advance for the references that we may have missed.} Our 
approach will be mostly qualitative and heuristic.

\subsubsection*{A history of BMS symmetry}

BMS symmetry was discovered in the sixties by Bondi, Van der Burg, Metzner 
\cite{Bondi:1960jsa,Bondi:1962px} and Sachs \cite{Sachs:1962zza,Sachs:1962wk}, as a group (\ref{BMS}) of 
globally well-defined diffeomorphisms of asymptotically flat space-times.\i{BMS group}\i{global BMS group} 
The 
notion of 
asymptotic flatness 
was put on firmer ground shortly thereafter, thanks to the notion of conformal compactifications\i{conformal 
compactification} 
\cite{Penrose:1962ij}. Arguably, at the time of the discovery, the presence of an infinite-dimensional 
group of supertranslations was seen as something of a pathology. Nevertheless it was quickly suggested that 
full BMS symmetry (with the supertranslations turned 
on) could be used to discuss aspects of both quantum gravity \cite{Komar:1965zz} and $S$-matrix physics 
\cite{Sachs:1962zza}. This led to the study of unitary representations of BMS.\\

The first suggestion that BMS representations might be relevant to particle physics appeared in 
\cite{Sachs:1962zza}. A little later McCarthy and collaborators set out to study the 
representation theory of the global BMS group in full detail, with scattering amplitudes as a motivation 
\cite{Mccarthy:1972ry}. Owing to the semi-direct product structure of (\ref{BMS}) the strategy was to build 
induced representations \`a la Wigner in terms of orbits and little groups, as described in this thesis 
in chapter \ref{c2bis}. In particular it was shown in \cite{McCarthy01,McCarthy00} that all little groups are 
compact, leading to the conclusion that BMS particles\i{BMS particle!in 4D} in four dimensions 
cannot have continuous spin, in 
contrast to their Poincar\'e counterparts. This was followed by the observation that the restriction of a BMS 
representation to its Poincar\'e subgroup is reducible and consists of a tower of Poincar\'e particles with 
different spins \cite{McCarthy317}; in \cite{McCarthy301} this spin mixing was interpreted as being due to 
the 
presence of the gravitational field. It was also shown in \cite{Crampin55} that certain BMS representations 
studied earlier in 
\cite{Cantoni1966,Cantoni1967} were in fact reducible induced representations.\\

Along the way it was realized that the absence of continuous-spin particles exhibited in 
\cite{McCarthy01,McCarthy00} was due to a delicate choice of topology, and that different topologies lead to 
radically different conclusions, including particles with continuous spin 
\cite{Girardello:1974sq}.\i{continuous spin!for BMS particles} These 
continuous-spin particles were then interpreted as scattering states in \cite{Crampin:1974aw,McCarthy489}. 
Finally, the whole construction was put on firm mathematical ground in \cite{Piard01}, where the theory 
of induced representations was extended to groups of the type $G\ltimes A$ with an infinite-dimensional 
Abelian group $A$. As a corollary, it was shown in \cite{Piard:1977vh} that induced representations \`a la 
Wigner exhaust all irreducible unitary representations of the global BMS group (\ref{BMS}). This analysis 
was later completed by the proof that the global BMS group has no non-trivial central extensions, and 
therefore admits no projective representations other than those originating from its non-trivial topology 
\cite{Cattaneo:1979tj}.

\subsubsection*{BMS symmetry and holography}

For about two decades after McCarthy's work on BMS representations, the study of BMS symmetry as such appears 
to have slowed down, with the exception of the discovery of its supersymmetric version in 
\cite{Awada:1985by}. Nevertheless, substantial progress was made during that period in closely related areas, 
particularly in the study of the structure of gravity near null infinity (see e.g.\ 
\cite{Geroch1977,Geroch:1978us,Ashtekar:1978zz,Ashtekar:1981bq}). This led for instance to the idea of 
asymptotic quantization \cite{Ashtekar:1981sf,Ashtekar:1987tt},\i{asymptotic quantization} which can roughly 
be thought of as a 
quantization of bulk degrees of freedom obtained by quantizing a suitable radiative phase space on the 
boundary --- an idea that sounds prophetic given the development of holography about ten years later.\\

As mentioned in the introduction of this thesis, holography emerged from general considerations on the nature 
of quantum gravity \cite{'tHooft:1993gx,Susskind:1994vu} guided by the seminal observation by Bekenstein and 
Hawking that black holes have entropy \cite{Bekenstein:1973ur,Hawking:1974sw}. Its best known realization 
occurs in the AdS/CFT duality \cite{Maldacena:1997re,Witten:1998qj,Gubser:1998bc}, but it was soon suggested 
that a suitable notion of holography should hold for all families of space-times (see 
e.g.\ \cite{Bousso:1999cb,Bousso:1999dw}), and in particular for asymptotically flat gravity. In 
that context interest in BMS symmetry slowly re-emerged 
\cite{Arcioni:2003xx,Arcioni:2003td,Dappiaggi:2004cp,Dappiaggi:2004kv,Arcioni:2004ba}, leading 
in particular 
to the proposal \cite{Barnich:2009se,Barnich:2010eb,Banks:2003vp} that the globally well-defined BMS group 
(\ref{BMS}) should be extended to include a Virasoro-like semi-group of local conformal transformations of 
celestial 
spheres. This proposal is the origin of the terminology of ``superrotations'' for asymptotic symmetry 
transformations that extend Lorentz transformations. The study of asymptotically flat holography then took 
off, both in four dimensions 
\cite{Virmani:2011aa,Barnich:2013axa,Barnich:2013sxa,Lambert:2014poa,Fujisawa:2015asa} and in three 
dimensions \cite{Barnich:2011ct,Barnich:2012aw,Gonzalez:2012nv,Barnich:2012xq,Barnich:2012rz,Barnich:2013yka, 
Schulgin:2013xya,Krishnan:2013wta,Barnich:2015uva,Oblak:2015sea,Hartong:2015usd,Bonzom:2015ans,Barnich:2015dvt
,Troessaert:2015syk}, although in the latter case a substantial part of the literature is written in the 
language of Galilean conformal symmetry 
\cite{Bagchi:2010eg,Bagchi:2010zz,Bagchi:2012cy,Bagchi:2012yk,Bagchi:2012xr,Afshar:2013bla,Fareghbal:2013ifa,
Bagchi:2014ava,Fareghbal:2014qga,Fareghbal:2014kfa,Bagchi:2015wna,Hosseini:2015uba,Fareghbal:2015bxd,
Banerjee:2015kcx,Fareghbal:2016hqr}.\\

Along the way, it was realized by Strominger and collaborators that BMS 
symmetry does have highly 
non-trivial implications for the $S$-matrix, in the form of soft graviton theorems\i{soft graviton} 
\cite{Strominger:2013jfa,He:2014laa,Cachazo:2014fwa,Kapec:2014opa,Kapec:2015vwa}. This discovery 
sparked a flurry of papers discussing the applications of the BMS group (and its gauge-theoretic 
generalizations 
\cite{Strominger:2013lka,He:2014cra,Lysov:2014csa,Kapec:2014zla,He:2015zea,Kapec:2015ena,Strominger:2015bla,
Dumitrescu:2015fej}) to scattering amplitudes \cite{Campiglia:2015qka,Campiglia:2015kxa}, memory effects 
\cite{Strominger:2014pwa,Pasterski:2015tva,Pasterski:2015zua} and more recently to black holes 
\cite{Hawking:2016msc}.

\subsubsection*{BMS$_4$ particles?}

Despite recent progress, we still seem to be quite far from having truly understood BMS symmetry in four 
dimensions. Representation theory provides an easy way to illustrate the 
problem. Indeed, one of the cornerstones of the relation between BMS symmetry and soft theorems is the fact 
that supertranslations generate soft graviton states when acting on the vacuum. Accordingly, if the 
global BMS group (\ref{BMS}) is correct, then the representations considered by McCarthy in 
\cite{McCarthy01,McCarthy00} should 
account for this effect: they should represent Poincar\'e particles dressed with soft gravitons.\\

However, 
it is easy to see that this is not the case. To illustrate this point, consider the analogue of the 
global BMS group (\ref{BMS}) in three dimensions,\i{global BMS group!in 3D}
\be
\text{gBMS}_3
\equiv
\text{PSL}(2,\RR)\ltimes\Vect_{\text{Ab}}\,,
\label{defgbms}
\ee
where $\PSL$ is the Lorentz subgroup of $\Diff$ consisting of projective 
transformations (\ref{protophi}) and acting on superrotations according to (\ref{advecis}). The 
representations of this group would be induced exactly in the same way as for the standard BMS$_3$ group 
(\ref{defbms}), but there would be two crucial differences:
\begin{itemize}
\item There would be no non-trivial central extensions.
\item The supermomentum orbits would all be finite-dimensional (since $\PSL$ is finite-dimensional).
\end{itemize}
In particular the Hilbert space of any irreducible unitary representation of the group (\ref{defgbms}) 
coincides with the space of states of a Poincar\'e particle; it consists of wavefunctions on a 
finite-dimensional momentum orbit. In fact, the only difference between the representations of this group and 
those of Poincar\'e would be that translations are paired with supermomenta according to the functional 
formula (\ref{denpar}) rather than a finite-dimensional product $p_{\mu}\alpha^{\mu}$. In particular the 
vacuum representation would be trivial and there would be no way for quantum supertranslations to create soft 
graviton states upon acting on the vacuum.\\

These observations exhibit an important point: the global BMS group (\ref{BMS}) 
cannot be the end of the story. It is too small to account for soft graviton degrees of freedom, 
and it must be extended in some way. Unfortunately the argument does not tell us \it{how} BMS symmetry should 
be extended. To our knowledge, two 
proposals have been formulated so far, both suggesting a superrotational 
extension of the Lorentz group.\i{extended BMS group} The first is the aforementioned idea of turning 
superrotations into a 
semi-group of local conformal transformations of celestial spheres 
\cite{Barnich:2009se,Barnich:2011ct,Barnich:2010eb,Barnich:2011mi,Barnich:2016lyg}; the second 
suggests that superrotations should instead span a group $\text{Diff}(S^2)$ of diffeomorphisms of the 
sphere \cite{Campiglia:2014yka,Campiglia:2015yka}. 
It appears that there are currently no definitive arguments for selecting one 
proposal over the other; 
see however \cite{Compere:2016jwb}, where it is argued that finite singular conformal transformations of 
celestial spheres in four dimensions are pathological.\\

The fact that BMS symmetry in four dimensions is ill-defined is a call for further developments. This thesis 
is one of them: it aims at understanding a three-dimensional toy model and using it as a guide for the 
realistic problem. Indeed, many properties that we have encountered in our investigation of BMS$_3$ should 
remain true in BMS$_4$. In particular the semi-direct product structure $G\ltimes A$ appears to be a robust 
feature and implies that BMS particles are classified by supermomentum orbits that coincide with orbits of 
the Bondi mass aspect under asymptotic symmetry transformations. Furthermore the occurrence of a central 
extension pairing superrotations with supertranslations has also been observed in BMS$_4$ 
\cite{Barnich:2011mi}. However, a sharp difference between BMS$_3$ and BMS$_4$ is that, in the former, 
supertranslations do \it{not} create new states when acting on the vacuum. A possibly related difference is 
that the language suited to the study of BMS$_4$ appears to be that of \it{groupoids} rather than groups. We 
will not have much more to say about this here, and return now to our study of BMS$_3$.

\section{\ BMS modules and flat limits}
\label{sebomodule}

In the previous pages we have described irreducible, unitary representations of BMS$_3$. In 
order to make contact with the representation theory of the Virasoro algebra it is useful to reformulate 
these representations in Lie-algebraic language, in the form of so-called \it{induced modules}.\i{induced 
module} This 
reformulation will also allow us to discuss the flat limit of dressed particles in AdS$_3$ and to understand 
the difference between unitarity in BMS$_3$ and unitarity for Galilean conformal symmetry. The plan is as 
follows: we first describe induced modules for the Poincar\'e algebra in three dimensions and interpret them 
as ultrarelativistic limits of highest-weight representations of $\sl$, before applying the same construction 
to the $\bms$ algebra. The 
presentation is adapted from \cite{Campoleoni:2016vsh}.

\subsection{\ Poincar\'e modules in three dimensions}
\label{COPsec:poincare}

Our goal here is to rewrite the three-dimensional relativistic particles of section \ref{sePoTri}
in Lie-algebraic language. This will allow us to relate Poincar\'e representations with the ultrarelativistic 
limit of highest-weight 
representations of $\sl\oplus\sl$.

\subsubsection*{Poincar\'e algebra}

In three dimensions, the Lie algebra of the Poincar\'e group is spanned by three Lorentz generators 
$j_m$ and three translation generators $p_m$ ($m=-1,0,1$) with Lie brackets (\ref{yebem}). As in the case 
(\ref{ellemok}) of $\sl$, it is more convenient to use a different complexified basis $J_m=ij_m$, 
$P_m=ip_m$, in terms of which the brackets become\i{Poincar\'e algebra}
\be
[J_m,J_n]=(m-n)\,J_{m+n}\,,\qquad
[J_m,P_n]=(m-n)\,P_{m+n}\,,
\qquad
[P_m,P_n]=0\,.
\label{COPpal}
\ee
These conventions are such that, in any 
unitary representation, the 
operators representing Poincar\'e generators satisfy Hermiticity conditions of the type 
(\ref{hermiconw}):\i{Hermiticity conditions!for Poincar\'e}
\be
\label{COPherm}
(P_m)^{\dagger}=P_{-m}\,,
\qquad
(J_m)^{\dagger}=J_{-m}\,.
\ee
As in section \ref{sevirep}, we abuse notation by denoting with the same letter both the abstract generators 
$J_m,P_n$ and the operators that represent them.\\

The Poincar\'e algebra has two quadratic Casimir operators: the 
mass squared\i{mass squared}\i{Casimir operator}
\be
\cM^2=P_0^2-P_1P_{-1}
\label{COPmsq}
\ee
and the three-dimensional analogue of the square of the Pauli-Lubanski vector,
\be
\cS
=
P_0J_0-\frac{1}{4}\left(J_1P_{-1}+J_{-1}P_1+P_1J_{-1}+P_{-1}J_1\right).
\label{COPcasij}
\ee
The eigenvalues of these operators classify irreducible representations according to mass and spin, exactly 
as in section \ref{sePoTri}. See e.g.\ \cite{barut1986theory} for the proof of the fact that the operator 
(\ref{COPmsq}) actually takes the value $M^2$ in the space of states of a relativistic particle with mass $M$.

\subsubsection*{Induced modules}

Irreducible unitary representations of the Poincar\'e group are obtained 
by 
considering the Lorentz orbit of a 
momentum $p$ and building a Hilbert space of wavefunctions on that orbit. A 
basis of this space is provided by plane waves (\ref{pawawest}), where the Dirac 
distribution is determined by the choice of measure on the orbit. (For instance one can take the 
Lorentz-invariant measure (\ref{lom}).)
Their transformation laws are given by (\ref{Tiga}).
Here, in order to 
make the link with the standard Dirac notation, we denote such plane waves by $\Psi_k\equiv|k,s\rangle$
for any $k\in\cO_p$, where $s$ is the spin of the representation.\\

For future comparison with $\bms$, we focus on a relativistic particle with mass $M>0$. 
Its little group $\un$ consists of spatial rotations generated by $J_0$. If we call $p$ the momentum of the 
representation 
in the rest frame, then the corresponding plane wave $\Psi_p\equiv|M,s\rangle$ satisfies\i{induced 
module!for Poincar\'e algebra}
\be
\label{COPrest1}
P_0 |M,s\rangle = M |M,s\rangle\,, \quad 
P_{-1} |M,s\rangle = P_1 |M,s\rangle = 0\,,\quad 
J_0 |M,s\rangle = s |M,s\rangle\,.
\ee
From now on we call $|M,s\rangle$ the \it{rest frame state}\i{rest frame} of the representation. Any other 
plane wave $\Psi_k=|k,s\rangle$ 
with boosted momentum $k\in\cO_p$ can be obtained by acting on $|M,s\rangle$ with a Lorentz 
transformation $g_k$, where $g_k$ is a standard boost. In this sense the rest frame state determines all the 
properties of the representation, in the same way that highest-weight representations are determined by 
their highest-weight state. Note, however, that the conditions (\ref{COPrest1}) that define $|M,s\rangle$ 
are \it{not} of the same form as the highest-weight conditions (\ref{hiwesta}) in that they involve both 
positive and negative modes.\\

Let us now understand how the conditions (\ref{COPrest1}) induce a representation of the Poincar\'e 
algebra. They define a one-dimensional 
representation of 
the subalgebra generated by $\{P_m,J_0\}$. This subalgebra consists of infinitesimal translations and spatial 
rotations, i.e.\ it is a semi-direct sum $\mathfrak{u}(1)\inplus\RR^3$ where $\mathfrak{u}(1)$ is generated 
by $J_0$ while $\RR^3$ is generated by the $P_m$'s. Thus the prescription (\ref{COPrest1}) is a 
Lie-algebraic version of the spin representation (\ref{rspin}) for the case of a little group $\un$ with 
$\cR[\theta]=e^{is\theta}$. Guided by our experience 
of induced representations, we can attempt to induce a representation of the full Poincar\'e algebra out 
of the one-dimensional representation (\ref{COPrest1}); the result is known as an \it{induced 
module}\i{induced module} (see e.g.\ section 10.7 of 
\cite{schottenloher}). We thus declare that the carrier space $\sH$ of the representation is spanned by all 
states obtained by acting on the rest frame state with operators that do \it{not} appear in the conditions 
(\ref{COPrest1}):
\be
\label{COPboost1}
|k,l\,\rangle
=
(J_{-1})^k(J_{1})^l|M,s\rangle\,,
\ee
where $k,l$ are non-negative integers. Such states are infinitesimally boosted states analogous to the 
descendant states (\ref{s72}) that span Verma modules for the Virasoro algebra. By definition, they form a 
basis of the space $\sH$. The latter provides a Poincar\'e representation as it should, since acting from the 
left on the states (\ref{COPboost1}) yields linear operators on $\sH$ whose 
commutators coincide with (\ref{COPpal}).
Moreover, the Casimir operators (\ref{COPmsq}) and (\ref{COPcasij}) have the same eigenvalue on each state 
(\ref{COPboost1}), since they commute by construction with all elements of the algebra. This readily implies 
that the representation is irreducible.\\

Note that unitarity is not obvious in this picture: 
if one did not know that the induced 
module follows from a 
manifestly unitary representation of the Poincar\'e group in terms of wavefunctions, there would be no 
straightforward way to define a scalar product on 
the 
space $\sH$ spanned by the states (\ref{COPboost1}), even after enforcing the Hermiticity 
conditions 
(\ref{COPherm}). In fact, the norm squared of any plane wave state is strictly infinite because of 
the delta function in (\ref{PlaSca}).
This is strikingly different from the highest-weight representations of section 
\ref{sevirep}, where the highest-weight conditions were enough to evaluate the norm squared (\ref{normaD}) of 
all descendant states.

\paragraph{Remark.}\label{pagindumodu} The definition of the infinitesimally boosted states (\ref{COPboost1}) 
follows from the general construction of induced modules,\i{induced module} as follows. Let $\mg$ be a Lie 
algebra with some 
subalgebra $\mh$. Let $\sS$ be a one-dimensional representation of $\mh$. If $\cU(\mg)$ denotes the universal 
enveloping algebra of $\mg$, then the $\mg$-module 
$\sT=\text{Ind}_{\mh}^{\mg}(\sS)$ induced by $\sS$ is the representation of $\mg$ that acts in the space 
$\cU(\mg)\otimes\CC$ quotiented by the relations $Y\otimes\lambda=1\otimes\sS[Y]\lambda$ for all $Y\in\mh$ 
and all $\lambda\in\RR$. For any Lie algebra element $X\in\mg$, the operator $\sT[X]$ acts on the carrier 
space by hitting on vectors from the left, in such a way that commutators of operators $\sT[X]$ reproduce the 
Lie brackets of the Lie algebra $\mg$. In this language the Poincar\'e module above is 
induced by the representation (\ref{COPrest1}) of the semi-direct sum $\mg_p\inplus\RR^3$, where $\RR^3$ is 
the Lie algebra of translations generated by the $P_m$'s while $\mg_p$ is the Lie algebra of the little group 
generated by $J_0$. The conditions (\ref{COPrest1}) define a one-dimensional representation $\sS$ of 
$\mg_p\inplus\RR^3$, analogous to the spin representation (\ref{rspin}).

\subsubsection*{Ultrarelativistic limit of $\mathfrak{sl}(2,\mathbb{R})$ modules}

In addition to being convenient for generalizations to infinite-dimensional extensions of the Poincar\'e 
algebra, Poincar\'e modules can be seen as a limits of  unitary representations of the AdS$_3$ 
isometry algebra $\mathfrak{so}(2,2)\cong 
\mathfrak{sl}(2,\mathbb{R}) \oplus \mathfrak{sl}(2,\mathbb{R})$. The generators of the latter can be 
divided 
in two groups, $L_m$ and $\bar{L}_m$ with $m=-1,0,1$, whose Lie brackets are two commuting copies of 
(\ref{wittibi}). In terms of these basis elements the quadratic Casimir of each copy of $\sl$ is 
(\ref{silcasimir}). As usual our conventions are such that, in any unitary representation, the Hermiticity 
conditions (\ref{hermiconw}) hold in both sectors.\\

The Poincar\'e algebra (\ref{COPpal}) can be recovered from $\sl\oplus\sl$ as a flat limit of the type 
described in section \ref{sebmsmod}. Thus we introduce a length scale $\ell$ (to be identified with the AdS 
radius) and define new 
generators as in (\ref{hajim}):\i{sl2R@$\sl$!flat limit}\i{flat limit!of sl2R@of $\sl$}
\be
J_m\equiv L_m-\bar L_{-m}\,,
\qquad
P_m\equiv\frac{1}{\ell}(L_m+\bar L_{-m})\,.
\label{COPlpj}
\ee
The resulting algebra is (\ref{haHHa}) without $i$'s on the left-hand side, and its 
limit $\ell\rightarrow+\infty$ reproduces the Poincar\'e algebra (\ref{COPpal}). In addition 
the quadratic Casimir (\ref{silcasimir}) can be combined with its barred counterpart 
$\bar\cC$, producing
\be
\frac{2}{\ell^2} \left( \cC+\bar\cC \right)
= \cM^2 + \cO(\ell^{-2}) ,
\qquad
\frac{1}{\ell} \left(\cC-\bar\cC\right)
= \cS ,
\label{COPcasilim}
\ee
where $\cM^2$ and $\cS$ are the Poincar\'e Casimirs (\ref{COPmsq}) and (\ref{COPcasij}).\\

The matching of Casimir operators suggests that the contraction also relates Poin\-ca\-r\'e modules to 
$\sl\oplus\sl$ representations. Concretely, consider the tensor product of two highest-weight representations 
(\ref{hiwesta}) of $\sl$ with weights $h,\bar h$:\i{highest-weight representation!flat limit}\i{flat limit!of 
highest-weight reps}
\be
L_1|h,\bar h\rangle=\bar L_1|h,\bar h\rangle=0\,,
\qquad
L_0|h,\bar h\rangle=h|h,\bar h\rangle\,,
\qquad
\bar L_0|h,\bar h\rangle=\bar h|h,\bar h\rangle\,.
\label{yahve}
\ee
This yields an irreducible representation of 
$\sl\oplus\sl$ whose carrier space is spanned by descendant states $(L_{-1})^m 
(\bar{L}_{-1})^n|h,\bar{h}\rangle$. 
Let us rewrite this representation in terms of the operators (\ref{COPlpj}). First we define the numbers
\be
M\equiv\frac{h+\bar h}{\ell}\,,
\qquad
s\equiv h-\bar h\,,
\label{COPms}
\ee
which are eigenvalues of energy and angular momentum:
\be
P_0|h, \bar h\rangle
=
\frac{h+\bar h}{\ell}|h, \bar h\rangle\,,
\qquad
J_0|h,\bar h\rangle
=
(h-\bar h)|h, \bar h\rangle
\ee
in terms of operators (\ref{COPlpj}). Similarly,
in terms of $J$'s and $P$'s,
the condition that $L_1$ and $\bar L_1$ annihilate the 
highest-weight state becomes
\be
\label{COPhwcPoincare}
\Big(
P_{\pm1}\pm\frac{1}{\ell}J_{\pm1}
\Big)
|h, \bar h\rangle
= 0\,.
\ee
This allows us to reformulate the whole representation of $\sl\oplus\sl$ in 
terms of operators $J_m,P_n$; it results in 
expressions of the form
\be
\label{COPmatrix-p}
P_n |k,l\rangle=\sum_{k',l'} \mathsf{P}^{(n)}_{k',l';\,k,l}(M,s,\ell) |k',l'\rangle\,,
\qquad
J_n |k,l\rangle=\sum_{k',l'} \mathsf{J}^{(n)}_{k',l';\,k,l}(M,s) |k',l'\rangle
\ee
where the states $|k,l\rangle$ take the form (\ref{COPboost1})
with the identification $|M,s\rangle\equiv|h,\bar h\rangle$,
while $\mathsf{P}^{(n)}$ and $\mathsf{J}^{(n)}$ are infinite matrices. Owing to the definition 
(\ref{COPlpj}) and property (\ref{COPhwcPoincare}), only negative powers of 
$\ell$ appear in (\ref{COPmatrix-p}). It follows that the matrix elements $\mathsf{P}^{(n)}_{k',l';\,k,l}$ 
and 
$\mathsf{J}^{(n)}_{k',l';\,k,l}$ have a 
well-defined limit $\ell \to \infty$. This limit coincides with the result that one would find in a 
Poincar\'e module spanned by states (\ref{COPboost1}), provided that the 
conformal 
weights 
scale as
\be
\label{COPscaleh}
h= \frac{M\ell+s}{2}+\lambda+\cO(1/\ell),
\qquad
\bar h= \frac{M\ell-s}{2}+\lambda+{\cO}(1/\ell) ,
\ee 
where $\lambda$ is an arbitrary parameter independent of $\ell$. Thus, in the flat limit, the 
$\mathfrak{sl}(2,\RR)$ highest-weight 
conditions (\ref{COPhwcPoincare}) are turned into a
rest frame condition (\ref{COPrest1}) and the 
Poincar\'e Casimirs $\cM^2$ and $\cS$ take the values $M^2$ and $Ms$, respectively.\i{induced module!flat 
limit} In short, Poincar\'e 
modules are flat limits of $\sl\oplus\sl$ modules.\\

Note again that unitarity is subtle: starting from scalar products of $\sl\oplus\sl$ descendants, 
the 
flat limit gives rise to scalar products of states (\ref{COPboost1}) that diverge like positive powers of 
$\ell$. 
Indeed the norms (\ref{normaD}) diverge when $\ell\rightarrow+\infty$, owing to the fact
that $h$ is proportional to $\ell$ in  (\ref{COPscaleh}). Equivalently, the wavefunctions
corresponding to states (\ref{COPboost1})
become (derivatives of)
delta functions in the flat limit;\i{AdS radius!as infrared regulator} from this 
point of 
view $\ell$ is an infrared regulator. Nevertheless, upon recognizing these divergent scalar products as delta 
functions (\ref{PlaSca}), one concludes that the Poincar\'e module is a unitary 
representation in disguise.\\

Relation (\ref{COPscaleh}) shows that the contraction defined by (\ref{COPlpj}) is an 
ul\-tra\-re\-la\-ti\-vis\-tic/\-high\--e\-ner\-gy limit\i{flat limit!as ultrarelativistic 
limit}\i{ultrarelativistic limit}\i{sl2R@$\sl$!ultrarelativistic limit} from the viewpoint of 
AdS$_3$. 
Poincar\'e modules are thus remnants of $\mathfrak{so}(2,2)$ representations whose energy becomes large in 
the limit $\ell \to \infty$. In section \ref{segalical} we shall see that the
\it{non}-relativistic contraction from 
$\mathfrak{so}(2,2)$ to $\mathfrak{iso}(2,1)$ gives rise to 
representations of a different type, that have been discussed in \cite{Bagchi:2009pe,Grumiller:2014lna}.\\

We should mention that highest-weight representations of $\mathfrak{sl}(2,\RR)$ can also be 
interpreted as induced modules. Indeed eq.\ (\ref{hiwesta}) defines a one-dimensional representation of 
the 
subalgebra spanned by $\{L_0,L_1\}$, while the vector space of descendant states can be identified with a 
quotient of $\,\cU(\mathfrak{sl}(2,\RR))\otimes\CC$ as discussed in the remark of page 
\pageref{pagindumodu}. 
The main 
difference with respect to Poincar\'e is the splitting of the algebra as $\mathfrak{n^-} \oplus 
\mathfrak{h} \oplus \mathfrak{n}^+$, where $\mathfrak{n}^\pm$ are nilpotent subalgebras, which
allows one to evaluate scalar products by enforcing the Hermiticity conditions (\ref{hermiconw}).

\subsection{\ Induced modules for $\mathfrak{bms}_3$}
\label{COPsec:bms}

Let us now apply the considerations of the previous pages to three-dimensional BMS symmetry.
The centrally extended $\bms$ algebra\footnote{Recall that we use the same notation for both the $\bms$ 
algebra and its central extension.} is spanned by superrotation generators $\cJ_m$ 
and supertranslation generators $\cP_m$ ($m \in \ZZ$) 
together with central charges $\cZ_1$, $\cZ_2$,
whose Lie brackets take the form (\ref{bammex}). Following the Virasoro convention (\ref{operatul}), we 
change the normalization and define
\be
J_m\equiv i\cJ_m+i\frac{\cZ_1}{24}\delta_{m,0}\,,
\qquad
P_m\equiv i\cP_m+i\frac{\cZ_2}{24}\delta_{m,0}\,,
\label{GxP}
\ee
as well as $Z_1\equiv i\cZ_1$ and $Z_2\equiv i\cZ_2$.
The constant shifts in $P_0$ and $J_0$ ensure that the vacuum state has zero eigenvalues under these 
operators. According to 
this definition the operators representing $J_m$ and $P_m$ in any unitary representation satisfy the 
Hermiticity conditions (\ref{COPherm}).\i{Hermiticity conditions!for BMS$_3$} Furthermore, in any irreducible 
representation the central charges $Z_1$ and $Z_2$ take definite values $c_1,c_2$, so we can write the 
commutation relations of the $\bms$ algebra in a form analogous to (\ref{serge}):
\begin{align}
{}[J_m,J_n] & = (m-n)J_{m+n}+\frac{c_1}{12}\,m(m^2-1)\,\delta_{m+n,0}\,,\nn\\
\label{COPjp}
{}[J_m,P_n] & = (m-n)P_{m+n}+\frac{c_2}{12}\,m(m^2-1)\,\delta_{m+n,0}\,,\\
{}[P_m,P_n] & = 0\,.\nn
\end{align}
\vspace{.1cm}

In contrast to Poincar\'e, the quadratic operators 
(\ref{COPmsq})-(\ref{COPcasij}) no longer commute with the algebra (\ref{COPjp}). Nevertheless, the 
classification of BMS$_3$ representations in section \ref{sebmspar} provides at least one obvious, yet 
non-trivial, Casimir operator.\footnote{I am 
indebted to Axel Kleinschmidt for this observation.}\i{bms3 algebra@$\bms$ 
algebra!Casimir operators}\i{Casimir operator!for $\bms$}\i{monodromy!as bms3 Casimir@as $\bms$ Casimir} 
Indeed, in the Hilbert space of any BMS$_3$ particle, the ``mass operator'' (\ref{bamonod})\i{mass 
operator}\i{mass squared!for BMS$_3$} 
takes a definite value when $\sfM$ is the monodromy matrix whose trace is given by the Wilson loop 
(\ref{wilson}), with $c$ replaced by $c_2$ and $p(\phii)$ replaced by the ``supermomentum operator''
\be
\hp(\phii)=\sum_{m\in\ZZ}P_me^{-im\phii}-\frac{c_2}{24}
\nn
\ee
where the $P_m$'s are the supertranslation generators appearing in (\ref{COPjp}).
Accordingly, upon writing the right-hand side of (\ref{bamonod}) in terms of $P_m$'s, one 
obtains a highly non-linear combination of operators that commutes, by construction, with 
the entire $\bms$ algebra. (That it commutes with $P_m$'s is trivial, since 
all supertranslations commute; that it commutes with $J_m$'s follows from the fact 
that (\ref{bamonod}) is invariant under superrotations.) The value of that Casimir operator can 
be used to classify BMS$_3$ particles, as we have done in section \ref{sebmspar}.\\

Aside from the mass operator (\ref{bamonod}), any function of the BMS$_3$ central charges $c_1$, $c_2$ is 
clearly a Casimir. To our knowledge, whether this list exhausts all possible $\bms$ Casimirs is an open 
question, though it seems plausible that it does since the only Casimirs of the Virasoro algebra are 
functions of its central charges \cite{feigin1983}. In particular, it is not clear whether there exists 
a $\bms$ Casimir whose value specifies the spin of a BMS$_3$ particle, analogously to the Poincar\'e 
combination (\ref{COPcasij}).\\

We now describe induced modules for the $\bms$ algebra (\ref{COPjp}), built analogously to the Poincar\'e 
modules above and classified by their mass and spin (and central charges). We discuss separately generic 
massive modules and the vacuum module, and end by showing how they can all be obtained as ultrarelativistic 
limits of highest-weight representations of Virasoro algebras.

\subsubsection*{Massive modules}

Consider a BMS$_3$ particle with mass $M>0$ and spin $s$. Its supermomentum orbit contains a constant 
$p=M-c_2/24$; the corresponding plane wave state $\Psi_p\equiv|M,s\rangle$ is such that\i{bms3 
module@$\bms$ module}
\be
\label{COPrest-bms}
P_0|M,s\rangle=M|M,s\rangle,
\qquad
P_m|M,s\rangle=0
\;\text{for }m\neq0,
\qquad
J_0|M,s\rangle=s|M,s\rangle.
\ee
Thus $|M,s\rangle$ is a supermomentum eigenstate with vanishing eigenvalues under $P_m$, $m\neq0$. In 
analogy 
with (\ref{COPrest1}), we call $|M,s\rangle$ the \it{rest frame state}\i{rest frame} of the module.\\

As in the Poincar\'e case, the conditions (\ref{COPrest-bms}) define a one-dimensional representation of the 
subalgebra of 
(\ref{COPjp}) spanned by $\{P_n,J_0,c_1,c_2\}$. This representation can be used to define an induced module 
$\sH$ with basis vectors analogous to (\ref{COPboost1}),\i{induced module}
\be
\label{COPeq:BoostedStatesDef}
J_{n_1}J_{n_2} \cdots J_{n_N}|M,s\rangle ,
\ee
where the $n_i$'s are non-zero integers such that $n_1\leq n_2\leq...\leq n_N$.\i{induced module!for $\bms$ 
algebra}\i{bms3 algebra@$\bms$ algebra!induced module}\i{massive bms3 module@massive $\bms$ module} With 
this ordering, states 
(\ref{COPeq:BoostedStatesDef}) with different combinations of $n_i$'s are linearly independent within the 
universal enveloping algebra of $\bms$, and acting on them from the left with the generators of 
the algebra provides linear operators on $\sH$ whose commutators coincide with (\ref{COPjp}). Thus one 
readily obtains a representation of the $\bms$ algebra.\\

As in the Poincar\'e case above, unitarity is hidden in this picture 
because there is no straightforward way to compute scalar products of states (\ref{COPeq:BoostedStatesDef}). 
In fact, since $|M,s\rangle$ is a delta function, all such states strictly have infinite norm. This 
is because realistic states of BMS$_3$ particles are smeared wavefunctions that consist of 
infinite linear combinations of plane waves. Unitarity can then be recognized in the fact that acting with 
(finite) superrotations on $|M,s\rangle$ produces a ``basis'' of plane waves that generate a space of 
square-integrable wavefunctionals on the supermomentum orbit. In particular the representation is 
automatically irreducible in the sense that all basis states are obtained by acting with symmetry 
transformations on the single state $|M,s\rangle$.

\subsubsection*{Vacuum module}

Recall that the BMS$_3$ vacuum is the scalar representation whose supermomentum orbit 
$\Diff/\PSL$ 
contains the vacuum configuration $p=p_{\text{vac}}=-c_2/24$. The corresponding induced module can be 
described 
similarly to massive ones. Owing to the normalization (\ref{GxP}), the Hilbert space of the vacuum 
representation contains a plane wave $\Psi_p\equiv|0\rangle$ such that\i{vacuum 
module!for bms3 algebra@for $\bms$ algebra}\i{bms3 module@$\bms$ module!vacuum}
\be
P_m|0\rangle=0\;\text{ for all }m\in\mathbb{Z}
\qquad\text{and}\qquad
J_n|0\rangle=0\;\text{ for }n=-1,0,1.
\label{COPeq:BMSVac}
\ee
Here the condition $P_0|0\rangle=0$ says that the vacuum has zero mass for the normalization (\ref{GxP}), 
while the extra conditions $J_{\pm1}|0\rangle=0$ enforce Lorentz-invariance. They reflect the fact that the 
little group of the vacuum is the whole Lorentz group, rather than the group of spatial rotations that occurs 
for massive particles.\\

If we were dealing with the Poincar\'e algebra, the requirements (\ref{COPeq:BMSVac}) would produce a trivial 
representation. Here, 
by contrast, there exist non-trivial ``boosted vacua'' of the form (\ref{COPeq:BoostedStatesDef}), where now 
the
$n_i$'s are integers different from $-1,0,1$. These vacua are Lie-algebraic analogues of the boundary 
gravitons described 
earlier. The fact that the vacuum is not invariant under the full BMS$_3$ 
symmetry, but only under its Poincar\'e subgroup, suggests that the boosted states 
(\ref{COPeq:BoostedStatesDef}) 
(with all $n_i$'s $\neq-1,0,1$) can be interpreted as Goldstone-like states associated with broken symmetry 
generators; see the discussion surrounding (\ref{Ledes}).
Note that, in contrast to the 
realistic four-dimensional case, BMS$_3$ supertranslations do \it{not} create new states when acting on the 
vacuum.

\subsubsection*{Ultrarelativistic limit of Virasoro modules}

In analogy with the observations of section \ref{COPsec:poincare}, $\mathfrak{bms}_3$ modules may be seen as 
limits of tensor products 
of highest-weight representations of Virasoro. Let therefore $L_m,\bar L_m$ be 
generators 
of two commuting copies of the Virasoro algebra (\ref{serge}) with definite central charges $c,\bar c$. 
Highest-weight representations are then obtained starting from a primary state $|h,\bar h\rangle$
which satisfies (\ref{yahve}) together with\i{primary state}\i{highest-weight representation}\i{Virasoro 
algebra!highest-weight rep}
\be
L_m|h,\bar h\rangle=\bar L_m|h,\bar h\rangle=0
\quad\text{ for }m>0.
\label{COPhighest-weight}
\ee
The carrier space is spanned by descendant states
\be
\label{COPeq:VirasoroVermaModule}
L_{-n_1}...L_{-n_k} \bar{L}_{-\bar{n}_1}...\bar{L}_{-\bar{n}_l}|h,\bar{h}\rangle  
\ee
with $1\leq n_1 \leq n_2 \leq...\leq n_k$ and $1\leq\bar n_1\leq...\leq\bar n_l$.
Since we eventually wish to take the ultrarelativistic limit of this representation, we will be interested in 
large values of $h$ 
and 
$\bar{h}$, where the representation is irreducible and unitary thanks to the standard 
Hermiticity conditions 
(\ref{hemingway}).\\

As in the Poincar\'e case, one can define new generators (\ref{COPlpj}), now including also the central 
charges 
$c_1,c_2$ defined by (\ref{CiCCoh}). In particular, the space of Virasoro descendants can be rewritten in 
the basis (\ref{COPeq:BoostedStatesDef}) with the identification $|M,s\rangle\equiv|h,\bar h\rangle$, where 
$M$ and $s$ are the eigenvalues of $P_0$ and $J_0$ related to 
$h$ 
and $\bar{h}$ by (\ref{COPms}). The change of basis from descendant 
states (\ref{COPeq:VirasoroVermaModule}) 
to infinitesimally boosted states (\ref{COPeq:BoostedStatesDef}) is invertible because none of the $J_n$'s 
annihilate the highest-weight state. The resulting Virasoro representation takes a form analogous to 
(\ref{COPmatrix-p}), 
where now each state is labelled by the quantum numbers $n_i$ of (\ref{COPeq:BoostedStatesDef})
and the matrices 
$\mathsf{P}^{(n)}$ 
and $\mathsf{J}^{(n)}$ also depend on the central charges (\ref{CiCCoh}).
As before, only negative powers of $\ell$ enter $\mathsf{P}^{(n)}$ via the highest-weight conditions 
(\ref{COPhighest-weight}) written in the new basis:\i{flat limit!of Virasoro 
reps}\i{bms3 module@$\bms$ module!as flat limit}
\be
\Big( P_{\pm n} \pm \frac{1}{\ell} J_{\pm n} \Big) | h, \bar{h} \rangle = 0 .
\ee
A limit $\ell \to \infty$ performed at fixed $M$, $s$ and $c_1$, $c_2$ (rather than fixed $h,\bar{h}$ say) 
then yields a massive $\bms$ module of the type described above. In particular, the limit maps the 
highest-weight state (\ref{COPhighest-weight}) on the rest frame 
state 
(\ref{COPrest-bms}). In this sense $\bms$ modules are high-energy limits of
tensor products of Virasoro modules, since $h$ and $\bar h$ go to infinity in the flat limit. By the way, 
this provides an intuitive picture of why the energy spectrum\i{BMS$_3$ particle!energy 
spectrum}\i{energy spectrum} of $\BMS$ 
particles is continuous: the typical distance between two consecutive eigenvalues of $P_0=(L_0+\bar 
L_0)/\ell$ is $1/\ell$, which shrinks to zero when $\ell$ goes to infinity.

\subsection{\ Representations of the Galilean conformal algebra}
\label{segalical}

Here we revisit the Galilean conformal algebra introduced in section \ref{sebmsmod}. As 
explained there, $\mathfrak{gca}_2$ coincides with $\bms$,
which effectively makes them classically interchangeable. We now argue that this 
equivalence does \it{not} hold at the quantum level; this observation will be the basis of our arguments in 
section \ref{COPsec:hs}, when explaining why the quantization of asymptotically flat gravity cannot be a 
Galilean conformal field theory. The 
highest-weight representations that we shall describe here were first obtained in \cite{Bagchi:2009pe}, but 
their identification with induced representations of BMS$_3$ is new.

\subsubsection*{Highest-weight representations of $\mathfrak{gca}_2$}

The $\mathfrak{gca}_2$ algebra is isomorphic to $\mathfrak{bms}_3$, but their interpretations differ: in 
$\bms$ the non-Abelian generators generalize the angular 
momentum operator and span superrotations, while the Abelian ones generalize the Hamiltonian and span 
supertranslations. By contrast, in $\mathfrak{gca}_2$,\i{gca2@$\gca$} the non-Abelian generators are the ones 
that 
generalize the Hamiltonian, and the Abelian ones generalize (angular) momentum. Due to this 
difference, one is naturally led to look for unitary representations of $\gca$ where the operator $J_0$ 
of (\ref{COPjp}) is bounded from below. From the BMS$_3$ perspective this is an awkward choice (since it 
explicitly
breaks 
parity by forcing all states to have the same sign of angular 
momentum), but from the Galilean viewpoint it is perfectly well motivated.\\

To describe these representations we use the method of induced modules applied to the algebra 
(\ref{COPjp}), with the Hermiticity conditions (\ref{COPherm}). To stress that we are dealing with 
\it{Galilean} rather than \it{relativistic} representations, we denote all $\gca$ generators with a 
tilde on top, such as $\tilde J_m$, $\tilde P_m$, plus central charges $\tilde c_1$, 
$\tilde c_2$. In order to obtain a representation where the spectrum of $\tilde J_0$ is bounded from below, we
start from a state $|\tilde M,\tilde s\rangle$ which has highest weight for the Virasoro subalgebra 
generated by $\tilde J$'s in the sense that\i{Galilean highest-weight reps}\i{highest-weight 
representation!of gca2@of $\gca$}
\be
\tilde J_0|\tilde M,\tilde s\rangle=\tilde s|\tilde M,\tilde s\rangle,
\qquad
\tilde P_0|\tilde M,\tilde s\rangle=\tilde M|\tilde M,\tilde s\rangle
\label{hws1}
\ee
and
\be
\tilde J_m|\tilde M,\tilde s\rangle
=
\tilde P_m|\tilde M,\tilde s\rangle=0
\qquad\text{for }m>0\,.
\label{hws}
\ee
We stress that $\tilde s$ is now interpreted as a (dimensionless) energy while $\tilde M$ is a 
(dimensionful) momentum. In analogy with Virasoro representations, one can then define descendant states of 
the form
\be
\tilde P_{-k_1}...\tilde P_{-k_n}\tilde J_{-l_1}...\tilde J_{-l_m}|\tilde M,\tilde s\rangle
\label{COPeq:GCAStates}
\ee
with $1\leq k_1\leq...\leq k_n$, $1\leq l_1\leq...\leq l_m$,
and declare that they form a basis of the carrier space. The conditions (\ref{hws}) allow one to evaluate the 
would-be scalar products of such descendants upon using the Hermiticity conditions (\ref{COPherm}), and one 
finds that the representation is \it{non-unitary}\i{non-unitary representation}\i{representation!non-unitary} 
whenever $\tilde M\neq0$ or $\tilde c_2\neq0$ 
\cite{Bagchi:2009pe}. When $\tilde M=c_2=0$, unitarity requires in addition that $\tilde s>0$ and that the 
superrotation central charge $\tilde c_1$ be non-negative. Thus, unitary representations of $\gca$ boil down 
to highest-weight representations of its Virasoro subalgebra generated by the $\tilde J$'s.\\

A similar construction can be applied to a vacuum-like highest-weight representation 
of $\gca$, whose highest-weight state $|\tilde 0\rangle$ is annihilated by all Poincar\'e generators 
$\tilde J_{-1},\tilde J_0,\tilde J_1$, $\tilde P_{-1},\tilde P_0,\tilde P_1$ and by all positive modes as in 
(\ref{hws1}):\i{Galilean vacuum}
\be
\tilde P_m|\tilde 0\rangle=0\,,
\qquad
\tilde J_m|\tilde 0\rangle=0
\qquad\text{ for }m\geq-1.
\label{GaVa}
\ee
Again, one concludes in that 
case that the representation is unitary if and only if $\tilde c_2=0$ and $\tilde c_1\geq0$.\\

The highest-weight representations of the type just described which are \it{unitary} (i.e.\ have $\tilde M=0$ 
and $\tilde c_2=0$) are special cases of induced representations of BMS$_3$ as described in section 
\ref{sebmspar}. Indeed, consider the vanishing supermomentum $(\tilde p,\tilde c_2)=(0,0)$. Its orbit under 
superrotations is trivial and its little group is the whole Virasoro group, so the corresponding induced 
representation is entirely determined by its spin $\tilde s$. The latter labels a unitary highest-weight 
representation of Virasoro, with central charge $\tilde c_1$ say. At the Lie-algebraic level this 
spin representation takes the form of a highest-weight representation (\ref{hws1})-(\ref{hws}) 
with $\tilde M=0$ and $\tilde c_2=0$. There is an analogue of this construction in the Poincar\'e group: the 
vanishing momentum vector $p=0$ has a trivial orbit and its little group is the whole Lorentz group, so the 
corresponding induced representation of Poincar\'e is just a unitary representation of the Lorentz group; it 
is a ``vacuum with spin''\i{spinning vacuum!for BMS$_3$}\i{BMS$_3$ 
vacuum!with spin} of the type mentioned in section \ref{relagroup}.\\

The difference betwen the BMS$_3$ vacuum (\ref{COPeq:BMSVac}) and the 
Galilean vacuum (\ref{GaVa}) implies sharp differences for all quantum systems enjoying such symmetries, 
since 
it affects the definition of normal ordering.\i{normal ordering} For example, the normal-ordered product 
${{:}\!\mathrel{J_2P_{-3}}\!{:}}$ equals $J_2P_{-3}$
in a BMS$_3$-invariant theory, while in a Galilean conformal field theory 
one has
${{:}\!\mathrel{\tilde J_2\tilde P_{-3}}\!{:}}=\tilde P_{-3}\tilde J_2$. We shall see explicit illustrations 
of this 
phenomenon 
in section \ref{COPsec:hs} below, when dealing with non-linear higher-spin symmetry algebras. It suggests in 
particular that theories enjoying $\bms$ symmetry or $\gca$ symmetry differ greatly at the quantum level, 
despite the isomorphism $\bms\cong\gca$.

\subsubsection*{Galilean limit of Virasoro modules}

We now recover the (generally non-unitary) Galilean highest-weight representations defined by 
(\ref{hws1})-(\ref{hws}) as a non-relativistic limit of Virasoro modules. As 
before we let the generators $L_m$, $\bar L_n$ satisfy the algebra (\ref{serge}) with central charges $c,\bar 
c$ 
respectively, and we consider a highest-weight representation of the type 
(\ref{yahve})-(\ref{COPhighest-weight}). In order to take the non-relativistic limit (\ref{limini}), we 
introduce a length scale $\ell$ and define\i{Virasoro algebra!non-relativistic limit}\i{non-relativistic 
limit}
\be
\label{COPeq:GCALinearCombinations}  
\tilde J_n\equiv\bar L_n+L_n\,,
\qquad
\tilde P_n\equiv\frac{1}{\ell}\left(\bar L_n-L_n\right)\,.
\ee
We stress that the combinations of $L_m$'s appearing here differ from those of the 
ultrarelativistic limit (\ref{COPlpj}). In particular, $\tilde J_0$ now generates time translations while 
$\tilde P_0$ generates spatial translations; the parameter $\ell$ should no longer be interpreted as the 
AdS$_3$ radius, and there is no mixing between positive and negative modes. In these terms, the 
limit $\ell\rightarrow+\infty$ of the direct sum of two Virasoro algebras reduces to a 
$\gca$ algebra (\ref{COPjp}) with tildes on top of all generators, including the central charges
\be
\tilde c_1=\bar c+c\,,
\qquad
\tilde c_2=\frac{\bar c-c}{\ell}\,.
\label{COPc-galileo}
\ee
Note that, up to central charges, the same redefinitions applied to $\sl\oplus\sl$ reproduce the Poincar\'e 
algebra $\mathfrak{iso}(2,1)$; this is a \it{non}-relativistic limit to be contrasted with the 
\it{ultra}relativistic limit described in section \ref{COPsec:poincare}.\\

In analogy with section \ref{COPsec:bms}, let us rewrite the tensor product of two highest-weight 
representations of Virasoro in terms of the operators (\ref{COPeq:GCALinearCombinations}). Given the weights 
$h,\bar h$ we define
\be
\tilde s\equiv\bar h+h\,,
\qquad
\tilde M\equiv\frac{\bar h-h}{\ell}
\nn
\ee
which we stress is radically different from the ultrarelativistic redefinition (\ref{COPms}).
Upon identifying $|\tilde M,\tilde s\rangle\equiv|h,\bar h\rangle$,
the 
highest-weight state satisfies (\ref{hws1}) and (\ref{hws}).
These conditions hold for any value of $\ell$, including the limit $\ell\rightarrow+\infty$. The 
descendant states (\ref{COPeq:GCAStates}) then provide a representation of the sum of two Virasoro algebras, 
which in the non-relativistic limit $\ell\rightarrow+\infty$ becomes a generically \it{non-unitary} 
representation of $\gca$. Unitarity is recovered if $\tilde M=\tilde c_2=0$ and $\tilde s,\tilde c_1\geq 
0$. Again, this is strikingly different from the ultrarelativistic contraction described above.

\paragraph{Remark.} The difference between $\mathfrak{gca}_2$ modules and $\bms$ modules has 
been 
known, albeit in disguise, ever since the nineties. Namely, the tensionless limit of string theory 
gives rise to so-called \it{null strings}\i{null string} \cite{Schild:1976vq}, whose worldsheet is a null 
surface and thus 
provides a stringy generalization of null geodesics. It was observed in \cite{Lizzi:1986nv} that the algebra 
of constraints arising from worldsheet reparameterization invariance of null strings is the $\bms$ algebra, 
although the name ``BMS'' was not used at the time.\footnote{This occurrence of the 
$\bms$ 
algebra predates its gravitational description \cite{Ashtekar:1996cd} by a decade!} In the same paper the 
authors 
observed that a suitable normal-ordering prescription gives rise to a consistent quantization of the null 
string in \it{any} space-time dimension, and systematically results in a continuous mass spectrum. This 
result is the stringy analogue of the $\bms$ modules described in section \ref{COPsec:bms}. Only later was it 
realized that a different normal ordering prescription \cite{Gamboa:1989zc,Gamboa:1989px} gives rise to the 
same critical dimension as in standard string theory (26 for the bosonic string and 10 for the superstring), 
but that the resulting spectrum is massless and discrete; in fact, the spectrum then coincides with the 
massless part of the spectrum of standard string theory. The latter result is the stringy 
analogue of the $\mathfrak{gca}_2$ modules described here and the critical dimension is analogous to the 
requirement $\tilde c_2=0$ that ensures unitarity for non-relativistic modules; see 
\cite{Bagchi:2013bga,Bagchi:2015nca,Casali:2016atr} for a recent account of these results.

\section{\ Characters of the BMS$_3$ group}
\label{sebmschar}

Now that we are acquainted with BMS$_3$ particles, we can start using them as a computational tool. In this 
section we use the Frobenius formula (\ref{fropo}),\footnote{Here
$\delta$ denotes the delta function associated by (\ref{didi}) with the measure $\mu$.}\i{Frobenius formula}
\be
\chi[(f,\alpha)]
=
\Tr\big(\cT[(f,\alpha)]\big)
=
\int_{\cO_p}d\mu(k)\,\delta(k,f\cdot k)\,
e^{i\langle k,\alpha\rangle}
\chi_{\cR}[g_k^{-1}fg_k]\,,
\label{fropokad}
\ee
to evaluate characters of rotations $f$ and 
supertranslations $\alpha$ in induced representations of the (centrally extended) BMS$_3$ group. Remarkably, 
the localization effect due to the delta function will allow us to 
compute characters despite the fact that we do not know explicit measures on supermomentum orbits. We focus 
on 
massive BMS$_3$ particles and on the BMS$_3$ vacuum, and compare the results to the 
ultrarelativistic 
limit of Virasoro characters. The 
results 
reviewed here were first reported in \cite{Oblak:2015sea}; their application to partition functions 
\cite{Barnich:2015mui,Campoleoni:2015qrh} will be exposed in the next chapter.

\subsection{\ Massive characters}
\label{CHARcharrBMS}

We consider a BMS$_3$ particle with mass $M>0$ and spin $s\in\RR$, so that the little group representation 
is $\cR[\theta]=e^{is\theta}$. Our goal is to evaluate the character (\ref{fropokad}) for arbitrary BMS$_3$ 
transformations $(f,\alpha)$, along the lines described in section \ref{relagroup} for the Poincar\'e group. 
The 
only subtlety is that now $\cO_p\cong\Diff/S^1$ is an infinite-dimensional supermomentum orbit; the $g_k$'s 
of eq.\ (\ref{fropokad}) are standard boosts and the pairing $\langle k,\alpha\rangle$ is given by 
(\ref{denpar}). We assume as before that there exists a quasi-invariant measure $\mu$ on the supermomentum 
orbit $\cO_p$, but we stress again that different measures yield equivalent representations so that the end 
result will be independent of $\mu$. We shall verify this point explicitly below.\\

The character (\ref{fropokad}) vanishes if $f$ is not conjugate to an element of the little group $\un$; 
furthermore it is a class function, so we may take $f(\phii)=\phii+\theta$ to be a pure 
rotation by some angle $\theta$, without loss of generality. We assume for simplicity that $\theta$ is 
non-zero; the case $\theta=0$ is radically different, and we shall briefly comment about it below. When 
$\theta\neq0$, the delta function $\delta(k,f\cdot k)$ of (\ref{fropokad}) localizes the supermomentum 
integral to the unique point of $\cO_p$ that is left invariant by rotations, namely the supermomentum at rest 
$p=M-c_2/24$.\i{localization} This allows us to pull the little group character 
$\chi_{\cR}[f]=e^{is\theta}$ out of the integral
and to reduce
the pairing $\langle k,\alpha\rangle$ to 
a product $p\alpha^0=M\alpha^0-c_2\alpha^0/24$, so the whole character (\ref{fropokad}) boils down to a 
BMS$_3$ analogue of eq.\ (\ref{Xixi}):
\be
\chi[(f,\alpha)]
=
e^{is\theta}e^{i\alpha^0(M-c_2/24)}
\int_{\cO_p}d\mu(k)\,
\delta(k,f\cdot k)\,.
\label{CHARlocChar}
\ee
\vspace{.1cm}

To evaluate the character it only remains to integrate the delta function. This requires local coordinates 
on the orbit in a neighbourhood of $p$, which can be obtained by Fourier-expanding each supermomentum 
$k(\phii)$ as in eq.\ (\ref{surfou}). The Fourier modes $k_n=k_{-n}^*$ then transform under rotations 
$f(\phii)=\phii+\theta$ according to\i{complex rotation}\i{iepsilon regularization@$i\epsilon$ 
regularization}\i{rotation!action on supermomenta}
\be
k_n\mapsto\left[f\cdot k\right]_n=k_n\,e^{in\theta}.
\nn
\ee
As we shall see, the character that follows from this transformation is divergent due to the fact that the 
group is infinite-dimensional. To cure this divergence we consider complex rotations rather than real ones 
and introduce a complex parameter
\be
\tau\equiv\frac{1}{2\pi}(\theta+i\epsilon)
\label{CHARmodPar}
\ee
where $\epsilon>0$. We then define the transformation of supermomentum Fourier modes under complex rotations 
to be
\be
k_n\mapsto\left[f\cdot k\right]_n=
\left\{\begin{array}{lcc}
k_ne^{2\pi in\tau} & \text{if} & n>0,\\
k_0 & \text{if} & n=0,\\
k_ne^{2\pi in\bar\tau} & \text{if} & n<0.\end{array}\right.
\label{CHARtFour}
\ee
We will see below that this modification can be justified by thinking of BMS$_3$ representations as 
high-energy limits of Verma modules. Note that this prescription leaves room for ``Euclidean'' rotations 
(i.e.\ rotations by an imaginary angle) while 
preserving the reality condition $(k_n)^*=k_{-n}$.\\

The problem now is to express the measure $\mu$ and the corresponding delta function $\delta$ in terms of 
Fourier modes. On the massive supermomentum orbit $\cO_p\cong\Diff/S^1$, the non-zero Fourier modes of 
$k(\phii)$ determine its energy $k_0$. This is analogous to the statement that the energy of a relativistic 
particle is determined by its momentum according to $E=\sqrt{M^2+\bbk^2}$. Let us prove this in a 
neighbourhood of the supermomentum at rest, $p=M-c_2/24$, by acting on it with an infinitesimal superrotation 
$X$ that we Fourier-expand as\i{Fourier series}
\be
X(\phii)=i\sum_{n\in\ZZ}X_ne^{-in\phii}.
\nn
\ee
Since the action of superrotations on supermomenta is the coadjoint representation (\ref{covinf}) of the 
Virasoro algebra, we find a variation
\be
\left(\delta_Xp\right)(\phii)
=
\sum_{n\in\ZZ}2n\left(M+\frac{c_2}{24}(n^2-1)\right)X_ne^{-in\phii}
\equiv
\sum_{n\in\ZZ}\delta p_n\,e^{-in\phii}.
\label{CHARinfim}
\ee
Here the variation of the zero-mode, $\delta p_0$, vanishes for any choice of $X$. By contrast, all other 
Fourier modes are acted upon in a non-trivial way and can therefore take arbitrary values by a suitable 
choice of $X$.\i{orbit!coordinates}\i{supermomentum orbit!coordinates} This implies that (at least in a 
neighbourhood of $p$) the non-zero modes of supermomenta 
provide local coordinates on $\cO_p$. In terms of the Fourier decomposition (\ref{surfou}), this is to say 
that when $k(\phii)=p+\varepsilon(\delta_Xp)(\phii)$, the non-zero modes $k_n$ coincide with 
$\varepsilon\delta p_n$ (while $k_0=p$ to first order in $\varepsilon$).\\

It follows that in terms of 
$k_n$'s, the 
supermomentum measure $\mu$ of (\ref{CHARlocChar}) reads\i{measure!on 
Virasoro orbit}
\be
d\mu(k)=\text{(Some $k$-dependent prefactor)}\times\prod_{n\in\ZZ^*}dk_n
\label{CHARmeasOp}
\ee
where the prefactor is unknown. In quantum mechanics one would write the infinite product
$\prod_{n\in\ZZ^*}dk_n$ as a path integral measure $\cD k$, with the extra rule that the zero-mode of $k$ is 
not to be integrated over. The definition of the delta function (\ref{didi}) associated 
with 
$\mu$ ensures that\i{delta function}
\be
\delta(q,k)
=
\text{(Some $k$-dependent prefactor)}^{-1}\times\prod_{n\in\ZZ^*}\delta(q_n-k_n),
\nn
\ee
where the $\delta$ on the right-hand side is the usual Dirac distribution in one dimension. Crucially, the 
prefactor appearing in front of the delta function is the inverse of the prefactor of the measure 
(\ref{CHARmeasOp}). As in eq.\ (\ref{Cancel}) this implies that the combination $d\mu(k)\delta (k,\cdot)$ is 
invariant under changes of measures, and it allows us to rewrite the character (\ref{CHARlocChar}) as
\begin{eqnarray}
\chi[(f,\alpha)]
& \!\!= &
\!\!e^{is\theta}e^{i\alpha^0(M-c_2/24)}
\int_{\RR^{2\infty}}\prod_{n\in\ZZ^*}dk_n\prod_{n\in\ZZ^*}
\delta\left(k_n-\left[f\cdot k\right]_n\right)\nn\\
\label{CHARstar12}
& \!\!\stackrel{\text{(\ref{CHARtFour})}}{=} &
\!\!e^{is\theta}e^{i\alpha^0(M-c_2/24)}
\bigg|
\int_{\RR^{\infty}}\prod_{n=1}^{+\infty}dk_n\prod_{n=1}^{+\infty}
\delta\left(k_n(1-e^{2\pi in\tau})\right)
\bigg|^2,
\end{eqnarray}
where we have replaced the real angle $\theta$ by its complex counterpart $2\pi\tau$ given by 
(\ref{CHARmodPar}). Denoting $q\equiv\exp[2\pi i\tau]$ and evaluating the integral, the 
character of a massive $\BMS$ particle finally reduces to\i{BMS$_3$ character}\i{character!for 
BMS$_3$}\i{massive BMS$_3$ particle!character}
\be
\boxed{
\bigg.
\chi[(f,\alpha)]
=
e^{is\theta}e^{i\alpha^0(M-c_2/24)}
\frac{1}{\prod_{n=1}^{+\infty}|1-q^n|^2}\,.
}
\label{CHARcharBMS}
\ee
We stress that this holds only provided $f$ is conjugate to a rotation by $\theta$. We recognize here the 
ubiquitous factor $(1-q)^{-1}$ arising 
from the Atiyah-Bott fixed point theorem (\ref{detif}). The result can also be rewritten in terms of the 
Dedekind eta function (\ref{ss86}),
\be
\chi[(f,\alpha)]
=
\frac{|q|^{1/12}}{|\eta(\tau)|^2}
e^{is\theta}e^{i\alpha^0(M-c_2/24)},
\nn
\ee
with $|q|=1$ in the (pathological) limit $\epsilon\rightarrow0$.

\paragraph{Remark.} At this stage, and in contrast to conformal field theory, 
the coefficient $\tau$ should not be seen as a modular parameter. The small parameter 
$\epsilon$ in 
(\ref{CHARmodPar}) was merely introduced to ensure convergence of the determinant arising from the 
integration 
of the delta function in (\ref{CHARstar12}).\i{iepsilon regularization@$i\epsilon$ regularization} This being 
said, the occurrence of the Dedekind eta function is 
compatible 
with the modular transformations\i{modular transformation}\i{flat space holography!modular transformations} 
used in 
\cite{Bagchi:2012xr,Barnich:2012xq} to derive a Cardy-like formula reproducing the entropy of 
flat space cosmologies.

\subsection{\ Comparison to Poincar\'e and Virasoro}
\label{CHARsubsecCompChar}

Formula (\ref{CHARcharBMS}) extends the Poincar\'e character (\ref{ChRot}) in three dimensions. Indeed, 
taking $\epsilon=0$ in 
(\ref{CHARcharBMS}) and forgetting about all convergence issues, one finds
\be
\chi[(f,\alpha)]
=
e^{is\theta}e^{i\alpha^0(M-c_2/24)}
\prod_{n=1}^{+\infty}\frac{1}{4\sin^2(n\theta/2)}\,.
\nn
\ee
Here the term $n=1$ coincides (\ref{ChRot}), while the 
contribution 
of higher Fourier modes is due, loosely speaking, to the infinitely many Poincar\'e subgroups of $\BMS$. This 
is analogous to the
fact that the Virasoro
character (\ref{s86}) may be seen as a product of infinitely many $\SL$ characters (\ref{silchar}) labelled 
by an integer $n$.\\

The 
divergence of the $\BMS$ character (\ref{CHARcharBMS}) as $\epsilon\rightarrow0$ is identical to that
of the Virasoro character (\ref{s86}) as $\tau$ becomes real. In this sense, the divergence is not a 
pathology of $\BMS$,\i{flat limit!of Virasoro character} but rather a general phenomenon to be expected from 
infinite-dimensional 
groups; here we have cured this divergence by adding an imaginary part $i\epsilon$ to the angle. The origin 
of this imaginary part can be 
traced back to the fact that $\BMS$ representations are ultrarelativistic limits of Virasoro representations, 
as discussed at length in section \ref{COPsec:bms}. Indeed, suppose we are given a tensor product of two 
Virasoro representations with highest weights $h,\bar h$ and central charges $c,\bar c$. The corresponding 
character generalizes the partition function (\ref{Vanouche}) as
\be
\text{Tr}\left(q^{L_0-c/24}\bar q^{\bar L_0-\bar c/24}\right)
\refeq{s86}
\frac{q^{h-c/24}\bar q^{\bar h-\bar c/24}}{\prod_{n=1}^{+\infty}|1-q^n|^2}\,,
\qquad q=e^{2\pi i\tau}.
\label{s86Bis}
\ee
Writing the modular parameter in the form (\ref{TaKatak}) with an $\ell$-independent $\beta$ and introducing 
a mass $M$ and a spin $s$ defined by (\ref{COPms}), the large $\ell$ limit of the quantities 
appearing in the right-hand side of (\ref{s86Bis}) is
\be
\tau\sim\frac{1}{2\pi}(\theta+i\epsilon),
\qquad
q^{h-c/24}\bar q^{\bar h-\bar c/24}\sim e^{i\theta(s-c_1/24)}e^{-\beta(M-c_2/24)}.
\label{CHARmolim}
\ee
(Here the imaginary part of $\tau$ goes to zero, but we keep writing it as $\epsilon>0$ to reproduce the 
regularization used in (\ref{CHARcharBMS}).) Thus the flat limit of 
(\ref{s86Bis}) is
\be
\lim_{\ell\rightarrow+\infty}\text{Tr}\left(q^{L_0-c/24}\bar q^{\bar L_0-c/24}\right)
=
e^{i\theta(s-c_1/24)}e^{-\beta(M-c_2/24)}
\frac{1}{\prod_{n=1}^{+\infty}|1-q^n|^2},
\nn
\ee
and coincides (up to a redefinition of spin) with the $\BMS$ character (\ref{CHARcharBMS}) for a 
supertranslation whose 
zero-mode is a Euclidean time translation, $\alpha=i\beta$. The 
left-hand side of this expression can be 
interpreted as a trace
\be
\lim_{\ell\rightarrow+\infty}\text{Tr}\left(q^{L_0-c/24}\bar q^{\bar L_0-c/24}\right)
\stackrel{\text{(\ref{CHARmolim})}}{=}
\text{Tr}\left(e^{i\theta(J_0-c_1/24)}e^{-\beta(P_0-c_2/24)}\right)
=
\chi[(f,\alpha)],
\label{trachar}
\ee
where $f$ is a rotation by $\theta$ and the operators $J_m$, $P_n$ are normalized so as to satisfy the 
commutation relations (\ref{COPjp}). In this form, the matching between the flat limit of 
the 
Virasoro character (\ref{s86Bis}) 
and the $\BMS$ character (\ref{CHARcharBMS}) is manifest.

\subsubsection*{Universality of BMS$_3$ characters}

Even though $\BMS$ and Virasoro characters are related by the limit just 
described, they are strikingly different in that the result (\ref{CHARcharBMS}) holds for 
any value of the central charge $c_2$, any mass $M$, and any spin $s$. By 
contrast, the characters of irreducible, unitary highest weight representations of the Virasoro 
algebra depend heavily on the values of the central charge $c$ and the highest weight $h$: when $c\leq1$, 
only 
certain 
discrete values of $c$ and $h$ lead to unitary representations, and the resulting character is not given 
by (\ref{s86Bis}) 
\cite{FeiginFuchs01,Wassermann01,Wassermann02}.\i{universality}\i{BMS$_3$ character!universal} In that sense, 
induced representations of 
the 
$\BMS$ group are less intricate than highest weight representations of the Virasoro algebra. Since the former 
are 
high-energy, high central charge limits of 
the latter, this could have been expected: all complications occurring at small $c$ vanish when $\ell$ goes 
to 
infinity, since $c$ scales linearly with $\ell$ by assumption.\\

We could also have guessed that such a simplification would occur thanks to dimensional 
arguments.\i{dimensional analysis} 
Indeed, both $M$ and $c_2$ are dimensionful parameters labelling BMS$_3$ representations, so their values can 
be tuned at will by a suitable 
choice of units. Accordingly, in 
contrast to Virasoro highest weight representations, one should not expect to find sharp 
bifurcations in the structure of BMS$_3$ particles as $M$ and $c_2$ vary. In this 
sense formula (\ref{CHARcharBMS}) is a universal character.

\subsection{\ Vacuum character}
\label{CHARssecVac}

We now turn to the character of the BMS$_3$ vacuum, that is, the scalar representation whose supermomentum 
orbit is that of $p_{\text{vac}}=-c_2/24$. 
The computation is identical to that of section \ref{CHARcharrBMS}, save for the fact that the little 
group is the Lorentz group $\text{PSL}(2,\RR)$ rather than $\text{U}(1)$, so the orbit 
$\cO_{\text{vac}}\cong\Diff/\PSL$ has codimension three rather than one in $\Diff$.\\

As before, the quantity we wish to compute is $\chi[(f,\alpha)]$, where $\alpha$ is any supertranslation. 
Since the little group is now larger than $\mathrm{U}(1)$, one can obtain non-trivial characters even when 
$f$ is not conjugate to a rotation. We will not consider such cases here and stick instead to our earlier 
convention that $f(\phii)=\phii+\theta$ is a rotation by $\theta\neq0$. (Equivalently we may take $f$ to be 
merely conjugate to a rotation since the character is a class function.) Then the integral of the
Frobenius formula (\ref{fropokad}) localizes to the unique rotation-invariant point $p_{\text{vac}}$ on the 
orbit and the character can be written as\i{supermomentum 
orbit!coordinates}\i{localization}\i{orbit!coordinates}\i{coordinates!on orbit}
\be
\chi_{\text{vac}}[(f,\alpha)]
=
e^{-i\alpha^0c_2/24}
\int_{\cO_{\text{vac}}}d\mu(k)\,
\delta(k,f\cdot k).
\label{CHARvacChar}
\ee
Here $\mu$ is some quasi-invariant measure on the vacuum orbit. Using Fourier expansions (\ref{surfou}), we 
can think of Fourier modes as redundant coordinates on the orbit; the subtlety is to understand which of 
these modes should be modded out so as to provide genuine, non-redundant local coordinates on 
$\cO_{\text{vac}}$.\\

As in the case of massive characters we work in a neighbourhood of the rest frame supermomentum 
$p_{\text{vac}}$ and rely on the action (\ref{CHARinfim}) of infinitesimal 
superrotations. Taking $M=0$ in that equation, we now find that all three modes $\delta p_1$, $\delta p_0$ 
and $\delta p_{-1}$ vanish for any choice of $X$. This is an infinitesimal restatement of the fact that the 
little group is $\PSL$.  Thus, in a neighbourhood of $p_{\text{vac}}$, we can use the higher Fourier modes 
$p_n$ with $|n|\geq 2$ as local coordinates. In particular the measure $\mu$ now takes the form\i{measure!on 
Virasoro orbit}
\be
d\mu(k)=\text{(Some $k$-dependent prefactor)}\times\prod_{n=2}^{+\infty}dk_ndk_{-n}\,,
\nn
\ee
where the prefactor is again unknown, but eventually irrelevant since it is cancelled by the prefactor of the 
corresponding delta function. The vacuum character (\ref{CHARvacChar}) thus boils down to
\begin{eqnarray}
\chi_{\text{vac}}[(f,\alpha)]
& \!\!= &
\!\!e^{-i\alpha^0c_2/24}
\int_{\RR^{2\infty-2}}\prod_{n=2}^{+\infty}dk_ndk_{-n}\prod_{n=2}^{+\infty}
\delta(k_n-[f\cdot k]_n)
\delta(k_{-n}-[f\cdot k]_{-n})\nn\\
& \!\!\refeq{CHARtFour} &
\!\!e^{-i\alpha^0c_2/24}
\bigg|
\int_{\RR^{\infty-1}}\prod_{n=2}^{+\infty}dk_n\prod_{n=2}^{+\infty}
\delta\left(k_n(1-q^n)\right)
\bigg|^2,
\nn
\end{eqnarray}
where $q\equiv\exp[2\pi i\tau]$ and $\tau=(\theta+i\epsilon)/2\pi$ contains an imaginary part $\epsilon$ 
that regularizes the divergence of the infinite product. Integrating the delta functions and 
taking into account the determinant, we finally obtain\i{iepsilon regularization@$i\epsilon$ 
regularization}\i{vacuum character!for BMS$_3$}\i{BMS$_3$ character!for vacuum 
representation}\i{character!for BMS$_3$}
\be
\boxed{\bigg.
\chi_{\text{vac}}[(f,\alpha)]
=
e^{-i\alpha^0c_2/24}
\frac{1}{\prod_{n=2}^{+\infty}|1-q^n|^2}\,.
}
\label{CHARcharBMSvac}
\ee
Note the truncated product starting at $n=2$, which reflects Lorentz-invariance.
As in the massive case above, this expression can be interpreted as a trace (\ref{trachar}), now taken in 
the Hilbert space of the vacuum representation. It can also be recovered as a flat limit of the product of 
two Virasoro vacuum characters (\ref{s88}). 

\paragraph{Remark.} In this section we have systematically assumed that $f$ is a rotation by some non-zero 
angle $\theta$. In doing so we have left aside the interesting problem of computing characters of pure 
supertranslations. This includes for instance Euclidean time translations, whose characters coincide with 
canonical partition functions of BMS$_3$ particles. Analogously to the Poincar\'e results 
(\ref{AHA}) or (\ref{TaBou}), all such characters are infrared-divergent and rely on an integral taken over 
the \it{whole} supermomentum orbit, due to the lack of a localizing delta function. We will not attempt to 
evaluate these characters here.

\newpage
~
\thispagestyle{empty}

\chapter{Partition functions and characters}
\label{c8}
\markboth{}{\small{\chaptername~\thechapter. Partition functions and characters}}

The asymptotic symmetries described in chapter \ref{c6} suggest that the quantization of asymptotically flat 
gravitational fields in three dimensions provides unitary representations of BMS$_3$. In particular, it 
should be possible to identify BMS$_3$ particles with quantized gravitational fluctuations around suitable 
background metrics.\i{BMS$_3$ particle!as quantized metric} The purpose of this chapter is to confirm 
this identification by matching one-loop 
partition functions of gravity with BMS$_3$ characters. As a by-product, the method of heat 
kernels that we shall use for this computation 
also allows us to evaluate partition functions for fields with arbitrary spin, which will lead us to 
higher-spin extensions of BMS$_3$ symmetry. As we will show, the resulting irreducible unitary 
representations can be classified analogously to the BMS$_3$ particles of chapter \ref{c7}, and their 
characters match one-loop partition functions of combinations of higher-spin fields in three-dimensional 
Minkowski space.\\

The plan is as follows. We start in section \ref{seRROT} by evaluating one-loop partition functions of free 
fields with arbitrary mass and spin in $D$-dimensional Minkowski space at finite temperature and angular 
potentials. We show that the result is an exponential of Poincar\'e characters which, for spin two in $D=3$, 
coincides with the vacuum BMS$_3$ character (\ref{CHARcharBMSvac}). In section \ref{CAPsec:3DW} we extend 
this matching to higher-spin theories in three dimensions by describing a method for obtaining induced 
irreducible unitary representations of the corresponding asymptotic symmetry groups. Section 
\ref{COPsec:hs} is devoted to the Lie-algebraic counterpart of that method, which we compare to earlier 
proposals in the literature \cite{Grumiller:2014lna}. We show in particular that ultrarelativistic and 
non-relativistic limits of quantum $\cW$ algebras differ, which singles out induced representations as the 
correct approach to flat space holography. Finally, in section \ref{CAPsupersec} we define supersymmetric 
extensions of the BMS$_3$ group, describe their irreducible unitary representations and show that their 
characters coincide with one-loop partition functions of asymptotically flat hypergravity. Sections 
\ref{LABEKKO} and \ref{CAPITOL} are technical appendices that summarize computations related to 
$\text{SO}(n)$ 
characters which are useful for sections \ref{seRROT} and \ref{CAPsupersec}, respectively.\\

The results described in this chapter first appeared in 
\cite{Barnich:2015mui,Campoleoni:2015qrh,Campoleoni:2016vsh}. They are Min\-kow\-skian analogues of earlier 
observations on partition functions in AdS$_3$ \cite{Giombi:2008vd,David:2009xg,Gaberdiel:2010ar} that we 
already referred to in section \ref{sevirep}. Note that our language in this chapter will be somewhat 
different than in the previous ones, as we will rely much more heavily on quantum field theory. On the other 
hand the group-theoretic tools that we will be using are essentially the same as in chapter \ref{c7}.

\section{\ Rotating canonical partition functions}
\label{seRROT}

We wish to study one-loop partition functions of higher-spin fields in 
$D$\--di\-men\-sio\-nal 
Minkowski space at 
finite temperature $1/\beta$, and with non-zero angular potentials. As in section \ref{relagroup}, we denote 
these potentials by 
$\vec\theta=(\theta_1,...,\theta_r)$,\i{angular potential} where $r=\lfloor(D-1)/2\rfloor$ is the rank of 
$\text{SO}(D-1)$, 
that 
is, the maximal number of independent rotations in $(D-1)$ space dimensions; we assume $D\geq3$. The 
computation 
involves a functional integral over fields living on a quotient of $\RR^D$, where the easiest way to 
incorporate one-loop effects is the heat kernel method. Accordingly we now briefly review this 
approach, before applying it to bosonic fields and rewriting the resulting partition function as an 
exponential of Poincar\'e characters; for spin two and $D=3$, the result coincides with the 
vacuum BMS$_3$ character. Fermions will be treated separately in section \ref{CAPsubsec2.3}.

\subsection{\ Heat kernels and method of images}
\label{suseHEAT}

Our goal is to evaluate partition functions of the form\i{partition function}\i{path integral}
\be
\label{CAPs2.5}
Z(\beta,\vec\theta\,)
=
\int\cD\phi\,e^{-S[\phi]}
\ee
where $\phi$ is some collection of fields (bosonic or fermionic) defined on a thermal quotient $\RR^D/\ZZ$ of 
flat Euclidean space,\i{thermal quotient}\i{quotient space!thermal} satisfying suitable (anti)periodicity 
conditions. (The explicit action of $\ZZ$ on 
$\RR^D$, with its dependence on $\beta$ and $\vec\theta$, will be displayed below --- see 
eq.~(\ref{CAP2.6}).) 
The functional $S[\phi]$ is a Euclidean action for these fields. 
Expression (\ref{CAPs2.5}) can be evaluated perturbatively around a saddle point $\phi_c$ of $S$,\i{saddle 
point approximation} leading to 
the 
semi-classical (one-loop) result\i{functional determinant}
\be
\label{CAPss2.5}
Z(\beta,\vec\theta\,)
\sim
e^{-S[\phi_c]}\left[\text{det}\!\left.\left(
\frac{\delta^2 S}{\delta\phi\delta\phi}
\right)\right|_{\phi_c}\right]^{\#}
\ee
where the exponent $\#$ depends on the nature of the fields that were integrated out. The quantity 
$\delta^2S/\delta\phi(x)\delta\phi(y)$ appearing in this expression is a differential operator acting on 
sections of a suitable vector bundle over $\RR^D/\ZZ$. The evaluation of the one-loop 
contribution to the partition function thus boils down to that of a functional determinant.\\

After gauge-fixing, such determinants reduce to expressions of the form $\text{det}(-\Delta+M^2)$, where 
$\Delta$ is a Laplacian operator on $\RR^D/\ZZ$. These, in 
turn, can be evaluated thanks to the method of heat kernels. In short (see 
e.g.\ \cite{Vassilevich:2003xt,Giombi:2008vd} for details), one can express 
$\text{det}(-\Delta+M^2)$ on $\RR^D$ as an integral\i{heat kernel}
\be
\label{CAP2.5}
-\log\det(-\Delta+M^2)
=
\int_0^{+\infty}
\frac{dt}{t}
\int_{\RR^D}d^Dx\,\text{Tr}\left[K(t,x,x)\right],
\ee
up to an ultraviolet divergence that can be regularized with standard methods. Here $K(t,x,x')$ is a 
matrix-valued bitensor 
known as the \it{heat kernel}\i{heat kernel} associated with $(-\Delta+M^2)$. It satisfies the heat 
equation
\be
\label{CAPs4}
\frac{\partial}{\partial t}\,K(t,x,x')-(\Delta_x-M^2)\,K(t,x,x')=0\,,
\ee
with the initial condition
\be
\label{CAPss4}
K(t=0,x,x')=\delta^{(D)}(x-x')\,\II
\ee
where $\II$ is an identity matrix having the same tensor structure as $K$ (here omitted for brevity) while 
$\delta^{(D)}$ is the Dirac delta function associated with the translation-invariant Lebesgue measure on 
$\RR^D$.\\

Heat kernels are well suited for the computation of functional determinants on quotient spaces. Indeed, 
suppose $\Gamma$ is a discrete subgroup of the isometry group of $\RR^D$, acting freely on $\RR^D$. 
Introducing the 
equivalence relation $x\sim y$ if there exists a $\gamma\in\Gamma$ such
that $\gamma(x)=y$, we define the quotient manifold $\RR^D/\Gamma$\i{quotient space} as the corresponding set 
of equivalence 
classes. Given a differential operator $\Delta$ on $\RR^D$, it naturally induces a 
differential operator on $\RR^D/\Gamma$, acting on fields that satisfy suitable (anti)periodicity conditions. 
Because the heat equation 
(\ref{CAPs4}) is linear, the heat kernel on the quotient space can be obtained from the heat kernel on 
$\RR^D$ 
by 
the method of images:\i{method of images}
\be
\label{CAPt4}
K^{\mathbb{R}^D/\Gamma}(t,x,x')=\sum_{\gamma\,\in\,\Gamma }K\big(t,x,\gamma(x')\big) \, .
\ee
Here, abusing notation slightly, $x$ and $x'$ denote points both in $\RR^D$ and in its quotient. In writing 
(\ref{CAPt4}) we are assuming, for simplicity, that 
the tensor structure of $K$ is trivial, but as soon as $K$ carries tensor or spinor indices 
(i.e.~whenever the fields under consideration have non-zero spin), the right-hand side involves Jacobians 
that account for the non-trivial transformation law of $K$. Once $K^{\RR^D/\ZZ}$ is known, the determinant 
of the operator $-\Delta+M^2$ is given by (\ref{CAP2.5}) with $K$ replaced by $K^{\RR^D/\ZZ}$ and $\RR^D$ 
replaced by $\RR^D/\ZZ$.\\

\begin{figure}[h]
\centering
\includegraphics[width=0.35\textwidth]{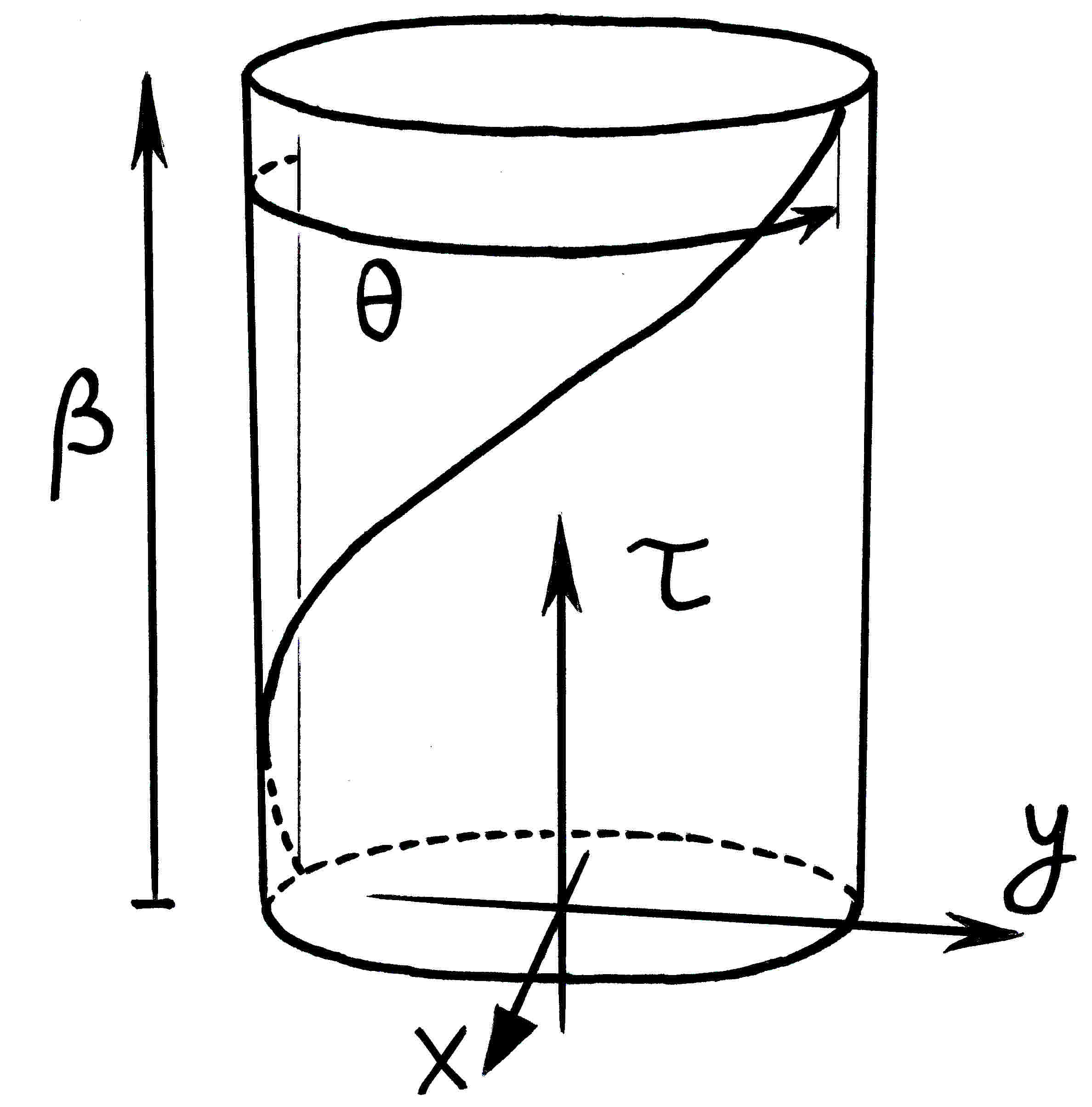}
\caption{The quotient space $\RR^3/\ZZ$ defined by identifications of $\RR^3$ generated by the group action 
(\ref{CAP2.6}); $\beta$ is an inverse temperature while $\theta$ is an angular potential.\label{KALET}}
\end{figure}

We shall be concerned with thermal quantum field theories on rotating Minkowski space, so we
define our fields on a quotient $\RR^D/\ZZ$ of Euclidean space 
with the action of $\ZZ$ obtained as follows. For odd $D$, we endow $\RR^D$ with Cartesian coordinates 
$(x_i,y_i)$ (where $i=1,...,r$) and a Euclidean time coordinate $\tau$,\i{Euclidean time} so that an integer 
$n\in\ZZ$ acts 
on 
$\RR^D$ according to\i{hot flat space} (see fig.\ \ref{KALET})
\be
\label{CAP2.6}
\gamma^n\begin{pmatrix} x_i \\ y_i \end{pmatrix}
=
\begin{pmatrix}
\cos(n\theta_i) & -\sin(n\theta_i)\\
\sin(n\theta_i) & \cos(n\theta_i)
\end{pmatrix} \cdot\begin{pmatrix} x_i \\ y_i \end{pmatrix}
\equiv
R(n\theta_i)\cdot\begin{pmatrix} x_i \\ y_i \end{pmatrix},
\qquad
\gamma^n(\tau)
=
\tau+n\beta\,.
\ee
For even $D$ we add one more spatial coordinate $z$, invariant under $\ZZ$. In terms of the 
coordinates $\{x_1,y_1,...,x_r,y_r,\tau\}$ (and also $z$ if $D$ is even), the 
Euclidean Lorentz transformation implementing the rotation (\ref{CAP2.6}) is the $n^{\text{th}}$ power of the 
rotation matrix
\be
\label{CAP2.7}
J
=
\begin{pmatrix}
 R(\theta_1) & 0 & \cdots & 0 \\
0 &   \ddots & 0 & \vdots \\
\vdots & 0 &R(\theta_r)   & 0 \\
0 & \cdots & 0 & 1
\end{pmatrix}
\quad\text{or}\quad
\begin{pmatrix}
 R(\theta_1) & 0 & \cdots & 0 & 0\\
0 &   \ddots & 0 & \vdots & 0\\
\vdots & 0 &R(\theta_r)   & 0 &0\\
0 & \cdots & 0 & 1&0\\
0 & \cdots & 0 & 0&1
\end{pmatrix}
\ee
for $D$ odd or $D$ even, respectively. Being isometries of flat space, these transformations are linear maps 
in Cartesian coordinates, so their $n^{\text{th}}$ power coincides with the Jacobian matrix 
$\partial\gamma^n(x)^{\mu}/\partial x^{\nu}$ that will be needed later for the method of images. Throughout 
this chapter we take all angles $\theta_1,...,\theta_r$ to be non-vanishing and combine them in a vector 
$\vec\theta=(\theta_1,...,\theta_r)$. We now display the computation 
of one-loop partition functions on $\RR^D/\ZZ$ for bosonic higher-spin fields.

\subsection{\ Bosonic higher spins}
\label{CAPsubsec2.2}

Here we study the rotating one-loop partition function of a free bosonic field with spin $s$ 
and mass $M$ (including the massless case).\i{higher spin} For $M>0$ its Euclidean action can be 
presented either (i) 
using a symmetric traceless field $\phi_{\mu_1...\mu_s}$ of rank $s$ together with a tower of auxiliary 
fields of 
ranks $s-2,s-3,..., 0$  that do not display any gauge symmetry \cite{Singh:1974qz}; or (ii) using a set of 
doubly traceless fields of ranks $s,s-1,...,0$ subject to a gauge symmetry generated by traceless gauge 
parameters of ranks $s-1,s-2,...,0$ \cite{Zinoviev:2001dt}. In the latter case, the action is 
a sum of Fronsdal actions \cite{Fronsdal:1978rb} for each of the involved fields, plus a set of 
cross-coupling terms with one derivative proportional to $M$, and a set of terms without derivatives 
proportional to $M^2$. In the massless limit, all these couplings vanish and one can consider independently 
the (Euclidean) Fronsdal action for the field of highest rank:\i{Fronsdal action}\i{action functional!for 
higher spin field}
\be
\label{CAPs3}
S[\phi_{\mu_1...\mu_s}]
=
-\frac{1}{2}\int d^Dx\;\phi^{\mu_1...\mu_s}
\left(
\cF_{\mu_1...\mu_s}
- 
\frac{1}{2}\,\delta_{(\mu_1\mu_2}{\cF_{\mu_3...\mu_s)\lambda}}^\lambda
\right),
\ee
where indices are raised and lowered thanks to the Euclidean metric, while
\be
\cF_{\mu_1 \cdots \mu_s}
\equiv
\Delta\,\phi_{\mu_1...\mu_s}
-\der_{(\mu_1|}\der^{\lambda}\phi_{|\mu_2...\mu_s)\lambda}
+\der_{(\mu_1}\der_{\mu_2}{\phi_{\mu_3...\mu_s)\lambda}}^{\lambda}\,.
\ee
Parentheses denote the symmetrization\i{symmetrization} of the indices they enclose, with the minimum number 
of terms 
needed 
and without any overall factor. The massless action (\ref{CAPs3}) has a gauge symmetry\i{gauge 
transformation} 
$\phi_{\mu_1...\mu_s}\mapsto\phi_{\mu_1...\mu_s}+\der_{(\mu_1}\xi_{\mu_2...\mu_s)}$, where 
$\xi_{\mu_2...\mu_s}$ is a symmetric tensor field. When $s=2$ the action reduces to that of a metric 
perturbation $h_{\mu\nu}$ around a flat background. We refer for instance to \cite{Rahman:2013sta} for many 
more details on this topic.

\subsubsection*{Massive case}

Applying e.g.\ the techniques of \cite{Gaberdiel:2010ar} to the presentation of the Euclidean action of a 
massive field of spin $s$ of \cite{Zinoviev:2001dt}, one finds that the partition function is given 
by\i{bosonic partition function}
\be
\label{CAPss5b}
\log Z
=
-\,\frac{1}{2}\log \det (-\Delta^{(s)}+M^2)+\frac{1}{2}\log \det (-\Delta^{(s-1)}+M^2) \, ,
\ee
where $\Delta^{(s)}$ is the Laplacian $\partial_{\mu}\partial^{\mu}$ acting on periodic,\footnote{More 
precisely, the field at time $\tau+\beta$ is rotated by $\vec\theta$ with respect to the field at 
time $\tau$.} symmetric, traceless 
tensor fields with $s$ indices on $\RR^D/\ZZ$. We denote the 
heat kernel associated with $(-\Delta^{(s)}+M^2)$ on $\RR^D$ by $K_{\mu_s,\nu_s}(t,x,x')$, where $\mu_s$ and 
$\nu_s$ are shorthands that denote sets of $s$ symmetrized indices.
The differential equation (\ref{CAPs4}) with initial condition (\ref{CAPss4}) for 
$K_{\mu_s,\nu_s}(t,x,x')$ then reads
\be
\label{CAP2.8}
(\Delta^{(s)}-M^2-\der_t)K_{\mu_s,\,\nu_{s}}=0\,,
\quad
K_{\mu_s,\,\nu_s}(t=0,x,x')=\II_{\mu_s,\,\nu_s} 
\delta^{(D)}(x-x')\,,
\ee
where $\II_{\mu_s,\nu_s}$ is an identity matrix with the same tensor structure as $K_{\mu_s,\nu_s}$. Sets of 
repeated covariant or contravariant indices denote sets of indices that are symmetrized with the 
minimum number of terms required and without multiplicative factors, while contractions involve as usual a 
covariant and a contravariant index. For instance the tracelessness condition on the heat kernel amounts 
to\i{traceless heat kernel}\i{heat kernel!traceless}
\be
\label{CAP2.9}
\delta^{\mu \mu} {K}_{\mu_{s},\,\nu_s}= \delta^{\nu \nu} K_{\mu_s,\,\nu_{s}}=0 \, .
\ee
The unique solution of (\ref{CAP2.8}) fulfilling this condition is
\be
\label{CAPs6}
K_{\mu_s,\,\nu_s}(t,x,x')
=
\frac{1}{(4\pi t)^{D/2}}\,
e^{-M^2 t-\frac{1}{4t}|x-x'|^2}\;
\II_{\mu_s,\,\nu_s}
\ee
where $|x-x'|$ is the Euclidean distance between $x$ and $x'$, while the spin-$s$ identity matrix is
\be
\II_{\mu_s,\,\nu_s}
=
\sum^{\lfloor\frac{s}{2}\rfloor}_{n=0}
\frac{(-1)^n 2^n n!\,[D+2(s-n-2)]!!}
{s!\,[D+2(s-2)]!!}\, \delta_{\mu\mu}^n \delta_{\mu\nu}^{s-2n} \delta_{\nu\nu}^n\,.
\ee
Note that the dependence of this heat kernel on the space-time points $x$, $x'$ and on Schwinger proper time 
$t$ is that of a scalar heat kernel, and completely factorizes from its spin/index structure which is 
entirely 
accounted for by the matrix $\II$.\footnote{Note also that the scalar heat kernel coincides with the 
propagator of a free particle in $\RR^D$, whose expression for $D=2$ was written in eq.\ (\ref{galipipi}).} 
This simplification is the reason why heat kernel 
computations are simpler in flat space than in AdS or dS.\\

To determine the heat kernel associated with the operator $(-\Delta^{(s)}+M^2)$ on $\RR^D/\ZZ$, we use the 
method of images (\ref{CAPt4}), taking care of the non-trivial index structure. Denoting the matrix 
(\ref{CAP2.7}) 
by ${J_{\alpha}}^{\beta}$ (it is the Jacobian of the transformation $x\mapsto \gamma(x)$), the spin-$s$ heat 
kernel on the quotient space $\RR^D/\ZZ$ is\i{method of images}
\be
K^{\mathbb{R}^D/\mathbb{Z}}_{\mu_s,\,\nu_s}(t,x,x')
=
\sum_{n\,\in\,\ZZ}{(J^n)_{\alpha}}^{\beta}...{(J^n)_{\alpha}}^{\beta}
K_{\mu_s,\,\beta_s}\big(t,x,\gamma^n(x')\big) \, ,
\ee
where we recall again that repeated covariant or contravariant indices are meant to be symmetrized with the 
minimum number of terms required and without multiplicative factors, while repeating a covariant index in a 
contravariant position denotes a contraction.
Accordingly, eq.\ (\ref{CAP2.5}) gives the determinant of $(-\Delta^{(s)}+M^2)$ on $\RR^D/\ZZ$:\i{functional 
determinant}
\be
\begin{split}
& -\log\det(-\Delta^{(s)}+M^2) =
\int_0^{+\infty} \frac{dt}{t}
\int_{\RR^D/\ZZ} d^Dx\,
(\delta^{\mu \alpha})^s\,
K^{\mathbb{R}^D/\mathbb{Z}}_{\mu_s,\,\alpha_s} (t,x,x) \\[3pt]
& =
\sum_{n\,\in\,\ZZ}
(J^n)^{\mu\beta}\cdots(J^n)^{\mu\beta}\,
\II_{\mu_s,\beta_s}
\int_0^{+\infty} \frac{dt}{t}
\int_{\RR^D/\ZZ} d^Dx\,
\frac{1}{(4\pi t)^{D/2}}\,
e^{-M^2 t-\frac{1}{4t}|x-\gamma^n(x)|^2}.\quad\quad\label{CAPs5t}
\end{split}
\ee
In this series the term $n=0$ contains both an ultraviolet divergence (due to the singular 
behaviour of the integrand as $t\rightarrow 0$) and an infrared one (due to the integral of a constant over 
$\RR^D/\ZZ$), proportional to the product $\beta V$ where $V$ is the spatial volume of the system. This 
divergence is a quantum contribution to the vacuum energy, which we ignore from now on. The only non-trivial 
one-loop contribution then comes 
from the terms $n\neq0$ in (\ref{CAPs5t}). Using
\be
|x-\gamma^n(x)|^2
=
n^2\beta^2+\sum_{i=1}^r4\sin^2(n\theta_i/2)(x_i^2+y_i^2)
\nn
\ee
in terms of the coordinates introduced around (\ref{CAP2.6}), the integrals over $t$ and $x$ give rise to a 
divergent series\i{one-loop determinant}\i{functional determinant}
\be
\label{CAPADIV}
-\log\det(-\Delta^{(s)}+M^2)
=
\sum_{n\,\in\,\ZZ^*}
\frac{1}{|n|}
\frac{\chi_s[n\vec\theta\,]}{\prod\limits_{j=1}^r|1-e^{in\theta_j}|^2}
\times
\begin{cases}
e^{-|n|\beta M}  &\mbox{if } D \ \text{odd},\\[3pt]
\frac{ML}{\pi}K_1(|n|\beta M)& \mbox{if } D \  \text{even},
\end{cases}
\ee
where $K_1$ is the first modified Bessel function of the second 
kind,\i{Bessel function} $L\equiv\int_{-\infty}^{+\infty}dz$ is an infrared divergence (\ref{tikk}) that 
arises in even dimensions 
because the $z$ axis is left fixed by the rotation (\ref{CAP2.7}), and
\be
\label{CAPs5.5t}
\chi_s[n\vec\theta\,]
\equiv
(J^n)^{\mu\beta}...(J^n)^{\mu\beta}\,\II_{\mu_s,\,\beta_s}
\equiv
\left[(J^n)^{\mu\beta}\right]^s \II_{\mu_s,\,\beta_s}
\ee
is the full mixed trace of $\II_{\mu_s,\nu_s}$. As such, expression (\ref{CAPADIV}) makes no sense because 
the sum over $n$ diverges. To cure this problem, one needs to choose a regularization procedure. 
Motivated by the similar situation already encountered in eq.\ (\ref{CHARmodPar}), for now we choose to 
regulate the series by a naive replacement: we let $\epsilon_j$, $j=1,...,r$ be small positive parameters and 
replace $\theta_j$ by $\theta_j\pm i\epsilon_j$ in all positive powers of $e^{\pm i\theta_j}$. As a result, 
expression (\ref{CAPADIV}) is replaced by the convergent series\i{iepsilon regularization@$i\epsilon$ 
regularization}
\be
\label{CAPss5t}
-\log \det(-\Delta^{(s)}+M^2)
=
\sum_{n\,\in\,\ZZ^*}
\frac{1}{|n|}
\frac{\chi_s[n\vec\theta,\vec{\epsilon}\,]}{\prod\limits_{j=1}^r|1-e^{in(\theta_j+i\epsilon_j)}|^2}
\times
\begin{cases}
e^{-|n|\beta M}  &\mbox{if } D \ \text{odd},\\[3pt]
\frac{ML}{\pi}K_1(|n|\beta M)& \mbox{if } D \  \text{even},
\end{cases}
\ee
where $\chi_s[n\vec\theta,\vec{\epsilon}\,]$ is still given by (\ref{CAPs5.5t}), except that now all 
factors $e^{\pm i\theta_j}$ appearing in the Jacobians are replaced by $e^{\pm i(\theta_j\pm i\epsilon_j)}$.\\

The regularization described here is motivated by the fact that, for 
odd $D$, the resulting expressions look very much like flat limits of AdS one-loop determinants, in which 
case 
the parameters $\epsilon_j\propto\beta/\ell$ are remnants of the inverse temperature (with $\ell$ the AdS 
radius). The subtlety, however, is that the exact matching of the flat limit of AdS with combinations such as 
(\ref{CAPss5t}) requires some of the $\epsilon_j$'s to be multiplied by certain positive coefficients; thus 
eq.~(\ref{CAPss5t}) is not quite the same as the flat limit of its AdS counterpart --- we will illustrate 
this point for $D=3$ in section \ref{susepabako}. As for even values of $D$, the 
situation is even worse since the flat limit of the AdS result contains an infrared 
divergence; it is not obvious how this divergence can be regularized so as to reproduce the combination 
$L\cdot K_1$ of (\ref{CAPss5t}), though apart from this the other terms of the expression 
indeed coincide with the flat limit of their AdS counterparts. From now on we will use the $i\epsilon$ 
prescription systematically, often omitting to indicate it explicitly. We will keep it only in the final 
results, and in section \ref{susepabako} we will introduce a refined regularization such that 
partition functions in $D=3$ exactly reproduce characters of suitable asymptotic symmetry algebras, while 
also matching the flat limit of their AdS peers.\\

In eq.\ (\ref{CAPss5t}), the divergence as $\epsilon_j\rightarrow0$ is the same as in the BMS$_3$ 
character (\ref{CHARcharBMS}). The new ingredient is the angle-dependent trace (\ref{CAPs5.5t}); in 
appendices \ref{CAPAppA1} and \ref{CAPAppA2} we show that the latter is the character of an irreducible, 
unitary 
representation of $\text{SO}(D)$ with highest weight $\lambda_s\equiv(s,0,...,0)$.\i{highest-weight 
representation}\i{son@$\mathfrak{so}(n)$!highest weight} More precisely, let $H_i$ denote the generator of 
rotations in the plane $(x_i,y_i)$, in the coordinates defined around (\ref{CAP2.6}). Then the Cartan 
subalgebra\i{Cartan subalgebra} $\mathfrak{h}$ of $\mathfrak{so}(D)$ is generated by 
$H_1,...,H_r$, plus, if $D$ is even, a generator of rotations in the plane $(\tau,z)$. Denoting the dual 
basis of $\mathfrak{h}^*$ by $L_1,...,L_r$ (plus possibly $L_{r+1}$ if $D$ is even), we can consider the 
weight $\lambda_s=sL_1$ whose only non-zero component (in the basis of $L_i$'s) is the first 
one. The 
character of the corresponding highest-weight representation of $\mathfrak{so}(D)$ coincides with expression 
(\ref{CAPs6.5}):\i{character!for SOn@for $\text{SO}(n)$}\i{son@$\mathfrak{so}(n)$!character}
\be
\label{CAPs6b}
\chi_s[n\vec\theta\,]
=
\chi_{\lambda_s}^{(D)}[n\theta_1,...,n\theta_r]
\qquad\text{or}\qquad
\chi_{\lambda_s}^{(D)}[n\theta_1,...,n\theta_r,0] \, ,
\ee
for $D$ odd or even, respectively. From now on, $\chi_{\lambda}^{(n)}$ denotes a character of $\text{SO}(n)$ 
with highest weight $\lambda$.\\

We can now display the one-loop partition function (\ref{CAPss5b}). Using 
expression (\ref{CAPss5t}) for the one-loop determinant together with property (\ref{CAPs6b}), 
we find
\be
Z(\beta,\vec\theta\,)= 
\exp\!
\left[
\sum_{n=1}^{+\infty}
\frac{n^{-1}}{\prod\limits_{j=1}^r|1-e^{in\theta_j}|^2} \times
\begin{cases}
\left(
\chi_{\lambda_s}^{(D)}[n\vec\theta\,]
-
\chi_{\lambda_{s-1}}^{(D)}[n\vec\theta\,]
\right)
e^{-n\beta M}\\[10pt]
\left(
\chi_{\lambda_s}^{(D)}[n\vec\theta,0]
-
\chi_{\lambda_{s-1}}^{(D)}[n\vec\theta,0]
\right)\!
\frac{ML}{\pi}K_1(n\beta M)
\end{cases}
\!\!\!\!\!\!\right]
\label{CAPs6.5b}
\ee
where the upper (resp.\ lower) line corresponds to the case where $D$ is odd (resp.\ even). Remarkably, the 
differences of $\text{SO}(D)$ characters appearing here can be simplified: according to 
eqs.~(\ref{CAPapp:SO(D)DifferenceProofRelations1}) and (\ref{CAPs37.5}), the difference of two $\text{SO}(D)$ 
characters with weights $(s,0,...,0)$ and $(s-1,0,...,0)$ 
is a (sum of) character(s) of $\text{SO}(D-1)$:
\be
\label{CAPt6.5b}
\begin{rcases*}
\chi_{\lambda_s}^{(D)}[\vec\theta\,]
-
\chi_{\lambda_{s-1}}^{(D)}[\vec\theta\,] & (D\text{ odd})\\[5pt]
\chi_{\lambda_s}^{(D)}[\vec\theta,0]
-
\chi_{\lambda_{s-1}}^{(D)}[\vec\theta,0] & (D\text{ even})
\end{rcases*}
=
\chi_{\lambda_s}^{(D-1)}[\vec\theta\,] \, .
\ee
Since the rank of $\text{SO}(D-1)$ is $r=\lfloor(D-1)/2\rfloor$, the right-hand side of this equality makes 
sense regardless of the parity of $D$, and the partition function (\ref{CAPs6.5b}) boils down to\i{bosonic 
partition function}
\be
\label{CAPss6.5b}
Z(\beta,\vec\theta\,)
=
\exp\!
\left[\,
\sum_{n=1}^{+\infty}
\frac{1}{n}
\frac{\chi_{\lambda_s}^{(D-1)}[n\vec{\theta},\vec{\epsilon}\,]}
{\prod\limits_{j=1}^r|1-e^{in(\theta_j+i\epsilon_j)}|^2}
\times
\begin{cases}
e^{-n\beta M} & (D\text{ odd})\\[5pt]
\frac{ML}{\pi}K_1(n\beta M) & (D\text{ even})
\end{cases}
\,\right].
\ee
Note that the function 
of $n\vec\theta$ and $n\beta$ appearing here in the sum over $n$ is essentially the character 
(\ref{Chamouss})-(\ref{Xufy}) of a Poincar\'e particle with mass $M$ and spin $\lambda_s$; we will return to 
this observation in section \ref{CAPsec:poincare}. An 
analogous result holds in Anti-de Sitter space \cite{Gibbons:2006ij,Gopakumar:2011qs,Beccaria:2015vaa}.

\subsubsection*{Massless case}

We now turn to the one-loop partition function associated with the Euclidean Fronsdal action (\ref{CAPs3}),
describing a \it{massless} field with spin $s$. The extra ingredient with respect to the massive case is the 
gauge 
symmetry
$\phi_{\mu_1...\mu_s}\mapsto\phi_{\mu_1...\mu_s}+\der_{(\mu_1}\xi_{\mu_2...\mu_s)}$.\i{gauge transformation} 
This 
forces one to fix a 
gauge and introduce ghost\i{ghost} fields that absorb the gauge redundancy \cite{Gaberdiel:2010ar}, which 
adds two more functional determinants to the massive result (\ref{CAPss5b}) and
leads to the 
following expression for the one-loop term of the partition function:
\be
\label{CAP2.4bos}
\log Z
=
-\frac{1}{2}\log \det (-\Delta^{(s)})+\log\det(-\Delta^{(s-1)})-\frac{1}{2}\log\det(-\Delta^{(s-2)}) \, .
\ee
As before, $\Delta^{(s)}$ is the Laplacian on $\RR^D/\ZZ$ acting on periodic, traceless, symmetric fields 
with $s$ indices. The functional determinants can be evaluated exactly as in the massive case, upon setting 
$M=0$. 
In particular, using $\lim_{x\rightarrow0}xK_1(x)=1$, the massless version of the functional determinant 
(\ref{CAPss5t}) is
\be
\label{CAPs6.5t}
-\log \det(-\Delta^{(s)})
=
\sum_{n\,\in\,\ZZ^*}
\frac{1}{|n|}
\frac{\chi_s[n\vec\theta,\vec{\epsilon}\,]}{\prod\limits_{j=1}^r|1-e^{in(\theta_j+i\epsilon_j)}|^2}
\times
\begin{cases}
1  &\mbox{if } D \ \text{odd},\\
\frac{L}{\pi|n|\beta}&\mbox{if } D \  \text{even},
\end{cases}
\ee
which has been regularized as in the massive case. The matching (\ref{CAPs6b}) between $\chi_s$ and 
a 
character of $\text{SO}(D)$ remains valid, but a sharp 
difference arises upon including all three functional determinants in 
(\ref{CAP2.4bos}). Indeed, the combination of $\chi_s$'s now is
\be
\label{CAPss6.5t}
\chi_s[n\vec\theta\,]-2\chi_{s-1}[n\vec\theta\,]+\chi_{s-2}[n\vec\theta\,]
\stackrel{\text{(\ref{CAPs6b})-(\ref{CAPt6.5b})}}{=}
\chi_{\lambda_s}^{(D-1)}[n\vec\theta\,]
-
\chi_{\lambda_{s-1}}^{(D-1)}[n\vec\theta\,]\,.
\ee
It is tempting to use (\ref{CAPt6.5b}) once more to rewrite this as a character of $\text{SO}(D-2)$, and 
indeed 
this is exactly what happens for even $D$ because in that case the rank of $\text{SO}(D-1)$ equals that of 
$\text{SO}(D-2)$:
\be
\label{CAPt6.5t}
Z(\beta,\vec\theta\,)
=
\exp
\left[
\sum_{n=1}^{+\infty}
\frac{1}{n}
\frac{\chi_{\lambda_s}^{(D-2)}[n\vec{\theta},\vec{\epsilon}\,]}
{\prod\limits_{j=1}^r|1-e^{in(\theta_j+i\epsilon_j)}|^2}
\frac{L}{\pi n\beta}
\right]
\qquad(\text{even }D).
\ee
If $D$ is \it{odd}, however, the rank 
decreases by one unit in going from $\text{SO}(D-1)$ to $\text{SO}(D-2)$, so expression 
(\ref{CAPss6.5t}) contains one angle too much to be a character of 
$\text{SO}(D-2)$. In fact, when $D=3$, the right-hand side of (\ref{CAPss6.5t}) is the best we can hope to 
get; for $s\geq2$ it takes the form
\be
\chi_{\lambda_s}^{(1)}[n\theta]
\equiv
\chi_{\lambda_s}^{(2)}[n\theta]
-
\chi_{\lambda_{s-1}}^{(2)}[n\theta]
=
e^{isn\theta}-e^{i(s-1)n\theta}+\text{c.c.}
\label{CAPILOBO}
\ee
where we have used the character $\chi_s[\theta]=e^{is\theta}+e^{-is\theta}$ for parity-invariant 
unitary representations of $\text{SO}(2)$ and ``c.c.'' means ``complex conjugate''. (For lower spins one has 
$\chi_{\lambda_0}^{(1)}[\theta]\equiv1$ and $\chi_{\lambda_1}^{(1)}[\theta]\equiv2\cos\theta-1$.) Hence the 
partition function given by (\ref{CAP2.4bos}) becomes
\be
Z(\beta,\theta)
=
e^{-S^{(0)}}
\exp
\left[
\sum_{n=1}^{+\infty}\frac{1}{n}\frac{1}{|1-e^{in(\theta+i\epsilon)}|^2}
\left(
e^{isn(\theta+i\epsilon)}-e^{i(s-1)n(\theta+i\epsilon)}+\text{c.c.}
\right)
\right]
\qquad
(D=3)
\label{CAPOEIRA}
\ee
upon using the crude regularization described below eq.~(\ref{CAPss5t}). For the sake of generality we have 
included a spin-dependent classical 
action $S^{(0)}$, whose value is a matter of normalization and is generally taken to vanish, except for spin 
two (see below). In the more general case where $D$ is odd and larger than three, a simplification does occur 
on the 
right-hand side of (\ref{CAPss6.5t}): as we show in appendix \ref{CAPAppA3}, the 
difference (\ref{CAPss6.5t}) can be written as a sum of $\text{SO}(D-2)$ characters with angle-dependent 
coefficients (see eq.~(\ref{CAPapp:SO(D)DifferenceProofRelations2})). Indeed, let us define 
\be
\label{CAPs6q}
\cA_k^r(\vec\theta\,)
\equiv
\frac{|\cos((r-i)\theta_j)|_{\theta_k=0}}{|\cos((r-i)\theta_j)|}\,,
\qquad k=1,...,r,
\ee
where $|A_{ij}|$ denotes the determinant of an $r\times r$ matrix. Then the rotating one-loop partition 
function for a massless field with spin $s$ in odd space-time dimension $D\geq5$ reads
\be
\label{CAPss6q}
Z(\beta,\vec\theta\,)
=
\exp
\left[
\sum_{n=1}^{+\infty}
\frac{1}{n}
\frac{\sum\limits_{k=1}^r\cA_k^r(n\vec\theta,\vec{\epsilon}\,)\,
\chi_{\lambda_s}^{(D-2)}[n\theta_1,...,\widehat{n\theta_k},...,n\theta_r, \vec{\epsilon}\,]}
{\prod\limits_{j=1}^r|1-e^{in(\theta_j+i\epsilon_j)}|^2}
\right]
\qquad(\text{odd }D\geq5),
\ee
where the hat on top of an argument denotes omission.\\

Note that the massless partition functions (\ref{CAPt6.5t}) and (\ref{CAPss6q}) are related to the massless 
limit of 
(\ref{CAPss6.5b}). Indeed, as we show in appendix \ref{CAPAppA4}, it turns out that\i{bosonic partition 
function!massless limit}
\be
\label{CAPt6q}
\chi_{\lambda_s}^{(D-1)}[\vec\theta\,]
=
\sum_{j=0}^s
\begin{cases}
\sum_{k=1}^r\cA^r_k(\vec\theta\,)\,
\chi_{\lambda_j}^{(D-2)}[\theta_1,...,\widehat{\theta_k},...,\theta_r]
& \text{ for odd }D,
\\[8pt]
\chi_{\lambda_j}^{(D-2)}[\vec\theta\,]
& \text{ for even }D.
\end{cases}
\ee
Accordingly, the massless limit of a massive partition function with spin $s$ is a product of massless 
partition functions with spins ranging from $0$ to $s$,
\be
\label{CAPs6.5q}
\lim_{M\rightarrow0}Z_{M,s}
=
\prod_{j=0}^sZ_{\text{massless},j}\,,
\ee
consistently with the structure of the massive action \cite{Zinoviev:2001dt}. This result stresses again the 
role of 
the functions 
$\cA_k^r(\vec\theta\,)$ defined in (\ref{CAPs6q}): when the space-time dimension is odd, one 
needs 
angle-dependent coefficients because the rank of the little group of massless particles is smaller than the 
maximum 
number of angular velocities, so that a single $SO(D-2)$ character cannot account for all of them. By the 
way, the results (\ref{CAPt6q}) and (\ref{CAPs6.5q}) also hold in dimension $D=3$, provided one takes the 
``characters'' $\chi^{(1)}_{\lambda_s}[\theta]$ to be of the form (\ref{CAPILOBO}) with 
$\chi^{(1)}_{\lambda_0}[\theta]=1$ and $\chi^{(1)}_{\lambda_1}[\theta]=2\cos\theta-1$.

\subsection{\ Partition functions and BMS$_3$ characters}
\label{susepabako}

Let us rewrite the three-dimensional partition function (\ref{CAPOEIRA}) in a form more convenient for the 
group-theoretic discussion of the remainder of this chapter. As mentioned above, the only non-trivial step 
will be to slightly modify the $i\epsilon$ regularization. Namely, instead of the combination of exponentials 
appearing in (\ref{CAPOEIRA}), consider the expression\i{iepsilon regularization@$i\epsilon$ regularization}
\be
e^{isn(\theta+i\epsilon)}-e^{i(s-1)n\theta-(s+1)n\epsilon}+\text{c.c.}
\label{KAPELLEKE}
\ee
Writing $q\equiv e^{i(\theta+i\epsilon)}$ and plugging (\ref{KAPELLEKE}) into the sum over $n$ of 
eq.\ (\ref{CAPOEIRA}), one obtains the series
\be
\sum_{n=1}^{+\infty}\frac{1}{n}\frac{q^{ns}-q^{ns}\bar q^{n}+\text{c.c.}}{|1-q^n|^2}
=
\sum_{n=1}^{+\infty}\frac{1}{n}\left(\frac{q^{ns}}{1-q^n}+\text{c.c.}\right)
=
-\sum_{j=s}^{+\infty}\log(1-q^j)+\text{c.c.}
\label{Storm}
\ee
where the new regularization (\ref{KAPELLEKE}) has ensured that the summand decomposes as the sum of a chiral 
and an anti-chiral piece in $q$. (This was \it{not} the case with the rough regularization of 
eq.~(\ref{CAPss5t})!) In order to write down the full partition function, it only remains to assign a value to 
the classical action 
$S^{(0)}$; a convention that has come to be standard in the realm of three-dimensional gravity is to set 
$S^{(0)}=0$ for any spin $s\neq2$ (vacuum expectation values are assumed to vanish), while 
$S^{(0)}=-\beta/8G$ for spin two (with $G$ the Newton constant in three dimensions). This choice ensures 
covariance of the on-shell action under modular 
transformations \cite{Barnich:2015mui,Bagchi:2013lma}, in analogy with the similar choice in AdS$_3$ 
\cite{Giombi:2008vd}. All in all, combining the value of $S^{(0)}$ with the series (\ref{Storm}) and renaming 
$j$ into $n$, one 
finds that the three-dimensional partition function (\ref{CAPOEIRA}) can be written as\i{higher 
spin!partition function}\i{bosonic partition function!in 3D}\i{partition 
function!for 3D gravity}\i{3D gravity!partition function}\i{boundary graviton!partition function}
\be
\label{CAPss6.5q}
Z(\beta,\theta)
=
e^{\delta_{s,2} \frac{\beta c_2}{24}}
\prod_{n=s}^{+\infty}\frac{1}{|1-e^{in(\theta+i\epsilon)}|^2}\,,
\quad
c_2=3/G.
\ee
This expression is the flat limit of the analogous higher-spin partition function in AdS$_3$ 
\cite{Gaberdiel:2010ar}. But most importantly for our purposes, taking $s=2$ in this formula, we recognize 
the vacuum BMS$_3$ character 
(\ref{CHARcharBMSvac}). This is our first key conclusion in this chapter: it confirms that boundary 
gravitons in flat space form an irreducible unitary representation of the BMS$_3$ group of the type described 
in chapter \ref{c7}. The case $s>2$ will be studied in section \ref{CAPsec:3DW}, with similar conclusions.\\

The result (\ref{CAPss6.5q}) can be generalized to orbifolds in flat space:\i{orbifold} upon declaring that 
the angular 
coordinate $\phii$ 
of (\ref{bominkow}) is identified as $\phii\sim\phii+2\pi/N$ with some integer $N>1$, one obtains a flat 
three-dimensional conical deficit. One can then evaluate heat kernels on that background by computing a sum 
over images (\ref{CAPt4}), where $\Gamma$ is now a group $\ZZ\times\ZZ_N$ whose two generators enforce (i) 
the 
thermal identifications (\ref{CAP2.6}), and (ii) the orbifolding $\phii\sim\phii+2\pi/N$. An important 
technical subtlety is that, in order to evaluate a partition function with temperature $1/\beta$ and angular 
potential 
$\theta$ on that background, the angle appearing in (\ref{CAP2.6}) must be $\theta/N$ rather than $\theta$. 
The rest of the computation is straightforward, and one finds that the one-loop partition 
function of gravity can be written as
\be
Z_N(\beta,\theta)
=
e^{-\beta p_0}\prod_{n=1}^{+\infty}\frac{1}{|1-e^{in(\theta+i\epsilon)}|^2}\,,
\nn
\ee
where $p_0=-c_2/(24N^2)$. Comparing with (\ref{CHARcharBMS}), we recognize the (Euclidean) character of a 
BMS$_3$ particle with mass $M=\frac{c_2}{24}(1-1/N^2)$, which is indeed the mass one would obtain by writing 
the conical deficit metric in BMS form (\ref{piment}). Note in particular that the sum over images of the 
orbifolding group $\ZZ_N$ converts the truncated product of (\ref{CAPss6.5q}) into a full product 
$\prod_{n=1}^{+\infty}(\cdots)$.\\

The result (\ref{CAPss6.5q}) first appeared in \cite{Barnich:2015mui} and parallels 
earlier observations in AdS$_3$ 
\cite{Giombi:2008vd}. It is tempting to conjecture that formula (\ref{CAPss6.5q}) is one-loop 
exact,\i{one-loop exactness} since it is the only expression compatible with BMS$_3$ symmetry. This being 
said we will not need to 
assume one-loop exactness in this thesis and we will not attempt to prove it. The remainder of this chapter 
is devoted to various extensions of this matching.

\paragraph{Remark.} In \cite{Bonzom:2015ans} it was shown that the one-loop partition function 
(\ref{CAPss6.5q}) with $s=2$, and hence the vacuum BMS$_3$ character (\ref{CHARcharBMSvac}), can be 
reproduced using quantum Regge calculus.\i{Regge calculus} In that context the truncation of the product 
over $n=2,3,...$ is a consequence of triangulation-invariance in the bulk.

\subsection{\ Relation to Poincar\'e characters}
\label{CAPsec:poincare}

We now show that all one-loop partition functions displayed above can be written as 
exponentials of (sums of) the Poincar\'e characters of section \ref{relagroup}. Recall 
in particular that massive characters are given by eq.\ (\ref{Chamouss}), where $f$ is the rotation 
(\ref{CAP2.7}) and $\chi_{\lambda}$ is the character of an irreducible representation of the little group 
$\SO(D-1)$ with highest weight $\lambda$.\\

In order to represent a massive relativistic particle with spin $s$, we choose the weight $\lambda$ to be 
$\lambda_s=(s,0,...,0)$ in terms of the dual basis of the Cartan subalgebra of $\mathfrak{so}(D-1)$ described 
above (\ref{CAPs6b}). With 
this choice, expressions (\ref{Chamouss}) and (\ref{Xufy}) actually appear in the exponent of 
(\ref{CAPss6.5b}): 
taking $\alpha^0=i\beta$, we can rewrite the rotating one-loop partition function for a massive field with 
spin $s$ as the exponential of a sum of Poincar\'e characters:\i{character!within partition 
function}\i{partition function}\i{Poincar\'e character!in a partition 
function}
\be
\label{CAPZexp}
Z_{M,s}[\beta,\vec\theta\,]
=
\exp\left[\,
\sum_{n=1}^{+\infty}\frac{1}{n}\,\chi_{M,s}[n\vec{\theta},in\beta]
\,\right].
\ee
The series in the exponent diverges for real $\theta_i$'s, which can be cured by 
adding suitable imaginary parts to these angles as explained above. This result holds for 
any space-time dimension $D$ (along with the infrared regularization (\ref{tikk})). From a  physical 
perspective, it is the statement that a free field is a collection of 
harmonic oscillators, one for each value of momentum: the index $n$ then labels the oscillator modes, while 
the integral over momenta is the one in the Frobenius formula (\ref{fropokad}). In particular, 
standard, non-rotating one-loop partition functions are exponentials of sums of characters of (Euclidean) 
time translations. This relation has also been observed in AdS 
\cite{Dolan:2005wy,Gibbons:2006ij,Gopakumar:2011qs}; our partition functions are flat limits of these earlier 
results, up to the even-dimensional regularization subtlety mentioned below eq.~(\ref{CAPs5.5t}). Note that 
this 
issue already occurs at the level of characters: although most of (\ref{Xufy}) is a flat limit of an 
$\text{SO}(D-1,2)$ character, it is not clear how to regularize the divergences that pop up when one of the 
angles vanishes so as to recover the regulators (\ref{tikk}). This problem also appears for odd $D$ when 
one or more angles are set to zero.\\

For massless fields, the situation is a bit more complicated. For even $D$ the massless 
Poincar\'e character (\ref{FACH}) is the limit $M\rightarrow0$ of its massive counterpart (\ref{Xufy}), and 
the one-loop partition function (\ref{CAPt6.5t}) can again be written as an exponential (\ref{CAPZexp}). But 
in odd space-time dimensions, $\text{SO}(D-2)$ has lower rank than $\text{SO}(D-1)$, so the 
rotation (\ref{CAP2.7}) is \it{not}, in general, conjugate to an element of 
the massless little group: it has one angle too much, and whenever all angles $\theta_1,...,\theta_r$ 
are non-zero, the character (\ref{fropo}) vanishes. The only 
non-trivial irreducible character arises when at least one of the angles $\theta_1,...,\theta_r$ vanishes, 
say $\theta_r=0$, in which case the massless character takes the form (\ref{ChD}). However, comparison with 
(\ref{CAPss6q}) reveals a 
mismatch: the partition function does \it{not} take the form (\ref{CAPZexp}) in terms of the massless 
characters (\ref{ChD}); in field theory, all $r$ angles $\theta_i$ may be 
switched on simultaneously! To accommodate for this one can resort to the angle-dependent coefficients 
$\cA_k^r(\vec\theta)$ introduced in (\ref{CAPs6q}), whose origin can again be understood through 
the massless limit of the 
character (\ref{Chamouss}). Using relation (\ref{CAPt6q}), the product of massless partition 
functions 
with spins ranging from zero to $s$ can be written as (\ref{CAPZexp}), where the characters on the right-hand 
side are massless limits of massive Poincar\'e characters. However, it is unclear whether the quantities 
appearing in the exponent of (\ref{CAPss6q}) can be related directly to Poincar\'e characters \it{without} 
invoking a massless limit.\\

The occurrence of Poincar\'e characters in (\ref{CAPZexp}) illuminates certain aspects of gravity and 
higher-spin theories in three 
dimensions. Indeed, recall expression (\ref{CAPOEIRA}) for the partition function of a field with spin 
$s$ in three-dimensional thermal, rotating Minkowski space. (The regularization is unimportant for 
our present argument.) Since the space-time dimension is odd, the terms of the series in the exponential are 
not quite massless 
Poincar\'e characters, but they can still be interpreted as contributions of specific field excitations. 
Indeed, the terms $e^{isn\theta}$ are due to the heat kernel with spin $s$, while the terms 
$-e^{i(s-1)n\theta}$ 
come from ghosts,\i{ghost}\i{3D gravity!no local degrees of freedom} with a minus sign due to their 
fermionic statistics. The difference 
$e^{isn\theta}-e^{i(s-1)n\theta}$ vanishes when $\theta=0$, in accordance with the fact that ghosts cancel 
all would-be local degrees of freedom.\i{no gravitons in 
3D}\i{degrees of freedom} However, when $\theta\neq0$, 
the cancellation is incomplete because would-be local field excitations and ghosts have different 
spins ($s$ and $s-1$, respectively). As a result the one-loop partition function is non-trivial 
despite the absence of physical local degrees of freedom.

\section{\ Representations and characters of flat $\cW_N$}
\label{CAPsec:3DW}

As an application of the results of the previous section, we now explain how certain 
combinations of one-loop partition functions in three-dimensional flat space reproduce characters of 
higher-spin asymptotic symmetry algebras at null infinity. As it turns out, the coadjoint representation 
of standard $\cW_N$ algebras \cite{Balog:1990mu,Khesin:1991ra,Bajnok:2000nb} will play a key role in the 
analysis, so we first review briefly the analogous situation of higher-spin fields in AdS$_3$. We then turn 
to 
the case of spin $3$ in flat space and describe certain irreducible unitary representations of the 
corresponding 
asymptotic symmetry group.
Upon evaluating their characters thanks to the Frobenius formula, we find that they
match suitable products of partition functions. We also extend these observations to arbitrary spin $N$. 
The description of the induced modules and quantum algebras that correspond to this construction are 
relegated to section \ref{COPsec:hs}.

\subsection{\ Higher spins in AdS$_3$ and $\cW_N$ algebras}

As a preparation for flat space computations, we review here the asymptotic symmetries of higher-spin 
theories in AdS$_3$. We also describe the corresponding quantum symmetry algebras, their unitary 
representations and their characters, which match field-theoretic one-loop partition functions.

\subsubsection*{Asymptotic symmetries}

Asymptotic symmetries of higher-spin theories in three dimensions were first studied in 
AdS$_3$ \cite{Henneaux:2010xg,Campoleoni:2010zq,Gaberdiel:2011wb,Campoleoni:2011hg},\i{higher spin!asymptotic 
symmetries}\i{asymptotic symmetry!for higher spins}
and are similar to the Brown-Henneaux asymptotic symmetries of gravity described in chapter \ref{AdS3}. Here 
we focus on models 
including fields with spin ranging 
from 2 to $N$; this setup can be described as an 
$\mathfrak{sl}(N,\RR)\oplus\mathfrak{sl}(N,\RR)$ Chern-Simons action with a principally embedded 
$\mathfrak{sl}(2,\RR)\oplus\mathfrak{sl}(2,\RR)$ gravitational subalgebra. When 
$N=3$, the asymptotic symmetries are generated by gauge transformations specified by 
four arbitrary, $2\pi$-periodic functions $\big(X(x^+),\xi(x^+)\big)$ and $\big(\bar 
X(x^-),\bar\xi(x^-)\big)$ 
of the 
light-cone coordinates $x^{\pm}$ on the boundary of AdS$_3$. In particular, the functions 
$X(x^+)$ and $\bar X(x^-)$ generate Brown-Henneaux conformal transformations of the type described in section 
\ref{sebohad}. Since 
the results are 
left-right symmetric, we focus on 
the left-moving sector. The surface charge associated with a transformation $(X,\xi)$ then generalizes the 
(left-moving half of the)
gravitational expression (\ref{s212}) and takes 
the form \cite{Campoleoni:2010zq}\i{surface charge!for higher spins}\i{higher spin!surface 
charge}
\be
Q_{(X,\xi)}[p,\rho]
=
\frac{1}{2\pi}\int_0^{2\pi}d\varphi\big[X(x^+)p(x^+)+\xi(x^+)\rho(x^+)\big]
\label{CAPchargeW}
\ee
when the normalization is chosen so that pure AdS$_3$ with all higher-spin fields switched off has vanishing 
higher-spin charges and negative mass $-1/8G$. Here $\varphi=(x^+-x^-)/2$, while $p(x^+)$ and $\rho(x^+)$ 
are two arbitrary, $2\pi$-periodic functions 
specifying a solution of the field equations. In particular $p(x^+)$ is one of the two functions 
$\big(p(x^+),\bar p(x^-)\big)$ that determine an on-shell AdS$_3$ metric (\ref{s209}) while $\rho(x^+)$, 
together with its anti-chiral counterpart $\bar\rho(x^-)$, specifies an on-shell higher-spin field 
configuration. The vacuum field configuration (\ref{s209b}) corresponds to pure AdS$_3$ with all higher-spin 
fields set to zero, and is given by $\rho=\bar\rho=0$, 
$p=\bar p=-\ell/16G$.\\

One can think of the pair $(X,\xi)$ as an element 
of the asymptotic symmetry algebra, so the charge (\ref{CAPchargeW}) is the pairing between this algebra and 
its dual space. This generalizes the pairing (\ref{denpar}) of the Virasoro algebra with CFT stress tensors, 
and $(p,\rho)$ may be seen as a coadjoint vector of the symmetry algebra. Its 
infinitesimal transformation law extends (\ref{s210}) and turns out to be 
\cite{Campoleoni:2010zq}\i{coadjoint representation!of W3 algebra@of $\cW_3$ 
algebra}\i{W4 algebra@$\cW_N$ algebra!coadjoint representation}
\begin{subequations}
\label{CAPdeltaP}
\begin{align}
\delta_{(X,\xi)}p
& =
Xp'+2X'p-\frac{c}{12}\,X'''+2\,\xi\rho'+3\,\xi'\rho\,,\\[5pt]
\delta_{(X,\xi)}\rho
& =
X\rho'+3X'\rho+2\xi p'''+9\xi'p''+15\xi''p'+10\xi'''p\nn\\
\label{CAPdeltaPi}
&
\quad-\frac{c}{12}\xi^{(5)}-\frac{192}{c}\left(\xi pp'+\xi'p^2\right),
\end{align}
\end{subequations}
where prime 
denotes differentiation with respect to $x^+$, 
and $c=3\ell/2G$ is the Brown-Henneaux central charge. 
Analogous formulas hold in the anti-chiral sector.
Since $X$ generates conformal transformations, this implies that $p$ is a (chiral) quasi-primary field with 
weight 2
while $\rho$ is a primary with weight 3. Together with the surface charges (\ref{CAPchargeW}), these 
transformation laws yield the Poisson 
bracket (\ref{dekuphi}):
\be
\label{CAPss25}
\left\{Q_{(X,\xi)}[\,p,\rho],Q_{(Y,\zeta)}[\,p,\rho]\right\}
=
-\,\delta_{(X,\xi)}Q_{(Y,\zeta)}[p,\pi]\,.
\ee
Formula (\ref{CAPdeltaP}) turns out to coincide with the coadjoint representation of a Poisson 
algebra known as the \it{$\cW_3$ algebra}, and indeed one finds that the bracket (\ref{CAPss25}) reproduces 
the non-linear bracket of a $\cW_3$ algebra with central charge 
$c$ (see eq.\ (\ref{CAPSULE}) below). Similar considerations apply to models including fields with spin 
ranging from 2 to $N$ \cite{Campoleoni:2010zq,Campoleoni:2011hg}; the 
resulting asymptotic symmetry algebra is the direct sum of two copies of $\cW_N$.

\subsubsection*{Quantum $\cW_3$ algebra}

Owing to the fact that (\ref{CAPdeltaP}) is the coadjoint representation of $\cW_3$, the orbit of the AdS$_3$ 
vacuum $\big(p=\bar p=-c/24,\rho=\bar\rho=0\big)$ under asymptotic symmetry transformations 
is a direct product of two vacuum coadjoint orbits of $\cW_N$; these orbits are well-defined 
infinite-dimensional manifolds even though the definition of finite symmetry transformations is more 
intricate than in the pure Virasoro case corresponding to $N=2$ \cite{Bajnok:2000nb}. Putting all 
mathematical subtleties under the rug, one thus expects the quantization of that orbit to produce the vacuum 
highest-weight representation of the quantum algebra $\cW_N\oplus\cW_N$. Accordingly, we now describe the 
quantum version of the $\cW_3$ algebra.\\

As in the purely gravitational case (\ref{ViCha}), the classical asymptotic symmetry 
algebra given by the 
surface charges (\ref{CAPchargeW}) can be written in terms of modes
\be
\cL_m\equiv Q_{(e^{imx^+},0)}\,,
\qquad
\cW_m\equiv Q_{(0,e^{imx^+})}
\label{LAVA}
\ee
and their barred counterparts in the right-moving sector.
The normalization is such that pure AdS$_3$ has 
all charges vanishing except $\cL_0=\bar\cL_0=-c/24$.
Using (\ref{CAPdeltaP}), one finds that
the Poisson brackets (\ref{CAPss25}) of the charges (\ref{LAVA}) 
take the form of a classical $\cW_3$ algebra:\i{classical W3 algebra@classical $\cW_3$ 
algebra}\i{W3 algebra@$\cW_3$ algebra}
\begin{align}
i\{\cL_m,\cL_n\} & = (m-n)\cL_{m+n}+\frac{c}{12}m^3\delta_{m+n,0}\,,\nn\\
\label{CAPSULE}
i\{\cL_m,\cW_n\} & = (2m-n)\cW_{m+n}\,,\\
i\{\cW_m,\cW_n\} & = 
(m-n)(2m^2+2n^2-mn)\cL_{m+n}+\frac{96}{c}\Lambda_{m+n}+\frac{c}{12}m^5\delta_{m+n,0}\,,\nn
\end{align}
where $\Lambda_m\equiv\sum_{p\in\ZZ}\cL_{m-p}\cL_p$ is a non-linear term and the first line is the usual 
Virasoro algebra. The same brackets hold in the right-moving sector,
so as announced the asymptotic symmetry algebra is a direct sum of two classical $\cW_3$ algebras. 
Under quantization the Poisson brackets 
are 
turned into commutators according to $i\widehat{\{\cdot,\cdot\}}=[\hat\cdot,\hat\cdot]$ and the charges 
$\cL_m,\cW_n$ become operators $L_m,W_n$ which, in any unitary representation, satisfy the Hermiticity 
conditions\i{Hermiticity conditions!for W3@for $\cW_3$}
\be
L_m^{\dagger}=L_{-m}\,,
\qquad
W_m^{\dagger}=W_{-m}\,.
\nn
\ee
It is also customary to normalize the Virasoro generators $L_m$ so that the vacuum state has vanishing 
eigenvalue under $L_0$, i.e.\ to rename $L_m+\frac{c}{24}\delta_{m,0}$ into $L_m$. As a result the 
commutators of the operators $L_m,W_n$ yield the \it{quantum} $\cW_3$ algebra\i{quantum W3 algebra@quantum 
$\cW_3$ algebra}\i{W3 algebra@$\cW_3$ algebra}
\begin{subequations}
\label{COPeq:QuantumW3}
\begin{align}
[L_m,L_n]
&= 
(m-n)L_{m+n}+\frac{c}{12}(m^3-m)\delta_{m+n,\,0}\,,\\
[L_m,W_n]
&=
(2m-n)W_{m+n}\,,\\
[W_m,W_n]
&=
(m-n)(2m^2+2n^2-mn-8)L_{m+n}
+
\frac{96}{c+22/5}(m-n)\,{{:}\!\mathrel{\Lambda_{m+n}}\!{:}}\nn\\
\label{HABANA}
&
\quad+\frac{c}{12}(m^2-4)(m^3-m)\delta_{m+n,0}\,,
\end{align}
\end{subequations}
whose non-linear terms are normal-ordered according to the prescription\i{normal ordering}
\be
\label{COPeq:VirasoroNO}
{{:}\!\mathrel{\Lambda_m}\!{:}}
\equiv
\sum_{p\geq-1}L_{m-p}L_p 
+
\sum_{p<-1}L_pL_{m-p}
-
\frac{3}{10}(m+3)(m+2)L_m\,.
\ee
Here the term linear in $L_m$ ensures that the operator 
${{:}\!\mathrel{\Lambda_m}\!{:}}$ is quasi-primary with 
respect to the action of $L_m$'s. Note how the denominator of the structure constant of the non-linear term 
in (\ref{HABANA}) involves a shifted central charge $c+22/5$ instead of the classical $c$ in the last line of 
(\ref{CAPSULE}). Analogous commutation relations hold in the barred sector.

\subsubsection*{Unitary representations and characters}

Unitary representations of the quantum $\cW_3$ algebra are obtained analogously to the Virasoro 
highest-weight representations of section \ref{sevirep}, and are spanned by the descendants of a 
highest-weight state annihilated by operators $L_m,W_n$ with $m,n>0$. At large $c$, such representations are 
irreducible. The same is true of $\cW_N$ algebras for any finite $N$, and irreducible unitary representations 
of $\cW_N\oplus\cW_N$ are tensor products of individual highest-weight representations of the two $\cW_N$ 
algebras (at large $c,\bar c$). Characters of such representations can be evaluated by adapting the counting 
argument that led to (\ref{s86}). In particular the character of a highest-weight representation of 
$\cW_N\oplus\cW_N$ with central charges $(c,\bar c)$, generic highest weights $(h,\bar h)$, vanishing 
higher-spin weights and vanishing higher-spin chemical potentials, is\i{W4 algebra@$\cW_N$ 
algebra!character}\i{character!for WN@for $\cW_N$}
\be
\text{Tr}\left(
q^{L_0-c/24}\bar q^{\bar L_0-\bar c/24}
\right)
=
q^{h-c/24}\bar q^{\bar h-\bar c/24}
\left(
\prod_{n=1}^{+\infty}\frac{1}{|1-q^n|^2}
\right)^{N-1}\,.
\label{HASSAN}
\ee
This reduces to the product of Virasoro characters (\ref{s86Bis}) when $N=2$. The vacuum character of 
$\cW_N\oplus\cW_N$ similarly reads\i{vacuum character!for WN@for $\cW_N$}
\be
\text{Tr}_{\text{vac}}\left(
q^{L_0-c/24}\bar q^{\bar L_0-\bar c/24}
\right)
=
\prod_{s=2}^N\left(\prod_{n=s}^{+\infty}\frac{1}{|1-q^n|^2}\right),
\label{VACANT}
\ee
where the truncated product arises because the vacuum state is left invariant by the wedge algebra 
$\mathfrak{sl}(N,\RR)$.\i{wedge algebra} This reduces to the vacuum character (\ref{provaco}) when $N=2$.\\

As mentioned earlier, it was shown in \cite{Giombi:2008vd} that the one-loop 
partition function of gravitons in AdS$_3$ at temperature $1/\beta$ and angular potential $\theta$ is a 
vacuum character (\ref{provaco}) with modular parameter (\ref{TaKatak}).\i{higher spin!partition 
function}\i{partition function!for higher spins} 
This result was later extended to 
higher-spin theories \cite{David:2009xg,Gaberdiel:2010ar}, whose one-loop partition functions on thermal 
AdS$_3$ coincide with vacuum $\cW_N\oplus\cW_N$ characters (\ref{VACANT}) upon including the contribution of 
fields with spins $s=2,3,...,N$. These results confirm the interpretation of irreducible unitary 
representations of asymptotic symmetry groups as particles dressed with boundary degrees of freedom. The 
purpose of the remainder of this chapter is to describe the similar matching that occurs in asymptotically 
flat theories.

\subsection{\ Flat $\cW_3$ algebra}
\label{CAPPOU}

The asymptotic symmetries of higher-spin theories at null infinity in 
three\--di\-men\-sio\-nal flat space were described in 
\cite{Afshar:2013vka,Gonzalez:2013oaa,Grumiller:2014lna}.\i{higher spin!asymptotic 
symmetries}\i{asymptotic symmetry!for higher spins} Here we focus on the model describing the 
gravitational coupling 
of a field of spin 3, which is a three-dimensional Chern-Simons theory whose gauge algebra
$\mathfrak{sl}(3,\RR) \inplus\mathfrak{sl}(3,\RR)_{\text{Ab}}$ is the flat limit of 
$\mathfrak{sl}(3,\RR)\oplus\mathfrak{sl}(3,\RR)$. The associated asymptotic symmetry generators turn out to 
be labelled by 
four arbitrary, $2\pi$-periodic functions 
$X(\varphi)$, $\xi(\varphi)$, $\alpha(\varphi)$ and $a(\varphi)$ on the celestial circle at (future) null 
infinity. Of 
these, $X(\varphi)$ and $\alpha(\phii)$ generate standard BMS$_3$ superrotations and supertranslations 
(respectively), while $\xi$ and $a$ are their higher-spin extensions. The corresponding surface charges 
extend the gravitational formula (\ref{bokka}) and read\i{higher spin!surface charge}\i{surface charge!for 
higher spins}
\be
\label{CAPs25}
Q_{(X,\xi,\alpha,a)}[j,\kappa,p,\rho]
=
\frac{1}{2\pi}
\int_0^{2\pi}d\varphi
\Big[
X(\varphi)j(\varphi)+\xi(\varphi)\kappa(\varphi)+\alpha(\varphi)p(\varphi)+a(\varphi)\rho(\varphi)
\Big]\,,
\ee
where the $2\pi$-periodic functions $j$, $\kappa$, $p$ and $\rho$ determine a solution of the equations of 
motion. In particular $p(\varphi)$ is the Bondi mass aspect (supermomentum) and $j(\varphi)$ is 
the 
angular momentum aspect (angular supermomentum)
appearing in an asymptotically flat metric (\ref{piment}). The functions $\rho$ and $\kappa$ are analogous
quantities 
for a spin-3 field. As usual, the quadruple $(j,\kappa,p,\rho)$ may be seen as an element of the 
dual space of the asymptotic symmetry algebra. In particular, the higher-spin supermomentum $(p,\rho)$ 
transforms under 
higher-spin superrotations $(X,\xi)$ as a coadjoint vector of the $\cW_3$ algebra, i.e.\ according to 
(\ref{CAPdeltaP}), albeit with a central charge $c_2=3/G$ instead of $c$.\\

The Poisson brackets satisfied by the surface charges (\ref{CAPs25}) are given as usual by (\ref{CAPss25}) 
and are most easily expressed in terms of generators
\be
\cJ_m\equiv Q_{(e^{im\phii},0,0,0)}\,,
\quad
\cK_m\equiv Q_{(0,e^{im\phii},0,0)}\,,
\quad
\cP_m\equiv Q_{(0,0,e^{im\phii},0)}\,,
\quad
\cQ_m\equiv Q_{(0,0,0,e^{im\phii})}\,.
\nn
\ee
Note that with this normalization pure Minkowski space has all its charges vanishing, except $\cP_0=-1/8G$. 
One then finds that the $\cJ_m$'s and $\cP_m$'s close according to a $\bms$ algebra (\ref{surall}) with 
central charge $c_2=3/G$, while brackets involving higher-spin charges take the form\i{flat W3 algebra@flat 
$\cW_3$ algebra}\i{classical flat W3 algebra@classical flat $\cW_3$ algebra}
\begin{subequations}
\label{ClaSH}
\begin{align}
i\{\cJ_m,\cK_n\}
&=
(2m-n)\cK_{m+n}\,,\\
i\{\cJ_m,\cQ_n\}
&=
(2m-n)\cQ_{m+n}\,,\\
i\{\cP_m,\cK_n\}
&=
(2m-n)\cQ_{m+n}\,,\\
i\{\cP_m,\cQ_n\}
&=
0\,,\\
i\{\cK_m,\cK_n\}
&=
(m-n)(2m^2+2n^2-mn)\cJ_{m+n}
+\frac{96}{c_2}(m-n)\Omega_{m+n}\,,\\
i\{\cK_m,\cQ_n\}
&=
(m-n)(2m^2+2n^2-mn)\cP_{m+n}+\frac{96}{c_2}\Theta_{m+n}+\frac{c_2}{12}m^5\delta_{m+n,0}
\end{align}
\end{subequations}
where the non-linear terms $\Omega$ and $\Theta$ are given by
\be
\Omega_m\equiv\sum_{p\in\ZZ}\left(\cP_{m-p}\cJ_p+\cJ_{m-p}\cP_p\right)\,,
\qquad
\Theta_m\equiv\sum_{p\in\ZZ}\cP_{m-p}\cP_p\,.
\label{COPeq:FW3NonLinTermsUR}
\ee
These formulas show that, up to central terms, the brackets of $(\cJ,\cK)$'s with 
$(\cP,\cQ)$'s take the same form as the brackets of $(\cJ,\cK)$'s with themselves. This is the situation 
described in (\ref{jappah}) and it implies that, similarly to (\ref{haBOP}), the asymptotic symmetry 
algebra is an exceptional semi-direct sum\i{exceptional semi-direct product/sum}
\be
\text{``flat $\cW_3$ algebra''}
\equiv
\cW_3\inplus_{\text{ad}}(\cW_3)_{\text{Ab}}\,,
\label{CAPWw}
\ee
where $\cW_3$ is the classical $\cW_3$ algebra (\ref{CAPSULE}) and $(\cW_3)_{\text{Ab}}$ denotes an Abelian 
Lie algebra 
isomorphic, as a vector space, to $\cW_3$. This algebra is centrally extended, as the bracket 
between generators of $\cW_3$ and those of $(\cW_3)_{\text{Ab}}$ includes a central charge $c_2$; there is 
also a central charge $c_1$ specific to the left (non-Abelian) $\cW_3$ subalgebra of (\ref{CAPWw}), but 
it is not switched on in parity-preserving theories. We shall return to this structure in section 
\ref{COPsec:hs}, upon describing its quantum version.

\subsubsection*{Induced representations and unitarity}

Since the flat $\cW_3$ algebra (\ref{CAPWw}) has the exceptional form $\mg\inplus\mg_{\text{Ab}}$, with $\mg$ 
the 
standard 
$\cW_3$ algebra, its unitary representations should be induced representations labelled by orbits of 
supermomenta under the coadjoint action of elements of a group whose tangent space at the identity is 
the $\cW_3$ algebra.\i{flat W3 algebra@flat $\cW_3$ algebra!induced representation}\i{induced 
representation!of flat W3 algebra@of flat $\cW_3$ algebra}\i{supermomentum orbit!for higher 
spins}\i{symplectic leaf} 
However, the non-linearities that appear in $\cW$ algebras make this step subtle, so one can 
bypass the need to control the group as follows. Generic $\cW$ algebras define a Poisson manifold through 
(\ref{CAPss25}) and one can classify the submanifolds on which the Poisson structure is invertible, i.e.\ 
their symplectic leaves in the terminology of section \ref{seSymPa}. In the case of the Virasoro algebra 
(which corresponds to $\cW_N$ with $N=2$) this concept coincides with that of a coadjoint orbit of 
the Virasoro group. For higher $N$ the symplectic leaves of $\cW_N$ algebras are still well defined 
\cite{Bajnok:2000nb} despite the lack of a straightforward definition of the group that corresponds to 
the $\cW_N$ algebra. 
These leaves may be seen as intersections of the coadjoint orbits of $\mathfrak{sl}(N)$-Kac 
Moody algebras with the constraints that implement the Hamiltonian reduction to $\cW_N$ algebras.
Accordingly, it should be possible to build unitary representations of flat $\cW_N$ algebras as 
Hilbert spaces of wavefunctions 
defined on their symplectic leaves, which we assume as usual to admit quasi-invariant measures.
In most of the remainder of this chapter, we test that proposal by showing how it can be used to evaluate 
characters that 
coincide with higher-spin one-loop partition functions. Note that the non-linearities 
appearing in the brackets of the algebra (\ref{CAPWw}) imply an extra 
complication for representation theory in that one has to devise a suitable normal-ordering prescription. In 
the standard $\cW_3$ case we displayed this normal ordering in (\ref{COPeq:VirasoroNO}). In the flat case 
we will address this issue in section \ref{COPsec:hs}.\\

The complete classification of the symplectic leaves of the $\cW_3$ algebra has been worked out in 
\cite{Khesin:1991ra,Bajnok:2000nb}; according to our proposal this settles the classification of 
irreducible unitary representations of the flat $\cW_3$ algebra, in the same way that 
Virasoro coadjoint orbits classify BMS$_3$ particles. Instead of describing the details of this 
classification, we focus from now on on orbits of constant supermomenta, which can be 
classified thanks to the infinitesimal transformation laws (\ref{CAPdeltaP}) given by the algebra. 
To describe such an orbit, let us pick a pair $(p,\rho)$ where $p(\varphi)=p_0$ and 
$\rho(\varphi)=\rho_0$ are constants, and act on it with an infinitesimal higher-spin superrotation 
$(X,\xi)$. Then, all terms involving derivatives of $p$ or 
$\rho$ in eq.\ (\ref{CAPdeltaP}) vanish, and we find
\begin{subequations}
\label{CAPdeltaP0}
\begin{align}
\delta_{(X,\xi)}p_0
& =
2\,X'p_0-\frac{c_2}{12}\,X'''+3\,\xi'\rho_0\,,\\[5pt]
\label{CAPdeltaPi0}
\delta_{(X,\xi)}\rho_0
& =
3\,X'\rho_0+10\,\xi'''p_0-\frac{c_2}{12}\,\xi^{(5)}-\frac{192}{c_2}\,\xi'p_0^2\,.
\end{align}
\end{subequations}
The little group for $(p_0,\rho_0)$ consists of higher-spin superrotations leaving it invariant. 
Its Lie algebra is therefore spanned by pairs $(X,\xi)$ such that the right-hand sides of 
eqs.~(\ref{CAPdeltaP0}) vanish:
\begin{subequations}
\label{CAPlilp}
\begin{align}
2\,X'p_0-\frac{c_2}{12}X'''+3\,\xi'\rho_0
& = 0\,,\\
\label{CAPlilpi}
3\,X'\rho_0+10\,\xi'''p_0-\frac{c_2}{12}\,\xi^{(5)}-\frac{192}{c_2}\,\xi'p_0^2
& = 0\,.
\end{align}
\end{subequations}
The solutions of these equations depend on the values of $p_0$ and $\rho_0$. Here we take $\rho_0=0$ for 
simplicity, i.e.~we only consider cases where all higher-spin charges are switched off. 
\label{CAPOff} Then, given $p_0$, eqs.~(\ref{CAPlilp}) become two decoupled differential 
equations for the functions $X(\phii)$ and $\xi(\phii)$, leading to three different 
cases:\i{flat W3 algebra@flat 
$\cW_3$ algebra!little groups}\i{little group!for flat W3@for flat $\cW_3$}
\begin{itemize}
\item For generic values of $p_0$, the only pairs $(X,\xi)$ leaving $(p_0,0)$ invariant are constants, and 
generate a little group $\text{U}(1)\times\RR$. 
\item For $p_0=-n^2c_2/96$ where $n$ is a positive odd integer, the pairs $(X,\xi)$ leaving $(p_0,0)$ 
invariant take the form
\be
X(\varphi)=A,
\qquad
\xi(\varphi)=B+C\cos(n\phii)+D\sin(n\phii),
\ee
where $A$, $B$, $C$ and $D$ are real numbers. The corresponding little group is the $n$-fold cover of 
$\text{GL}(2,\RR)$.
\item For $p_0=-n^2c_2/24=-(2n)^2c_2/96$ where $n$ is a positive integer, the Lie algebra of the 
little group is spanned by
\be
\begin{split}
X(\varphi)= &
\,A+B\cos(n\phii)+C\sin(n\phii),\\
\xi(\varphi)= &
\,D+E\cos(n\phii)+F\sin(n\phii)+G\cos(2n\phii)+H\sin(2n\phii),
\end{split}
\label{CAPexlil}
\ee
where $A,B,...,H$ are real coefficients. The little group is thus an $n$-fold cover of 
$\text{SL}(3,\mathbb{R})$. In particular, $p_0=-c_2/24$ realizes the absolute minimum of energy among all 
supermomenta belonging to orbits with energy bounded from below. It is thus the supermomentum of the vacuum 
state, and indeed, upon using $c_2=3/G$, the field configuration that corresponds to it is the metric of 
Minkowski space (with the spin-3 field set to zero on account of $\rho_0=0$).
\end{itemize}
These results extend our earlier observations on Virasoro orbits in section \ref{secovobi}.

\subsection{\ Flat $\cW_3$ characters}
\label{susepabaka}

The information on little groups turns out to be sufficient to evaluate certain characters along the 
lines of section \ref{sebmschar}.
For instance, consider an induced representation based on the orbit $\cO_p$ of a generic pair $(p_0,0)$, and 
call 
$(s,\sigma)$ the spin of the 
representation $\cR$ of the little group 
$\text{U}(1)\times\RR$. Then take a superrotation which is an element of the $\text{U}(1)$ subgroup (i.e.\ 
a rotation 
$f(\varphi)=\phii+\theta$). The only point on the orbit that is left invariant by the rotation 
is $(p_0,0)$, and the whole integral over the orbit in (\ref{fropokad}) localizes to that 
point. Therefore, in analogy with the BMS$_3$ example, the detailed knowledge of the orbit is irrelevant to 
compute the character. In particular,
including a higher-spin supertranslation $\big(\alpha(\phii),a(\phii)\big)$,
the only components of 
$\alpha(\varphi)$ and $a(\varphi)$ that survive the integration are their zero-modes $\alpha^0$ and $a^0$. 
The character thus takes the form 
\be
\chi[(\text{rot}_{\theta},\alpha,a)]
=
e^{is\theta}e^{ip_0\alpha^0}\int_{\cO_p}\!
d\mu(k)\,\delta(k,\text{rot}_{\theta}\cdot k)\,
\ee
where the little group character $e^{is\theta}$ factors out as in (\ref{Xixi}).
In writing this we assume the existence of a quasi-invariant measure $\mu$ on the orbit, whose precise 
expression is unimportant since different measures give representations that are unitarily equivalent. Our 
remaining task is to 
integrate the delta 
function. To do so, we use local coordinates on the orbit, which we choose to be the Fourier modes of 
higher-spin 
supermomenta as we did in section \ref{sebmschar}. Since $p_0$ is generic, the non-redundant 
coordinates 
on the 
orbit are the non-zero modes. As in (\ref{CHARstar12}) the integral is thus
\be
\int_{\cO_p}\!\!
d\mu(k)\delta(k,\text{rot}_{\theta}\cdot k)
=
\prod_{n\,\in\,\mathbb{Z}^*}\!\left(\int dk_n\delta(k_n-e^{in\theta}k_n)\right)\!
\prod_{m\,\in\,\mathbb{Z}^*}\!\left(\int d\rho_m\delta(\rho_m-e^{im\theta}\rho_m)\right),
\label{CAPintFourier}
\ee
where we call $k_n$ the Fourier modes of the standard (spin 2) supermomentum, while $\rho_m$ are the modes of 
its higher-spin counterpart. Performing the integrals over Fourier modes and adding small 
imaginary parts $i\epsilon$ to $\theta$ as in (\ref{CHARmodPar}), we obtain\i{flat W3 algebra@flat 
$\cW_3$ algebra!character}\i{character!for flat WN@for flat $\cW_N$}
\be
\chi[(\text{rot}_{\theta},\alpha,a)]
=
e^{is\theta}e^{ip_0\alpha^0}
\left(\prod_{n=1}^{+\infty}\frac{1}{|1-e^{in(\theta+i\epsilon)}|^2}
\right)^2.
\ee
This is a natural spin-3 extension of the spin-2 (BMS$_3$) massive character 
(\ref{CHARcharBMS}), in the same way that (\ref{HASSAN}) generalizes Virasoro characters. It is also a flat 
limit of (\ref{HASSAN}) for $N=3$.\\

A similar computation can be performed for orbits of other higher-spin supermomenta $(p_0,0)$. The only 
subtlety is that, for the values of $p_0$ for which the little 
group is larger than $\text{U}(1)\times\RR$, the orbit has higher codimension in $\cW_3^*$ than the 
generic orbit just discussed. Accordingly, there are fewer coordinates on the orbit and the 
products of integrals (\ref{CAPintFourier}) are truncated. For instance, when $p_0=-n^2c_2=/24$ with $n$ a 
positive integer, the little group is generated by pairs 
$(X,\xi)$ of the form (\ref{CAPexlil}), so that the Fourier modes providing non-redundant local coordinates 
on 
the orbit (in a neighbourhood of $(p_0,0)$) are the modes $k_m$ with $m\notin\{-n,0,n\}$ and the higher-spin 
modes $\rho_m$ with $m\notin\{-2n,-n,0,n,2n\}$. Assuming that the representation $\cR$ of the little group is 
trivial, this produces a character\i{vacuum character!for flat W3@for flat $\cW_3$}
\be
\chi[(\text{rot}_{\theta},\alpha,a)]
=
e^{-in^2c_2\alpha^0/24}
\Bigg(
\prod_{\substack{m=1,\\ m\neq n}}^{+\infty}
\frac{1}{|1-e^{im(\theta+i\epsilon)}|^2}
\Bigg)
\cdot
\Bigg(
\prod_{\substack{m=1,\\ m\neq n,\\ m\neq2n}}^{+\infty}
\frac{1}{|1-e^{im(\theta+i\epsilon)}|^2}
\Bigg)
\,.
\ee
The choice $n=1$ specifies the vacuum representation of the flat 
$\cW_3$ algebra; taking $\alpha$ to be a Euclidean time translation by $i\beta$, we get
\be \label{CAPvac3}
\chi_{\text{vac}}[(\text{rot}_{\theta},\alpha=i\beta,a=0)]
=
e^{\beta c_2/24}
\Bigg(
\prod_{n=2}^{+\infty}
\frac{1}{|1-e^{in(\theta+i\epsilon)}|^2}
\Bigg)
\cdot
\Bigg(
\prod_{n=3}^{+\infty}
\frac{1}{|1-e^{in(\theta+i\epsilon)}|^2}
\Bigg).
\ee
This is one of our key results in this chapter.\i{partition function!for higher spins} Indeed, comparing with 
eq.~(\ref{CAPss6.5q}), we recognize 
the product of the (suitably regularized) rotating one-loop partition functions 
of massless fields with spins two and three in three-dimensional flat space. It provides a first non-trivial 
check of our proposal for the construction of unitary representations of flat $\cW_N$ algebras.\\

All the induced representations described above are unitary by construction, provided one can 
define (quasi-invariant) measures on the corresponding orbits. In analogy with representations of the 
$\mathfrak{bms}_3$ algebra, they can also be described as induced modules that generalize those of 
section \ref{sebomodule}; we will turn to them in section \ref{COPsec:hs}.

\subsection{\ Flat $\cW_N$ algebras}
\label{susepabaki}

The considerations of the previous pages can be generalized to higher-spin theories in flat 
space with spins ranging from 2 to $N$. In AdS$_3$ the asymptotic symmetries of models with this field 
content consist of two copies of a $\cW_N$ algebra, so it is natural to anticipate that the corresponding 
theory in 
flat space will have an asymptotic symmetry algebra of the exceptional form\i{flat WN algebra@flat $\cW_N$ 
algebra}
\be
\text{``flat $\cW_N$ algebra''}
\equiv
\cW_N\inplus_{\text{ad}}(\cW_N)_{\text{Ab}}\,,
\label{CAPs29.5}
\ee
in analogy with (\ref{CAPWw}).
The surface charges generating these symmetries should coincide with the pairing of the Lie algebra of 
(\ref{CAPs29.5}) with its dual space, and they are likely to satisfy a centrally extended algebra. Since the 
presence 
of higher-spin fields does not affect the value of the central charge in three-dimensional AdS gravity 
\cite{Henneaux:2010xg,Campoleoni:2010zq}, one expects the central charge in this case to be the usual 
$c_2=3/G$ 
appearing in mixed brackets. This structure was indeed observed for $N=4$ in 
\cite{Grumiller:2014lna}. We now argue that this proposal must hold for any $N$ by showing that the 
vacuum character of (\ref{CAPs29.5}), computed along the lines followed above for flat $\cW_3$, reproduces 
the 
product of one-loop partition functions of fields of spin $2,3,...,N$.\\

According to our proposal for the characterization of the representations of semi-direct sums of the type 
(\ref{CAPs29.5}), unitary representation of flat $\cW_N$ algebras are classified by their symplectic leaves, 
that is, by orbits of higher-spin supermomenta $(p_1,...,p_{N-1})$.\i{supermomentum!for higher 
spins}\i{higher spin!supermomentum} (Here $p_1(\phii)$ is the 
supermomentum 
that we used to write as $p(\phii)$, while 
$p_2(\phii)$ is what we called $\rho(\phii)$ for $N=3$.) The infinitesimal transformations that generalize 
(\ref{CAPdeltaP}) and that define these orbits locally can 
be found for instance in \cite{Campoleoni:2011hg}. Here we focus on the vacuum orbit where we set all 
higher-spin 
charges to zero and take only $p_1=-c_2/24$ to be non-vanishing.\i{little group!for higher spin vacuum} This 
particular supermomentum is left fixed 
by higher-spin superrotations of the form
\be\label{CAPeq:LastEqFinal}
X_i(\phii)
=
A_i+
\sum_{j=1}^i\big(B_{ij}\cos(j\phii)+C_{ij}\sin(j\phii)\big),
\qquad i=1,...,N-1,
\ee
where the coefficients $A_i$, $B_{ij}$, $C_{ij}$ are real. In principle one can obtain such symmetry 
generators by looking for the little group of the vacuum as in (\ref{CAPlilp}), using for instance the 
explicit 
formulas of \cite{Campoleoni:2011hg}. Yet, a simpler way to derive the same result is to look for 
the 
higher-spin isometries of the vacuum in the first-order formulation, in which the theory is described by a 
Chern-Simons action with gauge 
algebra $\mathfrak{sl}(N,\mathbb{R})\inplus_{\text{ad}}(\mathfrak{sl}(N,\mathbb{R}))_{\text{Ab}}$ (see e.g.\ 
\cite{Campoleoni:2011tn,Gonzalez:2013oaa,Grumiller:2014lna}). In this language, and in terms of retarded 
Bondi coordinates $(r,\phii,u)$, 
the 
vacuum field configuration takes the form\i{higher spin!vacuum configuration}
\be
A_{\mu}(x)
=
b(r)^{-1}g(u,\phii)^{-1}
\partial_{\mu}
\left[g(u,\phii)b(r)\right],
\qquad
b(r)=
\exp\left[\frac{r}{2}\,P_{-1}\right],
\ee
where $g(u,\phii)$ is a field valued in $\text{SL}(N,\RR)\ltimes\mathfrak{sl}(N,\RR)$, given by
\be
\label{CAPggg}
g(u,\phii)=
\exp\left[
\left(P_1+\frac{1}{4}\,P_{-1}\right)u+
\left(J_1+\frac{1}{4}\,J_{-1}\right)\varphi
\right]
\ee
in terms of Poincar\'e generators that satisfy the commutation relations 
(\ref{COPpal}). The 
isometries of this field configuration are 
generated by gauge parameters of the form $(g\cdot b)^{-1}T_a(g\cdot b)$, where $T_a$ is any of the basis 
elements of the gauge algebra. Upon expanding $g^{-1}T_ag$ as a position-dependent linear combination of 
gauge 
algebra generators, the function multiplying the lowest weight generator coincides with the corresponding 
asymptotic symmetry parameter (see e.g.\ \cite{Campoleoni:2010zq} for details). The latter can be 
obtained as follows.\\

For convenience, we diagonalize the Lorentz piece of the group element (\ref{CAPggg}) as
\be
\exp\left[
\Big(J_1+\frac{1}{4}J_{-1}\Big)\varphi
\right]
=Se^{iJ_0\varphi}S^{-1}
\ee
where $S$ is some $\text{SL}(2,\RR)$ matrix. Then the gauge parameters that generate the little group of the 
vacuum configuration can be written as
\begin{subequations}
\begin{align}
& \exp\left[-\Big(J_1+\frac{1}{4}J_{-1}\Big)\varphi
\right]
\sum_{m=-\ell}^{\ell}\alpha^mW_m^{(\ell)}
\exp\left[
\Big(J_1+\frac{1}{4}J_{-1}\Big)\varphi
\right]\label{CAPfgfg}\\
\label{CAPtwiddle}
& =
S\,e^{-iJ_0\phii}\,\sum_{m=-\ell}^{\ell}
\alpha^m\,S^{-1}\,W_m^{(\ell)}\,S\,e^{iJ_0\phii}\,S^{-1},
\end{align}
\end{subequations}
where the $\alpha^m$'s are certain real coefficients, while the $W_m^{(\ell)}$'s (with $2 \leq \ell \leq N$ 
and 
$- \ell \leq m \leq \ell$) generate the $\mathfrak{sl}(N,\mathbb{R})$ algebra (including $J_m 
\equiv W^{(2)}_m$). Note that the matrix $S$ preserves the conformal weight since it is an exponential 
of $\mathfrak{sl}(2,\RR)$ generators, so that 
\be
\sum_{m=-\ell}^{\ell}\alpha^m\,S\,W_m^{(\ell)}\,S^{-1}
=
\sum_{m=-\ell}^{\ell}\tilde\alpha^mW_m^{(\ell)}
\ee
for some coefficients $\tilde\alpha^j$ obtained by acting on the $\alpha^m$'s with an invertible linear map. 
Since each generator $W_m^{(\ell)}$ has weight $m$ under $J_0$, eq.\ (\ref{CAPtwiddle}) can be 
rewritten as 
\be
\sum_{m={-\ell}}^{\ell}
e^{im\phii}
\tilde\alpha^m\,
S\,W_m^{(\ell)}\,S^{-1}
=
\sum_{m,n={-\ell}}^{\ell}\beta^{mn}W_n^{(\ell)}
e^{ij\phii}
=
\sum_{m=-\ell}^{\ell}e^{im\phii}\beta^{m\ell}W_{\ell}^{(\ell)}
+
\cdots
\ee
for some coefficients $\beta^{mn}$. In the last step we omitted all terms proportional to $W_m^{(\ell)}$'s 
with $m<\ell$; the important piece is the term that multiplies the highest-weight generator 
$W_{\ell}^{(\ell)}$: it is the function on the circle that generates the asymptotic symmetry corresponding to 
the generator $\sum_{m=-\ell}^{\ell}\alpha^mW_m^{(\ell)}$ that we started with in (\ref{CAPfgfg}). Since the 
$\beta^{m\ell}$'s are related to the $\alpha^m$'s by an invertible linear map, and since there are $2\ell+1$ 
linearly independent generators of this type, the isometries of the vacuum exactly span the set of functions 
of the form (\ref{CAPeq:LastEqFinal}). This is what we wanted to prove; there are $N^2-1$ linearly 
independent asymptotic symmetry generators of this form, and they span the Lie algebra 
$\mathfrak{sl}(N,\RR)$.\\

The character associated with the vacuum 
representation of (\ref{CAPs29.5}) can then be worked out exactly as in the cases $N=2$ and $N=3$ discussed 
above: using the Fourier modes of the $N-1$ components of supermomentum as coordinates on the orbit, we need 
to mod out the redundant modes. For the vacuum orbit, these are the modes ranging from $-(s-1)$ to $(s-1)$ 
for 
the $s^{\text{th}}$ component. The integral over the localizing delta function in the Frobenius formula 
(\ref{fropo}) then produces a character\i{vacuum character!for flat WN@for flat $\cW_N$}\i{flat 
WN algebra@flat 
$\cW_N$ algebra!vacuum character}\i{character!for flat WN@for flat $\cW_N$}
\be
\boxed{
\bigg.
\chi[(\text{rot}_{\theta},a_1=i\beta)]
=
e^{\beta c_2/24}
\prod_{s=2}^N
\left(\prod_{n=s}^{+\infty}\frac{1}{|1-e^{in(\theta+i\epsilon)^2}|}\right)\,}
\ee
where we implicitly set to zero all higher-spin supertranslations except the gravitational one, 
$a_1=i\beta$. Comparing with (\ref{CAPss6.5q}), we recognize the product of one-loop partition functions of 
massless 
higher-spin fields 
with spins ranging from 2 to $N$,\i{partition function!for higher spins} including a classical contribution. 
This result confirms, on the one hand, 
our conjecture (\ref{CAPs29.5}) for the asymptotic symmetry algebras of generic higher-spin theories in 
three-dimensional flat 
space, and on the other hand it provides another consistency check of our proposal for the characterization 
of unitary representations of flat $\cW_N$ algebras. It is also a flat limit of the vacuum $\cW_N\oplus\cW_N$ 
character displayed in (\ref{VACANT}).

\section{\ Flat $\cW_3$ modules}
\label{COPsec:hs}

We now turn to the algebraic analogue of the above considerations, i.e.\ we describe induced modules of flat 
$\cW_N$ algebras along the lines of section \ref{sebomodule}. For simplicity we focus on the case $N=3$ but 
the construction also applies to other higher-spin extensions of $\bms$. Due to the non-linearities of 
$\cW_3$, our plan in this 
section is slightly different from that of section \ref{sebomodule}. Namely, we start by describing the 
quantum flat $\cW_3$ algebra as an ultrarelativistic limit of the direct sum of two quantum $\cW_3$ 
algebras, which produces a specific ordering of operators in the non-linear terms of commutators. We then 
move on to the description of induced modules of the ultrarelativistic quantum flat $\cW_3$ algebra, and show 
that the ultrarelativistic normal ordering is defined with respect to a 
rest frame vacuum. Along the way we compare our results to those of the non-relativistic limit described in 
\cite{Grumiller:2014lna}, and point out that the two limits lead to different quantum algebras.

\subsection{\ Ultrarelativistic and non-relativistic limits of $\cW_3$}

The flat $\cW_3$ algebra (\ref{CAPWw}) can 
be obtained as an In\"on\"u-Wigner contraction of the direct sum of two $\cW_3$ algebras. This flat limit 
was discussed at the semiclassical level in \cite{Afshar:2013vka,Gonzalez:2013oaa}, and a Galilean 
limit of the quantum algebra was described in \cite{Grumiller:2014lna}. Here we are interested instead 
in an ultrarelativistic limit of $\cW_3 \oplus \cW_3$. 
The key difference between the Galilean and ultrarelativistic contractions is that the latter mixes 
generators with positive and negative mode numbers, while the former does not. For linear algebras such 
as Virasoro, this makes no difference and the two contractions yield identical quantum algebras, namely 
$\bms\cong\gca$. When non-linear terms are involved in the contraction, however, Galilean and 
ultrarelativistic limits generally give different quantum algebras, as we now explain.

\subsubsection*{Ultrarelativistic contraction}

The quantum $\cW_3$ algebra is spanned by two sets of generators $L_m$ and $W_m$ ($m\in\mathbb{Z}$) whose 
commutation relations were displayed in (\ref{COPeq:QuantumW3}). Consider now a direct sum 
$\cW_3\oplus\cW_3$, where the generators and the central charge of the other copy of 
$\mathcal{W}_3$ will be denoted with a bar on top ($\bar L_m,\,\bar W_m$ and 
$\bar{c}$). 
Introducing a length scale $\ell$ to be interpreted as the AdS$_3$ radius, we define new generators $P_m$ 
and $J_m$ as in (\ref{COPlpj}), as well as\i{flat limit!of W3 algebra@of $\cW_3$ algebra}\i{W3 
algebra@$\cW_3$ algebra!flat limit}
\be
\label{COPeq:fshsg12}
K_m\equiv W_m-\bar W_{-m}\,,
\qquad
Q_m\equiv\frac{1}{\ell}\left(W_m+\bar W_{-m}\right).
\ee
We also define central charges $c_1$ and $c_2$ as in (\ref{CiCCoh}). In the limit $\ell \to \infty$, and 
provided the central charges scale in such a way that both $c_1$ and $c_2$ are finite, one finds that $J_m$ 
and 
$P_m$ satisfy the $\bms$ brackets (\ref{COPjp}) together with
\begin{subequations}
\label{COPcomm-W}
\begin{alignat}{5}
[J_m,\, K_n] & = (2m-n) K_{m+n}\,, \qquad &
[J_m,\, Q_n] & = (2m-n) Q_{m+n}\,, \\  
[P_m,\, K_n] & = (2m-n) Q_{m+n}\,, \qquad &
[P_m,\, Q_n] & = 0\,.
\end{alignat}
The remaining brackets involving higher-spin generators are\i{quantum flat W3 algebra@quantum flat $\cW_3$ 
algebra}
\begin{align}
[K_m,K_n]
&=
(m-n)(2m^2+2n^2-mn-8)J_{m+n} 
+\frac{96}{c_2}(m-n)\Omega_{m+n}\\
&
\quad-\frac{96\,c_1}{c_2^2}\,(m-n)\Theta_{m+n} 
+\frac{c_1}{12}(m^2-4)(m^3-m)\delta_{m+n,\,0}\,,
\label{COP[W,W]}\\
[K_m,Q_n]
&= 
(m-n)(2m^2+2n^2-mn-8)P_{m+n} 
+\frac{96}{c_2}(m-n)\Theta_{m+n}\nn\\
& \quad +\frac{c_2}{12}(m^2-4)(m^3-m)\delta_{m+n,\,0} \,, \\
[Q_m,\, Q_n] &=0 \, ,
\end{align}
\end{subequations}
where the non-linear terms $\Omega_m$ and $\Theta_m$ are quadratic operators given by 
(\ref{COPeq:FW3NonLinTermsUR}), with the exact same ordering (and calligraphic letters replaced 
by usual capital letters):
\be
\Omega_m\equiv
\sum_{p\in\ZZ}\left(P_{m-p}J_p+J_{m-p}P_p\right)\,,
\qquad
\Theta_m\equiv\sum_{p\in\ZZ}P_{m-p}P_p\,.
\label{AMATHEO}
\ee
The commutation relations (\ref{COPcomm-W}) are quantum analogues of the Poisson brackets 
(\ref{ClaSH}),\i{flat W3 algebra@flat $\cW_3$ algebra}
including a 
central charge $c_1$ and with operators normalized so that the vacuum has zero eigenvalue under $P_0$.
One can check that with the definition (\ref{COPeq:FW3NonLinTermsUR}), the brackets given by (\ref{COPjp}) 
and (\ref{COPcomm-W}) satisfy Jacobi 
identities, so the generators $J_m,K_m,P_n,Q_n$ span a well-defined non-linear Lie algebra. We call 
it the \it{quantum flat $\cW_3$ algebra}. In any unitary 
representation, its 
generators satisfy the Hermiticity conditions\i{Hermiticity conditions!for flat W3@for flat $\cW_3$}
\be
(Q_m)^{\dagger}=Q_{-m}\,,
\qquad
(K_m)^{\dagger}=K_{-m}\,.
\ee
supplemented with (\ref{COPherm}) for $m\in\mathbb{Z}$.\\

The expressions (\ref{AMATHEO}) for the quadratic terms follow from the identities\i{flat limit!of normal 
ordering}\i{normal ordering!flat limit}
\be
{{:}\!\mathrel{\Lambda_m}\!{:}}
+
{{:}\!\mathrel{\bar\Lambda_m}\!{:}}
=
\frac{\ell^2}{2}\,\Theta_m+\cO(\ell)\,, 
\qquad
{{:}\!\mathrel{\Lambda_m}\!{:}}
-
{{:}\!\mathrel{\bar\Lambda_m}\!{:}}
=
\frac{\ell}{2}\,\Omega_m+\cO(1)\,
\ee
where ${{:}\!\mathrel{\Lambda_m}\!{:}}$ is the normal-ordered quadratic term (\ref{COPeq:VirasoroNO}) of the 
quantum $\cW_3$ 
algebra, while ${{:}\!\mathrel{\bar\Lambda_m}\!{:}}$ is its right-moving counterpart. Note, in particular, 
that both 
the linear term in (\ref{COPeq:VirasoroNO}) and the mixing between positive 
and 
negative modes in (\ref{COPlpj})-(\ref{COPeq:fshsg12}) are necessary to reorganize the sum of quadratic terms 
with 
the precise order  of (\ref{AMATHEO}). We shall see in section \ref{COPsufawa} that 
(\ref{COPeq:FW3NonLinTermsUR}) is a normal-ordered polynomial with respect to the natural vacuum in induced 
modules of the quantum flat $\cW_3$ algebra.

\subsubsection*{Galilean contraction}

In order to compare the quantum flat $\cW_3$ algebra (\ref{COPcomm-W}) with other results in the literature 
\cite{Grumiller:2014lna}, we now consider the non-relativistic limit of the quantum direct sum 
$\cW_3\oplus\cW_3$. It 
is obtained by 
defining central charges $\tilde c_1$ 
and $\tilde c_2$ as in (\ref{COPc-galileo}), introducing new generators $\tilde J_m$ and $\tilde P_m$ as in 
(\ref{COPeq:GCALinearCombinations}) and writing\i{W3 algebra@$\cW_3$ algebra!non-relativistic 
limit}\i{non-relativistic limit}
\be
\tilde K_m\equiv\bar W_m+W_m \,,
\qquad
\tilde Q_m\equiv \frac{1}{\ell}\left(\bar W_m-W_m\right).
\ee
Note the difference with respect to (\ref{COPeq:fshsg12}).
In the limit $\ell\rightarrow+\infty$ one obtains brackets of the same form as in (\ref{COPcomm-W}) 
upon putting tildes on top of all generators, but there are two important differences: 
(i) the coefficient in front of $\Theta_{m+n}$ in (\ref{COP[W,W]}) contains a shifted central charge $\tilde 
c_1+44/5$,
and (ii)
the quadratic term $\tilde\Omega_m$ reads
\be
\label{COPnoligal}
\tilde\Omega_m
=
\sum_{p\geq-1}
\left(\tilde P_{m-p}\tilde J_p+\tilde J_{m-p}\tilde P_p\right)
+
\sum_{p<-1}
\left(\tilde P_p\tilde J_{m-p}+\tilde J_p\tilde P_{m-p}\right)
-\,
\frac{3}{5}\,(m+3)(m+2)\tilde P_m
\ee
instead of (\ref{AMATHEO}). The non-linear term $\tilde\Theta_m$ remains the same (up to 
tildes) since the generators $\tilde P_m$ commute in the large $\ell$ limit. The quadratic combinations 
$\tilde\Theta_m$ and $\tilde\Omega_m$ can then be interpreted as normal-ordered operators with 
respect to a Galilean highest-weight vacuum defined by conditions of the type 
(\ref{hws}) with 
$\tilde M=\tilde s=0$. These differences show that the two contractions lead to different \emph{quantum} 
algebras, 
despite the fact that the corresponding classical algebras coincide.\footnote{An interesting problem is to 
understand if these algebras are merely different because of an unfortunate choice of basis, or if they are 
genuinely distinct in the sense that they are not isomorphic. We will not address this issue here.} Thus, in 
the presence of higher-spin fields, the difference between ultrarelativistic and Galilean limits manifests 
itself directly in the symmetry algebras and not only at the level of the representations that survive in 
the 
limit.\\

In the following we restrict attention to irreducible unitary representations of the ultrarelativistic 
quantum algebra (\ref{COPcomm-W}), built once again according to the induced module prescription of section 
\ref{sebomodule}. On the 
other 
hand, highest-weight representations of Galilean\i{Galilean flat W3 algebra@Galilean flat $\cW_3$ 
algebra}\i{gca2@$\gca$!higher spin version}\i{flat W3 algebra@flat $\cW_3$ algebra!Galilean version} 
contractions of two copies of non-linear $\mathcal{W}$ 
algebras were discussed in \cite{Grumiller:2014lna}, where it was shown that unitary representations with 
higher-spin states do not exist.

\subsection{\ Induced modules for the flat $\cW_3$ algebra}
\label{COPsufawa}

According to our proposal of section \ref{CAPPOU}, the Hilbert space of any unitary representation of the 
flat $\cW_3$ algebra consists of wavefunctions on the orbit of a higher-spin supermomentum 
$\big(p(\phii),\rho(\phii)\big)$. Assuming that the orbit admits a quasi-invariant measure, a basis 
of the Hilbert space is provided by plane waves (\ref{pawawest}) with definite supermomentum. For 
definiteness, let us focus on an orbit containing a constant higher-spin supermomentum 
$(p_0,\rho_0)$; this is to say that the representation admits a rest frame. There is 
a corresponding plane wave $\Psi_{(p_0,\rho_0)}$, and any other plane wave can be obtained by acting on 
$\Psi_{(p_0,\rho_0)}$ with a higher-spin superrotation.

\subsubsection*{Massive modules}

Let us take $p_0=M-c_2/24$ with $M>0$; this corresponds to a massive representation of the flat $\cW_3$ 
algebra. Assuming also that $\rho_0$ is generic, the little group is $\un\times\RR$ and the spin of the 
representation is therefore a pair $(s,\sigma)\in\RR^2$. Now, the plane wave at rest 
$\Psi_{(p_0,\rho_0)}\equiv|M,\rho_0\rangle$ is a state that satisfies\i{rest 
frame!for flat W3 algebra@for flat $\cW_3$ algebra}
\begin{subequations}
\label{COPrestW}
\be 
P_m |M,\rho_0 \rangle = 0 \, , \qquad 
Q_m |M,\rho_0 \rangle = 0 \, \qquad \mathrm{for}\ m \neq 0 \, ,
\ee
and is an eigenstate of zero-mode charges:
\begin{alignat}{5}
P_0 |M,\rho_0 \rangle & = M |M,\rho_0 \rangle \, , \qquad 
& J_0 |M,\rho_0 \rangle & = s |M,\rho_0 \rangle \, , \\
Q_0 |M,\rho_0 \rangle & = \rho_0 |M,\rho_0 \rangle \, , \qquad 
& K_0 |M,\rho_0 \rangle & = \sigma |M,\rho_0 \rangle \, .
\end{alignat}
\end{subequations}
Here $M$ and $s$ are the mass and spin labels encountered in (\ref{COPrest-bms}), while $\rho_0$ and $\sigma$ 
are their spin-3 
counterparts. As before we call $|M,\rho_0 \rangle$ the \it{rest frame state} of the representation and we 
normalize the operator $P_0$ so that the vacuum has vanishing $P_0$ 
eigenvalue.\\

The conditions (\ref{COPrestW}) define a one-dimensional representation of the subalgebra spanned by 
$\{P_m,Q_m,J_0,W_0\}$. They can be used to define an induced module $\sH$ with basis elements\i{induced 
module!for flat W3 algebra@for flat $\cW_3$ algebra}\i{flat W3 algebra@flat $\cW_3$ algebra!induced 
representation}
\be
\label{COPWmodule}
K_{k_1}...K_{k_m}J_{l_1}...J_{l_n}|M,\rho_0\rangle \, ,
\ee
where $k_1\leq...\leq k_m$ and $l_1\leq...\leq l_n$ are non-zero integers. This provides an 
explicit representation of the quantum flat $\cW_3$ algebra. Note that the presence of non-linearities in the 
commutators (\ref{COPcomm-W}) does not affect the construction of the induced module, which involves the 
universal enveloping algebra anyway.\\

As usual, unitarity is somewhat hidden in the induced module picture 
but can be recognized in the fact that the state $|M,\rho_0\rangle$ is a plane wave, and that acting on it 
with finite higher-spin superrotations generates an orthonormal basis of plane wave states for the carrier 
space of 
the representation. Irreducibility\i{irreducible representation}\i{representation!irreducible} can be 
inferred from the same argument that we used for 
$\mathfrak{bms}_3$: by 
construction, a 
supermomentum orbit is a homogeneous space for the action of superrotations, and this carries over to the 
higher-spin setting. This implies that $\cW_3$ superrotations can map any plane wave state on any other one, 
which in turn implies that the space of the representation has no non-trivial invariant subspace.

\subsubsection*{Vacuum module}

The vacuum module of the flat $\cW_3$ algebra can be built in direct analogy to its $\mathfrak{bms}_3$ 
counterpart discussed around (\ref{COPeq:BMSVac}). The only subtlety is the enhancement of the little group, 
which leads to additional conditions on superrotations. Indeed the vacuum state $|0\rangle$ is now an 
eigenstate 
of all modes $P_m$ and $Q_m$ with zero eigenvalue, and satisfies in addition\i{flat W3 algebra@flat 
$\cW_3$ algebra!vacuum module}\i{vacuum module!for flat W3 algebra@for flat $\cW_3$ algebra}
\be
J_n|0\rangle=0\;\;\text{ for }n=-1,0,1,
\qquad
K_m|0\rangle=0\;\;\text{ for }m=-2,-1,0,1,2.
\label{JAKTA}
\ee
These conditions ensure that the vacuum is invariant under the $\mathfrak{sl}(3,\RR)$ wedge algebra\i{wedge 
algebra} of the 
$\cW_3$ subalgebra (which includes in particular the Lorentz algebra). The corresponding module can then be 
built as usual by acting with higher-spin superrotation generators on the vacuum state and producing states 
of 
the form (\ref{COPWmodule}), where now all $l_i$'s must be different from $-1,0,1$ and all $k_i$'s must be 
different from $-2,-1,0,1,2$. We stress that the $l_i$'s and $k_i$'s can be positive or negative, in sharp 
contrast to the non-relativistic modules investigated in \cite{Grumiller:2014lna}.\\

The definition of the flat $\cW_3$ vacuum allows us to interpret the quadratic 
terms (\ref{AMATHEO}) as being normal-ordered. Indeed, their expectation values vanish in the vacuum 
$|0\rangle$:\i{normal ordering}
\be
\label{COPeq:RestFrameNO}
\langle0|\Theta_n|0\rangle
=
\langle0|\Omega_n|0\rangle
=0.
\ee
These considerations appear to be a robust feature of ``flat $\cW$ algebras'': ultrarelativistic 
contractions 
of $\cW_N \oplus \cW_N$ algebras always take the form
\be
\text{``flat $\cW_N$''}
=
\cW_N\inplus_{\text{ad}}(\cW_N)_{\text{Ab}}
\ee
and therefore contain an Abelian ideal, where the semi-direct sum ensures that the structure constants 
of the non-linear terms are 
always proportional to inverse powers of the central charge. Indeed, for a non-linear operator of 
$n^\textnormal{th}$ order the structure constants are of order $\frac{1}{c^{n-1}}$ at large $c$. When 
expanding them in powers of the contraction parameter $\ell$, this implies that the leading term is 
proportional to $\ell^{1-n}$ thanks to (\ref{CiCCoh}). In order to obtain a finite expression, it is thus 
necessary that the resulting non-linear operator consists of at least $n-1$ Abelian generators. Terms of this 
kind always have a vanishing expectation value in the rest frame vacuum state, although the precise ordering 
in the 
polynomial should be fixed by other means, e.g.\ by defining the algebra via a contraction of the quantum 
algebra or by imposing Jacobi identities. Thus the conditions (\ref{COPrestW}) with $M=\rho_0=s=\sigma=0$, 
together with (\ref{JAKTA}), provide a valid definition of the vacuum for all quantum flat $\cW_N$ algebras.\\

By contrast, for a highest-weight vacuum of the type (\ref{hws}), the quadratic operators $\Omega_m$ given by 
(\ref{AMATHEO}) generally have non-vanishing vacuum
expectation values. Thus the extra non-linear structure introduced by higher spins exhibits the fact
that the natural representations in the ultrarelativistic limit are the induced ones 
discussed above, rather than the highest-weight ones of \cite{Bagchi:2009pe,Grumiller:2014lna}.
This difference emphasizes the physical distinction between ultrarelativistic and Galilean limits: the 
former is adapted to gravity, and more generally to models of fundamental interactions, 
where unitarity is a key requirement. In particular, flat space holography (at least in the framework of 
Einstein gravity) is expected to rely on the unitary construction described in this thesis. By contrast, the 
Galilean viewpoint is suited to condensed matter applications, and more 
generally to situations where unitarity need not hold --- as was indeed argued in \cite{Bagchi:2009pe}. We 
stress that this 
difference is a genuine quantum higher-spin effect: it is not apparent at the classical level, and it does 
not occur in pure gravity either.

\section{\ Super-BMS$_3$ and flat supergravity}
\label{CAPsupersec}

This section is devoted to supersymmetric extensions of the BMS$_3$ group, to their representations, and to 
their characters. Accordingly we start by describing
rotating one-loop partition functions of fermionic fields in flat space, along the same 
lines as in section \ref{seRROT}. Upon confirming that they take the form of 
exponentials of Poincar\'e characters (\ref{CAPZexp}), we specialize to $D=3$ space-time dimensions. There we 
describe supersymmetric BMS$_3$ groups and their unitary representations, and note that super BMS$_3$ 
multiplets contain towers of infinitely many particles with increasing spins. Finally, we show that the 
resulting 
characters match suitable combinations of bosonic and fermionic one-loop partition functions.

\subsection{\ Fermionic higher spin partition functions}
\label{CAPsubsec2.3}

We wish to evaluate the partition function \ref{CAPs2.5} of a free fermionic field $\psi$ with spin 
$s+1/2$ (where $s$ is a non-negative integer) and mass $M > 0$.\i{higher spin}\i{fermion!higher 
spin} Its Euclidean action can be presented either (i) using a symmetric, 
$\gamma$-traceless spinor field with $s$ 
space-time indices and a set of auxiliary fields with no gauge symmetry \cite{Singh:1974rc} or (ii) using a 
set of symmetric spinor fields with $s,s-1,...,0$ space-time indices and vanishing triple $\gamma$-trace, 
subject to a gauge symmetry generated by $\gamma$-traceless parameters with $s-1,...,0$ space-time indices 
\cite{Metsaev:2006zy}. In the latter case, just as for bosons, the action is given by a sum of actions 
for massless fields 
of each of the involved spins, plus a set of cross-coupling terms proportional to the mass. In the limit $M 
\to 0$ the quadratic couplings vanish and one is left with a sum of decoupled Fang-Fronsdal actions 
\cite{Fang:1978wz}\i{Fang-Fronsdal action}\i{action functional!for higher spin field}
\be
\label{CAPss8.5}
S[\psi,\bar\psi]
=
\int d^Dx\,
\bar{\psi}^{\mu_1...\mu_s}
\left(
\cS_{\mu_1...\mu_s}
-
\frac{1}{2}\,\gamma_{(\mu_1}{\not {\!\cS}}_{\mu_2 \cdots \mu_s)}
-
\frac{1}{2}\,\delta_{(\mu_1\mu_2 }{\cS_{\mu_3 \cdots \mu_s)\lambda}}^\lambda
+
\text{h.c.}
\right),
\ee
where space-time indices are raised and lowered thanks to the Euclidean metric (and ``$\text{h.c.}$'' means 
``Hermitian conjugate'').
We use the same symmetrization conventions as in section \ref{CAPs2.5} and
\be
\cS_{\mu_1...\mu_s}
=
\left(
{\not {\!\der}}\,\psi_{\mu_1...\mu_s}-\der_{(\mu_1}{\not {\!\! \psi}}{}_{\mu_2...\mu_s)}
\right)\,.
\ee
The slash notation means ${\not{\!\cV}}\equiv\gamma^{\mu}\cV_{\mu}$, where the $\gamma^{\mu}$'s are Dirac 
matrices satisfying the anticommutation relations $\{\gamma^{\mu},\gamma^{\nu}\}=\delta^{\mu\nu}$.\\

To compute the partition function for $\psi$, $\bar\psi$ one has to evaluate a path integral (\ref{CAPs2.5}) 
with the integration measure $\cD\psi\cD\bar\psi$ and $S$ the action (\ref{CAPss8.5}) or its massive 
analogue. 
The fermionic fields live on $\RR^D/\ZZ$ as defined by the group action (\ref{CAP2.6}), but in contrast to 
bosons, they satisfy \it{anti}periodic\i{antiperiodic} boundary conditions along the thermal cycle. For a 
massive field, 
one thus finds that the partition function is given by\i{functional determinant}\i{partition 
function!for fermions}\i{fermion!partition function}
\be
\label{CAP2.18}
\log Z
=
\frac{1}{2}
\log\det(-\Delta^{(s+1/2)}+M^2)
-\frac{1}{2}\log\det(-\Delta^{(s-1/2)}+M^2)\,,
\ee
where $\Delta^{(s+1/2)}$ is the Laplacian acting on antiperiodic, symmetric, $\gamma$-traceless spinor fields 
with $s$ indices on $\RR^D/\ZZ$. For massless fields, the gauge symmetry enhancement requires gauge-fixing 
and ghosts,\i{ghost} leading to \cite{Creutzig:2011fe}
\be
\label{CAP2.4fer}
\log Z
=
\frac{1}{2}
\log\det(-\Delta^{(s+1/2)})
-
\log\det(-\Delta^{(s-1/2)})
+
\frac{1}{2}
\log\det(-\Delta^{(s-3/2)})\,.
\ee
Eqs.\ (\ref{CAP2.18}) and (\ref{CAP2.4fer}) are fermionic analogues of the bosonic formulas (\ref{CAPss5b}) 
and (\ref{CAP2.4bos}).
To evaluate the functional determinants, we rely once more on heat kernels and the method of 
images described in section \ref{suseHEAT}.\\

The heat kernel ${\cK^{AB}}_{\mu_s,\nu_s}$ associated with the operator $(-\Delta^{(s+1/2)}+M^2)$ on $\RR^D$ 
is the unique solution of\i{fermion!heat kernel}\i{heat kernel}
\be
\label{CAP2.14}
({\Delta_{(s+1/2)}}-M^2-\der_t){\cK^{AB}}_{\mu_s,\nu_s}=0\,,
\quad
{\cK^{AB}}_{\mu_s,\,\nu_s}(t=0,x,x')
= 
\II^{(F)}_{\mu_s,\,\nu_s}{\text{\bfseries{1}}}^{AB}\delta^{(D)}(x-x')\,.
\ee
Here ${\cK^{AB}}_{\mu_s,\nu_s}$ is a bispinor in the indices $A$ and $B$, and a symmetric bitensor 
in the indices $\mu_s$ and $\nu_s$. (We use again the shorthand $\mu_s$ to denote a set of $s$ symmetrized 
indices.) It is also $\gamma$-traceless in the sense that\i{gamma tracelessness@$\gamma$-tracelessness}
\be
\label{CAP2.15}
\gamma^{\mu}{\cK}_{\mu_{s},\,\nu_s}={\cK}_{\mu_s,\,\nu_{s}}\gamma^{\nu}=0 \, .
\ee
The 
solution of (\ref{CAP2.14}) satisfying this requirement is
\be
\label{CAPs9}
{\cK}_{\mu_s,\,\nu_s}(t,x,x')
=
\frac{1}{(4\pi t)^{D/2}}\,
e^{-M^2 t-\frac{1}{4t}|x-x'|^2}\;\II^{(F)}_{\mu_s,\,\nu_s}\,,
\ee
where $\II^{(F)}_{\mu_s,\nu_s}$ is the following bisymmetric, $\gamma$-traceless tensor:
\be
\II^{(F)}_{\mu_s,\,\nu_s}
=
\sum_{k=0}^{\lfloor\frac{s}{2}\rfloor}
\frac{(-1)^k 2^k k!\, [D+2(s-k-1)]!!}{s!\,[D+2(k-1)]!!}
\left(
\delta^k_{\mu \mu}  
\delta^{s-2k}_{\mu\nu}
\delta^s_{\nu\nu}
-
\frac{\delta^s_{\mu\mu}\delta^{s-2k-1}_{\mu \nu}  
\delta^s_{\nu\nu}\gamma_{\mu}\gamma_{\nu}}{D+2(s-k-1)}
\right).
\ee
Up to the replacement of $\II$ by $\II^{(F)}$, the fermionic heat kernel (\ref{CAPs9}) is the same as the 
bosonic one in eq.~(\ref{CAPs6}). In particular, $\II^{(F)}$ carries all its tensor and spinor indices.\\

To evaluate the determinant of $(-\Delta^{(s+1/2)}+M^2)$ on $\mathbb{R}^D/\ZZ$, we use once more the method 
of 
images (\ref{CAPt4}). As before, we need to keep track of the non-trivial index structure of 
${\cK^{AB}}_{\mu_s,\nu_s}$, which leads to\i{method of images}
\be
\label{CAPss9}
\cK^{\mathbb{R}^D/\mathbb{Z}}_{\mu_s,\, \alpha_s}(t,x,x')
=
\sum_{n\,\in\,\mathbb{Z}}(-1)^n
{(J^n)_{\alpha}}^{\beta}...{(J^n)_{\alpha}}^{\beta}
\,
U^n\,
\cK_{\mu_s,\,\beta_s}\big(t,x,\gamma^n(x')\big) \, ,
\ee
where the factor $(-1)^n$ comes from antiperiodic boundary conditions, $J$ is the matrix (\ref{CAP2.7}), and 
$U$ 
is a $2^{\lfloor D/2\rfloor}\times2^{\lfloor D/2\rfloor}$ matrix acting on spinor indices 
in such a way that
\be
\label{CAP2.16}
{J^\alpha}_{\beta}\gamma^{\beta}
=
U \gamma^{\alpha} U^{-1}\,.
\ee
In other words, $U$ is the matrix corresponding to the transformation (\ref{CAP2.7}) in the spinor 
representation of $\text{SO}(D)$, and it can be written as
\be
U
=
\exp\left[\frac{1}{4}\sum_{j=1}^{\lfloor(D-1)/2\rfloor}\theta_j[\gamma_{2j-1},\gamma_{2j}]\right].
\nn
\ee
In particular, a rotation by $2\pi$ around any given axis maps the field $\psi$ on $-\psi$, in accordance 
with the fact 
that spinors represent $\text{SO}(D)$ up to a sign. Note that, using an explicit $D$-dimensional 
representation of the $\gamma$ matrices, one gets
\be
\label{CAPss9.5}
\text{Tr}(U^n)
=
2^{\lfloor D/2\rfloor}\prod_{i=1}^r\cos(n\theta_i/2)\,.
\ee
Now, plugging (\ref{CAPss9}) into formula (\ref{CAP2.5}) for the determinant of 
$-\Delta^{(s+1/2)}$, one obtains a sum of integrals which can be evaluated exactly as in the bosonic case. 
The only difference with respect to bosons comes from the spin structure, and the end result is\i{one-loop 
determinant}\i{functional determinant}
\be
-\log \det(-\Delta^{(s+1/2)}+M^2)
=
\sum_{n\,\in\,\mathbb{Z}^*}\frac{(-1)^n}{|n|}
\frac{\chi^{(F)}_s[n\vec\theta,\vec{\epsilon}\,]}{\prod\limits_{j=1}^r|1-e^{in(\theta_j+i\epsilon_j)}|^2}
\times
\begin{cases} 
e^{-|n|\beta M}  &\!\!D\,\text{odd}\\[3pt] 
\frac{ML}{\pi}K_1(|n|\beta M)&\!\!D\,\text{even}\end{cases}
\ee
where we have discarded a volume divergence independent of all chemical potentials (as in eq.\ 
(\ref{CAPss5t})), and where
\be
\label{CAPs10}
\chi^{(F)}_s[n\vec\theta,\vec{\epsilon}\,]
=
(J^{\mu\alpha})^s\,
\text{Tr}\left[
\II^{(F)}_{\mu_s, \alpha_s} 
\right]
\ee
is the fermionic analogue of (\ref{CAPs5.5t}), with the same rough regularization as in eq.\ (\ref{CAPss5t}) 
(a more careful regularization will be described below for $D=3$). This result takes 
the same form as (\ref{CAPss5t}), up to 
the replacement of $\chi_s$ by $\chi_s^{(F)}$ and the occurrence of $(-1)^n$ due to antiperiodicity. In 
appendices \ref{CAPAppB1} and \ref{CAPAppB2}, we show that
\be
\label{CAPs9.5}
\chi_s^{(F)}[n\vec\theta]
\stackrel{\text{\text{\ref{CAPAppB1}}\&\ref{CAPAppB2}}}{=}
\begin{cases}
\chi_{\lambda_s^{(F)}}^{(D)}[n\vec \theta\,] & \text{ for odd }D,\\[3pt]
\chi_{\lambda_s^{(F)}}^{(D)}[n\vec \theta,0] & \text{ for even }D,
\end{cases}
\ee
where the term on the right-hand side is the character of an 
irreducible representation of $\text{SO}(D)$ with highest weight 
$\lambda_s^{(F)}=(s+1/2,1/2,...,1/2)$, written here in the dual basis of the Cartan subalgebra of 
$\mathfrak{so}(D)$ described above 
(\ref{CAPs6b}).\\

Having computed the required functional determinants on $\RR^D/\ZZ$, we can now write down the partition 
functions given by (\ref{CAP2.18}) and (\ref{CAP2.4fer}). In the massive case, the difference of Laplacians 
acting 
on fields with spins $(s+1/2)$ and $(s-1/2)$ produces the difference of two factors (\ref{CAPs9.5}), with 
labels 
$s$ and $s-1$. It turns out that formula (\ref{CAPt6.5b}) still holds if we replace $\lambda_s$ and 
$\lambda_{s-1}$ by their fermionic counterparts, $\lambda^{(F)}_s$ and 
$\lambda^{(F)}_{s-1}$. (The proof of this statement follows the exact same steps as in the bosonic case 
described in appendix \ref{CAPAppA3}, up to obvious replacements that account for the change in the highest 
weight vector.) Accordingly, the rotating one-loop partition function of a massive field with spin 
$s+1/2$ is\i{fermion!partition function}\i{higher spin!partition function}\i{partition function!for 
fermions}\i{partition function!for higher spins}
\be
\label{CAPmassivefermion}
Z(\beta,\vec\theta\,)
=
\exp\left[
\sum_{n=1}^\infty\frac{(-1)^{n+1}}{n}
\frac{\chi_{\lambda_s^{(F)}}^{(D-1)}[n\vec\theta,\vec{\epsilon}\,]}
{\prod\limits_{j=1}^r|1-e^{in(\theta_j+i\epsilon_j)}|^2}
\times
\begin{cases}
e^{-n\beta M} & \text{($D$ odd)}\\
\frac{ML}{\pi}K_1(n\beta M) & \text{($D$ even)}
\end{cases}
\right].
\ee
In the massless case we must take into account one more difference of characters, namely (\ref{CAPss6.5t}) 
with 
$\lambda_s$ replaced by $\lambda^{(F)}_s$. For $D\geq4$, this difference can be written as a combination of 
$\text{SO}(D-2)$ characters (the proof is essentially the same as in appendix \ref{CAPAppA3}), and the 
partition 
function of a massless field with spin $s+1/2$ exactly takes the form (\ref{CAPt6.5t}) or (\ref{CAPss6q}) 
(for $D$ even or odd, respectively) with an additional factor of $(-1)^{n+1}$ in the sum over $n$, and the 
replacement of $\lambda_s$ by $\lambda^{(F)}_s$. One can also verify that relation 
(\ref{CAPs6.5q}) remains true for fermionic partition functions.\\

For $D=3$, differences of 
$\text{SO}(2)$ characters cannot be reduced any further (recall the discussion surrounding (\ref{CAPILOBO})), 
so the best one can do is to write the partition function of a massless field with spin $s+1/2$ as
\be
Z(\beta,\theta)
=
\exp\left[
\sum_{n=1}^{+\infty}
\frac{(-1)^{n+1}}{n}
\frac{1}{|1-e^{in(\theta+i\epsilon)}|^2}
\left(
e^{i(s+1/2)n(\theta+i\epsilon)}-e^{i(s-1/2)n(\theta+i\epsilon)}+\text{c.c.}
\right)
\right]
\label{SAMBAPATI}
\ee
provided $s\geq1$. (For $s=0$ the exponentials in the summand reduce to 
$e^{in(\theta+i\epsilon)/2}+\text{c.c.}$, without any negative contribution.) Here we are using once more the 
crude regularization described around (\ref{CAPss5t}); a more careful prescription, motivated by the bosonic 
combination (\ref{KAPELLEKE}), consists in regulating the sum of exponentials in the summand according 
to\i{iepsilon regularization@$i\epsilon$ regularization}
\be
e^{i(s+1/2)n(\theta+i\epsilon)}-e^{i(s-1/2)n\theta-(s+3/2)n\epsilon}+\text{c.c.}
\label{BAKTOTO}
\ee
Upon using this expression in the summand of (\ref{SAMBAPATI}) instead of the naive combination of 
exponentials written there, the series in the exponential becomes
\begin{align}
\sum_{n=1}^{+\infty}\frac{(-1)^{n+1}}{n}
\frac{q^{n(s+1/2)}-q^{n(s+1/2)}\bar q^n+\text{c.c.}}{|1-q^n|^2}
&=
\sum_{n=1}^{+\infty}
\left(
\frac{(-1)^{n+1}}{n}
\frac{q^{n(s+1/2)}}{1-q^n}
+\text{c.c.}
\right)\nn\\
&=
\sum_{j=s}^{+\infty}\log(1+q^{j+1/2})+\text{c.c.}
\nn
\end{align}
in terms of $q=e^{i(\theta+i\epsilon)}$. As in the bosonic case (\ref{Storm}), the regularization 
(\ref{BAKTOTO}) has ensured that $\log Z$ splits as the sum of a chiral and an anti-chiral function of $q$. 
After renaming $j$ into $n$, the end result is the following expression for the partition function of a field 
with spin $s+1/2$ in three dimensions:
\be
\label{CAPs9.5b}
Z
=
\prod^{+\infty}_{n=s}
\big|1+e^{i\left(n+1/2\right)(\theta+i\epsilon)}\big|^2,
\ee
which can also be recovered as 
the flat limit of the corresponding AdS result \cite{Creutzig:2011fe}. The remainder of this chapter is 
devoted to relating this partition function to the vacuum characters of various supersymmetric extensions of 
the BMS$_3$ group.

\subsection{\ Supersymmetric BMS$_3$ groups}
\label{CAPsec:3DSugra}

The supersymmetric BMS$_3$ groups describe the symmetries of three\--di\-men\-sio\-nal, 
a\-symp\-to\-ti\-cal\-ly flat 
supergravity \cite{Barnich:2014cwa,Barnich:2015sca,Fuentealba:2015jma,Fuentealba:2015wza,Mandal:2010gx}. Here 
we 
briefly review some background on super Lie groups and the super Virasoro algebra, which we then use to 
provide a definition of various supersymmetric extensions of BMS$_3$. The corresponding unitary 
representations and characters will be investigated in section \ref{SAMBOR}.

\subsubsection*{Supersymmetric induced representations}

A super Lie group\i{super Lie group} is a pair
$(\Gamma_0,\gamma)$ 
where $\Gamma_0$ is a Lie group in the 
standard sense, while $\gamma$ is a super Lie algebra whose even part coincides with the Lie algebra of 
$\Gamma_0$, and whose odd part is a $\Gamma_0$-module such that the differential of the $\Gamma_0$ action be 
the bracket 
between even and odd elements of $\gamma$ \cite{deligne1999quantum}. Then a \it{super 
semi-direct product} is a super Lie group of the form 
\cite{Carmeli:2005rg,CarmeliThesis}
\be
\big(
G\ltimes_{\sigma}A
,
\mathfrak{g}\inplus(A+\cA)
\big),
\label{CAPsupergroup}
\ee
where $G\ltimes A$ is a standard (bosonic) semi-direct product group with Lie algebra $\mathfrak{g}\inplus 
A$, while $\mathfrak{g}\inplus(A+\cA)$ is a super Lie algebra whose odd subalgebra $\cA$ 
is a $G$-module such 
that
the bracket between elements of $\mathfrak{g}$ and elements of $\cA$ be the differential of the action of $G$ 
on $\cA$, and such that $[A,\cA]=0$ and $\{\cA,\cA\}\subseteq A$. By virtue of this definition, the action of 
$G$ on $\cA$ is compatible with the super Lie bracket:
\be
\left\{g\cdot S,g\cdot T\right\}
=
\sigma_g\left\{S,T\right\}
\qquad
\forall\,S,T\in\cA\,,
\label{CAPcompa}
\ee
where $\sigma$ is the action of $G$ on $A$.\\

It was shown in \cite{Carmeli:2005rg,CarmeliThesis} that all irreducible, unitary representations 
of a super 
semi-direct product are induced in essentially the same sense as for standard, bosonic groups.\i{induced 
representation!of super semi-direct product}\i{representation!of super semi-direct product} 
In 
particular, 
they are classified by the orbits and little groups of $G\ltimes_{\sigma}A$, as explained in section 
\ref{sesemi}. However, there are two important 
differences with 
respect to the purely bosonic case:
\begin{enumerate}
\item Unitarity rules out all orbits on which energy can be negative, so that the
momentum orbits giving rise to unitary representations of the supergroup form a subset of the full menu of 
orbits available in the purely bosonic case. More precisely, 
given a momentum $p\in A^*$, it must be such that\i{admissible momentum}\i{momentum!admissible}
\be
\langle p,\{S,S\}\rangle\geq0\qquad\forall S\in\cA\,.
\label{CAPadmiss}
\ee
When this condition is not satisfied, the representations of (\ref{CAPsupergroup}) 
associated 
with the orbit $\cO_p$ are not unitary. The momenta satisfying condition (\ref{CAPadmiss}) are said to be 
\it{admissible}. Note that admissibility is a $G$-invariant statement: if $f\in G$ and 
if $p$ is admissible, 
then so is $f\cdot p$, by virtue of (\ref{CAPcompa}). For instance, the only admissible 
momenta for the super Poincar\'e group
are those of massive or massless particles with positive energy (and the trivial momentum $p=0$).
\item Given an admissible momentum $p$, the odd piece $\cA$ of the supersymmetric translation algebra 
produces a (generally degenerate) Clifford algebra\i{Clifford algebra}
\be
\cC_p
=
T(\cA)/\left\{S^2-\langle p,\{S,S\}\rangle\;|\;S\in\cA\right\},
\label{CAPcliff}
\ee
where $T(\cA)$ is the tensor algebra of $\cA$.\i{tensor algebra} Quotienting this algebra by its ideal 
generated by the radical 
of $\cA$, one obtains a non-degenerate Clifford algebra $\bar\cC_p$. Since $\cA$ is a $G$-module, there 
exists 
an action of the little group $G_p$ on $\bar\cC_p$; let 
us denote this action by $a\mapsto g\cdot a$ for $a\in\bar\cC_p$ and $g\in G_p$. To obtain a representation 
of the full supergroup (\ref{CAPsupergroup}), one must find an irreducible representation $\tau$ of 
$\bar\cC_p$ and a representation $\cR_0$ of $G_p$ acting in the same space, and compatible with $\tau$ in 
the 
sense that\i{little group!fundamental representation}\i{fundamental representation}
\be
\tau[g\cdot a]
=
\cR_0[g]\cdot\tau[a]\cdot(\cR_0[g])^{-1}.
\label{CAPcompat}
\ee
For finite-dimensional groups, the pair $(\tau,\cR_0)$ turns out to be unique up to multiplication of $\cR_0$ 
by a character of $G_p$ (and possibly up to parity-reversal). Given such a pair, we call it the 
\it{fundamental representation} of the 
supersymmetric little group.
\end{enumerate}

The Clifford algebra (\ref{CAPcliff}) leads to a replacement of the irreducible, ``spin'' representations of 
the 
little group, by generally \it{reducible} representations $\cR_0\otimes\cR$. This is the multiplet 
structure of supersymmetry:\i{supermultiplet}\i{multiplet} the restriction of 
an irreducible unitary 
representation of a supergroup to its 
bosonic subgroup is generally reducible, and the various irreducible components account for the combination 
of 
spins that gives rise to a {\scshape{susy}} multiplet. In the Poincar\'e group, an irreducible supermultiplet 
contains finitely 
many spins; by contrast, we will see below that super-BMS$_3$ multiplets contain infinitely many spins. 
Apart from this difference, the structure of induced representations of super semi-direct products is 
essentially the same as in the bosonic case: they consist of wavefunctions on an orbit, taking their values 
in the space of the representation $\cR_0\otimes\cR$.\label{CAPROR} In particular, the Frobenius formula 
(\ref{fropokad}) 
for characters remains valid, up to the replacement of $\cR$ by $\cR_0\otimes\cR$.

\subsubsection*{Supersymmetric Virasoro algebra}

As a preparation for super BMS$_3$, let us first recall the definition of the super Virasoro algebra. The 
latter is 
built by adding to 
$\text{Vect}(S^1)$ an odd subalgebra $\cF_{-1/2}(S^1)$ of $-1/2$-densities on the circle 
\cite{Dai,guieu2007algebre}. This produces a Lie superalgebra, isomorphic to 
$\text{Vect}(S^1)\oplus\cF_{-1/2}(S^1)$ as a vector space, which we shall write as 
$\mathfrak{s}\text{Vect}(S^1)$. Its 
elements are pairs $(X,S)$, where $X=X(\varphi)\partial/\partial\varphi$ and 
$S=S(\varphi)(d\varphi)^{-1/2}$, and the super Lie bracket is 
defined as\i{super Witt algebra}\i{Witt algebra!supersymmetric version}\i{VectS1@$\Vect$!supersymmetric 
version}
\be
\label{CAP3.28}
\big[
(X,S),(Y,T)
\big\}
\equiv
\Big(
[X,Y]+S\otimes T,X\cdot T-Y\cdot S
\Big).
\ee
Here $[X,Y]$ is the standard Lie bracket of vector fields and the dot denotes the natural action of 
vector fields on $\cF_{-1/2}(S^1)$, so that $X\cdot T$ is the $-1/2$-density with component
\be
X\cdot T
\equiv
XT'-\frac{1}{2}X'T.
\ee
(This is formula (\ref{infident}) with $h=-1/2$.)
Upon expanding the functions $X(\varphi)$ and $S(\varphi)$ in Fourier modes, one recovers the standard 
$\cN=1$ supersymmetric extension of the Witt algebra. Choosing $S(\phii)$ to be periodic or antiperiodic 
leads to the Ramond\i{Ramond sector} or the Neveu-Schwarz\i{Neveu-Schwarz sector} sector of the superalgebra, 
respectively.\\

The central extension of $\mathfrak{s}\text{Vect}(S^1)$ is the super 
Virasoro algebra, $\mathfrak{svir}$. Its elements are triples $(X,S,\lambda)$ where 
$(X,S)\in\mathfrak{s}\text{Vect}(S^1)$ and $\lambda\in\RR$, with a super Lie bracket\i{super Virasoro 
algebra}\i{Virasoro algebra!supersymmetric version}
\be
\big[
(X,S,\lambda),(Y,T,\mu)
\big\}
\equiv
\Big(
[X,Y]+S\otimes T,X\cdot T-Y\cdot S,\sfc(X,Y)+\sfh(S,T)
\Big),
\label{CAPsuperbra}
\ee
where $\sfc$ is the Gelfand-Fuks cocycle (\ref{gefuks}) while $\sfh$ is its supersymmetric cousin,
\be
\sfh(S,T)\equiv\frac{1}{12\pi}\int_0^{2\pi}d\varphi\, S'T'\,.
\label{CAPcocycle}
\ee
By expanding the functions $X$ and $S$ in Fourier modes, one 
obtains the usual commutation relations of $\cN=1$ super Virasoro. Explicitly, defining the 
generators
\be
\cL_m\equiv\big(e^{im\phii}\der_{\phii},0,0\big),
\qquad
\cQ_r\equiv\big(0,e^{ir\phii}(d\phii)^{-1/2},0\big),
\qquad
\cZ\equiv(0,0,1),
\nn
\ee
one finds that (\ref{CAPsuperbra}) yields the super Lie brackets
\begin{align}
i[\cL_m,\cL_n\} & = (m-n)\cL_{m+n}+\frac{\cZ}{12}m^3\delta_{m+n,0}\,,\nn\\
i[\cL_m,\cQ_r\} & = \left(\frac{m}{2}-r\right)\cQ_{m+r}\,,\nn\\
{}[\cQ_r,\cQ_s\} & = \cL_{r+s}+\frac{\cZ}{6}r^2\delta_{r+s,0}\,.
\end{align}
\vspace{.1cm}

The super Virasoro algebra is (half of) the asymptotic symmetry algebra of 
three-dimensional supergravity\i{supergravity}\i{asymptotic symmetry!for supergravity}\i{asymptotic Killing 
spinor}
with Brown-Henneaux boundary conditions \cite{Banados:1998pi,Henneaux:1999ib} (see also 
\cite{Coussaert:1993jp}). 
In that context the vector field $X$ is one of the components of an asymptotic Killing vector field 
(\ref{s208}), while $S$ is one of the components of an asymptotic Killing spinor. The fact that the 
quantization of three-dimensional supergravity produces super Virasoro representations was verified in 
\cite{David:2009xg} by showing that the one-loop partition 
function of supergravity on thermal AdS$_3$ coincides with the vacuum character of two super Virasoro 
algebras. In the remainder of this section our goal is to describe the flat analogue of these results.

\subsubsection*{Supersymmetric BMS$_3$ groups}

Equipped with the definition of super semi-direct products and that of the super Virasoro algebra, 
we can now define the $\cN=1$ super BMS$_3$ group \cite{Barnich:2014cwa,Barnich:2015sca}: it is a super 
semi-direct 
product (\ref{CAPsupergroup}) whose even piece is the BMS$_3$ group (\ref{defbms}), and whose odd subspace 
is the space of densities $\cF_{-1/2}(S^1)$ with the bracket $\{S,T\}=S\otimes T$. In other words, the 
(centreless) super 
$\mathfrak{bms}_3$ algebra is a super semi-direct sum\i{super bms3 algebra@super $\bms$ 
algebra}\i{bms3 algebra@$\bms$ algebra!supersymmetric version}
\be
\label{CAPsuperbms3}
\mathfrak{sbms}_3
=
\text{Vect}(S^1)\inplus
\left(\text{Vect}(S^1)_{\text{Ab}}\oplus\cF_{-1/2}\right),
\ee
where $\text{Vect}(S^1)_{\text{Ab}}\oplus\cF_{-1/2}$ may be seen as an Abelian version of 
$\mathfrak{s}\text{Vect}(S^1)$. Again, choosing periodic/antiperiodic boundary conditions for $\cF_{-1/2}$ 
yields the 
Ra\-mond/\-Ne\-veu-\-Schwarz sector of the theory (respectively). 
Central extensions can be included as in 
(\ref{hibiscus}) and lead to a supersymmetric version of the centrally extended algebra (\ref{haBOP}). The 
elements of the resulting super Lie algebra $\mathfrak{s}\widehat{\mathfrak{bms}}{}_3$ are $5$-tuples 
$(X,\lambda;\alpha,S,\mu)$, where $(X,\alpha,S)$ belongs to $\mathfrak{sbms}_3$ and $\lambda,\mu$ are 
real 
numbers, with a super Lie bracket that extends (\ref{wisig}):
\be
\begin{split}
& \Big[
(X,\lambda;\alpha,S,\mu),(Y,\kappa;\beta,T,\nu)
\Big\}
=\\
& =
\Big(
[X,Y],\sfc(X,Y);
[X,\beta]-[Y,\alpha],X\cdot T-Y\cdot S;
\sfc(X,\beta)-\sfc(Y,\alpha)+\sfh(S,T)
\Big)\,.
\end{split}
\label{CAPsuperbms}
\ee
Here $\sfc$ is again the Gelfand-Fuks cocycle (\ref{gefuks}) while $\sfh$ is given by (\ref{CAPcocycle}). 
Upon introducing generators analogous to (\ref{JimmA}), the central charges (\ref{zaza}) 
and $\cQ_r\equiv(0,0;0,e^{ir\phii}(d\phii)^{-1/2},0)$, one finds the brackets (\ref{bammex}) 
supplemented with
\begin{subequations}
\label{CAPs32.5}
\begin{align}
i[\cJ_m,\cQ_r\} & = \left(\frac{m}{2}-r\right)\cQ_{m+r}\,,\\[1pt]
i[\cP_m,\cQ_r\} & = 0\,,\\[1pt]
[\cQ_r,\cQ_s\} & = \cP_{r+s}+\frac{\cZ_2}{6}\,r^2\delta_{r+s,0}\,.
\end{align}
\end{subequations}
The indices $r$, $s$ are integers/half-integers in the 
Ramond/Neveu-Schwarz sector, respectively. Note that the centrally extended bracket of supercharges only 
involves the 
central charge $\cZ_2$ that pairs superrotations with supertranslations.\\

In the gravitational context, the functions $X$ and $\alpha$ generate superrotations and supertranslations, 
while $S(\phii)$ generates local supersymmetry transformations that become global symmetries upon enforcing 
suitable boundary conditions on the fields. The surface charge associated with $(X,\alpha,S)$ then takes the 
form \cite{Barnich:2014cwa}\i{surface charge!for 3D supergravity}
\be
\label{CAPchargesusy}
Q_{(X,\alpha,S)}[j,p,\psi]
=
\frac{1}{2\pi}\int^{2\pi}_0 d\varphi
\Big[
X(\phii)j(\phii)+\alpha(\phii)p(\phii)+S(\phii)\psi(\phii)
\Big]\,,
\ee
where $j$ and $p$ are the angular momentum and Bondi mass aspects of (\ref{piment}), 
while $\psi(\phii)$ is one of the subleading components of the gravitino\i{gravitino} at null infinity.
The triple $(j,p,\psi)$ is a coadjoint vector for the (centrally extended) super BMS$_3$ group. In particular 
$(j,p)$ are quadratic densities, while $\psi(\phii)$ has 
weight $3/2$ on the circle.
Upon using formula 
(\ref{CAPss25}), the charges (\ref{CAPchargesusy}) satisfy the algebra (\ref{CAPsuperbms}) with definite 
values $\cZ_1=0$, 
$\cZ_2=c_2=3/G$ for the central charges. Note 
that the 
gravitino naturally satisfies Neveu-Schwarz boundary conditions on the celestial circle, as it represents 
Lorentz transformations up to a sign.\\

The construction of the super BMS$_3$ group can be generalized in a straightforward 
way. Indeed, let $G$ be a (bosonic) group, $\mathfrak{g}$ its Lie algebra, 
$\mathfrak{sg}$ a super Lie algebra whose even subalgebra is $\mathfrak{g}$. Then one can associate with $G$ 
a (bosonic) exceptional semi-direct product $G\ltimes\mathfrak{g}$ --- the even $\hBMS$ group 
(\ref{hibiscus}) is of that 
form, with $G$ 
the Virasoro group. Now let $\mathfrak{sg}_{\text{Ab}}$ denote the ``Abelian'' super Lie algebra which is 
isomorphic to 
$\mathfrak{sg}$ as a vector space, but where all brackets involving elements of $\mathfrak{g}$ are set to 
zero. One may then define a super semi-direct product
\be
\big(
G\ltimes\mathfrak{g},
\mathfrak{g}\inplus\mathfrak{sg}_{\text{Ab}}
\big)
\ee
where we use the notation (\ref{CAPsupergroup}). This structure appears to be ubiquitous in 
three-dimensional, asymptotically flat supersymmetric higher-spin theories.

\subsection{\ Supersymmetric BMS$_3$ particles}
\label{SAMBOR}

Unitary representations of the super BMS$_3$ group can be classified along the lines explained in chapter 
\ref{c7}. In the remainder of this section we describe this classification and use it to evaluate characters 
of the centrally extended super BMS$_3$ group. We conclude with the 
observation that these characters reproduce one-loop partition functions of three-dimensional asymptotically 
flat supergravity and hypergravity.

\subsubsection*{Admissible super BMS$_3$ orbits}

The unitary representations of super BMS$_3$ are classified by the same supermomentum orbits as in the 
purely bosonic case, i.e.\ coadjoint orbits of the Virasoro group. However, supermomenta that do not satisfy 
condition (\ref{CAPadmiss}) are forbidden, so 
our 
first task is to understand which 
orbits are admissible. To begin, recall that the admissibility condition (\ref{CAPadmiss}) is invariant under 
superrotations. Thus, if we consider a supermomentum orbit containing a constant $p_0$ say,
the supermomenta on the orbit will be admissible if and only if $p_0$ is. 
Including the central charge $c_2$, we ask: which pairs $(p_0,c_2)$ are such that
\be
\langle (p_0,c_2),\{S,S\}\rangle\geq 0\qquad\text{for any }\,S\in\cF_{-1/2}(S^1)\,?
\ee
Here $\langle\cdot,\cdot\rangle$ is the pairing (\ref{virpar}) of centrally extended supermomenta with 
centrally extended supertranslations. Using the super Lie 
bracket (\ref{CAPsuperbms}), we find
\be
\langle (p_0,c_2),\{S,S\}\rangle
=
\frac{1}{2\pi}\int_0^{2\pi}d\varphi
\left(
p_0(S(\varphi))^2+\frac{c_2}{6}(S'(\varphi))^2
\right).
\label{CAPpss}
\ee
Since the 
term involving $(S')^2$ can be made arbitrarily large while keeping $S^2$ arbitrarily small, a necessary 
condition for $(p_0,c_2)$ to be admissible is that $c_2$ be non-negative. 
Already note that this condition did not arise in the bosonic BMS$_3$ group.
The admissibility condition on 
$p_0$, 
on the other hand, depends on the sector under 
consideration:\i{admissible momentum}\i{supermomentum!admissible}
\begin{itemize}
\item In the Ramond sector, $S(\varphi)$ is a periodic function on the circle. In particular, 
$S(\varphi)=\text{const.}$ is part of the supersymmetry algebra, so for expression (\ref{CAPpss}) to be 
non-negative for any $S$, we must impose $p_0\geq0$.
\item In the Neveu-Schwarz sector, $S(\varphi)$ is antiperiodic (i.e.~$S(\varphi+2\pi)=-S(\varphi)$) and can 
be expanded in Fourier modes as
\be
S(\varphi)=\sum_{n\,\in\,\mathbb{Z}}s_{n+1/2}\,e^{i(n+1/2)\varphi}.
\ee
Then expression (\ref{CAPpss}) becomes
\be
\label{CAPs35}
\langle (p_0,c_2),\{S,S\}\rangle
=
\sum_{n\,\in\,\mathbb{Z}}
\left[
p_0+\frac{c_2}{6}(n+1/2)^2
\right]|s_{n+1/2}|^2,
\ee
and the admissibility condition amounts to requiring all coefficients in this series to be non-negative, 
which gives
\be
p_0\geq-\frac{c_2}{24}\,.
\label{CAPbound}
\ee
\end{itemize}
These bounds are consistent with earlier observations in three-dimensional supergravity 
\cite{Barnich:2014cwa}, according to which Minkowski space-time 
(corresponding to $p_0=-c_2/24$) realizes the Neveu-Schwarz vacuum, while the Ramond vacuum is realized by 
the 
null orbifold (corresponding to $p_0=0$). Analogous results hold in AdS$_3$ \cite{Coussaert:1993jp}. More 
general admissibility conditions can presumably be worked out for \it{non-constant} supermomenta by adapting 
the proof 
of the positive energy theorem of section \ref{senepost}, but we will not address this 
question here.

\subsubsection*{Super BMS$_3$ multiplets}

As explained around (\ref{CAPcliff}), any unitary representation of super BMS$_3$ based on a supermomentum 
orbit $\cO_p$ 
comes 
equipped with a 
representation $\tau$ of the Clifford algebra\i{Clifford algebra}
\be
\label{CAPs36}
\cC_p=T\left(\cF_{-1/2}(S^1)\right)/
\left\{
S^2-\langle(p,c_2),\{S,S\}\rangle
\right\}.
\ee
Let us build such a representation. For definiteness we work in the Neveu-Schwarz sector and take $p$ to 
be a 
constant admissible supermomentum $p_0=M-c_2/24$ with $M>0$, whose little group is $\text{U}(1)$. Then the 
bilinear form (\ref{CAPs35}) is non-degenerate and the representation $\tau$ of the Clifford algebra 
(\ref{CAPs36}) 
must be such that
\be
\label{CAPs35.5}
\tau[\cQ_r]\cdot\tau[\cQ_s]+\tau[\cQ_s]\cdot\tau[\cQ_r]=\left(
\frac{c_2}{6}(r^2-1/4)+M\right)\delta_{r+s,0}\,,
\quad r,s\in\ZZ+1/2.
\ee
In order to make $\tau$ irreducible, we start with a highest-weight state $|0\rangle$ such that 
\mbox{$\tau[\cQ_r]|0\rangle=0$} for $r>0$, and generate the space of the representation by its 
``descendants'' 
$\tau[\cQ_{-r_1}]...\tau[\cQ_{-r_n}]|0\rangle$, with $0<r_1<...<r_n$.\i{supermultiplet}\i{super BMS3 
multiplet@super BMS$_3$ multiplet} It follows from the Lie brackets 
(\ref{CAPs32.5}) 
that each descendant state has spin $s+\sum_{i=1}^nr_i$, where $s$ is the spin of the state $|0\rangle$; this 
observation uniquely determines the little group representation $\cR_0$ satisfying (\ref{CAPcompat}). Thus, a 
super BMS$_3$ particle consists of infinitely many particles with spins increasing from $s$ to infinity.\\

A similar construction can be carried out for the vacuum supermomentum at $M=0$, with the subtlety that the 
Clifford algebra (\ref{CAPs36}) (or equivalently (\ref{CAPs35.5})) is degenerate. As explained below 
(\ref{CAPcliff}), 
one needs to quotient (\ref{CAPs36}) by the radical of the bilinear form (\ref{CAPs35}), resulting in a 
non-degenerate Clifford algebra $\bar\cC_p$. In the case at hand this algebra is generated by supercharges 
$\cQ_r$ with $|r|>1$, and the representation $\tau$ must satisfy (\ref{CAPs35.5}) with $M=0$ and $|r|,|s|>1$. 
The 
remainder of the construction is straightforward: starting from a state $|0\rangle$ with, say, vanishing 
spin, one generates the space of the representation by acting on it with $\tau[\cQ_{-r}]$'s, where $r>1$. The 
vacuum representation of super BMS$_3$ thus contains infinitely many ``spinning vacua'' with increasing spins.

\subsubsection*{Characters}

The Fock space representations just described can be used to evaluate characters. For example, in the massive 
case with spin $s$ one finds the supersymmetric little group character
\be
\label{CAPs35b}
\text{tr}\left[e^{i\theta J_0}\right]
=
e^{is\theta}\left[
1+e^{i\theta/2}+e^{3i\theta/2}+e^{2i\theta}+\cdots
\right]
=
e^{is\theta}\prod_{n=1}^{+\infty}\left(1+e^{i(n-1/2)(\theta+i\epsilon)}\right),
\ee
where we have added a small imaginary part to $\theta$ to ensure convergence of the product; the trace is 
taken in the fermionic Fock space associated with the ``highest-weight state'' $|0\rangle$. The vacuum case 
is similar, except that the product would start at $n=2$ rather than $n=1$ (and $s=0$). Note that 
(\ref{CAPs35b}) explicitly breaks parity invariance; this can be fixed by replacing the parity-breaking Fock 
space representations $\tau$ described above by parity-invariant tensor products $\tau\otimes\bar\tau$, where 
$\bar\tau$ is the same as $\tau$ with the replacement of $\cQ_r$ by $\cQ_{-r}$. The trace of a rotation 
operator 
in the space of $\tau\otimes\bar\tau$ then involves the norm squared of the product appearing in 
(\ref{CAPs35b}).\\

As explained at the beginning of section \ref{CAPsec:3DSugra}, the character of an induced representation of 
a super semi-direct 
product 
takes the same form (\ref{fropo}) as in the bosonic case, but with the character of $\cR$ replaced by that 
of 
a (reducible) representation $\cR_0\otimes\cR$ compatible with the Clifford algebra representation $\tau$. We 
thus find that the character of a rotation by $\theta$ (together with a Euclidean time translation by 
$\beta$), in the parity-invariant vacuum representation of the $\cN=1$, Neveu-Schwarz super BMS$_3$ group, 
reads\i{character!for supersymmetric BMS$_3$}\i{super BMS3 character@super BMS$_3$ character}
\begin{align}
\chi_{\text{vac}}^{\text{super BMS}}[(\text{rot}_{\theta},i\beta)]
& =
\chi_{\text{vac}}^{\text{BMS}}[(\text{rot}_{\theta},i\beta)]\cdot
\prod_{n=2}^{+\infty}|1+e^{i(n-1/2)(\theta+i\epsilon)}|^2\nn\\
& =
e^{\beta c_2/24}
\prod_{n=2}^{+\infty}\frac{|1+e^{i(n-1/2)(\theta+i\epsilon)}|^2}{|1-e^{in(\theta+i\epsilon)}|^2}\,.
\end{align}
Comparing with (\ref{CAPss6.5q}) and (\ref{CAPs9.5b}), we recognize the product of the (suitably regularized) 
partition functions of 
two 
massless fields with spins 2 and 3/2,\i{partition function!for supergravity} that is, the one-loop partition 
function of $\cN=1$ supergravity in 
three-dimensional flat space.

\subsubsection*{Higher-spin supersymmetry and hypergravity}

In \cite{Fuentealba:2015jma,Fuentealba:2015wza}, the authors considered a three-dimensional hypergravity 
theory\i{hypergravity} consisting of a metric coupled to a single field with half-integer spin $s+1/2$, with 
$s$ larger than 
one. Upon imposing suitable asymptotically flat boundary conditions, they found that the asymptotic symmetry 
algebra spans a superalgebra that extends the bosonic $\mathfrak{bms}_3$ algebra by generators $\cQ_r$ of 
spin 
$s+1/2$.  The one-loop partition function of that system is the product of the graviton partition function 
(see eq.~(\ref{CAPss6.5q}) for $s=2$) with the fermionic partition function (\ref{CAPs9.5b}). We now show 
that 
this partition function coincides with the vacuum character of the corresponding asymptotic symmetry group 
(in 
the Neveu-Schwarz sector).\\

The irreducible, unitary representations of the asymptotic symmetry group of \cite{Fuentealba:2015wza} are 
classified by the same orbits and little groups as for the standard BMS$_3$ group. In particular, we can 
consider the orbit of a constant supermomentum $p_0=M-c_2/24$; the associated Clifford algebra representation 
$\tau$ mentioned below (\ref{CAPcliff}) then satisfies a natural generalization of eq.~(\ref{CAPs35.5}) (see 
eq.~(7.23) in \cite{Fuentealba:2015wza}):
\be
\label{CAPQQ}
\tau[\cQ_r]\tau[\cQ_{\ell}]+\tau[\cQ_{\ell}]\tau[\cQ_r]
=
\prod_{j=0}^{s-1}
\left(
\frac{c_2}{6}
\Big(
r^2-\frac{(2j+1)^2}{4}
\Big)
+M
\right)\delta_{r+\ell,0}\,,
\ee
where $r$ and $\ell$ are integers or half-integers, depending on the sector under consideration (Ramond or 
Neveu-Schwarz, respectively). In order for the orbit to be admissible in the sense of (\ref{CAPadmiss}), the 
value of $M$ must be chosen so as to ensure that all coefficients on the right-hand side of (\ref{CAPQQ}) are 
non-negative. In particular, the vacuum value $M=0$ is admissible in the Neveu-Schwarz sector, in which case 
the anticommutators $\{\tau[\cQ_r],\tau[\cQ_{-r}]\}$ vanish for $|r|=1/2,...,s-1/2$. Thus, in the 
Neveu-Schwarz 
vacuum, the Clifford algebra (\ref{CAPQQ}) degenerates and $\tau$ must really be seen as a representation of 
the non-degenerate subalgebra generated by the $\cQ_r$'s with $|r|\geq s$. The corresponding Fock space 
representation can be built as explained below (\ref{CAPs35.5}), and the spins of the basis states in this 
representation are uniquely determined by the fact that the $\cQ_r$'s have spin $s+1/2$. The corresponding 
Fock 
space character is thus
\be
\text{tr}\left[e^{i\theta J_0}\right]
=
\prod_{n=s+1}^{+\infty}
\left(1+e^{i(n-1/2)(\theta+i\epsilon)}\right),
\ee
which generalizes (\ref{CAPs35b}). The character for $\tau\otimes\bar\tau$ is the squared norm of this 
expression, and the resulting vacuum character of the hypersymmetric BMS$_3$ group is\i{character!for 
hypersymmetric BMS$_3$}
\be
\chi_{\text{vac}}^{\text{hyper BMS}}
[(\text{rot}_{\theta},i\beta)]
=
e^{\beta c_2/24}\;
\frac
{\prod\limits_{n=s}^{+\infty}
|1+e^{i(n+1/2)(\theta+i\epsilon)}|^2}
{\prod\limits_{m=2}^{+\infty}
|1-e^{im(\theta+i\epsilon)}|^2}\,.
\ee
As announced earlier, this coincides with the (suitably regularized) one-loop partition function of 
asymptotically flat gravity 
coupled to a massless field with spin $s+1/2$.\i{partition function!for hypergravity} We have thus completed 
our overview of the relation between 
BMS$_3$ characters and one-loop partition functions in three dimensions.

\begin{subappendices}

\begin{advanced}
\section{\ From mixed traces to bosonic characters}
\label{LABEKKO}
\end{advanced}

This section and the next one are technical appendices that describe various computations concerned with 
characters of highest-weight representations of $\text{SO}(n)$. These considerations are useful for sections 
\ref{CAPsubsec2.2} and \ref{CAPsubsec2.3}. Other than that, they may be skipped on a first reading.\\

\subsection{\ Mixed traces and symmetric polynomials}
\label{CAPAppA1}

In this part of the appendix we prove that the mixed trace (\ref{CAPs5.5t}) of $\II_{\mu_s,\alpha_s}$ in 
$D$ 
dimensions coincides with a certain difference of complete homogeneous symmetric polynomials\i{complete 
homogeneous symmetric polynomial} in the traces of 
$J^n$ as given by
\be
\label{CAPs6.5}
\chi_s[n\vec\theta\,]
=
h_s(J^n)-h_{s-2}(J^n)\,,
\ee
where
\be
\label{CAPss6.5}
h_s(J^n)
=
\sum_{\substack{m_1,...,m_s\in\,\NN\\
m_1+2m_2+...+sm_s=\,s}}\!
\left[\,
\prod_{k=1}^s
\frac{\left(
\text{Tr}[(J^n)^k]
\right)^{m_k}}{m_k!k^{m_k}}
\,\right].
\ee
By definition, the \it{complete homogeneous symmetric polynomial} of degree $s$ in $D$ complex 
variables $\lambda_1,...,\lambda_D$ is
\begin{equation}
\label{CAPs43.5}
h_s(\lambda_1,...,\lambda_D)
=
\sum_{\substack{\ell_1,...,\ell_D=\,0\\
\ell_1+...+\ell_D=\,s}}^s
\lambda_1^{\ell_1}\lambda_2^{\ell_2}...\lambda_D^{\ell_D}
=
\sum_{1\leq\ell_1\leq\ell_2\leq...\leq\ell_s\leq D}
\lambda_{\ell_1}\lambda_{\ell_2}...\lambda_{\ell_s}.
\end{equation}
Using the variant of Newton's identities\i{Newton identities}
\be
h_s(\lambda_1,...,\lambda_D)
=
\frac{1}{s}\sum_{N=1}^sh_{s-N}(\lambda_1,...,\lambda_D)(\lambda_1^N+...+\lambda_D^N)\,,
\ee
one can show by recursion (see e.g. \cite[p.\ 24f]{Macdonald1995}) that the polynomial (\ref{CAPs43.5}) can 
equivalently be written as in (\ref{CAPss6.5}):
\be
\label{CAPss43.5}
h_s(\lambda_1,...,\lambda_D)
=
\sum_{\substack{m_1,...,m_s\in\,\NN \\ m_1+2m_2+...+sm_s=\,s}}
\prod_{k=1}^s\frac{(\lambda_1^k+...+\lambda_D^k)^{m_k}}{m_k!\,k^{m_k}}\,.
\ee
We shall use this relation later. To prove (\ref{CAPs6.5}), we start with the 
following:

\paragraph{Lemma.} Let $J$ be a complex $D\times D$ matrix with eigenvalues $\lambda_1,...,\lambda_D$. 
Then,
\be
\label{CAPeq:ProofPart1}
\left(\delta^{\mu\alpha}\right)^s\frac{1}{s!}\left(J_{\mu\alpha}\right)^s
=
h_s(\lambda_1,\lambda_2,...,\lambda_D)\,,
\ee
where we use the same notation for contracting symmetrized indices as in (\ref{CAP2.9}).

\paragraph{Proof.} The left-hand side of (\ref{CAPeq:ProofPart1}) can be seen as a trace over symmetric 
tensor 
powers of $J$. 
Indeed, $\delta^{\mu\alpha}J_{\mu\alpha}=\text{Tr}(J)$ is clear; as 
for $\frac{1}{2}\left(\delta^{\mu\alpha}\right)^2\left(J_{\mu\alpha}\right)^2$, one gets
\begin{equation}
\frac{1}{2}\left(\delta^{\mu\alpha}\right)^2\left(J_{\mu\alpha}\right)^2
=
\frac{1}{2}\left(
\textnormal{Tr}\!
\left(J\right)^2+\textnormal{Tr}\!\left(J^2\right)
\right)
=
\textnormal{Tr}\!
\left(
S^2\!\left(J\right)
\right)
=
\frac{1}{2}\sum_{i=1}^2
\textnormal{Tr}\!\left(J^i\right)\textnormal{Tr}\!
\left(S^{2-i}\!\left(J\right)\right),
\end{equation}
where $S^k(J)$ is the $k^{\textnormal{th}}$ symmetric tensor power of $J$. One then defines 
recursively
\begin{equation}
\frac{1}{s!}\left(\delta^{\mu\alpha}\right)^s\left(J_{\mu\alpha}\right)^s=\textnormal{Tr}
\left(S^s\!\left(J\right)\right)=\frac{1}{s}\sum_{i=1}^s\textnormal{Tr}\left(J^i\right)\textnormal{Tr}\left(S^
{
s-i}\!\left(J\right)\right),
\end{equation}
so that $\frac{1}{s!}\left(\delta^{\mu\alpha}\right)^s\left(J_{\mu\alpha}\right)^s$ is just a 
trace in the $s^\textnormal{th}$ symmetric tensor power of the $D$-dimensional vector space $V$ on which 
$J_{\mu\alpha}$ acts as a linear operator. Now consider an eigenbasis $\{e_1,...,e_D\}$ for 
$J_{\mu\alpha}$, with $J\cdot e_k=\lambda_ke_k$. Since $\frac{1}{s!}\left(J_{\mu\alpha}\right)^s$ is the 
$s^\textnormal{th}$ symmetric tensor power of $J_{\mu\alpha}$ one can construct an eigenbasis for 
$\frac{1}{s!}\left(J_{\mu\alpha}\right)^s$ by symmetrizing $e_{k_1}\otimes e_{k_2}\otimes...\otimes 
e_{k_D}$, with $k_1\leq k_2\leq...\leq k_D$. These eigenvectors have eigenvalues 
$\lambda_{l_1}\lambda_{l_2}...\lambda_{l_D}$, and 
since $\left(\delta^{\mu\alpha}\right)^s\frac{1}{s!}\left(J_{\mu\alpha}\right)^s$ 
is the trace of $\frac{1}{s!}\left(J_{\mu\alpha}\right)^s$, relation (\ref{CAPeq:ProofPart1}) follows upon 
using 
the second expression of $h_s(\lambda_1,...,\lambda_D)$ in (\ref{CAPs43.5}).\hfill$\blacksquare$\\

We can now turn to the proof of (\ref{CAPs6.5}). To this end we fix conventionally the number of terms 
entering the contraction of two symmetrized expressions as follows. Objects with \it{lower} indices are 
symmetrized with the minimum number of terms required and 
without
overall normalization factor, while objects with \it{upper} indices are not symmetrized at all, since the 
symmetrization is induced by the contraction. This specification is needed because 
terms with lower and upper indices in a contraction may have a different index structure and therefore the 
number of terms needed for their symmetrization may be different. For instance
\be
\begin{split}
& A^{\mu} B^{\mu} C^{\mu}D_{\mu\mu} E_{\mu}\equiv A^{\mu}B^{\nu}C^{\rho}
\left(
D_{\mu\nu} E_{\rho}+ D_{\nu\rho} E_{\mu}+ D_{\rho\mu} E_{\nu} 
\right) \\
& = \frac{1}{2}
\left(
A^{\mu}B^{\nu}C^{\rho}+ A^{\nu}B^{\rho}C^{\mu}+ A^{\rho}B^{\mu}C^{\nu}+A^{\mu}B^{\rho}
C^{\nu}+ A^{\rho}B^{\nu}C^{\mu}+ A^{\nu}B^{\mu}C^{\rho}
\right) D_{\mu\nu} E_{\rho}\, .  
\end{split}
\ee

In order to simplify computations, we define
\be
\label{CAPeq:TDefinition}
T_{\mu_s,\,\alpha_s}\equiv J_{\mu\alpha}... J_{\mu\alpha}\,,\qquad 
T^{[s]}\equiv 
T_{\mu_s,\,\alpha_s}\left(\delta^{\mu\alpha}\right)^s,
\ee
which implies the contraction rules
\be
\label{CAPeq:TContractionRules}
\delta^{\mu\mu}T_{\mu_s,\,\alpha_s}=2\,\delta_{\alpha\alpha}T_{\mu_{s-2},\,\alpha_{s-2}}\,,
\quad
\delta^{\alpha\alpha}T_{\mu_s,\alpha_s}=2\,\delta_{\mu\mu}T_{\mu_{s-2},\alpha_{s-2}}\,.
\ee
In terms of the tensors $T_{\mu_s,\alpha_s}$, the mixed trace (\ref{CAPs5.5t}) can be written as 
\begin{align}
& \chi_s[n\vec\theta]
=
\frac{1}{s!}\,T_{\mu_s,\beta_s}\!
\bigg[\!
\left(\delta^{\mu\beta}\right)^s
\!+\sum_{m=1}^{\lfloor\frac{s}{2}\rfloor}
\frac{\left(-1\right)^ms!\left[D+2\left(s-m-2\right)\right]!!}{2^{m}m!\left(s-2m\right)!
\left[
D+2\left(s-2\right)
\right]!!}\times\\
&
\qquad\qquad\qquad\qquad
\times
(\delta^{\mu\mu})^m
(\delta^{\mu\beta})^{s-2m}
(\delta^{\beta\beta})^m
\bigg]\nn\\
& \!\refeq{CAPeq:TContractionRules}\!
\frac{1}{s!}\,T^{[s]}+\sum_{m=1}^{\left[\frac{s}{2}\right]}\frac{\left(-1\right)^m\left[
D+2\left(s-m-2\right)\right]!!}{2^{m-1}m!\left(s-2m\right)!\left[D+2\left(s-2\right)\right]!!}\times\\
&
\qquad\qquad\qquad\qquad
\times
(\delta^{
\mu\mu})^m(\delta^{\mu\beta})^{s-2m}(\delta^{\beta\beta})^{m-1}\delta_{\mu\mu}T_{
\mu_{s-2},\beta_{s-2}}\,.
\end{align}
To compute the trace of 
the  
$(\delta^{\mu\mu})^m(\delta^{\mu\beta})^{s-2m}(\delta^{\beta\beta})^{m-1}$ 
terms, we first change our symmetrization from $\delta_{\mu\mu}T_{\mu_{s-2},\beta_{s-2}}$ (which 
contains $\frac{s!}{2(s-2)!}$ terms) to the aforementioned product of $\delta$'s. In doing so one has to 
introduce a factor accounting for the number of terms in each structure as
\begin{subequations}
\begin{align}
\delta_{\mu\mu}T_{\mu_{s-2},\,\beta_{s-2}}&
\leadsto\frac{s!}{2(s-2)!}\ \text{terms},\\
\left(\delta^{\mu\mu}\right)^{mu}
\left(\delta^{\mu\beta}\right)^{s-2m}
\left(\delta^{\beta\beta}\right)^{m-1}
&\leadsto\frac{s!}{2^mm!}\times\frac{(s-2)!}{2^{m-1}(m-1)!(s-2m)!}\ \text{terms},
\end{align}
\end{subequations}
which implies
\be
\begin{split}
\chi_s[n\vec\theta]
=
\frac{1}{s!}T^{[s]}+
\sum_{m=1}^{\lfloor\frac{s}{2}\rfloor}
\frac {
\left(-1\right)^m2^{m-1}(m-1)!\left[D+2\left(s-m-2\right)\right]!!}{\left[(s-2)!\right]^2\left[
D+2\left(s-2\right)\right]!!}\times\\
\qquad\qquad\qquad\times\delta_
{\mu\mu}^m\delta_{\mu\beta}^{s-2m}\delta_{\beta\beta}^{m-1}\delta^{\mu\mu}T^{\mu_{s-2},\beta_{s-2}}.
\end{split}
\ee
Taking into account the correct combinatorial factors one obtains
\be
\delta_{\mu\mu}^m\delta_{\mu\beta}^{s-2m}\delta_{\beta\beta}^{m-1}\delta^{\mu\mu}=\left[D+2(s-m-1)\right]
\delta_{\mu\mu}^{m-1}\delta_{\mu\beta}^{s-2m}\delta_{\beta\beta}^{m-1}+2m\,\delta_{\mu\mu}^m\delta_{\mu\beta}^
{
s-2m-2}\delta_{\beta\beta}^{m}\,,
\ee
which then yields
\begin{align}
&\chi_s[n\vec\theta]=\frac{1}{s!}\,T^{[s]}+
\left(\sum_{m=1}^{\lfloor\frac{s}{2}\rfloor}
\frac{\left(-1\right)^m2^{m-1}(m-1)!\left[D+2\left(s-m-1\right)\right]!!}{\left[D+2\left(s-2\right)\right]!!}
\delta_{\mu\mu}^{m-1}\delta_{\mu\beta}^{s-2m}\delta_{\beta\beta}^{m-1}\right.\nn\\
\label{CAPs47}
&\left. + \sum_{m=1}^{\lfloor\frac{s}{2}\rfloor-1}
\frac{\left(-1\right)^m2^{m}m!\left[D+2\left(s-m-2\right)\right]!!}
{\left[D+2\left(s-2\right)\right]!!}\delta_{\mu\mu}^m\delta_{\mu\beta}^{s-2m-2}\delta_{\beta\beta}^{m}
\right)\!\frac{1}{\left[(s-2)!\right]^2}\,T^{\mu_{s-2},\beta_{s-2}}.
\end{align}
Shifting $m\rightarrow m+1$ in the upper sum one can see that both sums are identical apart from the overall 
sign and the lower extremum. Thus (\ref{CAPs47}) boils down to
\begin{equation}
\chi_s[n\vec\theta]
=
\frac{1}{s!}\,T^{[s]}-\frac{1}{\left[(s-2)!\right]^2}\,\delta_{\mu\beta}^{s-2}T^{\mu_{s-2},\beta_{s-2}}
=
\frac{1}{s!}\,T^{[s]}-\frac{1}{(s-2)!}\,T^{[s-2]}\,.
\end{equation}
Now using (\ref{CAPeq:TDefinition}) and (\ref{CAPeq:ProofPart1}) one obtains
\begin{equation}
\label{CAPss47}
\chi_s[n\vec\theta]
=
\frac{1}{s!}\,T^{[s]}-\frac{1}{(s-2)!}\,T^{[s-2]}
=
h_s(\lambda_1, \lambda_2,...,\lambda_D)-h_{s-2}(\lambda_1, \lambda_2,...,\lambda_D)\,,
\end{equation}
where $\lambda_1,...,\lambda_D$ are the eigenvalues of $J^n$. (These eigenvalues are $e^{\pm in\theta_j}$ 
for 
$j=1,...,r$, and one or two unit eigenvalues depending on whether $D$ is odd or even, respectively.) This 
leads to the desired result: since traces of powers of $J^n$ can be written as
\be
\text{Tr}[(J^n)^k]
=
\lambda_1^k+...+\lambda_D^k
\ee
in terms of the eigenvalues of $J^n$, the complete homogeneous symmetric polynomials expressed as 
(\ref{CAPss43.5}) exactly coincide with the combination (\ref{CAPss6.5}), and equation (\ref{CAPss47}) 
coincides with 
(\ref{CAPs6.5}).

\subsection{\ Symmetric polynomials and $\text{SO}(D)$ characters}
\label{CAPAppA2}

In this part of the appendix we review the relation between complete homogeneous symmetric polynomials and 
characters of orthogonal groups. Most of the explicit proofs can be found in \cite{fulton1991}, 
chapter 24, to which we refer for details on our arguments below. We study separately the cases 
of odd and even $D$ and let $r\equiv\lfloor(D-1)/2\rfloor$, with $\theta_1,...,\theta_r$ the 
non-vanishing angles appearing in the rotations (\ref{CAP2.7}).

\subsubsection*{Odd $D$}

We consider the Lie algebra $\mathfrak{so}(D)=\mathfrak{so}(2r+1)$, with rank $r$. Choosing a basis of 
$\CC^{2r+1}$ such that the Lie algebra $\mathfrak{so}(2r+1)_{\CC}$ can be written in terms of complex 
matrices, we may choose the Cartan subalgebra\i{son@$\mathfrak{so}(n)$!Cartan 
subalgebra}\i{Cartan subalgebra} to be the subalgebra 
$\mathfrak{h}$ of $\mathfrak{so}(2r+1)_{\CC}$ consisting of diagonal matrices. As a basis of $\mathfrak{h}$ 
we choose the matrices $H_i$ whose entries all vanish, except the $(i,i)$ and $(r+i,r+i)$ entries which are 
$1$ and $-1$, respectively (with $i=1,...,r$). In our convention (\ref{CAP2.6}), the operator $H_i$ 
generates 
rotations in the plane $(x_i,y_i)$. Then, calling $L_i$ the elements of the dual basis (such that $\langle 
L_i,H_j\rangle=\delta_{ij}$), a dominant weight is one of the form 
$\lambda=\lambda_1L_1+...\lambda_rL_r\equiv(\lambda_1,...,\lambda_r)$ with 
$\lambda_1\geq...\geq\lambda_r\geq0$.\\

Let $\lambda$ be a dominant weight for $\mathfrak{so}(2r+1)$. According to formula (24.28) in 
\cite{fulton1991}, the character of the irreducible representation of $\mathfrak{so}(2r+1)$ with 
highest weight $\lambda$ is\i{son@$\mathfrak{so}(n)$!character}\i{character!for SOn@for 
$\text{SO}(n)$}
\begin{equation}
\label{CAPeq:SO2n+1CharacterFormula}
\chi^{(2r+1)}_\lambda[q_1,...,q_r]
=
\text{Tr}_{\lambda}\left[q_1^{H_1}\cdots q_r^{H_r}\right]
=
\frac
{\left|
q_j^{\lambda_i+r-i+\frac{1}{2}}-q_j^{-\left(\lambda_i+r-i+\frac{1}{2}\right)}
\right|}
{\left|
q_j^{r-i+\frac{1}{2}}-q_j^{-\left(r-i+\frac{1}{2}\right)}
\right|}\,,
\end{equation}
where $q_1,\cdots q_r$ are arbitrary complex numbers\footnote{Eventually these numbers will be exponentials 
of 
angular potentials, so they are fugacities associated with the rotation generators 
$H_i$.}, $\text{Tr}_{\lambda}$ denotes a trace taken in the space of the representation, and 
$\left|A_{ij}\right|$ denotes the determinant of the matrix $A$ with rows $i$ and columns $j$. This 
expression 
is a corollary of the Weyl character formula. Using proposition A.60 and Corollary A.46 of  
\cite{fulton1991}, it can be rewritten as
\begin{equation}
\label{CAPeq:PropA60+CorA46}
\chi^{(2r+1)}_\lambda[q_1,...,q_r]
=
\left|h_{\lambda_i-i+j}-h_{\lambda_i-i-j}\right|,
\end{equation}
where $h_j=h_j\left(q_1,...,q_n,q_1^{-1},...,q_n^{-1},1\right)$ is a complete homogeneous symmetric 
polynomial of degree $j$ in $2r+1$ variables. In particular, for a highest weight 
$\lambda_s=(s,0,...,0)$ (where $s$ is a non-negative integer), the matrix appearing on the right-hand 
side of (\ref{CAPeq:PropA60+CorA46}) is upper triangular, with the entry at $i=j=1$ given by $h_s-h_{s-2}$ 
and 
all other entries on the main diagonal equal to one. Accordingly, the determinant in 
(\ref{CAPeq:PropA60+CorA46}) boils down to $h_s-h_{s-2}$ in that simple case. For the rotation (\ref{CAP2.7}) 
we 
may identify $q_j=e^{in\theta_j}$, and we conclude that
\be
\label{CAPeq:SO2n+1CharacterProofFinish}
\chi^{(2r+1)}_{\lambda_s}[n\vec\theta]
=
\frac{\left|
\sin\left[
\left(\lambda_i+r-i+\frac{1}{2}\right)n\theta_j\right]\right|}
{\left|\sin\left[\left(r-i+\frac{1}{2}\right)n\theta_j\right]\right|}
=
h_s(J^n)-h_{s-2}(J^n)\,,
\end{equation}
where $\lambda_i=s\,\delta_{i1}$. Thus for odd $D$ the difference of symmetric polynomials in (\ref{CAPs6.5}) 
is just a 
character of $\text{SO}(D)$.

\subsubsection*{Even $D$}

We now turn to the Lie algebra $\mathfrak{so}(2r+2)$, with rank $r+1$. As in the odd case we choose a basis 
of 
$\mathbb{C}^{2r+2}$ such that we can write the Lie algebra $\mathfrak{so}(2r+2)$ in terms of complex matrices 
and the Cartan subalgebra\i{son@$\mathfrak{so}(n)$!Cartan 
subalgebra}\i{Cartan subalgebra} is generated by $r+1$ diagonal matrices $H_i$ whose entries all vanish, 
except 
$(H_i)_{ii}=1$ and $(H_i)_{r+1+i,r+1+i}=-1$. We call $L_i$ the elements of the dual basis, and with these 
conventions a weight $\lambda=\lambda_1L_1+...+\lambda_{r+1}L_{r+1}\equiv(\lambda_1,...,\lambda_{r+1})$ 
is 
dominant if $\lambda_1\geq\lambda_2\geq...\geq\lambda_r\geq|\lambda_{r+1}|$.\\

Let $\lambda$ be a dominant weight for $\mathfrak{so}(2r+2)$. Then formula (24.40) in 
\cite{fulton1991} gives the character of the associated highest-weight representation 
as
\be 
\begin{split}
& \chi^{(2r+2)}_\lambda[q_1,...,q_{r+1}]
 =
\text{Tr}_{\lambda}\left[q_1^{H_1}\cdots q_{r+1}^{H_{r+1}}\right]\\
\label{JESUS}
& =
\frac
{\left|
q_j^{\lambda_i+r+1-i}+q_j^{-\left(\lambda_i+r+1-i\right)}
\right|
+
\left|
q_j^{\lambda_i+r+1-i}-q_j^{-\left(\lambda_i+r+1-i\right)}
\right|}
{\left|
q_j^{r+1-i}+q_j^{-\left(r+1-i\right)}
\right|}\,,\quad\quad\quad\quad
\end{split}
\ee
where we use the same notations as in (\ref{CAPeq:SO2n+1CharacterFormula}), except that now 
$i,j=1,...,r+1$. 
Note that the second term in the numerator of this expression vanishes whenever $\lambda_{r+1}=0$ (because 
the $(r+1)^{\text{th}}$ row of the matrix $q_j^{\lambda_i+r+1-i}-q_j^{-\left(\lambda_i+r+1-i\right)}$ 
vanishes). Since this is the case that we will be interested in, we may safely forget about that second term 
from now on. Alternatively, for the mixed traces (\ref{CAPs5.5t}) that we need, we may take 
$q_j=e^{in\theta_j}$ 
for $j=1,...,r$ and $q_{r+1}=1$ without loss of generality, so that this second term vanishes again. 
Using proposition A.64 of \cite{fulton1991}, one can then rewrite 
(\ref{JESUS}) as
\begin{equation}
\label{CAPeq:PropA64}
\chi^{(2r+2)}_\lambda[q_1,...,q_r,1]
=
\left|h_{\lambda_i-i+j}-h_{\lambda_i-i-j}\right|,
\end{equation}
where $h_j=h_j\left(q_1,...,q_r,1,q_1^{-1},...,q_r^{-1},1\right)$. Finally, using the same arguments as 
for 
odd $D$, one easily verifies that the determinant on the right-hand side of (\ref{CAPeq:PropA64}) reduces 
once 
more to $h_s-h_{s-2}$ for a highest weight $\lambda_s=(s,0,...,0)$. Writing again $q_j=e^{in\theta_j}$, 
one concludes that, for even $D$,
\begin{equation}
\label{CAPeq:SO2nCharacterProofFinish}
\chi^{(2r+2)}_{\lambda_s}
\left[n\theta_1,...,n\theta_r,n\theta_{r+1}=0\right]
=
\frac
{\left|
\cos\left[\left(\lambda_i+r+1-i\right)n\theta_j\right]
\right|}
{\left|
\cos\left[\left(r+1-i\right)n\theta_j\right]
\right|}
\Bigg|_{\theta_{r+1}=0}
\!\!\!\!\!\!\!\!\!=
h_s(J^n)-h_{s-2}(J^n),
\end{equation}
where $\lambda_i=s\,\delta_{i1}$. This concludes the proof of (\ref{CAPs6b}). Note that, for \it{
non-vanishing} 
$\theta_{r+1}$, the quotient of denominators in the middle of (\ref{CAPeq:SO2nCharacterProofFinish}) is 
actually 
the character $\chi^{(2r+2)}_{\lambda_s}\left(n\theta_1,...,n\theta_r,n\theta_{r+1}\right)$. This 
detail will be useful in appendix \ref{CAPAppA3}.

\subsection{\ Differences of $\text{SO}(D)$ characters}
\label{CAPAppA3}

In this part of the appendix we prove the following relations between characters of orthogonal groups:
\begin{subequations}
\label{CAPapp:SO(D)DifferenceProofRelations}
\begin{align}
\chi^{(2r+1)}_{\lambda_s}[\vec\theta\,]-\chi^{(2r+1)}_{\lambda_{s-1}}[\vec\theta\,]
=
&\ \chi^{(2r)}_{\lambda_s}[\vec\theta\,]\,,
\label{CAPapp:SO(D)DifferenceProofRelations1}\\
\chi^{(2r)}_{\lambda_s}[\vec\theta\,]-\chi^{(2r)}_{\lambda_{s-1}}[\vec\theta\,]
=
&\ \sum_{k=1}^r\mathcal{A}^r_k[\vec\theta\,]
\chi^{(2r-1)}_{\lambda_{s}}[\theta_1,...,\widehat{\theta_k},...,\theta_r]\,.
\label{CAPapp:SO(D)DifferenceProofRelations2}
\end{align}
\end{subequations}
Here $\vec\theta=(\theta_1,..,\theta_r)$, $\lambda_s$ is the weight with components $(s,0,...,0)$ in the 
basis defined above equations (\ref{CAPeq:SO2n+1CharacterFormula}) and (\ref{JESUS}), 
and the 
hat denotes omission of an argument, while the 
coefficients $\cA^r_k$ are the quotients of determinants defined in (\ref{CAPs6q}). Note that, when one of 
the 
angles $\theta_1,...,\theta_r$ vanishes, say $\theta_{\ell}=0$, then $\cA^r_k=\delta_{k\ell}$ and relation 
(\ref{CAPapp:SO(D)DifferenceProofRelations2}) reduces to
\be
\label{CAPs37.5}
\left.\chi^{(2r)}_{\lambda_s}[\vec\theta\,]\right|_{\theta_{\ell}=0}
-
\left.\chi^{(2r)}_{\lambda_{s-1}}[\vec\theta\,]\right|_{\theta_{\ell}=0}
=
\chi^{(2r-1)}_{\lambda_s}[\theta_1,...,\widehat{\theta_{\ell}},...,\theta_r]\,.
\ee

\paragraph{Proof of (\ref{CAPapp:SO(D)DifferenceProofRelations}).} We start by defining the matrices
\be
\label{CAPA.2a}
(A^r)_{ij}=\sin\left[(r-i+\tfrac{1}{2})\theta_j\right],\qquad 
(B^r)_{ij}=\cos\left[(r-i)\theta_j\right],
\ee
so that in particular
\be
\label{CAPs38}
\cA^r_k(\vec\theta)
=
\frac{|B^r|_{\theta_k=0}}{|B^r|}.
\ee
We shall also use the shorthand notation
\be
\label{CAPnotation}
M^{r}[\theta_k]\equiv|M_{ij}(\theta_1,...,\theta_{k-1},\theta_{k+1},...,\theta_{r+1})|
\ee
to denote the determinant of the $r\times r$ matrix missing the angle $\theta_k$ of any of 
the matrices defined in (\ref{CAPA.2a}). As a preliminary step towards the proof, we list the 
four following identities:
\begin{subequations}
\label{CAPapp:SO(D)DifferenceProofRelationsToProveBosonic}
\begin{align}
\frac{|A^r|}{\prod_{j=1}^r\sin\left(\theta_j/2\right)}
=&\,2^{r-1}|B^r|\,,
\label{CAPapp:SO(D)DifferenceProofRelationsToProveBosonic1}\\[3pt]
|\cos\left[(r-i)\theta_j\right]|
=&\,2^\frac{(r-1)(r-2)}{2}
\prod_{1\leq i<j\leq r}(\cos(\theta_i)-\cos(\theta_j))\,,
\label{CAPapp:SO(D)DifferenceProofRelationsToProveBosonic2}\\
\frac{|B^r|_{\theta_k=0}}{A^{r-1}[\theta_k]}
=&\,2^{r-1}(-1)^{k+1}\prod_{\substack{j=1\\j\neq k}}^r\sin\left(\theta_j/2\right),
\label{CAPapp:SO(D)DifferenceProofRelationsToProveBosonic3}\\
|B^r|
=&\,\sum_{k=1}^r|B^r|_{\theta_k=0}\,.
\label{CAPapp:SO(D)DifferenceProofRelationsToProveBosonic4}
\end{align}
\end{subequations}
Here (\ref{CAPapp:SO(D)DifferenceProofRelationsToProveBosonic1}) can be proven by induction on $r$ upon 
expanding the determinant $|A^r(\vec\theta)|$ along the first line of the matrix $A^r$. Property 
(\ref{CAPapp:SO(D)DifferenceProofRelationsToProveBosonic2}) can be shown by observing that
\begin{equation}
\cos[(r-i)\theta_j]
=
2^{r-i-1}\cos^{r-i}(\theta_j)+\sum_{k=1}^{r-i-1}c_k\cos(k\theta_j)
\end{equation}
with some irrelevant real coefficients $c_k$, and that the contribution of the second term of this expression 
to the determinant $|\cos[(r-i)\theta_j]|$ vanishes by linear dependence. Equation 
(\ref{CAPapp:SO(D)DifferenceProofRelationsToProveBosonic3}) then follows from 
(\ref{CAPapp:SO(D)DifferenceProofRelationsToProveBosonic1}) and 
(\ref{CAPapp:SO(D)DifferenceProofRelationsToProveBosonic2}), while property 
(\ref{CAPapp:SO(D)DifferenceProofRelationsToProveBosonic4}) can again be proved by induction on $r$.\\

Thanks to eqs.~(\ref{CAPapp:SO(D)DifferenceProofRelationsToProveBosonic}), we can tackle the proof of 
(\ref{CAPapp:SO(D)DifferenceProofRelations}). Equation (\ref{CAPapp:SO(D)DifferenceProofRelations1}) is easy: 
using 
expression (\ref{CAPeq:SO2n+1CharacterProofFinish}) for the character $\chi_{\lambda_s}^{(2r+1)}$, 
we 
can write the difference of characters on the left-hand side of (\ref{CAPapp:SO(D)DifferenceProofRelations1}) 
as
\be
\chi^{(2r+1)}_{\lambda_s}-\chi^{(2r+1)}_{\lambda_{s-1}}
=
\frac
{\sum\limits_{k=1}^r(-1)^{k+1}
2\cos[(s+r-1)\theta_k]\sin\left(\theta_k/2\right)A^{r-1}[\theta_k]}
{|A^r|}\,.
\ee
Property (\ref{CAPapp:SO(D)DifferenceProofRelationsToProveBosonic1}) then allows us to reduce this expression 
to 
the quotient of denominators appearing in the middle of eq.~(\ref{CAPeq:SO2nCharacterProofFinish}) (with the 
replacement of $r+1$ by $r$ and all angles non-zero), which is indeed the sought-for character 
$\chi_{\lambda_s}^{(2r)}[\vec\theta\,]$.\\

Equation (\ref{CAPapp:SO(D)DifferenceProofRelations2}) requires more work. Using once more the expression in 
the 
middle of (\ref{CAPeq:SO2nCharacterProofFinish}), we first rewrite the left-hand side of 
(\ref{CAPapp:SO(D)DifferenceProofRelations2}) as
\begin{equation}
\label{CAPapp:SO(D)DifferenceProofRelationsOddDRHS}
\chi^{(2r)}_{\lambda_s}-\chi^{(2r)}_{\lambda_{s-1}}
=
\frac
{\sum\limits_{k=1}^r(-1)^{k+1}
(-2\sin[(s+r-\tfrac{3}{2})\theta_k]
\sin\left(\theta_k/2\right)B^{r-1}[\theta_k]}
{|B^r|}\,.
\end{equation}
Let us now recover this expression as a combination of characters of $\text{SO}(2r-1)$: using formula 
(\ref{CAPeq:SO2n+1CharacterProofFinish}) and the identities 
(\ref{CAPapp:SO(D)DifferenceProofRelationsToProveBosonic}), one finds
\begin{eqnarray}
&   & \sum_{k=1}^r\chi^{(2r-1)}_{\lambda_s}[\theta_1,...,\widehat{\theta_k},...,\theta_r]
|B^r|_{\theta_k=0}\nn\\
& \stackrel{\text{(\ref{CAPapp:SO(D)DifferenceProofRelationsToProveBosonic3})}}{=} &
\sum_{k=1}^r(-1)^{k+1}2^{r-1}
\prod_{\substack{j=1\\j\neq k}}^r
\sin\left(\theta_j/2\right)
\times\nn\\
&   &
\times\bigg[
\sum_{j=1}^{k-1}(-1)^{j+1}\sin[(s+r-\tfrac{3}{2})\theta_j]A^{r-2}[\theta_j,\theta_k]\nn\\
&   &
+\sum_{j=k+1}^{r}(-1)^{j}\sin[(s+r-\tfrac{3}{2})\theta_j]A^{r-2}[\theta_j,\theta_k]\bigg] \nn \\
& \stackrel{\text{(\ref{CAPapp:SO(D)DifferenceProofRelationsToProveBosonic1})}}{=} &
\sum_{k=1}^r(-1)^{k+1}2^{2r-4}\sin[(s+r-\tfrac{3}{2})\theta_k]
\sin\left(\theta_k/2\right)\times\nn\\
&   &
\times\bigg[
\sum_{j=1}^{k-1}(-1)^{j}B^{r-2}[\theta_j,\theta_k]
\prod_{\substack{i=1\\i\notin\{j,k\}}}^r\sin^2\left(\theta_i/2\right)\nn\\
&   &
+\sum_{j=k+1}^r(-1)^{j+1}B^{r-2}[\theta_j,\theta_k]\prod_{\substack{i=1\\i\notin\{j,k\}}}^r
\sin^2\left(\theta_i/2\right)
\bigg]\nn\\
& \stackrel{\text{(\ref{CAPapp:SO(D)DifferenceProofRelationsToProveBosonic2})}}{=} &
\sum_{k=1}^r(-1)^{k+1}(-2)\sin[(s+r-\tfrac{3}{2})\theta_k]
\sin\left(\theta_k/2\right)\nn\\
&   &
\left[
\sum_{j=1}^{k-1}(-1)^{j}
\left.B^{r-1}[\theta_k]\right|_{\theta_j=0}
+\sum_{j=k+1}^r\left.B^{r-1}[\theta_k]\right|_{\theta_j=0}
\right]\nn\\
& \stackrel{\text{(\ref{CAPapp:SO(D)DifferenceProofRelationsToProveBosonic4})}}{=} &
\sum_{k=1}^r(-1)^{k+1}(-2)\sin[(s+r-\tfrac{3}{2})\theta_k]
\sin\left(\theta_k/2\right)
B^{r-1}[\theta_k].
\end{eqnarray}
This coincides with the numerator of the right-hand side of 
(\ref{CAPapp:SO(D)DifferenceProofRelationsOddDRHS}), 
so identity (\ref{CAPapp:SO(D)DifferenceProofRelations2}) follows with $\cA^r_k$ given by 
(\ref{CAPs38}).\hfill$\blacksquare$

\subsection{\ From $\text{SO}(D)$ to $\text{SO}(D-1)$}
\label{CAPAppA4}

In this appendix we prove relation (\ref{CAPt6q}) between characters of $\text{SO}(D)$ and $\text{SO}(D-1)$:

\paragraph{Lemma.} One has the following relations:
\begin{subequations}\label{CAPapp:SO(D)SO(D-1)ProofRelations}
\begin{align}
\chi^{(2r+1)}_{\lambda_s}[\theta_1,...,\theta_r]
=&
\sum_{j=0}^s\chi^{(2r)}_{\lambda_j}(\theta_1,...,\theta_r),
\label{CAPapp:SO(D)SO(D-1)ProofRelations1}
\\
\chi^{(2r)}_{\lambda_s}[\theta_1,...,\theta_r]
=&
\sum_{j=0}^s\sum_{k=1}^r
\mathcal{A}^r_k(\vec\theta)
\chi^{(2r-1)}_{\lambda_j}[\theta_1,...,\widehat{\theta_k},...,\theta_r].
\label{CAPapp:SO(D)SO(D-1)ProofRelations2}
\end{align}
\end{subequations}
Here $\lambda_j$ is the weight $(j,0,...,0)$ as explained above (\ref{CAPs6b}) or below 
(\ref{CAPeq:PropA60+CorA46}), and $\cA^r_k(\vec\theta)$ is the quotient (\ref{CAPs6q}) or (\ref{CAPs38}). 
Since the 
proofs of these two identities are very similar, we will only display the proof of 
(\ref{CAPapp:SO(D)SO(D-1)ProofRelations1}).

\paragraph{Proof of (\ref{CAPapp:SO(D)SO(D-1)ProofRelations1}).} 
Eq.~(\ref{CAPapp:SO(D)SO(D-1)ProofRelations1}) can be written as
\be
\frac{\sum\limits_{k=1}^r(-1)^{k+1}\sin[(s+r-\tfrac{1}{2})\theta_k]
A^{r-1}[\theta_k]}{|A^r|}
\stackrel{\text{(\ref{CAPapp:SO(D)DifferenceProofRelationsToProveBosonic1})}}{=}
\sum_{j=0}^s
\frac{\sum\limits_{k=1}^r(-1)^{k+1}\cos[(j+r-1)\theta_k]B^{r-1}[\theta_k]}{|B^r|},
\ee
where we used formulas (\ref{CAPeq:SO2n+1CharacterProofFinish}) and (\ref{CAPeq:SO2nCharacterProofFinish}) 
for the 
characters, as well as the definition (\ref{CAPA.2a}) of $A^r$ and $B^r$. One can then use identities 
(\ref{CAPapp:SO(D)DifferenceProofRelationsToProveBosonic1}) and 
(\ref{CAPapp:SO(D)DifferenceProofRelationsToProveBosonic3}) to match the right-hand side of 
this expression with the left-hand side, proving the desired identity.\hfill$\blacksquare$

\begin{advanced}
\section{\ From mixed traces to fermionic characters}
\label{CAPITOL}
\end{advanced}

This appendix is the fermionic (half-integer spin) analogue of section \ref{LABEKKO}. It may be skipped in a 
first reading.\\

\subsection{\ Mixed traces and symmetric polynomials}
\label{CAPAppB1}

Our goal here is to prove the first equality of (\ref{CAPs9.5}), following the same method as in appendix 
\ref{CAPAppA1} for the bosonic case. First, using the definition (\ref{CAP2.16}) of $U$ and the contraction 
rules 
(\ref{CAPeq:TContractionRules}), one can write (\ref{CAPs10}) as
\begin{align}
\label{CAPapp:FermionicCHSPStartingEqSimple1}
\chi_s^{(F)}[n\vec\theta]
=&
\Bigg[
\frac{1}{s!}T^{[s]}
+
\sum_{m=1}^{\lfloor\frac{s}{2}\rfloor}
\frac
{(-1)^m[D+2(s-m-1)]!!}
{2^{m-1}m!(s-2m)![D+2(s-1)]!!}\times\nn\\
&
\times
(\delta^{\mu\mu})^m(\delta^{\mu\beta})^{s-2m}(\delta^{\beta\beta})^{m-1}\delta_{\mu\mu}
T_{\mu_{s-2},\beta_{s-2}}
\Bigg]\!
\textnormal{Tr}[U^n]\nn\\
+&\sum_{m=0}^{\lfloor\frac{s-1}{2}\rfloor}
\frac{(-1)^{m+1}[D+2(s-m-2)]!!}{2^mm!(s-2m-1)![D+2(s-1)]!!}\times\nn\\
&
\times
\textnormal{Tr}
[T_{\mu_{s-1},\beta_{s-1}}
\gamma_\mu\gamma^\mu(\delta^{\mu\mu})^m(\delta^{\mu\beta})^{s-2m-1}(\delta^{\beta\beta})^m U^n],
\end{align}
where $T^{[s]}$ is the notation (\ref{CAPeq:TDefinition}). In the first term of this expression, we shift the 
symmetrization on the $\delta$'s i.e. we exchange upper and lower indices while taking into account the 
change 
in multiplicities of the terms involved; in all other terms, we compute one contraction with 
$\delta^{\beta\beta}$. Eq.~(\ref{CAPapp:FermionicCHSPStartingEqSimple1}) then 
simplifies to
\begin{align}
\label{CAPapp:FermionicCHSPStartingEqSimple2}
&\chi_s^{(F)}[n\vec\theta]
=
\frac{1}{s!}T^{[s]}\textnormal{Tr}[U^n]
-
\frac{1}{[(s-1)!]^2[D+2(s-1)]}\textnormal{Tr}
[T^{\mu_{s-1},\beta_{s-1}}
\gamma_\mu\gamma^\mu\delta_{\mu\beta}^{s-1}U^n]\\
&+
\sum_{m=1}^{\lfloor\frac{s}{2}\rfloor}
\left[
\frac{(-1)^m2^{m-1}(m-1)![D+2(s-m-1)]!!}{[(s-2)!]^2[D+2(s-1)]!!}
T^{\mu_{s-2},\beta_{s-2}}\delta^{\mu\mu}\delta_{\mu\mu}^m\delta_{\mu\beta}^{s-2m}\delta_{\beta\beta}^{m-1}
\textnormal{Tr}[U^n]
\right.\nn\\
&+
\left.
\frac{(-1)^{m+1}2{m-1}(m-1)![D+2(s-m-2)]!!}{[(s-2)!]^2[D+2(s-1)]!!}
\textnormal{Tr}
[T^{\mu_{s-2},\beta_{s-2}}
\delta^{\mu\mu}\delta_{\mu\mu}^m\delta_{\mu\beta}^{s-2m-1}
\delta_{\beta\beta}^m\gamma_\mu\gamma_\beta U^n]
\right].\nn
\end{align}
The $\gamma$ traces and mixed traces can now be evaluated using
\begin{subequations}
\begin{align}
\gamma^\mu\gamma_\mu\delta_{\mu\beta}^{s-1}
=&
\,[D+2(s-1)]\delta_{\mu\beta}^{s-1}-\gamma_\mu\gamma_\beta\delta_{\mu\beta}^{s-2},\\[5pt]
\delta^{\mu\mu}\delta_{\mu\mu}^m\delta_{\mu\beta}^{s-2m}\delta_{\beta\beta}^{m-1}
=&
\,[D+2(s-m-1)]\delta_{\mu\mu}^{m-1}\delta_{\mu\beta}^{s-2m}
\delta_{\beta\beta}^{m-1}\nn\\
&+2m\,\delta_{\mu\mu}^m\delta_{\mu\beta}^{s-2m-2}\delta_{\beta\beta}^m,\\[5pt]
\delta^{\mu\mu}\delta_{\mu\mu}^m\delta_{\mu\beta}^{s-2m-1}\delta_{\beta\beta}^{m-1}\gamma_\mu\gamma_\beta
=&
\,[D+2(s-m-1)]\delta_{\mu\mu}^{m-1}\delta_{\mu\beta}^{s-2m-1}\delta_{\beta\beta}^{m-1}
\gamma_\mu\gamma_\beta\nn\\
&
+4m\,\delta_{\mu\mu}^m\delta_{\mu\beta}^{s-2m-2}\delta_{\beta\beta}^{m}+2m\,\delta_{\mu\mu}^m\delta_{\mu\beta}
^{s-2m-3}\delta_{\beta\beta}^{m}\gamma_\mu\gamma_\beta,
\end{align}
\end{subequations}
which yields
\begin{align}
\chi_s^{(F)}[n\vec\theta]
=&
\left[
\frac{1}{s!}T^{[s]}-\frac{1}{(s-1)!}T^{[s-1]}
\right]\textnormal{Tr}[U^n]\nn\\
&
+
\frac{1}{[(s-1)!]^2
[D+2(s-1)]}
\textnormal{Tr}[T^{\mu_{s-1},\beta_{s-1}}\delta_{\mu\beta}^{s-2}\gamma_\mu\gamma_\beta U^n]\nn\\
&
-\frac{D+2(s-2)}{[(s-2)!]^2[D+2(s-1)]}T^{\mu_{s-2},\beta_{s-2}}\delta_{\mu\beta}^{s-2}\textnormal{Tr}[U^n]
\nn\\
&
+\frac{1}{[(s-2)!]^2[D+2(s-1)]}\textnormal{Tr}[T^{\mu_{s-2},\beta_{s-2}}\delta_{\mu\beta}^{s-3}
\gamma_\mu\gamma_\beta U^n].
\end{align}
Using (\ref{CAPss47}) and the definition (\ref{CAP2.16}) of $U$, together with some careful counting, one 
verifies 
that this expression matches $\left[h_s(J^n)-h_{s-1}(J^n)\right]\textnormal{Tr}[U^n]$, which was to be proven.

\subsection{\ Symmetric polynomials and $\text{SO}(D)$ characters}\label{CAPAppB2}

In this part of the appendix we prove the second equality in (\ref{CAPs9.5}), following essentially the same 
steps as in appendix \ref{CAPAppA2}. We refer again to \cite{fulton1991} for details, and we write the 
components of weights in the dual basis of the Cartan subalgebra described above 
(\ref{CAPeq:SO2n+1CharacterFormula}) and  
(\ref{JESUS}). We will consider separately odd and even space-time dimensions.

\subsubsection*{Odd $D$}

The character of a half-spin representation of $\mathfrak{so}(2r+1)$ with a dominant highest weight
$\lambda=(\lambda_1+\tfrac{1}{2},\lambda_2+\tfrac{1}{2},...,\lambda_r+\tfrac{1}{2})$ is 
\cite[p.258f]{Littlewood:1950}\i{character!for SOn@for $\text{SO}(n)$}\i{son@$\mathfrak{so}(n)$!character}
\begin{equation}
\chi^{(2r+1)}_{\lambda}\left[\theta_1,...,\theta_r\right]
=
\frac
{\left|\sin\left[\left(\lambda_i+r-i+1\right)\theta_j\right]\right|}
{\left|\sin\left[\left(r-i+\frac{1}{2}\right)\theta_j\right]\right|}
=
\left(\prod_{i=1}^r2\cos\left(\tfrac{\theta_i}{2}\right)\right)
\frac{\left|\sin\left[\left(\lambda_i+r-i+1\right)\theta_j\right]\right|}
{\left|\sin\left[\left(r-i+1\right)\theta_j\right]\right|}.
\end{equation}
Owing to expression (\ref{CAPss9.5}) for the trace of $U^n$, the second equality in (\ref{CAPs9.5}) is 
equivalent to
\begin{equation}
\label{CAPeq:SO2n+1FermionicCharacterProofStart}
h_s(J)-h_{s-1}(J)
=
\frac
{\left|\sin\left[\left(\lambda_i+r-i+1\right)\theta_j\right]\right|}
{\left|\sin\left[\left(r-i+1\right)\theta_j\right]\right|}
\end{equation}
for $\lambda_i=s\delta_{i1}$. To prove this, consider the difference of the bosonic character 
(\ref{CAPeq:SO2n+1CharacterProofFinish}) and the right-hand side of 
(\ref{CAPeq:SO2n+1FermionicCharacterProofStart}):
\be
\label{CAPss50b}
\frac{\left|
\sin[(\lambda_i+r-i+\tfrac{1}{2})\theta_j]
\right|}{\left|
\sin[(r-i+\tfrac{1}{2})\theta_j]
\right|}
-
\frac{\left|
\sin[(\lambda_i+r-i+1)\theta_j]
\right|}{\left|
\sin[(r-i+1)\theta_j]
\right|}\,.
\ee
Introducing the notation
\be
\label{CAPA.2b}
(\cA^r)_{ij}=2\sin\left[(r-i+1)\theta_j\right],
\quad 
(\cB^r)_{ij}=2\cos\left[(r-i+\tfrac{1}{2})\theta_j\right]
\ee
and in terms of (\ref{CAPA.2a}), this difference can be written as
\be
\label{CAPs50b}
\frac
{\sum_{k=1}^r(-1)^{k+1}\sin[(s+r-\tfrac{1}{2})\theta_k]A^{r-1}[\theta_k]}
{|A^r|}
-
\frac
{\sum_{k=1}^r(-1)^{k+1}2\sin[(s+r)\theta_k]\cA^{r-1}[\theta_k]}
{|\cA^r|}
\ee
upon expanding the determinants along the first row. Now it turns out that\footnote{See 
e.g.~\cite[p.259]{Littlewood:1950}.}
\be
2^r\left|A^r\right|
\prod_{i=1}^r2\cos\left(\theta_i/2\right)
=
|\cA^r|,
\quad
2^{r-1}
\left|B^r\right|
\prod_{i=1}^r2\cos\left(\theta_i/2\right)
=
|\cB^r|,
\ee
and plugging this property in (\ref{CAPs50b}) one sees that (\ref{CAPss50b}) is just $h_{s-1}(J)-h_{s-2}(J)$. 
Since the 
first term of (\ref{CAPss50b}) equals $h_s(J)-h_{s-2}(J)$ by 
virtue of (\ref{CAPeq:SO2n+1CharacterProofFinish}), this proves 
(\ref{CAPeq:SO2n+1FermionicCharacterProofStart}).

\subsubsection*{Even $D$}

The character of an irreducible representation of $\mathfrak{so}(2r+2)$ with (dominant) highest-weight 
$\lambda=(\lambda_1+1/2,...,\lambda_{r+1}+1/2)$ can be written as 
\cite[p.258-259]{Littlewood:1950}\i{character!for SOn@for $\text{SO}(n)$}\i{son@$\mathfrak{so}(n)$!character}
\be
\chi^{(2r+2)}_{\lambda}
\left[\theta_1,...,\theta_r\right]
=
\frac{\left|\cos\left[\left(\lambda_i+r-i+\tfrac{3}{2}\right)\theta_j\right]\right|}
{\left|\cos\left[\left(r-i+1\right)\theta_j\right]\right|}
=
\prod_{i=1}^{r+1}2\cos\left(\tfrac{\theta_i}{2}\right)
\frac{\left|\cos\left[\left(\lambda_i+r-i+\tfrac{3}{2}\right)\theta_j\right]\right|}
{\left|\cos\left[\left(r-i+\tfrac{3}{2}\right)\theta_j\right]\right|},
\ee
where we are including the possibility of a non-zero angle $\theta_{r+1}$ (while in (\ref{CAPs9.5}) we take 
$\theta_{r+1}=0$). Taking into account (\ref{CAPss9.5}), proving the second equality in (\ref{CAPs9.5}) 
amounts to 
showing that
\be
\label{CAPeq:SO2nFermionicCharacterProofStart}
h_s(J)-h_{s-1}(J)
=
\frac{\left|\cos\left[\left(\lambda_i+r-i+\tfrac{3}{2}\right)\theta_j\right]\right|}
{\left|\cos\left[\left(r-i+\tfrac{3}{2}\right)\theta_j\right]\right|}
\Bigg|_{\theta_{r+1}=0}
\ee
for $\lambda_i=s\delta_{i1}$. To prove this we proceed as in the odd-dimensional case: the difference of the 
bosonic character (\ref{CAPeq:SO2nCharacterProofFinish}) and the right-hand side of 
(\ref{CAPeq:SO2nFermionicCharacterProofStart}),
\be
\left.\frac
{\left|
\cos[(\lambda_i+r-i+1)\theta_j]
\right|}
{\left|
\cos[(r-i+1)\theta_j]
\right|}
\right|_{\theta_{r+1}=0}
-
\left.\frac
{\left|
\cos[(\lambda_i+r-i+\tfrac{3}{2})\theta_j]
\right|}
{\left|
\cos[(r-i+\tfrac{3}{2})\theta_j]
\right|}
\right|_{\theta_{r+1}=0},
\ee
can be written as
\begin{equation}
\left[
\frac{\sum\limits_{k=1}^{r+1}(-1)^{k+1}\cos[(s+r)\theta_k]B^r[\theta_k]}
{|B^{r+1}|}
-
\frac{\sum\limits_{k=1}^{r+1}(-1)^{k+1}2\cos[(s+r+\tfrac{1}{2})\theta_k]\cB^r[\theta_k]}
{|\cB^{r+1}|}
\right]_{\theta_{r+1}=0}
\end{equation}
upon expanding the determinants along the first row and using the notation (\ref{CAPA.2a})-(\ref{CAPA.2b}). 
One can 
then verify that this reduces to $h_{s-1}(J)-h_{s-2}(J)$ by the same argument as in the odd-dimensional case. 
By 
virtue of the second equality in (\ref{CAPeq:SO2nCharacterProofFinish}), this proves 
(\ref{CAPeq:SO2nFermionicCharacterProofStart}).

\end{subappendices}

\newpage
~
\thispagestyle{empty}

\chapter{Conclusion}
\label{c13}
\markboth{}{\small{\chaptername~\thechapter. Conclusion}}

We have now completed our survey of the group-theoretic aspects of three-dimensional gravity, and in 
particular of BMS symmetry in three dimensions. In this conclusion we take one last look at what we have 
achieved.

\subsection*{Quantum symmetries}

The overarching theme of this thesis has been group theory and its application to quantum systems with 
symmetries. Accordingly, parts I and II of this thesis were devoted to a broad overview of group 
representations and geometry. In particular we have motivated and introduced central extensions, defined 
induced representations, applied them to semi-direct products and relativistic symmetry groups, and explained 
how classical mechanical systems with symmetries become symmetric quantum systems upon ``replacing Poisson 
brackets by commutators''. We have also applied some of these tools to the Virasoro group --- the symmetry 
group of two-dimensional conformal field theories --- and used it to analyse certain properties of 
three-dimensional gravity on Anti-de Sitter backgrounds.

\subsection*{BMS$_3$ particles}

Part III of the thesis was devoted to the application of group theoretic methods to the study of 
asymptotically flat quantum gravity in three dimensions. In that context our weapon of choice has been the 
BMS group in three dimensions, which we have introduced as an asymptotic symmetry group in chapter \ref{c6}, 
before working out its abstract definition independently of gravity. We have seen 
in particular that it enjoys an exceptional structure of the type $G\ltimes\mg$, where $G$ is the Virasoro 
group spanned by superrotations while $\mg$ is its Lie algebra, spanned by supertranslations. A crucial 
implication of this structure was that irreducible unitary representations of the BMS$_3$ group, i.e.\ 
BMS$_3$ particles, have supermomenta that span coadjoint orbits of the Virasoro group. This observation has 
allowed us to classify all such representations thanks to the classification of Virasoro orbits exposed 
earlier, in chapter \ref{c4bis}.\\

We also observed that supermomentum orbits have a straightforward interpretation in gravity, since 
supermomenta coincide with Bondi mass aspects of asymptotically flat space-time metrics. As a result we 
interpreted BMS$_3$ particles in two equivalent ways: (i) as quantizations of orbits of 
asymptotically flat metrics under BMS$_3$ transformations, and (ii) as relativistic particles dressed with 
gravitational boundary degrees of freedom. These topological degrees of freedom are the three-dimensional 
analogue of soft gravitons, so a BMS$_3$ particle is effectively a particle dressed with soft gravitons.\\

As a confirmation of this interpretation, we evaluated BMS$_3$ characters and showed that they coincide 
with gravitational one-loop partition functions. On the group-theoretic side this computation involves the 
Frobenius character formula (\ref{fropo}), which is essentially an integral of little group characters over a 
supermomentum orbit. Remarkably, for non-zero angular potentials, we found that the integral localizes 
to a single point on the supermomentum orbit, which allowed us to evaluate characters exactly. On the 
field-theoretic side the one-loop partition function was evaluated using heat kernel methods which turn 
out, unsurprisingly, to be more tractable in flat space than in Anti-de Sitter space.

\subsection*{Higher spins and supergravity}

The study of thermodynamics has led us into higher-spin theories, whose partition functions in flat space 
could be computed with little extra effort compared to the gravitational case. We have used this 
computation as an excuse to investigate the unitary representations of asymptotic symmetry algebras that 
occur in that context, with methods and results very similar to those of the purely gravitational setting. A 
striking aspect of these considerations was the fact that it enabled us to compare ultrarelativistic and 
non-relativistic limits of conformal higher spins in a way that pure gravity does not allow. In doing so we 
uncovered a sharp difference between the two limits at the quantum level. As a corollary we 
concluded that flat space holography should not be described as a Galilean conformal field theory.\\

We have similarly studied the supersymmetric generalization of BMS$_3$ symmetry, whose unitary 
representations are essentially super BMS$_3$ multiplets consisting of an infinite tower of BMS$_3$ particles 
with 
increasing spins.

\subsection*{Is this quantum gravity?}

In the introduction of the thesis we motivated the study of BMS symmetry by presenting it as a way to tackle 
the quantization of gravity.\i{quantum gravity} This is a good moment to ask to what extent this proposal has 
succeeded. To 
begin, we should realize that what we have done \it{is}, indeed, a partial quantization of gravity: we 
have described almost explicitly a family of Hilbert spaces endowed with operator algebras inherited from 
gravitational symmetries, and we have used these Hilbert spaces to compute concrete quantities such as 
partition functions. In this sense we have actually worked with a partial version of quantum gravity.\\

This being said, one should not be overly enthusiastic about what we have achieved: in essence we have worked 
out the flat space analogue of results that were mostly already known in the framework of AdS$_3$/CFT$_2$. In 
fact, in many cases our results were flat limits of their AdS peers, although the 
BMS$_3$ approach often led us to consider slightly different questions and use somewhat different methods 
than those suggested by conformal symmetries. Thus we have indeed made progress in our understanding of flat 
space holography, but whether this opens new doors towards quantum gravity is a whole other matter.\\

There are two simple arguments that show why our work is not \it{quite} quantum gravity. First, what 
we have studied are \it{irreducible} unitary representations of asymptotic symmetry groups, while realistic 
gravitational systems are expected to form highly reducible representations. In essence, saying that we have 
studied quantum gravity by studying irreducible representations of BMS$_3$ would be tantamount to saying that 
relativistic one-particle quantum mechanics is the same as quantum field theory, which is of 
course untrue. Secondly, one should realize that our study of the symmetries of gravity hasn't taught 
us anything about the microscopic details of gravity itself. For instance, the fact that the phase space of 
gravity forms the coadjoint representation of the asymptotic symmetry group is merely a restatement of the 
fact that momentum maps belong to the coadjoint representation, which is a robust feature of all symmetric 
phase spaces; it does not tell us anything about the details of gravity.

\subsection*{A look forward}

Our observations on BMS particles in three dimensions have allowed us to describe dressed particles in a 
group-theoretic framework. While we haven't described interacting particles in this thesis, it is likely that 
our methods do apply to such cases as well. In particular, describing scattering phenomena in terms of BMS 
particles instead of standard (naked) particles should incorporate soft graviton contributions. In three 
dimensions this could presumably be used to describe, say, the merger of two particles into a flat space 
cosmology; optimistically, BMS$_3$ representations might then even account for gravitational quantum 
corrections to such 
amplitudes! In four dimensions the situation is much less well understood, for reasons that we alluded to 
earlier. In that case the problem is much more basic, since the very definition of BMS symmetry is elusive 
--- let alone its quantum representations.\\

A related project is the description of BMS world lines.\i{world line!for BMS$_3$ particle}\i{BMS world 
line} The reader may recall that we described in chapter 
\ref{c3} a general procedure for building world line actions associated with arbitrary Lie groups, and that 
the application of this method to the Poincar\'e group resulted in relativistic world lines. It is tempting 
to ask what happens when that approach is applied to the BMS$_3$ group; the answer is very natural: 
the resulting action principle describes world lines propagating in the space of supertranslations, which can 
equivalently be seen as relativistic world lines dressed with gravitational degrees of freedom. Yet another 
way to think of these world-lines is to interpret them as two-dimensional field theories dual to 
three-dimensional 
asymptotically flat gravity, and indeed one finds that their partition functions coincide with gravitational 
(one-loop) partition functions. These considerations should appear 
soon in a separate publication \cite{Barnich:2050zq}. Note that, as before, the application of these ideas to 
BMS in \it{four} dimensions is much more problematic due to the lack of a proper definition of BMS$_4$ 
symmetry.\\

There is undoubtedly much more to be 
done in the future, both in toy models such as three-dimensional gravity and in real-world, four-dimensional 
systems. In the wake of the experimental observation of gravitational waves 
\cite{Abbott:2016blz},\i{gravitational wave} it is 
likely that new tools and methods will soon be required to understand and study gravity, both 
classically and quantum-mechanically. On a more philosophical note, it is remarkable that a question 
seemingly as simple as ``what is a particle?'' has an answer as 
intricate and rich as what we have been attempting to describe in this thesis. We hope to have 
contributed to a partial solution to the problem, and look forward to investigating some of its future 
applications.\\

~\\
~\\
~\\
~\\
\begin{center}
\it{All nature is but art, unknown to thee;\\
All chance, direction, which thou canst not see;\\
All discord, harmony not understood.}
\end{center}
\vspace{.2cm}
\hfill Alexander Pope, \it{An Essay on Man}, 1733.

\i{adjoint vector!for BMS$_3$|see{$\bms$ algebra}}
\i{AdS$_3$|seealso{Brown-Henneaux}}
\i{asymptotic flatness|seealso{BMS$_3$ fall-offs}}
\i{Banados metric@Ba\~{n}ados metric|seealso{Brown-Henneaux metric}}
\i{BMS$_3$ momentum|see{supermomentum}}
\i{BMS group!in 3D|see{BMS$_3$ group}}
\i{boost|seealso{standard boost}}
\i{boundary conditions|see{fall-off cond.}}
\i{CFT vacuum|see{Virasoro vacuum}}
\i{character|seealso{Frobenius formula}}
\i{circle|see{$S^1$}}
\i{conformal field theory|seealso{AdS/CFT}}
\i{conformal transformation|seealso{M\"obius transformation}}
\i{coset space|see{quotient space}}
\i{DiffS1@$\Diff$!central extension|see{Virasoro group}}
\i{Dirac distribution|see{delta function}}
\i{extended BMS group!in 3D|see{BMS$_3$}}
\i{Frobenius formula|seealso{character}}
\i{fundamental vector field|see{infinitesimal generator}}
\i{GA@$G\ltimes A$|see{semi-direct product}}
\i{Galilean conformal algebra|see{$\gca$}}
\i{Galilean limit|see{non-relativistic limit}}
\i{Galilei group|seealso{Bargmann group}}
\i{induced representation!of BMS$_3$|see{BMS$_3$ particle}}
\i{induced representation!of Lie algebra|see{induced module}}
\i{inhomogeneous Lorentz group|see{Poincar\'e group}}
\i{Lie algebra!of BMS$_3$|see{$\bms$}}
\i{light-like infinity|see{null infinity}}
\i{little group|seealso{stabilizer}}
\i{massive BMS$_3$ particle|seealso{elliptic Virasoro orbit}}
\i{massless BMS$_3$ particle|seealso{parabolic Virasoro orbit}}
\i{metric|see{AdS$_3$, Minkowski space, on-shell metric}}
\i{momentum!for BMS$_3$|see{supermomentum}}
\i{momentum!eigenstate|see{plane wave}}
\i{one-loop partition function|see{partition function}}
\i{one-particle state|see{particle}}
\i{PGL2R@$\text{PGL}(2,\RR)$|see{$\PSL$}}
\i{relativistic particle|see{particle}}
\i{representation!of BMS$_3$|see{BMS$_3$ particle}}
\i{representation!of Poincar\'e|see{particle}}
\i{representation!of sl2R@of $\sl$|see{highest-weight representation}}
\i{representation!of Virasoro|see{highest-weight representation}}
\i{retarded Bondi coordinates|see{Bondi coordinates}}
\i{stabilizer|seealso{little group}}
\i{superboost|seealso{superrotation}}
\i{symmetry!asymptotic|see{asymptotic symmetry}}
\i{tachyonic BMS$_3$ particle|seealso{hyperbolic Virasoro orbit}}
\i{Thomas precession|seealso{Wigner rotation}}
\i{three-dimensional gravity|see{3D gravity}}
\i{transitive action|see{homogeneous space}}
\i{two-cocycle|see{central extension}}
\i{ultrarelativistic limit|seealso{flat limit}}
\i{unitary representation|seealso{induced representation}}
\i{unitary representation!of BMS$_3$|see{BMS$_3$ particle}}
\i{unitary representation!of Poincar\'e|see{particle}}
\i{vector field!generating asymptotic symmetries|see{asymptotic Killing vector}}
\i{Virasoro cocycle|see{Gelfand-Fuks}}

\backmatter
\pagestyle{Bibliography}

\markboth{}{\small{\chaptername~\thechapter. Bibliography}}

\clearpage
\pagestyle{Index}
\printindex

\end{document}